\documentclass[10pt]{elsarticle}

\usepackage{titlesec}
\titleformat{\subsection}[runin]{\normalfont\bfseries}{\thesubsection.}{.5em}{}[.]\titlespacing{\subsection}{0pt}{2ex plus .1ex minus .2ex}{.8em}
\titleformat{\subsubsection}[runin]{\normalfont\itshape}{\thesubsubsection.}{.3em}{}[.]\titlespacing{\subsubsection}{0pt}{1ex plus .1ex minus .2ex}{.5em}
\titleformat{\paragraph}[runin]{\normalfont\bfseries}{\theparagraph.}{.3em}{}[.]\titlespacing{\paragraph}{0pt}{1ex plus .1ex minus .2ex}{.5em}

\usepackage[labelfont=sc,font=small,labelsep=period]{caption}
\usepackage{amsmath} %[intlimits]
\usepackage{amssymb}
\usepackage{amsfonts}
\usepackage{latexsym}
\usepackage{amsthm}
\usepackage{amsxtra}
\usepackage{amscd}
\usepackage{bbm}
\usepackage{mathrsfs}
\usepackage{bm}
\usepackage{longtable}
\usepackage{mathtools} %contains useful extra commands, such as \bigtimes
\usepackage{subcaption}
\usepackage{algorithm}	%write algo
\usepackage{algpseudocode} 	%write algo
\usepackage{multirow}
\usepackage{geometry}

\usepackage{graphicx, color}

\definecolor{darkred}{rgb}{0.9,0,0.3}
\definecolor{darkblue}{rgb}{0,0.3,0.9}

\newcommand{\nc}{\normalcolor}

\usepackage[pdftex, colorlinks, linkcolor=blue,citecolor=red]{hyperref}

\usepackage{natbib}
\numberwithin{equation}{section}
\numberwithin{figure}{section}
\theoremstyle{plain} %plain, definition, remark

\newtheorem*{theorem*}{Theorem}

\newtheorem*{lemma*}{Lemma}

\newtheorem*{corollary*}{Corollary}

\newtheorem*{proposition*}{Proposition}

\newtheorem*{definition*}{Definition}

\newtheorem*{conjecture*}{Conjecture}

\theoremstyle{definition} %plain, definition, remark

\newtheorem*{example*}{Example}

\newtheorem*{remark*}{Remark}

\renewcommand{\b}[1]{\boldsymbol{\mathrm{#1}}} %bold
\renewcommand{\cal}{\mathcal} 
 
\newcommand{\fra}{\mathfrak} 
\newcommand{\ul}[1]{\underline{#1} \!\,} %underline
\newcommand{\wh}{\widehat}
\newcommand{\wt}{\widetilde}

\newcommand{\R}{\mathbb{R}}
\newcommand{\N}{\mathbb{N}}

\newcommand{\ii}{\mathrm{i}}
\newcommand{\dd}{\mathrm{d}}
\newcommand*{\deq}{\mathrel{\vcenter{\baselineskip0.65ex \lineskiplimit0pt \hbox{.}\hbox{.}}}=}

\renewcommand{\leq}{\leqslant}
\renewcommand{\geq}{\geqslant}
\renewcommand{\epsilon}{\varepsilon}

\newcommand{\qq}[1]{[\![{#1}]\!]}

\newcommand{\pb}[1]{\bigl({#1}\bigr)}
\newcommand{\pB}[1]{\Bigl({#1}\Bigr)}

\newcommand{\pBB}[1]{\Biggl({#1}\Biggr)}

\newcommand{\qb}[1]{\bigl[{#1}\bigr]}
\newcommand{\qB}[1]{\Bigl[{#1}\Bigr]}

\newcommand{\qBB}[1]{\Biggl[{#1}\Biggr]}

\newcommand{\hB}[1]{\Bigl\{{#1}\Bigr\}}

\newcommand{\abs}[1]{\lvert #1 \rvert}
\newcommand{\absb}[1]{\bigl\lvert #1 \bigr\rvert}
\newcommand{\absB}[1]{\Bigl\lvert #1 \Bigr\rvert}

\newcommand{\norm}[1]{\lVert #1 \rVert}
\newcommand{\normb}[1]{\bigl\lVert #1 \bigr\rVert}

\newcommand{\avg}[1]{\langle #1 \rangle}
\newcommand{\avgb}[1]{\bigl\langle #1 \bigr\rangle}

\newcommand{\avgBB}[1]{\Biggl\langle #1 \Biggr\rangle}

\newcommand{\scalar}[2]{\langle{#1} \mspace{2mu}, {#2}\rangle}

\newcommand{\scalarBB}[2]{\Biggl\langle{#1} \,\mspace{2mu},\, {#2}\Biggr\rangle}

\DeclareMathOperator{\diag}{diag}
\DeclareMathOperator{\tr}{Tr}

\DeclareMathOperator{\supp}{supp}
\DeclareMathOperator{\re}{Re}
\DeclareMathOperator{\im}{Im}

\renewcommand{\Re}{\text{Re}}

\newcommand{\argmin}{\operatornamewithlimits{argmin}}
\let\e=\varepsilon 

  \def\AAA{{\bf A}}
\def\BBB{{\bf B}}

\def\ie{{i.e.}}\def\eg{{e.g.}}

\newcommand{\G}{\b G}
\newcommand{\stj}{\fra g}
\newcommand{\hil}{\fra h}
\newcommand{\btr}{\cal B}
\newcommand{\rtr}{\cal R}
\newcommand{\wtr}{\cal W}
\newcommand{\ttr}{\cal T}
\newcommand{\str}{\cal S}

\newcommand{\mso}{\Phi}
\newcommand{\Tr}{\text{Tr}}
\newcommand \C {\textbf{C}}
\newcommand \E {\textbf{E}}
\newcommand \X {\textbf{X}}
\newcommand \Y {\textbf{Y}}
\newcommand \M {\textbf{M}}
\renewcommand \S {\textbf{S}}
\newcommand \B {\textbf{B}}
\newcommand \W {\textbf{W}}
\newcommand \Wishart {\b {\cal W}}

\newcommand \Supp {\text{Supp}}

\newcommand \In {\b I_{N}}
\newcommand \It {\b I_{T}}
\def\Xint#1{\mathchoice
{\XXint\displaystyle\textstyle{#1}}%
{\XXint\textstyle\scriptstyle{#1}}%
{\XXint\scriptstyle\scriptscriptstyle{#1}}%
{\XXint\scriptscriptstyle\scriptscriptstyle{#1}}%
\!\int}
\def\XXint#1#2#3{{\setbox0=\hbox{$#1{#2#3}{\int}$}
\vcenter{\hbox{$#2#3$}}\kern-.5\wd0}}
\def\dashint{\Xint-}

\makeatletter
\newcommand{\thickline}{%
	\noalign {\ifnum 0=`}\fi \hrule height 1pt
	\futurelet \reserved@a \@xhline
}
\makeatother

\journal{Physics Reports}

\begin{document}

\begin{frontmatter}

\title{Cleaning large Correlation Matrices: tools from Random Matrix Theory}

\author{Jo{\"e}l Bun}
\address{Capital Fund Management, 23--25, rue de l'Universit\'e, 75\,007 Paris}
\address{LPTMS, CNRS, Univ. Paris-Sud, Universit{\'e} Paris-Saclay, 91405 Orsay, France}
%\address{Leonard de Vinci P{\^o}le Universitaire, Finance Lab, 92916 Paris La D{\'e}fense, France}
\ead{joel.bun@gmail.com}

\author{Jean-Philippe Bouchaud}
\ead{jean-philippe.bouchaud@cfm.fr}
%\address{Capital Fund Management, 23--25, rue de l'Universit\'e, 75\,007 Paris}
\author{Marc Potters}
\ead{marc.potters@cfm.fr}
\address{Capital Fund Management, 23--25, rue de l'Universit\'e, 75\,007 Paris}



\begin{abstract}
This review covers recent results concerning the estimation of large covariance matrices using tools from Random Matrix Theory (RMT). We introduce several RMT methods and analytical techniques, such as the Replica formalism and Free Probability, with an emphasis on the Mar{\v c}enko-Pastur equation that provides information on the resolvent of multiplicatively corrupted noisy matrices. Special care is devoted to the statistics of the eigenvectors of the empirical correlation matrix, which turn out to be crucial for many applications. We show in particular how these results can be used to build consistent ``Rotationally Invariant'' estimators (RIE) for large correlation matrices when there is no prior on the structure of the underlying process. The last part of this review is dedicated to some real-world applications within financial markets as a case in point. We establish empirically the efficacy of the RIE framework, which is found to be superior in this case to all previously proposed methods. The case of additively (rather than multiplicatively) corrupted noisy matrices is also dealt with in a special Appendix. Several open problems and interesting technical developments are discussed throughout the paper.
\end{abstract}


\end{frontmatter}

%\linenumbers

% \title{Cleaning large Correlation Matrices: tools from Random Matrix Theory}

% \author{Joel Bun, Jean-Philippe Bouchaud, Marc Potters\footnote{CFM, 23 rue de l'Universit\'e, 75007 Paris, France}}
% %\address{CFM, 23 rue de l'Universit\'e, 75007 Paris, France}

% \sloppy
% \maketitle
% \begin{abstract} 
% This review covers recent results concerning the estimation of large covariance matrices using tools from Random Matrix Theory (RMT). We introduce several RMT methods and analytical techniques, such as the Replica formalism and Free Probability, with an emphasis on the Mar{\v c}enko-Pastur equation that provides information on the resolvent of multiplicatively corrupted noisy matrices. Special care is devoted to the statistics of the eigenvectors of the empirical correlation matrix, which turn out to be crucial for many applications. We show in particular how these results can be used to build consistent ``Rotationally Invariant'' estimators (RIE) for large correlation matrices when there is no prior on the structure of the underlying process. The last part of this review is dedicated to some real-world applications within financial markets as a case in point. We establish empirically the efficacy of the RIE framework, which is found to be superior in this case to all previously proposed methods. The case of additively (rather than multiplicatively) corrupted noisy matrices is also dealt with in a special Appendix. Several open problems and interesting technical developments are discussed throughout the paper.

% \end{abstract} 

\tableofcontents

%\newpage

%-----------------------------------------------------------------------------------------------------
%*****************************************************************************************************
%-----------------------------------------------------------------------------------------------------

\clearpage%!TEX root = RMT_Covariance_Review.tex
\section{Introduction}

\subsection{Motivations}

In the present era of ``Big Data'', new statistical methods are needed to
decipher
large dimensional data sets that are now routinely generated in almost all
fields -- physics, image analysis, genomics, epidemiology,
engineering, economics and finance, to quote only a few. It is
very natural to try to identify common causes
(or factors) that explain the joint dynamics of $N$ quantities. These
quantities might be daily returns of the different
stocks of the S\&P 500, temperature variations in different locations
around the planet, velocities of individual grains in a packed granular
medium,
or different biological indicators (blood pressure, cholesterol, etc.)
within a population, etc., etc. The simplest mathematical object that
quantifies the similarities between these observables is an $N \times N$
correlation matrix $\bf C$. Its eigenvalues and eigenvectors can then be
used to characterize the most important common dynamical ``modes'', i.e. linear 
combinations of the original variables with the largest variance. This is
the well known ``Principal Component Analysis'' (or PCA) method. More formally, 
let us denote by $\b y \in \R^{N}$ the set of demeaned and 
standardized\footnote{This apparently innocuous assumption will be discussed in Chapter \ref{chap:spectrum}.} variables which are thought to display some degree of interdependence. 
Then, one possible way to quantify the underlying interaction network between these variables is through the standard, Pearson correlations:
\begin{equation}
\label{eq:population_covariance_matrix}
\C_{ij} = \mathbb{E}\qb{y_{i} y_{j}}, \quad i,j \in [\![1,N]\!],
\end{equation}
%where the bracket notation will be used throughout this review to denote averages over the underlying (perhaps unknown) statistical process. 
We will refer to the matrix $\C$ as the \emph{population} correlation matrix throughout the following. 

The major concern in practice is that the expectation value in \eqref{eq:population_covariance_matrix} is rarely computable precisely because the underlying distribution 
of the vector $\b y$ is unknown and is what one is struggling to determine. 
Empirically, one tries to infer the matrix $\C$ by collecting a large number $T$ of realizations of these $N$ variables 
that defines the input sample data matrix $\b Y = (\b y_1, \b y_2, \dots, \b y_T) \in \R^{N\times T}$. 
Then, in the case of a sufficiently large number of realizations $T$, one tempting solution to estimate $\C$ is 
to compute that \emph{sample correlation matrix} estimator $\bf E$, defined as:
\begin{equation}
E_{ij} \;\deq\; \frac{1}{T} \sum_{t=1}^T \, Y_{it} \, Y_{jt} \;\equiv\; \frac1T \left
({\bf Y} {\bf Y}^* \right)_{ij},
\end{equation}
where $Y_{it}$ is the realization of the $i$th observable ($i=1, \dots, N$)
at ``time'' $t$
($t=1,\dots,T$) that will be assumed in the following to be demeaned and
standardized (see previous footnote). 

Indeed, in the case where $N \ll T$, it is well known using result of classical multivariate statistics that $\E$ converges (almost surely) to $\C$ \cite{van2000asymptotic}. 
However, when $N$ is large, the simultaneous estimation of all $N(N-1)/2$ the elements of $\bf C$ -- or in fact only of its $N$ eigenvalues -- 
becomes problematic when the total number $T$ of observations is not very large compared to $N$ itself.  In the example
of stock returns, $T$ is the total number of trading days in the sampled data; but in the biological example, $T$ would be the size of the population
sample, etc. Hence, in the modern framework of high-dimensional statistics, the empirical correlation matrix ${\bf E}$
(i.e. computed on a given realization) must be carefully distinguished from the ``true'' correlation matrix ${\bf C}$ of the underlying statistical
process (that might not even be well defined). In fact, the whole point of the present review is to characterize the difference between
${\bf E}$ and ${\bf C}$, and discuss how well (or how badly) one may reconstruct ${\bf C}$ from the knowledge of $\bf E$ in the case where $N$ and $T$ 
become very large but with their ratio $q=N/T$ not vanishingly small; this is often called the large dimension limit (LDL), or else the ``Kolmogorov regime''.
% Of course, if $N$ is small (say $N=4$) and the number of observations is
% huge (say $T=10^6$),
% then one can intuitively expect that any observable computed using ${\bf
% E}$ will be very close to its ``true'' value,
% computed using ${\bf C}$. 

There are numerous situations where the estimation of the high-dimensional covariance matrix is crucial. Let us give some well-known examples:
\begin{enumerate}

	\item Generalized least squares (GLS): Suppose we try to explain the vector $\b y$ using a linear model
		\begin{equation}
			\b y = X \b \beta + \b \e,
		\end{equation}
		where $X$ is a $N \times k$ design matrix ($k \geq 1$), $\b \beta$ denotes the regression coefficients to these $k$ factors, and $\b \e$ denotes the residual. Typically, one seeks to find $\b \beta$ that best explains the data and this exactly the purpose of GLS. Assume that $\E[\b \e | X] = 0$ and $\mathbb{V}[\b \e | X] = \C$ the covariance matrix of the residuals. Then GLS estimates $\b \beta$ as (see \cite{amemiya1985advanced} for a more detailed discussion): 
		\begin{equation}
			\wh {\b \beta} = \left(X^* \C X\right)^{-1} X^* \C^{-1} \b y.
		\end{equation}
		We shall investigate this estimator in Section \ref{chap:application}.
		
        \item Generalized methods of moments (GMM): Suppose one wants to calibrate the parameters $\Theta$ of a model on some data set. The idea is to compute the empirical average of a set of $k$ functions (generalized moments) of the data, which should all be zero for the correct values of the parameters, $\Theta = \Theta_0$. The distance to zero is measured using the covariance of these functions. A precise measurement of this $k \times k$ covariance matrix increases the efficiency of the GMM -- see \cite{hansen1982large}. Note that GLS is a special form of GMM. 
        
	\item Classification \cite{friedman2001elements}: Suppose that we want to classify the variables $\b y$ between two Gaussian populations with different mean $\b \mu_1$ and $\b \mu_2$, priors $\pi_1$ and $\pi_2$, but same covariance matrix $\C$. The \emph{Linear Discriminant Analysis} rule classifies $\b y$ to class 2 if
			\begin{equation}
				\b x^* \C^{-1}(\b \mu_1 - \b \mu_2) > \frac12 (\b \mu_2 + \b \mu_1)^{*}\C^{-1}(\b \mu_2 - \b \mu_1) - \log(\pi_2/\pi_1)
	\end{equation}
		
	\item Large portfolio optimization \cite{markowitz1952portfolio}: Suppose we want to invest on a set of financial assets $\b y$ in such a way that the overall risk of the portfolio is minimized, for a given performance target $\nu$. According to Markowitz's theory, the optimal investment strategy is a vector of weights $\b w \deq (w_1, \dots,w_p)^*$ that can be obtained through a quadratic optimization program where we minimize the variance of the strategy $\scalar{\b w}{\C \b w}$ subject to a constraint on the expectation value $\scalar{\b w}{\b g} \geq \mu$, with $\b g$ a vector of predictors and $\mu$ fixed. (Other constraints can also be implemented). The optimal strategy reads
	\begin{equation}
		\b w = \nu \, \frac{\C^{-1} \b g}{\b g^{*} \C^{-1} \b g}.
	\end{equation}
\end{enumerate}
As we shall see in Chapter \ref{chap:application}, a common measure of the ``risk'' of estimation in high-dimensional problems like 
(i) and (iv) above is given by  $\Tr {\bf E}^{-1}/\Tr {\bf C}^{-1}$, which turns out to be very close to
unity $T$ is large enough for a fixed $N$, i.e. when $q=N/T \to 0$. However, when the number of observables $N$ is also large, such that
the ratio $q$ is not very small, we will find below that $\Tr {\bf E}^{-1} =  \Tr {\bf C}^{-1}/(1-q)$ for a wide   class
of processes. In other words, the out-of-sample risk $\Tr {\bf E}^{-1}$ can excess by far the true optimal risk $\Tr {\bf C}^{-1}$ when $q > 0$, and even diverge when $q \to 1$. Note that for a similar scenario when Value-at-Risk is minimized in-sample was elicited in \cite{caccioli2015portfolio} and in \cite{ciliberti2007feasibility} for the Expected Shortfall. Typical number in the case of stocks is $N=500$ and $T=2500$,
corresponding to 10 years of daily data, already quite a long
strand compared to the lifetime of stocks or the expected structural
evolution time of markets, but that corresponds to $q=0.2$. For macroeconomic indicators -- say inflation,
20 years of monthly data produce a meager $T=240$, whereas the number of sectors of activity for which inflation is
recorded is around $N=30$, such that $q=0.125$. Clearly, effects induced by a non zero value of $q$ are expected to be highly relevant in 
many applications.

\subsection{Historical survey}

The rapid growth of RMT (Random Matrix Theory) in the last two decades is due both to the increasing complexity of the data in many fields of science (the ``Big Data'' phenomenon) and to many new, groundbreaking mathematical results that challenge classical results of statistics. In particular, RMT has allowed a very precise study of large sample covariance matrices and also the design of estimators that are consistent in the large dimensional limit (LDL) presented above. The aim of this review is to provide the reader an introduction to the different RMT inspired techniques that allow one to investigate problems of high-dimensional statistics, with the estimation of large covariance matrices as the main thread.

The estimation of covariance matrices is a very old problem in multivariate statistics and one of the most influential work goes back to 1928 with John Wishart \cite{wishart1928generalised} who investigated the distribution of the sample covariance matrix $\E$ in the case of i.i.d Gaussian realizations $\b y_1, \b y_2, \dots, \b y_T$. In particular, Wishart obtained the following explicit expression for the distribution of $\E$ given $\C$ \cite{wishart1928generalised}:
\begin{equation}
\label{eq:wishart_distribution}
{\cal P}_{W}(\E | \C) = \frac{T^{NT/2}}{2^{NT/2}  \Gamma_{N}(T/2)} \frac{ \det(\E)^{\frac{T-N-1}{2}}}{\det(\C)^{T/2}} e^{-\frac{T}{2} \Tr \C^{-1} \E},
\end{equation}
where $\Gamma_{N}(\cdot)$ is the multivariate Gamma function with parameter $N$.\footnote{$\Gamma_{N}(u) = \pi^{N(N-1)/4} \prod_{j=1}^N \Gamma(u + (1-j)/2)$.} In Statistics, one says that $\E$ follows a $\text{Wishart}(N,T,\C/T)$ distribution and it is often referred to as one of the first result in RMT. Note that for a finite $N$ and $T$, the marginal probability density distribution of the eigenvalues is known \cite{anderson1984introduction}:
\begin{equation}
	\label{eq:wishart_marginal_PDF}
	\rho_N(\lambda) = \frac{1}{N} \sum_{k=0}^{N-1} \frac{k!}{T-N+k} \qb{L_k^{T-N}(\lambda)}^2 \lambda^{T-N} e^{-\lambda},
\end{equation}
where we assumed that $T > N$ and $L_{k}^{l}$ are the Laguerre polynomials\footnote{$L_{k}^{l}(\lambda) = \frac{e^{\lambda}}{k! \lambda^l} \frac{d^k}{d\lambda^k} (e^{-\lambda} \lambda^{k+l}).$}.

Even though the Wishart distribution gives us many important properties concerning $\E$, the behavior of the sample estimator as a function of $N$ was understood much later with the pioneering work of Charles Stein in 1956 \cite{stein1956inadmissibility}. The most important contribution of Stein can be summarized as follows: when the number of variables $N \geq 3$, there exist combined estimators more accurate in terms of \emph{mean squared error} than any method that handles the variables separately (see \cite{efron1977stein} for an elementary introduction). This phenomenon is called \emph{Stein's paradox} and establishes in particular that the sample  matrix $\E$ becomes more and more inaccurate as the dimension of the system $N$ grows. The idea of ``combined'' estimators has been made precise with the James-Stein estimator \cite{james1961estimation} for the mean of a Gaussian vector that outperforms traditional methods such as maximum likelihood or least squares whenever $N \geq 3$. To achieve this, the authors used a \emph{Bayesian} point of view, i.e. by assuming some \emph{prior} probability distribution on the parameters that we aim to estimate. For sample covariance matrices, Stein's paradox also occurs for $N \geq 3$ as shown by using properties of the Wishart distribution and the so-called \emph{conjugate} prior technique (see Chapter \ref{chap:bayes}). This was first shown for the \emph{precision} matrix $\C^{-1}$ in \cite{efron1976multivariate,haff1977minimax} and then for the covariance matrix $\C$ in \cite{haff1980empirical} and lead to the famous {\it linear shrinkage} estimator
\begin{equation}
	\label{eq:intro_linear_shrinkage}
	\b \Xi = \alpha_s \E + (1 -\alpha_s) \b I_N,
\end{equation}
where $\Xi$ denotes, here and henceforth, an estimator of $\C$ and $\alpha_s \in (0,1)$ is the shrinkage intensity parameter. In \cite{haff1980empirical}, Haff proposed to estimate $\alpha_s$ using the marginal probability distribution of the observed matrix $\b Y$ as advocated in the so-called \emph{empirical} Bayes framework. We see that this shrinkage estimator interpolates between the empirical ``raw'' matrix $\E$ (no shrinkage, $\alpha_s=1$) and the null hypothesis $\b I_N$ (extreme shrinkage, $\alpha_s=0$). This example illustrates the idea of a combined estimator, not based only on the data itself, that offers better performance when the dimension of the system grows. The improvement made by using the simple estimator \eqref{eq:intro_linear_shrinkage} rather than the sample covariance matrix $\E$  has been precisely quantified much later in 2004 \cite{ledoit2004well} in the asymptotic regime $N \to \infty$, with an explicit and observable estimator for the shrinkage intensity $\alpha_s$. To summarize, the Bayesian approach turns out to be a cornerstone in estimating high dimensional covariance matrices and will be 
discussed in more details in the Section \ref{chap:bayes}. 

\begin{figure}[h]
% \begin{subfigure}{.5\textwidth}
%   \centering
%   \includegraphics[width=1.\linewidth]{Figures/introduction/dirac}
%   \caption{Spectrum of $\C = \b I_N$.}
%   %\label{fig:multiple}
% \end{subfigure}%
% \begin{subfigure}{.5\textwidth}
%   \centering
%   \includegraphics[width=1.\linewidth]{Figures/introduction/MP_den_intro}
%   \caption{Spectrum of $\E$ for $q = 0.25$ (blue) and $q = 0.5$ (red).}
%   %\label{fig:dGOE}
% \end{subfigure}\\
\centering
\includegraphics[scale=0.45]{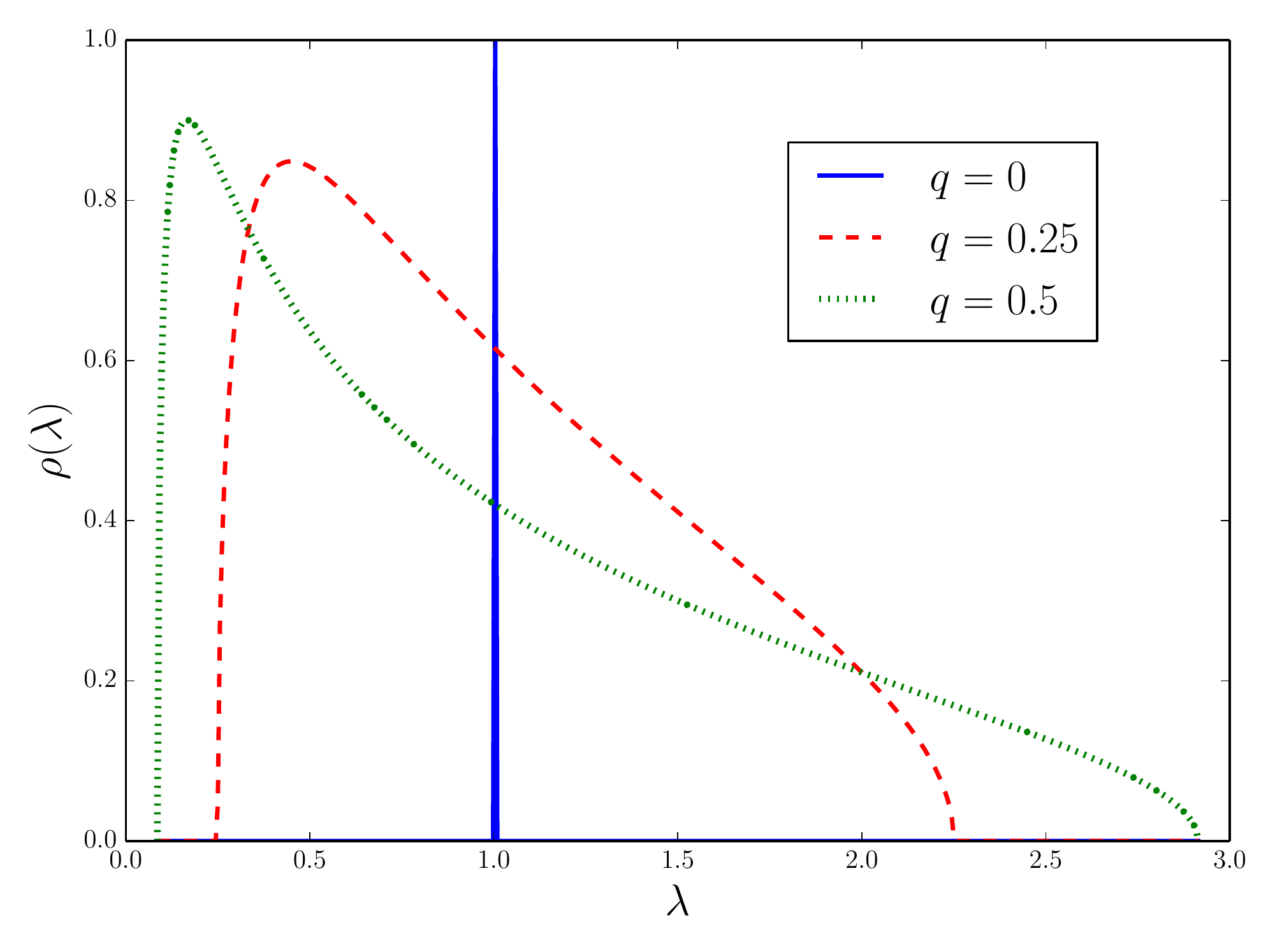}
  %\caption{Spectrum of $\E$ for $q = 0.25$ (blue) and $q = 0.5$ (red).}
\caption{Plot of the sample eigenvalues and the corresponding sample eigenvalues density under the null hypothesis with $N = 500$. The blue line ($q=0$) corresponds to a perfect estimation of the population eigenvalues. The larger is the observation ratio $q$, the wider is the sample density. We see that even for $T = 4N$, the deviation from the population eigenvalues is significant. }
\label{fig:intro_MP}
\end{figure}

Interestingly, the first result on the behavior of sample covariance matrices in the LDL did not come from the statistics community. It is due to the seminal work of Mar{\v c}enko and Pastur in 1967 \cite{marchenko1967distribution} where they obtained a self-consistent equation for the spectrum of $\E$ \emph{given} $\C$ as $N$ goes to infinity. In particular, the influence of the quality ratio $q$ appears precisely. Indeed, it was shown in the classical limit $T \to \infty$ and $N$ fixed in 1963 by Anderson that the sample eigenvalues converge to the population eigenvalues \cite{anderson1963asymptotic}, a result indeed recovered by the Mar{\v c}enko-Pastur formula for $q=0$. However, when $q = \cal O(1)$, the same formula shows that all the sample eigenvalues become noisy estimators of the ``true'' (population) ones no matter how large $T$ is. This is also called the \emph{curse of dimensionality}. More precisely, the distortion of the spectrum of $\E$ compared to the ``true'' one becomes more and more substantial as $q$ becomes large (see Figure \ref{fig:intro_MP}). The heuristic behind this phenomenon is as follows. When the sample size $T$ is very large, each individual coefficient of the covariance matrix $\b C$ can be estimated with negligible error (provided one can assume that 
$\b C$ itself does vary with time, i.e. that the observed process is stationary). But if $N$ is also large and of
the order of $T$, as is often the case in many situations, the sample estimator $\E$ becomes ``inadmissible''. More specifically, the large number of simultaneous noisy variables creates important systematic errors in 
the computation of the eigenvalues of the matrix.

The Mar{\v c}enko-Pastur result had a tremendous impact on the understanding the ``curse of dimensionality''. Firstly, it was understood in 1995 that this result is to a large degree \emph{universal} when $N \to \infty$ and $q = \cal O(1)$, much as the Wigner semi-circle law is universal: the  Mar{\v c}enko-Pastur  equation is valid for a very broad range of random measurement processes and for general population covariance matrix $\C$ \cite{yin1986limiting,silverstein1995strong,sengupta1999distributions}. This property is in fact at the core of RMT which makes this theory particularly appealing. 
At the same time, some empirical evidence of the relevance of these results for sample covariance matrices weres provided in \cite{laloux1999noise,plerou2002random} using financial data sets, which are known to be non-Gaussian \cite{bouchaud2003theory}. More precisely, these works suggested that most of the eigenvalues (the \emph{bulk}) of financial correlation matrices agrees, to a first approximation, with the null hypothesis $\C = \b I$, while a finite number of ``spikes'' (\emph{outliers}) reside outside of the bulk. This observation is the very essence of the \emph{spiked covariance matrix} model named after the celebrated paper of Johnstone in 2001 with many applications in \emph{principal components analysis} (PCA)  \cite{johnstone2001distribution}. Indeed, the author showed another manifestation of universal properties of RMT, namely the Tracy-Widom distribution for the top bulk eigenvalues in the spiked covariance matrix \cite{tracy1994level,johnstone2001distribution}. This result suggest that the edge of the bulk of eigenvalues is very \emph{rigid} in the sense that the 
position of the edge has very small fluctuations of order $T^{-2/3}$. This provides a very simple recipe to distinguish meaningful eigenvalues (beyond the edge) from noisy ones (inside the bulk) \cite{laloux2000random,plerou2002random}. This method is known as ``eigenvalue \emph{clipping}'': all eigenvalues in the bulk of the Mar{\v c}enko-Pastur spectrum are deemed as noise and thus replaced by a constant value whereas the principal components outside of the bulk (the spikes) are left unaltered. This very simple method provides robust out-of-sample performance \cite{bouchaud2009financial} and emphasizes that the notion of regularization -- or cleaning -- is very important in high-dimension. 

Even if the spiked covariance matrix model provides quite satisfactory results in many different contexts \cite{bouchaud2009financial}, one may want to work without such an assumption on the structure of $\C$ using the Mar{\v c}enko-Pastur equation to reconstruct numerically the spectrum of $\C$ \cite{silverstein1995analysis}. However, this is particularly difficult in practice since the Mar{\v c}enko-Pastur equation is easy to solve in the other direction, i.e. knowing the spectrum of $\C$, we easily get the spectrum of $\E$. In that respect, many studies attempting to ``invert'' the Mar{\v c}enko-Pastur equation appeared since 2008 \cite{bouchaud2009financial,mestre2008improved,yao2012eigenvalue,el2008spectrum}. The first one consists in finding a parametric ``true'' spectral density that fits the data \cite{bouchaud2009financial}. The method of \cite{mestre2008improved}, further improved in \cite{yao2012eigenvalue}, is completely different. Under the assumption that the spectrum of $\C$ consists of a finite number of eigenvalues, an exact analytical estimator of each population eigenvalue is provided. However, this method requires some very strong assumptions on the structure of the spectrum of $\C$. The last approach can be considered as a \emph{nonparametric} method and seems to be very appealing. Indeed, El Karoui proposed a ``consistent'' numerical scheme to invert the Mar{\v c}enko-Pastur equation using the observed sample eigenvalues \cite{el2008spectrum}. Nevertheless, while the method is very informative, it turns out that the algorithm also needs prior knowledge on the location of the true eigenvalues which makes the implementation difficult in practice. 

% The first one adopted an ``empirical Bayes'' approach in the sense that they consider the eigenvalues of $\C$ belongs to a rotationally invariant prior distribution that can be estimated from the data itself \cob **??really??**\nc (\cor By empirical bayes approach, I mean that the prior is fitted by maximizing the likelihood. Maybe this is not the correct term !\nc). 

\begin{figure}[h]
	\begin{center}
   \includegraphics[scale = 0.5]{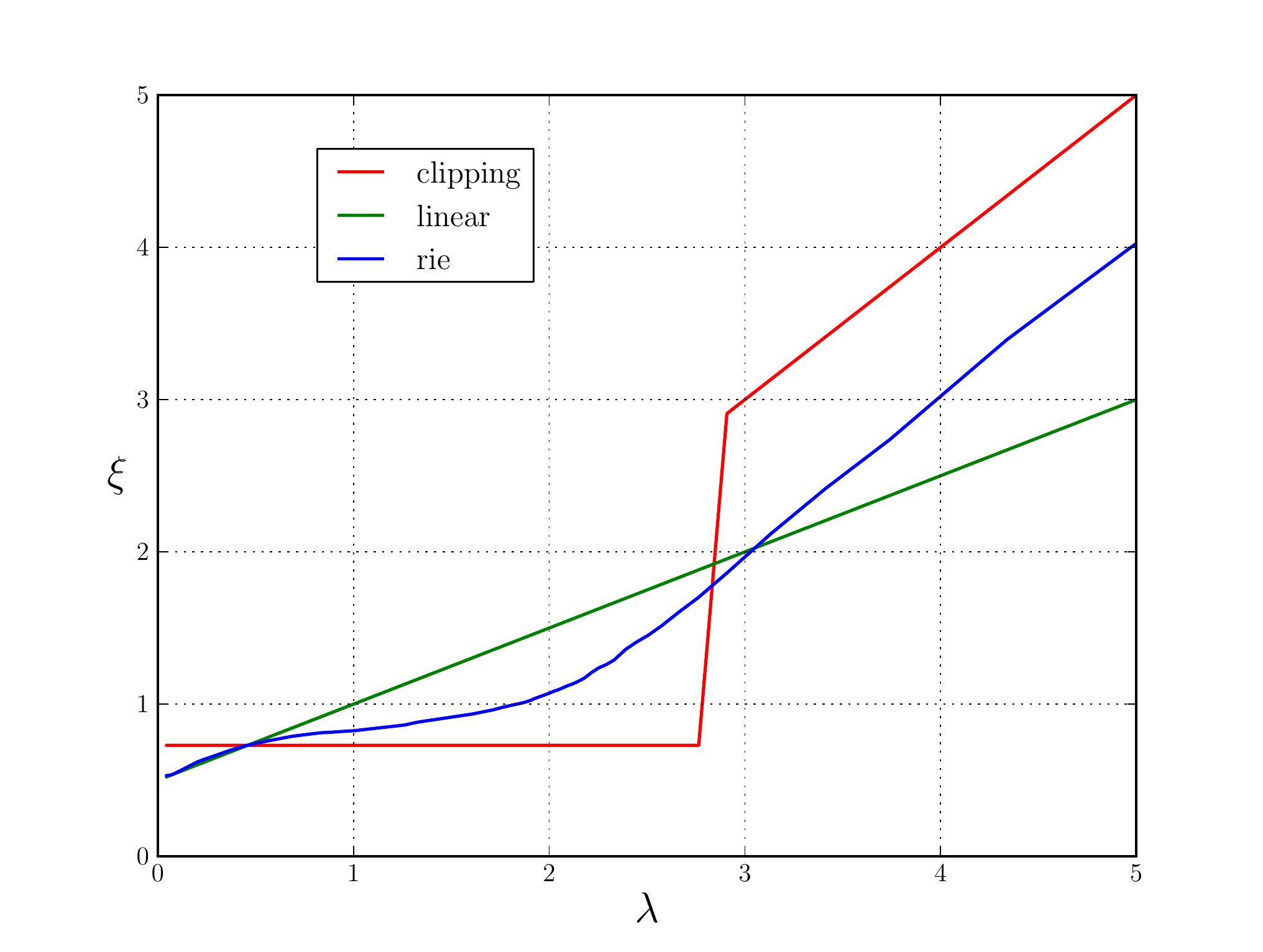} 
   \end{center}
   \caption{(Color online). Three shrinkage transformations: ``cleaned'' eigenvalues on the y-axis  as a function of the sample eigenvalues (see Chapter \ref{chap:numerical} for more details). This figure is a quick summary the evolution of shrinkage estimators starting with the linear method (green), then the heuristic eigenvalues clipping method (red) to the optimal RIE (blue).}
   \label{fig:intro_RIE_other}
\end{figure}

These inversion schemes thus allow in principle to retrieve the spectrum of $\C$ but as far as estimating high-dimensional covariance matrices is concerned, merely substituting the sample eigenvalues by the estimated ``true'' ones does not give a satisfactory answer to our problem. Indeed, the Mar{\v c}enko-Pastur equation only describes the spectrum of eigenvalues of large sample covariance matrices but does not yield any information about the \emph{eigenvectors} of $\E$. In fact, except for some work by Jack Silverstein around 1990 \cite{silverstein1986eigenvalues,silverstein1989eigenvectors}, most RMT results about sample covariance matrices were focused on the eigenvalues, as discussed above. The first fundamental result on the eigenvectors of $\E$ was obtained in \cite{paul2007asymptotics} in the special case of the spiked covariance matrix model, but is somehow disappointing for inference purposes. Indeed, Paul noticed that outliers' eigenvectors obey a cone concentration phenomenon with respect to the true eigenvectors whereas all other ones retain very little information \cite{paul2007asymptotics}. Differently said, the eigenvectors of $\E$ are not consistent estimators of the eigenvectors of $\C$ in the high-dimensional framework. A few years later, these observations were generalized to general population covariance matrices $\C$ \cite{ledoit2011eigenvectors,bun2015rotational,bun2016optimal,benaych2011eigenvalues,monasson2015estimating}. When dealing with the estimation of $\C$, information about eigenvectors has to be taken into account somehow in the inference problem. Clearly, the above ``eigenvalue substitution'' method cannot be correct as it
proposes to take the best estimates of the eigenvalues of $\C$ but in an unknown eigenvalue basis. Consequently, a different class of estimators flourished very recently that we shall refer to as \emph{rotational invariant estimators}\footnote{This is sometimes called rotation-equivariant estimators} (RIE) \cite{ledoit2011eigenvectors,bun2015rotational,bun2016optimal}. In this particular class of estimators, the main assumption is that any estimator $\Xi$ of $\C$ must share the same eigenvectors as $\E$ itself. This hypothesis has a very intuitive interpretation in practice as it amounts to posit that one has no prior insights on the structure of $\C$, i.e. on the particular directions in which the eigenvectors of $\C$ must point. It is easy to see that the linear shrinkage estimator \eqref{eq:intro_linear_shrinkage} falls into this class of estimators. Compared to the aforementioned RMT-based methods, RIE explicitly uses the information on the eigenvectors of $\E$, in particular their average overlap with the true eigenvectors. It turns out that one can actually obtain an optimal estimator of $\C$ in the LDL for any general population covariance matrix $\C$ \cite{bun2016optimal}. Note that the optimal estimator is in perfect agreement with Stein's paradox, that is to say, the optimal cleaning recipe takes into account about the information of all eigenvectors and all eigenvalues of $\E$. The conclusion is therefore that combining all the information's about $\E$ always provide more accurate prediction than any method that handles the parameters separately within the modern era of ``Big Data''. We summarize the above long journey concerning the estimation of large sample 
covariance matrices in Figure \ref{fig:intro_RIE_other}, which can be seen as a thumbnail picture of the present review. Note that a very recent work \cite{monasson2015estimating} attempts to incorporate prior information on the true components. While it remains unclear how to use this framework for the estimation of correlation, this may allows one to construct ``optimal'' non-rotational invariant estimators.  We shall address this issue at the end of this review.  

\subsection{Outline}

Our aim is to review several Random Matrix Theory (RMT) results that take advantage of the high-dimensionality of the problem to estimate covariance matrices consistently, spanning nearly fifty years of research from the result of Mar{\v c}enko and Pastur \cite{marchenko1967distribution} to the very recent ``local'' optimal RIE for general population covariance matrices \cite{bun2016optimal}. We emphasize that this review is not intended to provide detailed proofs (in the mathematical sense) but we will include references to this mathematical literature as often as possible for those who might be interested. 

In Chapter \ref{chap:RMT}, we begin with a detailed but still incomplete introduction to RMT and some of the analytical methods available to study the behavior of large random matrices in the asymptotic regime. In fact, most of the computations in Chapter \ref{chap:RMT} will be performed under very general model of random matrices and will be used throughout the following. The first method is arguably the most frequently used in the Physics literature known as the Coulomb gas analogy \cite{brezin1978planar}. This is particularly useful to  deal with invariant ensembles, leading to Boltzmann-like weights that allows one to recover very easily well-known results such as Wigner's semicircle law \cite{wigner1951statistical} or Mar{\v c}enko-Pastur density \cite{marchenko1967distribution}. This is the main purpose of Section \ref{sec:potential}. The second method is Voiculescu's \emph{free probability theory} which was originally proposed in 1985 to understand a special class of von Neumann algebras through the concept of \emph{freeness} \cite{voiculescu1985symmetries}. Loosely speaking, two matrices $\AAA$ and $\BBB$ are mutually free if their eigenbasis are related to one another by a random rotation, or said differently if the eigenvectors of $\AAA$ and $\BBB$ are almost surely orthogonal. Voiculescu discovered in 1991 \cite{voiculescu1991limit} that some random matrices do satisfy asymptotically the freeness relation, which considerably influenced RMT. We present in Section \ref{sec:free_probability} a precise definition of the concept of freeness and then provide some applications for the computations of the spectral density of a large class of random matrices. In Section \ref{sec:replica}, we present a more formal tool known as the Replica method in statistical physics of disordered systems \cite{edwards1976eigenvalue,mezard1987spin}. While being less rigorous, this method turns out to be very powerful to compute the average behavior of large complex systems (see \cite{morone2014replica} for a recent review). In our case, we shall see how this method allows us to compute the \emph{resolvent} of a large class of random matrices which will be especially useful to deal with the statistics of eigenvectors. 

In Chapters \ref{chap:spectrum} and \ref{chap:eigenvectors}, we study in details the different properties of large sample covariance matrices. Chapter \ref{chap:spectrum} is dedicated to the statistics of the eigenvalues of $\E$, and in particular we propose a very simple derivation of the Mar{\v c}enko-Pastur equation using tools from free probability theory. Then, we review different properties that we can learn about $\C$ using $\E$ such as the moment generating functions, or the edges of the support of the spectral density of $\E$. We discuss the properties of the edges of the distribution for finite $N$ and also the outliers. In Chapter \ref{chap:eigenvectors}, we focus the recent results concerning the eigenvectors of $\E$ for a general $\C$. We distinguish two different cases. The first one is the angle between the true and estimated eigenvectors and we shall see that the initial results of \cite{paul2007asymptotics} hold for a general $\C$. The second case is the angle between two \emph{independent} sample eigenvectors, a result that allows one to infer interesting properties about the structure of $\C$. 

After these three relatively technical sections, we then turn on the main theme of this review which is the estimation of large sample covariance matrices. In Chapter \ref{chap:bayes}, we formalize the Bayesian method for covariance matrices. We present the class of conjugate prior from which we re-obtain the linear 
shrinkage \eqref{eq:intro_linear_shrinkage} initially derived by Haff \cite{haff1980empirical}. Next, we consider the class of Boltzmann-type, rotational invariant prior distributions. We then relate the Bayes optimal estimator with the least squares optimal oracle estimator of $\C$. The so-called oracle estimator is the main quantity of interest in the following Chapter \ref{chap:RIE}. In particular, we show that this estimator converges to a limiting and -- remarkably -- fully observable function in the limit of large dimension using the results on eigenvectors obtained in Chapter \ref{chap:eigenvectors}. Hence, there exists an optimal estimator of large population covariance $\C$ depending only on $\E$ inside the class of RIEs. The rest of the Chapter \ref{chap:RIE} is dedicated to some theoretical and numerical applications of the optimal RIE. 

Chapter \ref{chap:application} concerns the applications of the optimal RIE for Markowitz optimal portfolio. In particular, we characterize explicitly, under some technical assumptions, the danger of using the sample covariance matrix $\E$ in a large scale and out-of-sample framework. As alluded to above, we shall see that if $\E$ has no exact zero mode (i.e. when $q=N/T < 1$), the realized risk associated to this ``naive'' estimator overestimate the true risk by a factor $(1-q)^{-1}$. Also, we shall see that the best we can do in order to minimize the out-of-sample risk is actually given by the optimal RIE of the Chapter \ref{chap:RIE}. Several alternative cleaning ``recipes'', proposed in previous work, are also reviewed in that Chapter.  

Finally, Chapter \ref{chap:numerical} contains empirical results using real financial data sets. We give further evidence that using a correctly regularized estimator of $\C$ is highly recommended in real life situations. Moreover, we discuss about the implementation of the optimal RIE in the presence of finite size effects, to wit, when $N$ is large but finite. 

The appendices contain auxiliary results which are mentioned in the paper. The first appendix copes with the so-called Harish-Chandra--Itzykson-Zuber (HCIZ) integral which routinely appears in calculations involving sums or products of free random matrices. The HCIZ is an integral over the group of orthogonal matrices for which explicit and analytical results are scarce. The second appendix is a reminder on some results of linear algebra which are particularly useful for the study of eigenvectors. The third appendix is another analytical tool in RMT to establish self-consistent equations for the resolvent (or the Stieltjes transform) of large random matrices. This technique is very convenient when working with independent entries and it provides a nice illustration of the Central Limit Theorem for random matrices. However, the formalism is not as synthetic as the method provided in Chapter \ref{chap:RMT} but is now standard in the RMT literature, which is why we relegate its presentation to an appendix. Finally, we devote a full appendix to the case where the noise in the matrix is {\it additive}, rather than {\it multiplicative} for correlation matrices. Although not directly relevant to the main issue discussed in the present review, the additive noise model is interesting in itself and finds many applications in different fields of science.

%---------  Theoretical part ---------------- %

\clearpage%!TEX root = RMT_Covariance_Review.tex
\section{Random Matrix Theory: overview and analytical tools}
%\section{RMT in a nutshell}
\label{chap:RMT}

\subsection{RMT in a nutshell}
\label{section:RMT_nutshell}

\subsubsection{Large dimensional random matrices}
\label{section:RMT_definition}

As announced in the introduction, the main analytical tool that we
shall review in this article is Random Matrix Theory (RMT).  In order
to be as self-contained as possible, we recall in this section some of
the basic results and techniques of RMT.  The study of random matrices
began with the work of Wishart in 1928, who was interested in the
distribution of the so-called empirical (or sample) covariance
matrices, which ultimately lead to the Mar{\v c}enko-Pastur
distribution in 1967.  RMT was also introduced by Wigner in
the 1950's as a statistical model for the energy levels of heavy
nuclei, and lead to the well-known Wigner semi-circle distribution, as
well as Dyson's Brownian motion (see
e.g. \cite{weidenmuller2009random}, \cite{akemann2011oxford} for
comprehensive reviews). Branching off from these early physical and
statistical applications, RMT has become a vibrant research field of
its own, with scores of beautiful results in the last decades -- one
of the most striking being the discovery of the Tracy-Widom
distribution of extreme eigenvalues, which turns out to be related to
a large number of topics in statistical mechanics and probability
theory \cite{dean2006large,majumdar2006random}.  Here, we will only
consider the results of RMT that pertain to statistical inference, and
leave aside many topics -- see e.g. \cite{akemann2011oxford}, \cite{anderson2010introduction},
\cite{terence2012topics}, \cite{tulino2004random},
\cite{bai2009spectral} or
\cite{couillet2011random} for more detailed and rigorous introductions
to RMT. We will also restrict to square, symmetric correlation
matrices, even though the more general problem of rectangular
correlation matrices (measuring the correlations between $M$ input
variables and $N$ output variables) is also extremely
interesting. This problem leads to the so-called Canonical Component
Analysis \cite{hotelling1936relations} and can be dealt with the
Singular Value Decomposition, for which partial results are available,
see e.g. \cite{wachter1980limiting,bouchaud2007large}.

We begin with a formal definition of ``large'' random matrices. A common assumption in RMT is that the matrix under scrutiny is of infinite size. However, this is obviously not a realistic assumption for practical problems where one rather deals with \emph{large} but \emph{finite} $N$ dimensional matrices. Nonetheless, we shall see that working in the $N \to \infty$ limit leads to very precise approximations of the properties of large but finite matrices. More precisely, it is well known that probability distributions describing the fluctuations of macroscopic observables often converge to limiting laws in the limit of large sizes. Hence, we expect that the statistical properties (say the distribution of eigenvalues) of a random matrix $\M$ of dimension $N$ shows, to a certain extent, a deterministic or self-averaging behavior\footnote{i.e. independent of the specific realization of the matrix itself} when the dimension $N$ goes to infinity. These deterministic features can be used to characterize the matrix under scrutiny, provided it is large enough. This is why we consider the limit $N \rightarrow \infty$ from now on. 

The limiting behavior of ``large'' random matrices is in fact at the
heart of RMT, which predicts that infinite dimensional matrices do
display \emph{universal} features, both at the macroscopic and at the
microscopic levels.  To be more precise, we define a $N \times N$
random matrix\footnote{Boldface letters will refer throughout this
  paper to matrices.}  $\M$ with a certain probability measure ${\cal
  P}_{\beta}(\M)$, where $\beta$ is the Dyson's threefold way index
and specifies the symmetry properties of the ensemble ($\beta = 1$ for
Orthogonal, $\beta = 2$ for Unitary and $\beta = 4$ for Symplectic
ensembles). A property is said to be \emph{universal} if it does not
depend on the specific probability measure ${\cal P}_{\beta}(\M)$. One
well known example of universality pertains to the distribution of the
distance $s$ between two successive eigenvalues (see
\cite{tao2011random} for an extended discussion).

The ensemble most relevant for our purpose is the Orthogonal one, which deals with real symmetrical matrices. In this case, the matrix $\M$ is said to be rotationally invariant if the probability is invariant under the transformation $\M \rightarrow \b\Omega \M \b\Omega^{\dag}$ for any matrix $\b\Omega$ belonging to the Orthogonal group $\b O(N)$, \ie~  ${\cal P}_{\beta}(\M) = {\cal P}_{\beta}(\b\Omega \M \b\Omega^{\dag}) ,\, \forall \b \Omega \in \b O(N)$. A typical example of invariant measure in the physics literature is that ${\cal P}_{\beta}(\M)$ is of the form of a Boltzmann distribution:
\begin{equation} \label{eq:Boltz}
{\cal P}_{\beta}(\M) {\cal D}\M \propto e^{- \frac{\beta N}{2} \Tr V(\M)} {\cal D}\M
\end{equation}
with $V$ the so called \emph{potential} function and ${\cal D}\M = \prod_{i=1}^{N} d\M_{ii} \prod_{i < j}^{N} d\M_{ij}$ denotes the (Lebesgue) flat measure. The rotational invariant property is evident since the above parametrization only involves the trace of powers of $\M$. Already at this stage, it is interesting to notice that the distribution (\ref{eq:Boltz}) can alternatively be rewritten in terms of the eigenvalues and eigenvectors of $\M$ as:
\begin{equation} \label{eq:Boltz2}
{\cal P}_{\beta}(\M) {\cal D}\M \propto  e^{- \frac{\beta N}{2} \sum_{i=1}^N V(\nu_i) }\prod_{i<j}^{N} \abs{\nu_i - \nu_j}^\beta \pB{\prod_{i=1}^{N} \dd\nu_i} \pB{\dd \Omega},
\end{equation}
where the Vandermonde determinant ($\prod \abs{\nu_i - \nu_j}^\beta$)comes from the change of variables (from the $\M_{ij}$ to the $\nu_i$ and $\Omega_{ij}$). This representation is extremely useful, as will be illustrated below.

What kind of universal properties can be of interest in practice? Let us consider a standard problem in multivariate statistics. Suppose that we have a very large dataset with correlated variables. A common technique to deal with this large dataset is to reduce the dimension of the problem using for instance a \emph{principal component analysis} (PCA), obtained by diagonalizing the covariance matrix of the different variables. But one can wonder whether the obtained eigenvalues $\nu_i$ and their associated eigenvectors are reliable or not (in a statistical sense). Hence, the characterization of eigenvalues (and eigenvectors) is an example of features that one would like to know a priori. In that respect, RMT provided (and continues to provide) many groundbreaking results on the eigenvalues and the eigenvectors of matrices belonging to specific invariant ensembles (Unitary, Orthogonal and Symplectic). The distribution of the eigenvalues $\{\nu_i\} : i = \{1,\dots, N\}\}$ can be characterized through the \emph{Empirical Spectral Distribution} (ESD) (also known as the ``Eigenvalue Distribution''):
\begin{equation}
\label{ESD}
\rho_{\M}^{N}(x) = \frac1N \sum_{i=1}^{N} \delta(x - \nu_{i})
\end{equation}
with $\delta$ the Dirac delta function. Note that the symmetry of the considered matrices ensures that the eigenvalues of $\M$ are defined on the real line (complex eigenvalues are beyond the scope of this review, but see 
\cite{bai2009spectral,couillet2011random,bordenave2012} for more on this). One of the most important property of large random matrices is that one expects the ESD to converge (almost surely in many cases) to a unique and  \emph{deterministic} limit $\rho_{\M}^N \to \rho_{\M}$ as $N \to \infty$. Note that it is common to refer to this deterministic density function $\rho_{\M}$ as the \emph{Limiting Spectral Density} (LSD), or else the ``Eigenvalue Spectrum'' of the matrix. 
An appealing feature of RMT is the predicted \emph{self-averaging} (sometimes call \emph{ergodicity} or \emph{concentration}) property of the LSD: when the dimension $N$ becomes very large, a single sample of $\M$ spans the whole eigenvalue density function, independently of the specific realization of $\M$. The consequence of this self-averaging property is that we can replace the computation of the ESD (\ref{ESD}) for a specific $\M$ by the average according to the probability measure of $\M$ (\eg\  over the measure (\ref{eq:Boltz})):
\begin{equation}
\label{average_ESD}
\rho_{\M}(x) = \lim_{N \rightarrow \infty} \rho_{\M}^{N}(x) ,\quad\text{with }\quad \rho_{\M}^{N}(x) = \left\langle \frac1N \sum_{i=1}^{N} \delta(x - \nu_{i}) \right\rangle_{\M}.
\end{equation}

For real life data-sets, it is often useful to distinguish the eigenvalues that lie within the spectrum of $\rho_{\M}$ from those that are well separated from it. We will refer to the first category as the \textbf{bulk} of the eigenvalues with a slight abuse of notation. We will call the second type of eigenvalues \textbf{outliers} or \textbf{spikes}. Throughout this work, we assume the LSD that describes the bulk of $\rho_\M$ to be a non-negative continuous function, defined on an unique compact support -- denoted $\supp[\rho_{\M}]$ -- meaning that $\supp[\rho_{\M}]$ consists of a single ``bulk'' component (often called the \emph{one-cut} assumption). Moreover, we allow the presence of a finite number $r \ll N$ of outliers, which are of crucial importance in many fields. Throughout this chapter, we shall denote by $\nu_1 \ge \nu_2 \ge \dots \ge \nu_N $ the eigenvalues of $\M$. We furthermore define the associated eigenvectors by $\b w_1 , \b w_2 , \dots , \b w_N$. For $N$ that goes to infinity, it is often convenient to index the eigenvectors by their corresponding eigenvalues, \ie~ $\b w_i \equiv \b w_{\nu_i}$ for any integer $1 \le i \le N$, and this is the convention that we adopt henceforth.

\subsubsection{Various RMT transforms}
\label{section:RMT_transforms}

We end this section with an overview of different transforms that appear in the RMT literature. These transforms are especially useful to study the spectral properties of random matrices in the limit of large dimension, and to deal with sums and products of random matrices.

\paragraph{Resolvent and Stieltjes transform}
\label{section:RMT_resolvent_stieltjes}

We start with the resolvent of $\M$ which is defined as\footnote{Note that in the mathematical and statistical literature, the resolvent differs from ours by a minus sign.}
\begin{equation}
\label{resolvent}
\b G_{\M}(z) := (z\In - \M)^{-1},
\end{equation}
with $z \deq x - \ii \eta \in \mathbb{C}^{-}$, where $\mathbb{C}^{-} = \{ z \in \mathbb{C} : \im(z) < 0\}$. We define accordingly $\mathbb{C}^{+} = \{ z \in \mathbb{C} : \im(z) > 0\}$. This quantity displays several interesting properties, making it the relevant object to manipulate. First, it is a continuous function of $z$ and is easy to differentiate (compared to working directly on the ESD), 
providing a well-defined tool for mathematical analysis. Furthermore, it contains the complete information about the eigenvalues $\{\nu_i\}$ {\it and} the eigenvectors $\{\b w_i\}$ 
since it can be rewritten as:
\begin{equation}
\label{eq:resolvent_decomposition}
\b G_{\M}(z) = \sum_{i=1}^{N} \frac{\b w_i \b w_i^*}{z - \nu_i}.
\end{equation}
It is easy to see that the number of singularities of the resolvent is equal to the number of eigenvalues of $\M$. Suppose that $z \to \nu_i$ for any $i \in \qq{N}$, then the residue of the pole defines a projection operator onto 
the eigenspace associated to the eigenvalues $\nu_i$. We will show in chapter \ref{chap:eigenvectors} how this property can be used to study the statistics of the eigenvectors. 

% Another nice property in the large $N$ limit is that when the real of $z$ is in the bulk component of $\rho_{\M}$, the resolvent is self-averaging for $z$ not too close of the real axis. By ``$z$ not too close to the real axis'', we mean that the imaginary part of $z$ has to be much larger than $N^{-1}$. Indeed, in the case where $\im(z) \sim N^{-1}$, it is not hard to see from the eigendecomposition of the resolvent that the value of $\b G_{\M}$ will be dominated by the fluctuations of individual eigenvalue, and the law of large number would not apply anymore. We therefore adopt the following definition for the complex parameter $z := x - \ii\eta$, with $x \in \mathbb{R}^+$ and $\eta \gg N^{-1}$.

While the statistics of the eigenvectors is an interesting and non-trivial subject in itself, we focus for now on the statistics of the eigenvalues through the ESD \eqref{average_ESD}. 
For this aim, we define the normalized trace of Eq. \eqref{resolvent} as
\begin{equation}
	\label{eq:stieltjes_emp}
	\stj_{\M}^{N}(z) := \frac1N \Tr\left[\b G_{\M}(z)\right],
\end{equation}
We shall skip the index $_{\M}$ as soon as there is no confusion about the matrix we are dealing with. In the limit of large dimension, one has
\begin{equation}
	\label{eq:stieltjes}
	\stj^N(z) \underset{N \rightarrow \infty}{\sim} \stj(z), \qquad \stj(z) \deq \int \frac{\rho(u)}{z - u}  \dd u.
\end{equation}
which is known as the \emph{Stieltjes} (or \emph{Cauchy}) transform of $\rho$. The Stieltjes transform has a lot of appealing properties. For instance, if the density function $\rho$ does not contain Dirac masses, then this is the unique solution of the so-called \emph{Riemann-Hilbert} problem, i.e :
\begin{enumerate}
	\item $\stj(z)$ is analytic in $\mathbb{C}^{+}$ except on its branch cut on the real axis inside $\supp[\rho_{\M}]$;
	\item $\lim_{|z| \to \infty} z \stj(z) = 1$;
	\item $\stj(z)$ is real for $z \in \mathbb{R} \backslash \supp[\rho_{\M}]$;
	\item When near the branch cut, two different values for $\stj(z)$ are possible, depending on whether the cut is approached from above or from below, i.e.: 
	\begin{equation}
		\label{eq:stieltjes_inversion_formula}
		\lim_{\eta \rightarrow 0^{+}} \stj(x \pm \ii \eta) = \hil(x) \mp i \pi \rho(x), \qquad x \in \supp[\rho] \text{ and } \rho(x) \in \R^+,
	\end{equation}
	where the function $\hil$ denotes the \emph{Hilbert} transform of $\rho$ defined by
	\begin{equation}
		\label{eq:hilbert_transform}
		\hil(x) \deq  \dashint_{\supp[\rho]} \frac{\rho(u)}{x - u} \dd u
	\end{equation}
	with $\dashint$ denoting Cauchy's principal value.
\end{enumerate}
It is now immediate to see that if one knows $\stj(z)$ in the complex plane, the density $\rho$ can be retrieved by inverting the last property of the Riemann-Hilbert problem:
\begin{equation}
\label{eq:stieltjes_inversion}
\rho(x) \equiv \frac{1}{\pi} \lim_{\eta \rightarrow 0^+} \im(\stj(x - \ii \eta)),\qquad x \in \supp[\rho].
\end{equation}
The continuous limit of $\stj(z)$ in the large $N$ limit thus allows to investigate the distribution of the eigenvalues that lie in the bulk component. 

Another interesting property is to study the asymptotic expansion of $\stj(z)$ when $z$ is large (and outside of \Supp[$\rho$]). Expanding $\stj(z)$ in powers of 
$z^{-1}$ yields:
\begin{equation*}
\stj(z) \underset{z \rightarrow \infty}{=} \frac{1}{z} \int \rho(u) \sum_{k=0}^{\infty} \left(\frac{u}{z}\right)^{k} \dd u .
\end{equation*}
To leading order, we get, in agreement with property (ii) above: 
\begin{equation*}
\stj(z) \sim \frac1z \int \rho(u) \dd u \equiv \frac1z,
\end{equation*}
where the last equality comes from the fact that the ESD is normalized to unity. 
The other terms of the expansion are also of particular interest. Indeed, we see that 
\begin{equation}
\label{eq:stieltjes_Moment}
\stj(z) \underset{z \rightarrow \infty}{=}  \frac1z + \frac1N \sum_{k=1}^{\infty} \frac{\Tr \M^{k}}{z^{k+1}} \equiv \frac1z + \sum_{k=1}^{\infty} \frac{\varphi(\M^{k})}{z^{k+1}},
\end{equation}
where we defined the $k$-th moment of the ESD by $\varphi(\M^{k}) := N^{-1} \Tr \M^{k}$. We see that the Stieltjes transform is related to the \textit{moment generating function} of the random matrix $\M$. This is another illustration of the fact that the Stieltjes transform contains the complete information about the eigenvalues density. Inversely, if one can measure the moments of the eigenvalues distribution, it is possible reconstruct a parametric eigenvalues density function that matches the empirical data. This nice property is an important feature of the Stieltjes transform for statistical inference purposes. Note that we will sometimes abbreviate $\varphi(\M^{k}) \equiv \varphi_{k}$ when there is no confusion about the matrix we are studying.  

Last but not least, it is easy to check the following scaling property
\begin{equation}
	\label{eq:stieltjes_scale_prop}
	\stj_{a\M}(z) = \frac{1}{a} \stj_{\M}\left(\frac{z}{a}\right),
\end{equation}
for any $a \in \mathbb{R}\backslash \{0\}$. Moreover, suppose that $\M$ is invertible, then using \eqref{eq:stieltjes_emp} we also have
\begin{equation}
	\label{eq:stieltjes_M_Minv}
	z \stj_\M(z) + \frac{1}{z} \stj_{\M^{-1}}\pBB{\frac{1}{z}} = 1,
\end{equation}
so that we are able to compute the Stieltjes transform of $\M^{-1}$ given the Stieltjes transform of $\M$. 

\paragraph{Blue function and $\rtr$-transform}
\label{section:RMT_R_transform}

There are many other useful RMT transforms, some that will turn out to be important in the next chapter. We begin with the \emph{free cumulant} generating function which is known as the $\rtr$-transform in the literature \cite{voiculescu1992free,tulino2004random,speicher2009free}. To define this quantity, it is convenient to introduce the functional inverse of the Stieltjes transform, also known as the \emph{Blue} transform \cite{zee1996law}
\begin{equation}
\label{eq:blue}
\btr(\stj(z)) = z,
\end{equation}
and the \emph{$\rtr$-transform} is simply defined by
\begin{equation}
\label{eq:r_transform}
\rtr(\omega) = \btr(\omega) - \frac1\omega.
\end{equation}
Note that one may deduce from \eqref{eq:stieltjes_scale_prop} the following property
\begin{equation}
	\rtr_{a\M}(\omega) = a \rtr_\M(a\omega),
\end{equation}
for any $a \in \mathbb{R}$. One very nice property is that the $\rtr$-transform admits a Taylor expansion in the limit $\omega \rightarrow 0$. Indeed, by plugging $\omega = \stj(z)$ into Eq.\ \eqref{eq:r_transform}, we obtain the formula 
\begin{equation}
\rtr(\stj(z)) + \frac{1}{\stj(z)} = z.
\end{equation}
Then, one can find after expanding the Stieltjes transform in powers of $z^{-1}$ that $\rtr(\omega)$  can be expanded as
\begin{equation}
	\label{eq:r_transform_series}
\rtr(\omega) = \sum_{\ell=1}^{\infty} \kappa_{\ell}(\M) \omega^{\ell-1}
\end{equation}
where the sequence $\{ \kappa_{\ell} \}_{\ell \ge 0}$ denotes the \textit{free cumulant} of order $\ell$ which are expressed as a function of the moments of the matrix. For completeness, we give the first four free cumulants: 
\begin{eqnarray}
	\label{eq:first_cumulants}
	\kappa_1 & = & \varphi_1 \nonumber \\
	\kappa_2 & = & \varphi_2 - \varphi_1^2 \nonumber \\
	\kappa_3 & = & \varphi_3 - 3 \varphi_2 \varphi_1 + 2 \varphi_1^3 \nonumber \\
	\kappa_4 & = & \varphi_4 - 4 \varphi_3 \varphi_1 - 2 \varphi_2^2 + 10 \varphi_2 \varphi_1^2 - 5 \varphi_1^4.
\end{eqnarray}
Note that the first three cumulants are equivalent to the `standard' cumulants of ordinary random variables and only differ from $\ell \geq 4$. Note for example that when $\varphi_1=0$, one finds $\kappa_4 = \varphi_4 - 2 \varphi_2^2$, whereas the standard kurtosis would read $\varphi_4 - 3 \varphi_2^2$. It will turn out that the free cumulants of the sum of independent -- in a sense specified below -- random matrices are given by the sum of the cumulants of these random matrices, i.e.\  $ \kappa_{\ell}(\M) = \kappa_{\ell}(\AAA) + \kappa_{\ell}(\BBB)$, see section \ref{sec:free_probability} below.

\paragraph{Moment generating function and S-transform}
\label{section:RMT_S_transform}

The moment generating function of the LSD $\rho$ is obtained by considering 
\begin{equation}
\label{eq:T_transform}
\ttr(z) \deq z \stj(z) - 1 = \int \frac{\dd u\rho(u) u}{z- u},	
\end{equation}
frequently known as the $\cal T$ (or sometimes $\eta$ \cite{tulino2004random}) transform \cite{benaych2011eigenvalues}. Indeed, by taking $z\to\infty$, one readily finds 
\begin{equation}
\label{eq:T_transform_moment}
\ttr_{\M}(z) = \sum_{k=1}^{\infty} \frac{\varphi(\M^{k})}{z^{k}}.
\end{equation}
We can then introduce the so-called $\str$-transform as \cite{voiculescu1992free}:
\begin{equation}
\label{eq:s_transform}
\str(\omega) \;\deq\; \frac{\omega+1}{\omega \ttr^{-1}(\omega)} 
\end{equation}
where $\ttr^{-1}(\omega)$ is the functional inverse of the $\ttr$-transform. Using the series expansion of $\ttr_{\M}(z)$ in powers of $z^{-1}$ and Eq.\ \eqref{eq:first_cumulants}, one finds that the $\str$-transform also admits a Taylor series which reads:
\begin{eqnarray}
\label{S_transform_taylor}
\str_{\M}(\omega) & = & \frac{1}{\varphi_{1}} + \frac{\omega}{\varphi_1^3} (\varphi_1^2 - \varphi_2) + \frac{\omega^2}{\varphi_1^5} (2\varphi_2^2 - \varphi_2 \varphi_1^2 - \varphi_3 \varphi_1) + {\cal O}(\omega^3) \nonumber \\
& = & \frac{1}{\kappa_{1}} - \frac{\kappa_2}{\kappa_1^3} \omega + \frac{2\kappa_2^2 - \kappa_1 \kappa_3}{\kappa_1^5} \omega^2 + {\cal O}(\omega^3).
\end{eqnarray}
From this last equation, it is not hard to see that the $\str$-transform of a matrix $\M$ which has a zero trace is ill-defined. Hence, the $\str$-transform of a Wigner matrix does not make sense, but it will be very useful when manipulating positive definite covariance matrices (see Section \ref{subsubsec:freemult}) 

Note finally that there exists a relation between the $\rtr$-transform and the $\str$-transform
\begin{equation}
	\label{eq:r_s_transform}
	\rtr(\omega) = \frac{1}{\str(\omega\rtr(\omega))}, \qquad \str(\omega) = \frac{1}{\rtr(\omega\str(\omega))}
\end{equation}
which allows one to deduce $\rtr(z)$ from $\str(z)$ and vice versa. Other properties on the $\rtr$ and $\str$ transforms can be found e.g.\ in \cite{burda2013free}.

\begin{changemargin}{1.0cm}{1.0cm} 
\footnotesize
Let us show the second equality of \eqref{eq:r_s_transform} for the sake of completeness. The derivation of the first identity is similar and we omit details. Using \eqref{eq:r_transform} and \eqref{eq:s_transform}, one obtains 
\begin{equation}
	\rtr(\omega\str(\omega)) = \btr\pBB{\frac{\omega+1}{\ttr^{-1}(\omega)}} - \frac{\ttr^{-1}(\omega)}{\omega+1}.
\end{equation}
Next, by setting $z = \ttr^{-1}(\omega)$, we can rewrite \eqref{eq:T_transform} as
\begin{equation}
	\label{eq:T_transform_inv}
	 \frac{\omega+1}{\ttr^{-1}(\omega)} = \stj\pb{\ttr^{-1}(\omega)}.
\end{equation}
Hence, we conclude that
\begin{equation}
	\rtr(\omega\str(\omega)) = \ttr^{-1}(\omega) - \frac{1}{\stj\pb{\ttr^{-1}(\omega)}} = \frac{\omega}{\stj\pb{\ttr^{-1}(\omega)}}.
\end{equation}
The conclusion then follows from \eqref{eq:T_transform_inv}. 

\end{changemargin} 
\normalsize

% \subsection{Analytical tools for spectrum analysis}
% \label{chap:analytical}

\subsection{Coulomb gas analogy}
\label{sec:potential}

There exists several techniques to compute the limiting value of the
Stieltjes transform: (i) Coulomb gas methods, (ii) method of moments,
(iii) Feynman diagrammatic expansion, (iv) Dyson's Brownian motion,
(v) Replicas, (vi) Free probability, (vii) recursion formulas, (viii) supersymmetry... We
devote the rest of this section to provide the reader with a brief
introduction to (i), (v) and (vi). Dyson's Brownian motion (iv) and
the recursion method (vii) are mentioned in appendices
\ref{app:recursion} and \ref{app:DBM}. We refer to
\cite{anderson2010introduction} for the moment methods (ii), to
\cite{brezin1978planar,burda2004signal} for Feynman diagrams
(iii) or to \cite{guhr2010supersymmetry} and references therein for summetry applied to RMT. Again, we emphasize that this presentation is not intended to
be rigorous in a mathematical sense, and we refer to standard RMT
textbooks such as
\cite{akemann2011oxford,anderson2010introduction,terence2012topics,couillet2011random}
for more details.

We begin with the \emph{Coulomb gas analogy} that, loosely speaking, consists in considering the eigenvalues of $\M$ as the positions of fictitious charged particles, repelling each other via a $2$-d Coulomb (logarithmic) potential (see \cite{mehta2004random} for a self-contained introduction or to e.g. \cite{brezin1978planar,dean2006large,dean2008extreme} for concrete applications). We shall highlight in this section the strong link between the potential function and the Stieltjes transform $\stj(z)$ whenever the probability measure over the matrix ensemble is rotationally invariant, i.e. of the form Eq.\ (\ref{eq:Boltz}). 

\subsubsection{Stieltjes transform and potential function}

\begin{figure}[!ht]
	\begin{center}
   \includegraphics[scale = 0.4]{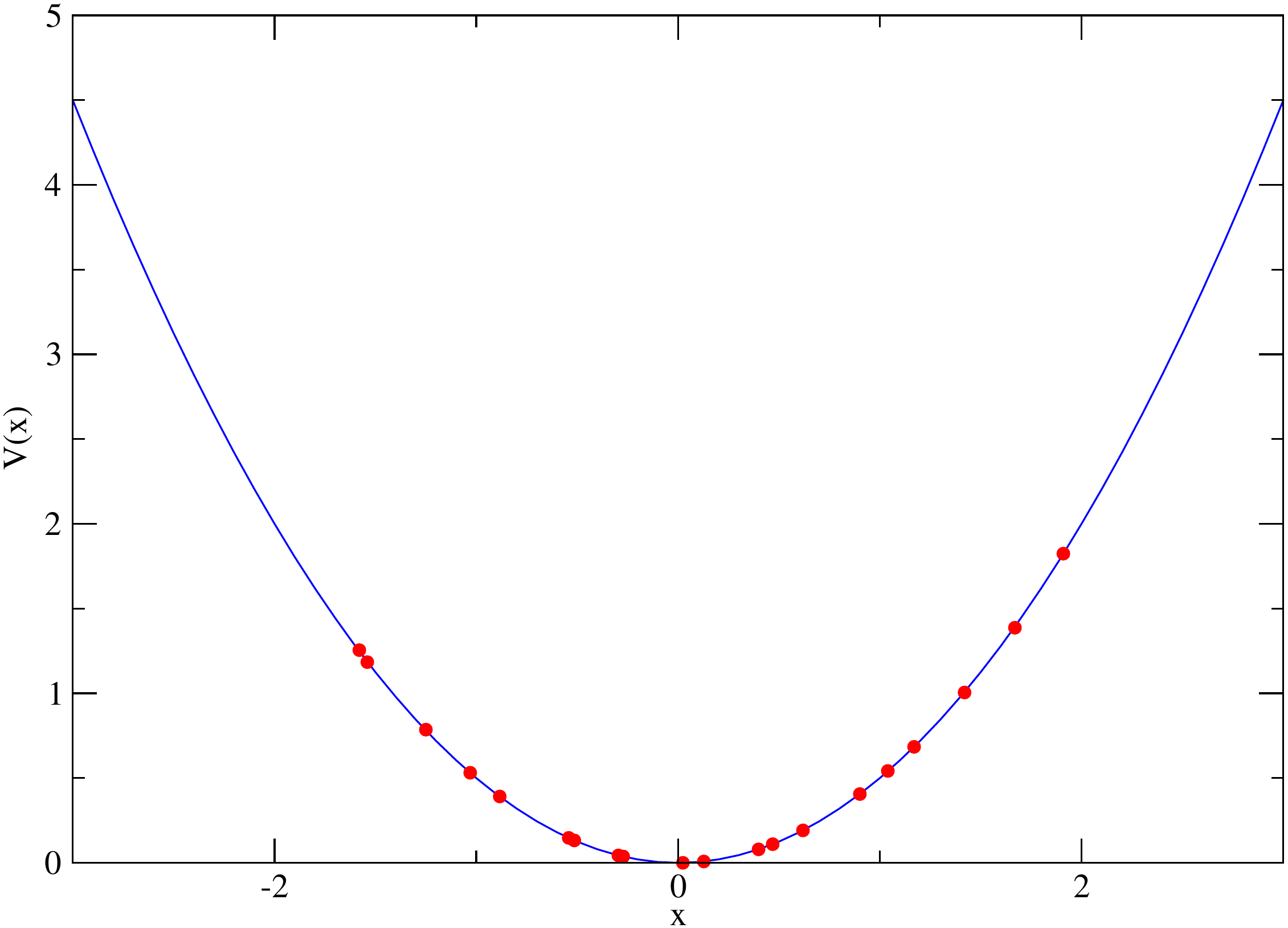} 
   \end{center}
   \caption{Typical configuration of a repulsive Coulomb gas with $N = 20$ particles (red dots) in the potential $V(x) = x^2/2$ as a function of x.}
   \label{potential_quadratic}
\end{figure}

First, we write from \eqref{eq:Boltz} the \emph{partition function} of the model as 
\begin{equation*}
{\cal Z} \propto \int e^{-\frac{\beta N}{2} \Tr V(\M)} {\cal D}\M,
\end{equation*}
and this can be used as a starting point to obtain the LSD -- or rather its Stieltjes transform -- using a saddle point method. This relation has first been obtained in the seminal paper of Br{\'e}zin-Itzykson-Parisi-Zuber \cite{brezin1978planar} and we repeat here the main idea of the derivation (see also \cite[Section 2.1]{zuber2012introduction}). Let us first 
express the partition function in terms of the eigenvalues and eigenvectors of $\M$, using (\ref{eq:Boltz2}):
\begin{equation*}
{\cal Z} \propto \int \pB{\prod_{i=1}^{N} \dd\nu_i} \exp\left\{- N \sum_{i=1}^{N}\left[ V(\nu_i) - \frac{\beta}{2N} \sum_{i \neq j} \log \lvert \nu_i - \nu_j \lvert \right] \right\},
\end{equation*}
up to a constant factor that comes from integrating over the Haar measure $d\Omega$.  It is then customary to introduce the \textit{action} $S(\{\nu_i\}) \equiv S(\nu_1, \nu_2, \dots, \nu_N)$ such that we can rewrite the partition function as:
\begin{equation}
\label{Partition_RMM}
{\cal Z} \propto  \int \prod_{i=1}^{N} d\nu_i e^{- N^2 S(\{\nu_i\}) } \quad\text{with}\quad S(\{\nu_i\}) = \frac1N \sum_{i=1}^{N} V(\nu_i) - \frac{\beta}{2N^2} \sum_{i \neq j} \log \lvert \nu_i - \nu_j \lvert,
\end{equation}
Note that the action is normalized so that its large $N$ limit is of order $1$. The eigenvalues can be seen as a thermal gas of one-dimensional particles in an external potential $V(z)$ and subject to a (logarithmic) ``electrostatic'' repulsive interaction: this is the Coulomb gas analogy. At thermal equilibrium, the eigenvalues typically gather in potential well(s), but cannot accumulate near the minimum due to the repulsive force, which keeps them at distance of order $\cal O(N^{-1})$. For instance, if we take a quadratic potential function $V(x) = x^2/2$, then all the particles tend to gather around zero as it is shown in the Fig. \ref{potential_quadratic}. We recall that we consider only densities which are defined on an unique compact support (\textit{one-cut} assumption) and we thus require that the fictitious particles evolve in a confining convex potential $V(z)$. The class of potential function that we consider is such that its derivative gives a Laurent polynomial, i.e., $V'(z) = \sum_{k} c_k z^k$ with $k$ integers that can be negative. Since we can always rewrite $V'(z) = z^{-\ell} P(z)$, with the ``order'' $\ell$ the lowest (negative) power of $V'(z)$ and $P(z)$ a polynomial, we define by $d$ the ``degree'' of $V'(z)$ which corresponds to the degree of $P(z)$. In particular, if $V'(z)$ is a polynomial, then $\ell = 0$. 

In the large $N$ limit, the integral over eigenvalues can be computed by the saddle-point method which yields the following ``force equilibrium'' condition:\footnote{The reader might wonder why a system in thermal equilibrium ends up being described by simple mechanical equilibrium, as at zero temperature. It turns out that the system is effectively at very low temperatures and that entropy effects are of order $N^{-1}$ compared to interaction effects, see e.g. \cite{dean2008extreme} for a detailed discussion. Entropy effects start playing a role for extended $\beta$ ensembles where $\beta = c/N$ where $c$ is finite, see \cite{allez2012invariant}.}
\begin{equation}
\label{eq:potential_sp}
V'(\nu_i) = \frac{\beta}{N} \sum_{j=1; j\neq i}^{N} \frac{1}{\nu_i - \nu_j}, \quad \forall \, i = 1,\dots,N.
\end{equation}
It seems hopeless to find the eigenvalues $\{\lambda_{i}\}$ that solve these $N$ equations. However, we may expect to find the LSD $\rho_{\M}$ in the limit $N \to \infty$, corresponding to configuration of the eigenvalues that satisfies these saddle-point equations. In the case of the one-cut assumption, the result reads \cite{brezin1978planar}:
\begin{equation}
\label{eq:potential_stieltjes}
\stj(z) = V'(z) - Q(z) \sqrt{ (z - \nu_{+})} \sqrt{(z - \nu_{-})},
\end{equation}
where $\nu_- < \nu_+$ denote the edges of $\supp[\rho]$ and $Q(z)$ is
also a Laurent polynomial with degree $d-1$ and order
$\ell$. Therefore, we see that we have $d+1$ unknowns to determine,
namely the coefficients of $Q(z)$, $\nu_-$ and $\nu_+$ which are
determined using the series expansion \eqref{eq:stieltjes_Moment}. We
shall give a detailed illustration of this procedure in Section
\ref{sec:MP_law} below\footnote{In the case of positive definite
  covariance matrices, we can use the series
  \eqref{eq:MP_equation_mgf_inverse} that corresponds to the limit $z
  \to 0$}.

We observe that as soon as we can characterize the potential function of $V(z)$ that governs the entries of $\M$, we are then able to find the corresponding LSD $\rho_{\M}$. We will show in the rest of this section that this Coulomb gas analogy allows one to retrieve some important laws in RMT.  \\

\begin{changemargin}{0.5cm}{0.5cm} 
\footnotesize
Let us show how to obtain \eqref{eq:potential_stieltjes}. In the following we set $\beta = 1$. First, we introduce the normailzed trace of the resolvent $\stj(z)$ in \eqref{eq:potential_sp} by multiplying on both sides by $N^{-1} (z-\nu_i)^{-1}$ and summing over all $i$, which yields
\begin{equation}
\label{eq:stieltjes_BIPZ_sp}
\frac1N \sum_{i=1}^{N} \frac{V'(\nu_i)}{z-\nu_i} = \frac{1}{N^2} \sum_{i=1}^{N} \sum_{j=1; j\neq i}^{N} \frac{1}{(z-\nu_i)(\nu_i - \nu_j)}.
\end{equation}
Notice that this last equation is indeed an analytical function for $z \in \mathbb{C} \backslash \Supp[\rho_{\M}]$. Then, we rewrite the LHS using  some algebraic manipulations that leads to
\begin{equation*} 
\frac1N \sum_{i=1}^{N} \frac{V'(\nu_i)}{z-\nu_i} = V'(z) \stj (z) - \frac1N \sum_{i=1}^{N} \frac{V'(z) - V'(\nu_i)}{z-\nu_i}, 
\end{equation*}
and for the RHS, we obtain
\begin{equation*} 
\frac{1}{N^2} \sum_{i=1}^{N} \sum_{j=1; j\neq i}^{N} \frac{1}{(z-\nu_i)(\nu_i - \nu_j)} \equiv \frac12 \left[ \stj^2(z) + \frac1N \stj'(z) \right].
\end{equation*}
Regrouping these last two equations into the saddle-point equation \eqref{eq:stieltjes_BIPZ_sp} gives
\begin{equation*}
\frac{1}{2}\left[ \stj^{2}(z) + \frac1N \stj'(z) \right] = V'(z) \stj(z) - \frac1N \sum_{i=1}^{N} \frac{V'(z) - V'(\nu_i)}{z-\nu_i}.
\end{equation*}
Since we are interested in the limit of large $N$, we thus have to solve for $\stj(z)$ the following quadratic equation
\begin{equation}\label{eq:stieltjes_BIPZ_quadratic}
\stj^2(z) - 2 V'(z) \stj(z) + \frac2N \sum_{i=1}^{N} \frac{V'(z) - V'(\nu_i)}{z-\nu_i} = 0.
\end{equation}
The most difficult term is the last one because the sum is not
explicit. For the sake of simplicity, we consider the case where
$V'(z)$ is a polynomial of degree $d > 0$ as the extension to Laurent
polynomial, i.e. polynomial with negative powers, is immediate. For
$V'(z)$ a polynomial function in $z$, we have that
\begin{equation*}
P(z) \deq \frac1N \sum_{i=1}^{N} \frac{V'(z) - V'(\nu_i)}{z-\nu_i}
\end{equation*}
is also a polynomial but with a degree $d-1$ whose coefficients can be
determined later by the normalization constraint, or by matching some
moments. Then, the solution of Eq.\ \eqref{eq:stieltjes_BIPZ_quadratic} is such
that:
\begin{equation*}
% \label{stieltjes_potential}
\stj(z) = V'(z) \pm \sqrt{V'(z)^{2} - 2P(z)}.
\end{equation*}
The nice property in the one-cut framework (i.e., a unique compact support for $\rho$) is that the above expression can be simplified to (when $d \geq 1$):
\begin{equation*}
\stj(z) = V'(z) \pm Q(z) \sqrt{ (z - \nu_{+})(z - \nu_{-})}
\end{equation*}
where $\nu_-$ and $\nu_+$ denote the edges of $\supp[\rho]$ and $Q(z)$ is a polynomial with degree $d-1$ and this gives \eqref{eq:potential_stieltjes}. \\
% It is sometimes more convenient to write the above equation in the following form:
% \begin{equation}
% \label{Stieltjes_potential_onecut_2}
% G(z) = V'(z) \pm Q(z) \sqrt{z^2 - 2az + b^2}
% \end{equation}
% with $a = (\nu_{+} + \nu_{-})/2$ and $b = \nu_{+} \nu_{-}$. 
\end{changemargin} 
\normalsize

\subsubsection{Wigner's semicircle law}
\label{sec:wigner}

\begin{figure}[!ht]
	\begin{center}
   \includegraphics[scale = 0.5]{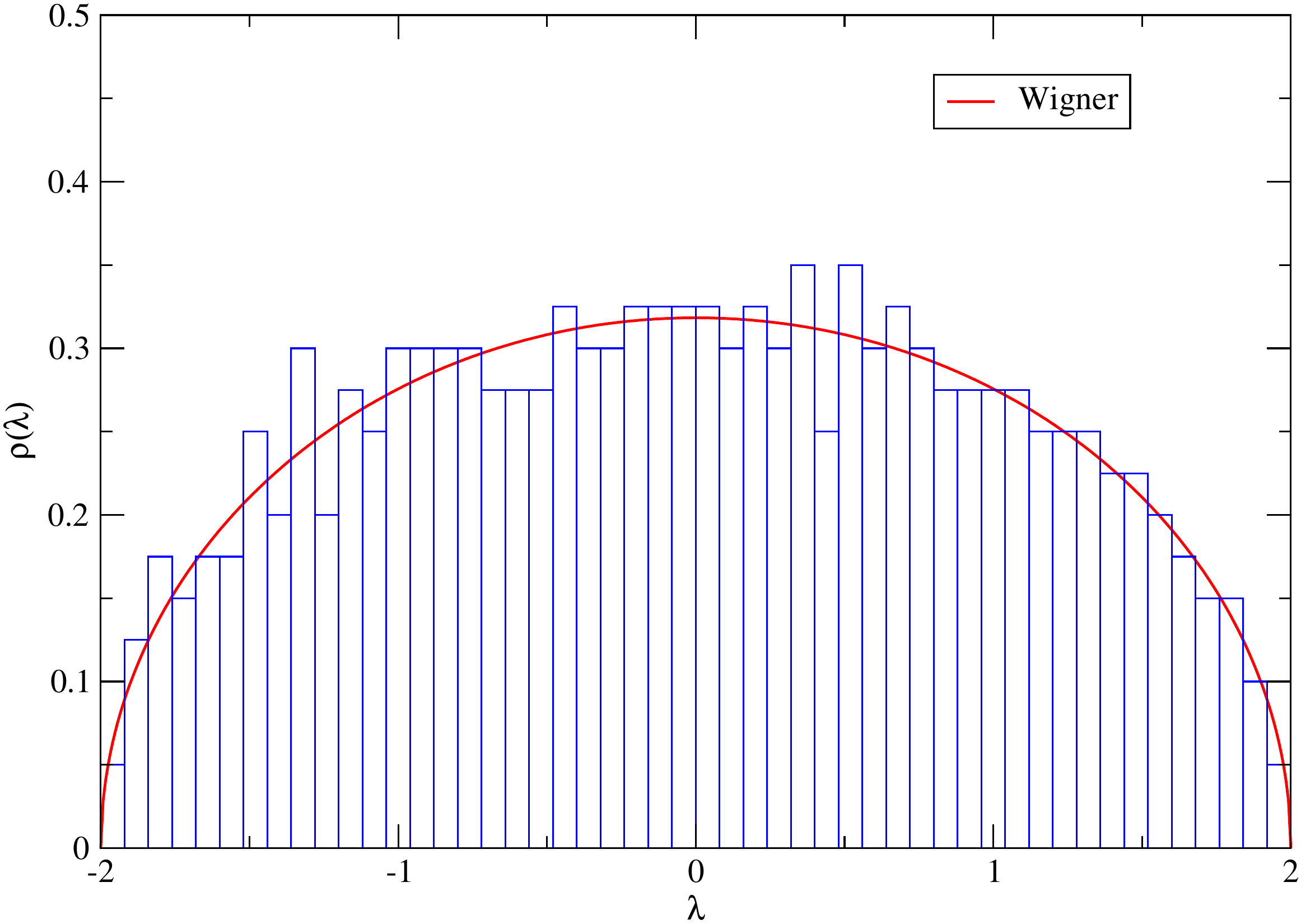} 
   \end{center}
   \caption{Wigner semi-circle density \eqref{density_Wigner} compared with empirical results with $N = 500$ (histogram) from one sample, illustrating the convergence of the ESD at finite $N$ to the asymptotic LSD. }
   \label{chap1_wigner}
\end{figure}

As a warm-up exercise, we begin with Wigner's semi-circle law  \cite{wigner1951statistical}, one of the most important result in RMT. Note that this result has first been obtained in the case of Gaussian matrix with independent and identically distributed entries (while preserving
the symmetry of the matrix). For real entries, we refer to this class of random matrices as the Gaussian Orthogonal Ensemble (GOE). It has been proved, see e.g.\ \cite{anderson2010introduction}, that the semi-circle law can be extended to a broader class of random matrices, known as the \textit{Wigner Ensemble} that deals with a matrix $\M$ with independent and identically distributed entries such that:\footnote{The case where the variance of the 
matrix elements diverge corresponds to {\it L\'evy matrices}, introduced in \cite{cizeau1994theory}. For a rigorous approach, we refer the readers to \cite{arous2008spectrum}. For recent developments, see \cite{tarquini2016level} .}
\begin{equation}
	\label{eq:def_Wigner_matrix}
\mathbb{E} \qb{\M_{ij}} = 0, \quad\text{and}\quad \mathbb{E} \qb{\M_{ij}^2}  = \sigma^2/N.
\end{equation}
Let us consider here the specific case of a GOE matrix. For Gaussian entries, it is not hard to see that the associated probability measure ${\cal P}_{\beta}(\M)$ is indeed of the Boltzmann type with a potential function $V(\M) = \M^2/2\sigma^2$. From Eq.\ (\ref{eq:potential_stieltjes}), we remark that the unknown polynomial $Q(z)$ is simply a constant because the derivative of the potential has degree $d=1$. To determine this constant, we enforce the property (ii) of the Riemann-Hilbert problem which enable us to get by identification: $Q(z) = 1$, $\nu_{\pm} = \pm 2 \sigma$. We thus finally obtain: 
\begin{equation}
\label{Stieltjes_Wigner}
\stj_{W}(z) = \frac{z - \sqrt{z + 2 \sigma}\sqrt{z - 2 \sigma}}{2\sigma^2},
\end{equation}
where $\sqrt{\cdot}$ denotes throughout the following the principal square root, that is the non-negative square root of a non-negative real number. Equation \eqref{Stieltjes_Wigner} is indeed the Stieltjes transform of Wigner's semi-circle law. Note that it is frequent to see the above result written as
\begin{equation*}
\stj_{W}(z) = \frac{z \pm \sqrt{z^2 - 4 \sigma^2} }{2\sigma^2},
\end{equation*}
where the convention ``$\pm$'' refers to the fact that we have to chose the correct sign such that $\stj(z) \sim z^{-1}$ for large $|z|$ (property (ii) of the Riemann-Hilbert problem). The density function is then retrieved using the 
inversion formula \eqref{eq:stieltjes_inversion} that yields the celebrated \emph{Wigner's semicircle law}:
\begin{equation}
\label{density_Wigner}
\rho_{W}(x) = \frac{1}{2\pi \sigma^2}\sqrt{4\sigma^2 - x^2}, \, \qquad |x| < 2\sigma.
\end{equation}
We plot in Fig \ref{chap1_wigner} the density of the semi-circle and compared with the ESD obtained from a GOE matrix of size $N = 500$. As stated at the beginning of this section, we see that the limiting density agrees well with the ESD of the large but finite size matrix. In fact, one can rigorously estimate the expected difference between the ESD at finite $N$ and the asymptotic LSD for $N = \infty$, which vanishes as $N^{-1/4}$ as soon as the 
$\M_{ij}$'s have a finite fourth moment, and as $N^{-2/5}$ if all the moments of the $\M_{ij}$ are finite (see \cite{bai1993convergence}).

Due to the relative simplicity of the expression of Eq.\ \eqref{Stieltjes_Wigner}, one can easily invert this expression to find the Blue transform to find that the $\rtr$-transform of the semicircle law reads
\begin{equation}
	\label{eq:R_transform_Wigner}
		\rtr_{W}(z) = \sigma^2 z.
\end{equation}
Since the average trace $\varphi_1$ is exactly 0, the $\str$-transform of a Wigner matrix is an ill-defined object. 

\subsubsection{The Mar{\v c}enko-Pastur law}
\label{sec:MP_law}

As stated in the introduction, the study of random matrices began with
John Wishart \cite{wishart1928generalised}. More precisely, let us
consider the $N \times T$ matrix $\b Y$ consisting of $T$ independent
realizations of random centered Gaussian vectors of size $N$ and
covariance $\C$, then the Wishart matrix is defined as the $N \times
N$ matrix $\M$ as $\M \deq T^{-1} \Y \Y^{*}$. In multivariate
statistics, this matrix $\M$ is better known as the sample covariance
matrix (see Chapter \ref{chap:spectrum}). For \emph{any} $N$ and $T >
N$, Wishart derived the exact PDF of the entries $\M$ which reads:
\begin{equation}
\label{eq:wishart_distribution}
{\cal P}_{\text{w}}(\M|\C) = \frac{1}{2^{NT/2}  \Gamma_{N}(T/2)} \frac{ \det(\M)^{\frac{T-N-1}{2}}}{\det(\C)^{T/2}} e^{-\frac{T}{2} \Tr \C^{-1} \M}.
\end{equation}
As alluded in the introduction, we say that $\M$ (given $\C$) follows a Wishart$(N,T, \C/T)$ distribution. In the ``isotropic'' case, i.e., when $\C = \b I_N $, we can deduce from \eqref{eq:wishart_distribution} 
\begin{equation}
\label{eq:wishart_distribution_isotropic}
{\cal P}_{\text{w}}(\M|\b I_N) \propto \det(\M)^{\frac{T-N-1}{2}} e^{-\frac{T}{2} \Tr \M} := e^{-\frac{T}{2} \Tr \M + \frac{T-N-1}{2} \Tr \log \M},
\end{equation}
which clearly belongs to the class of Boltzmann ensembles \eqref{eq:Boltz}. Throughout the following, we shall denote by $\Wishart$ the $N \times N$ matrix whose distribution is given by \eqref{eq:wishart_distribution_isotropic}. Ignoring sub-leading terms, the corresponding potential function is given by:
\begin{equation}
V(z) = \frac{1}{2q} \left[ z - (1-q) \log z \right], \qquad\text{with}\qquad q := N/T.
\end{equation}
It is easy to see that the derivative indeed gives a Laurent polynomial in $z$ as we have
\begin{equation*}
V'(z) = \frac{1}{2qz}\left[ z - (1-q) \right].
\end{equation*}
Following our convention, $V'(z)$ is a Laurent polynomial of degree $1$ and order $\ell = -1$ so that we deduce $Q(z)$ in \eqref{eq:stieltjes_BIPZ_sp} is of the form $c/z$ with $c$ a constant to be determined using \eqref{eq:stieltjes_Moment}. We postpone the computation of the Stieltjes transform $\stj(z)$ to the end of this section. The final result reads:\nc
\begin{equation}
\label{eq:stieltjes_isotropic_wishart}
%\stj(z) = \frac{ (z+q-1) \pm \sqrt{ (z-1)^2 - 2q(z+1) + q^2} }{2qz},
\stj(z) = \frac{ (z+q-1) - \sqrt{z-\nu_-}\sqrt{z-\nu_+} }{2qz}, \qquad \nu_\pm\;\deq\; (1\pm\sqrt{q})^2,
\end{equation}
and this is the solution found by Mar{\v c}enko and Pastur in \cite{marchenko1967distribution} in the special case $\C = \b I_N$. We can now use the inversion formula \eqref{eq:stieltjes_inversion} to find the celebrated Mar{\v c}enko-Pastur (MP) law (for $q \in (0,1)$)
\begin{equation}
\label{eq:MP_density}
\rho_{\text{MP}}(\nu) = \frac{\sqrt{ 4\nu q - (\nu +q - 1)^2}}{2q\pi \nu} , \qquad \forall \, \nu\in \qb{\nu_-, \nu_+} .
\end{equation}
Note that for $q\geq 1$, it is plain to see that $\M$ has $N-T$ zero
eigenvalues that contribute $(1-q)\delta_0$ to the density
Eq.\ \eqref{eq:MP_density}. Note that the convergence of the ESD
towards the asymptotic MP law occurs, for $q < 1$, at the same speed
as in the Wigner case, i.e. as $N^{-2/5}$ in the present case where
the random elements of $\b Y$ are Gaussian (for a full discussion of
this issue, see \cite{bai2003convergence}).

Again, the expression of $\stj(z)$ is simple enough to obtain a closed formula for the Blue transform, and deduce from Eq.\ \eqref{eq:stieltjes_isotropic_wishart} the $\rtr$-transform of the MP law:
\begin{equation}
	\label{eq:R_transform_MP}
		\rtr_{\text{MP}}(\omega) = \frac{1}{1-q\omega}.
\end{equation}
One can compute the $\str$-transform of the MP law using the relation \eqref{eq:r_s_transform}: 
\begin{equation}
	\label{eq:S_transform_MP}
		\str_{\text{MP}}(\omega) = \frac{1}{1+q\omega}.
\end{equation}

\begin{changemargin}{1.0cm}{1.0cm} 
\footnotesize
We now derive the Stieltjes transform \eqref{eq:stieltjes_isotropic_wishart} through a complete application of the BIPZ formalism introduced in Eq.\ \eqref{eq:stieltjes_BIPZ_sp}. As alluded to above, the Stieltjes transform \eqref{eq:stieltjes_BIPZ_sp} for the isotropic Wishart matrix has the form
\begin{equation}
	\label{eq:stieltjes_isotropic_wishart_tmp}
	\stj(z) = \frac{1}{2q} \qBB{1 - \frac{1-q}{z}} - \frac{c}{z} \sqrt{z-\nu_+}\sqrt{z-\nu_-},
\end{equation}
and the constants that we have to determine are $c, \nu_+$ and $\nu_-$. To that end, we use \eqref{eq:stieltjes_Moment} that tells us that when $\abs{z} \to \infty$
\begin{equation}
	\label{eq:stieltjes_z_infty_tmp}
	\stj(z) = \frac{1}{z} + \frac{\varphi(\M)}{z^2} + \cal O(z^{-3}).
\end{equation}
On the other hand, one finds by taking the limit $z \to \infty$ into \eqref{eq:stieltjes_isotropic_wishart_tmp} that
\begin{eqnarray}
	\stj(z) = \frac{1}{2q} \qBB{1 - \frac{1-q}{z}} - c\qBB{1 - \frac{\nu_+ + \nu_-}{2z} - \frac{(\nu_+ - \nu_-)^2}{8z^2}} + \cal O(z^{-3}),
\end{eqnarray}
Then, by comparing this last equation to \eqref{eq:stieltjes_z_infty_tmp}, we may fix $c$ by noticing that we have a leading order
\begin{equation*}
	\frac{1}{2q} - c = 0,
\end{equation*}
since $\stj(z)$ behave as $\cal O(z^{-1})$ for very large $z$ and therefore we have
\begin{equation}
	\label{eq:MP_alpha_cst}
	c = \frac{1}{2q}. 
\end{equation}
Next, we find at order $\cal O(z^{-1})$:
\begin{equation}
	1 = - \frac{(1-q)}{2q} + \frac{\nu_+ + \nu_-}{4q},
\end{equation}
that is to say
\begin{equation}
	\label{eq:MP_nu_plus}
	\nu_+ = 2(1+q) - \nu_-.
\end{equation}
Finally, the last constant is determined with the condition at order $\cal O(z^{-2})$, 
\begin{equation}
	\varphi(\M) = \frac{(\nu_+ - \nu_-)^2}{16q}, 
\end{equation}
which is equivalent to
\begin{equation}
	\label{eq:MP_nu_minus}
	\nu_- = \nu_+ - 4\sqrt{q\varphi(\M)} = (1+q) - 2\sqrt{q} = (1-\sqrt{q})^2,
\end{equation}
where we used \eqref{eq:MP_nu_plus} and $\varphi(\M) = 1$ in the third step. Consequently, we deduce from \eqref{eq:MP_nu_plus} that $\nu_+ = (1+\sqrt{q})^2$ and the result \eqref{eq:stieltjes_isotropic_wishart} follows from the equations \eqref{eq:MP_alpha_cst}, \eqref{eq:MP_nu_plus} and \eqref{eq:MP_nu_minus}. \nc

\end{changemargin} 
\normalsize

\subsubsection{Inverse Wishart matrix}
\label{sec:IMP_law}
\begin{figure}[!ht]
	\begin{center}
   \includegraphics[scale = 0.5]{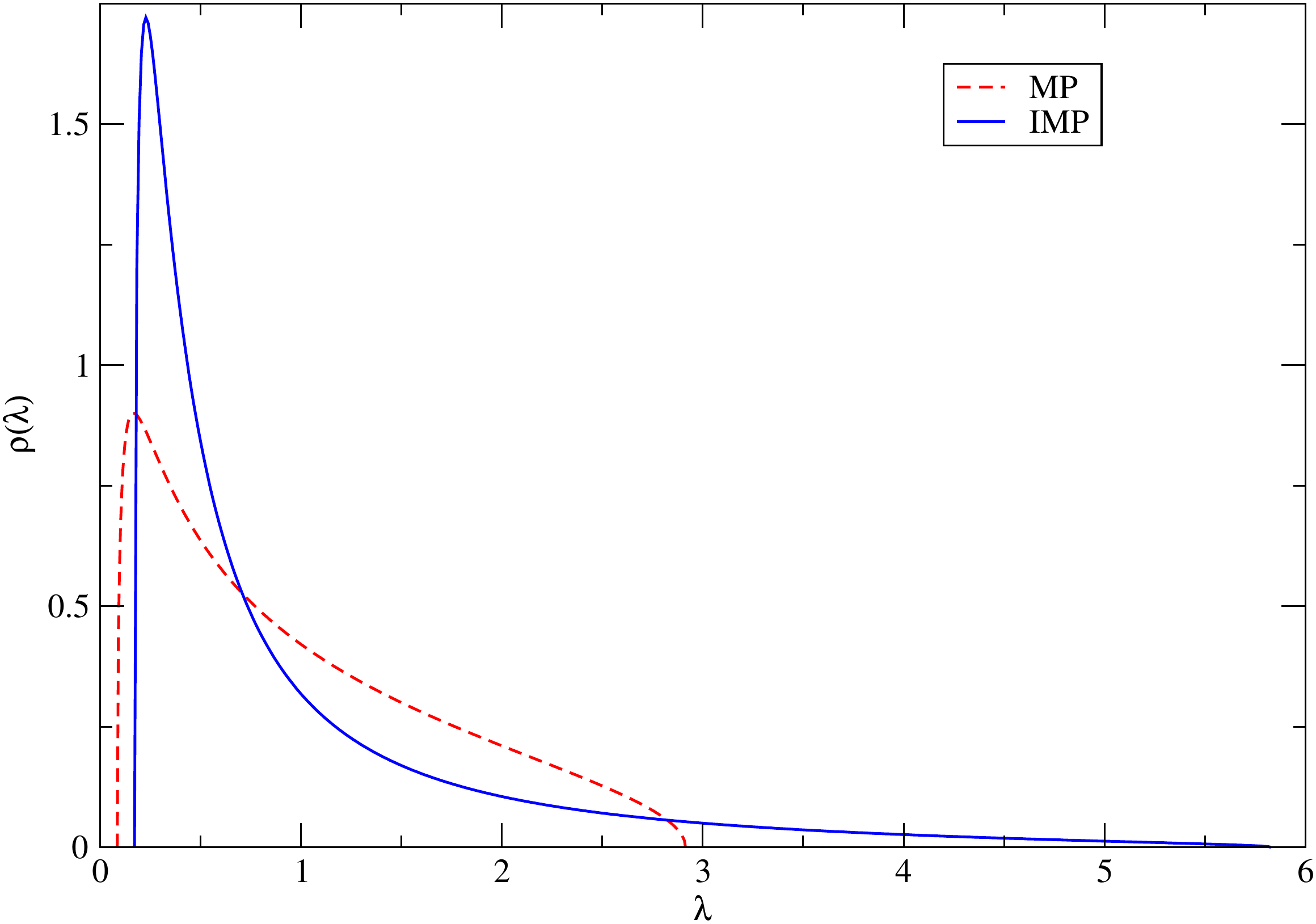} 
   \end{center}
   \caption{The red dotted curve corresponds to the Mar{\v c}enko-Pastur density \eqref{eq:MP_density} with $q=0.5$. We repeat the experiment with the Inverse Wishart matrix still with $q=0.5$ (plain blue curve).}
   \label{chap1_MP_IMP}
\end{figure}

%{\bf Inverse Mar{\v c}enko-Pastur density}. 
Another very interesting case is the inverse of a Wishart matrix, simply named the ``inverse Wishart'' matrix. The derivation of the corresponding eigenvalue density is straightforward from the Mar{\v c}enko-Pastur law \eqref{eq:MP_density}. Indeed, one just needs to make the change of variable $u = ((1-q)\nu)^{-1}$ into Eq.\ \eqref{eq:MP_density} to obtain:\footnote{The factor $(1-q)^{-1}$ is introduced to keep the mean at one as will be explained below.}
% One can again deduce the potential function and its derivative:
% \begin{equation}
% 	\label{eq:invw_potential}
% V'(z) = \frac{1}{2qz^2} \left[ (1+q) z + 1 \right].
% \end{equation}
% Proceeding like in the Wishart case, one can compute exactly the Stieltjes transform 
\begin{equation}
\label{eq:IMP_density}
\rho_{\text{IMP}}(u) = \frac{\kappa}{\pi u^2} \sqrt{ (u_+ - u)(u - u_-)}, \qquad u_{\pm} \;\deq\; \frac{1}{\kappa} \left[ \kappa + 1 \pm \sqrt{2\kappa + 1} \right],
\end{equation}
where the subscript $\text{IMP}$ stands for ``Inverse Mar{\v c}enko-Pastur'' and $\kappa$ is related to $q$ through
\begin{equation}
\label{eq:kappa_invWishart}
q = \frac{1}{2\kappa + 1} \in (0,1)\,.
\end{equation}
In particular, one notices that $u_{\pm} = (1-q)/\nu_{\mp}$ where
$\nu_{\mp}$ is defined in
Eq.\ \eqref{eq:stieltjes_isotropic_wishart}. We plot in
Fig.\ \ref{chap1_MP_IMP} the density of the Mar{\v c}enko-Pastur
\eqref{eq:MP_density} and of its inverse \eqref{eq:IMP_density} both
with parameter $q=0.5$.

In addition to the eigenvalue density \eqref{eq:IMP_density}, one can also derive explicit expressions for the other transforms presented in Section \ref{section:RMT_transforms}. For the Stieltjes transform, it suffices to apply the same change of variable $u = ((1-q)z)^{-1}$ and to use the properties \eqref{eq:stieltjes_scale_prop} and \eqref{eq:stieltjes_M_Minv} to obtain:
% One can again deduce the potential function and its derivative:
% \begin{equation}
% 	\label{eq:invw_potential}
% V'(z) = \frac{1}{2qz^2} \left[ (1+q) z + 1 \right].
% \end{equation}
% Proceeding like in the Wishart case, one can compute exactly the Stieltjes transform 
\begin{equation}
\label{eq:stieltjes_invWishart}
\stj_{\text{iw}}(u) = \frac{ u(\kappa + 1) - \kappa - \kappa \sqrt{ u-u_-} \sqrt{u-u_+}}{u^2}\,,
\end{equation}
where the bounds $u_{\pm}$ are given in Eq.\ \eqref{eq:IMP_density}. One can easily check with the inversion formula \eqref{eq:stieltjes_inversion_formula} that we indeed retrieve the density of states \eqref{eq:IMP_density} as expected. \nc
%  of the Inverse Wishart matrices:
% \begin{equation}
% \label{eq:IMP_density}
% \rho_{\text{IMP}}(\nu) = \frac{\kappa}{\pi \nu^2} \sqrt{ (\nu_+ - \nu)(\nu - \nu_-)}, \qquad \forall \; \nu \in [\nu_-, \nu_+],
% \end{equation}
% where the subscript $\text{IMP}$ stands for ``Inverse Mar{\v c}enko-Pastur''.   
% %The reason why we introduce the inverse Wishart matrix is because this family plays an important role in Bayesian statistics. We will see that most of the results concerning this family of matrices can be retrieved with a RMT framework (see Section \ref{??}). 

Using the Stieltjes transform \eqref{eq:stieltjes_invWishart}, one can then compute the $\rtr$-transform of the Inverse Mar{\v c}enko-Pastur density to find 
\begin{equation}
	\label{eq:R_transform_IMP}
		\rtr_{\text{IMP}}(\omega) = \frac{\kappa - \sqrt{\kappa(\kappa - 2\omega)}}{\omega}, \qquad \kappa > 0,
\end{equation}
and then, from \eqref{eq:r_s_transform}, the $\str$-transform reads
\begin{equation}
	\label{eq:S_transform_IMP}
		\str_{\text{IMP}}(\omega) = 1 - \frac{\omega}{2\kappa}. 
\end{equation}

% Finally, the inversion formula \eqref{eq:stieltjes_inversion_formula} gives the density of states of the Inverse Wishart matrices:
% \begin{equation}
% \label{eq:IMP_density}
% \rho_{\text{IMP}}(\nu) = \frac{\kappa}{\pi \nu^2} \sqrt{ (\nu_+ - \nu)(\nu - \nu_-)}, \qquad \forall \; \nu \in [\nu_-, \nu_+],
% \end{equation}

%The reason why we introduce the inverse Wishart matrix is because this family plays an important role in Bayesian statistics. We will see that most of the results concerning this family of matrices can be retrieved with a RMT framework (see Section \ref{??}). 
%We plot in Fig.\ \ref{chap1_MP_IMP} the density of the Mar{\v c}enko-Pastur \eqref{eq:MP_density} and of its inverse \eqref{eq:IMP_density} both with parameter $q=0.5$. \nc

In statistics, the derivation of the inverse Wishart distribution is slightly different.  Let $\M$ be a $N \times N$ real symmetric matrix that we assume to be invertible and suppose that $\M^{-1}$ follows a Wishart$(N,T, \C^{-1})$  and $\C$ is a $N \times N$ real symmetric positive definite ``reference'' matrix and $T > N - 1$. In that case, it turns out that the PDF of $\M$ is also explicit. More precisely, we say that $\M$ is distributed according to an Inverse-Wishart$(N,T, \C)$ whose PDF is given by \cite{anderson1984introduction}:
\begin{equation}
	\label{eq:inverse_wishart_distribution}
	\cal P_{\text{iw}}(\M^{-1} |\C) \;=\; \frac{1}{2^{NT/2} \Gamma_N(T/2)} \frac{\det(\C)^{T/2}}{\det(\M)^{(T+N+1)/2)}} e^{-\frac{1}{2} \tr \C \M^{-1}}\,.
\end{equation}
In order to get that distribution, one should note that the Jacobian of the transformation $\M \to \M^{-1}$ is equal to $(\det \M)^{-N-1}$, as can be derived by using the eigenvalue/eigenvector representation of the measure, see Eq. (\ref{eq:Boltz2}). A detailed derivation of this change of variable may be found e.g.\ in \cite[Eq.\ (15.15)]{dwyer1967some}. 

An important property of the Inverse-Wishart distribution is the following closed formula for the expectation value:
\begin{equation}
	\label{eq:inverse_wishart_mean}
	\avgb{\M}_{\cal P_{\text{iw}}} = \frac{\C}{T - N -1}.
\end{equation}
The derivation of this result can be obtained using the different identities of \cite{haff1979identity}.

We may now explain the factor $(1-q)$ in the above change of variable. If we consider $\C = \In/T$, we deduce from \eqref{eq:inverse_wishart_mean} that
\begin{equation}
	\avgb{\M}_{\cal P_{\text{iw}}} = \frac{T}{T-N-1} \In \underset{\text{LDL} }{\sim} \frac{1}{1-q} \In.
\end{equation}
In order to have a normalized spectral density, i.e. $N^{-1} \Tr\M = 1$, we see that we need to apply $\tilde\M = (1-q)\M$ so that $\avg{\tilde \M}=\In$. This was exactly the purpose of the change of variable $u = ((1-q)\nu)^{-1}$ in Eq.\ \eqref{eq:IMP_density}. 
%Roughly, We say that $\M$ is an inverse Wishart matrix if and only if $\M^{-1}$ is a Wishart matrix defined as above (see Appendix \ref{app:invW} for the exact definition and parametrization). Note that in order to have that $\M^{-1}$ is invertible, the parameter of $q\deq N/T$ of Wishart matrix must obey $q < 1$. 

We conclude this section by stating that one can characterize entirely the eigenvalue density function of a broad class of random matrices $\M$ through a potential function. This allows one to reproduce a 
large variety of empirical spectral densities by adequately choosing the convex confining potential.

\subsection{Free probability}
\label{sec:free_probability}

We saw in the previous two examples that one can derive, from the potential function, some analytical results about the ESD which can be very interesting for statistical purposes ($\eg$ the inverse Wishart density). However, the Coulomb gas method does not allow one to investigate the spectrum of a matrix that is perturbed by some noise source. This is a classical problem in Statistics where one is often interested in extracting the ``true'' signal from noisy observations. Standard models in statistics deal with either an additive or multiplicative noise (as will be the case for empirical correlation matrices). Unless one can write down exactly the PDF of the entries of the corrupted matrix, which is rarely the case, the Coulomb gas analogy is not directly useful.

This section is dedicated to a short introduction to free probability theory, which is an alternative method to study the asymptotic behavior of some large dimensional random matrices. More precisely, free probability provides a robust way to investigate the LSD of either sums or products of random matrices with specific symmetry properties. We will only give here the basic notions of free probability applied to symmetric real random matrices and we refer 
to e.g.\  \cite{speicher2009free} or \cite{burda2013free} for a more exhaustive presentation. 

\subsubsection{Freeness}
\label{subsubsec:freeness}

Free probability theory was initiated in 1985 by Dan Voiculescu in order to understand special classes of von Neumann algebras \cite{voiculescu1985symmetries}, by establishing calculus rules for non commutative operators relying on the notion of \textbf{freeness}, defined below for the special case of matrices. A few years later,  Voiculescu \cite{voiculescu1992free} and Speicher \cite{speicher1994multiplicative}  found that rotationally invariant random matrices asymptotically satisfy the freeness criteria, and this has had a tremendous impact on RMT. 

Roughly speaking, two matrices $\AAA$ and $\BBB$ are mutually \textit{free} if their eigenbasis are related to one another by a random rotation, \ie~ when their eigenvectors are almost surely orthogonal. For random matrices, we rather use the notion of ``asymptotic'' freeness. The precise statement is as follows \cite{voiculescu1992free}: let $\AAA$ and $\BBB$ be two independent self-adjoint matrices of size $N$. If the spectral density of each matrix converges almost surely in the large $N$ limit and if $\BBB$ is invariant under rotation, then $\AAA$ and $\BBB$ are asymptotically free. This statement can also be found in a different context in \cite{speicher1994multiplicative}. 

The notion of freeness for random matrices is the counterpart of independence for random variables. Indeed, recall that the normalized trace operator, defined as
\begin{equation}
\label{eq:trace_matrix}
\varphi(\M) := \frac1N \Tr \M,
\end{equation}
is equal to the first moment of $\rho_\M$. Then, provided that $\varphi(\AAA) = \varphi(\BBB) = 0$, we say that $\AAA$ and $\BBB$ are free if the so-called \emph{freeness} property is satisfied, to wit:
\begin{equation}
	\label{eq:freeness}
		\varphi\left( \AAA^{n_1} \BBB^{m_1}\AAA^{n_2} \BBB^{m_2}\dots\AAA^{n_k} \BBB^{m_k}  \right) = \varphi (\AAA^{n_1}) \varphi (\BBB^{m_1}) \varphi (\AAA^{n_2}) \varphi (\BBB^{m_2}) \dots \varphi (\AAA^{n_k}) \varphi (\BBB^{m_k}), 
\end{equation}
for any integers $n_{1}, \dots, n_{k}$ and $m_{1}, \dots, m_{k}$ with $k \in \N^+$. Note that if $\varphi(\AAA) \neq 0$ and $\varphi(\BBB) \neq 0$, then it suffices to consider the centered matrices $\AAA - \varphi(\AAA) \b I_N$ and $\BBB - \varphi(\BBB) \b I_N$. 

Let us explore \eqref{eq:freeness} in the simplest case. For any free matrices $\AAA$ and $\BBB$ defined as above, one has 
\begin{equation}
\label{asymptotic_freeness}
\varphi\left( (\AAA - \varphi(\AAA))(\BBB - \varphi(\BBB)) \right) = 0,
\end{equation}
from which we deduce $\varphi \left( \AAA \BBB  \right) = \varphi (\AAA) \varphi (\BBB)$. Hence, if one thinks of the trace operator \eqref{eq:trace_matrix} as the analogue of the expectation value for non commutative random variables, the freeness property is the analogue of the moment factorization property. More generally, freeness allows the computation of mixed moments of products of matrices from the knowledge of the moments of $\AAA$ and $\BBB$, similar to classical independence in probability theory. For example, from 
\begin{equation}
	\label{eq:mixed_moment1}
	\varphi\left( (\AAA - \varphi (\AAA))( \BBB - \varphi (\BBB))(\AAA - \varphi (\AAA))\right) = 0,
\end{equation}
we can deduce that
\begin{equation}
\varphi\left( \AAA \BBB \AAA \right) =  \varphi(\AAA^2 \BBB) = \varphi(\AAA^2) \varphi(\BBB).
\end{equation}

One typical example of free pairs of matrices is when $\AAA$ is a fixed 
matrix and when $\BBB$ is a random matrix belonging to a rotationally invariant ensemble, i.e. $\BBB = \Omega \BBB_{\text{diag}} \Omega^{*}$, where $\BBB_{\text{diag}}$ is diagonal and $\Omega$ distributed according to the Haar (flat) measure over the orthogonal group, in the limit where $N$ is infinitely large. This concept of asymptotic freeness is also related to the notion of vanishing non-planar diagrams \cite{hooft1974planar}. As we shall see in Chapter \ref{chap:application}, the computation of mixed moments will be used to derive some useful relations for estimating over-fitting for statistical estimation problems.

\subsubsection{Sums of free matrices}
\label{sec:freeadd}

In addition to the computation of mixed moments such as
Eq.\ \eqref{eq:mixed_moment1}, free probability theory allows us to
compute the LSD of sums and products of invariant random
matrices, as we discuss now.

Let us look at the additive case first. Suppose that we observe a matrix $\M$ which is built from the addition of a fixed ``signal'' matrix $\AAA$ and a noisy (or random) matrix $\BBB$ that we assume to be invariant under rotation, i.e.,
\begin{equation*}
\M = \AAA + \b\Omega \BBB \b\Omega^{*},
\end{equation*}
for any $N \times N$ matrix $\b\Omega$ that belongs to the orthogonal group $\b O(N)$. A typical question is to evaluate the LSD of $\M$ and estimate the effect of the noise on the signal in terms of the modification of its eigenvalues. Assuming that the ESD of $\AAA$ and $\BBB$ converge to a well defined limit, the spectral density of $\M$ can be computed using the law of addition for non commutative operators, namely Voiculescu's \emph{free addition}
\begin{equation}
\label{eq:free_add_formula}
%\rtr_{\AAA \bxp \BBB}(\omega) = 
\rtr_{\M}(\omega) = \rtr_{\AAA}(\omega) + \rtr_{\BBB}(\omega).
\end{equation}
Hence, we can interpret the $\rtr$-transform \eqref{eq:r_transform} as
the analogue in RMT of the logarithm of the Fourier transform for
standard additive convolution. It is possible to rewrite
Eq.\ \eqref{eq:free_add_formula} as a function of the Stieltjes
transform of $\M$ that contains all the information about the spectral
density of $\M$. Equation \eqref{eq:free_add_formula} is equivalent to
\begin{equation*}
\btr_{\M}(\omega) = \btr_{\AAA}(\omega) + \rtr_{\BBB}(\omega).
\end{equation*}
Next, we introduce $\omega = \stj_{\M}(z)$ that yields
\begin{equation*}
\btr_{\AAA}(\stj_{\M}(z)) = z - \rtr_{\BBB}(\stj_{\M}(z)).
\end{equation*}
It now suffices to apply the function $\stj_{\AAA}$ on both sides to obtain
\begin{equation}
\label{eq:stieltjes_free_addition}
\stj_{\M}(z) = \stj_{\AAA}(z - \rtr_{\BBB}(\stj_{\M}(z))).
\end{equation}
This last relation establishes the influence of the additive noise coming from the matrix $\BBB$ on the ``signal'' (or true) eigenvalues of $\AAA$. 

To gain more insight on this result, let us assume that the noise matrix $\BBB$ is a simple GOE matrix with centered elements of variance $\sigma^2$/N. 
We know from Eq.\ \eqref{eq:R_transform_Wigner} that $\rtr_{\BBB}(z) = \sigma_{\BBB}^2 z$. Hence, the spectrum of the sample matrix $\M$ is characterized by the following fixed-point equation:\footnote{This equation can also be interpreted as the solution of a Burgers equation, that appears within the Dyson Brownian 
motion interpretation of the same problem -- see Appendix \ref{app:addition} for more about this.}
\begin{equation}
\stj_{\M}(z) = \stj_{\AAA}(z - \sigma_{\BBB}^2 \stj_{\M}(z)).
\end{equation}
This is the Stieltjes transform of the deformed GOE matrix\footnote{This result can be generalized to the class of deformed Wigner matrices, i.e. where the noise is given by \eqref{eq:def_Wigner_matrix} but not necessarily Gaussian, see e.g. \cite{khorunzhy1994eigenvalue}.} which is a well-known model in statistical physics of disordered systems. Indeed, this model can be seen as a Hamiltonian that consists of a fixed source subject to an external additive perturbation $\BBB$ \cite{brezin1995universalb}. Taking $\AAA$ to be a GOE as well, we find that $\M$ is a GOE with variance $\sigma_{\AAA}^2 + \sigma_{\BBB}^2$, as expected. In a inference theory context, this model might be useful to describe general linear model where the signal we try to infer is corrupted by an additive noise. 

Another interesting application is when the matrix $\BBB$ has low rank, frequently named a \emph{factor model}. In the example of stocks market, this model can be translated into the fact that there exist few common factors to all stocks such as global news about the economy for instance. For the sake of simplicity, we consider the rank-$1$ case but the following argument can be easily generalized to a finite rank $r \ll N$. Let us denote the unique nontrivial eigenvalue of $\BBB$ as $\beta > 0$ and ask ourselves how adding a (randomly oriented) rank-1 matrix affects the spectrum of $\M$. This problem can be solved explicitly using free matrix tools in the LDL. Indeed, as we show below, the largest eigenvalue pops out of the spectrum of $\AAA$ whenever there exists $z \in \mathbb{R}\backslash \supp[\rho_\AAA]$ such that
\begin{equation}
	\label{eq:outlier_condition_factor_model}
	\stj_\AAA(z) = \frac1\beta.
\end{equation}
For instance, if $\AAA$ is a Wigner matrix with variance $\sigma^2 > 0$, one can easily check from \eqref{eq:outlier_condition_factor_model} and \eqref{eq:R_transform_Wigner} that the largest eigenvalue $\nu_1$ of $\M$ is given by
\begin{equation}
	\nu_1 = 
	\begin{cases}
		\beta + \sigma^2/\beta & \text{if } \beta > \sigma \\
		2\sigma & \text{otherwise}\,.
	\end{cases}
\end{equation}
When $\beta > \sigma$, we say that $\nu_1$ is an \emph{outlier}, i.e. it lies outside the spectrum of $\rho_\AAA$. Hence, we see that free probability allows one to find a simple criterion for the possible presence of outliers. 

\begin{changemargin}{1cm}{1cm}
\footnotesize
Let us now derive the criterion \eqref{eq:outlier_condition_factor_model}. First we need to compute the $\rtr$-transform of the rank one matrix $\BBB$ in order to use \eqref{eq:free_add_formula}. From \eqref{eq:stieltjes}, we easily find that
\begin{equation}
	\stj_\BBB(u) = \frac{1}{N} \frac{1}{u-\beta} + \pBB{1-\frac1N} \frac1u = \frac{1}{u}\qBB{1+\frac1N\frac{\beta}{1-u^{-1} \beta} }.
\end{equation}
Using perturbation theory, we can invert this last equation to find the Blue transform, and this yields at leading order,
\begin{equation}
	\btr_\BBB(\omega) = \frac{1}{\omega} + \frac{\beta}{N(1-\omega\beta)} + \cal O(N^{-2}).
\end{equation}
We may therefore conclude from \eqref{eq:r_transform} that
\begin{equation}
	\rtr_\BBB(\omega) = \frac{\beta}{N(1-\beta\omega)} + \cal O(N^{-2}).
\end{equation}
Hence, we obtain by applying \eqref{eq:free_add_formula} and \eqref{eq:r_transform} that
\begin{equation}
	\btr_\M(\omega) = \btr_\AAA(\omega) + \frac{\beta}{N(1-\beta\omega)} + \cal O(N^{-2}).
\end{equation}
Next, we set $\omega = \stj_\M(z)$ so that this latter equation becomes
\begin{equation}
	\label{eq:factor_model_tmp1}
	z = \btr_\AAA(\stj_\M(z)) + \frac{\beta}{N(1-\beta\stj_\M(z))} + \cal O(N^{-2}).
\end{equation}
From this equation, we expect the Stieltjes transform of $\rho_\M$ to be of the form
\begin{equation}
	\stj_\M(z) = \stj_0(z) + \frac{\stj_1(z)}{N} + \cal O(N^{-2}). 
\end{equation}
By plugging this ansatz into \eqref{eq:factor_model_tmp1}, we see that $\stj_0(z)$ and $\stj_1(z)$ satisfies
\begin{eqnarray}
	z & = & \btr_\AAA(\stj_0(z)) \nonumber \\
	\stj_1(z) & = & - \frac{\beta}{\btr_\AAA'(\stj_0(z))(1-\stj_0(z) \beta)}\,.
\end{eqnarray}
It is easy to find that $\stj_0(z) = \stj_\AAA(z)$ as expected. We now focus on the $1/N$ correction term and using that $\btr_\AAA'(\stj_\AAA(z)) = 1/\stj_\AAA(z)$, we conclude that
\begin{equation}
	\stj_1(z) = -\frac{\beta \stj_\AAA'(z)}{1-\stj_\AAA(z) \beta}\,.
\end{equation}
Finally, we obtained that
\begin{equation}
	\stj_\M(z) \approx \stj_\AAA(z) - \frac1N \frac{\beta \stj_\AAA'(z)}{1-\stj_\AAA(z) \beta},
\end{equation}
and we see that the correction term only survive in the large $N$ limit if $\stj_\AAA(z) = \beta^{-1}$ has a non trivial solution. Differently said, $z$ is an eigenvalue of $\M$ and not of $\AAA$ if there exists $z \in \mathbb{R}\backslash \supp[\rho_\AAA]$ such that $\stj_\AAA(z) = \beta^{-1}$ and this leads to the criterion \eqref{eq:outlier_condition_factor_model}.
\end{changemargin}
\normalsize
\nc

\subsubsection{Products of free matrices}
\label{subsubsec:freemult}

Similar results are available for free multiplicative convolution. Before showing how to obtain the LSD of the product of free matrices, we first emphasize that one has to carefully define the product of free matrices. Indeed, the naive analogue of the free addition would be to define $\M = \AAA \BBB $. However the product $\AAA\BBB$ is in general not self-adjoint when $\AAA$ and $\BBB$ are self-adjoint but not commuting. In the case where $\AAA$ is positive definite, we can see that the product $\AAA^{1/2} \BBB \AAA^{1/2}$ makes sense and share the same moments than the product $\AAA\BBB$. Therefore, we define the product of free matrices by 
\begin{equation}
	\label{eq:model_freemult}
	\M : = \sqrt{\AAA} \BBB \sqrt{\AAA}. 
\end{equation} 
Note that in this case, $\BBB$ need not be necessarily positive definite but must have a trace different from zero (see the Taylor expansion below). For technical reason, we need the LSD of ${\bf B}$ to be well-defined. Under this assumption, the free multiplicative convolution rule for random matrices is given by 
\begin{equation}
\label{eq:free_mult}
%\str_{\AAA \bxt \BBB}(z) := 
\str_{\M}(\omega)  \;=\; \str_{\AAA}(\omega) \str_{\BBB}(\omega).
\end{equation}
This is the so-called \emph{free multiplication}, which has been first obtained by Voiculescu \cite{voiculescu1992free} and then by \cite{zinn1999adding} in 
a physics formalism. 
%The $\str$-transform is therefore the analogue of the Fourier transform for free multiplicative convolution. 

Again, if one is interested in the limiting spectral density of $\M$, one would like to write \eqref{eq:free_mult} in terms of its Stieltjes transform. 
Using the very definition of the $\str$-transform, we rewrite \eqref{eq:free_mult} as
\begin{equation*}
\frac{1}{\ttr^{-1}_{\M}(\omega)} = \frac{\str_{\BBB}(\omega)}{\ttr^{-1}_{\AAA}(\omega) }.
\end{equation*}
The trick is the same as above so we therefore set $\omega = \ttr_{\M}(z)$ to find
\begin{equation}
\ttr^{-1}_{\AAA}(\ttr_{\M}(z)) = z \str_{\BBB}(\ttr_{\M}(z)).
\end{equation}
It is now immediate to get the analogue of \eqref{eq:stieltjes_free_addition} for the multiplicative case 
\begin{equation}
\ttr_{\M}(z) = \ttr_{\AAA}\left( z \str_{\BBB}(\ttr_{\M}(z)) \right),
\end{equation}
that gives in terms of the Stieltjes transform
\begin{equation}
\label{eq:stieltjes_free_multiplication}
z \stj_{\M}(z) = Z(z) \stj_{\AAA}\left( Z(z) \right), \qquad Z(z) \;\deq\; z \str_{\BBB}(z \stj_{\M}(z) - 1).
\end{equation}
This is certainly one the most important results of RMT for statistical inference. It allows one to generalize the Mar{\v c}enko-Pastur law for sample covariance matrices to arbitrary population covariance matrices $\b C$ (see next section), and obtain results on the eigenvectors as well. We emphasize that the literature on free products can be adapted to non Hermitian matrices, see \cite{burda2013free} or \cite{burda2011multiplication} for a recent review on the multiplication of random matrices.

\subsection{Replica analysis}
\label{sec:replica}

\subsubsection{Resolvent and the Replica trick}

As we noticed above (Eq.\ \ref{eq:resolvent_decomposition}), information about the eigenvectors can be studied through the resolvent. However, both the Coulomb gas analogy and free probability tools are blind to the structure of eigenvectors since these only give information about the normalized trace of the resolvent. In order to study the resolvent matrix, we need to introduce other tools, for example one borrowed from statistical physics named the \emph{Replica} method. To make it short, the Replica method allows one to rewrite the expectation value of a logarithm in terms of moments, expressed as expectation values of many copies, named the \emph{replicas}, of the initial system. This method has been extremely successful in various contexts, including RMT and disordered systems, see e.g.\  \cite{mezard1987spin,edwards1976eigenvalue}, or \cite{morone2014replica} for a more recent review. We stress that even if this method turns out to be a very powerful heuristic, it is not rigorous mathematically speaking (see below). Therefore, it is essential to verify the result obtain from the Replica method using other methods, for example numerical simulations. Note that a rigorous but more difficult way to deal with resolvent is the recursion technique that uses linear algebra results, as explained in Appendix \ref{app:recursion}. Other available techniques include Feynman diagrams \cite{burda2004signal,burda2004spectral}. 

As a warm-up exercise, we present briefly the approach for the Stieltjes transform and then explain how to extend it to the study of full resolvent. We notice that any Stieltjes transform can be expressed as
\begin{equation}
	\label{eq:stieltjes_log_tmp}
	\stj(z) = \sum_{i=1}^{N} \frac{1}{z - \nu_i} = \frac{\partial}{\partial z} \log \prod_{i=1}^{N} (z - \nu_i) = \frac{\partial}{\partial z} \log \det(z I - \M).
\end{equation}
Then, using the Gaussian representation of $\det(z I - \M)^{-1/2}$, we have that 
\begin{equation}
	{\cal Z}(z) \equiv \left(\det(z I - \M)\right)^{-1/2} = \int \exp \qBB{ - \frac12 \sum_{i,j=1}^{N} \eta_i (z I - \M)_{ij} \eta_j} \prod_{j=1}^{N} \pBB{\frac{\dd \eta_j}{\sqrt{2\pi}}}.
\end{equation}
Plugging this last equation into \eqref{eq:stieltjes_log_tmp} and assuming that the Stieltjes transform is self-averaging, we see that we need to compute 
the average of the logarithm of ${\cal Z}(z)$:
\begin{equation}
	\label{eq:stieltjes_log}
	\stj(z) = -2 \frac{\partial}{\partial z} \mathbb{E} \log {\cal Z}(z),
\end{equation}
where the average is taken over the probability distribution $\cal P_\M$. However, it would be easier to compute the moments $\mathbb{E} {\cal Z}^n(z)$ instead of $\mathbb{E} \log {\cal Z}(z)$ and this is precisely the purpose of the \emph{Replica trick} which was initially formulated as the following identity
\begin{equation}
	\label{eq:replica_trick_log}
	\log {\cal Z} = \lim_{n\to 0} \frac{{\cal Z}^{n} - 1}{n},
\end{equation}
so that one formally has 
\begin{equation}
	\label{eq:stieltjes_replica}
	\stj(z) = \lim_{n\to0} \frac{\partial}{\partial z} \frac{\mathbb{E} {\cal Z}^{n} - 1}{n}.
\end{equation}
We have thus transformed the problem \eqref{eq:stieltjes_log} into the computation of $n$ replicas of the system involved in ${\cal Z}^n(z)$. 
The non-rigorous part of this method is quite obvious at this stage. While the integer moments of ${\cal Z}$ can indeed be expressed as an average
of the replicated system, the identity \eqref{eq:replica_trick_log} requires vanishingly small, \emph{real} values of $n$. Typically, one works with integer $n$'s and then perform an analytical continuation of the result to real values of $n$ before taking the limit $n \rightarrow 0$ (after, as it turns out, sending the size 
of the matrix $N$ to infinity!). Therefore, the main concern of this method is that we assume that the analytical continuation poses no problem, which is not necessarily the case. It is precisely this last step that could lead to uncontrolled approximations in some cases \cite{parisi1980sequence}, which is why numerical (or other) checks are mandatory. Nonetheless, the Replica trick gives a simple heuristic to compute the Stieltjes transform $\stj(z)$ which, as shown below, is 
exact for the quantities considered in this review.

For our purposes, we need to extend the above Replica formalism for the entire resolvent and not only its normalized trace. In that case, we will need a slightly different \emph{Replica identity}, extending \eqref{eq:replica_trick_log}, that we shall now present. The starting point is to rewrite the entries of the resolvent matrix ${\b G}(z)$ using the Gaussian integral representation of an inverse matrix
\begin{equation}
\label{eq:resolvent_entries}
(z\In - \M)^{-1}_{ij} = \frac{\int \left(\prod_{k=1}^{N} d\eta_k\right) \eta_i \eta_j \exp\left\{-\frac12 \sum_{k,l=1}^{N} \eta_k (z\delta_{kl} - \M_{kl}) \eta_l \right\}}{\int \left(\prod_{k=1}^{N} d\eta_k\right) \exp\left\{-\frac12 \sum_{k,l=1}^{N} \eta_k (z\delta_{kl} - \M_{kl}) \eta_l \right\}}.
\end{equation}
%Using the spectral decomposition of the resolvent \eqref{eq:resolvent_decomposition}, 
As explained in Appendix \ref{app:recursion}, we expect that \eqref{eq:resolvent_entries} is self-averaging in the LDL thanks to the Central Limit Theorem, so that:
\begin{equation}
\label{eq:resolvent_entries_avg}
(z\In - \M)^{-1}_{ij} = \avgBB{\frac{1}{\cal Z} \int \left(\prod_{k=1}^{N} d\eta_k\right) \eta_i \eta_j \exp\left\{-\frac12 \sum_{k,l=1}^{N} \eta_k (z\delta_{kl} - \M_{kl}) \eta_l \right\}}_{\cal P_\M},
\end{equation}
where ${\cal Z}$ is as above the partition function, i.e. the denominator in Eq. (\ref{eq:resolvent_entries}). The replica identity for resolvent is given by 
\begin{eqnarray}
\label{eq:resolvent_replica}
{G}_{ij}(z)  & = & \underset{n \rightarrow 0}{\lim} \avgBB{ {\cal Z}^{n-1} \int \left(\prod_{k=1}^{N} d\eta_k\right) \eta_i \eta_j \exp\left\{-\frac12 \sum_{k,l=1}^{N} \eta_k (z\delta_{kl} - \M_{kl}) \eta_l \right\} }_{\cal P_\M} \nonumber \\
& = & \underset{n \rightarrow 0}{\lim} \int \left(\prod_{k=1}^{N} \prod_{\alpha=1}^{n} d\eta_k^{\alpha}\right) \eta_i^{1} \eta_j^{1} \avgBB{ \prod_{\alpha=1}^{n} \exp\left\{-\frac12 \sum_{k,l=1}^{N} \eta_k^{\alpha} (z\delta_{kl} - \M_{kl}) \eta_l^{\alpha} \right\} }_{\cal P_\M}.
\end{eqnarray}
Again, we managed to rewrite the initial problem \eqref{eq:resolvent_entries_avg} as the computation of $n$ replicas. We emphasize that \eqref{eq:resolvent_replica} is valid for any random matrix $\M$, and is useful provided that we are able to compute the average over the probability density $\cal P_\M$. The identity \eqref{eq:resolvent_replica} is the central tool of this section. In particular, it allows one to study the asymptotic behavior of the resolvent entry-wise, which contains more information about the spectral decomposition of $\M$ than just the normalized trace \cite{bun2015rotational}. As will become apparent below, we consider a model of random matrices inspired by Free Probability theory, i.e. $\M = \AAA + \b\Omega \BBB \b\Omega^* $ and $\M = \AAA^{1/2}\b\Omega \BBB \b\Omega^*  \AAA^{1/2}$ (see Section \ref{sec:free_probability} above for a more details). We shall focus on the model of free multiplication since the arguments below may be repeated almost verbatim for the free additive case (see Appendix \ref{app:addition}).

\subsubsection{Matrix multiplication using replicas}

We reconsider the model \eqref{eq:model_freemult} and assume without loss of generality that $\AAA$ is diagonal. In that case, we see that $\cal P_{\M}$ is simply the Haar measure over the orthogonal group $\b O(N)$. We specialize the replica identity \eqref{eq:resolvent_replica} to $\M = \AAA^{1/2}\b\Omega \BBB \b\Omega^*  \AAA^{1/2}$ so that we get
\begin{equation}
	\label{eq:resolvent_replica_freemult}
	{G}_{ij}(z)  = \underset{n \rightarrow 0}{\lim} \int \left(\prod_{k=1}^{N} \prod_{\alpha=1}^{n} d\eta_k^{\alpha}\right) \eta_i^{1} \eta_j^{1} e^{-\frac{z}{2} \sum_{\alpha = 1}^{n} \sum_{k=1}^{N} (\eta_k^{\alpha})^2 } \cal I_{1}\pBB{\sum_{\alpha=1}^{n} \pb{\eta^{\alpha} \AAA^{1/2}} \pb{\eta^{\alpha} \AAA^{1/2}}^*, \BBB},
\end{equation}
where 
\begin{equation}
	\label{eq:HCIZ}
	\cal I_{\beta}(\AAA',\BBB) \;\deq\; \int \exp\qB{-\frac{\beta N}{2}\tr\AAA'\Omega \BBB \Omega^* } \cal D \Omega,
\end{equation}
is the so-called \emph{Harish-Chandra--Itzykson-Zuber} integral \cite{harish1957differential,itzykson1980planar}. Explicit results for this integral are known for Hermitian matrices ($\beta = 2$) for any integer dimension $N$, but not for real orthogonal matrices. Even the study of \eqref{eq:HCIZ} in the limit $N \to \infty$ is highly non trivial (see Appendix \ref{app:HCIZ}). Nevertheless, in the case where $\AAA'$ is of finite rank, the leading contribution for $N \to \infty$ is known for any symmetry group. Fortunately, we see that $\AAA'$ in our case is of rank $n$ and the result is obtained from Eq.\ \eqref{eq:HCIZ_rankn} in Appendix \ref{app:HCIZ}:\footnote{Recall that we work with $n$ as an integer throughout the intermediate steps of the computation.} 
\begin{equation}
	\cal I_{1}\pB{\sum_{\alpha=1}^{n} \pb{\eta^{\alpha} \AAA^{1/2}} \pb{\eta^{\alpha} \AAA^{1/2}}^*, \BBB} \underset{N\to\infty}{\sim} \exp\qBB{\frac{N}{2} \sum_{\alpha=1}^{n} \cal W_{\BBB}\left( \frac1N  \sum_{i=1}^{N} (\eta_i^{\alpha})^2 a_i \right)},
\end{equation}
with 
\begin{equation}
	\label{eq:W_transform}
	\wtr_{\BBB}^{\,\prime}(.) = \cal R_{\BBB}(.)\,,
\end{equation} 
and where we assume that the vectors $[\eta^{\alpha}]_{\alpha=1}^{n}$ are orthogonal to each other, which is generically true provided $n \ll N$. We then plug this result into \eqref{eq:resolvent_replica_freemult} and introduce an auxiliary variable $p^{\alpha} = \frac1N \sum_{i=1}^{N} (\eta_i^{\alpha})^2 a_i $ that we enforce using the exponential representation of a Dirac delta function
\begin{equation}
	\delta\pB{ p^{\alpha} - \frac1N \sum_{i=1}^{N} (\eta_i^{\alpha})^2 a_i } = \int \frac{1}{2\pi} \exp\qBB{\ii\zeta^{\alpha}\pB{ p^{\alpha} - \frac1N \sum_{i=1}^{N} (\eta_i^{\alpha})^2 a_i} } \dd\zeta^{\alpha},
\end{equation}
for each $\alpha = 1,\dots,n$. This allows to retrieve a Gaussian integral on $\eta^\alpha$. Renaming $\zeta^\alpha =- 2\ii\zeta^\alpha/N$ yields the result
\begin{equation}
	\label{eq:global_law_freemult_tmp1}
{G}_{ij}(z) \propto \int \int \left(\prod_{\alpha = 1}^{n} \dd p^{\alpha} \dd\zeta^{\alpha}\right) \frac{\delta_{ij}}{z - \zeta^1 a_i} \exp\qBB{ - \frac{Nn}{2} F_{0}(p^{\alpha}, \zeta^{\alpha})}
\end{equation}
where $F_0$ is the free energy given by
\begin{equation}
F_{0}(p^{\alpha}, \zeta^{\alpha}) = \frac{1}{n} \sum_{\alpha=1}^{n} \left[ \frac{1}{N} \sum_{k=1}^{N} \log(z - \zeta^{\alpha} a_k ) + \zeta^{\alpha} p^{\alpha} - \cal W_{\BBB}(p^{\alpha}) \right].
\end{equation}
Now, one sees that the integral over $\dd p^{\alpha} \dd \zeta^{\alpha}$ involves the exponential of $Nn/2$ times the free energy, which is of order unity. Provided that $n$ 
is non-zero, one can estimate this integral via a saddle point method (but of course $n$ will be sent to zero eventually...). We assume a \textit{replica symmetric} ansatz for the saddle point, i.e. $p^{\alpha}=p^*$ and $\zeta^{\alpha} = \zeta^*$, $\forall \alpha  = 1, \dots, n$.  This is natural since $F_0$ is invariant under the permutation
group $P_n$. Note however that the replica symmetric ansatz can lead to erroneous results and this phenomenon is known as \emph{replica symmetry breaking}, see e.g.\ \cite{mezard1987spin,parisi1980sequence} or \cite{talagrand2006parisi} and references therein for a mathematical formalism. The rest of the calculation relies on a saddle-point analysis whose details we postpone below, and we finally obtain a so-called ``global law'' for the resolvent of $\M$:\footnote{The term ``global'' assumes that the imaginary part of $z$ is much larger than $N^{-1}$, in contrast to many different studies of the resolvent at a ``local'' scale (see \cite{benaych2016lectures} for a detail presentation of this concept for Wigner matrices).}
\begin{equation}
	\label{eq:global_law_freemult}
	z {\b G}_{\M}(z)_{i,j} \underset{N\to\infty}{\sim} Z(z) \b G_{\AAA}(Z(z))_{i,j}, \qquad Z(z) \deq z \cal S_{\BBB}(z \stj_{\M}(z) - 1),
\end{equation}
which is often referred to as a \emph{subordination} relation between the resolvent of $\M$ and $\AAA$. Taking the trace of both sides of the above equation, one notices that \eqref{eq:global_law_freemult} is a generalization of the formula \eqref{eq:stieltjes_free_multiplication} as a matrix. We should emphasize that Eq.\ \eqref{eq:global_law_freemult} is self-averaging element by element for the matrix ${\b G}_{\M}(z)$, i.e. ${G}_{ij}(z)=\langle{G}_{ij}(z)\rangle + \cal O (N^{-1/2})$. The matrix ${\b G}_{\M}(z)$ taken as a whole cannot be considered deterministic, for example $\langle{\b G}_{\M}(z)\rangle^2$ is in general different from $\langle{\b G}^2_{\M}(z)\rangle$. When considering the whole matrix ${\b G}_{\M}(z)$ one should rather write:
\begin{equation}
	\label{eq:global_law_freemult_avg}
	z \left\langle{\b G}_{\M}(z)\right\rangle \underset{N\to\infty}{\sim} Z(z) \b G_{\AAA}(Z(z)), \qquad Z(z) \deq z \cal S_{\BBB}(z \stj_{\M}(z) - 1),
\end{equation}
Note that the average resolvent $\langle{\b G}_{\M}(z)\rangle$ is diagonal in the eigenbasis of $\AAA$, as expected by symmetry. 
%Nevertheless in what follows we will write deterministic equations for ${\b G}_{\M}(z)$ which should be interpreted as element by element self-averaging equations.

We can redo the exact same calculations for the free addition model $\M = \AAA +\Omega \BBB \Omega^* $, still with $\AAA = \diag(a_1, a_2, \dots, a_N)$ (see Appendix \ref{app:addition}). Starting from the replica identity \eqref{eq:resolvent_replica} and then applying \eqref{eq:HCIZ_rankn}, we obtain the following expression \cite{bun2015rotational}:
\begin{equation}
{G}_{ij}(z) \propto \int \int \left( \prod_{\alpha=1}^{n}  dp^{\alpha} d\zeta^{\alpha}\right) \frac{\delta_{ij}}{z - \zeta^{1} - a_i} \exp\left\{ -\frac{Nn}{2} F^{a}_0(p^{\alpha}, \zeta^{\alpha}) \right\},
\end{equation}
where the `free energy' $F^{a}_0$ is given by
\begin{equation}
\label{free_convolution_energy}
F_0^a(p, \zeta) \deq \frac{1}{Nn} \sum_{\alpha = 1}^{n} \left[ \sum_{k=1}^{N} \log(z - \zeta^{\alpha} - a_k) - \cal W_{\BBB}(p^{\alpha}) + p^{\alpha}\zeta^{\alpha} \right].
\end{equation}
Invoking once again the replica symmetric ansatz, the subordination for the resolvent under the free addition model follows from a saddle-point analysis \cite{bun2015rotational}
\begin{equation}
	\label{eq:global_law_free_addition}
	{\b G}_{\M}(z)_{i,j} \underset{N\to\infty}{\sim} \b G_{\AAA}(Z_a(z))_{i,j}, \qquad Z_a(z) \deq z - \cal R_{\BBB}(\stj_{\M}(z)),
\end{equation}
which is exactly the result obtained in \cite{kargin2015subordination} in a mathematical formalism. Again taking the trace of both sides of this equation allows 
one to recover the relation \eqref{eq:stieltjes_free_addition} between Stieltjes transforms.

\subsubsection{Free multiplication: replica saddle-point analysis}

\begin{changemargin}{1cm}{1cm}
\footnotesize
We now present the derivation of \eqref{eq:global_law_freemult} from \eqref{eq:global_law_freemult_tmp1}. We shall that it actually provides an elementary derivation of the free multiplication formula \eqref{eq:free_mult}. Under the replica symmetric ansatz, the free energy becomes 
\begin{equation*}
F_{0}(p^{\alpha}, \zeta^{\alpha}) \equiv F_{0}(p, \zeta) = \frac{1}{N} \sum_{k=1}^{N} \log(z - \zeta a_k ) + \zeta p - \wtr_{\BBB}(p),
\end{equation*}
which needs to be extremized. We first consider the first order condition with respect to $p$ which leads to
\begin{equation}
\label{eq:zeta_star_mult}
\zeta^{*} = \rtr_{\BBB}(p^*).
\end{equation}
The other derivative with respect to $\zeta$ gives: 
\begin{equation}
\label{eq:p_star_mult}
p^{*} = \frac{1}{\zeta^* N} \sum_{k=1}^{N} \frac{a_k}{z/\zeta^* -  a_k} = \frac{\ttr_{\AAA} \left( \frac{z}{\rtr_{\BBB}(p^*)} \right)}{\rtr_{\BBB}(p^*)}.
\end{equation}
Hence, plugging \eqref{eq:zeta_star_mult} and \eqref{eq:p_star_mult} into \eqref{eq:global_law_freemult_tmp1}, we get in the large $N$ limit and then the limit $n \rightarrow 0$ by 
\begin{equation}
\label{eq:global_law_tmp}
{G}_{ij}(z)_{ij} = \frac{\delta_{ij}}{z - \rtr_{\BBB}(p^*) c_i}.
\end{equation}
We can find a genuine simplification of the last expression using the connection with the free multiplication convolution. By taking the normalized trace of ${\b G}_{\M}(z)$, we see that we have
\begin{equation}
\label{eq:free_mult_tmp}
z \stj_{\M}(z) = Z \stj_{\AAA}(Z), \quad\text{with }\quad Z \equiv Z(z) = \frac{z}{\rtr_{\BBB}(p^*)},
\end{equation}
which can rewrite as 
\begin{equation*}
\ttr_{\M}(z) = \ttr_{\AAA}(Z).
\end{equation*}
Let us define 
\begin{equation}
	\label{eq:x_T}
	\omega = \ttr_{\M}(z) = \ttr_{\AAA}(Z).
\end{equation} 
Using Eq.\ \eqref{eq:p_star_mult}, this latter equation implies $p^* = \omega/\cal R_{\BBB}(p^*)$. Let us now show how to retrieve the free multiplicative convolution \eqref{eq:free_mult} from \eqref{eq:free_mult_tmp} in the large $N$ limit. Indeed, let us rewrite \eqref{eq:x_T} as
\begin{equation}
\label{tmp}
z \ttr_{\M}(z) = Z \ttr_{\AAA}(Z) \rtr_{\BBB}(p^*),
\end{equation}
and it is trivial to see that using \eqref{eq:x_T} that this last expression can be rewritten as $\omega \ttr_{\M}^{-1}(\omega) = \omega \ttr_{\AAA}^{-1}(\omega) \rtr_{\BBB}(p^*)$. Finally, using the definition of the $\str$-transform \eqref{eq:s_transform}, this yields
\begin{equation}
\label{eq:S_transform_tmp}
\str_{\M}(\omega) = \str_{\AAA}(\omega) \frac{1}{\rtr_{\BBB}(p^*)}.
\end{equation}
Using \eqref{eq:r_s_transform}, we also have
\begin{equation}
\label{R_S_transform_p_star}
\frac{1}{\rtr_{\BBB}(p^*)} = \str_{\BBB}(p^* \rtr_{\BBB}(p^*)),
\end{equation}
But recalling that $p^* = \omega/\rtr_{\BBB}(p^*)$, we conclude from \eqref{eq:zeta_star_mult}, \eqref{eq:x_T} and \eqref{R_S_transform_p_star} that
\begin{equation}
	\label{eq:zeta_star_mult_sol}
	\frac{1}{\zeta^{*}} = \rtr_{\BBB}(p^*) = \str_{\BBB}(\ttr_{\M}(z)).
\end{equation}
Going back to \eqref{eq:S_transform_tmp}, we see that the spectral density of $\M$ is given by Voiculescu's free multiplication formula
\begin{equation}
\str_{\M}(\omega) = \str_{\AAA}(\omega) \str_{\BBB}(\omega),
\end{equation}
confirming that the replica symmetry ansatz is indeed valid in this case. Finally, by plugging \eqref{eq:zeta_star_mult_sol} into \eqref{eq:global_law_tmp}, we get the result \eqref{eq:global_law_freemult}.
\end{changemargin}
\normalsize

\clearpage%!TEX root = RMT_Covariance_Review.tex
\section{Spectrum of large empirical covariance matrices}
\label{chap:spectrum}

\subsection{Sample covariance matrices} 

\subsubsection{Setting the stage}

After a general introduction to RMT and to some of the many different analytical tools, we are now ready to handle the main issue of this review, which is the 
statistics of sample covariance matrices. As a preliminary remark, note that we assume that the variance of each variable can be estimated independently with great accuracy given that we have $T \gg 1$ observations for each of them. Consequently, all variables will be considered to have unit variance in the following and we
will not distinguish further covariances and correlations henceforth. 

As stated in the introduction, the study of correlation matrices has a long history in statistics. Suppose we consider a (random) vector $\b y = (y_1, y_2, \dots, y_N)$. One standard way to characterize the underlying interaction network between these variables is through their correlations. Hence, the goal is to measure as 
precisely as possible the \textit{true} (or \textit{population}) covariance matrix, defined as 
\begin{equation}
\label{eq:population_covmat}
\C_{ij} = \mathbb{E} \qb{y_{i} y_{j}}, \quad i,j \in \qq{1,N}
\end{equation}
where we assumed that the $\{ y_{i}\}_{i \in \qq{1,N}}$ have zero mean without loss of generality (see below). It is obvious from the definition of $\C$ that the covariance matrix is symmetric. Throughout the following, we shall define the spectral decomposition of $\C$ as
\begin{equation}
\label{eq:population_covariance_spectral}
	\C \; = \; \sum_{i=1}^{N} \mu_i \b v_i \b v_i^*,
\end{equation}
with $\mu_1 \geq \mu_2\geq\dots\geq\mu_N$ the real eigenvalues and $\b v_1, \dots,\b v_N$ the corresponding eigenvectors.

As illustrated in the introduction, the concept of covariance is of crucial importance in a wide range of applications. For instance, let us consider an example that stems from financial applications. The probability of a large loss of a diversified portfolio is dominated by the correlated moves of its different constituents (see section \ref{sec:markowitz} for more details).  In fact, the very notion of diversification depends on the correlations between the assets in the portfolio. Hence, the estimation of the correlations between the price movements of these assets is at the core of risk management policies. 

The major concern in practice is that the \textit{true} covariance matrix $\C$ is in fact unknown. To bypass this problem, one often relies on a large number $T$ of \textit{independent} measurements, namely the ``samples'' $\b y_1, \dots, \b y_T$, to construct empirical estimates of $\C$. We thus define the $N \times T$ matrix $\Y_{it} \in \R^{N\times T}$, whose elements are the $t$-th measurement of the variable $y_i$. Within our example from finance, the random variable $Y_{it}$ would be the return of the asset $i$ at time $t$. Eq. (\ref{eq:population_covmat}) is then approximated by an average value over the whole sample data of size $T$, leading to the \textit{sample} (or \textit{empirical}) covariance matrix estimator:
\begin{equation}
\label{eq:SCM}
\E_{ij} = \frac1T (\Y \Y^{*})_{ij} = \frac1T \sum_{\tau = 1}^{T} \Y_{it} \Y_{jt}.
\end{equation}
In the statistical literature, this estimator is known as \emph{Pearson} estimator and in the RMT community, the resulting matrix sometimes referred to as the Wishart Ensemble. Whereas the Wigner Ensemble has been the subject of a large amount of studies in physics \cite{akemann2011oxford}, results on the Wishart Ensemble mostly come from mathematics \& statistics \cite{marchenko1967distribution,bai2009spectral,paul2014random}, telecommunication \cite{couillet2011random} or the financial/econophysics literature \cite{plerou2002random,bouchaud2009financial,burda2004signal}, 
although some work in the physics literature also exists \cite{vivo2007large, majumdar2012number, perret2015finite,wirtz2013distribution} -- to cite a few. 

In what we call the ``classical'' statistical limit, i.e. $T \rightarrow \infty$ with $N$ fixed, the law of large numbers tells us that $\E$ converges to the true covariance $\C$. However, as recalled in the introduction, in the present ``Big Data'' era where scientists are confronted with large data-sets such that the sample size $T$ {\it and} the number of variables $N$ are both very large, specific issues arise when the observation ratio $q = N/T$ is of order unity. This setting is known in the literature as the high-dimensional limit or Kolmogorov regime (or more commonly called the Big Data regime). This regime clearly differs from the traditional large $T$, fixed $N$ situation (i.e. $q \rightarrow 0$), where classical results of multivariate statistics apply. The setting $q \sim O(1)$ is precisely where tools from RMT can be helpful to make precise statements on the empirical covariance matrix \eqref{eq:SCM}. 

A typical question would be to study the ESD of $\E$ in order to quantify its deviation from the true covariance matrix $\C$. More precisely, does the ESD converges to an explicit LSD? If it does, can we get a tractable expression for this LSD? In the case where the samples $\{\b y_{t}\}_{t = 1}^{T}$ are given by a multivariate Gaussian distribution with zero mean and covariance $\C$, the distribution of the matrix $\E$ is exactly known since Wishart \cite{wishart1928generalised}, and is given by Eq. \eqref{eq:wishart_distribution} above, with $\M \to \E$. In the case where $\C = T^{-1} \In$, we retrieve the isotropic Wishart matrix above that we fully characterized in the previous chapter. The aim is now to provide the LSD of $\E$ for an \textit{arbitrary} true covariance matrix $\C$. More specifically, we shall look at linear models where the data matrix $\Y$ can be decomposed as 
\begin{equation}
	\label{eq:SCM_entries}
	\Y = \sqrt{\C} \X,
\end{equation} 
where $\X$ is a $N \times T$ random matrix with uncorrelated entries satisfying
\begin{equation}
\label{eq:SCM_entries_X}
\mathbb{E} [X_{it}] = 0, \qquad \mathbb{E} [X_{it}^2] = \frac1T.
\end{equation}
The above decomposition is always possible for multivariate Gaussian variables. Otherwise, the above framework assumes that our correlated random variables $y_i$ are 
obtained as linear combinations of uncorrelated random variables. In addition, we also require that the random variables $\sqrt{T} X_{it}$ have a bounded $4$-th moment, in other words that the 
distribution cannot be extremely fat-tailed. 

Next, we introduce the spectral decomposition of $\E$,
\begin{equation}
\label{eq:SCM_eigen}	
\E = \sum_{i=1}^{N} \lambda_i \b u_i \b u_i^*,
\end{equation} 
with $\lambda_1 \geq \lambda_2\geq\dots\geq\lambda_N$ the eigenvalues and $\b u_1, \dots,\b u_N$ the corresponding eigenvectors. Let us now list the main assumptions on the spectrum of $\E$, that we shall suppose to hold throughout this review:
\begin{enumerate}
	\item The support of $\rho_\E$ consists of $r+1$ (connected) components with $r \geq 0$. We call the $r$ largest components the \textit{outliers} and the smallest component the {\it bulk}. The boundary points of the bulk component are labeled $\lambda_-$ and $\lambda_+$ (with $\lambda_- \leq \lambda_+$).
	\item We suppose that the outliers are separated from each other and from the bulk (non-degeneracy).
	\item We suppose that the bulk is \textit{regular} in the sense that the density of $\rho_\E$ vanishes as a square root at the boundary points  $\lambda_-, \lambda_+$.
\end{enumerate}
In this chapter, we will look at the statistics of the eigenvalues of this model and the following one will be devoted to the eigenvectors. 

We end up this short introduction with two different remarks. 
The first one comments the zero-mean assumption made above, while the second one is concerned with the possible fat-tailed nature of the random variables
under scrutiny.

\subsubsection{Zero-mean assumption}

In real data-sets, sample vectors $\b y_t$ usually have a non-zero mean (even if the true underlying distribution is of zero mean). One can therefore
choose to shift the sample vectors in such a way that the empirical mean is exactly zero. This leads to the following definition of the empirical correlation 
matrix, often found in the literature:
\begin{equation}
	\breve{E}_{ij} \;=\; \frac{1}{T-1} \sum_{t=1}^{T} \pB{Y_{it} - \overline{Y_i}}\pB{Y_{jt} - \overline{Y_j}}, \qquad \overline{Y_i} = \frac{1}{T} \sum_{\tau=1}^{T} Y_{i\tau}.
\end{equation}
which is clearly unbiased as for $T \rightarrow \infty$ with $N$ fixed. This can be rewritten as:
\begin{equation*}
\breve{\E} = \frac{1}{T-1} \Y \left(\It - \b e \b e^{*}\right) \Y^*,\qquad \b e \deq (1,1, \dots, 1)^*/\sqrt{T} \in \mathbb{R}^T.
\end{equation*}
Still, the asymptotic properties of the eigenvalues (and eigenvectors) of $\E$ and of $\breve{\E}$ are identical, up to a possible extra outlier eigenvalue located at zero when $q > 1$. The simplest way to understand that the outlier has no influence on the asymptotic behavior of the spectrum is when $\Y$ is a Gaussian matrix. In this case, we know that a Gaussian matrix is statistically invariant under rotation so one can always rotate the vector $\b e$ in the $T$ dimensional space such that it becomes, say, $(1,0, \dots, 0)$. Then one has:
\begin{equation*}
\breve{E}_{ij} \sim \frac{1}{T-1} \sum_{t=2}^{T} \Y_{it} \Y_{jt}
\end{equation*}
which means that $\breve{\E}$ and $\E$ share identical statistically properties when $N, T \rightarrow \infty$ up to a rank one perturbation of eigenvalue $\sim T^{-1} \to 0$ (see Section \ref{sec:freeadd} for a related discussion). For $q < 1$, this has no influence at all 
on the spectrum since the corresponding eigenvalue is reabsorbed in the bulk. The possible spike associated to the rank-one perturbation only survives when $N  \geq T$, and it leads to an extra zero eigenvalue 
from the last equation. But in the case where $q \geq 1$, we know that there are $(N - T)$ additional zero eigenvalues, meaning that the extra spike at the origin is harmless. 
The case where $\Y$ is not rotationally invariant is harder to tackle and needs more sophisticated arguments for which we refer the reader to \cite[Section 9]{bloemendal2014principal} for more details.  
As a consequence, all the results concerning the statistics of the eigenvalues of $\E$ that we shall review below hold for $\breve{\E}$ as well. From a practical point of view, it is indifferent to consider raw data or demeaned data. We will henceforth assume that the samples data $(\b y_1, \dots, \b y_T)$ has exactly zero mean and will work with the corresponding $\E$ in the next sections.

\subsubsection{Distribution of the data entries}
\label{sec:SCM_entries}

The second remark deals with the distribution of the entries of the matrix $\Y$ given in Eq.\ \eqref{eq:SCM_entries_X}. It is well-known for instance that financial returns are strongly non-Gaussian, with power-law tails \cite{bouchaud2003theory}, and hence, the condition of a sufficient number of bounded moments can be seen as restrictive. What can be said in the case of entries that possess extremely fat tails? This is the main purposes of the theory of \emph{robust} estimators \cite{huber2011robust,maronna2006robust} where the RMT regime $N \asymp T$ has been subject to a lot studies in the past few years, especially in the case of elliptical distributions \cite{couillet2011random,biroli2007student,el2009concentration,chicheportiche2013non}. In particular, the so-called \emph{Maronna} robust $M$-estimator of $\C$ is the (unique) solution of the fixed point equation
\begin{equation}
	\label{eq:Maronna_SCM}
		\M \deq \frac{1}{T} \sum_{t=1}^{T} U\pB{\frac1N \b y_t^* \M^{-1} \b y_t} \b y_t \b y_t^*,
\end{equation}
where $U$ is a non-increasing function. It was shown recently \cite{couillet2015random} that the matrix $\M$ converges to a matrix of the form encountered in Eq.\ \eqref{eq:model_freemult} and thus different from $\E$. However, tractable formula are scarce except for the multivariate Student distribution where $U(x) \sim x^{-1}$ \cite{biroli2007student,el2009concentration,tyler1987distribution,zhang2014marchenko}. In that case, we have from \cite{couillet2016second} that the LSD of $\M$ converges (almost surely) to that of standard Wishart matrix $\E$ as $N \to \infty$. Therefore, all the results that we will present below holds for the robust estimator of $\C$ under a multivariate Student framework (see also 
\cite{biroli2007student}). We postpone discussions about other class of distributions to Chapter \ref{chap:conclusion}.

\subsection{Bulk statistics}
\label{sec:bulk statistics}

% We have studied the case where the true matrix $\C$ is the identity and its generalization in the presence of several special eigenvalues, called the outliers. While the spiked covariance matrix model has some appealing properties for many real life statistical problem, it is not necessarily the most realistic hypothesis. 
% An alternative way of thinking to the eigenvalues that spread outside the Mar{\v c}enko-Pastur sea, like in Fig. \ref{eigen_justMP}, is to assume that there is an \textit{anisotropic}\footnote{not proportional to the identity matrix.} true correlation matrix $\C$ that drives the whole system. Differently said, 

\subsubsection{Mar{\v c}enko-Pastur equation}
\label{sec:MP}

As we alluded to in the introduction, the fundamental tool to analyze the spectrum of large sample covariance matrices is the Mar{\v c}enko-Pastur equation \cite{marchenko1967distribution}. We actually have already encountered a special case of this equation in Section \ref{sec:MP_law} where we consider the LSD of $\E$ under the null hypothesis $\C = \b I_N$ (isotropic case). In this section, we allow the population correlation matrix $\C$ to be \emph{anisotropic}, that is to say not proportional to the identity matrix. As we shall see, the final result is not as simple as Eq.\ \eqref{eq:stieltjes_isotropic_wishart} but many properties can be inferred from it.  

The Mar{\v c}enko-Pastur (MP) equation dates back to their seminal paper \cite{marchenko1967distribution} which gives an exact relation between the limiting Stieltjes transforms of $\E$ and $\C$. This result is at the heart of many advances in statistical inference in high dimension (see Chapter \ref{chap:application} for some examples or \cite{paul2014random} and references therein). There are several ways to obtain this result, using e.g. recursion techniques \cite{silverstein1995empirical}, Feynman diagram expansion \cite{burda2004signal}, replicas (see \cite{sengupta1999distributions} or Section \ref{sec:replica} above for a generalization) or free probability. We will present this last approach, which is perhaps the simplest way to derive the MP equation. 

The key observation is that, for linear models, we can always rewrite $\E$ using Eq.\ \eqref{eq:SCM_entries} as
\begin{equation*}
	\E = \sqrt{\C} \Wishart \sqrt{\C}, \qquad \Wishart \;\deq\; \X \X^*,
\end{equation*}
where the matrix $\X$ satisfies Eq.\ \eqref{eq:SCM_entries_X} and is independent from $\C$. The model falls into the model of free multiplication encountered in Section \ref{sec:free_probability} since $\E$ is the free multiplicative convolution of $\C$ with a white Wishart kernel for $N \to \infty$ \cite{mingo2004annular}. Therefore, the Stieltjes transform of $\E$ is exactly given by Eq.\ \eqref{eq:stieltjes_free_multiplication} that we specialize to 
\begin{equation}\label{eq:MP_equation_stieltjes}
z \stj_{\E}(z) \;=\; Z(z) \stj_{\C}\left(Z(z)\right), \qquad\text{with}\qquad Z(z) \;\deq\; z \str_{\Wishart}(z \stj_{\E}(z) - 1).
\end{equation}
Moreover, the $\str$-transform of $\Wishart$ was obtained in Eq.\ \eqref{eq:S_transform_MP}, i.e. $\str_{\Wishart}(z) = (1+qz)^{-1}$ for any $q > 0$. Thus, we can re-express $Z(z)$ as:
\begin{equation}
Z(z) \;=\; \frac{z}{1-q+qz\stj_{\E}(z)},
\end{equation}
which is exactly the Mar{\v c}enko-Pastur self-consistent equation which relates the Stieltjes transforms of $\E$ and $\C$. 
The remarkable thing is that the RHS of Eq.\ \eqref{eq:MP_equation_stieltjes} is ``deterministic'' as $\C$ is fixed in this framework. 
Note that this equation is often written in the mathematical and statistical literature in an equivalent form as:
\begin{equation}
\label{eq:MP_equation_int}
\stj_{\E}(z) = \int \frac{\rho_{\C}(\mu) \dd\mu}{z - \mu(1-q+qz \stj_{\E}(z))} .
\end{equation}
There are two ways to interpret the above Mar{\v c}enko-Pastur equation:
\begin{itemize}
	\item[1.] the `direct' problem: we know $\C$ and we want to compute the expected eigenvalues density $\rho_{\E}$ of the empirical correlation matrix;
	\item[2.] the `inverse' problem: we observe $\E$ and try to infer the true $\C$ that satisfies equation (\ref{eq:MP_equation_stieltjes}).
\end{itemize}
Obviously, the inverse problem is the one of interest for many statistical applications, but is much more difficult to solve than the direct one as the mapping between $\stj_{\C}$ from $\stj_{\E}$ is numerically unstable.  
Still, the work of El-Karoui \cite{el2008spectrum} and, more recently, of Ledoit \& Wolf \cite{ledoit2013spectrum} allows one to make progress in this direction with a numerical scheme that solves a discretized version of the inverse problem Eq.\ \eqref{eq:MP_equation_int}. On the other hand, the direct problem leads to a self-consistent equation, which can be exactly solved numerically and sometimes analytically for some special forms of $\stj_{\C}$  (see next section). 

Let us finally make a remark that we have not seen in the literature before. Enhancing $Z(z)$ to $Z(z,q)$ to emphasize its dependence on $q$, one can check that this object obeys the following simple PDE \cite{bun2016dyson}:
\begin{equation}
\label{eq:MP_equation_diff}
q \frac{\partial Z(z,q)}{\partial q} = (Z(z,q)-z) \frac{\partial Z(z,q)}{\partial z},
\end{equation}
with initial condition $Z(z,q \to 0) = z + q z (1 - z\stj_{\C}(z))$. This representation can be given a direct interpretation but whether it is useful numerically or analytically remains to be seen.  

\subsubsection{Spectral statistics of the sample covariance matrix}
\label{sec:MP_spectrum_statistics}

For statistical purposes, the Mar{\v c}enko-Pastur equation provides an extremely powerful framework to understand the behavior of large dimensional sample covariance matrices, despite the fact that the inverse problem is not numerically stable. As we shall see in this section, one can infer many properties of the spectrum of $\E$ knowing that of $\C$, using the moment generating function. Recall the definition of the $\ttr$-transform in Eq.\ \eqref{eq:T_transform}, it is easy to see that we can rewrite Eq.\ \eqref{eq:MP_equation_stieltjes} as
\begin{equation}
	\label{eq:MP_equation_Ttransform}
		\ttr_\E(z) \;=\; \ttr_\C(Z(z)), \qquad Z(z) \;=\; \frac{z}{1+q\ttr_\E(z)}.
\end{equation}
We know from Eq. \eqref{eq:T_transform_moment} that the $\ttr$-transform can be expressed as power series for $z\to\infty$, hence we have
\begin{equation}
	\ttr_\E(z) \underset{z\to \infty}{=} \sum_{k=1}^{\infty} \varphi(\E^k) z^{-k},
\end{equation}
where $\varphi(.)=N^{-1} \Tr(.)$ is the normalized trace operator. We thus deduce that
\begin{equation*}
	Z(z)  \underset{z\to \infty}{=} \frac{z}{1+q \sum_{k=1}^{\infty} \varphi(\E^k) z^{-k}}.
\end{equation*}
Therefore we have for $z\to\infty$
\begin{equation}
	\ttr_\C(Z(z)) \underset{z\to \infty}{=} \sum_{k=1}^{\infty} \frac{\varphi(\C^k)}{z^k} \pBB{1+q \sum_{\ell=1}^{\infty} \varphi(\E^\ell) z^{-\ell}}^k.
\end{equation}
All in all, one can thus relate the moments of $\rho_\E$ with the moments of $\rho_\C$ by taking $z \to \infty$ in Eq.\ \eqref{eq:MP_equation_Ttransform} which yields
\begin{equation}
	\label{eq:MP_equation_mgf}
	\sum_{k=1}^{\infty} \frac{\varphi(\E^k)}{z^{k}} = \sum_{k=1}^{\infty} \frac{\varphi(\C^k)}{z^k} \pBB{1+q \sum_{\ell=1}^{\infty} \varphi(\E^\ell) z^{-\ell}}^k,
\end{equation}
which was first obtained in \cite{burda2004signal}. In particular, we infer from Eq.\ \eqref{eq:MP_equation_mgf} that the first three moments of $\rho_\E$ satisfy
\begin{eqnarray}
	\label{eq:SCM_moments}
	\varphi(\E)   & = & \varphi(\C) \quad = 1 \nonumber \\
	\varphi(\E^2) & = & \varphi(\C^2) + q  \nonumber \\
	\varphi(\E^3) & = & \varphi(\C^3) + 3q \varphi(\C^2) + q^2.
\end{eqnarray}
We thus see that the variance of the LSD of $\E$ is equal to that of $\C$ plus $q$, i.e. the spectrum of the sample covariance matrix $\E$ is always wider (for $q > 0$) than the spectrum of the population covariance matrix $\C$. This is an alternative way to convince ourselves that $\E$ is a noisy estimator of $\C$ in the high-dimensional regime. 

Note that we can also express the Mar{\v c}enko-Pastur equation in terms of a cumulant expansion. Indeed, we can rewrite Eq.\ \eqref{eq:MP_equation_stieltjes} in terms of the $\rtr$-transform (see below for a derivation) 
\begin{equation}
	\label{eq:MP_equation_Rtransform}
	\omega \rtr_\E(\omega) = \zeta(\omega) \rtr_\C(\zeta(\omega)), \qquad \zeta(\omega) = \omega\pb{1+q\omega \rtr_\E(\omega)}.
\end{equation}
Using the cumulants expansion of the $\rtr$-transform, given in Eq.\ \eqref{eq:r_transform_series}, we obtain for $\omega \to 0$
\begin{equation}
	\omega \rtr_\E(\omega) = \sum_{\ell=1}^{\infty} \kappa_\ell(\E) \omega^\ell,
\end{equation}
and 
\begin{equation}
	\zeta(\omega) \rtr_\C(\zeta(\omega)) = \sum_{\ell=1}^{\infty} \kappa_\ell(\C) \omega^\ell\pBB{1+q\sum_{m=1}^{\infty} \kappa_m(\E) \omega^m}^\ell.
\end{equation}
By regrouping these last two equations into Eq.\ \eqref{eq:MP_equation_Rtransform}, the analogue of Eq.\ \eqref{eq:MP_equation_mgf} in terms of free cumulants reads:
\begin{equation}
	\label{eq:MP_equation_cgf}
	\sum_{\ell=1}^{\infty} \kappa_\ell(\E) \omega^\ell = \sum_{\ell=1}^{\infty} \kappa_\ell(\C) \omega^\ell\pBB{1+q\sum_{m=1}^{\infty} \kappa_m(\E) \omega^m}^\ell,
\end{equation}
which would allow one to express the cumulants of $\E$ in terms of the cumulants of $\C$. 

Another interesting expansion is the case where $q < 1$, meaning that $\E$ is invertible. Hence $\stj(z)$ for $z \to 0$ is analytic and one can readily find
\begin{equation}
	\label{eq:stieltjes_expand_zero}
	\stj(z) \underset{z \to 0}{=} \; - \sum_{k=1}^{\infty} \varphi(\E^{-k}) z^{k-1}.
\end{equation}
This allows one to study the moment of the LSD of $\E^{-1}$ and this turns out to be an important quantity many applications (see Chapter \ref{chap:application}). Using Eq.\ \eqref{eq:MP_equation_stieltjes}, we can actually relate the moments of the spectrum $\E^{-1}$ to those of $\C^{-1}$ as one has, for $z \to 0$:
\begin{equation*}
	Z(z) = \frac{z}{1-q-q \sum_{k=1}^{\infty} \varphi(\E^{-k}) z^{k}}.
\end{equation*}
Hence, we obtain the following expansion for Eq.\ \eqref{eq:MP_equation_stieltjes} at $z \to 0$ and $q \in (0,1)$:
\begin{equation}
	\label{eq:MP_equation_mgf_inverse}
	\sum_{k=1}^{\infty} \varphi(\E^{-k}) z^{k} = \sum_{k=1}^{\infty} \varphi(\C^{-k}) \pBB{\frac{z}{1-q}}^k  \pBB{\frac{1}{1-\frac{q}{1-q} \sum_{\ell=1}^{\infty} \varphi(\E^{-\ell}) z^{\ell} }}^k,
\end{equation}
that is a little bit more cumbersome than the moment generating expansion Eq.\ \eqref{eq:MP_equation_mgf} or the cumulant expansion \eqref{eq:MP_equation_cgf}. Still, we get at leading order that
\begin{equation}
	\label{eq:SCM_inverse_moment_1}
	\varphi(\E^{-1}) = \frac{\varphi(\C^{-1})}{1-q}, \qquad \varphi(\E^{-2}) = \frac{\varphi(\C^{-2})}{(1-q)^2} + \frac{q\varphi(\C^{-1})^2}{(1-q)^3}.
\end{equation}
We will see in Section \ref{sec:markowitz} that the first relation (that can be found in \cite{burda2004signal}) has direct consequences for the out-of-sample risk of optimized portfolios. 

\begin{changemargin}{0.5cm}{0.5cm}
\footnotesize
Let us now give a formal derivation of Eq.\ \eqref{eq:MP_equation_Rtransform}. Let us define 
\begin{equation}
	\omega = \stj_\E(z), \qquad \zeta = \stj_\C(Z),
\end{equation}
which allows us to rewrite Eq.\ \eqref{eq:MP_equation_stieltjes} as
\begin{equation}
	\omega \btr_\E(\omega) = \zeta \btr_\C(\zeta), \qquad Z \equiv \btr_\C(\zeta) = \frac{\btr_\E(\omega)}{1-q+q \omega \btr_\E(\omega)}.
\end{equation}
Then, using the definition \eqref{eq:r_transform} of the $\rtr$-transform, we can rewrite this last equation as
\begin{equation}
	\omega \rtr_\E(\omega) = \zeta \rtr_\C(\zeta), \qquad \rtr_\C(\zeta) + \frac1\zeta = \frac{\rtr_\E(\omega) + 1/\omega}{1+q \omega \rtr_\E(\omega)}.
\end{equation}
We deduce that 
\begin{equation}
	\label{eq:tmp_MP_eq_rtr}
	\rtr_\C(\zeta) = \frac{\rtr_\E(\omega) + 1/\omega}{1+q \omega \rtr_\E(\omega)} - \frac1\zeta,
\end{equation}
which yields
\begin{equation}
	\omega \rtr_\E(\omega) = \zeta \pBB{ \frac{\rtr_\E(\omega) + 1/\omega}{1+q \omega \rtr_\E(\omega)} - \frac1\zeta }.
\end{equation}
By re-arranging the terms in this last equation, we obtain
\begin{equation}
	\omega \rtr_\E(\omega) + 1 = \frac{\zeta}{\omega} \pBB{ \frac{\omega\rtr_\E(\omega) + 1}{1+q \omega \rtr_\E(\omega)} },
\end{equation}
that is to say 
\begin{equation}
	\zeta \equiv \zeta(\omega)  = \omega\pb{1+q \omega \rtr_\E(\omega)},
\end{equation}
and Eq.\ \eqref{eq:MP_equation_Rtransform} immediately follows by plugging this last equation into Eq.\ \eqref{eq:tmp_MP_eq_rtr}.
\end{changemargin}
\normalsize

\subsubsection{Dual representation and edges of the spectrum}
\label{sec:MP_dual_representation}

Although a lot of information about the spectrum of $\E$ can be gathered from the Mar{\v c}enko-Pastur equation \eqref{eq:MP_equation_stieltjes}, the equation itself is not easy to solve analytically. 
In particular, what can be said about the edges of the spectrum of $\E$? We shall see that one can answer some of these questions by using a dual representation of Eq.\ \eqref{eq:MP_equation_stieltjes}. 

The ``dual'' representation that we are speaking about comes from studying the $T \times T$ matrix $\S$:
\begin{equation}
\label{eq:SCM_dual}
\S := \frac1T \Y^{*} \Y \equiv \X^{*} \C \X,
\end{equation}
where we used Eq.\ \eqref{eq:SCM_entries} in the last equation.  The dual matrix $\S$ can also be interpreted as a correlation matrix. In a financial context, $\E$ tells us how similar is the movement of two stocks
over time, while $\S$ tells us how similar are two dates in terms of the overall movements of the stocks on these two particular dates.  
Using a singular value decomposition, it is not difficult to show that $\S$ and $\E$ share the same non-zero eigenvalues -- hence the ``duality''. In the case where $T > N$, 
the matrix $\S$ has a zero eigenvalue with multiplicity $T-N$ in addition to the eigenvalues $\{\lambda_i\}_{i \in [\!1,N\!]}$ of $\E$. Therefore, it is easy to deduce the Stieltjes transform of $\S$:
\begin{equation}
\label{eq:stieltjes_E_dual}
\stj_{\S}(z) = \frac{1}{T} \left[ \frac{T-N}{z} + N \stj_{\E}(z) \right] = \frac{1-q}{z} + q \stj_{\E}(z) = \frac{1}{Z(z)}.
\end{equation}
The introduction of this dual representation of the empirical matrix allows one to get the following expression from Eq. \eqref{eq:MP_equation_int}:
\begin{equation*}
\stj_{\S}(z) = \frac1z \left( 1 - q + q \int \frac{\rho_{\C}(\mu) d\mu}{1 - \mu \stj_{\S}(z)}  \right).
\end{equation*}
After some manipulations, we can rewrite this last equation as 
\begin{equation}
	\label{eq:MP_equation_BaiSilverstein}
	z = \frac{1}{\stj_{\S}(z)} + q \int \frac{\rho_{\C}(\mu) d\mu}{\mu^{-1} - \stj_{\S}(z)}.
\end{equation}
Writing $\omega = \btr_{\S}(\stj_{\S}(z))$ in the above equation, we obtain a characterization of the functional inverse of $\stj_{\S}$ as 
\begin{equation}
\label{eq:MP_equation_dual_inverse} 
\btr_{\S}(\omega)\;:=\;  \frac{1}{\omega} + q \int \frac{\rho_{\C}(\mu) d\mu}{\mu^{-1} - \omega},
\end{equation}
and this is the dual representation of the Mar{\v c}enko-Pastur equation \eqref{eq:MP_equation_stieltjes}. 
The analytic behavior of this last equation has been the subject of several studies, especially in \cite{silverstein1995analysis}. In particular, it was proved that there exists a \textit{unique} $\omega \in \mathbb{C}_+$ that solves the equation \eqref{eq:MP_equation_dual_inverse}. This yields the Stieltjes transform of $\S$ from which we re-obtain the Stieltjes transform of $\E$ using Eq.\ \eqref{eq:stieltjes_E_dual}. We will see in the next section that the dual representation \eqref{eq:MP_equation_dual_inverse} of the Mar{\v c}enko-Pastur equation is particularly useful when we will try to solve the direct problem.

\begin{figure}[h]
	\begin{center}
   \includegraphics[scale = 0.55]{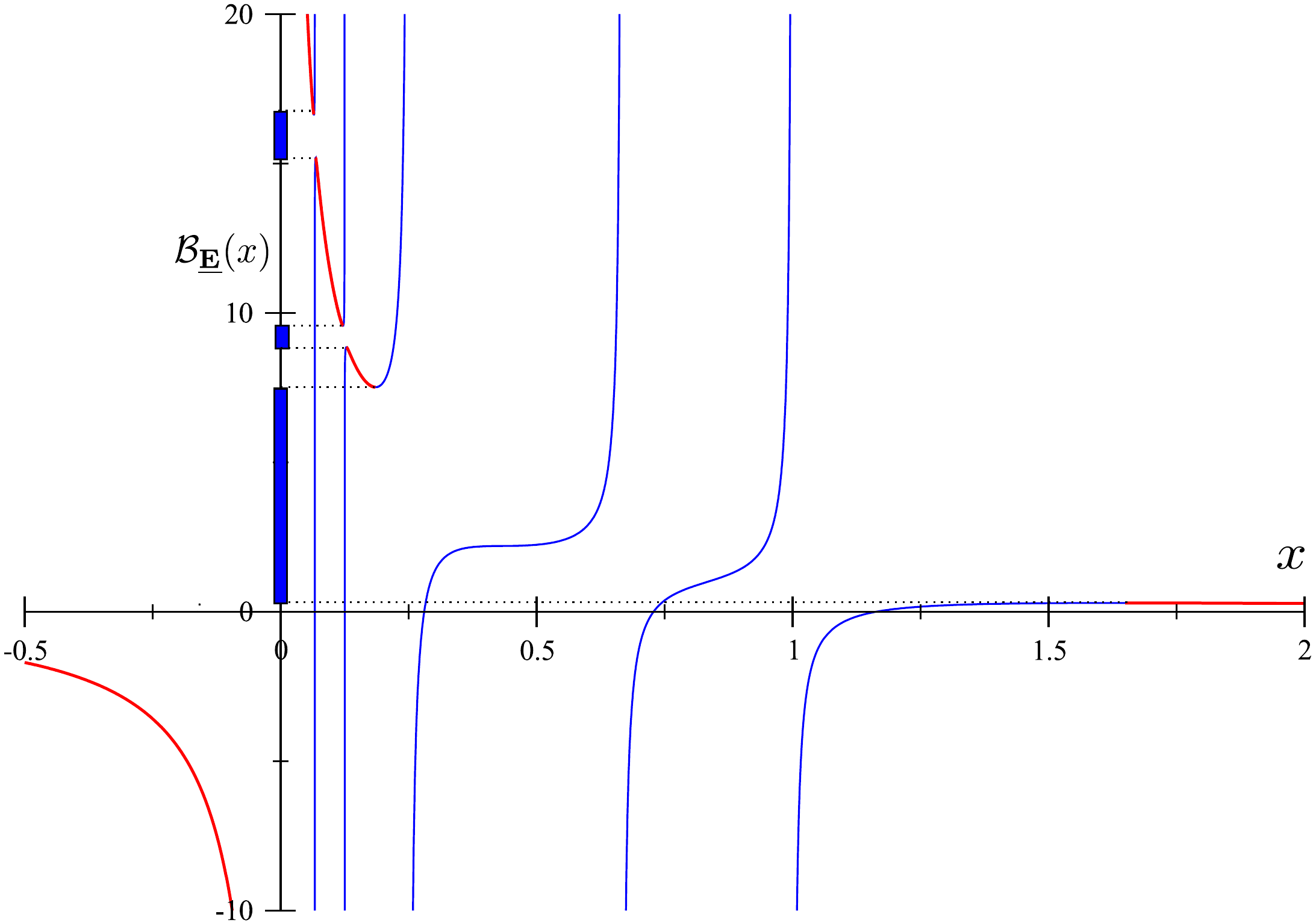} 
   \end{center}
   \caption{The function $\btr_{\underline \E}(x)$ with population eigenvalues density given by $0.002 \, \delta_{15} + 0.002 \, \delta_{8} + 0.396 \, \delta_{3} + 0.3 \, \delta_{1.5} + 0.3 \, \delta_{1} $. Here $T = 1000$, $N = 500$ \nc and we have $3$ connected components. The vertical asymptotes are located at each $- x^{-1}$ for $x \in \{1, 1.5, 3, 8, 15\}$. The support of $\rho_\S$ is indicated with thick blue lines on the vertical axis. The inverse of $\stj_\S \vert_{\R \setminus \supp \rho_\S}$ is drawn in red.}
   \label{f_sigma}
\end{figure}

In addition, the position of the edges of the LSD of $\E$ can be inferred from Eq.\ \eqref{eq:MP_equation_dual_inverse}. Within a one cut-assumption, the edges of the support of $\rho_\E$ are given by:
\begin{equation}
\label{bounds_empircal}
\lambda_{\pm}^{\E} = \btr_{\S}(\omega_{\pm}) \quad\text{where $\omega_{\pm} \in \mathbb{R}^+$ is such that } \btr_{\S}'(\omega_{\pm}) = 0.
\end{equation}
Indeed, knowing the spectral density of $\S$ allows us to get the spectral density of $\E$ since from Eq.\ \eqref{eq:stieltjes_E_dual} one gets:
\begin{equation}
	\rho_\S(\lambda) = q \rho_\E(\lambda) + (1-q)^{+} \delta_0,
\end{equation}
for any $\lambda \in \supp\rho_\S$. Next, one easily obtains
\begin{equation}
	\stj_\S'(z) = - \int \frac{\rho_\S(x) \dd x}{(z-x)^2} < 0,
\end{equation}
for any $z \not\in \supp[\rho_\S]$, meaning that it is strictly decreasing outside of the support. We saw in Section \ref{section:RMT_transforms} that the Stieltjes transform $\stj(z)$ is analytical and positive for any $z \in \mathbb{R}$ outside of the support. Moreover, for $z \to \infty$, we have $\stj_\S(z) \sim z^{-1} + \cal O(z^{-2})$ so that we deduce $\stj_\S(z)$ is a bijective decreasing function. Its inverse function $\btr_\S$ therefore also decreases in those same intervals. Consequently, the union of intervals where $\btr_\S(x)$ is decreasing  will lead to the complement of the support and the edges of the support of $\rho_\S$ are thus given by the critical points of $\btr_{\S}$, as in Eq. \eqref{bounds_empircal} above. If one assumes that there are a finite number $r$ of (non-degenerate) spikes, we can readily generalize the above arguments and find that there will be $2(r+1)$ critical points (see Figure \ref{f_sigma} for an illustration with two non-degenerate spikes).

\subsubsection{Solving Mar{\v c}enko-Pastur equation}
\label{sec:solving_MP}
%Practical considerations on the 

In this section, we investigate the direct problem of solving the Mar{\v c}enko-Pastur equation Eq.\ \eqref{eq:MP_equation_stieltjes} for $\stj_\E$ given $\stj_\C$. 
We will discuss briefly the inverse problem at the end of this section. 

\paragraph{Exactly solvable cases} As far as we know, there are only a few cases where we can find an explicit expression for the LSD of $\E$. The first one is trivial: it is when one considers the ``classical'' limit 
in statistics where $T \to \infty$ for a fixed value of $N$. In this case $q = 0$ in \eqref{eq:MP_equation_int}, and obviously $\stj_{\E}(z) = \stj_{\C}(z) $ in this case, as expected. 

\begin{figure}[!]
	\begin{center}
   \includegraphics[scale = 0.45]{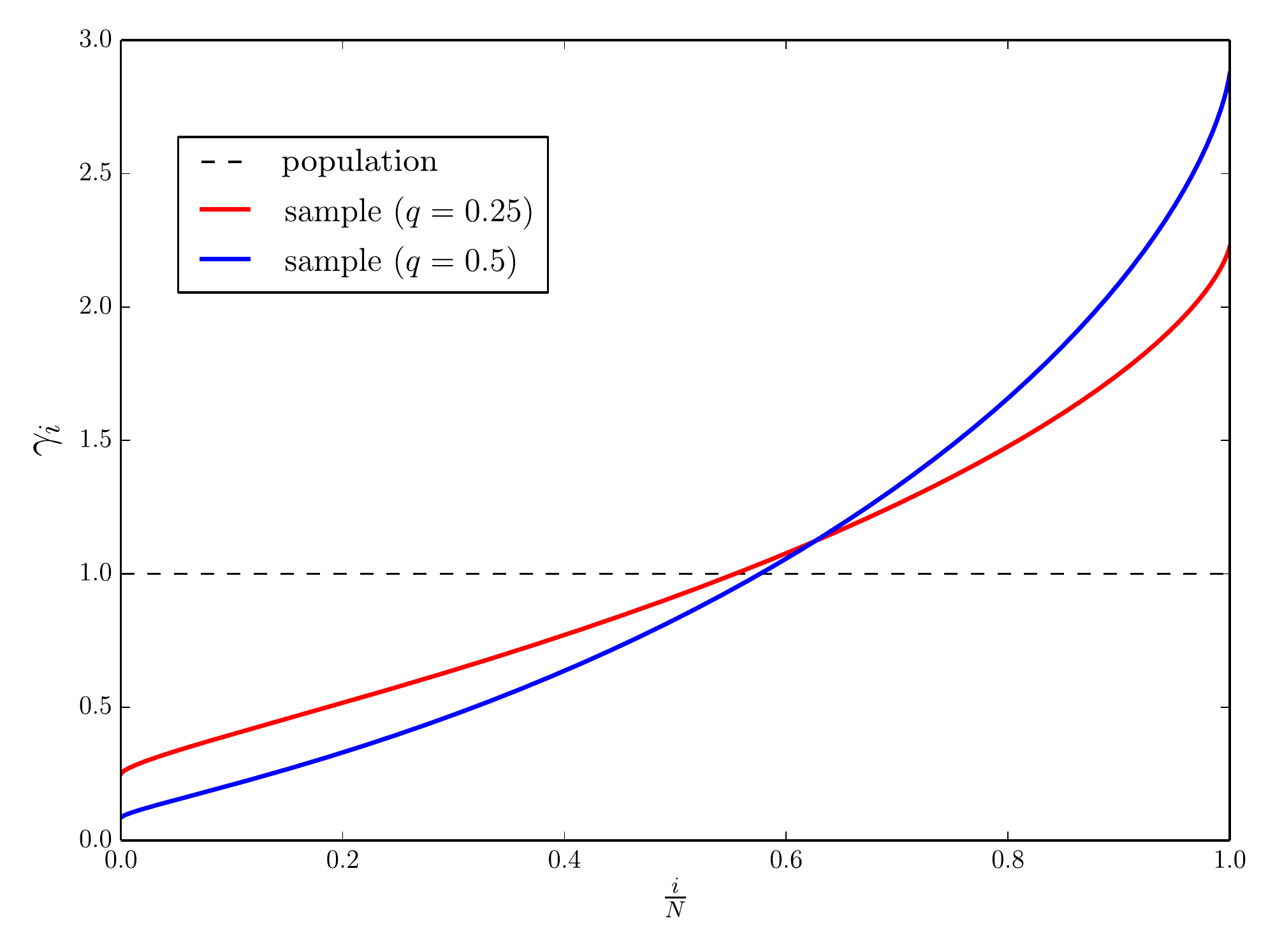} 
   \end{center}
   \caption{Typical position of the sample eigenvalues under the Mar{\v c}enko-Pastur law \eqref{eq:MP_density} with a finite observation ratio $q = 0.25$ (red line) and $q = 0.5$ (blue line). The dotted line corresponds to the locations of the population eigenvalues and we see a significant deviation. }
   \label{fig:MP_den_quantile}
\end{figure}

However, for any finite observation ratio $q > 0$, we anticipate from the discussion of Section \ref{sec:MP_spectrum_statistics} above that the LSD of $\E$ will be significantly different from that of $\C$. 
The influence of $q$ can be well understood in the simple case where $\C = \In$. We know from Section \ref{sec:MP_law} that this case is exactly solvable and the LSD of $\E$ is the well-known Mar{\v c}enko-Pastur law \eqref{eq:MP_density}, that we recall here:
\begin{equation}
	\stj_\E(z) = \frac{z+1-q - \sqrt{z -\lambda_-^{\text{mp}}}\sqrt{z - \lambda_+^{\text{mp}}}}{2qz}, \qquad \lambda_{\pm}^{\text{mp}} = (1\pm\sqrt{q})^2
\end{equation}
In words, the sample eigenvalues spans the interval $[(1-\sqrt{q})^2, (1+\sqrt{q})^2]$ while the population eigenvalues are all equal to unity. We therefore deduce that the variance of the sample eigenvalue distribution is order ${q}$, highlighting the systematic bias in the estimation of the eigenvalues using $\E$ when $q = \cal O(1)$. This effect can be visualized using the quantile representation of the spectral distribution. Indeed, it is known since \cite{bloemendal2014principal,knowles2014anisotropic} that the bulk eigenvalues $[\lambda_i]_{i\in \qq{r+1,N}}$ converge in the high-dimensional regime to their ``quantile positions'' $[\gamma_i]_{i \in \qq{r+1,N}}$. More precisely, this reads:
\begin{equation}
	\label{eq:classical_location_bulk}
	 \lambda_i \approx \gamma_i, \quad\text{where}\quad \frac{i}{N} = \int^{\gamma_i} \rho_\E(\lambda)\dd\lambda, \qquad i \geq r+1\,.
\end{equation}
We plot the $\gamma_i$'s of the Mar{\v c}enko-Pastur law in Fig.\ \ref{fig:MP_den_quantile} for $q=1/4$ and $q=1/2$, and observe systematic and significant deviations from the ``classical'' positions $\gamma_i^{q=0} \equiv 1$. This again illustrates that $\E$ is an untrustworthy estimator when the sample size is of the same order of magnitude as the number of variables. 

\begin{figure}[!]
	\begin{center}
   \includegraphics[scale = 0.45]{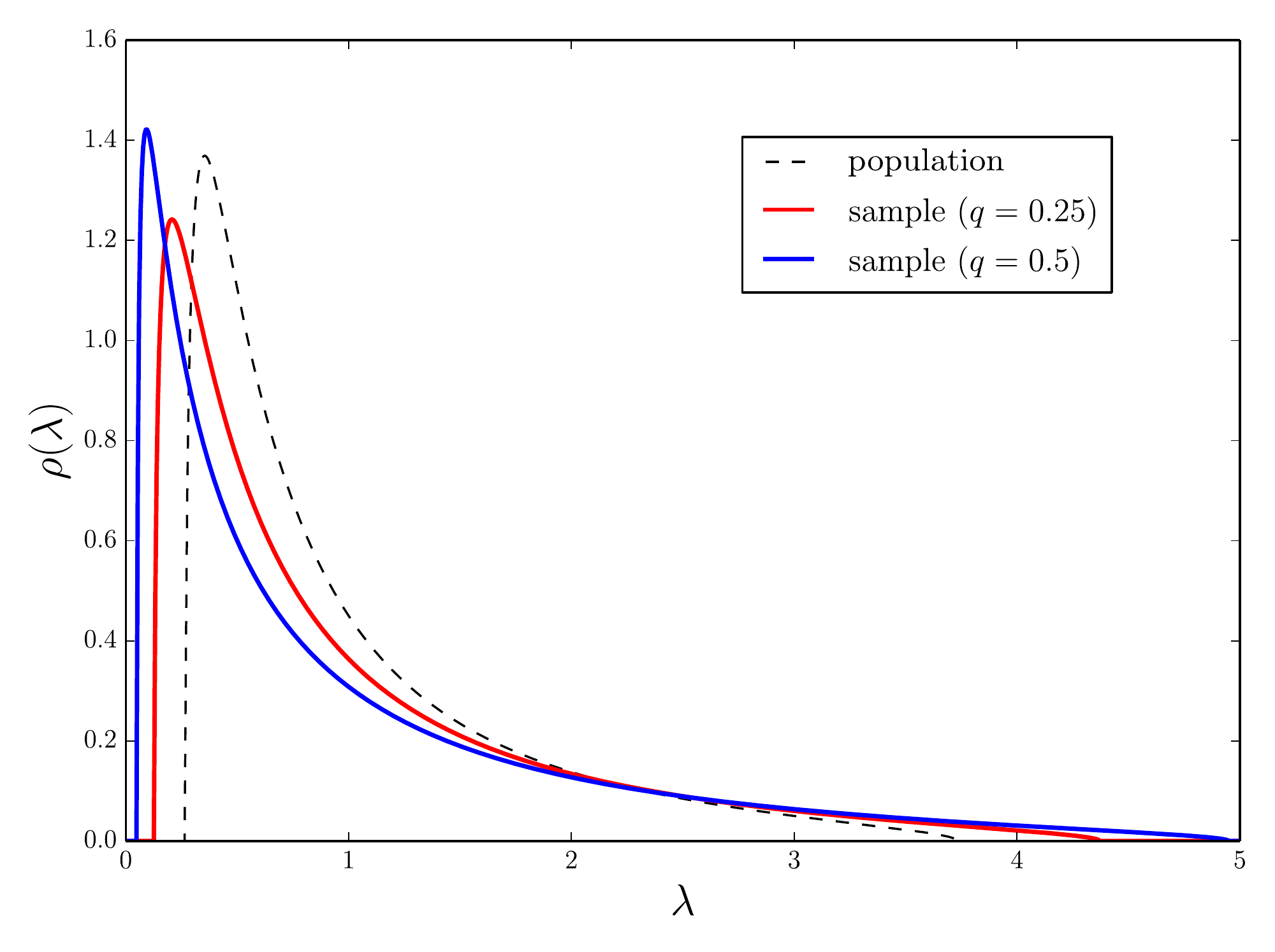} 
   \end{center}
   \caption{Solution of the Mar{\v c}enko-Pastur equation for the eigenvalue distribution of $\E$ when $\C$ is an inverse Wishart matrix with parameter $\kappa = 1.0$ for $q = 0.25$ (red line) and $q = 0.5$ (blue line). 
   The black dotted line corresponds to the LSD $\rho_\C$. }
   \label{fig:IMP_density}
\end{figure}

Now that the qualitative impact of the observation ratio $q$ is well understood, a natural extension would be to examine the Mar{\v c}enko-Pastur equation for a non trivial correlation matrix $\C$. To this aim, we 
now consider another interesting solvable case, especially for statistical inference, which is the case and of an (isotropic) inverse Wishart matrix with hyper-parameter $\kappa > 0$. 
From Section \ref{sec:IMP_law}, we recall that
\begin{equation*}
\str_{\C}(\omega) = 1 - \frac{\omega}{2\kappa},
\end{equation*}
for $\kappa > 0$. Then, using the free multiplication formula \eqref{eq:free_mult}, we have $\str_\E(\omega) = \str_\C(\omega) \str_{\Wishart}(\omega)$ where $\str_{\Wishart}(\omega)$ is given in \eqref{eq:S_transform_MP}, which yields a quadratic equation in $\ttr_{\E}(z)$. This implies that $\stj_\E$ reads:
\begin{equation}
\label{eq:stieltjes_IW_MP}
\stj_{\E}(z) = \frac{ z(1+\kappa) - \kappa(1-q) \pm \sqrt{ (\kappa(1-q) - z(1+\kappa))^2 - z(z+2q\kappa)(2\kappa +1)}}{z(z+2q\kappa)},
\end{equation}
from which we can retrieve the edges of the support:
\begin{equation}
\label{eq:edges_IW_MP}
\lambda_{\pm}^{\text{iw}} = \frac{1}{\kappa}\left[ (1+q)\kappa + 1 \pm \sqrt{ (2\kappa+1)(2q\kappa+1)} \right].
\end{equation}
One can check that the limit $\kappa \to \infty$ recovers the null hypothesis case $\C = \In$; the lower $\kappa$, the wider the spectrum of $\C$. We plot in Figure \ref{fig:IMP_density} the spectral density $\rho_\C$ and $\rho_\E$ for $q = 0.25$ and $q=0.5$ as a function of the eigenvalues. Again, we see that the spectral density of $\E$ puts significant weights on regions of the real axis which are outside the support of $\rho_\C$, due to the measurement noise. %Hence, it is not surprising that we obtain the same conclusion as above whether we compare the typical locations of $\rho_\C$ and $\rho_\E$ (see Fig.\ \ref{fig:IMP_quantile_diff}). 
From an inference theoretic viewpoint, the interest of the Inverse-Wishart ensemble is to provide a parametric prior distribution for $\C$ where everything can be computed analytically (see Chapter \ref{chap:bayes} below for some applications).

\begin{changemargin}{0.5cm}{0.5cm}
\footnotesize

There exist several other examples where the Mar{\v c}enko-Pastur equation is exactly solvable even though the Stieltjes transform is not explicit. For instance, if we consider $\C$ to be a Wishart matrix of parameter $q_0$ independent from ${\cal \W}$, then we have from \eqref{eq:free_mult} that 
\begin{equation*}
	\str_\E(\omega) = \frac{1}{(1+q_0\omega)(1+q\omega)}. 
\end{equation*}
It is then easy to see from the definition \eqref{eq:s_transform} that $\ttr_\E(z) \equiv \omega(z)$ is solution of the cubic equation, 
\begin{equation}
	z (1+\omega(z))(1+q_0\omega(z))(1+q\omega(z)) - \omega(z) = 0,
\end{equation}
from which we obtain $\stj_\E(z)$ thanks to \eqref{eq:T_transform} and by choosing the unique solution of the latter equation in $\mathbb{C}^+$ (see the following section for details on this point). Another toy example that uses the Mar{\v c}enko-Pastur with the $\rtr$-transform formalism is when $\C$ is a GOE centered around the identity matrix. In this case we have
\begin{equation}
	\rtr_\C(\omega) = 1+\sigma^2\omega,
\end{equation}
where we add the constraint $\sigma \leq 0.5$ such that $\C$ remains a positive semi-definite matrix. Then, by plugging this formula into \eqref{eq:MP_equation_Rtransform}, we find that $\stj_\E(z) = \omega$ is the solution of quartic equation:
\begin{equation}
	\sigma^2 \omega^2 (1+q\omega\rtr_\E(\omega))^2 + \omega (1+q\omega\rtr_\E(\omega)) - \omega\rtr_\E(\omega) = 0,
\end{equation}
and as above, we take the unique solution in $\mathbb{C}^+$ in order to get the right Stieltjes transform. 
\end{changemargin}
\normalsize

% ****I would suppress this fig (not much new info)***
% \begin{figure}[!]
% 	\begin{center}
%    \includegraphics[scale = 0.4]{Figures/spectrum/IMP_quantile_diff} 
%    \end{center}
%    \caption{Comparison of the ``typical locations'' $[\mu_i]_{i\in\qq{1,N}}$ and $[\gamma_i]_{i\in\qq{1,N}}$ defined in Eq.\ \eqref{eq:classical_location_bulk}. We depict the difference between $[\gamma_i]_{i\in\qq{1,N}}$ and $[\mu_i]_{i\in\qq{1,N}}$, which explains why the population values lies at the origin. }
%    \label{fig:IMP_quantile_diff} \lambda_0
% \end{figure}

% The fact that we obtain an explicit formula for $G_{\E}(z)$ seems almost miraculous and to our knowledge, this is the only case aside from $\C = \In$ where we can solve exactly Mar{\v c}enko-Pastur equation in great generality. Nonetheless, from a statistician point of view, considering the inverse Wishart as the underlying distribution of the true $\C$ (``\textit{prior}'') is not that surprising as it turns out to be a cornerstone in Bayesian statistics on covariance matrices thanks to its analytical tractability. Hence, finding a close formula for $G_{\E}(z)$ is finally not that miraculous. Again, we stress that this result should be of particular interest in high-dimensional Statistics and might allows to adopt a Bayesian point of view regarding the Mar{\v c}enko-Pastur equation. We will come back to this last observation in much more details in section VII. \\

\textbf{The general case: numerical method} 
%In our attempt to estimate $\C$, a natural (but naive) solution is to find $\rho_{\C}$ (or $G_{\C}(z)$) such that we replace the sample eigenvalues $[\lambda_i]_{i\in\qq{1,N}}$ by the population ones. Since the indirect problem is %quite complicated to handle, adopting a ``Bayesian'' viewpoint is a possible way to overcome this problem. However, this means that we should be able to solve Eq.\ \eqref{eq:MP_equation_stieltjes} for an arbitrary true population %covariance matrix $\C$ in order to make it relevant.
Apart from the very specific cases discussed above, finding an explicit expression for $\stj_{\E}(z)$ is very difficult. This means that we have to resort to numerical schemes in order to solve the Mar{\v c}enko-Pastur equation. In that respect, the dual representation \eqref{eq:MP_equation_dual_inverse} of Eq.\ \eqref{eq:MP_equation_stieltjes} comes to be particularly useful. 
To solve the MP equation for a given $z$, we seek a certain $\stj \equiv \stj_{\S}$ such that\footnote{Recall that $\S$ is the $T\times T$ equivalent of $\E$ defined in Eq.\ \eqref{eq:SCM_dual}. }
\begin{equation}
z = \btr_{\S}(\stj), \qquad \stj \in \mathbb{C}_+,
\end{equation}
where the expression of $\btr_\S$ in terms of $\rho_\C$ is explicit and given in Eq.\ \eqref{eq:MP_equation_dual_inverse}. Numerically, the above equation is easily solved using
a simple gradient descent algorithm, i.e. find $\stj \in \mathbb{C}_+$ such that
\begin{equation}
	\label{eq:MP_numerical_scheme}
	\begin{cases}
		\re(z) = \re\qb{\btr_\S(\stj)} \\
		\im(z) = \im\qb{\btr_\S(\stj)}\,.
	\end{cases}
\end{equation}
It then suffices to use Eq.\ \eqref{eq:stieltjes_E_dual} in order to get $\stj_\E(z)$ for any $z \in \mathbb{C}_{-}$. Hence, if one wants to retrieve the eigenvalues density $\rho_\E$ at any point on the real line, we simply have to set $z = \lambda - \ii \e$ with $\lambda \in \Supp(\E)$ and $\e$ an arbitrary small real positive  number into Eq.\ \eqref{eq:MP_numerical_scheme}. Note that in the case where $\stj_{\C}$ is known, one can rewrite equation (\ref{eq:MP_equation_dual_inverse}) as
\begin{equation}
\label{eq:MP_equation_dual_inverse_v2}
\btr_{\S}(x) = \frac{1}{x} \left[ 1-q + \frac{q}{x} \stj_{\C}\left(\frac{1}{x}\right) \right],
\end{equation}
which is obviously more efficient since we avoid to compute the integral over eigenvalues. 

\begin{figure}[!ht]
	\begin{center}
   \includegraphics[scale = 0.5]{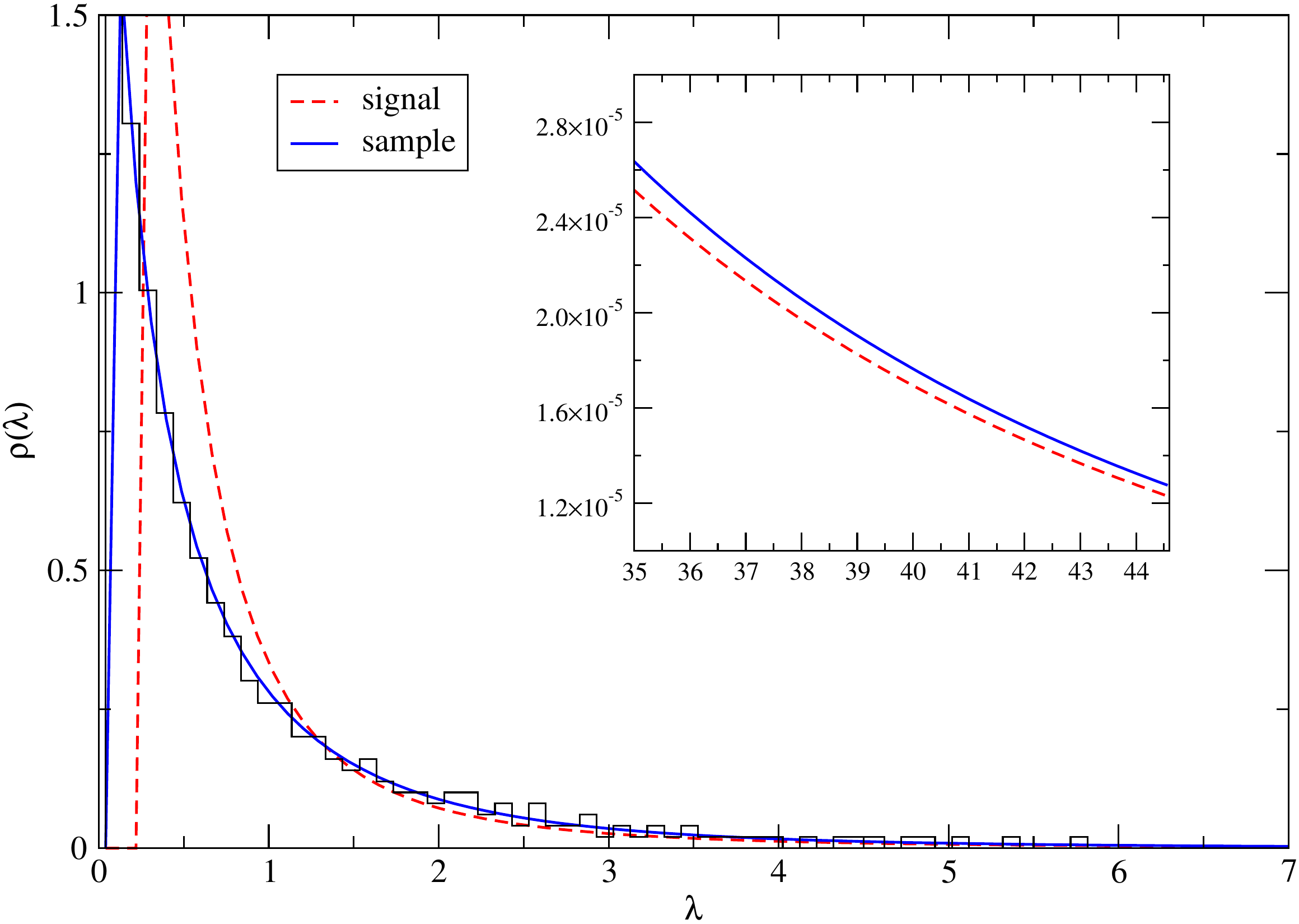} 
   \end{center}
   \caption{Resolution of the Mar{\v c}enko-Pastur equation when $\rho_{\C}$ is given a power law density with parameter $\lambda_0 = 0.3$ and a finite observation ratio $q = 0.5$ and $N=500$. The dotted line corresponds to the LSD of $\C$ while the plain line corresponds to the LSD of $\E$. The histogram is the ESD when we compute $\E$ from the definition \eqref{eq:SCM}.  The main figure covers the bulk of the eigenvalues while the inset zooms in the region of very large eigenvalues. }
   \label{fig:MP_powerlaw}
\end{figure}

In order to illustrate this numerical scheme, let us consider a covariance matrix whose LSD has a heavy right tail. One possible parametrization is to assume a power-law distribution of the form \cite{bouchaud2009financial}:
\begin{equation}
\label{eq:density_powerlaw}
\rho_{\C}(\lambda) = \frac{s A}{(\lambda + \lambda_{0})^{1+s}} \Theta(\lambda - \lambda_{\min}),
\end{equation}
where $\Theta(x)=x^+$ is the Heaviside step function, $s$ is an exponent that we choose to be $s=2$ \cite{bouchaud2009financial}, and $\lambda_{\min}$ the lower edge of the spectrum below which there are no eigenvalues of $\C$. 
$A,\lambda_{\min}$ are then determined by the two normalization constraints $\int \rho_{\C}(x)\dd x = 1$ and $\int x \rho_{\C}(x) \dd x = 1$. This leads to: $\lambda_{\min} = (1-\lambda_0)/2$ and $A = (1-\lambda_{\min})^2$. We restric to $\lambda_0 >-1 1$ such that $\lambda_{\min}< 1$. 
From the density Eq. \eqref{eq:density_powerlaw}, one can perform the Stieltjes transform straightaway to find
\begin{equation}
\label{eq:stieltjes_powerlaw}
\stj_{\C}(z) = \frac{1}{z + 1 - 2\lambda_0} + \frac{2(1-\lambda_0)}{(z + 1 - 2\lambda_0)^2} + \frac{2(1-\lambda_0)^2}{(z + 1 - 2\lambda_0)^3} \left[ \log \left( \frac{\lambda_0 - z}{1-\lambda_0} \right) \right],
\end{equation}
which allows one to solve Eq.\ \eqref{eq:MP_equation_dual_inverse_v2} for $\stj_{\E}(z)$ with only a few iterations. As we observe in Fig. \ref{fig:MP_powerlaw}, the theoretical value obtained from the numerical scheme \eqref{eq:MP_numerical_scheme} agrees perfectly with the empirical results, obtained by diagonalizing matrices of size $N=500$ matrices obtained as $\sqrt{\C} \Wishart \sqrt{\C}$, where $\Wishart$ is a Wishart matrix.  This illustrates the robustness of the above numerical scheme, even when the spectrum of $\C$ is fat-tailed. In addition, we can notice that the more we add structure in the true covariance $\C$, the wider is the empirical distribution as in the above case, where the spectrum of $\E$ embraces nearly all the positive real number line.

\subsection{Edges and outliers statistics}

\begin{figure}[!]
	\begin{center}
   \includegraphics[scale = 0.5]{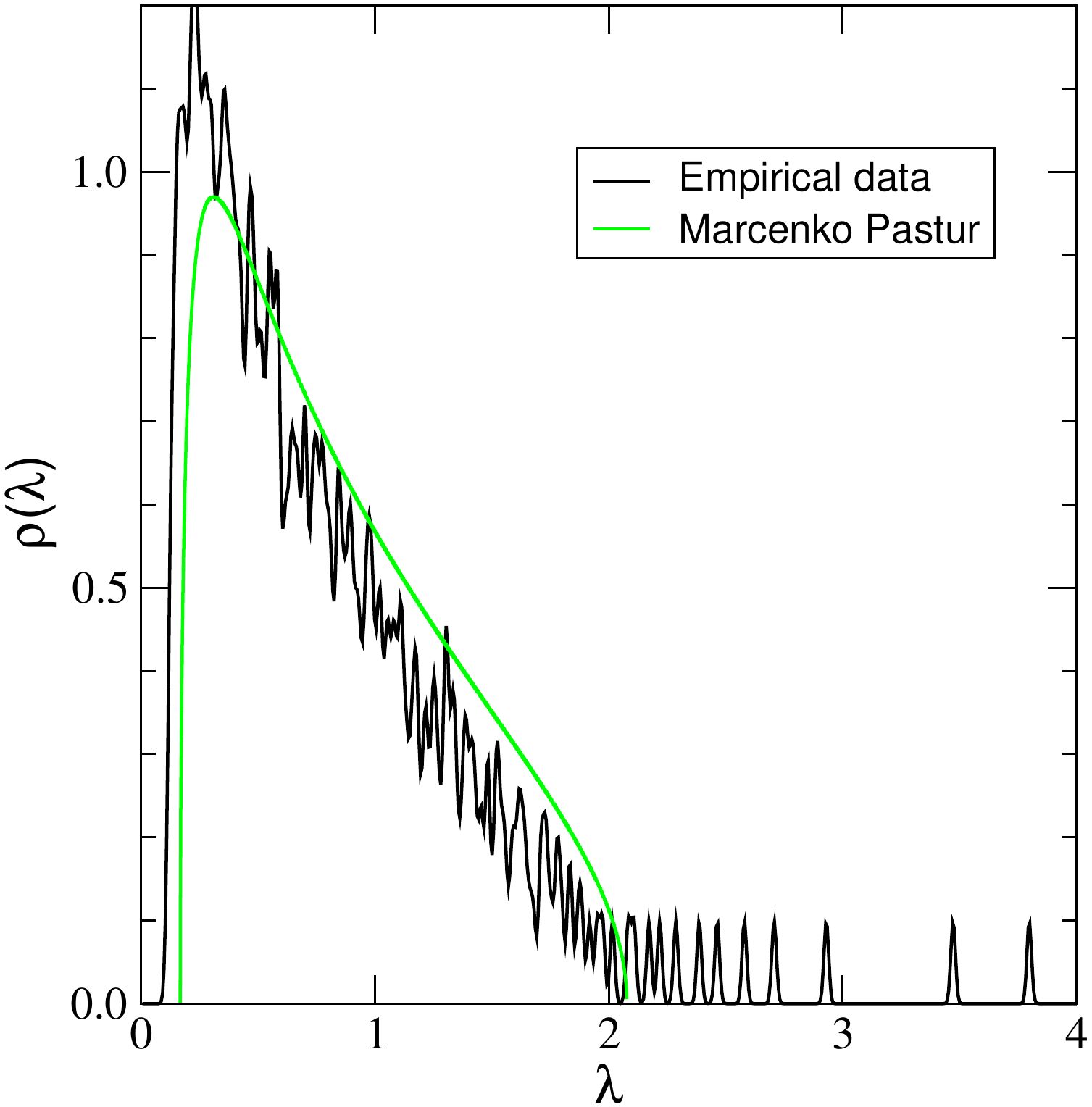} 
   \end{center}
   \caption{Test of the null hypothesis on the empirical correlation matrix $\E$ using US stocks' data with $N = 406$ and $T = 1300$.}
   \label{fig:eigen_justMP}
\end{figure}

As we alluded to several times above, the practical usefulness of the theoretical predictions for the eigenvalue spectra of
random matrices is (i) their universality with respect to the distribution of the underlying random variables and
(ii) the appearance of sharp edges in the spectrum, meaning that the existence of eigenvalues lying outside the 
allowed region is a possible indication against simple ``null hypothesis'' benchmarks. Illustrating the last point, Fig.\ \ref{fig:eigen_justMP}
shows the empirical spectral density of the correlation matrix corresponding to $N=406$ and $T = 1300$ so that $q \approx 0.31$, compared to the simplest 
Mar{\v c}enko-Pastur spectrum in the null hypothesis case $\C=\b I_N$. While the bulk of the distribution is roughly accounted for (but see Section
\ref{sec:past_cleaning} for a much better attempt), there seems to exist a finite number of eigenvalues lying outside the Mar{\v c}enko-Pastur sea, which may be called outliers
or spikes. However, even if there are no such spikes in the spectrum of $\C$, one expects to see, for finite $N$ some eigenvalues beyond the Mar{\v c}enko-Pastur upper edge. 
The next two subsections are devoted first to a discussion of these finite size effects, and then to a model with ``true'' outliers that survive in the
large $N$ limit. 

\subsubsection{The Tracy-Widom region}
\label{sec:edges}

This existence of sharp edges delimiting a region where one expects to see a non zero density of eigenvalues from a region where there should be none is only true in the asymptotic $N, T \to \infty$, and in the absence 
of ``fat-tails'' in the distribution of matrix elements (see \cite{arous2008spectrum,biroli2007top}). For large but finite $N$, on the other hand, one expects that the probability to find an eigenvalue beyond the Mar{\v c}enko-Pastur
sea is very small but finite. The width of the transition region, and the tail of the density of states was investigated already a while ago \cite{bowick1991universal}, culminating in the beautiful results by Tracy \& Widom on
the distribution of the {\it largest} eigenvalue of a random matrix \cite{tracy1994level}. The Tracy-Widom result is actually a very nice manifestation of the universality phenomenon that describes the fluctuations of macroscopic observables in many large dimensional systems (see the recent paper \cite{majumdar2014top} on this topic). The derivation of the Tracy-Widom distribution mainly relies on Orthogonal polynomials that we will not discuss in this review (see e.g. \cite{tracy1994level,nadal2011simple}) but there also exists an alternative approach \cite{ramirez2011beta}. The link between this limiting law and the largest eigenvalue of large sample covariance matrices has been subject to a large amount of studies that we will not attempt to cover here (see e.g. \cite{johnstone2001distribution,dean2006large,majumdar2006random,johansson2000shape,baik2005phase,peche2009universality} for details and references).

The Tracy-Widom result characterizes precisely the distance between the largest eigenvalue $\lambda_{1}$ of $\E$ and the upper edge of the spectrum that we denoted by
$\lambda_+$. This result can be (formally) stated as follows: the rescaled distribution of $\lambda_{1}-\lambda_+$
converges towards the Tracy-Widom distribution, usually noted $F_1$,
\begin{equation}
	\label{eq:TW}
	\cal P\left(\lambda_{1} \leq \lambda_+ + \gamma N^{-2/3} u\right)=F_1(u),
\end{equation}
where $\gamma$ is a constant that depends on the problem. For the isotropic Mar\v{c}enko-Pastur problem, $\lambda_+ =(1+\sqrt{q})^2$ and $\gamma=\sqrt{q}\lambda_+^{2/3}$, whereas for the Wigner problem, $\lambda_+=2$ and $\gamma=1$. We stress that this result holds for a large class of $N \times N$ matrices (e.g. symmetric random matrices with IID elements of a finite fourth
moment, see \cite{biroli2007top,arous2008spectrum}).

% The behaviour of the width of the transition region can be understood using a simple heuristic argument. Suppose that
% the $N=\infty$ density goes to zero near the upper edge $\lambda_+$ as $(\lambda_+-\lambda)^\theta$ (generically, 
% $\theta=1/2$ as is the case for the Wigner and the Mar\v{c}enko-Pastur distributions). For finite $N$, one expects
% not to be able to resolve the density when the probability to observe an eigenvalue is smaller than $1/N$. This criterion reads:
% \begin{equation}
% (\lambda_+-\lambda^*(N))^{\theta+1} \propto \frac1N \to \Delta \lambda^* \sim N^{-\frac{1}{1+\theta}},
% \end{equation}
% or a transition region that goes to zero as $N^{-2/3}$ in the generic case. More precisely, for Gaussian ensembles, the average 
% density of states at a distance $\sim N^{-2/3}$ from the edge behaves as:
% \begin{equation}
% \rho_N(\lambda \approx \lambda_+) = N^{-1/3} \Phi\left[N^{2/3}(\lambda-\lambda_+)\right],
% \end{equation}
% with $\Phi(x \to -\infty) \propto \sqrt{-x}$ as to recover the asymptotic density of states, and $\ln \Phi(x \to +\infty) \propto x^{3/2}$,
% showing that the probability to find an eigenvalue outside of the allowed band decays exponentially with $N$ and super exponentially with 
% the distance to the edge. 

Everything is known about the Tracy-Widom density $f_1(u)=F_1'(u)$, in particular its left and right far tails:
\begin{equation}
\ln f_1(u) \propto -u^{3/2}, \quad (u \to +\infty); \qquad \ln f_1(u) \propto -|u|^{3}, \quad (u \to -\infty);
\end{equation}
One notices that the left tail is much thinner:
pushing the largest eigenvalue inside the allowed band implies compressing the whole Coulomb gas of repulsive charges, which
is difficult. Using this analogy, the large deviation regime of the Tracy-Widom problem (i.e. for $\lambda_{1} - \lambda_+= \cal O(1)$)
can also be obtained \cite{dean2006large}.

Note that the distribution of the smallest eigenvalue $\lambda_{\min}$ around the lower edge $\lambda_-$ is also
Tracy-Widom, except in the particular case of Mar\v{c}enko-Pastur matrices with $q=1$. In this case, $\lambda_-=0$
which is a `hard' edge since all eigenvalues of the empirical matrix must be non-negative. This special case is treated
in, e.g. \cite{peche2003universality}.

\subsubsection{Outlier statistics} 
\label{sec:outlier_eigenvalues}

Now, there are cases where a finite number of eigenvalues genuinely reside outside the Mar{\v c}enko-Pastur sea (or more generally outside of the bulk region) even when $N \to \infty$. For example, the empirical data
shown in Fig.\ \ref{fig:eigen_justMP} indeed suggests the presence of true outliers, that have a real financial interpretation in terms of economic sectors of activity. Therefore, we need a framework to describe 
correlation matrices that contain both a bulk region and a finite number of spikes. The purpose of this section is to study the statistics of these eigenvalues from an RMT point of view. 

The standard way to treat outliers is to ``blow out'' a finite number of eigenvalues of a given (spikeless) correlation matrix $\underline\C$, that we construct as:
\begin{equation}
	%\label{eq:spikeless population eigenvalue}
	\underline\C = \sum_{i=1}^{N} \ul{\mu}_i \b v_i \b v_i^*,\quad\text{where}\quad \ul{\mu}_i = \begin{cases}
 		\mu_{0} & \text{if } i \leq r \\
 		\mu_i & \text{if } i \geq r+1 \,.
 	\end{cases}
\end{equation}
We choose the eigenvalue $\mu_{0}$ within the spectrum of $\underline\C$ such that there is no outliers initially. Here we fix $\mu_0 = \mu_{r+1}$ for simplicity, but any other choice in the set $[\mu_i]_{i\geq r+1}$ would do equally well. Then with this prescription, we may rewrite $\C$ as a small rank perturbation of $\ul\C$. Indeed, since each outlier $[\mu_i]_{i \leq r}$ are well separated from the bulk by assumption, 
we may parametrize each spike $\mu_i$ by a positive real number $d_i$ for any $i \leq r$ as follows:
\begin{equation}
	\label{eq:outlier_parametrization}
	\mu_i = \mu_{0}(1+d_i) \equiv \mu_{r+1}(1+d_i), \qquad d_i > 0\;,\; i \leq r.
\end{equation}
Hence, the population covariance matrix $\C$ is given by:
\begin{equation}
	\label{eq:spikeless population eigenvalue}
	\C = \sum_{i=1}^{N} {\mu}_i \b v_i \b v_i^*,\quad\text{where}\quad {\mu}_i = \begin{cases}
 		\mu_{0}(1+d_i)  & \text{if } i \leq r \\
 		\mu_i & \text{if } i \geq r+1 \,.
 	\end{cases}
\end{equation} 
More synthetically, one can write $\C$ as:
\begin{equation}
	\label{eq:spikeless_population_cov}
	\C = \underline \C \pb{\b I_N + \b V^{(r)} \b D \b V^{(r)*} },
\end{equation}
where $\b V^{(r)} \deq [\b v_1, \dots, \b v_r] \in \mathbb{R}^{N\times r}$ and $\b D \deq \diag(d_1, \dots, d_r)$ is a diagonal matrix that characterizes the spikes. 
We also define a fictitious spikeless sample covariance matrix as $\ul{\E} = \ul\C^{1/2} \X \X^* \ul\C^{1/2}$ and denote by $\underline{\S}  = \X^* \ul\C \X$ the $T \times T$ ``dual'' matrix. 
As noticed in \cite{bun2016optimal}, the statistics of the outliers of $\E$ can be investigated through that of $\ul{\E}$. Let us consider the rank-one $r=1$ case for the sake of simplicity 
(see \cite{bun2016optimal} for the general case). Then, we have
\begin{equation*}
\det(z \In - \E) = \det( z\In - \X^* \ul\C (\In + d_1 \b v_1 \b v_1^*) \X) = \det(z \In - \X\X^* \ul\C(\In + d_1 \b v_1 \b v_1^*)).
\end{equation*}
which can be transformed into: 
\begin{equation}
\det(z \In - \E) = \det( z\In - \ul\E) \det( \In - d_1 (z\In - \ul\E)^{-1} \b v_1 \b v_1^* \E) 
%\nonumber \\
\end{equation}
We can conclude that $\lambda_1$ in an eigenvalue of $\E$ and not of $\ul\E$ if and only if the second determinant vanishes, i.e. if $d_1 (\lambda_1\In - \ul\E)^{-1} \b v_1 \b v_1^* \ul \E$ has an eigenvalue equals to unity. To find $\lambda_1$, we remark that this second determinant is simply a rank-one update, meaning that it has only one non-trivial eigenvalue given by the equation:
\begin{equation}
\label{eq:clasical_location_equation}
d_1 \left[ \lambda_1 \langle \b v_1, \b G_{\ul\E}(\lambda_1) \b v_1 \rangle - 1 \right] = 1,
\end{equation}
where $\b G_{\ul\E}$ is the resolvent of $\ul \E$. The difficult part of \eqref{eq:clasical_location_equation} is to find an (asymptotic) expression for the scalar product $\langle \b v_1, \b G_{\ul\E} \b v_1 \rangle$. Let us assume without loss of generality\footnote{The extension to non-Gaussian entries can be done using standard comparison techniques, see e.g. \cite{knowles2014anisotropic} for details.} that $\C$ is Gaussian, which allows us to arbitrarily set $\b v_1 = (1,0,\dots,0)$. Then the equation we try to solve is:
\begin{equation}
	\label{eq:classical_location_equation_rank1}
\lambda_1 \b G_{\ul\E}(\lambda_1)_{11} = d_1^{-1}  + 1.
\end{equation}
As we shall see in the next section, the entries of $\G_{\ul\E}$ actually converges to a deterministic quantity for $N \to \infty$ and one obtains using Eq.\ \eqref{eq:global_law_SCM_diag} (see \eqref{eq:SCM_green_fct} for an alternative derivation). The result reads
\begin{equation*}
 	\b G_{\ul\E}(z)_{11} \approx \frac{1}{z-\ul\mu_1 (1-q+qz \stj_{\ul\E}(z))} = \frac{1}{z(1-\mu_{r+1} \stj_{\ul\S}(z))},
\end{equation*} 
where we used the identity \eqref{eq:stieltjes_E_dual} and that $\ul\mu_1 \equiv \mu_{r+1}$ by construction of \eqref{eq:spikeless_population_cov} in the last step. If $\lambda_1$ is not an eigenvalue of $\ul\E$, we find that Eq.\ \eqref{eq:classical_location_equation_rank1} becomes in the LDL
\begin{equation}
	\frac{1}{1-\mu_{r+1} \stj_{\ul\S}(\lambda_1)} = d_1^{-1} + 1,
\end{equation}
which is equivalent to:
\begin{equation}
	\stj_{\ul\S}(\lambda_1) = \frac{1}{\mu_{r+1}(1+d_1)} \equiv \frac{1}{\mu_1},
\end{equation}
where we used \eqref{eq:outlier_parametrization} in the last step. \nc Hence, we see that $\lambda_1$ is an outlier if it satisfies for large $N$:
\begin{equation}
\label{eq:classical_location_outlier}
\lambda_1 = \theta(\mu_1) \deq \btr_{\underline \S}\left( \frac{1}{\mu_1} \right),
\end{equation}
This result is very general and can be extended for any outlier $\lambda_i$ with $i \in \qq{1,r}$.  Moreover, we see that for $N \to \infty$, the (random) outlier $\lambda_1$ converges to a deterministic function of $\mu_1$. Hence, the function \eqref{eq:classical_location_outlier} depicts the ``classical location'' at which an outlier sticks and can therefore be interpreted as the analog of \eqref{eq:classical_location_bulk} for outliers. Note however that \eqref{eq:classical_location_outlier}  requires the knowledge of the spikeless matrix $\ul\S$ (or $\ul\E$). In practice, one should make some assumptions 
to decide whether a given empirical eigenvalue should be considered as a spike. 

The result \eqref{eq:classical_location_outlier} generalizes the result of Baik-Ben Arous-P{\'e}ch{\'e} for the spiked covariance matrix model \cite{baik2005phase}. Indeed, let us assume that the eigenvalues of the true covariance matrix $\C$ is composed of one outlier and $N-1$ eigenvalues at unity. Then, one trivially deduces that $\ul{\mu}_i = 1$ for all $i = 1, \dots, N$ which implies that the spectrum of $\ul\E$ is governed by the Mar{\v c}enko-Pastur law \eqref{eq:MP_density}. In fact, in the limit $N \to \infty$, the spectrum of $\ul\E$ and $\E$ are equivalent since the perturbation is of finite rank.  Therefore, we can readily compute the Blue transform of the dual matrix $\underline\S$ from \eqref{eq:MP_equation_dual_inverse} to find
\begin{equation}
\label{eq:blue transform spiked cov}
	\btr_{\underline \S}(x) = \frac{1}{x} + \frac{q}{1-x}.
\end{equation}
Applying this formula to Equation \eqref{eq:classical_location_outlier} then leads to the so-called BBP phase transition 
\begin{equation}
	\label{eq:spiked_cov_eigenvalue_phase_trans}
	\begin{cases}
		\lambda_{1} = \mu_1 + q \frac{\mu_1}{\mu_1-1} & \text{if } \mu_1>1+\sqrt{q};\\
		\lambda_{1} = \lambda_+ = (1+\sqrt{q})^2 & \text{if } \mu_1 \leq 1+\sqrt{q},
	\end{cases}
\end{equation}
where $\mu_1= \mu_0(1+d_1)$ is the largest eigenvalue of $\C$, which is assumed to be a spike. Note that in the limit $\mu_1 \to  \infty$, we get $\lambda_{1} \approx \mu_1 + q + \cal O(\mu_1^{-1})$. For rank $r$ perturbation, all eigenvalues such that $\mu_k>1+\sqrt{q}$, $1 \leq k \leq r$ will end up isolated above the Mar\v{c}enko-Pastur sea, all others disappear below $\lambda_+$. All these isolated eigenvalues have Gaussian fluctuations of order $T^{-1/2}$ \cite{baik2005phase}. The typical fluctuation of order $T^{-1/2}$ is also true for an arbitrary $\C$ \cite{bun2016optimal}, and is much smaller than the uncertainty in the bulk of the distribution, of order $\sqrt{q}$. Note that a naive application of Eq.\ \eqref{eq:MP_equation_stieltjes} to outliers would lead to a ``mini-Wishart'' distribution around the top eigenvalue, which incorrect (the distribution is Gaussian) except if the top eigenvalue has a degeneracy proportional to $N$.

\clearpage%!TEX root = RMT_Covariance_Review.tex
\section{Statistics of the eigenvectors}
\label{chap:eigenvectors}

We saw in the previous chapter that tools from RMT allow one to infer many properties of the (asymptotic) spectrum of $\E$, be it for the bulk or for more localized regions of the spectrum (edges and outliers). These results allow us to characterize in great detail the statistics of the eigenvalues of large sample covariance matrices. In particular, it is clear that in the high-dimensional limit, the use of sample covariance matrices is certainly not recommended 
as each sample eigenvalue $[\lambda_i]_{i\in\qq{N}}$ converges to a non-deterministic value, but this value is different from the corresponding ``true'' population eigenvalue $[\mu_i]_{i\in\qq{N}}$. Note that the results presented above only cover a 
small part of the extremely vast literature on this topic, including the study microscopic/local statistics (down to the $N^{-1}$ scale) \cite{perret2015finite,wirtz2013distribution,knowles2014anisotropic,hachem2015survey}. 

On the other hand, results concerning the eigenvectors are comparatively scarce. One reason is that most studies in RMT focus on rotationally invariant ensembles, such that the statistics of eigenvectors is featureless by definition. Notwithstanding, this question turns out to be very important for sample covariance matrices since in this case, the direction of the eigenvectors of the ``population'' matrix must somehow leave a trace. 
There are, at least, two natural questions about the eigenvectors of the sample matrix $\E$:
\begin{enumerate}
  \item How similar are sample eigenvectors $[\b u_i]_{i\in\qq{N}}$ and the true ones $[\b v_i]_{i\in\qq{N}}$?
  \item What information can we learn about the population covariance matrix by observing two independent realizations -- say $\E=\sqrt{\C} \b{\cal W} \sqrt{\C}$ and $\E'=\sqrt{\C} \b{\cal W}' \sqrt{\C}$ -- that remain correlated through $\C$?
\end{enumerate}

The aim of this chapter is to present some of the most recent results about the eigenvectors of large sample covariance matrices that will allow us to answer these two questions. 
More precisely, we will show how the tools developed in Section \ref{chap:RMT} can help us extract the statistical features of the eigenvectors $[\b u_i]_{i\in\qq{1,N}}$. 
Note that we will discuss these issues for a multiplicative noise model (see \eqref{eq:model_freemult} above), but the same questions can be investigated for additive noise as well, see \cite{benaych2011eigenvalues,knowles2014anisotropic,allez2014eigenvectors,allez2013eigenvectors,bun2016overlaps} and Appendix \ref{app:addition}. 

% In particular, we will ask ourselves two different questions:
% \begin{enumerate}
%   \item Can we characterize the effect of the noise on the eigenvectors? Differently said, how do the sample eigenvectors deviate from the population ones? 
%   \item Does the asymptotic relation between eigenvectors coming from of two independent realizations $\E$ and $\tilde \E$ of $\C$ converge to a deterministic quantity? 
% \end{enumerate}
% These two questions yields different consequences. For question (i), we ask ourself how much information is retained by the sample eigenvectors concerning the population ones. For question (ii), we rather want to know what is the relation between the eigenvectors of two independent realizations of $\C$ should satisfy. For instance, this would provide a statistical test regarding the stability of the population covariance matrix $\b C$ upon times. 

A natural quantity to characterize the similarity between two arbitrary vectors -- say $\b \xi$ and $\b \zeta$ -- is to consider the scalar product of $\b \xi$ and $\b \zeta$. More formally, we define the ``overlap'' as
$\scalar{\b \xi}{\b \zeta}$. Since the eigenvectors of real symmetric matrices are only defined up to a sign, we shall in fact consider the squared overlaps $\scalar{\b \xi}{\b \zeta}^2$. In the first problem alluded to above, we want to understand the relation between the eigenvectors of the population matrix 
$[\b v_i]_{i\in\qq{N}}$ and those of the sample matrix $[\b u_i]_{i\in\qq{N}}$. The matrix of squared overlaps is defined as $\scalar{\b u_i}{\b v_j}^2$, it forms a so-called bi-stochastic matrix (positive elements with the sums over both rows and columns all equal to unity). 

In order to study these overlaps, the central tool of this chapter will be the resolvent (and not its normalized trace as in the previous section). Indeed, if we choose the $\b v$'s to be our reference basis, we find from \eqref{eq:resolvent_decomposition}:
\begin{equation}
  \label{eq:gen_entries_G}
  \scalar{\b v}{\b G_\E(z) \b v} = \sum_{i=1}^{N} \frac{\scalar{\b v}{\b u_i}^2}{z - \lambda_i},
\end{equation}
for $\b v$ a deterministic vector in $\mathbb{R}^N$ of unit norm. Note that we can extend the formalism to more general entries of $\G_\E(z)$ of the form:
\begin{equation}
\label{eq:gen_entries_G2}
  \scalar{\b v}{\b G_\E(z) \b v'} = \sum_{i=1}^{N} \frac{\scalar{\b v}{\b u_i}\scalar{\b u_i}{\b v'}}{z - \lambda_i},
\end{equation}
for $\b v$ and $\b v'$ two unit norm deterministic vectors in $\mathbb{R}^N$. 

We see from Eqs.\ \eqref{eq:gen_entries_G} and \eqref{eq:gen_entries_G2} that each pole of the resolvent defines a projection onto the corresponding sample eigenvectors. This suggests that the techniques we need to apply are very 
similar to the ones used above to study of the density of states. However, one should immediately stress that contrarily to eigenvalues, each eigenvector $\b u_i$ for any given $i$ continues to fluctuate when $N \to \infty$,\footnote{Recall that we have indexed the eigenvectors by their associated eigenvalue.} and never reaches a deterministic limit. As a consequence, we will need to introduce some averaging procedure to obtain a well defined result. We will thus consider the following quantity,% \cor , \nc
\begin{equation}
\label{eq:overlap}
  %\cor \mso(\lambda, \mu) \;\deq\; \mathbb{E}\qBB{\frac1N \sum_{i,j=1}^N \scalar{\b u_i}{\b v_j}^2 \delta(\lambda - \lambda_i) \delta(\mu-\mu_j) } \nc
  \mso(\lambda_i, \mu_j) \;\deq\; N \mathbb{E}[ \langle \b u_i, {\b v}_j \rangle^2 ],
  %N \mathbb{E}[ \langle \b u_i, \b v_j \rangle^2 ],
\end{equation}
where the expectation $\mathbb{E}$ can be interpreted either as an average over different realizations of the randomness or, perhaps more meaningfully for applications, as an average {\it for a fixed sample} over small intervals of eigenvalues of width $\dd\lambda = \eta $ that we choose in the range $1 \gg \eta \gg N^{-1}$ (say $\eta = N^{-1/2}$) such that there are many eigenvalues in the interval $\dd\lambda$, while keeping $\dd\lambda$ sufficiently small for the spectral density to be constant. Interestingly, the two procedures lead to the same result for large matrices, i.e. the locally ``smoothed'' quantity $\mso(\lambda, \mu)$ is self averaging. We emphasize that we consider the population eigenvectors to be deterministic throughout this section. Only the sample eigenvectors are random. Note also the factor $N$ in the definition above, indicating that we expect typical square overlaps to be of order $1/N$, see below. 

For the second question, the main quantity of interest is, similarly, the (mean squared) overlap between two independent noisy eigenvectors
\begin{equation}
  \label{eq:mso}
  %\cor \smso(\lambda, \tilde \lambda) \;\deq\; \mathbb{E}\qBB{\frac1N \sum_{i,j=1}^N \scalar{\b u_i}{\tilde{\b u}_j}^2 \delta(\lambda - \lambda_i) \delta(\tilde\lambda-\tilde \lambda_j) } \nc
  \mso(\lambda_i, \tilde \lambda_j) \;\deq\; N \mathbb{E}[ \langle \b u_i, \tilde{\b u}_j \rangle^2 ],
\end{equation}
where $[\tilde \lambda_i]_{i\in\qq{N}}$ and $[\tilde{\b u}_i]_{i\in\qq{N}}$ are the eigenvalues and eigenvectors of $\tilde \E$, i.e. another sample matrix that is independent from $\E$ (but with the same underlying population matrix $\C$). 

We end this introduction with a short remark on the somewhat vague definitions \eqref{eq:overlap} and \eqref{eq:mso}. As explained above, we  index the eigenvectors by their corresponding eigenvalues and this allows us to consider the continuous limit of \eqref{eq:overlap}. However, a more precise definition 
should be that $\mso(\lambda, \mu) \deq \mathbb{E}\qb{N^{-1} \sum_{i,j=1}^{N} \scalar{\b u_i}{\tilde{\b b}_j}^2 \delta(\lambda-\lambda_i) \delta(\mu-\mu_i)}$ 
but we keep the notation \eqref{eq:overlap}, with a slight abuse of notation, as it will be more convenient to separate the analysis between an outlier or bulk eigenvalue. We emphasize that this remark also holds for the overlaps \eqref{eq:mso} as well. 

\subsection{Asymptotic eigenvectors deformation in the presence of noise}
\label{sec:eigenvectors_sample_true}

We consider in this section the first question, that is: can we characterize the effect of the noise on the eigenvectors? Differently said, how do the sample eigenvectors deviate from the population ones? In order to answer to this question, Eq.\ \eqref{eq:overlap} seems to be a good starting point since it allows one to extract exactly the projection of the sample eigenvectors onto the population ones.  We shall now show that Eq.\ \eqref{eq:overlap} converges to a deterministic quantity in the large $N$ limit; more precisely, we can summarize the main results of this section as follows:
\begin{enumerate}
  \item Any bulk sample eigenvectors is \emph{delocalized} in the population basis, i.e. $\mso(\lambda_i, \mu_j) \sim \cal O(1)$ (and not $\cal O(N)$) for any $i \in \qq{r+1,N}$ and $j \in \qq{N}$;
  \item For any outlier (i.e. $i \leq r$), $\b u_i$ is concentrated within a cone with its axis parallel to $\b v_i$ but is completely delocalized in any direction orthogonal to the spike direction $\b v_i$.
\end{enumerate} 
Therefore, these results look quite disappointing for a inference standpoint. Indeed, for the bulk eigenvectors, we discover that projection the estimated eigenvectors and their corresponding ``true'' directions converges almost surely to zero for large $N$; i.e. sample eigenvectors appear to contain very little information about the true eigenvectors (on this point, see however \cite{monasson2015estimating}). Still, as we will see below, the squared overlaps 
are not all equal to $1/N$ but some interesting modulations appear, that we compute below by extending the Mar{\v c}enko-Pastur equation to the full resolvent. For the outliers, on the other hand, the global picture is quite different. In particular, the phase transition phenomenon alluded in section \ref{chap:spectrum} above also holds for the projection of the sample spike eigenvector onto its parent population spike: as soon as an eigenvalue pops out from the bulk, 
the square overlap becomes of order $1$, as noticed in e.g. \cite{paul2007asymptotics,benaych2011eigenvalues,hoyle2003limiting}. In fact, the angle between the sample spike eigenvectors with the parent spike can be computed exactly, see below.

\subsubsection{The bulk}
\label{sec:eigenvectors_sample_true_bulk}

%\subsubsection{Resolvent and overlaps}

%\paragraph{Deterministic Equivalent for resolvents.} 

Let us focus on the bulk eigenvectors first, i.e. eigenvectors associated to eigenvalues lying in the bulk of the spectral density when the dimension of the empirical correlation matrix grows to infinity. This question has been investigated very recently in \cite{ledoit2011eigenvectors,bun2015rotational} and we repeat the different arguments here. The first step is to characterize the asymptotic behavior of the resolvent of sample covariance matrices. This can be done by specializing Eq. \eqref{eq:global_law_freemult} for the resolvent of the product of free matrices to the case where $\b A = \C$ and $\b B = \b X \b X^*$. In words, $\b A$ is the population matrix while $\b B$ is a white Wishart matrix that plays the role of the noisy multiplicative perturbations. Using \eqref{eq:S_transform_MP}, we know the $\str$-transform of white Wishart matrices explicitly so that one finds from Eq.\ \eqref{eq:S_transform_MP}, for $N \to \infty$:
\begin{equation}
\label{eq:global_law_SCM}
z {\b G}_{\E}(z)_{ij} = Z(z) {\b G}_{\C}(Z(z))_{ij}, \quad\text{with }\quad Z = \frac{z}{1-q+qz \stj_{\E}(z)}.
\end{equation}
In the literature, such a limiting result is referred to as  a ``deterministic equivalent'', as the RHS depends only on deterministic quantities\footnote{Recall that $\stj_\E(z)$ is the \emph{limiting} Stieltjes transform.}, and this is another evidence of the self-averaging property for large random matrices. 

\begin{figure}[!]
\begin{subfigure}{.5\textwidth}
  \centering
  \includegraphics[width= 1\linewidth]{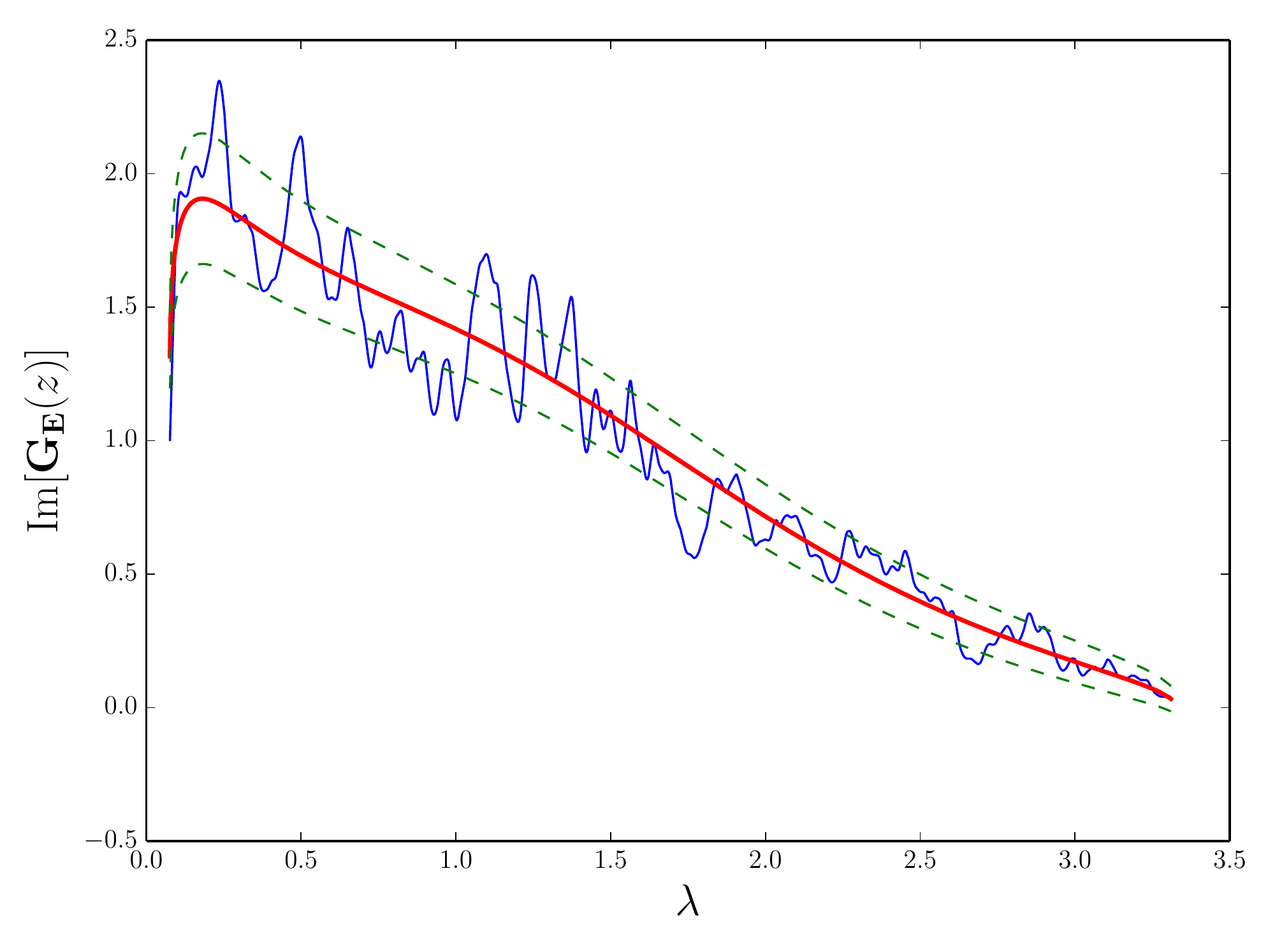}
  \caption{Diagonal entry of $\im[\G_\E(z)]$ with $i = j = 1000$. }
  \label{fig:multiple}
\end{subfigure}%
\begin{subfigure}{.5\textwidth}
  \centering
  \includegraphics[width= 1.0\linewidth]{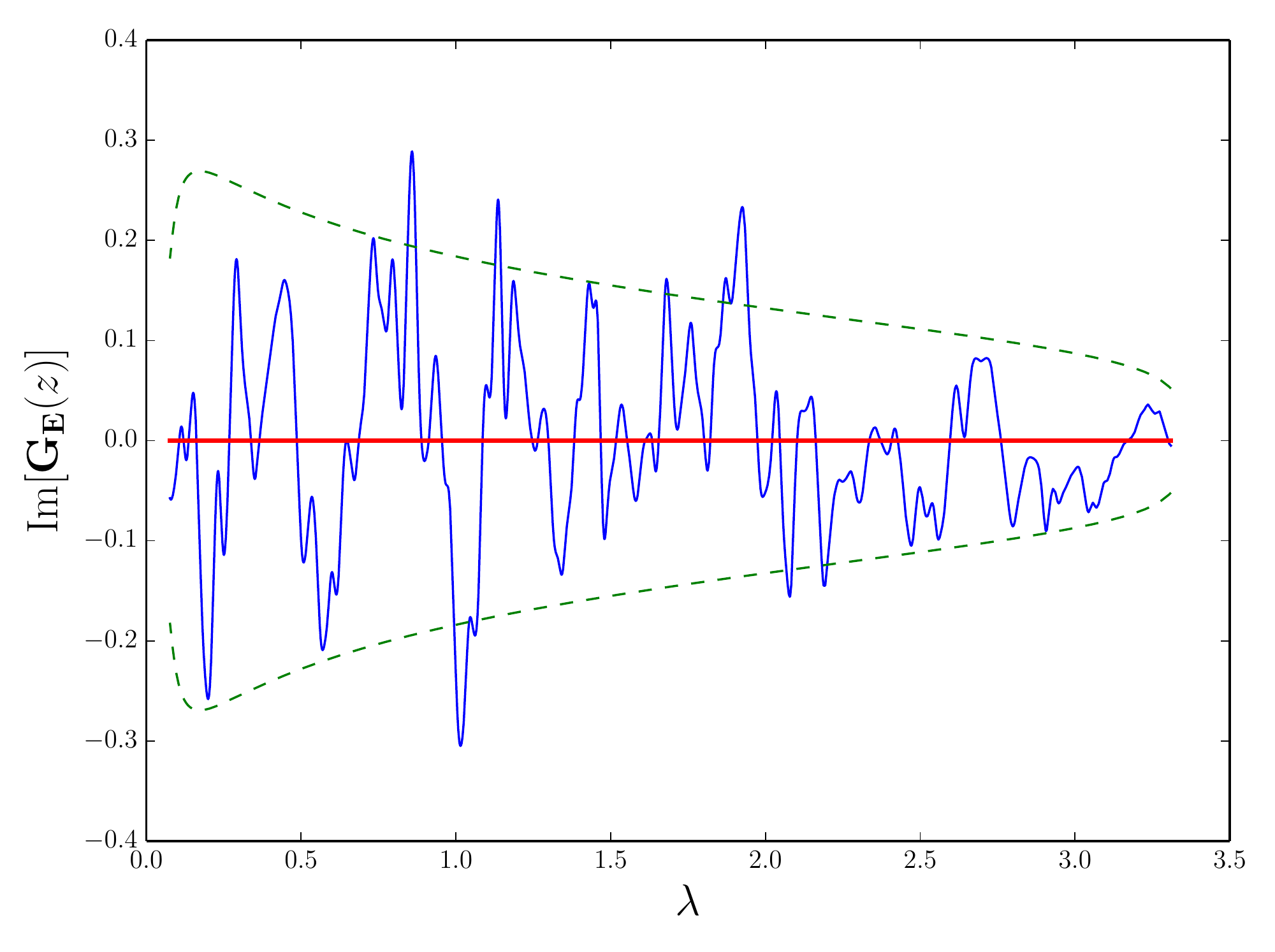}
  \caption{Off diagonal entry of $\im[\G_\E(z)]$ with $i = 999$ and $j = 1001$. }
  \label{fig:dGOE}
\end{subfigure}\\
\begin{subfigure}{.5\textwidth}
  \centering
  \includegraphics[width= 1.0\linewidth]{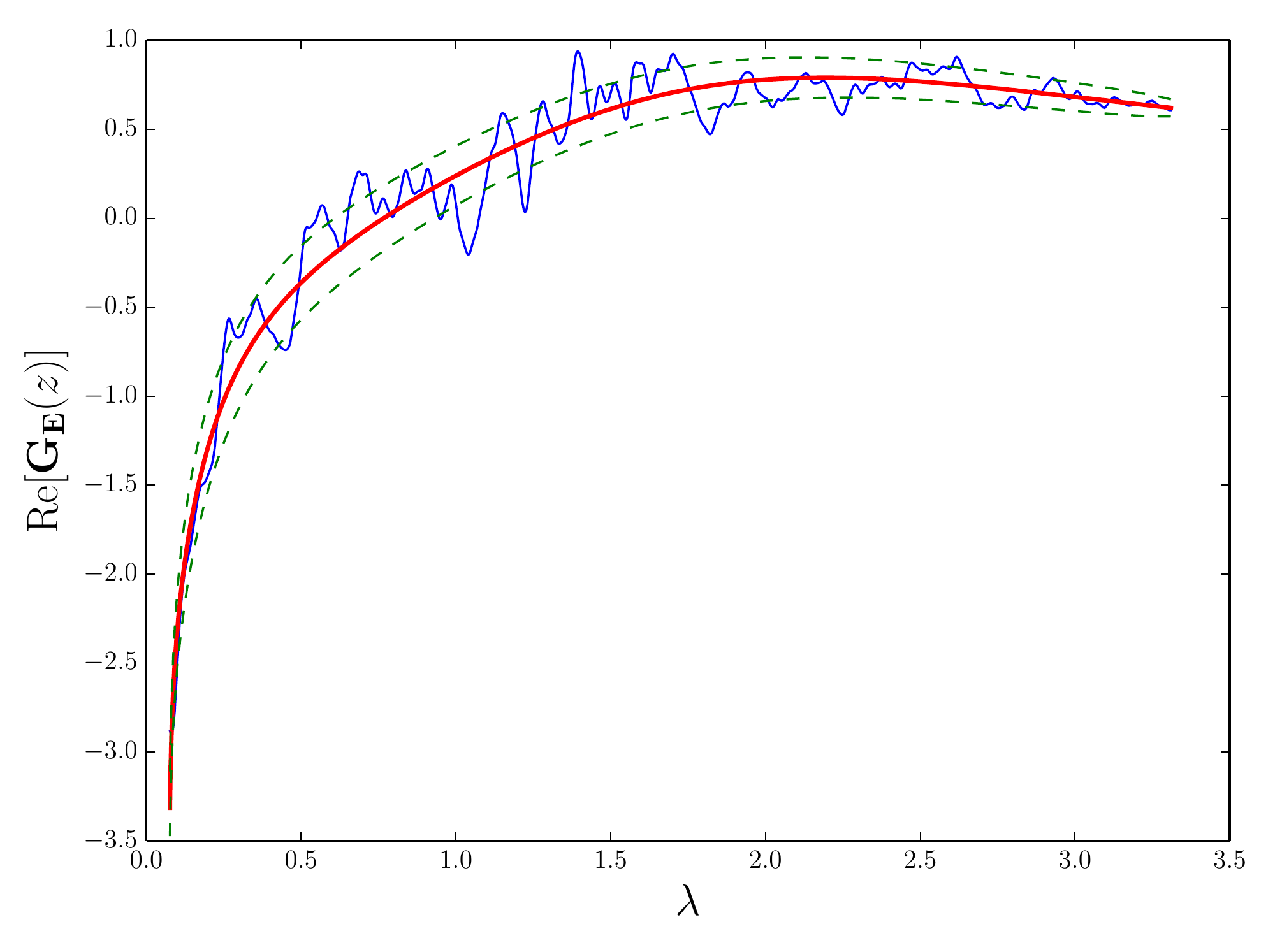}
  \caption{Diagonal entry of $\re[\G_\E(z)]$ with $i = j = 1000$.}
  \label{fig:multiple}
\end{subfigure}%
\begin{subfigure}{.5\textwidth}
  \centering
  \includegraphics[width= 1.0\linewidth]{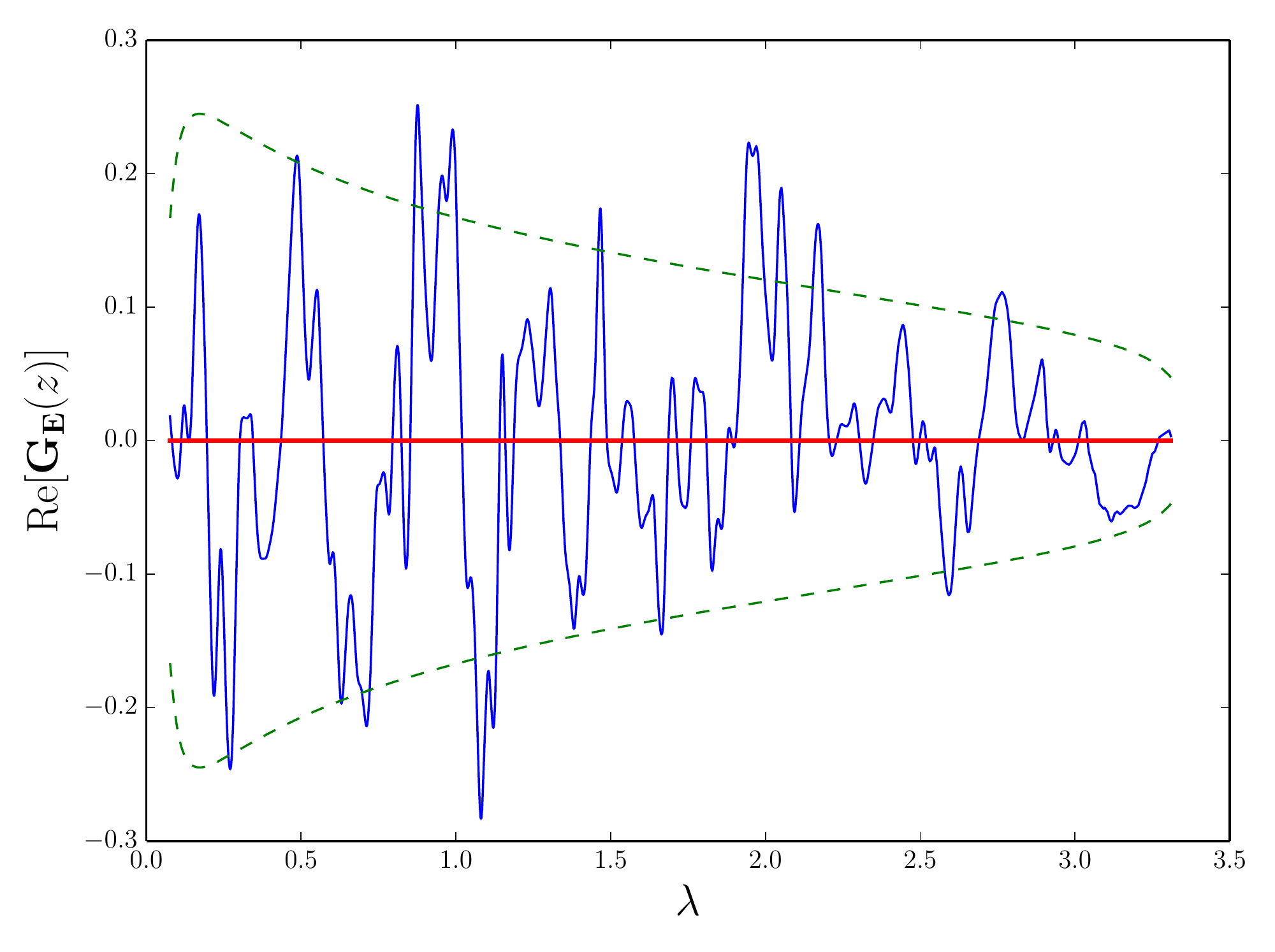}
  \caption{Off diagonal entry of $\re[\G_\E(z)]$ with $i = 999$ and $j = 1001$. }
  \label{fig:dGOE}
\end{subfigure}
\caption{Illustration of Eq.\ \eqref{eq:global_law_SCM_diag}. The population matrix is an Inverse Wishart matrix with parameter $\kappa = 5$ and the sample covariance matrix is generated using a Wishart distribution with $T = 2N$ and $N = 2000$. The empirical estimate of $\G_\E(z)$ (blue line) is computed for any $z = \lambda_i - \ii N^{-1/2}$ with $i \in \qq{1,N}$ comes from one sample and the theoretical one (red line) is given by the RHS of Eq.\ \eqref{eq:global_law_SCM}. The green dotted corresponds to the confidence interval whose formula is given by Eq.\ \eqref{eq:global_law_SCM_error_term}. 
%v The X-axis denotes the sample eigenvalues and the Y-axis the \emph{cleaned} eigenvalues.
}
\label{fig:resolvent_entries}
\end{figure}

One should notices that \eqref{eq:global_law_SCM} is a relation between resolvent matrices that generalizes the scalar Mar{\v c}enko-Pastur equation \eqref{eq:MP_equation_stieltjes} (which can be recovered by taking the trace on both sides of the equation). This relation first appeared in \cite{burda2004signal}, obtained using a planar diagram expansion valid for Gaussian entries. A few years later, that result was proven rigorously in Ref. \cite{knowles2014anisotropic} in a much more general framework, highlighting again the \emph{universal} nature of the resolvent of random matrices, down to the local scale.\footnote{Note that the Gaussian assumption is not needed either within the Replica method presented in Section \ref{chap:RMT}.} Choosing to work in the basis where $\C$ is diagonal, Eq.\ \eqref{eq:global_law_SCM} reduces to:
\begin{equation}
\label{eq:global_law_SCM_diag}
{\b G}_{\E}(z)_{ij} =  \frac{\delta_{ij}}{z - \mu_{i} (1-q+qz \stj_{\E}(z))}.
\end{equation}
This deterministic equivalent holds with fluctuations of order $N^{-1/2}$. This can be deduced e.g. from the Central Limit Theorem (CLT) (see Appendix \ref{app:recursion}). Quite interestingly, 
an explicit upper bound for the error term is provided in \cite{knowles2014anisotropic}. In particular, the authors showed that Eq.\ \eqref{eq:global_law_SCM} holds at a local scale $\eta = \widehat \eta N^{-1}$ with $\widehat \eta \gg 1$, with an error term bounded from above by:
\begin{equation}
  \label{eq:global_law_SCM_error_term}
  \Psi(z) \;\deq\; \sqrt{q\frac{\im \stj_\S(z)}{\widehat \eta}} + \frac{q}{\widehat \eta},
\end{equation}
provided that $N$ is large enough. We give an illustration of this ergodic behavior in Figure \ref{fig:resolvent_entries}, and we see the agreement is excellent. 

%\paragraph{Inversion formula for Eq. \eqref{eq:overlap}}

\begin{figure}[!ht]
  \begin{center}
   \includegraphics[scale = 0.45]{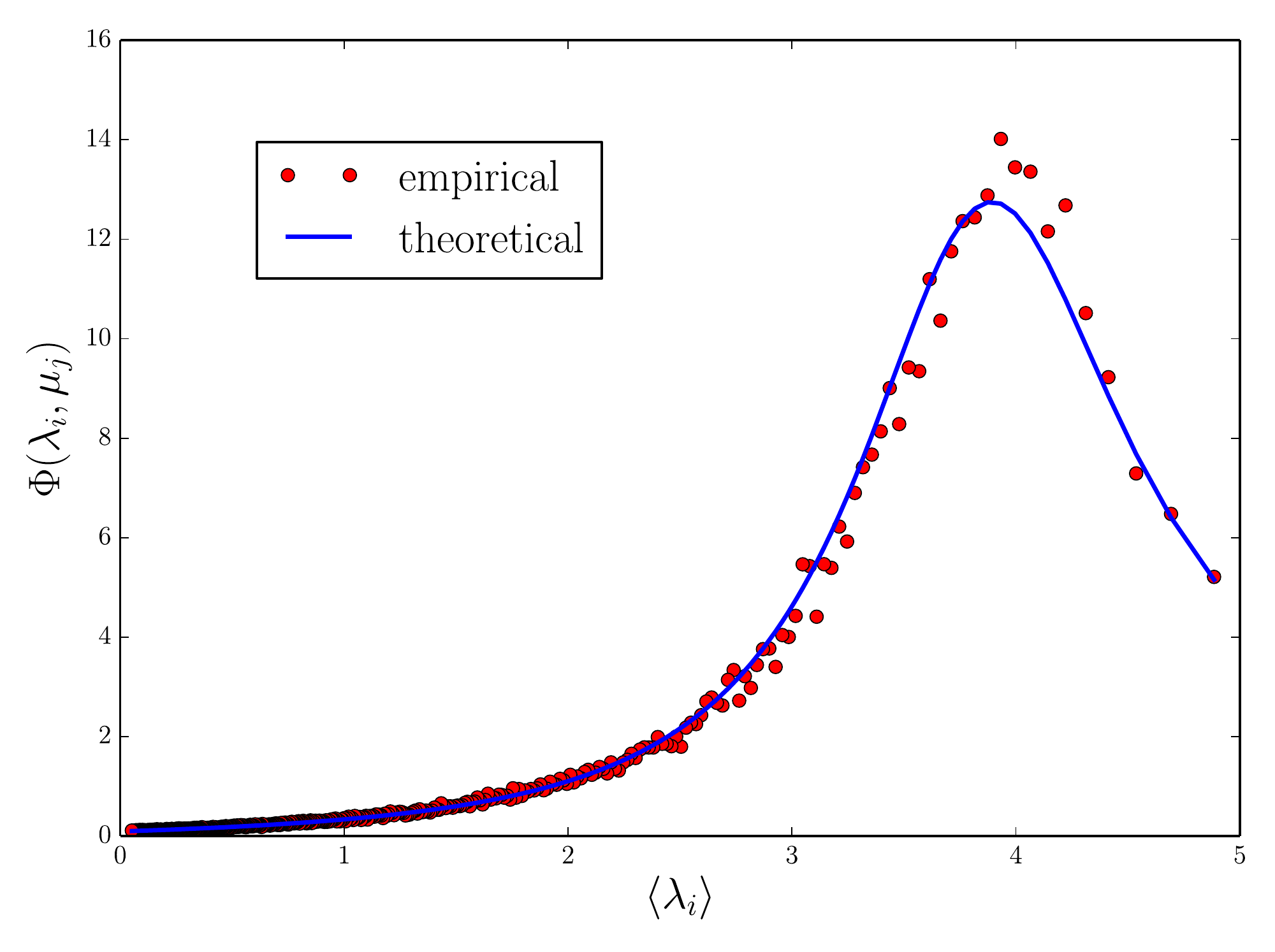} 
   \end{center}
   \caption{Rescaled mean squared overlaps $\Phi(\lambda_i, \mu_j)$ as a function of $\lambda_i$. We choose $\C$ as an inverse-Wishart matrix with parameter $\kappa = 1.0$ and set $N = 500$, $q = 0.5$. The empirical average (red points) comes from 500 independent realizations of $\b E$. The theoretical prediction (blue line) is given by Eq. \eqref{eq:overlap_SCM_bulk}. The peak of the mean squared overlap is in the vicinity of $\lambda_i \approx \mu_j\approx 4$.}
   \label{fig:overlap_cov}
\end{figure}

How can we compute the mean squared overlap using \eqref{eq:global_law_SCM}? The idea is to derive an inversion formula similar to \eqref{eq:stieltjes_inversion} for the full resolvent. More specifically, we start from \eqref{eq:resolvent_decomposition} for a given $\b v = \b v_j$ and notice that the true eigenvectors are deterministic. Therefore, the sum on the RHS of the latter equation is expected to converge in the 
large $N$ limit provided $z$ is outside of the support of the spectrum of $\E$. Moreover, the eigenvalues in the bulk converge to their classical position \eqref{eq:classical_location_bulk} so that we obtain for $N\to\infty$ that
\begin{equation}
  \scalar{\b v_j}{\b G_{\E}(z) \b v_j} \underset{N \uparrow \infty}{\sim} \int \frac{\mso(\lambda, \mu_j) \rho_{\b E}(\lambda) }{\lambda_i - \lambda - \ii \eta} \dd \lambda.
\end{equation}
where we have set $z = \lambda_i - \ii \eta$, $\eta \gg N^{-1}$ and $\mso(\lambda, \mu_j)$ is the smoothed squared overlap, averaged over a small interval of width $\eta$ around $\lambda$. Therefore, the final inversion formula is 
obtained using the Sokhotski-Plemelj identity as:
\begin{equation}
  \label{eq:resolvent_overlap}
  \mso(\lambda_i, \mu_j) = \frac{1}{\pi\rho_{\b E}(\lambda_i)} \lim_{\eta \to 0^{+}} \im \scalar{\b v_j}{ \b G_{\E}(\lambda_i - \ii \eta) \b v_j}\,,
\end{equation}
where the assumption that
$\lambda_i$ lies in the bulk of the spectrum is crucial here. This last identity thus allows us to compute the squared overlap $\mso(\lambda_i, \mu_j)$ from the full resolvent $\b G_{\b E}$, for any $i$ in the bulk ($i \geq r+1$) and a fixed $j \in \qq{1,N}$. Specializing to the explicit form of $\b G_{\E}(z)$ given in Eq. \eqref{eq:global_law_SCM_diag}, we finally obtain a beautiful explicit result for the (rescaled) average squared overlap:
\begin{equation}
\label{eq:overlap_SCM_bulk}
\mso(\lambda_i, \mu_j) = \frac{q \mu_j \lambda_i}{ (\mu_j(1-q) - \lambda_i + q\mu_j\lambda_i \hil_{\E}(\lambda_i))^2 + q^2 \mu_j^2 \lambda_i^2 \pi^2 \rho_{\E}^2(\lambda_i)},
\end{equation}
with $i \in \qq{r+1, N}, j \in \qq{1,N}$ and $\hil_{\E}(\lambda_i)$ denotes the real part of the Stieltjes transform $\stj_{\E}$ (see Eq.\ \eqref{eq:stieltjes_inversion_formula}). This relation is exact in the limit $N \to \infty$ 
and was first derived by Ledoit and P\'ech\'e in \cite{ledoit2011eigenvectors}. We emphasize again that this expression remains correct even if $\mu_j$ is an outlier. Since $\mso(\lambda_i, \mu_j)$ is of order unity whenever $q > 0$, 
we conclude that the dot product between any bulk eigenvector $\b u_i$ of $\E$ and the eigenvectors $\b v_j$ of $\C$ is of order $N^{-1/2}$, i.e vanishes at large $N$, and therefore non-outlier sample eigenvectors retain very little information about their corresponding true eigenvectors. This implies that any bulk eigenvector is a extremely poor estimator of the true one in the high-dimensional regime. We provide in Figure \ref{fig:overlap_cov} an illustration of Eq.\ \eqref{eq:overlap_SCM_bulk} for $N = 500$ and $\C$ an Inverse Wishart matrix with $\kappa = 1$. The empirical average comes from 500 independent realization of $\E$ and we see that it agrees perfectly with the asymptotic theoretical prediction, Eq. \eqref{eq:overlap_SCM_bulk}. Note that in the limit $q \to 0$, $\mso(\lambda_i, \mu_j)$ becomes more and more peaked around $\lambda_i \approx \mu_j$, with an amplitude that diverges for $q = 0$. Indeed, in this limiting case, one should find that $\b u_i \to \pm \b v_j \delta_{ij}$, i.e. the sample eigenvectors become equal to the population ones. 

\subsubsection{Outliers}
\label{sec:outlier_vec}

By construction, the spiked correlation model of Section \ref{eq:SCM} is such that the top $r$ eigenvalues $[\lambda_i]_{i\in\qq{1,r}}$ lie outside the spectrum of $\rho_\E$. What can be said about the statistics of the associated spike eigenvectors $[\b u_i]_{i\in\qq{1,r}}$? If we think of these outliers as a finite-rank deformation of a (fictitious) spikeless matrix ${\ul\E}$, then by Weyl's eigenvalue interlacing inequalities \cite{weyl1949inequalities},  the asymptotic density $\rho_\E$ is not influenced by the presence of non-macroscopic spikes, by which we mean that $\rho_\E(\lambda_i) = 0$ for any outlier eigenvalues. We saw in the previous section that for non-outlier eigenvectors, the main ingredients to compute the overlap are (i) the self-averaging property and (ii) the inversion formula \eqref{eq:resolvent_overlap}. Both implicitly rely on the continuous limit being valid, which is however not the case for outliers. Hence, we expect the statistics of outlier eigenvectors to be quite different from the bulk eigenvectors as confirmed for the null hypothesis case ${\ul\C}=\In$ \cite{hoyle2003limiting,paul2014random}. In this section, we present the analytical tools to analyze these overlaps for outliers in the case of an arbitrary population covariance, following the lines of \cite{bun2016optimal}. 

From Eq.\ \eqref{eq:classical_location_outlier} we saw that each outlier eigenvalues $[\lambda_i]_{i \in \qq{1,r}}$ of $\E$ converges to a deterministic limit $\theta(\mu_i)$, where $\mu_i$ is the corresponding population spike and $\theta$ is a certain function related to the Mar\v{c}enko-Pastur equation. Consequently, for isolated spikes $i \in \qq{1,r}$ we can define the closed disc $D_i$ in the complex plane, centered at $\theta(\mu_i)$ with radius chosen such that each it encloses no other point in the set $[\theta(\mu_j)]_{j\in\qq{1,r}}$ (see \cite{bun2016optimal} for details). Then, defining $\Gamma_i$ to be the boundary of the closed disc $D_i$, we can obtain the squared overlap for outlier eigenvectors using Cauchy's integral formula 
\begin{equation}
\label{eq:outlier_overlap}
\langle \b u _i, \b v_j \rangle^2 = \frac{1}{2 \pi \, \ii} \oint_{\Gamma_{i}} \langle \b v_j, {\b G}_{\E}(z) \b v_j \rangle \dd z, 
\end{equation}
for $i,j \in \qq{1,r}$. We emphasize there is no expectation value in Eq.\ \eqref{eq:outlier_overlap} (compare to our definition of the overlap in Eq.\ \eqref{eq:overlap}). The evaluation of the integral is highly non-trivial since $\b G_\E$ is singular in the vicinity of $\theta(\mu_j)$ for any $j \in \qq{1,r}$ and finite $N$. To bypass this problem, we reconsider the spikeless population covariance matrix $\ul\C$ defined in \eqref{eq:spikeless_population_cov} and the corresponding spikeless sample covariance matrix by $\ul\E$. Clearly, the resolvent $\b G_{\ul\E}$ is no longer singular in the vicinity of $\theta(\mu_j)$, by construction. Moreover, as we said above, the global statistics of the eigenvalues of $\ul\E$ and $\E$ are identical in the limit $N \to \infty$. Lastly, we can relate any projection of $\b G_{\E}$ onto the outlier population covariance eigenbasis using Schur complement formula (see Appendix \ref{app:linear_algebra} for a reminder):
\begin{equation}
  \label{eq:identity_resolvent}
    \b V^{(r)*} \b G_\E (z) \b V^{(r)} = - \frac1z \left[ \b D^{-1} - \frac{\sqrt{\In + \b D}}{\b D} \big( \b D^{-1} + \In - z \b V^{(r)*} \b G_{\ul\E} \b V^{(r)} \big)^{-1}\frac{\sqrt{\In + \b D}}{\b D} \right].
\end{equation}
This identity has been used in several studies that deal with related problems \cite{bloemendal2014principal, bun2016optimal} and references therein. Its derivation only needs linear algebra arguments and can be found in the section \ref{eq:linear_algebra_identity}. With this identity, the statistics of the outliers of $\E$ is seen to only rely on the spikeless matrix $\ul\E$. In particular, the integrand of \eqref{eq:outlier_overlap} can be rewritten using the spikeless resolvent which is analytic everywhere outside the spectrum of $\ul\E$. Since the global law of resolvent of $\ul\E$ is the same than $\E$  in the large $N$ limit, we can again use the estimate \eqref{eq:global_law_SCM}. By plugging \eqref{eq:global_law_SCM} into \eqref{eq:identity_resolvent}, one obtains
\begin{equation}
\label{eq:overlap_outlier_residue}
\langle \b u_i, \b v_j \rangle^2 = - \frac{1}{2 \pi \, \ii} \oint_{\theta(\Gamma_{i})} \frac1z \left[ \frac{1}{d_j} - \frac{1+d_j}{d_j^2} \frac{1}{d_j^{-1} + 1 - z \langle \b v_j, {\b G}_{\E_0}(z) \b v_j \rangle } \right] dz.
\end{equation}
Then, using Eq. \eqref{eq:clasical_location_equation} and Cauchy's theorem, one eventually finds \cite{bun2016optimal}
\begin{equation}
\label{eq:overlap_outlier_outlier}
\langle \b u_i, \b v_j \rangle^2 = \delta_{ij} \mu_i \frac{\theta'(\mu_i)}{\theta(\mu_i)} + \cal O(N^{-1/2})  = \delta_{ij} \mu_i \frac{\theta'(\mu_i)}{\lambda_i} + \cal O(N^{-1/2}),
\end{equation}
for any $i,j \in \qq{1,r}$ and where we used \eqref{eq:classical_location_outlier} in the denominator in the last step. Therefore, we conclude that the sample outlier eigenvector $\b u_i$ is concentrated on a cone around $\b v_i$ with aperture $2\arccos( \mu_i \theta'(\mu_i) / \theta(\mu_i))$. We also deduce from Eq.\ \eqref{eq:overlap_outlier_outlier} that $\b u_i$ is delocalized in all directions $\b v_j$ associated to different spikes $\mu_j \neq \mu_i$.  

An interesting application of  \eqref{eq:overlap_outlier_outlier} is to reconsider the spiked covariance matrix model introduce in the previous chapter. Let us assume for simplicity a single spike ($r =1$) and from equation \eqref{eq:blue transform spiked cov}, one gets, for $\mu_1 > 1 + \sqrt{q}$
\begin{equation*}
\theta(\mu_1) = \mu_1 + q + \frac{q}{\mu_1 - 1},
\end{equation*}
and plugging this result into equation \eqref{eq:overlap_outlier_outlier} yields 
\begin{equation}
  \label{eq:spiked_cov_eigenvector_phase_trans}
\langle \b u_1, \b v_1 \rangle^2 = \frac{\mu_1}{\theta(\mu_1)}\left( 1 - \frac{q}{(\mu_1 - 1)^2}\right) + \cal O(T^{-1/2})\,,
%\equiv \delta_{ij} \frac{\theta'(\mu_1) \mu_1}{\lambda_1},
\end{equation}
which is the expected result \cite{paul2007asymptotics,benaych2011eigenvalues,bloemendal2014principal,biroli2007top,monasson2015estimating}. This result shows that the coherence between the population spike and its sample counterpart becomes progressively lost when $\mu_1 \rightarrow 1+\sqrt{q}$ as it should be from the result \eqref{eq:spiked_cov_eigenvalue_phase_trans}. 

\begin{figure}
  \begin{center}
   \includegraphics[scale = 0.4]{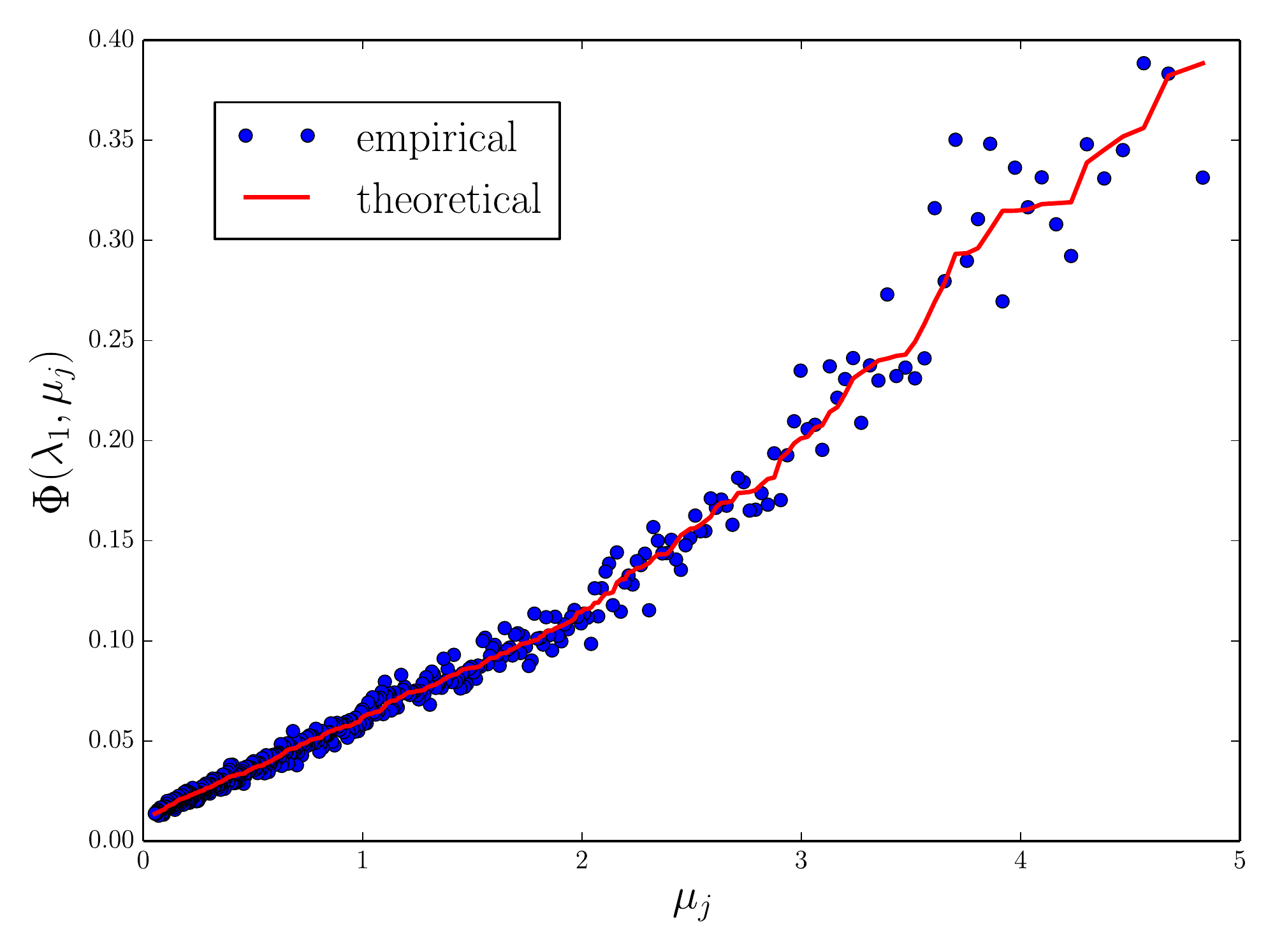} 
   \end{center}
   \caption{Rescaled mean squared overlap $\mso(\lambda_1, \mu_j)$ as a function of $\mu_j$ for $j > 1$. We chose the spikeless population matrix $\ul\C$ to be an Inverse-Wishart matrix with parameter $\kappa = 1.0$ and $N = 500$. We add a rank one perturbation such that $\lambda_1 \approx 10$ is isolated from the others. The sample matrix $\E$ is given by a Wishart matrix with $q = 0.5$. We compare the empirical average (blue points) comes from 200 independent realizations of $\E$. The theoretical prediction (red line) is given by Eq.\ \eqref{eq:overlap_outlier_bulk}.}
   \label{fig:mso_out_bulk}
\end{figure}

The same analysis can be applied for the overlap between the sample spikes and the population bulk eigenvalues $j > r$. The details can be found in \cite{bun2016optimal} and the final result reads 
\begin{equation}
  \label{eq:overlap_outlier_bulk}
    \mso(\lambda_i, \mu_j) = q \frac{\mu_j}{\lambda_i(1 - \mu_j/\mu_i)^2}, \qquad i \in \qq{1,r}, \, j \in \qq{r+1, N}.
\end{equation}
As expected, any outlier eigenvector $\b u_i$ has only $\sim N^{-1/2}$ overlap with any eigenvector of $\C$ except its ``parent'' from $\b v_i$. We illustrate Eq.\ \eqref{eq:overlap_outlier_bulk} in Figure \ref{fig:mso_out_bulk} as a function of the population eigenvalues $\mu_i$ with $i > 2$ in the case where $r=1$: in our example $\C$ is an Inverse Wishart matrix with parameter $\kappa = 1$ and we add a rank one perturbation such that $\lambda_1 \approx 10$. The empirical average comes from 200 realizations of $\E$ and we see that the agreement with the theoretical prediction in excellent.

\subsubsection{Derivation of the identity \eqref{eq:identity_resolvent}}
\label{eq:linear_algebra_identity}

\begin{changemargin}{1cm}{1cm}
\footnotesize

The derivation of the identity \eqref{eq:identity_resolvent} is the central tool in order to deal with the outliers of the sample covariance matrix $\E$. It relies purely on linear algebra arguments (see Appendix \ref{app:linear_algebra} for a reminder). In order to lighten the notations, let us rename $\b V \equiv \b V^{(r)}$ in this section. The first step is to write the following identity from Eq.\ \eqref{eq:spikeless_population_cov}:
\begin{eqnarray}
  \label{eq:identity_resolvent_tmp}
  \sqrt{\ul{\C}} \, \C^{-1} \, \sqrt{\ul{\C}} - \In  & = & (\In + \b V \b D \b V^*)^{-1} - \In  \nonumber \\
   & = & - (\In + \b V \b D \b V^*)^{-1} \b V \b D \b V^* \nonumber \\
   & = & - \b V \b D (\b I_r + \b D)^{-1} \b V^*
\end{eqnarray}
where we used the resolvent identity \eqref{eq:resolvent_identity} in the second line. This allows us to get (omitting the argument $z$)
\begin{eqnarray}
  \ul{\C}^{-1/2} \C^{1/2} \G_\E \C^{1/2} \ul{\C}^{-1/2}  & = & \ul{\C}^{-1/2} \pb{ z \C^{-1} - \X\X^{*}}^{-1} \ul{\C}^{-1/2}  \nonumber \\
   & = & \pb{ z (\ul{\C}^{1/2} \C^{-1}  \ul{\C}^{1/2} - \In) + z \In - \ul{\E}}^{-1} \nonumber\\
   & = & \pb{ -z \b V \b D (I + \b D)^{-1} \b V^* + \G_{\ul\E}^{-1} }^{-1},
\end{eqnarray}
where we invoked the previous identity  Eq.\ \eqref{eq:identity_resolvent_tmp} in the last step. From \eqref{eq:woodbury}, we have with $\AAA \equiv  z\In - \ul\E,  \; \BBB \equiv -z \b V, \; \b D \equiv \b D(\b I_r +\b D)^{-1}$ and $\C \equiv \b V^*$:
\begin{eqnarray}
  \ul{\C}^{-1/2} \C^{1/2} \G_\E \C^{1/2} \ul{\C}^{-1/2}  & = & \G_{\ul\E} + z \G_{\ul\E} \b V \pB{\b D^{-1} + \b I_r - z \b V^*  \G_{\ul\E} \b V}^{-1} \b V^* \G_{\ul\E}.
\end{eqnarray}
From there, one has
\begin{eqnarray}
  (\In + \b D)^{1/2} \b V^* \G_\E \b V (\In+\b D)^{1/2}  & = & \b V^* \G_{\ul\E} \b V + z \b V^* \G_{\ul\E} \b V \pB{\b D^{-1} + \b I_r - \b V^*  \G_{\ul\E} \b V}^{-1} \b V^* \G_{\ul\E} \b V. \nonumber\\
\end{eqnarray}
We then use the identity
\begin{equation}
  \AAA - \AAA(\AAA+\BBB)^{-1} \AAA = \BBB - \BBB(\AAA+\BBB)^{-1} \BBB,
\end{equation}
with $\AAA = V^* \G_{\ul\E} V$ and $\BBB = - (\b D^{-1} + \b I_r)/z$ to obtain
\begin{equation}
  (\b I_r + \b D)^{1/2} \b V^* \G_\E \b V (\b I_r+\b D)^{1/2}  = - \frac1z \qBB{ \frac{\b I_r+\b D}{\b D} + \frac{\b I_r+\b D}{\b D} \pb{-(\b D^{-1} + \b I_r) + z \b V^* \G_{\ul\E} \b V}^{-1} \frac{\b I_r+\b D}{\b D}}.
\end{equation}
By rearranging the terms, we finally get
\begin{equation}
\b V^* \G_\E \b V  = - \frac1z \qBB{ \b D^{-1} -  \frac{\sqrt{\b I_r+\b D}}{\b D} \pb{\b D^{-1} + \b I_r - z \b V^* \G_{\ul\E} \b  V }^{-1} \frac{\sqrt{\b I_r+\b D}}{\b D}},
\end{equation}
which is precisely Eq.\ \eqref{eq:identity_resolvent}.
\end{changemargin}
\normalsize

\subsection{Overlaps between the eigenvectors of correlated sample covariance matrices}
\label{sec:mso_two_sample}

We now consider the second problem of this chapter, that is to say how much information can we learn about the structure of $\b C$ from the sample eigenvectors? Differently said, imagine one measures the sample covariance matrix of the same process but on two independent time intervals, how close are the corresponding eigenvectors expected to be? To answer this question, let us denote by $\E$ and $\tilde{\E}$ the independent sample estimates of the same population matrix $\C$ defined as 
\begin{equation}
  \E \;\deq\; \sqrt{\C} \b{\cal W} \sqrt{\C}, \qquad \tilde{\E} \;\deq\; \sqrt{\C} \tilde{\b{\cal W}} \sqrt{\C},
\end{equation}
where $\b{\cal W}$ and $\tilde{\b{\cal W}}$ are two independent white Wishart matrix with parameter $q$ and $q'$ respectively. As in Section \ref{sec:eigenvectors_sample_true}, we can investigate this problem through the mean squared overlaps. 

In this section, we provide exact, explicit formulas for these overlaps in the high dimensional regime, and perhaps surprisingly, we will see that they may be evaluated without any prior knowledge on the spectrum of $\C$. 
%Differently said, we are able to determine whether $\E$ and $\tilde{\E}$ indeed correspond to the very same underlying true covariance matrix $\C$ without any prior knowledge on $\C$ itself. 
More specifically, we will show that Eq.\ \eqref{eq:mso} exhibits yet again a self-averaging behavior in the large $N$ limit, i.e. independent from the realization of $\E$ and ${\tilde{\E}}$.  We will moreover see that the overlaps \eqref{eq:mso} significantly depart from the trivial null hypothesis as soon as the population $\C$ has a non-trivial structure. Hence, this suggests that we might be able to infer the correlation structure of very large databases using empirical quantities only. 

%\subsubsection{An inversion formula}

All these results have been obtained in the recent work \cite{bun2016overlaps} and we shall only give here the main steps. For the sake of clarity, we use the notations $ \tilde\lambda_1 \geq \tilde\lambda_2 \geq \dots \geq \tilde\lambda_N$ to denote the eigenvalues of $\tilde{\b E}$ and by $ \tilde{\b u}_1, \tilde{\b u}_2, \dots, \tilde{\b u}_N$ the associated eigenvectors. Note that we will again index the eigenvectors by their corresponding eigenvalues for convenience. 

The central tool in this section is an inversion formula for \eqref{eq:mso} as it is usually done in RMT. To that end, we define the bivariate complex function
\begin{equation}
  \label{eq:psi_transform}
  \psi(z,  \tilde z) \;\deq\; \avgBB{\frac1N \tr \left[ (z-\b E)^{-1} ( \tilde z - \tilde {\b E})^{-1} \right]}_{\cal P},
\end{equation}
where $z,  \tilde z \in \mathbb{C}$ and $\avg{\cdot}_{\cal P}$ denotes the average with respect to probability measure associated to $\b E$ and $\tilde{\b E}$. Then, by a spectral decomposition of $\b E$ and $\tilde{\b E}$, one has
\begin{eqnarray}
\psi(z,\tilde z) & = & \avgBB{\frac1N \sum_{i,j=1}^{N} \frac{1}{z-\lambda_i} \frac{1}{\tilde z-\tilde \lambda_j} \scalar{\b u_i}{\tilde{\b u}_j}^2}_{\cal P}, %\nonumber \\
%& \underset{N\to\infty}{\sim} & \int \frac{\rho(\lambda)}{z-\lambda}\frac{\rho(\tilde\lambda)}{\tilde z-\tilde\lambda} \Phi(\lambda_i, \tilde \lambda_j) \dd\lambda \dd\tilde\lambda
\end{eqnarray}
where $\cal P$ denotes the probability density function of the noise part of $\b E$ and $\tilde{\b E}$. For large random matrices, we expect the eigenvalues of $[\lambda_i]_{i\in\qq{1,N}}$ and $[\tilde \lambda_i]_{i\in\qq{1,N}}$ stick to their \emph{classical} locations, i.e. smoothly allocated with respect to the quantile of the spectral density (see Section \ref{sec:MP}) so that the sample eigenvalues become deterministic in the large $N$ limit. Hence, we obtain after taking the continuous limit
\begin{equation}
\label{eq:zeta_to_overlap}
\psi(z,\tilde z) \sim \int \int \frac{\rho(\lambda)}{z-\lambda} \frac{\tilde\rho(\tilde\lambda)}{\tilde z -\tilde \lambda} \Phi(\lambda, \tilde\lambda) \dd\lambda \dd\tilde\lambda,
\end{equation}
where $\rho$ and $\tilde\rho$ are respectively the spectral density of $\b E$ and $\tilde{\b E}$, and $\Phi$ denotes the mean squared overlap defined in \eqref{eq:mso} above. Then, it suffices to compute
\begin{eqnarray}
\label{eq:m value 1}
\psi(x - \ii\eta, y \pm \ii\eta) & \sim &  \int \int \frac{(x - \lambda + \ii\eta)}{(x-\lambda)^2 + \eta^2}\frac{(y-\tilde \lambda \mp \ii\eta)}{(y -\tilde\lambda)^2 + \eta^2} \rho(\lambda)\tilde\rho(\tilde\lambda)\Phi(\lambda, \tilde\lambda) \dd\lambda \dd\tilde\lambda \nonumber \\
%& = &  \int \int \frac{(x - \lambda)(y-\tilde\lambda) \pm \eta^2 + \ii\eta(y - \tilde\lambda \mp(x-\lambda))}{((x-\lambda)^2 + \eta^2)((y -\tilde\lambda)^2 + \eta^2)} \rho(\lambda)\tilde\rho(\tilde\lambda)\Phi(\lambda, \tilde\lambda) \dd\lambda \dd\tilde\lambda,
\end{eqnarray}
from which, one deduces that
\begin{equation}
  %\underset{N\to\infty}{\sim}
  \re\qb{\psi(x - \ii\eta, y + \ii\eta) - \psi(x - \ii\eta, y - \ii\eta) } \sim 2 \int\int \frac{\eta \rho(\lambda)}{(x-\lambda)^2 + \eta^2}\frac{\eta \tilde\rho(\tilde\lambda)}{(y -\tilde\lambda)^2 + \eta^2}  \Phi(\lambda, \tilde\lambda) \dd\lambda \dd\tilde\lambda.
\end{equation}
Finally, the inversion formula follows from Sokhotski-Plemelj identity 
\begin{equation}
  \label{eq:mso_inversion_formula}
  \lim_{\eta\to 0^{+}} \re\qb{\psi(x - \ii\eta, y + \ii\eta) - \psi(x - \ii\eta, y - \ii\eta) } \sim 2\pi^2 \rho(x) \tilde \rho(y) \Phi(x,y).
\end{equation}
Note that the derivation holds for any models of $\b E$ and $\tilde{\b E}$ as long as its spectral density converges to a well-defined deterministic limit.

The inversion formula \eqref{eq:mso_inversion_formula} allows us to study the mean squared overlap \eqref{eq:mso} through the asymptotic behavior of the bivariate function $\psi(z,\tilde z)$. 
Moreover, since we are able control each entry of the resolvent of $\E$ and $\tilde{\E}$ (see Eq.\ \eqref{eq:global_law_SCM}), the evaluation of Eq.\ \eqref{eq:psi_transform} is immediate and leads to
\begin{equation}
\psi(z,\tilde z) \sim \frac{1}{z \tilde z} \frac1N \tr \qb{ Z(z) (Z(z) - \b C)^{-1} \tilde Z(\tilde z)( \tilde Z(\tilde z) - \b C)^{-1} },
\end{equation}
where $Z(z)$ is defined in \eqref{eq:global_law_SCM} and $\tilde Z(z)$ is obtained from $Z$ by replacing $q$ and $\stj_\E$ by $\tilde q$ and $\stj_{\tilde{\E}}$. Then, we use the identity 
\begin{equation}
  \label{eq:resolvent_identity}
  \pB{Z(z) - \b C}^{-1} \pB{\tilde Z(\tilde z) - \b C}^{-1} = \frac{1}{\tilde Z(\tilde z) - Z(z)} \qB{ \pB{Z(z) - \b C}^{-1} - \pB{\tilde Z(\tilde z) - \b C}^{-1} }
\end{equation}
to obtain
\begin{equation}
\psi(z,\tilde z) \sim  \frac{Z(z)\, \tilde Z(\tilde z)}{z \tilde z} \frac{1}{\tilde Z(\tilde z) - Z(z)} \frac1N \tr \qB{ \pB{Z(z) - \b C}^{-1} -  \pB{ \tilde Z(\tilde z) - \b C}^{-1} }.
\end{equation}
From this last equation and using Mar{\v c}enko-Pastur equation \eqref{eq:MP_equation_stieltjes}, we finally conclude that
\begin{equation}
\label{eq:psi_SCM}
\psi(z,\tilde z) \sim   \frac{1}{\tilde Z(\tilde z) - Z(z)} \left[\frac{\tilde Z(\tilde z)}{\tilde z} \stj_\E(z) - \frac{Z(z)}{z} \stj_{\tilde \E}(\tilde z) \right].
\end{equation}
One notices that Eq.\ \eqref{eq:psi_SCM} only depends on \emph{a priori} observable quantities, i.e. they do not involve explicitly the unknown matrix $\b C$. 
Once we characterized the asymptotic behavior of the bivariate function $\psi(z,\tilde z)$, we can then apply the inversion formula Eq.\ \eqref{eq:mso_inversion_formula} in order to retrieve the mean squared overlap \eqref{eq:mso}. Before stating the main result of this section, we first rewrite \eqref{eq:psi_SCM} as a function of the Stieltjes transform $\stj_{\b S}$ of the $T \times T$ dual matrix $\b S = T^{-1} \b X^{*} \C \b X$ that satisfies $\X \X^* = \b{\cal W}$ and Eq.\ \eqref{eq:stieltjes_E_dual}. Similarly, we define $\tilde{\b S} = T^{-1} \tilde{\b X}^{*} \C \tilde{\b X}$ with $\tilde{\X} \tilde{\X}^* = \tilde{\b{\cal W}}$.  Using \eqref{eq:stieltjes_E_dual} and omitting the argument $z$ and $\tilde z$, we can rewrite \eqref{eq:psi_SCM} as
\begin{eqnarray}
  \label{eq:m v2}
  \psi(z, \tilde z) & \sim & \frac{1}{q \tilde q z \tilde z} \left[ \frac{(\tilde q z - q \tilde z) \stj_{\tilde{\b S}}^2}{\stj_{\b S} - \stj_{\tilde{\b S}}} + \frac{(q - \tilde q) \stj_{\tilde{\b S}}}{\stj_{\b S} - \stj_{\tilde{\b S}}} \right] + \frac{\stj_{\b S}+\stj_{\tilde{\b S}}}{q\tilde z} - \frac{1-q}{qz\tilde z}.
\end{eqnarray}
We see from \eqref{eq:mso_inversion_formula} that it now suffices to consider the limit $\eta \to 0^{+}$ in order to get the desired result. To lighten the notations, let us define 
\begin{equation}
  \label{eq:m0}
  m_0(\lambda) \equiv \lim_{\eta\to 0^{+}} \stj_{\b S}(\lambda - \ii\eta) = m_R(\lambda) + \ii m_I(\lambda)
\end{equation} 
with 
\begin{equation}
  m_R(\lambda) = q \hil_\E(\lambda) + \frac{1-q}{\lambda}, \qquad m_I(\lambda) = q \rho_\E(\lambda) + (1-q)\delta_0,
\end{equation}
where $\hil_\E$ is the Hilbert transform of $\rho_\E$. Note that this relation follows from Eq.\ \eqref{eq:MP_equation_stieltjes}.  We also define $\tilde m_0(\lambda) = \lim_{\eta\to0} \stj_{\tilde{\b S}}(\lambda - \ii\eta)$ and denote by $\tilde m_R, \tilde m_I$ the real and imaginary part, respectively. Then, the asymptotic behavior of Eq. \eqref{eq:mso} for any $\lambda \in \supp \varrho$ and $\tilde\lambda \in \tilde\varrho$ is given by (see \cite{bun2016overlaps} for a detailed derivation) 
\begin{equation}
  \label{eq:mso_SCM_general}
  \Phi_{q,\tilde q}(\lambda, \tilde\lambda) = \frac{2(\tilde q \lambda - q \tilde\lambda) \qb{m_R \abs{\tilde m_0}^2 - \tilde m_R \abs{m_0}^2 } + (\tilde q - q) \qb{\abs{\tilde m_0}^2 - \abs{m_0}^2 }  }{\lambda\tilde\lambda \qB{(m_R - \tilde m_R)^2 + (m_I + \tilde m_I)^2 }\qB{(m_R - \tilde m_R)^2 + (m_I - \tilde m_I)^2}}.
\end{equation}
An interesting consistency check is when $\tilde q = 0$ in which case the sample eigenvalues coincide with the true ones for the tilde matrices, i.e.\ $\tilde\lambda \to \mu$. In this case we fall back on the framework of 
the previous section, i.e. obtaining the overlaps between the eigenvectors of $\E$ and $\C$. One can easily check that $\tilde m_R = 1/\mu$  and $\tilde m_I = 0$. Hence, we deduce from \eqref{eq:mso_SCM_general} that
\begin{eqnarray}
  \Phi_{q,\tilde q=0}(\lambda, \mu) =  \frac{q}{\lambda\mu \qb{ (m_R - 1/\mu)^2 + m_I^2}}  = \frac{q\mu}{\lambda \abs{1-\mu m_0(\lambda)}^2 },
\end{eqnarray}
which is another way to write \eqref{eq:overlap_SCM_bulk} after applying the formula \eqref{eq:stieltjes_E_dual} in the limit $\eta \to 0^{+}$. It therefore shows that the result \eqref{eq:mso_SCM_general} generalizes Eq.\ \eqref{eq:overlap_SCM_bulk} in the sense that we are able to study the mean squared overlaps between two possibly noisy sample estimates. Note that in the case $\tilde q = q$, Eq.\ \eqref{eq:mso_SCM_general} can be somewhat simplified to:
\begin{equation}
  \label{eq:mso_SCM_bulk}
  \Phi(\lambda, \tilde\lambda) = \frac{q(\lambda - \tilde\lambda)\pb{m_R(\lambda) \abs{ m_0(\tilde\lambda)}^2 -  m_R(\tilde\lambda) \abs{m_0(\lambda)}^2} }{\lambda\tilde\lambda \qB{(m_R - \tilde m_R)^2 + (m_I + \tilde m_I)^2 }\qB{(m_R - \tilde m_R)^2 + (m_I - \tilde m_I)^2}},
\end{equation}
that becomes when $\tilde\lambda = \lambda$ \cite{bun2016overlaps},
\begin{equation}
  \label{eq:mso_SCM_bulk_same_q_eig}
  \Phi(\lambda,\lambda) = \frac{q}{2 \lambda^2}  
  \frac{ \abs{m_0(\lambda)}^4 \partial_\lambda\qb{m_R(\lambda)/\abs{m_0(\lambda)}^2}}{m_I^2(\lambda) |\partial_\lambda m_0(\lambda)|^2}.
\end{equation}
This last ``self-overlap'' result quantifies the stability of the eigenvectors $\b u_i$ and $\tilde{\b u}_j$ associated to the very same eigenvalue $\lambda$ when they both come from the same population matrix $\b C$. Any statistically significant deviation between this predicted overlap and empirical results can be
interpreted as a violation of the hypothesis that the ``true'' population matrices corresponding to $\E$ and $\tilde{\E}$ are in fact different. This is 
extremely interesting from the point of view of applications, in particular to financial data where nothing ensures that $\b C$ is time independent.

% \begin{figure}[h]
%   \begin{center}
%    \includegraphics[scale = 0.45]{Figures/eigenvectors/mso_MP_N_500_q_05_M_200} 
%    \end{center}
%    \caption{Evaluation of the rescaled $N \E\scalar{\b u_i}{\tilde{\b u}_i}^2$ with $\b C = I_N$ with $N = 500$ and $q = 0.5$. The notations $[ \cdot ]_e$ denotes the empirical average over the 200 realizations. The blue points corresponds to the empirical average and the red line to Eq. \eqref{eq:mso_SCM_bulk_same_q_eig} evaluated at any $[\lambda_i]_e$.}
%    %as a function the typical $[\lambda_i]_e$
%    \label{fig:overlap_MP}
% \end{figure}

Now that we have all these theoretical results, let us now give some applications of the formula \eqref{eq:mso_SCM_bulk} as they will highlight that we can indeed find genuine information about the spectrum of $\C$ from the mean squared overlap \eqref{eq:mso}. We emphasize that all the following applications are performed in the case $q = \tilde q$ in order to give more insights about the results. As usual, we begin with the null hypothesis $\C = I_N$ as it will serve as the benchmark when we shall deal with more structured spectrum. As we shown in Section \eqref{sec:MP_law}, the Stieltjes transform $\stj_\E$, and thus $\stj_\S$ is explicit and obtained from the Mar{\v c}enko-Pastur density. More precisely, we deduce from Eq.\ \eqref{eq:stieltjes_isotropic_wishart} and \eqref{eq:stieltjes_E_dual} that $\stj_{\b S}$ is given by 
\begin{equation}
  \label{eq:m_MP}
  \stj_{\b S}(z) = \frac{z+q-1 - \ii\sqrt{4zq - (z+q-1)^2}}{2z}
\end{equation}
for any $z \in \mathbb{C}_-$. It is easy to see using the definition \eqref{eq:m0} that we have 
\begin{equation}
    m_R(\lambda) = \frac{\lambda+q-1}{2z}, \qquad m_I(\lambda) = \frac{\sqrt{4\lambda q - (\lambda +q-1)^2}}{2\lambda }.
\end{equation}
Hence, one obtains $\abs{m_0(\lambda)}^2 = \lambda^{-1}$ and $\abs{m_0'(\lambda)}^2 = q/(2\lambda^2)$, and by plugging this expressions into Eq. \eqref{eq:mso_SCM_bulk_same_q_eig}, we eventually get
\begin{equation}
  \label{eq:overlap_same_eig_MP}
  \Phi_{q,q}(\lambda,\lambda) = 1,
\end{equation}
for any $\lambda \in [ (1-\sqrt{q})^2, (1+\sqrt{q})^2]$. This simple result was expected as it corresponds to the case where the spectrum of $\C$ has \emph{no} genuine structure, so all the anisotropy in the problem is induced by the noise, which is independent in the two samples. 

% We plot in Figure \ref{fig:overlap_MP} an empirical check with $N = 500$, $q=0.5$ and $200$ independent realizations of the noise. **I think this graph is not needed**

\begin{figure}[h]
  \begin{center}
   \includegraphics[scale = 0.45]{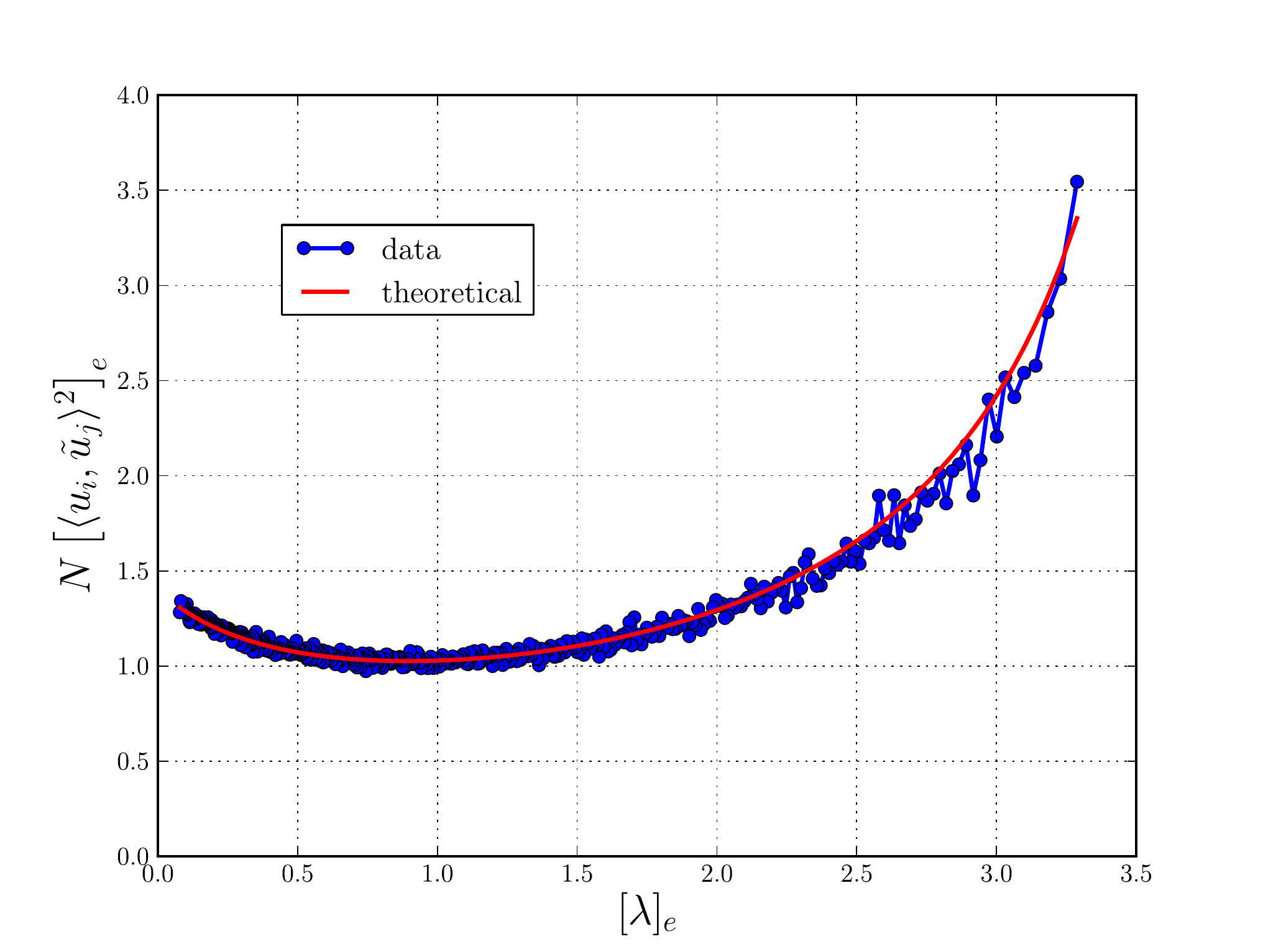} 
   \end{center}
   \caption{Evaluation of $N \mathbb{E}\scalar{\b u_i}{\tilde{\b u}_i}^2$ with $N = 500$ and $q = \tilde q = 0.5$. The population matrix $\b C$ is given by an Inverse-Wishart with parameter $\kappa$ and the sample covariance matrices $\b S$ and $\tilde{\b S}$ are generated from a multivariate Gaussian distribution. The empirical average (blue points) is taken over 200 realizations and the theoretical prediction Eq. \eqref{eq:mso_SCM_bulk_same_q_eig} (red line) is evaluated for all $[\lambda_i]_e$.}
   %as a function the typical $[\lambda_i]_e$
   \label{fig:overlap_invW}
\end{figure}

Next, we consider a more structured example of a population correlation matrix $\C$. A convenient case that can be treated analytically is when $\b C$ to be an inverse Wishart matrix, i.e. distributed according to \eqref{eq:inverse_wishart_distribution} with $\kappa > 0$ defined in Eq.\ \eqref{eq:kappa_invWishart}. As we saw in the previous chapter, the Stieltjes transform $\stj_\E(z)$ is explicit in this case (see Eq.\ \eqref{eq:stieltjes_IW_MP}). Going back to Eq. \eqref{eq:mso_SCM_bulk_same_q_eig}, one can readily obtain from Eq.\ \eqref{eq:stieltjes_IW_MP},
\begin{eqnarray}
  m_R(\lambda) \;=\; \frac{\lambda(1+q\kappa) + q\kappa(1-q)}{\lambda(\lambda + 2q\kappa)}, \qquad m_I(\lambda) \;=\; q \frac{\sqrt{\lambda - \lambda^{\text{iw}}_{-}} \sqrt{\lambda^{\text{iw}}_{+} - \lambda} }{\lambda(\lambda + 2q\kappa)},  
  %\\
  %\sqrt{2\kappa\lambda(1+\kappa(1+q)) - \kappa^2 \lambda^2 - \kappa^2(1-q)^2}
  %\lambda_{\text{iw}}^{\pm} & = & \frac{1}{k} \qB{ \kappa(1+q) + 1 \pm \sqrt{(2\kappa+1)(2q\kappa+1)}}, \nonumber
\end{eqnarray}
with $\lambda \in [\lambda^{\text{iw}}_{-},\lambda^{\text{iw}}_{+}]$ where $\lambda^{\text{iw}}_{\pm}$ is defined in \eqref{eq:edges_IW_MP}. Plugging these expressions into Eq.\ \eqref{eq:mso_SCM_bulk_same_q_eig} and after elementary computations, one finds 
\begin{equation}
  \label{eq:mso_same_q_eig_invW}  
  \Phi_{q,q}(\lambda,\lambda) = \frac{(1+q\kappa)(\lambda + 2q\kappa)^2}{2 q \kappa \qb{2\lambda(1+\kappa(1+q)) - \lambda^2\kappa+\kappa(-1+2q(1 + q\kappa))}}.
\end{equation}
The immediate consequence of this last formula is that in the presence of \emph{anisotropic} correlations, the mean squared overlap \eqref{eq:mso} clearly deviates from the null hypothesis $\Phi(\lambda,\lambda) = 1$. In the 
nearly isotropic limit $\kappa \to \infty$, that corresponds to the limit $\C \to \In$, one gets \cite{bun2016overlaps}
\begin{equation}
  \label{eq:mso_same_q_eig_invW_exp}  
  \Phi(\lambda,\tilde\lambda) \underset{\kappa\to\infty}{\sim} \qBB{1+\frac{(\lambda - 1)(\tilde\lambda-1)}{2q^2\kappa} + \cal O(\kappa^{-2})},
\end{equation}
which is in fact {\it universal} in this limit (i.e. independent of the precise statistical properties of the matrix $\C$), provided the eigenvalue spectrum of $\b C$ has a variance given by $(2\kappa)^{-1} \to 0^{+}$ \cite{bun2016overlaps}. In the general case, we provide a numerical illustration of this last statement in Figure \ref{fig:overlap_invW} with $\kappa = 5$, $N = 500$ and $q=0.5$. As we expect $\lambda_i \approx \tilde \lambda_i$ for any $i \in \qq{1,N}$, we compare our theoretical result \eqref{eq:mso_same_q_eig_invW} with the empirical average $[\scalar{\b u_i}{\tilde{\b u}_i}^2]_e$ taken over 200 realizations of $\E$ and we see that  the agreement is again excellent. We therefore conclude that a possible application of \eqref{eq:mso_SCM_general} is to estimate directly the statistical texture of $\C$ using only sample eigenvectors: see Section \ref{chap:application} for an interesting example. 

We now present an alternative derivation of $\Phi_{q,\tilde q}$ that uses the result of the Section \ref{sec:eigenvectors_sample_true}. The following argument is very general and might be useful when considering the overlaps between the eigenvectors of more general random matrices. The starting point is the orthonormality of the true eigenbasis, i.e. $\b V \b V^* = \In$  for $\b V \deq [\b v_1, \dots, \b v_N]$. Hence, we may always write
\begin{equation}
  \scalar{\b u_i}{\tilde{\b u}_j} = \scalarBB{\b u_i}{\pBB{\sum_{k=1}^{N} \b v_k \b v_k^*} \tilde{\b u}_j} = \sum_{k=1}^{N} \scalar{\b u_i}{\b v_k} \scalar{\b v_k}{\tilde{\b u}_j}
\end{equation}
Using the results of Section \ref{sec:eigenvectors_sample_true}, we rename the overlaps $\scalar{\b u_i}{\b v_k} = \sqrt{\mso_q(\lambda_i, \mu_k)/N}\times \e(\lambda_i, \mu_k)$ where $\mso_q(\lambda,\mu)$ is defined in \eqref{eq:overlap} and $\e(\lambda,\mu)$ are random variables of unit variance. Hence, we have
\begin{equation}
  \scalar{\b u_i}{\tilde{\b u}_j} = \frac{1}{N} \sum_{k=1}^{N} \sqrt{\mso_q(\lambda_i,\mu_k) \mso_{\tilde q}(\tilde\lambda_j, \mu_k)} \; \e(\lambda_i,\mu_k) \e(\tilde\lambda_j,\mu_k).
\end{equation}
As noticed in \cite{bun2016overlaps}, by averaging over the noise and making an ``ergodic hypothesis" \cite{deutsch1991quantum} -- according to which all signs $\epsilon(\mu,\lambda)$ are 
in fact independent from one another in the large $N$ limit -- one ends up with the following rather intuitive convolution result for the square overlaps:
\begin{equation}
\label{eq:overlap_convolution}
%N \langle \left({\bf u}_\lambda \cdot {\bf u}'_{\lambda'}\right)^2 \rangle
\Phi_{q,\tilde q}(\lambda_i,\tilde\lambda_j)  = \frac{1}{N} \sum_{k=1}^{N} \mso_q(\lambda_i,\mu_k) \mso_{\tilde q}(\tilde\lambda_j, \mu_k)
\end{equation}
It turns out that this expression is completely general and exactly equivalent to Eq.\ \eqref{eq:mso_SCM_bulk} if we replace the overlaps function $\mso$ by \eqref{eq:overlap_SCM_bulk}. However, whereas this 
expression still contains some explicit dependence on the structure of the pure matrix $\b C$, it has completely disappeared in Eq.\ \eqref{eq:mso_SCM_bulk}. An interesting application of the formula \eqref{eq:overlap_convolution} is when the spectrum of $\E$ (and $\tilde \E$) contains a finite number of outliers. Using the results \eqref{eq:overlap_outlier_outlier} and \eqref{eq:overlap_outlier_bulk} yields in the LDL and for $i \leq r$:
\begin{equation}
  \label{eq:mso_outlier}
  \Phi_{q,\tilde q}(\lambda_i,\tilde\lambda_i) \approx \mu_1^2 \frac{\theta'(\mu_1) \tilde\theta'(\mu_1)}{\theta(\mu_1)\tilde\theta(\mu_1)},
\end{equation}
where we recall that the function $\theta$ is defined in \eqref{eq:classical_location_outlier} and we define $\tilde\theta$ accordingly by replacing $q$ with $\tilde q$. Note that we can express \eqref{eq:mso_outlier} in terms of observable variables by noticing that 
\begin{equation}
  \mu_1 = \frac{1}{\stj_{\ul\S}(\lambda_1)}, \qquad \theta'(\mu_1) = \frac{-1}{\stj_{\ul\S}'(\theta(\mu_1)) \mu_i^2},
\end{equation}
that we plug into \eqref{eq:mso_outlier} to conclude that
\begin{equation}
  \label{eq:mso_outlier_obs}
  \Phi_{q,\tilde q}(\lambda_1,\tilde\lambda_1) \approx \frac{\stj_{\ul\S}(\lambda_1)}{ \lambda_1 \stj^{\prime}_{\ul\S}(\lambda_1)} \frac{\stj_{\tilde{\ul\S}}(\lambda_1)}{ \tilde\lambda_1 \stj^{\prime}_{\tilde{\ul\S}}(\lambda_1)}.
\end{equation}
This expression becomes even simpler when $q = \tilde q$ as it becomes
\begin{equation}
  \label{eq:mso_outlier_obs_same_q}
  \Phi_{q,q}(\lambda_1,\tilde\lambda_1) \approx \pBB{\frac{\stj_{\ul\S}(\lambda_1)}{ \lambda_1 \stj^{\prime}_{\ul\S}(\lambda_1)}}^2.
\end{equation}
One further deduces from \eqref{eq:overlap_outlier_outlier} and \eqref{eq:overlap_outlier_bulk} that for $i \leq r$, $\Phi_{q,\tilde q}(\lambda_i,\tilde\lambda_j) \sim \cal O(N^{-1})$ for any $j \neq i$.

%---------  Cleaning part ---------------- %

\clearpage%!TEX root = RMT_Covariance_Review.tex
\section{Bayesian Random Matrix Theory} 
\label{chap:bayes}

%%%%%%%%%%%%%%%%%%%%%%%%%%%%%%%%%%%%%%%%%%%%%%%%%%%%%%%%%%%%%%%%%%%%%%%%%%%%%%%%%%%%%%%%%%%%%%%%%%%%%%
%%%%%%%%%%%%%%%%%%%%%%%%%%%%%%%%%%%%%%%%%%%%%%%%%%%%%%%%%%%%%%%%%%%%%%%%%%%%%%%%%%%%%%%%%%%%%%%%%%%%%%

We saw in the previous chapters that RMT allows one to make precise statements about large empirical covariance matrices. In particular, we emphasized that the classical sample estimator $\E$ is not consistent in the high-dimensional limit as the sample spectral density $\rho_\E$ deviates significantly from the true spectrum whenever $q = \cal O(1)$. There has been many attempts in the literature to correct this ``curse of dimensionality'' using either heuristics or decision theoretic arguments (see Section \ref{sec:past_cleaning} for a summary of these attempts). Despite the strong differences in these approaches, all of them fall into the class of so-called \emph{shrinkage} estimators, to wit, one seeks the best way to ``clean'' the sample eigenvalues in such a way that the estimator 
is as robust as possible to the measurement noise.

In the previous chapter, we insisted that the bulk sample eigenvectors are delocalized, with a projection of order $N^{-1/2}$ in all directions, which means that 
they are extremely noisy estimators of the population eigenvectors. As a consequence, the naive idea of replacing the sample eigenvalues by the estimated true ones, obtained by inverting the Mar\v{c}enko-Pastur equation, will not necessarily lead to satisfactory results -- it would only be the optimal strategy if we had a perfect knowledge of the eigenvectors of $\C$. Hence, we are left with a very complicated problem: how can we estimate ``accurately'' the matrix $\C$ in the high-dimensional regime knowing that the eigenvalues are systematically biased and the eigenvectors nearly completely unknown? 

The aim of the present chapter and the following one is to answer this question by developing an optimal strategy to estimate $\C$, consistent with the quality ratio $q$. By optimal, we mean that the estimator we aim to construct has to minimize a given loss function. A natural optimality criteria is the squared distance between the estimator -- called $\Xi(\E)$ henceforth -- and the true matrix $\C$. As for the James-Stein estimator, we expect that ``mixed'' estimators provide better performance than ``classical'' ones (like the Pearson estimator) in high-dimension. In that respect, we introduce a Bayesian framework which, loosely speaking, allows one to introduce probabilistic models that encode the available data through the notion of \emph{prior} belief. 

The fact that probabilities represent degrees of belief is at the heart of Bayesian inference. As explained in the introduction to this review, this theory has enjoyed much success, especially in a high-dimensional framework. The central tool of this theory is the well known Bayes formula that allows one to introduce the concept of conditional probability. There are many different ways to make use of this formula and the corresponding schools of thought are referred to as empirical, subjective or objective Bayes (see e.g. \cite{gelman2014bayesian} for an exhaustive presentation). Here we shall not discuss these different points of view but  rather focus on the inference part of the problem. More precisely, our aim in this chapter is to construct a Bayesian estimator for $\Xi(\E)$. We therefore organize this chapter as follows. In the first part, we recall some basic results on Bayesian inference and introduce the estimator that will interest us. 
We then re-consider the famous ``linear shrinkage'' estimator, mentioned in Eq.\ \eqref{eq:intro_linear_shrinkage}, that interpolates linearly between the sample estimator and the identity matrix through the notion of \emph{conjugate priors}. Finally, we consider the class of rotational invariant prior where the RMT formalism introduced in the previous chapters is applied to derive an optimal estimator for $\C$, which will turn out to be more efficient that all past attempts -- see Chapter \ref{chap:numerical}.

\subsection{Bayes optimal inference: some basic results}

\subsubsection{Posterior and joint probability distributions}

Bayesian theory allows one to answer, at least in principle, the following question: given the observation matrix $\b Y$, how can we best estimate $\C$ if some prior knowledge of the statistics of $\C$ is available?  This notion of prior information has been the subject of many controversies but is a cornerstone to Bayes inference theory. More precisely, the main concept of Bayesian inference is the well-known Bayes formula
\begin{equation}
\label{eq:Bayes_rule}
{\cal P}(\C \lvert \Y) = \frac{{\cal P}(\Y \lvert \C) {\cal P}(\C)}{{\cal P}(\Y)}
\end{equation}
where
\begin{itemize}
	\item ${\cal P}(\C \lvert \Y)$ is the \emph{posterior} probability for $\C$ given the measurements $\Y$.
	\item ${\cal P}(\Y \lvert \C)$ is the \emph{likelihood} function, modeling the measurement process.
	\item ${\cal P}(\C)$ is called the \emph{prior} probability of $\b C$, that is to say the prior belief (or knowledge) about $\C$.
	\item ${\cal P}(\Y)$ is the marginal distribution, sometimes called the \emph{evidence}.
\end{itemize}
Note that the marginal distribution is often considered as a mere normalization constant (or partition function) since it is given by 
\begin{equation}
\label{eq:marginalPDF}
{\cal P}(\Y) = \int {\cal D} \C {\cal P}(\C) {\cal P}(\Y \lvert \C).
\end{equation}
Furthermore, we shall often use the concept of \emph{joint} probability distribution defined by
\begin{equation}
	\label{eq:joint_pdf}
	\cal P(\C, \Y) = \cal P(\Y | \C) \cal P(\C).
\end{equation}
Thus, the two crucial inputs in a Bayesian model are the likelihood process and the prior distribution. Learning using a Bayesian framework can actually be split in two different steps, which in our context are:
\begin{itemize}
  \item[1.] Set a joint probability distribution ${\cal P}(\C, \Y)$ defined as the product of the prior distribution and the likelihood function, i.e.
  \begin{equation}
	\label{jointPDF}
		{\cal P}(\C, \Y) = {\cal P}(\Y |\C) {\cal P}(\C).
  \end{equation}
  \item[2.] Test the consistency of the posterior distribution ${\cal P}(\C | \Y)$ on the available data.
\end{itemize}
We emphasize that the presence of a prior distribution does not imply that $\C$ is stochastic, it simply encodes the degree of belief about the structure of $\C$. 
%Hence, suppose that we have no prior on $\C$ (flat probability measure over all positive definite symmetric matrices), then the posterior distribution is simply given by the likelihood function. 
The main advantage of adopting this point of view is that it facilitates the interpretation of the statistical results. For instance, a Bayesian (probability) interval tells us how probable is the value of a parameter we attempt to estimate.  This is in contrast to the frequentist interval, which is only defined with respect to a sequence of similar realizations (confidence interval). We will discuss the difference between these points of view in the next 
paragraph.

\subsubsection{Bayesian inference}

The notion of Bayesian inference is related to the concept of the so-called \emph{Bayes risk}. In our problem, we want to estimate the true covariance matrix $\C$ given our sample data $\Y$; we shall denote by $\Xi(\Y)$ this estimator. There are two ways to think about this problem: the frequentist and the Bayesian approach. We will detail the difference between these two in this section.

Let us introduce a loss function ${\cal L}(\C, \Xi(\Y))$ that quantifies how far the estimator is from the true quantity $\C$. 
In general, this loss function is assumed to be a non-negative convex function with $\cal L(\C,\C) = 0$. 
The traditional \emph{frequentist} approach is to evaluate the performance of a given estimator by averaging the loss function over different sets of observations, for a fixed $\C$. 

An alternative point of view is to think that the precise nature of $\C$ is unknown. This change in the point of view has to be encoded in the inference problem and one way to do it is to look at the average value of the loss function over all the \emph{a priori} possible realizations of $\C$, and not on the realizations of 
$\Y$ itself. This is Bayes optimization strategy and the corresponding the decision rule is the so-called \textit{Bayes risk function} that is defined as:
\begin{equation}
\label{eq:Bayes_risk}
R^{\text{Bayes}}(\cal L(\C,\Xi(\Y))) \;\deq\; \bigg\langle \cal L(\C, \Xi(\Y)) \bigg\rangle_{{\cal P}(\C, \Y)},
\end{equation}
where, unlike the frequentist approach, the expectation value is taken over the joint probability of $\Y$ and $\C$. One of the most commonly 
used loss function is the squared Hilbert-Schmidt (or Euclidean) $L_2$ norm, i.e., 
\begin{equation}
\label{eq:bayes_squared_loss}
\cal L^{\text{$L_2$}}(\C, \Xi(\Y)) = \Tr\left[ (\C - \Xi(\Y))(\C - \Xi(\Y))^{*} \right].
\end{equation}
Using that covariance matrices are symmetric and applying Bayes rule, we see that 
\begin{eqnarray} 
\label{eq:bayes_risk_MMSE}
%\underset{ \Xi(\Y) \in \cal M(\Y)}{\argmin} 
R^{\text{Bayes}} & = & \bigg\langle \bigg\langle \Tr\left[ (\C -  \Xi(\Y))^2 \right] \bigg\rangle_{{\cal P}(\Y | \C)} \bigg\rangle_{{\cal P}(\C)} \nonumber \\
& = & \bigg\langle \bigg\langle \Tr\left[ (\C -  \Xi(\Y))^2 \right] \bigg\rangle_{{\cal P}(\C | \Y)} \bigg\rangle_{{\cal P}(\Y)},
\end{eqnarray}
where we have used that marginal distributions are positive in order to interchange the order of integration in the second line. 

The optimal Bayes estimator is defined as follows: let us denote by $\cal M_N(\Y)$ is the set of $N \times N$ positive definite matrices which are functions of $\Y$. This defines the set of admissible estimators of $\C$. Then the Bayes estimator associated to the loss function  \eqref{eq:bayes_squared_loss} is given by the \textit{minimum mean squared error} (MMSE) condition, i.e.
\begin{equation}
\label{eq:bayes_estimator}
\Xi^{\text{MMSE}} \;\equiv\; {\Xi}^{\text{MMSE}}(\Y) \;\deq\; \underset{\Xi(\Y) \in \cal M_N(\Y)}{\argmin} \bigg\langle \cal L^{\text{$L_2$}}(\C, \Xi(\Y)) \bigg\rangle_{{\cal P}(\C, \Y)},
\end{equation}
Expanding \eqref{eq:bayes_risk_MMSE}, it is readily seen that the MMSE estimator is given by the posterior mean:
\begin{equation}
\label{eq:bayes_MMSE}
\Xi^{\text{MMSE}} = \langle \C \rangle_{\cal P(\C | \Y)}.
\end{equation}
Note that the natural choice of the loss function may depend on the nature of the problem. Other loss functions often lead to different Bayes estimators, but we do not investigate such generalizations here. 

%Note that several loss functions have been considered in the literature \cite{donoho2013optimal} but as far as prediction (or out-of-sample estimation) is concerned, the MSE is the loss function that we have to consider in order to design optimal cleaning schemes (see section \ref{??}). 

%It is worth to mention at this point that the Bayesian framework is appropriate with the Mar{\v c}enko-Pastur formalism presented above. Indeed, the Mar{\v c}enko-Pastur equation tells us that if we know the true correlation matrix $\C$, then we know exactly the spectrum of the empirical eigenvalues of the matrix $\E$. In a Bayesian context this can be reformulated as follow: let us suppose that we can find a 'good' prior on $\C$, i.e., such that it allows to describe precisely the empirical eigenvalues spectrum, then we should be able to propose an estimator of $\C$ from this prior. So in some sense, the Mar{\v c}enko-Pastur equation is the key tool that permits us to find a good prior on the eigenvalues distribution of $\C$. Besides this observation, we will also show that this framework is in fact the right tool to consider for optimizing out-of-sample quadratic programs like the Markowitz problem. 

\subsection{Setting the Bayesian framework}

Now that we have derived the optimal estimator we are looking for, we still need to parametrize the joint probability function $\cal P(\C, \Y)$. There are thus two inputs in the Bayesian model: the likelihood function and the prior distribution, and we focus on the former quantity in this section. 

In a multivariate framework, the most common assumption (but not necessarily the most realistic) is that the measurement process $\Y$ is  Gaussian, that is to say,
\begin{equation}
\mathbb{P}( \Y | \C) = \frac{1}{(2\pi)^{\frac{NT}{2}} \det(\C)^{\frac{T}{2}}} \exp \left\{ - \frac{1}{2} \sum_{t=1}^{T} \sum_{i,j=1}^{N} Y_{it} \C_{i,j}^{-1} Y_{jt} \right\}.
\end{equation}
It is easy to see that this is of the Boltzmann type, as in Eq. \eqref{eq:Boltz}. More precisely, using the cyclic property of the trace operator one gets
\begin{equation*}
\sum_{t=1}^{T} \sum_{i,j=1}^{N} Y_{it} \C_{ij}^{-1} Y_{jt} = \Tr \left[ \Y \C^{-1} \Y^{*} \right] = T \Tr \left[ \E \C^{-1} \right].
\end{equation*}
Thus, the $N$-variate Gaussian likelihood function can be written as
\begin{equation}
\label{eq:gaussian_likelihood}
\cal{P}( \Y | \C) = \frac{1}{(2\pi)^{\frac{NT}{2}}} \exp\left\{-\frac{T}{2} \Tr \qb{\log(\C) + \E \C^{-1} } \right\} \equiv \cal{P}( \E | \C),
\end{equation}
where we used Jacobi's formula $\det(\AAA)  = \exp [\Tr \log \AAA]$ for any square matrix $\AAA$. 
As a result, we can rewrite the inference problem as a function of the sample covariance matrix $\E$, and in particular, the MMSE estimator becomes
\begin{equation}
\label{eq:bayes_MMSE_SCM}
	\Xi^{\text{MMSE}} \;\equiv\; \Xi^{\text{MMSE}}(\E) \deq  \avg{\C}_{\cal P(\C |\E)}.
\end{equation}

After a little thought, this set-up agrees perfectly with the framework developed in the Chapters \ref{chap:spectrum} and \ref{chap:eigenvectors} above. Indeed, in those sections we studied the spectral properties of the sample covariance matrix $\E$ given the limiting spectral distribution of $\C$ (the so-called ``direct problem'' introduced in Section \ref{sec:MP}). Differently said, the Mar{\v c}enko-Pastur equation \eqref{eq:MP_equation_stieltjes} has a natural Bayesian interpretation: it provides the (limiting) spectral density of $\E$ conditional to a population covariance matrix $\C$ that we choose within a specific prior probabilistic ensemble.

\subsection{Conjugate prior estimators} 

Once we have set the likelihood function, the next step is to focus on the prior distribution ${\cal P}(\C)$, keeping in mind that the ultimate goal is to compute the Bayes posterior mean estimator \eqref{eq:bayes_MMSE_SCM}. Unfortunately, the evaluation of the posterior distribution often leads to non trivial computations and closed-form estimators are thus  scarce. Nonetheless, there exists some classes of prior distributions where the posterior distribution can be computed exactly. The one that interests us is known as the class of `conjugate priors' in Statistics. Roughly speaking, suppose that we know the likelihood distribution ${\cal P}(\E \lvert \C)$, then the prior distribution ${\cal P}(\C)$ and the posterior distribution ${\cal P}(\C \lvert \E)$ are said to be conjugate if they belong to the same family of distributions. 

As an illustration, let us consider a warm-up example before going back to the estimation of the covariance. Suppose that we want to estimate the mean vector -- say $\b\mu$ -- given the $N$-dimensional vector data $\b y$ we observe. Moreover, assume that the likelihood function is a multivariate Gaussian distribution with a known covariance matrix $\sigma^2 \In$. Then, by taking a Gaussian prior on $\b \mu$ with zero ``mean'' and ``covariance'' matrix $\tau^2 \In$, one can easily check that 
\begin{equation}
	\cal P(\b \mu|\b y) = \cal N_N\pBB{ \frac{\tau^2}{\tau^2+\sigma^2} \b y, \frac{\tau^2\sigma^2}{\tau^2+\sigma^2} \In}.
\end{equation}
Therefore, the Bayes MMSE \eqref{eq:bayes_MMSE} of $\b\mu$ is given by
\begin{equation}
	\label{eq:JS_estimator}
		\avg{\b\mu}_{\cal P(\b \mu|\b y)} = \pBB{1 - \frac{\sigma^2}{\sigma^2 + \tau^2}} \b y,
\end{equation}
that is -- loosely speaking -- the celebrated James-Stein estimator \cite{james1961estimation}. In fact, the James-Stein estimator follows using the evidence $\cal P(\b y)$, and this approach is known as \emph{empirical Bayes} (see at the end of this section for more details).

\begin{figure}[!]
	 \begin{center}
	\includegraphics[scale = 0.4]{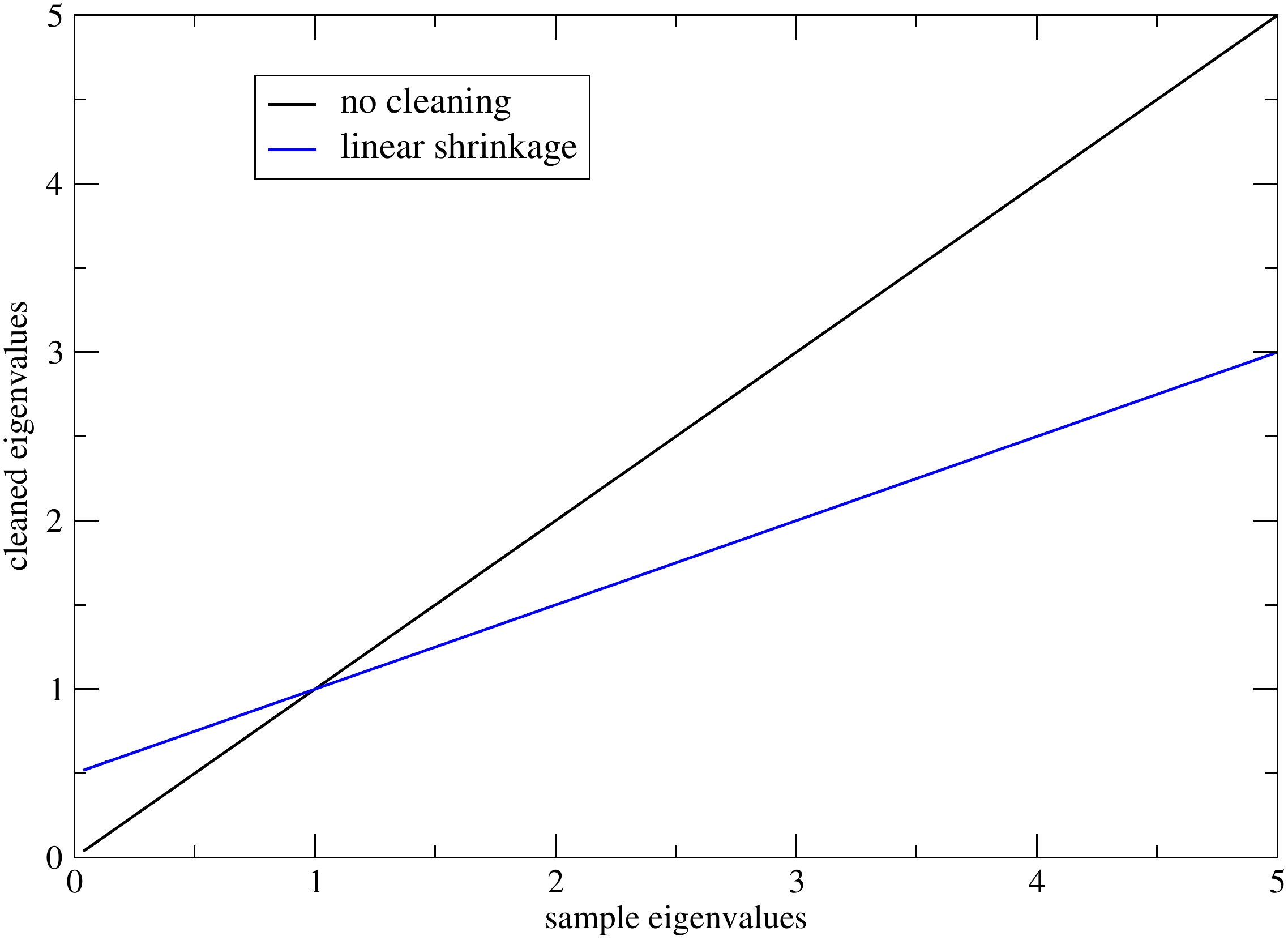} 
	\end{center}
	\caption{Impact of the linear shrinkage \eqref{eq:linear_shrinkage} with $\alpha_s = 0.5$ on the eigenvalues (blue line), compared to the sample eigenvalues (black line). We see that the small eigenvalues are shifted upward and the large ones are pulled downward. }
	\label{fig:linear_shrinkage}
\end{figure}

One can now wonder whether we can generalize this conjugate prior property to the case of covariance matrices under a measurement process characterized by the likelihood function ${\cal P}(\E | \C)$ given in Eq.\ \eqref{eq:gaussian_likelihood}. Again, we will see that conjugate prior approach yields a very interesting result. Using the potential theory formalism introduced in \eqref{eq:Boltz} and in Section \ref{sec:potential}, it is easy to see from Eq.\ \eqref{eq:gaussian_likelihood} that the potential function associated to a Gaussian likelihood function reads 
\begin{equation}
	\label{eq:gaussian_potential}
V_{q}(\E , \C) = \frac{1}{2q} \left[ \log(\C) + \E \C^{-1} \right],
\end{equation}
that is clearly the Inverse-Wishart distribution encountered in \eqref{eq:inverse_wishart_distribution} in the presence of an external field $\E$. Hence, let us 
introduce an inverse-Wishart ensemble with two hyper-parameters $\{\gamma, \kappa\}$ as a prior for $\C$:\footnote{More precisely, it is an inverse Wishart distribution ${\cal IW}_{N}(N, N(2\gamma - 1) - 1, 2N\kappa \In)$ defined in Eq.\ \eqref{eq:inverse_wishart_distribution}.}
\begin{equation*}
{\cal P}(\C) = Z \exp\left\{ - N \Tr \left[ \gamma \log \C + \kappa \C^{-1} \right] \right\},
\end{equation*}
with $Z$ a normalization constant that depends on $\gamma, \kappa$ and $N$. For simplicity, we impose that $\avg{\C}_{\cal P(\C)} = \b I_N$ and easily obtain (omitting term in $\cal O(N^{-1})$) that $\gamma = \kappa + 1$. This is the convention that we adopt henceforth. Using Bayes rule and the Gaussian likelihood function \eqref{eq:gaussian_likelihood}, we find that the posterior distribution is also an inverse-Wishart distribution of the form: 
\begin{equation}
\label{posterior_invW}
{\cal P}(\C \lvert \E) \propto \exp\left\{ - \frac12 \Tr \left[ (T+\nu+N+1) \log \C + T(2q\kappa \In + \E) \C^{-1} \right] \right\} ,
\end{equation}
where we defined $\nu := N(2\kappa + 1) - 1$. As a consequence, we expect the Bayes estimator to be explicit like the James-Stein estimator \eqref{eq:JS_estimator} and the final result for $\Xi^{\text{MMSE}}$ is obtained from \eqref{eq:inverse_wishart_mean}:
\begin{equation}
\Xi^{\text{MMSE}} =   \frac{T}{T+\nu-N-1} (2q\kappa \In + \E).
\end{equation}
This estimator is known as the \emph{linear shrinkage} estimator, first obtained in \cite{haff1980empirical},
\begin{equation}
\label{eq:linear_shrinkage_notcentered}
\Xi^{\text{lin}} \;\deq\; \frac{T}{T+\nu-N-1} (2q\kappa \In + \E) \approx \frac{1}{1+2q\kappa} \E + \frac{2q\kappa}{1+2q\kappa} \In + {\cal O}(T^{-1}),
\end{equation}
where we used that $T \rightarrow \infty$ with $q = N/T$ finite in the RHS. 
%We see that the estimator \eqref{eq:linear_shrinkage_notcentered} does not necessarily preserve the trace of the correlation matrix. One can fix one of the two parameters and determine the second one such that $\tr\Xi^{\text{lin}}/N = 1$, leading to
%the relation $\gamma = \kappa + 1$. 
%Therefore, we can simplify the linear estimator \eqref{eq:linear_shrinkage_notcentered} as 
All in all, we have derived the linear shrinkage estimator:
\begin{equation}
\label{eq:linear_shrinkage}
\Xi^{\text{lin}} = \alpha_{s} \E + (1-\alpha_{s}) \In  \quad\text{where}\quad  \alpha_s \;\deq\; \frac{1}{1+2q\kappa} \, \in [0,1], \; \kappa > 0\,.
\end{equation}
As for the James-Stein estimator, this estimator tells us to \emph{shrink} the sample covariance matrix $\E$ toward the identity matrix (our prior) with an intensity given by $\alpha_s$. We give a simple illustration of how this estimator transforms the eigenvalues in Figure \ref{fig:linear_shrinkage}. In particular, we see that small eigenvalues are lifted upwards while the top ones are pulled downwards. Furthermore, it is easy to see this estimator shares the same eigenvectors than the sample covariance matrix $\E$. This property will be important in the following. 

The remaining question is how can we consistently choose the parameter $\kappa$ (or directly $\alpha_s$) in order to use this estimator in practice? In \cite{haff1980empirical}, Haff promoted an empirical Bayes approach similar to the work of James and Stein \cite{james1961estimation}. In the high-dimensional regime, Ledoit \& Wolf \cite{ledoit2004well} noticed that this approach may suffer from the fact that classical estimators become unreliable and consequently proposed a consistent estimator of $\alpha_s$. There also exist more straightforward methods to estimate the parameter $\kappa$ directly from the data, using RMT tools. We summarize all these approaches in Section \ref{sec:linear_shrinkage}.

One may finally remark that the above derivation of the linear shrinkage estimator can be extended to the case where the prior is different from the identity matrix. Suppose that the prior distribution of $\C$ is a generalized inverse-Wishart distribution:
\begin{equation*}
{\cal P}(\C) = Z \exp\left\{ - N \Tr \left[ \gamma \log \C + \kappa \C_0 \C^{-1} \right] \right\},
\end{equation*}
where $\C_0$ is a certain matrix (referred as a \textit{fundamental} or \textit{prior} matrix) with a possibly non-trivial structure encoding what we believe about the problem at hand. In this case, it is easy to see that the above linear estimator still holds, with:
\begin{equation}
\label{eq:linear_shrinkage-bis}
\Xi^{\text{lin}} = \alpha_{s} \E + (1-\alpha_{s}) \C_0  \qquad  \alpha_s \in [0,1].
\end{equation}
Note that when $\C_0 \neq \In$, $\cal P(\C)$ is no longer rotationally invariant. A simple example is to choose $\C_0 = (1 - \rho) \In + \rho \b J$, where $\b J$ has all its elements equal to unity. This corresponds to a one-factor model in financial applications, where the correlations between any pair of stocks are constant. This 
can also be seen as a spike correlation model, as was shown in \eqref{eq:spikeless_population_cov} above, with $\ul{\C}= \In$, $r=1$, $v_1= (1,1,\dots,1)$ and $d_1 = (N-1) \rho$.

\begin{changemargin}{1.0cm}{1.0cm} 
\footnotesize
We now present the empirical Bayes approach through the ``non-observable'' James-Stein estimator \eqref{eq:JS_estimator}. This approach can be useful in order to estimate parameters directly from the data but it requires that the marginal distribution can be computed exactly. If we reconsider the framework of the estimator \eqref{eq:JS_estimator}, it is not hard to see that the evidence $\cal P(\b y)$, defined in \eqref{eq:marginalPDF}, is given by
\begin{equation}
	\label{eq:JS_marginal}
 	\cal P(\b y) \sim \cal N_N(0, (\sigma^2 + \tau^2) \b I_N)\,.
\end{equation}
Recall from \eqref{eq:JS_estimator} that our aim is to estimate the ratio $\sigma^2/(\sigma^2 +\tau^2)$ where $\sigma^2$ is known. To that end, we notice from \eqref{eq:JS_marginal} that 
\begin{equation}
	\normb{\b y}_{2}^{2} \sim (\sigma^2 + \tau^2) \chi_{N}^2\,,
\end{equation}
where $\normb{\cdot}_2$ is the $\mathbb{L}_2$ norm and $\chi_N^2$ is the chi-square distribution with $N$ degrees of freedom. Therefore, we can conclude by maximum likelihood estimation that
\begin{equation}
	\frac{\sigma^2 \times \max(N-2,0) }{\normb{\b y}_{2}^{2}} \approx \frac{\sigma^2}{\sigma^2 + \tau^2}\,,
\end{equation}
which yields an estimator of the unobservable term in Eq.\ \eqref{eq:JS_estimator}. Hence, if we plug this sample estimate into \eqref{eq:JS_estimator}, it yields the celebrated James-Stein estimator:
\begin{equation}
	\hat{\b \mu}_{\text{JS}} = \pBB{1 - \frac{\sigma^2 \times \max(N-2,0)}{\normb{\b y}_{2}^{2}}} \b y\,,
\end{equation}
that provides an improvement upon the maximum likelihood estimator of the mean of a Gaussian population whenever $N \geq 3$.

\end{changemargin} 
\normalsize

\subsection{Rotational invariant prior estimators}

The major drawback of the above conjugate prior class of estimator is that it does not make use of the enormous amount of information contained, for large $N$, 
in the observed spectral density of the sample correlation matrix $\E$. In fact, we know that its Stieltjes transform $\stj_{\E}(z)$ must obey the Mar{\v c}enko-Pastur equation relating it to $\stj_{\C}(z)$, and there is no guarantee whatsoever that this relation can be obeyed for any $\C$ belonging to an Inverse-Wishart ensemble. More precisely, the likelihood that $\stj_{\E}(z)$ indeed corresponds to a certain $\stj_{\C}(z)$ with $\C$ an Inverse-Wishart matrix is exponentially small in $N$, even for the optimal choice of the parameter $\kappa$. This is the peculiarity of the Bayesian approach in the large $N$ limit: the ensemble to which $\C$ belongs is in fact extremely strongly constrained by the Mar{\v c}enko-Pastur relation. In this section and in the next chapter, we discuss how these constraints can be implemented in practice, allowing us to construct a truly consistent estimator of $\C$. 

Let us consider a class of {\it rotationally invariant prior} distributions that belong to the Boltzmann class, Eq.\ \eqref{eq:Boltz}, i.e. 
\begin{equation}
	\label{eq:prior_boltzmann}
	\cal P(\C) \propto \exp[-N \tr V_0(\C)]
\end{equation}
where $V_0$ denotes the potential function. Therefore, it is easy to see that $\C \overset{\text{law}}{=} \b\Omega \C \b\Omega^*$ for any $N \times N$ orthogonal matrix $\b\Omega \in \b O(N)$. In other words, the eigenbasis of $\C$ is not biased in any specific direction. Moreover, using the Gaussian likelihood function \eqref{eq:gaussian_likelihood}, the posterior distribution reads:
\begin{equation}
	\label{eq:posterior_boltzmann}
	\cal P(\C | \E) = \frac1Z \exp\qB{- N \tr \cal V(\C, \E) }, \qquad \cal V(\C,\E) \;\deq\; V_q(\C,\E) + V_{0}(\C),
\end{equation}
where $V_q$ is defined in Eq.\ \eqref{eq:gaussian_potential}. As a result, one can derive the identity: 
\begin{equation}
	\label{eq:RI_posterior}
	\cal P(\C|\E) = \cal P(\b\Omega \C \b\Omega^* | \b\Omega \E \b\Omega^*),
\end{equation}
Therefore, the Bayes MMSE estimator Eq.\ \eqref{eq:bayes_MMSE} obeys the following property:
\begin{eqnarray}
	\label{eq:RIE_class_property}
	\avg{\C}_{\cal P(\C | \E)} & = & \int \b\Omega \C' \b\Omega^* \cal P(\b\Omega \C' \b\Omega^* | \E) \cal D\C' \nonumber \\
	& = & \b\Omega \left[\int  \C' \cal P(\C' |\b\Omega^* \E \b\Omega) \cal D\C' \right]\b\Omega^*  \equiv \b\Omega \avg{\C}_{\cal P(\C| \b\Omega^* \E \b\Omega)} \b\Omega^*  
\end{eqnarray}
where we changed variables $\C  \to \b\Omega \C' \b\Omega^*$ and used Eq.\ \eqref{eq:RI_posterior} in the last step. Now we can always choose $\b\Omega= \b U$ such that $\b U^* \E \b U$ is diagonal. In this case, it is not 
difficult to convince oneself using symmetry arguments that $\avg{\C}_{\cal P(\C| \b U^* \E \b U)}$ is then also diagonal. The above result then simply means that in general, the MMSE estimator of $\C$ is diagonal in the 
same basis as $\E$ --  see Takemura \cite{takemura1983orthogonally} and references therein:
\begin{equation}
	\label{eq:Bayes_estimator_eigen}
		\Xi^{\text{MMSE}} = \b U \b \Gamma(\Lambda) \b U^*,
\end{equation}
where $\b U \in \mathbb{R}^{N \times N}$ is the eigenvectors of $\E$ and $\b \Gamma(\b\Lambda) = \diag(\gamma_1(\b\Lambda), \dots, \gamma_N(\b\Lambda))$ is a $N \times N$ diagonal matrix whose entries are functions of the sample eigenvalues $\b\Lambda = \diag(\lambda_1,\lambda_2, \dots, \lambda_N)$. We see that assuming a rotationally invariant prior, the Bayesian estimation problem is reduced to finding a set of optimal eigenvalues $\gamma_i(\Lambda)$. This framework agrees perfectly with the linear shrinkage estimator \eqref{eq:linear_shrinkage}, for which $\gamma_i(\Lambda) := \alpha_{s} \lambda_i + (1 - \alpha_s)$, and can be seen as a generalized shrinkage estimator.

Before going into details on the explicit form of the $\b \Gamma(\b\Lambda)$, let us motivate the assumption of rotational invariance for the prior distribution of $\C$. Suppose that we have no prior information 
on possible privileged directions in the N-dimensional space that would allow one to bias the eigenvectors of the estimator $\Xi^{\text{MMSE}}$ in these special directions. In this case, it makes sense that the only 
reasonable eigenbasis for our estimator $\Xi^{\text{MMSE}}$ must be that the (noisy) observation $\E$ at our disposal. Any estimator satisfying Eq. \eqref{eq:RIE_class_property} will be referred to as a Rotational Invariant Estimator (RIE). However, we emphasize that such an assumption is not optimal when the components of $\E$ reveal some non-trivial structures. One example is the top eigenvector of financial correlation matrices, which is clearly biased in the 
$(1,1,\dots,1)$ direction. Dealing with such non-rotational invariant objects is however more difficult (see \cite{bun2016optimal,monasson2015estimating} and Chapter \ref{chap:conclusion} for a discussion on this topic).

We are now in a position to derive the explicit form of our optimal Bayes estimator within the class of RIEs. The eigen decomposition \eqref{eq:Bayes_estimator_eigen} of the estimator $\Xi^{\text{MMSE}}$ 
states that the eigenvalues of $\gamma_i \equiv \gamma_i(\Lambda)$ can be written as
\begin{equation*}
\gamma_i = \scalar{\b u_{i}}{\avg{\C}_{\cal P(\C | \E)} \b u_i},
\end{equation*}
where we have used the fact that $\avg{\C}_{\cal P(\C | \E)}$ is diagonal in the $\b U$ basis. After a little thought, one can see that the following identity holds:
\begin{equation}
\label{eq:stieltjes_F_RIE}
\frac1N \Tr \left[ (z\In - \E)^{-1} \avg{\C}_{\cal P(\C | \E)} \right] = \frac1N \sum_{i=1}^{N} \frac{ \gamma_i}{z - \lambda_i},
\end{equation}
%Recall that we indexed the eigenvectors by their associated eigenvalues, so that in the large $N$ limit, one may write
% \begin{equation}
% 	\label{eq:Bayes_RIE_1}
% \frac1N \Tr \left[ (z\In - \E)^{-1} \avg{\C}_{\cal P(\C | \E)} \right]  \sim \int \frac{\psi(\lambda) \rho_\E(\dd\lambda)}{z - \lambda},
% \end{equation}
which will allow us to extract the $\gamma_i$ we are looking for, i.e. determine the optimal shrinkage function of the Bayes estimator \eqref{eq:Bayes_estimator_eigen}. 
To that end, we invoke the usual self-averaging property that holds for very large $N$, so that we can take the average value over the marginal probability of $\E$ in the LHS of the last equation, yielding:
\begin{eqnarray}
\Tr \left[ (z\In - \E)^{-1} \avg{\C}_{\cal P(\C | \E)} \right] & = & \bigg\langle \Tr \left[ (z\In - \E)^{-1}  \avg{\C}_{\cal P(\C | \E)} \right] \bigg\rangle_{\cal P(\E)}, \nonumber\\
& = & \bigg\langle \bigg\langle \Tr \left[ (z\In - \E)^{-1} \C \right] \bigg\rangle_{\cal P(\C | \E)} \bigg\rangle_{\cal P(\E)}.
\end{eqnarray}
Using Bayes formula \eqref{eq:Bayes_rule}, we rewrite this last equation as
\begin{eqnarray}
\Tr \left[ (z\In - \E)^{-1} \avg{\C}_{\cal P(\C | \E)} \right]  & = & \bigg\langle \bigg\langle \Tr \left[ (z\In - \E)^{-1} \C \right] \bigg\rangle_{\cal P(\E | \C)} \bigg\rangle_{\cal P(\C)},  \nonumber\\
& = &  \bigg\langle \Tr \left[ \avgb{(z\In - \E)^{-1}}_{\cal P(\E|\C)} \C \right] \bigg\rangle_{\cal P(\C)}.
\end{eqnarray}
We recognize in the last line the definition of the Stieltjes transform of $\E$ for a given population matrix $\C$, which allows us to use the Mar{\v c}enko-Pastur formalism introduced in Chapters \ref{chap:spectrum} and \ref{chap:eigenvectors}. Therefore, since the eigenvalues $[\lambda_i]_i$ become deterministic in the limit $N \to \infty$ (see Chapter \ref{chap:spectrum}), we deduce that for large $N$
\begin{equation}
	\label{eq:Bayes_RIE_2}
\frac1N \Tr \left[ (z\In - \E)^{-1} \avg{\C}_{\cal P(\C | \E)} \right] \approx \int \frac{\rho_\E(\lambda) \dd\lambda}{z-\lambda} \bigg\langle \sum_{j=1}^{N} \mu_j\, \mso(\lambda, \mu_j)  \bigg\rangle_{\C},
\end{equation}
where $\mso(\lambda,\mu)$ is the mean squared overlap defined in Eq.\ \eqref{eq:overlap}. By comparing Eqs.\ \eqref{eq:stieltjes_F_RIE} and \eqref{eq:Bayes_RIE_2}, we can readily conclude that 
\begin{equation}
	\label{eq:Bayes_RIE}
	\gamma(\Lambda) \equiv \gamma(\lambda) = \bigg\langle \sum_{j=1}^{N} \mu_j\, \mso(\lambda, \mu_j) \bigg\rangle_{\C} \sim \int \mu \, \mso(\lambda, \mu) \rho_\C(\mu)\dd\mu,
\end{equation}
where we used again an ``ergodic hypothesis'' \cite{deutsch1991quantum} as $N \to \infty$ in the last step. Hence, we see that in the large $N$ limit, we are able to find a closed formula for the optimal shrinkage function $\gamma$ of the Bayes estimator \eqref{eq:Bayes_estimator_eigen} that depends on the mean squared overlap, studied in Chapter \ref{chap:eigenvectors}, and the prior spectral density $\rho_\C$. Said differently the final result Eq.\ \eqref{eq:Bayes_RIE} is explicit but still seems to depend on the prior we choose for $\C$. In fact, as we shall see in the next chapter, Eq.\ \eqref{eq:Bayes_RIE} can be estimated from the 
knowledge of $\E$ itself, i.e. without making any explicit choice for the prior! This is in line with our discussion at the beginning of this section: for large $N$, the observation of the spectral distribution of $\E$ is enough to 
determine the correct prior ensemble to which $\C$ must belong. 

\begin{figure}[!ht]
  \begin{center}
   \includegraphics[scale = 0.4]{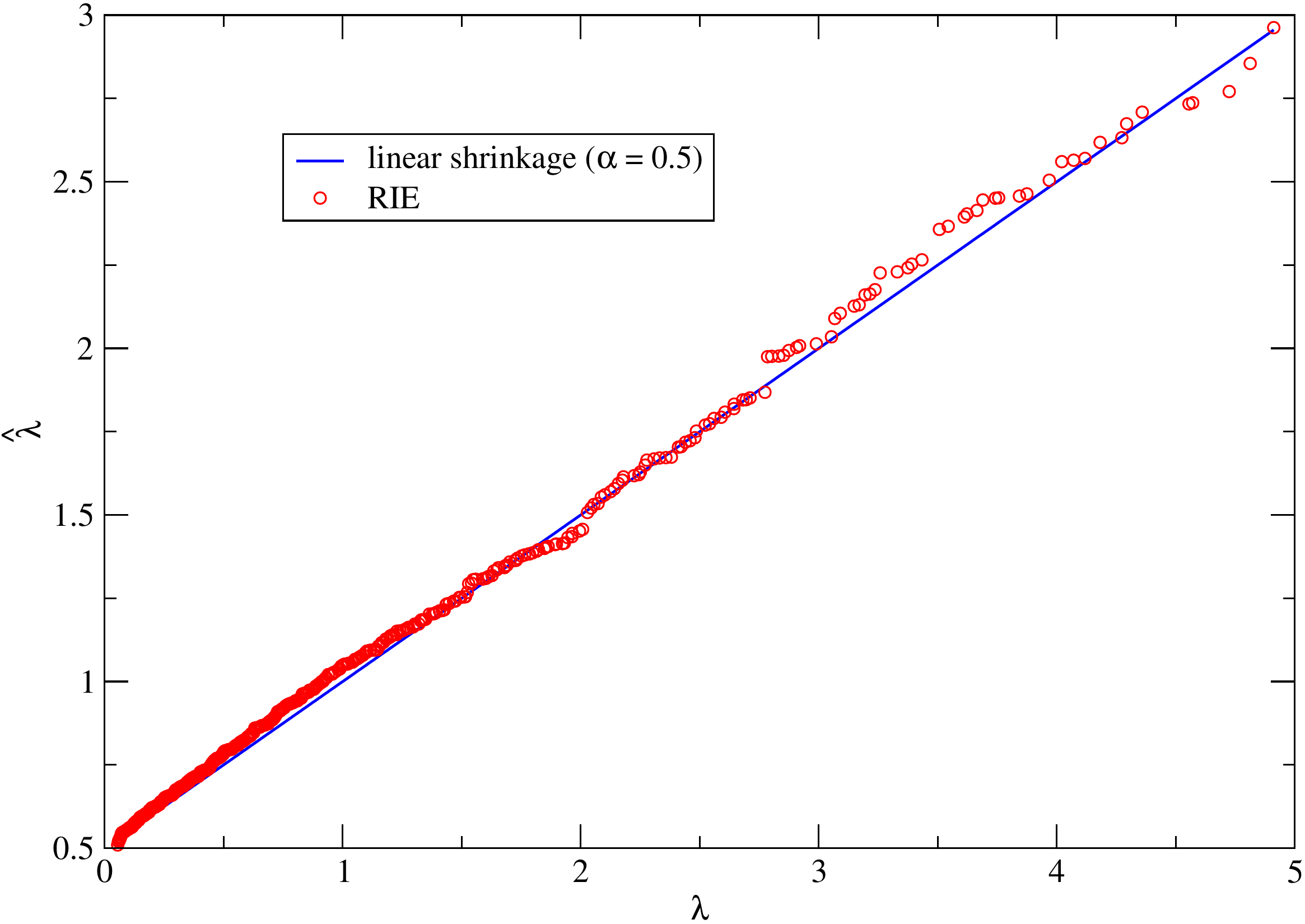} 
   \end{center}
   \caption{Comparison of our analytical RI-Bayes estimator \eqref{eq:Bayes_RIE} (red dots) with the theoretical result Eq.\ \eqref{eq:linear_shrinkage} (blue line) when the prior distribution is an inverse Wishart \eqref{eq:inverse_wishart_distribution}. The parameters are $N = 500$, $q = 0.5$ and $\alpha_s = 0.5$.}
   \label{fig:chap_Bayes_RIE_linear_N_500}
\end{figure}

We end this section with a self-consistency check in order to illustrate the result \eqref{eq:Bayes_RIE}. As alluded to above, the nonlinear shrinkage function \eqref{eq:Bayes_RIE} generalizes the linear shrinkage \eqref{eq:linear_shrinkage}. To highlight this, we assume that $\C$ is an isotropic Inverse Wishart matrices, such that the prior spectral density $\rho_\C$ is given by Eq.\ \eqref{eq:IMP_density}. We plot in Fig. \ref{fig:chap_Bayes_RIE_linear_N_500} the eigenvalues we obtain using our Bayes estimator \eqref{eq:linear_shrinkage} (red dots) coming from a single realization of $\E$ with $\C$ an inverse Wishart matrix of size $N = 500$. The parameter of the prior distribution has been chosen such that the shrinkage intensity is equal to one half. We see that the agreement is excellent, showing the validity of the ergodic hypothesis and at the same time, of the RI-Bayes estimator \eqref{eq:Bayes_RIE} in this particular case. In section \ref{sec:revisiting_linear}, we will show explicitly that Eq.\ \eqref{eq:Bayes_RIE_2} reproduces Eq.\ \eqref{eq:linear_shrinkage} when $\C$ is 
an isotropic Inverse Wishart matrix.

\clearpage%!TEX root = RMT_Covariance_Review.tex
\section{Optimal rotational invariant estimator for general covariance matrices}
\label{chap:RIE}

\subsection{Oracle estimator}

In the previous chapter, we introduced a Bayesian framework to build an estimator of the population correlation matrix $\C$ using the data $\Y$ at our disposal. 
We showed that using a conjugate prior assumption naturally leads to the class of linear shrinkage estimators, which is arguably among the most influential contributions to this topic. It was used successfully in many contexts as  a simple way to provide robustness against the noise in high dimensional settings (see e.g. \cite{stein1956inadmissibility,haff1980empirical} or \cite{karoui2011geometric} for a more recent review). However, the main concern regarding this estimator is that the conjugate prior ensemble is expected to be exponentially improbable (for large $N$) with the data at hand. In order to make full use of the information of 
the spectral density of the sample correlation matrix, we introduced a class of rotational invariant prior distributions. Within this framework, we have derived an
explicit formula for the \textit{minimum mean squared error} (MMSE) estimator valid in the limit of large dimension, which can be seen as a non-linear shrinkage
procedure. In this chapter, we want to show that the resulting estimator can be also understood as a so-called ``oracle'' estimator. This change of viewpoint 
is quite interesting as it shows that the above Bayes estimator has a much wider basis than anticipated. 

Imagine that one actually \emph{knows} the population matrix $\C$ -- hence the name ``oracle'' -- but that one 
decides to create an estimator of $\C$ that is constrained to have a predetermined eigenbasis $\b U$. (In practice, this eigenbasis will be that of the sample
correlation matrix $\E$). What is the best one can do to estimate the true matrix $\C$? The basic idea might look strange at first sight, since we do not know $\C$ 
at all! But as we shall see below, the oracle estimator will turn out to coincide with the MMSE estimator which is, for large $N$, entirely expressible in terms of observable quantities. More precisely, let us introduce the set $\cal M(\b U)$ of real symmetric definite positive $N \times N$ matrices that are diagonal in the 
basis $\b U = [\b u_i]_{i \in \qq{1,N}}$. The optimal estimator of $\C$ in $\cal M(\b U)$ in the $L_2$ sense is given by:
\begin{equation}
	\Xi^{\text{ora.}} \; = \; \underset{\Xi \in \cal M(\b U)}{\argmin} \normb{\Xi - \C}_{L_2}^2.
\end{equation}
It is trivial to find that the solution of this quadratic optimization problem, as:
\begin{equation}
	\label{eq:oracle}
	\Xi^{\text{ora.}} = \sum_{i=1}^{N} \xi^{\text{ora.}}_i \b u_i \b u_i^*, \qquad \xi^{\text{ora.}}_i = \scalar{\b u_i}{\C \b u_i}.
\end{equation}
This provides the best possible estimator of $\C$ given that we are ``stuck'' with the eigenbasis $[\b u_i]_{i \in \qq{1,N}}$. 
The meaning of this estimator is better understood if we rewrite it a function of the eigenvectors of $\C$, to wit: 
\begin{equation}\label{oracle_simple}
	\xi^{\text{ora.}}_i = \sum_{j=1}^{N} \mu_j \scalar{\b u_i}{\b v_j}^2.
\end{equation}
Indeed, we see from this last equation that the oracle estimator is given by a weighted average of the population eigenvalues with weights given by the transition from the imposed basis $\b u_i$ to the true basis $\b v_j$ with $j \in \qq{1,N}$. Hence, the ``oracle'' estimator \eqref{eq:oracle} explicitly uses the fact that 
the estimator lies in a wrong basis. 

Coming back to our estimation of $\C$ given a sample matrix $\E$, it is clear that if we have no information whatsoever on the true eigenbasis of $\C$, the only possibility is to use the eigenbasis of $\E$ itself as $\b U$. This is equivalent to the assumption of a rotationally invariant prior distribution for 
$\C$, but we do not rely on any Bayesian argument here. Now, one notices that in the limit $N \to \infty$, the oracle eigenvalues of $[\xi^{\text{ora.}}_i]_{i \in \qq{1,N}}$ are indeed equivalent to the RI-Bayes MMSE formula \eqref{eq:Bayes_RIE}, except that in Eq.\ \eqref{eq:oracle}, the population matrix $\C$ is a (deterministic) general covariance matrix. The equivalence between Bayes estimator \eqref{eq:Bayes_RIE} and unconditional estimator is not that surprising in the large $N$ limit and has been mentioned in different contexts \cite{karoui2011geometric,dicker2016ridge}. 

\subsection{Explicit form of the optimal RIE}

For practical purposes, the oracle estimator \eqref{eq:oracle} looks useless since it involves the matrix $\C$ which is exactly the quantity we wish to estimate. But in the high-dimensional limit a kind of ``miracle'' happens in the sense that the oracle estimator converges to a deterministic RIE that does not involve the matrix $\C$ anymore. Let us derive this formula first for bulk eigenvalues, then for outliers -- with the further surprise that the final expression is exactly the same in the two cases. 

\subsubsection{The bulk}

The derivation of the optimal nonlinear shrinkage function for the bulk eigenvalues in the limit of infinite dimension was considered in different recent works. The first one goes back to the work of Ledoit \& P{\'e}ch{\'e} \cite{ledoit2011eigenvectors}. More recently, this oracle estimator was considered in a more general framework \cite{bun2015rotational} (including the case of additive noise models, see Appendix \ref{app:addition}) with the conclusion was that the oracle estimator can be easily computed as soon as the convergence of the mean squared overlap $\mso(\lambda_i, \mu_j)$ defined in Eq.\ \eqref{eq:overlap} can be established. 

More precisely, let us fix $i \geq r+1$\footnote{Recall that the largest $r$ eigenvalues are assumed to be outliers.}, we expect that in the limit of large dimension, the squared overlaps $\scalar{\b u_i}{\b v_j}^2$ for any $j = 1, \dots, N$ will display asymptotic independence so that the law of large number applies, leading to a deterministic result for $\xi^{\text{ora.}}_i$. Hence, for large $N$, we have that for any $i > r$,
\begin{align}
\label{eq:RIE_resolvent_relation}
\xi^{\text{ora.}}_i = \sum_{j = 1}^{N}  \mu_j \, \mso(\lambda_i, \mu_j) \approx \frac{1}{N \pi \rho_{\E}(\lambda_i)} \lim_{\eta \rightarrow 0^+} \im \left[ \sum_{j=1}^{N}
\mu_j \left(z_i \In - \b E\right)^{-1}_{jj} \right],
\end{align}
where we have used the result Eq.\ \eqref{eq:resolvent_overlap} with $z_i =\lambda_i  - \ii\eta$. 
One finds using the Mar{\v c}enko-Pastur relation \eqref{eq:MP_equation_int} and after simple algebraic manipulations that 
\begin{equation*}
\xi^{\text{ora.}}_i \sim \frac{1}{q \pi \rho_{\E}(\lambda_i)}  \lim_{\eta \rightarrow 0^+} \im \left[ 1 - \frac{1}{1-q+qz_i\stj_{\E}(z_i)} \right],
\end{equation*} 
which can be further simplified to the final Ledoit-P\'ech\'e formula for the  oracle estimators $[\xi^{\text{ora.}}_i]_{i\in\qq{r,N}}$:
\begin{equation}
\label{eq:RIE_optimal}
\xi^{\text{ora.}}_i \sim \hat\xi(\lambda_i) \qquad\text{with}\qquad \hat\xi(\lambda) \;\deq\; \frac{\lambda}{ \absb{1-q+q \lambda \lim_{\eta \to 0^+} \stj_{\E}(\lambda - \ii\eta)}^2 }\,,
%\frac{\lambda_i}{ \left| 1-q+q \lambda_i \lim_{\eta \to 0^+} \stj_{\E}(\lambda_i - \ii \eta) \right|^2 },
\end{equation}
where $|\cdot|$ denotes the complex modulus. We notice that the RHS of this last equation does not involve the matrix $\C$ anymore and depends only 
on deterministic quantities. This is the ``miracle'' of the large $N$ limit we alluded to above: the {\it{a priori}} non-observable oracle estimator converges to a 
deterministic quantity that may be estimated directly from the data. 

\subsubsection{Outliers}
\label{sec:rie_outlier}

As usual, the arguments needed to derive the limiting value of the oracle estimator for outlier eigenvalues, i.e., $\xi^{\text{ora.}}_i$ for $i \leq r$, are a little bit different from those used above for bulk eigenvalues. Indeed, the latter explicitly needs the density of $\varrho_\E(\lambda_i)$ to be non-vanishing (for $N \to \infty$) and as we know from Chapter \ref{chap:spectrum}, this is not the case for outliers. Hence, the method of \cite{ledoit2011eigenvectors} and \cite{bun2015rotational} are not valid anymore. Surprisingly, though, the final result happens to be identical to Eq.\ \eqref{eq:RIE_optimal}! This has been established recently in \cite{bun2016optimal} and the starting point of the method is to rewrite the oracle solution as 
\begin{equation}
	\label{eq:oracle_outlier_decomp}
	\xi^{\text{ora.}}_i = \sum_{j=1}^{r} \mu_j \scalar{\b v_j}{\b u_i}^2 + \sum_{j=r+1}^{N} \mu_j \scalar{\b v_j}{\b u_i}^2,
\end{equation}
from which we conclude, using also the results of section \ref{chap:eigenvectors}, that if $r$ is finite both terms above will have a non-vanishing contribution for $i \leq r$. Roughly speaking, the first sum will contribute in $\cal O(1)$ for $j=i$ and the second sum gives a term of order $\cal O((N-r) \times 1/N) \sim \cal O(1)$.

We begin with the easy term which is the first one in the RHS of Eq.\ \eqref{eq:oracle_outlier_decomp}. Indeed, recall from Eq.\ \eqref{eq:overlap_outlier_outlier} that any outlier eigenvector $\b u_i$ is concentrated on a cone with its axis parallel to $\b v_i$ and completely delocalized in any direction orthogonal $\b v_j$ with $j \in \qq{1,N}$, $j \neq i$ fixed. Hence, the only term that contributes to leading order will be $\scalar{\b v_i}{\b u_i}^2$ and we therefore conclude that 
\begin{equation}
	\label{eq:oracle_outlier_1st_r}
	\sum_{j=1}^{r} \mu_j \scalar{\b v_j}{\b u_i}^2 \sim \mu_i^2 \frac{\theta'(\mu_i)}{\theta(\mu_i)} 
	%= \frac{\mu_i^2}{\lambda_i} \theta'(\mu_i),
\end{equation}
where we used Eq.\ \eqref{eq:classical_location_outlier} in the last step. The second term in Eq.\ \eqref{eq:oracle_outlier_decomp} is trickier to handle. As $r$ is finite and thus much smaller than $N$, we can assume that the second sum will concentrate around its mean value, i.e.
\begin{equation*}
	\sum_{j=r+1}^{N} \mu_j \scalar{\b v_j}{\b u_i}^2 \sim \sum_{j=r+1}^{N} \mu_j \mathbb{E} \scalar{\b v_j}{\b u_i}^2.
	%\underset{N \rightarrow \infty}{\sim}
\end{equation*}
The mean squared overlap in the RHS for $j \geq r+1$ and $i \leq r$ has been evaluated in section \ref{chap:eigenvectors} and the result is given in Eq.\ \eqref{eq:overlap_outlier_bulk} that we recall here for convenience:
\begin{equation*}
		\mathbb{E}[\scalar{\b u_i}{\b v_j}^2] =  \frac{\mu_i^2}{\theta(\mu_i)} \frac{\mu_j}{T(\mu_i - \mu_j)^2} , \qquad i \leq r, j \geq r+1.
\end{equation*}
Therefore we find for $r \ll N$ \cite{bun2016optimal}
\begin{equation}
	\label{eq:tmp_RIE_outlier_sum}
	\sum_{j=r+1}^{N} \mu_j \scalar{\b v_j}{\b u_i}^2 \sim \frac{\mu_i^2}{\theta(\mu_i)} \frac 1T \sum_{j=1}^{N} \frac{\mu_j^2}{(\mu_i - \mu_j)^2},
\end{equation}
where one notices that the sum of the RHS goes from $j = 1$ to $N$. 
We can simplify the sum in the RHS of this last equation by using the Mar{\v c}enko-Pastur equation \eqref{eq:MP_equation_dual_inverse}. Indeed, by setting $z = \theta(\mu_i)$ with $i \leq r$ and $\theta$ defined in Eq.\ \eqref{eq:classical_location_outlier}, Eq.\ \eqref{eq:MP_equation_dual_inverse}, becomes
\begin{equation}
	\theta(\mu_i) = \mu_i + \frac{1}{T}\sum_{j=1}^{N} \frac{1}{\mu_j^{-1} - \mu_i^{-1}}
\end{equation}
and by taking the derivative with respect to $\mu_i$, this yields
\begin{equation}
	\frac{1}{T}\sum_{j=1}^{N} \frac{\mu_j^2}{(\mu_i - \mu_j)^2} = 1 - \theta'(\mu_i),
\end{equation}
for any $i \leq r$. By plugging this identity into Eq.\ \eqref{eq:tmp_RIE_outlier_sum}, we then obtain 
\begin{equation}
	\label{eq:oracle_outlier_after_r}
	\sum_{j=r+1}^{N} \mu_j \scalar{\b v_j}{\b u_i}^2 \sim \frac{\mu_i^2}{\theta(\mu_i)} \pb{1 - \theta'(\mu_i)},
\end{equation}
for any $i \leq r$. All in all, we see by plugging Eqs. \eqref{eq:oracle_outlier_1st_r} and \eqref{eq:oracle_outlier_after_r} into Eq. \eqref{eq:oracle_outlier_decomp} that we finally get
% \begin{equation}
% 	\xi^{\text{ora.}}_i \sim - \frac{\mu_i^2 \theta'(\mu_i)}{\theta(\mu_i)} \frac{G_{\underline \E_0}'(\theta(\mu_i))}{G_{\underline \E_0}^2(\theta(\mu_i))}.
% \end{equation}
% Then, using the fact that $\theta(\mu_i) = \cal B_{\underline \E_0}(1/\mu_i)$\footnote{Recall that the Blue transform $\cal B$ is the functional inverse of the Stieltjes transform.}, we get
% \begin{equation}
% 	G_{\underline \E_0}^2(\theta(\mu_i)) = \mu_i^{-2} \qquad G_{\underline \E_0}'(\theta(\mu_i)) = - \frac{1}{\theta'(\mu_i) \mu_i^2},
% \end{equation}
% which allows to conclude for $i \leq r$ that
\begin{equation}
	\label{eq:RIE outlier nonobserv}
	\xi^{\text{ora.}}_i \sim \frac{\mu_i^2}{\theta(\mu_i)},
\end{equation}
i.e. the oracle estimator for outliers also converge to a deterministic value which is very simple, but depends on the population eigenvalues which are not observable. However, using Eq.\ \eqref{eq:classical_location_outlier}, we can rewrite the RHS of Eq.\ \eqref{eq:RIE outlier nonobserv} as a function of the sample eigenvalues. Firstly, one notices that $\theta(\mu_i) = \lambda_i$ for $N \to \infty$ thanks to Eq. \eqref{eq:classical_location_outlier}. Moreover, we can also invert Eq. \eqref{eq:classical_location_outlier} to find 
\begin{equation*}
	\mu_i \sim \frac{1}{\stj_{\ul\S}(\lambda_i)} = \frac{\lambda_i}{1-q+q \lambda_i \stj_{\ul\E}(\lambda_i)},
\end{equation*}
for any $i \leq r$ and where we use relation Eq.\ \eqref{eq:stieltjes_E_dual} in the last step. Therefore, we deduce that in the high dimensional limit, we can rewrite Eq.\ \eqref{eq:RIE outlier nonobserv} as
\begin{equation}
	\xi^{\text{ora.}}_i \sim \frac{\lambda_i}{\absb{1-q+q \lambda_i \stj_{\ul\E}(\lambda_i)}^2}.
\end{equation}
We see that the result is similar to the result for the bulk eigenvalues except that for outliers, we need the Stieltjes transform of the spikeless, fictitious sample covariance matrix $\ul\E$. But as we consider the limit $N \to \infty$, we easily deduce using Weyl's interlacing inequalities \cite{weyl1949inequalities} that we can replace it by the Stieltjes transform of $\E$ so that we finally conclude that for any outlier $i \leq r$,
\begin{equation}
	\label{eq:RIE_optimal_outlier}
	\xi^{\text{ora.}}_i \sim \hat\xi(\lambda_i)\,, 
	%\quad\text{with}\quad \xi(\lambda) \;\deq\; \frac{\lambda}{ \absb{1-q+q \lambda \lim_{\eta \to 0^+} \stj_{\E}(\lambda - \ii\eta)}^2 }\,,
\end{equation}
where the optimal shrinkage function $\hat\xi$ is defined in \eqref{eq:RIE_optimal}. We see that the outliers of oracle estimator also converge to a deterministic function which is exactly the same than for bulk eigenvalues \eqref{eq:RIE_optimal} in the large $N \to \infty$. 

To conclude, we found that the oracle estimator converges to a limiting function that does not explicitly require the knowledge of $\C$ and is identical to the Bayes-MMSE estimator obtained in the previous Chapter. Moreover, this function is ``universal'' in the sense that the optimal non linear shrinkage needed to clean bulk eigenvalues and outliers is given by the very same function in the limit $N \to \infty$, which is very appealing for practical applications. This function is defined in Eqs.\ \eqref{eq:RIE_optimal} or \eqref{eq:RIE_optimal_outlier} and only requires the knowledge of the Stieltjes transform of $\E$, which is observable -- see below.

\subsection{Some properties of the ``cleaned'' eigenvalues} 
\label{sec:RIE_prop}

Even though the optimal nonlinear shrinkage function \eqref{eq:RIE_optimal_obs} seems relatively simple, it is not immediately clear what is the effect induced by the transformation $\lambda_i \to \hat\xi(\lambda_i)$. In this section, we thus give some quantitative properties of the optimal estimator $\Xi^{\text{ora.}}$ to understand the impact of the optimal nonlinear shrinkage function $\hat\xi(\lambda)$. 

\begin{figure}[!ht]
  \begin{center}
   \includegraphics[scale = 0.4]{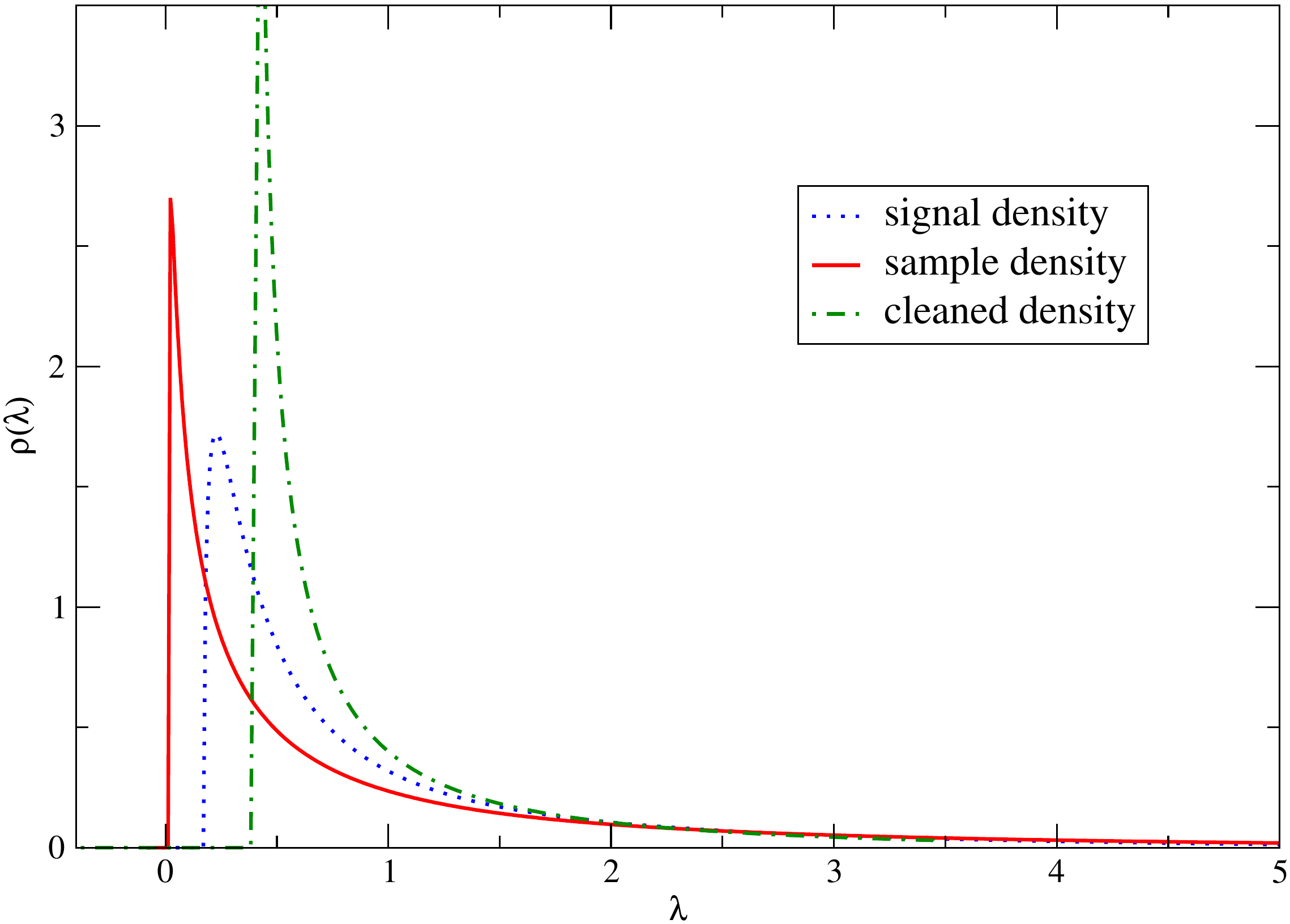} 
   \end{center}
   \caption{Evaluation of the eigenvalue density of the signal, sample and cleaned density for $q=0.5$ when the prior is an inverse Wishart of parameter $\kappa = 1$. We see that the cleaned density is the narrowest one, while the sample is the widest, as expected. }
   \label{fig:RIE_clean_density}
\end{figure}

First let us consider the moments of the spectrum of $\Xi^{\text{ora.}}$. From Eq. \eqref{oracle_simple} we immediately derive that:
\begin{equation}
	\label{eq:RIE_first_moment}
\Tr \Xi^{\text{ora.}} = \sum_{j=1} \mu_j \b v_j^* \left(\sum_{i=1} \b u_i \b u_i^*\right) \b v_j = \Tr \C,
\end{equation}
meaning that the cleaning operation preserves the trace of the population matrix $\C$, as it should be. For the moment of order 2 of the oracle estimator, we have:
\begin{equation*}
\Tr (\Xi^{\text{ora.}})^2 = \sum_{j,k=1}^{N} \mu_j \mu_k \sum_{i=1} \langle \b u_i, \b v_j \rangle^2 \langle \b u_i, \b v_k \rangle^2.
\end{equation*}
Now, if we define the matrix $\b P$ as $\{ \sum_{i=1} \langle \b u_i, \b v_j \rangle^2 \langle \b u_i, \b v_k \rangle^2 \}$ for ${j,k = 1, N}$, it is not hard to see that it is a square matrix with non-negative entries and whose rows all sum to unity. The matrix $\b P$ is therefore a (bi)stochastic matrix and the Perron-Frobenius theorem tells us that its largest eigenvalues is equal to unity. Hence, we deduce the following general inequality 
\begin{equation*}
\sum_{j,k=1}^{N} P_{j,k} \mu_j \mu_k \le \sum_{j=1}^{N} \mu_j^2,
\end{equation*}
which implies that
\begin{equation}
	\label{eq:RIE_second_moment}
	\Tr (\Xi^{\text{ora.}})^2 \leq \Tr \C^2 \leq \Tr \E^2,
\end{equation}
where the last inequality comes from Eq.\ \eqref{eq:SCM_moments}. In words, this result states that the spectrum of $\Xi^{\text{ora.}}$ is narrower than the spectrum of $\C$, which is itself narrower than the spectrum of $\E$. The optimal RIE therefore tells us that we better be even more ``cautious'' than simply bringing back the sample eigenvalues to their estimated ``true'' locations. This is because we have only partial information about the true eigenbasis of $\C$. In particular, one should always shrink downward (resp. upward) the top (resp. small) eigenvalues compared to their ``true'' locations $\mu_i$ for any $i\in\qq{1,N}$, except for the trivial case $\C = \In$. As a consequence, estimating the population eigenvalues $[\mu_i]_{i\in\qq{1,N}}$ is \emph{not} what one should do to obtain an optimal estimator of $\C$ when there is only partial information about its eigenvectors. We provide an illustration in Figure \ref{fig:RIE_clean_density} where we consider $\C$ to be an inverse-Wishart matrix with parameter $\kappa = 1$. 

%The reason why the clipping or factor models approaches are sub-optimal is now well understood: we have to shrink the largest eigenvalues downwards in order to be optimal even though they presumably correspond to true factors. This remark will be highlighted for the largest eigenvalue $\lambda_1$ in the next chapter. Hence, those two cleaning methods are not cautious enough regarding the extreme eigenvalues compared to the RIE \eqref{RIE_overlaps}.

Next, we consider the asymptotic behavior of the oracle estimator for which we recall from Eqs.\ \eqref{eq:RIE_optimal} and \eqref{eq:RIE_optimal_outlier} that
\begin{equation*}
	 \xi_i^{\text{ora.}} \sim \hat\xi_i\,, \qquad\text{with}\qquad \hat\xi_i \;\deq\; \frac{\lambda_i}{\abs{1-q+q\lambda_i \lim_{\eta\downarrow0}\stj_\E(\lambda_i - \ii\eta)}^2}\,.
\end{equation*}
Throughout the following, suppose that we have an outlier at the left of the lower bound of $\supp\rho_\E$ and let us assume $q < 1$ so that $\E$ has no exact zero mode\footnote{Recall that we assume $\C$ to be positive definite for the sake of simplicity.}. We know since Section \ref{sec:rie_outlier} that the estimator \eqref{eq:RIE_optimal} holds for outliers. Moreover, we have that $\lim_{\lambda \to 0^+} \stj_\E(\lambda)$ is real and analytic so that we have from Eq. \eqref{eq:MP_equation_mgf_inverse} that $\lambda  \stj_\E(\lambda) = \cal O(\lambda)$ for $\lambda \to 0^+$. This allows us to conclude from Eq.\ \eqref{eq:RIE_optimal} that for very small outliers, 
\begin{equation}
	\label{eq:RIE_near_zero}
	\lim_{\lambda \to 0^+} \hat\xi(\lambda) = \frac{\lambda}{(1-q)^2} + \cal O(\lambda^2),
\end{equation}
which is in agreement with Eq.\ \eqref{eq:RIE_second_moment}: small eigenvalues are enhanced for $q \in (0,1)$. 

The other asymptotic limit $\lambda \to \infty$ is also useful since it gives us the behavior of the nonlinear shrinkage function $\hat\xi$ for 
large outliers. In that case, we know from Eq.\ \eqref{eq:MP_equation_mgf} that $\lim_{\lambda \uparrow \infty} \lambda \stj_\E(\lambda)  \sim 1 + \lambda^{-1} \varphi(\E)$, where $\varphi$ denotes the normalized trace operator \eqref{eq:trace_matrix}. Therefore, we conclude that
\begin{equation}
	\label{eq:RIE_near_infty}
	\lim_{\lambda \to \infty} \hat\xi(\lambda) \approx \frac{\lambda}{\pB{1+q\lambda^{-1} \varphi(\E) + \cal O(\lambda^{-2})}^2} \sim \lambda - 2q\varphi(\E) + \cal O(\lambda^{-1}) ,
\end{equation}
and if we use that $\tr \E = \tr\C = N$, we simply obtain
\begin{equation}
	\label{eq:RIE_near_infty_normalized}
	\lim_{\lambda \to \infty} \hat\xi(\lambda) \approx \lambda - 2q + \cal O(\lambda^{-1}).
\end{equation}
It is interesting to compare this with the well-known ``Baik-Ben Arous-P{\'e}ch{\'e}'' (BBP) result on large outliers \cite{baik2005phase}, which reads (see Eq.\ \eqref{eq:spiked_cov_eigenvalue_phase_trans})  $\lambda \approx \mu + q$ for $\lambda\to\infty$. As a result, we deduce from Eq.\ \eqref{eq:RIE_near_infty_normalized} that $\hat\xi(\lambda) \approx 
\mu - q$ and we therefore find the following ordering relation
\begin{equation}
	\hat\xi(\lambda) < \mu < \lambda,
\end{equation}
for an isolated and large eigenvalues $\lambda$ and for $q > 0$. Again, this result is in agreement with Eq.\ \eqref{eq:RIE_second_moment}: large 
eigenvalues should be reduced for any $q > 0$, even below the ``true'' value of the outlier $\mu$. More generally, the non-linear shrinkage function $\hat\xi$ interpolates smoothly between $\lambda/(1-q)^2$ for small $\lambda$'s to $\lambda - 2q$ for large $\lambda$'s. Even though we did not manage to prove it, we believe that this is another manifestation of the fact that the limiting optimal nonlinear shrinkage function \eqref{eq:RIE_optimal} is monotonic with respect to the sample eigenvalues.

\subsection{Some analytical examples}

The above general properties of the oracle shrinkage procedure can be given more
flesh in some exactly solvable cases. In this section we provide 
two simple toy models where the function $\hat\xi(\lambda)$ can be characterized explicitly, before turning to numerical illustrations.  

\subsubsection{Null Hypothesis}

The first one is the null hypothesis $\C = \In$ where we shall see that, as expected $\xi^{\text{ora.}}(\lambda_i) = 1$ for any eigenvalues 
$[\lambda_i]_{i\geq r+1}$ in the bulk of the distribution. Outside of the spectrum, we observe a ``phase transition'' phenomena similar to the BBP transition \cite{baik2005phase}, that leads to a non-trivial shrinkage formula.

\begin{figure}[h]
  \begin{center}
   \includegraphics[scale = 0.45]{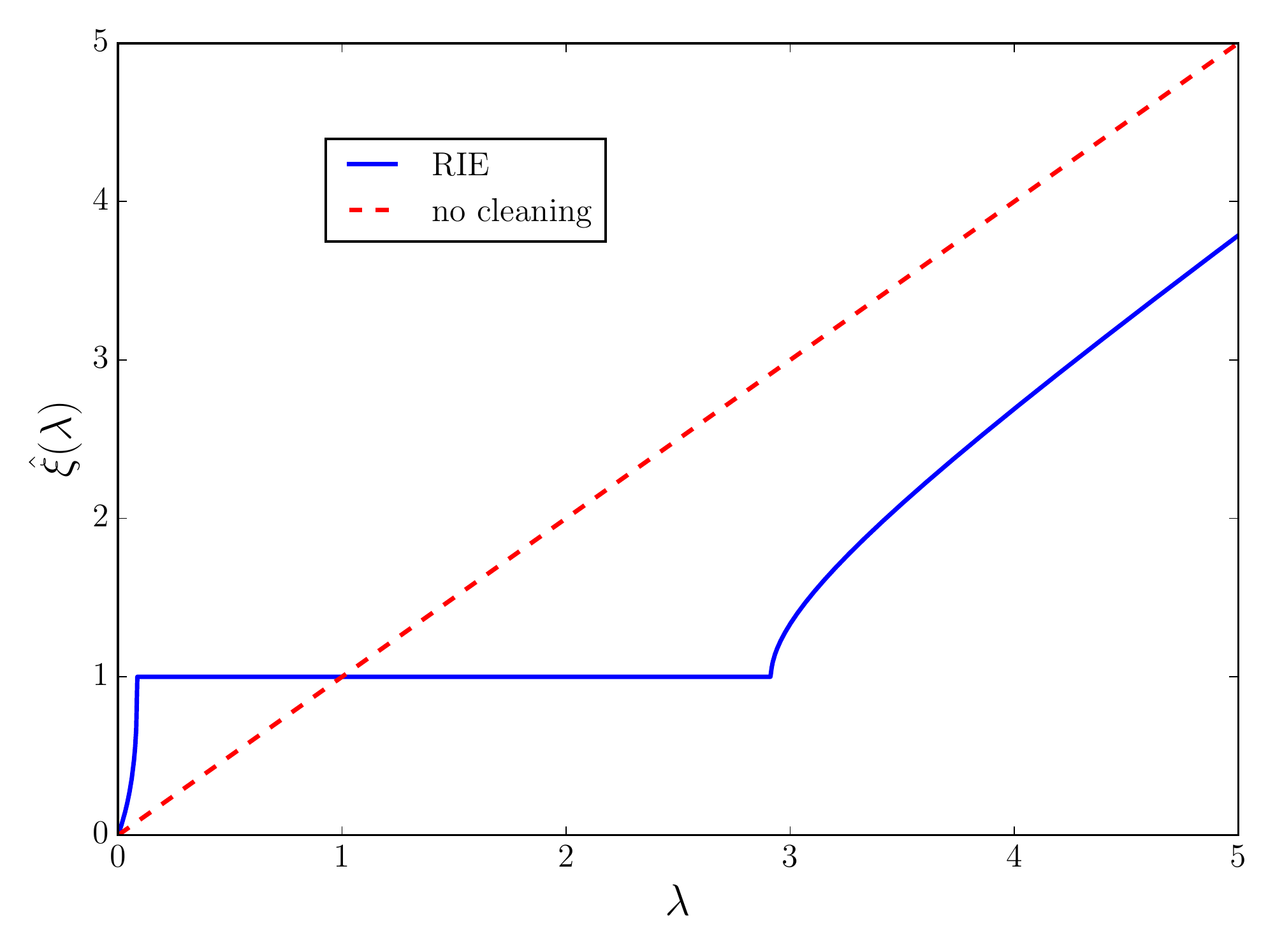} 
   \end{center}
   \caption{Evaluation of the optimal RIE's eigenvalues for $\C = I_N$ as a function of the sample eigenvalues $[\lambda_i]_{i\in\qq{1,N}}$ for $q=1/2$. The nonlinear shrinkage function is plotted with the plain blue line. We see that for $\lambda > (1+\sqrt{q})^2$, a phase transition occurs and the corresponding ``cleaned'' eigenvalues converge for large $\lambda$ to $\lambda - 2q$ (the red dotted line shifted down by $2q=1$). Note the square-root singularity of the estimator as one gets close to the edge of the spectrum. There is a similar phase transition for outliers $\lambda < (1-\sqrt{q})^2$ (see Figure \ref{fig:rie_left_edge_pt}). }
   \label{fig:rie_null}
\end{figure}

We begin with the outliers of $\E$. By assumption of our model, all the outliers have a contribution of order $N^{-1}$ so that in the limit $N \to \infty$, $\stj_\E$ is real and analytic for any $\lambda_i$ with $i \leq r$. Hence, the estimator is easily obtained by plugging the Stieltjes transform \eqref{eq:stieltjes_isotropic_wishart} into Eq.\ \eqref{eq:RIE_optimal}, with a result shown in Fig.\ \ref{fig:rie_null}. 

For bulk eigenvalues, the computation can be done more explicitly. First, using Eq.\ \eqref{eq:stieltjes_isotropic_wishart}, one finds
\begin{equation*}
	1-q + qz\stj_\E(z) = \frac{(z+1-q) \pm \sqrt{(z+q-1)^2 - 4zq}}{2}.
\end{equation*}
For $z = \lambda - \ii\eta$ with $\lambda \in \qb{(1-\sqrt{q})^2, (1+\sqrt{q})^2}$, we know that the square root in the latter equation becomes imaginary for $\eta\to 0^+$. Hence, if we take the square modulus, one gets
\begin{equation*}
	\lim_{\eta\to0}\absb{1-q + q\lambda \stj_\E(\lambda-\ii\eta)}^2 = \frac{(z+1-q)^2 + \pb{4\lambda q - (\lambda+q-1)^2 }}{4},
\end{equation*}
from which we readily find
\begin{equation*}
	\lim_{\eta\to0}\absb{1-q + q\lambda \stj_\E(\lambda-\ii\eta)}^2 = \lambda,
\end{equation*}
and this gives the expected answer
\begin{equation}
	\label{eq:RIE_nullH}
	\hat\xi(\lambda) = 1, \qquad \lambda \in \qb{(1-\sqrt{q})^2, (1+\sqrt{q})^2}.
\end{equation}
We provide an illustration of this phase transition in Figure \ref{fig:rie_null} in the case where $\C = \In$, corresponding to a matrix $\E$ is generated using an isotropic Wishart matrix with $q = 0.5$. It also confirms the asymptotic prediction for large and isolated eigenvalue Eq.\ \eqref{eq:RIE_near_infty_normalized}.

\subsubsection{Revisiting the linear shrinkage}
\label{sec:revisiting_linear}

In Chapter \ref{chap:bayes}, we saw that the linear shrinkage (towards the identity matrix) is equivalent to assuming that $\C$ itself belongs to an Inverse-Wishart ensemble with some parameter $\kappa$. We want to revisit this result within the framework of the present chapter, and we will see that in the presence of extra spikes, the optimal shrinkage function \eqref{eq:RIE_optimal} again shows a phase transition phenomenon and therefore differs from the linear estimator Eq.\ \eqref{eq:linear_shrinkage} for eigenvalues lying outside the spectrum of $\E$. 

\begin{figure}[h]
  \begin{center}
   \includegraphics[scale = 0.45]{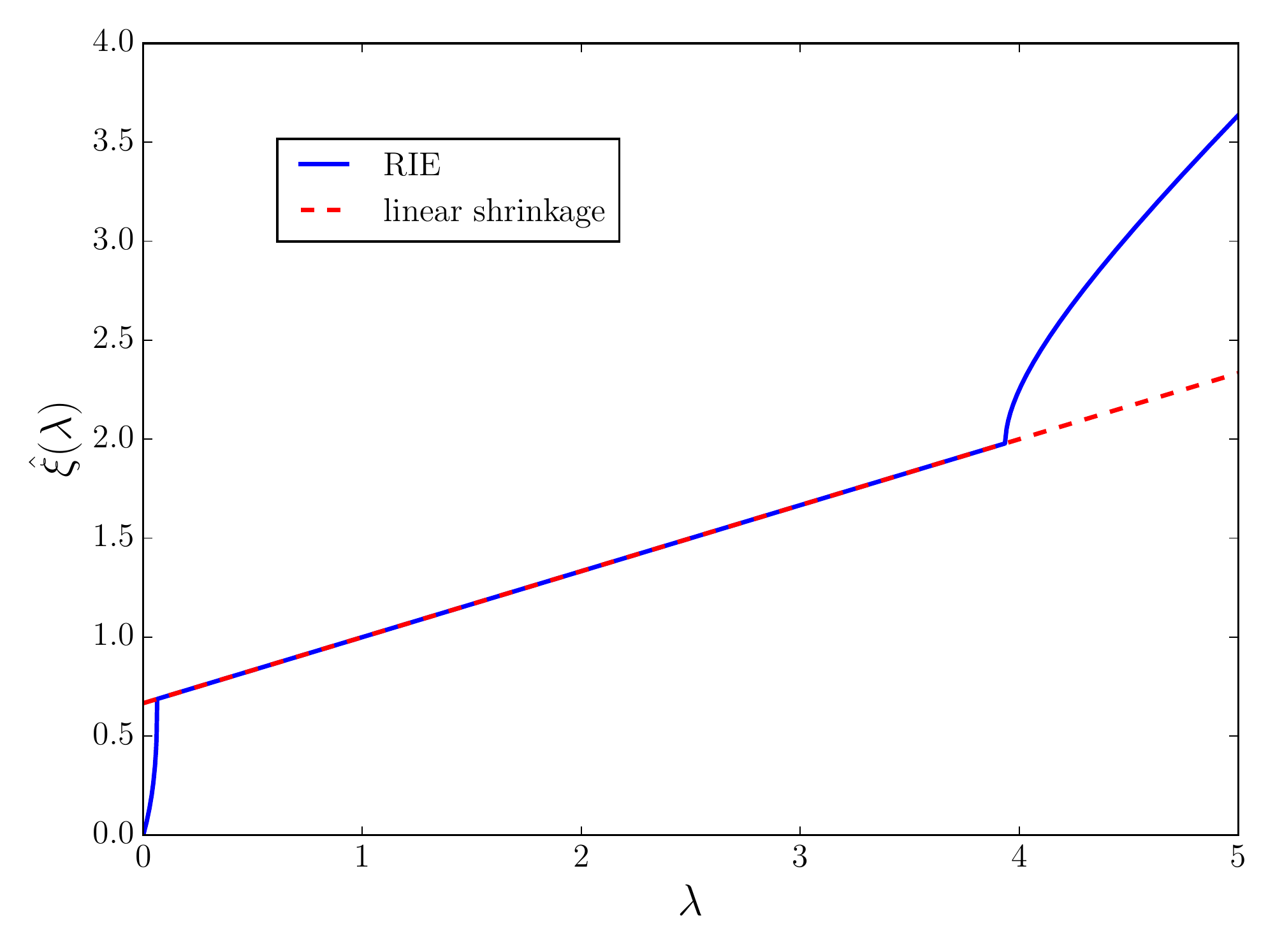} 
   \end{center}
   \caption{Evaluation of the optimal RIE's eigenvalues for an Inverse Wishart prior with $\kappa = 2$ as a function of the sample eigenvalues $[\lambda_i]_{i\in\qq{1,N}}$. The matrix $\E$ is generated using Wishart matrix with parameter $N = 500$ and $q = 0.5$. The nonlinear shrinkage function is plotted with the plain blue line and it coincides with the estimator Eq.~\eqref{eq:linear_shrinkage} (red dotted line). We nonetheless see that for $\lambda > \lambda_{+}^{\text{iw}}$, a phase transition occurs and the two estimators split up. The same phenomenon is observed for $\lambda < \lambda_{+}^{\text{iw}}$ (see Figure \ref{fig:rie_left_edge_pt}). }
   \label{fig:rie_iw}
\end{figure}

As for the null hypothesis case above, there is no particular simplifications for outliers and the numerical result is immediately obtained from Eq.\ \eqref{eq:RIE_optimal} and \eqref{eq:stieltjes_IW_MP}. For the bulk component, the square root term in Eq.\ \eqref{eq:stieltjes_IW_MP} becomes imaginary. Hence, setting $z = \lambda - \ii \eta$ into Eq.\ \eqref{eq:stieltjes_IW_MP} with $\lambda \in [\lambda_{-}^{\text{iw}}, \lambda_{+}^{\text{iw}}]$ and $\lambda_{\pm}^{\text{iw}}$, defined in Eq.\ \eqref{eq:edges_IW_MP}, one obtains
% \begin{equation*}
% \left| 1-q+q\lambda_{i}\underset{\varepsilon \rightarrow 0}{\lim} \stj_{\E}(\lambda_{i} - i\varepsilon) \right|^{2} = \left| 1-q+ q \lambda_{i} \left( \frac{\lambda_{i}(\kappa +1) - \kappa(1-q)}{\lambda_{i}(\lambda_{i}+2q\kappa)} + i \frac{\sqrt{2\lambda_{i}\kappa( (1+\kappa) + 1) - \kappa^2(1-q)^2 - \lambda_{i}^2\kappa^2} }{\lambda_{i}(\lambda_{i}+2q\kappa)} \right) \right|^{2},
% \end{equation*}
% that is to say
\begin{equation*}
\left| 1-q+q\lambda\underset{\eta \rightarrow 0^+}{\lim} \stj_{\E}(\lambda - \ii\eta) \right|^{2} = \frac{ \qb{\lambda(1+q\kappa) + \kappa q (1-q)}^2 + q^2 \qb{2\lambda \kappa( \kappa(1+q) + 1) - \kappa^2(1-q)^2 - \lambda^2\kappa^2}}{(\lambda+2q\kappa)^2},
\end{equation*}
with $\kappa > 0$. This can be rewritten after expanding the square as
\begin{equation}
\left| 1-q+q\lambda\underset{\eta \rightarrow 0^+}{\lim} \stj_{\E}(\lambda - \ii\eta) \right|^{2} = \frac{\lambda(1+2q\kappa)}{(\lambda+2q\kappa)}.
\end{equation}
By plugging this last equation into Eq.\ \eqref{eq:RIE_optimal} gives for any $\lambda \in [\lambda_{-}^{\text{iw}}, \lambda_{+}^{\text{iw}}]$
\begin{equation}
\xi^{\text{ora.}}(\lambda) = \frac{\lambda +2q\kappa}{1+2q\kappa},
\end{equation}
and if we recall the definition $\alpha_{s} = 1/(1+2q\kappa) \, \in [0, 1]$ of Eq.\ \eqref{eq:linear_shrinkage}, we retrieve exactly the linear shrinkage estimator \eqref{eq:linear_shrinkage},
\begin{equation}
	\label{eq:RIE_lin}
	\xi^{\text{ora.}}(\lambda) \sim \alpha_{s} \lambda + (1-\alpha_{s}), \qquad  \lambda \in [\lambda_{-}^{\text{iw}}, \lambda_{+}^{\text{iw}}].
\end{equation}
This last result illustrates in a particular case the genuine link between the optimal RIE $\Xi^{\text{ora.}}$ and Bayes optimal inference techniques in the LDL. In particular, we show that for an isotropic Inverse Wishart matrix, the estimator $\Xi^{\text{ora.}}$ gives the same result than the conjugate prior approach in the high dimensional regime. Nevertheless, this is valid \emph{only for the bulk component} as the presence of outliers induces a phase transition for the optimal RIE, which is absent within the conjugate prior theory that is blind to outliers. We illustrate this last remark in Figure \ref{fig:rie_iw} where $\C$ is an Inverse-Wishart matrix of parameter $\kappa = 2$. The link between Bayesian statistics and RIE in the high-dimensional regime has been noticed in \cite{bun2015rotational} where the case of an additive noise 
is also considered -- see Appendix \ref{app:addition}, yielding a generalization of the well-known Wiener's signal-to-noise ratio optimal estimator \cite{wiener1949extrapolation}. 

\begin{figure}[h]
\begin{center}
   \includegraphics[scale = 0.5]{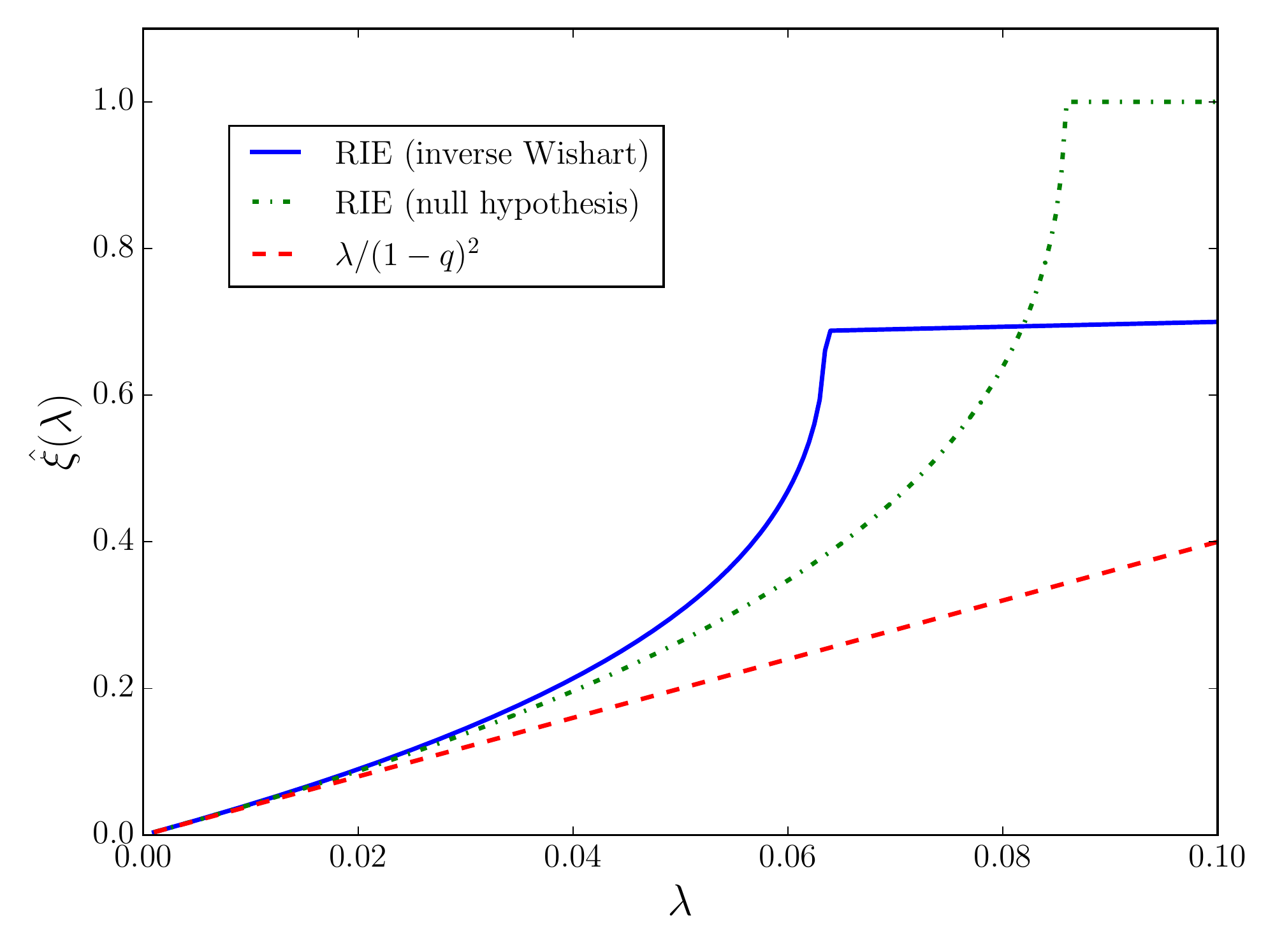} 
   \end{center}
\caption{Comparison of the prediction Eq.\ \eqref{eq:RIE_near_zero} (red dashed line) compared to the analytical solution of the null hypothesis \eqref{eq:RIE_nullH} (green dash-dotted line) and the Inverse Wishart prior \eqref{eq:RIE_lin} with parameter $\kappa = 2$ (blue plain line). In both cases, wet set $q = 0.5$. The asymptotic prediction \eqref{eq:RIE_near_zero} becomes less and less accurate as $\lambda$ moves closer to the left edge and the analytic solution (blue line) depicts a phase transition. 
%v The X-axis denotes the sample eigenvalues and the Y-axis the \emph{cleaned} eigenvalues.
}
\label{fig:rie_left_edge_pt}
\end{figure}

We also illustrate in Figure \ref{fig:rie_left_edge_pt} the phase transition observed for outliers at the left of the lower bound of the spectrum for both analytical examples. We see that for very small eigenvalues, the theoretical prediction \eqref{eq:RIE_near_zero} is pretty accurate. This prediction becomes less and less effective as $\lambda$ moves closer to the left edge. 

\subsection{Optimal RIE at work}
\label{sec:RIE_simulations}

In order to conclude this section, we now consider different cases where $\stj_\E(z)$ is not explicit, and where the problem must be solved numerically. In that case, the main question is to estimate the function $\stj_\E(z)$ without imposing any ``prior'' on $\C$. Indeed, even though the function $\xi^{\text{ora.}}$ only depends on observables quantities, we still need to estimate the function $\stj_\E(z)$ using only a finite (and random) set of sample eigenvalues. 

This question has been addressed recently in \cite{bun2016optimal}, where apart from extending the result of \cite{ledoit2011eigenvectors} to outliers (as reviewed above), the mathematical technique used in \cite{bun2016optimal} provides a derivation of Eq.\ \eqref{eq:RIE_optimal} at a \emph{local} scale and for any large but finite $N$. As alluded to in Chapter \ref{chap:eigenvectors}, the local scale can be understood as an average over small intervals of eigenvalues of width $\eta = \dd\lambda \geq N^{-1}$. The main result of \cite{bun2016optimal} can be summarized as follows: the limiting Stieltjes transform $\stj_\E(z)$ can be replaced by its discrete form
\begin{equation}
	\stj_\E^N(z) = \frac1N \sum_{i=1}^{N} \frac{1}{z - \lambda_i}\,,
\end{equation}
with \emph{high probability} (see e.g.\ \cite{knowles2014anisotropic} for the exact statement). Therefore, this yields a fully observable nonlinear shrinkage function and moreover, the choice $\eta = N^{-1/2}$ gives a sharp upper error bound for any finite $N$ and $T$. Precisely, for $z_i = \lambda_i - \ii N^{-1/2}$, there exists a constant $K$ such that for large enough $T$,
\begin{equation}
	\label{eq:RIE_optimal_obs}
 	\absB{\xi^{\text{ora.}}_i - \hat\xi_i^N } \leq \frac{K}{\sqrt{T}}, \qquad \hat\xi_i^N \;\equiv\; \hat\xi^N(\lambda_i) \deq \frac{\lambda_i}{\absb{1-q+qz_i  \stj_\E^{N}(z_i)}^2},
\end{equation} 
provided that $\lambda_i$ is not near zero \cite{bun2016optimal}. We see that Eq.\ \eqref{eq:RIE_optimal_obs} is extremely simple to implement numerically as it only requires to compute a sum over $N$ terms. 

\begin{figure}[!ht]
\begin{subfigure}{.5\textwidth}
  \centering
  \includegraphics[width=.9\linewidth]{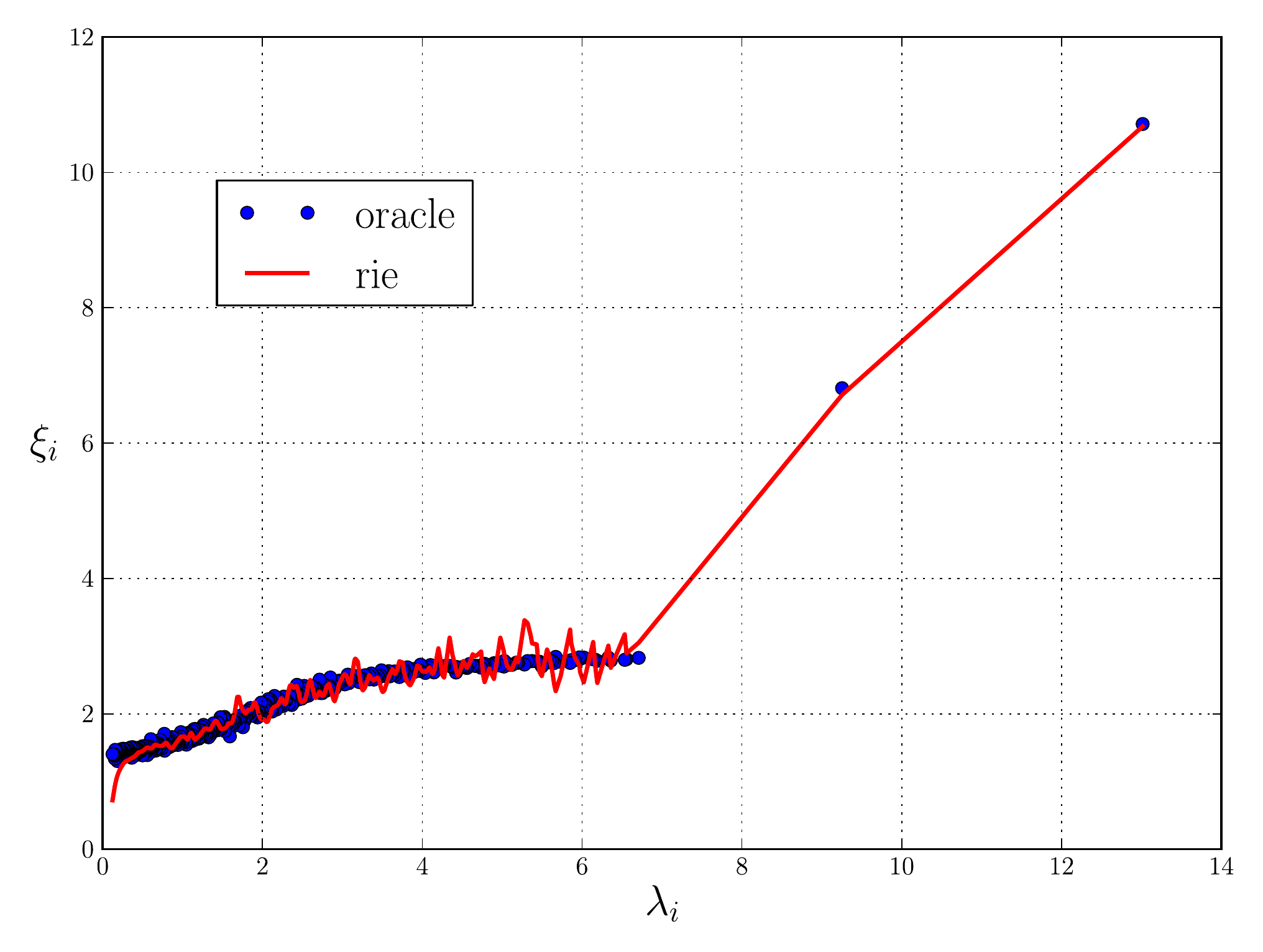}
  \caption{Multiple sources (case (i)).}
  \label{fig:multiple}
\end{subfigure}%
\begin{subfigure}{.5\textwidth}
  \centering
  \includegraphics[width=.9\linewidth]{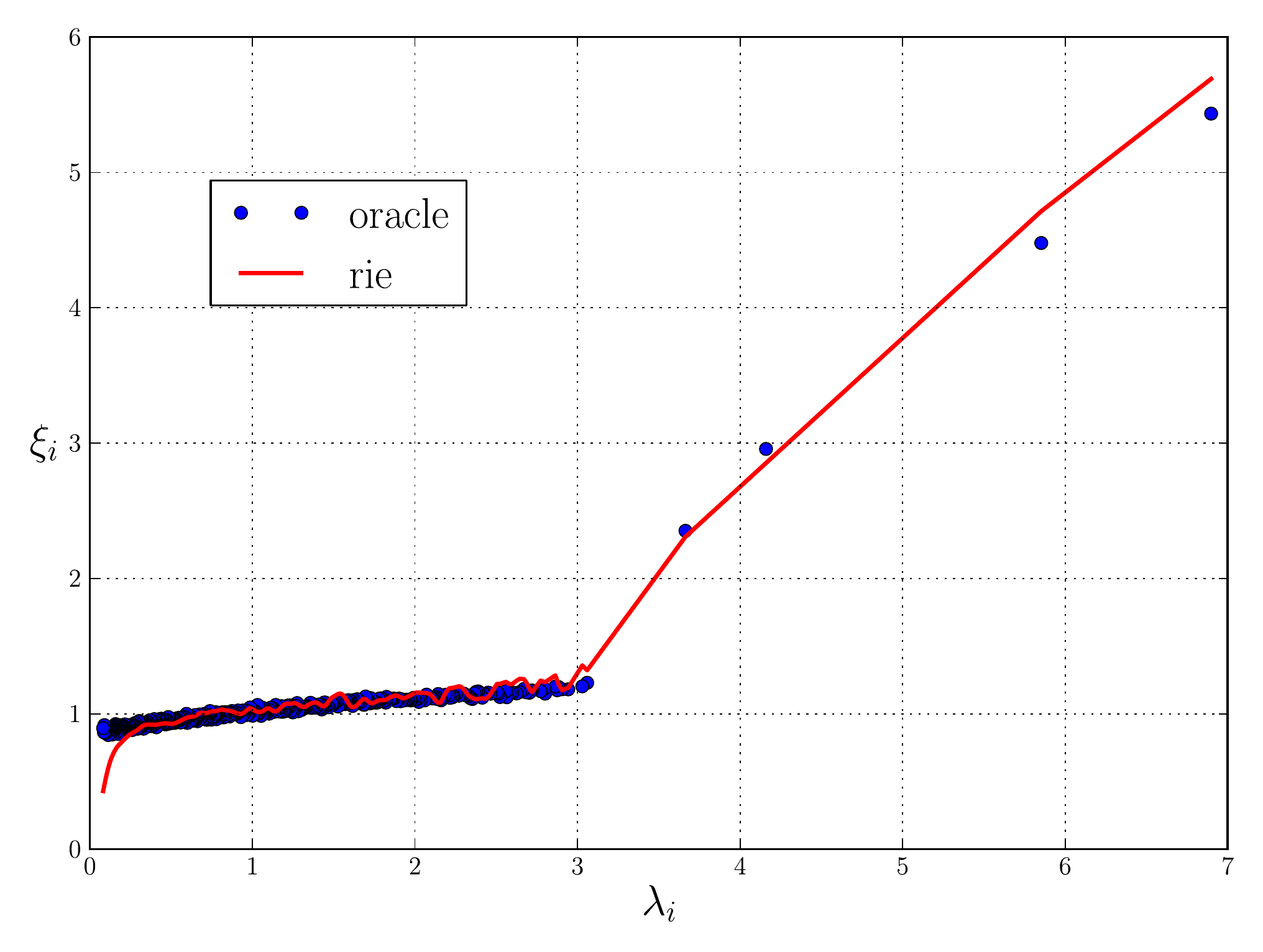}
  \caption{deformed GOE (case (ii)).}
  \label{fig:dGOE}
\end{subfigure}\\
\begin{subfigure}{.5\textwidth}
  \centering
  \includegraphics[width=.9\linewidth]{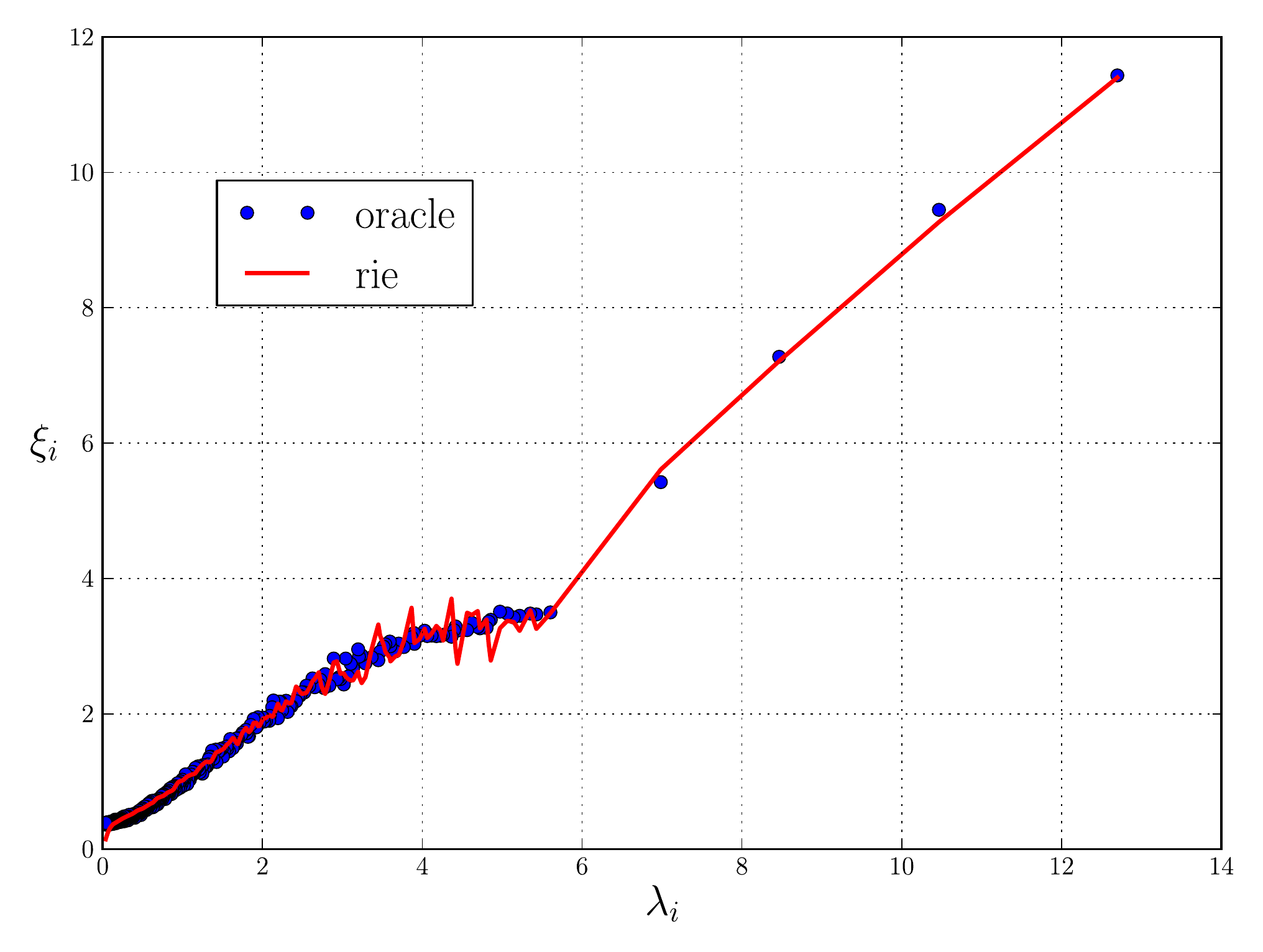}
  \caption{Toeplitz (case (iii))}
  \label{fig:toeplitz}
\end{subfigure}%
\begin{subfigure}{.5\textwidth}
  \centering
  \includegraphics[width=.9\linewidth]{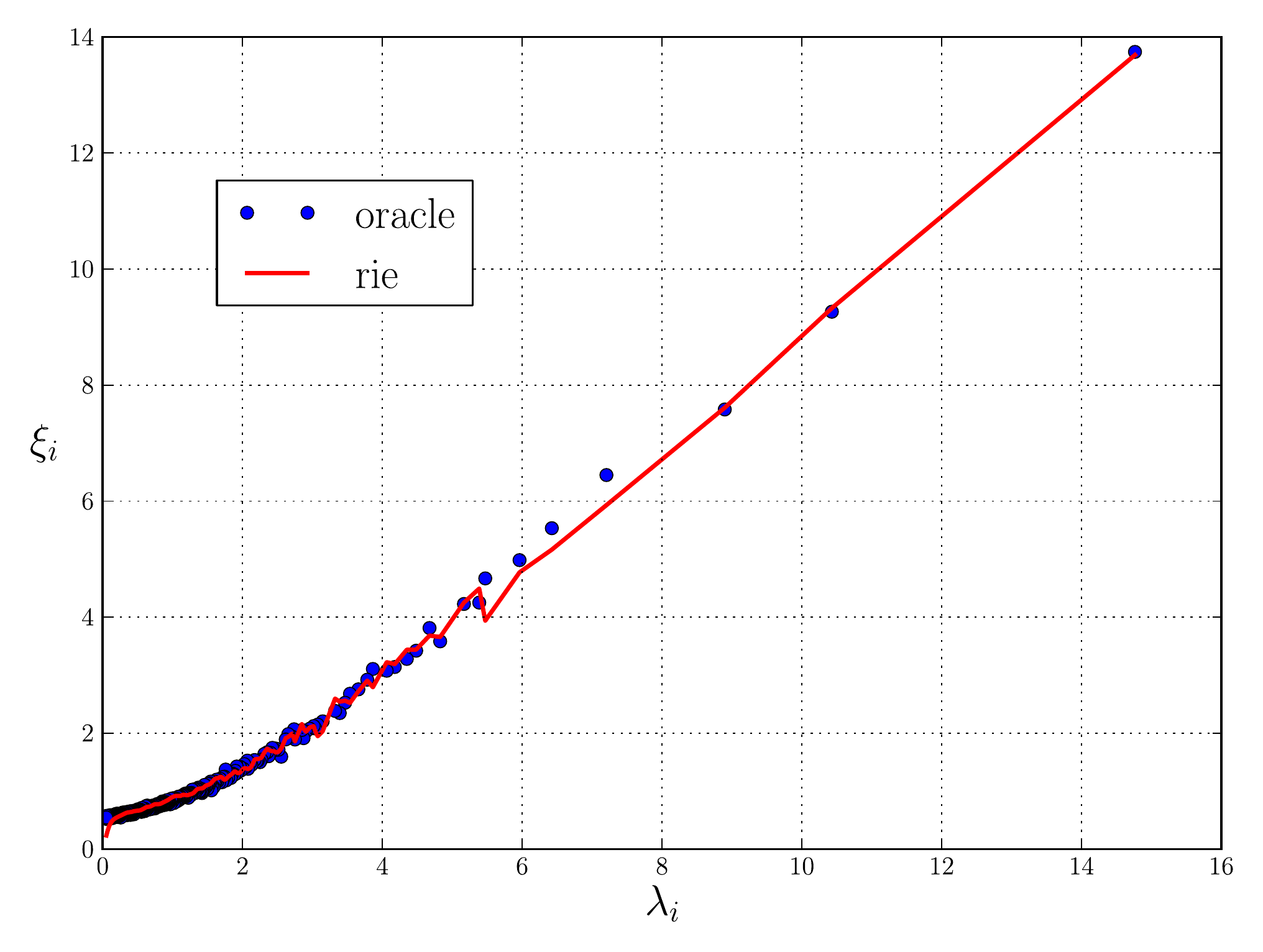}
  \caption{Power law (case (iv))}
  \label{fig:powerlaw}
\end{subfigure}
\caption{Comparison of numerically estimated oracle estimator \eqref{eq:RIE_optimal_obs} (red line) with the exact oracle RIE estimator \eqref{eq:oracle} (blue points) for the four cases presented at the beginning of Section \ref{sec:RIE_simulations} with $N = 500$ and $T = 1000$. The results come from a single realization of $\E$ using a multivariate Gaussian measurement process.
%v The X-axis denotes the sample eigenvalues and the Y-axis the \emph{cleaned} eigenvalues.
}
\label{fig:rie_examples}
\end{figure}

We now test numerically the accuracy of the finite $N$, observable optimal nonlinear shrinkage function \eqref{eq:RIE_optimal_obs} in four different settings for
the population matrix $\C$. We choose $N = 500$, $T = 1000$ (which are quite reasonable numbers in real cases, not too small nor too large) and consider the following four different cases:
\begin{enumerate}
	\item Diagonal matrix whose ESD is composed of multiple sources with ``spikes'',
	\begin{equation}
		\label{eq:rie_multiplie}
		\rho_\C = 0.002 \delta_{15} + 0.002 \delta_{8} + 0.396 \delta_{3} + 0.3 \delta_{1.5}  + 0.3 \delta_{1}.
	\end{equation}
	\item Deformed GOE, i.e $\C = I_N + \text{GOE}$ (of width $\sigma = 0.2$) with extra spikes located at \{3, 3.5, 4.5, 6\}.
	\item Toeplitz matrix with entries $\C_{ij} = 0.6^{\abs{i-j}}$ with spikes located at \{7, 8, 10, 11\};
	\item Power-law distributed eigenvalues (see \cite{bouchaud2009financial} and Chapter \ref{chap:spectrum}) with $\lambda_0 = - 0.6$ (or $\lambda_{\text{min}} = 0.8$. Using a large $N$ proxy for the 
	classical positions of the $\mu_i$, one gets \cite{bouchaud2009financial}:
		\begin{equation}
			\label{eq:power law proxy}
			%\mu_i = 2\lambda_{\text{min}} - 1 + (1 - \lambda_{\text{min}}) \sqrt{\frac{N}{i}} \qquad i \in \qq{1,N}.
			\mu_i = -\lambda_0 + \frac{(1 + \lambda_{0})}{2} \sqrt{\frac{N}{i}} \qquad i \in \qq{1,N}\,.
		\end{equation}
\end{enumerate}
Note that the last power law distribution automatically generates a bounded number of outliers. Moreover, since we work with $N$ and $T$ bounded, the largest eigenvalue of $\C$ is remains bounded. We plot the results obtained with the estimator Eq.\ \eqref{eq:RIE_optimal_obs} and the oracle estimator Eq.\ \eqref{eq:oracle} in Figure \ref{fig:rie_examples}. 

Overall, the estimator \eqref{eq:RIE_optimal_obs} gives accurate predictions for both the bulk eigenvalues and outliers. We have considered several configurations of outliers. For the case (i), we see that the two isolated outliers are correctly estimated. For the deformed GOE or the Toeplitz case, the outliers are chosen to be a little bit closer to one another and again, the results agree well with the oracle estimator. For the more complex case of a power law distributed spectrum, where there is no sharp right edge, we see that \eqref{eq:RIE_optimal_obs} matches again well with the oracle estimator. We nevertheless notice that the small eigenvalues are \emph{systematically} underestimated by the empirical optimal RIE \eqref{eq:RIE_optimal_obs}. This effect will be investigated in more details in Chapter \ref{chap:numerical}.

%\begin{document}

\begin{figure}[h]
  \begin{center}
   \includegraphics[scale = 0.6]{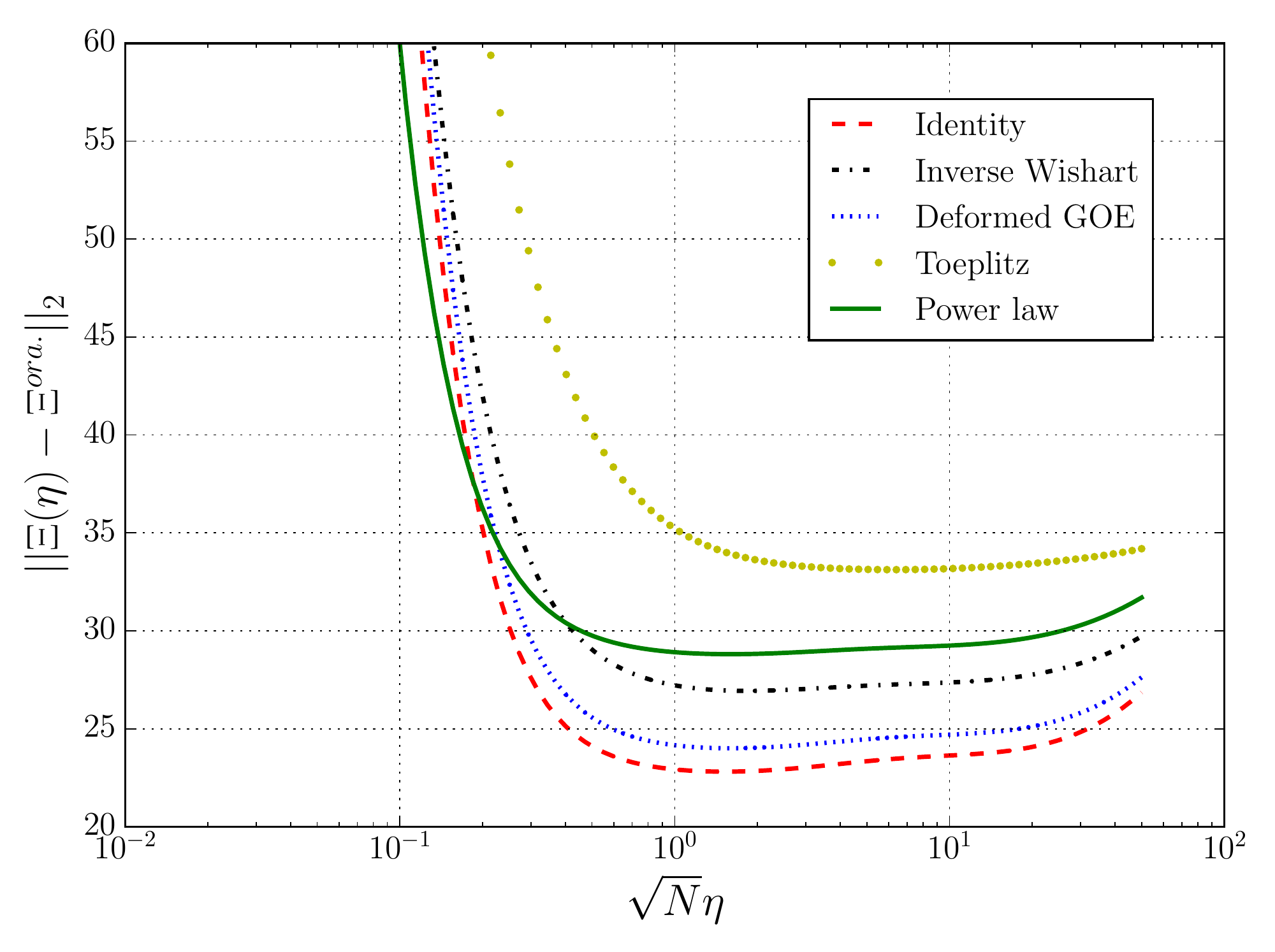} 
   \end{center}
   \caption{Mean squared difference between the optimal estimator \eqref{eq:RIE_optimal_obs} and the oracle estimator. The estimator \eqref{eq:RIE_optimal_obs} is now studied as a function of $\eta$. The x-axis (in logarithm scale) shows the value of $\alpha = \sqrt{N}\eta$ for the sake of clarity. We consider five different examples for $\C$ (same configuration as in Figure \ref{fig:rie_examples} and the identity matrix). For each example, we generate $100$ independent realizations of $\E$ with $N = 500$ and $T=1000$. 
   }
   \label{fig:eta_semilog_N_500_T_1000}
\end{figure}

As a further check, we provide here a numerical test of the ``optimal'' scale $\eta$. As explained above, it was shown in \cite{bun2016optimal} that the value $\eta = N^{-1/2}$ gives the upper bound in \eqref{eq:RIE_optimal_obs}. However, one might wonder if this value is indeed optimal with real (or synthetic) data. To test this, we study the estimator \eqref{eq:RIE_optimal_obs} as a function of $\eta$ and compute the corresponding mean squared error with respect to the oracle estimator $\Xi^{\text{ora.}}$ for $\eta = \alpha  N^{-1/2}$ and $\alpha \in [0.01, 50]$. For each $\C$, we evaluate the error for $100$ different realizations of $\E$ using a multivariate Gaussian process. The results are reported in Figure \ref{fig:eta_semilog_N_500_T_1000}. The optimal value of $\alpha \approx 1.5$ for all the examples except when $\C$ is a Toeplitz matrix (yellow dots) where the optimal value of $\alpha \approx 8.4$.

\subsection{Extension to the free multiplicative model}
\label{sec:RIE_freemult}

\begin{changemargin}{1.0cm}{1.0cm} 
\footnotesize

As highlighted in \cite{bun2015rotational}, the evaluation of the optimal RIE for bulk eigenvalues can be extended to more general multiplicative random matrix models (for additive noise models, see Appendix \ref{app:addition}). In particular, it is possible to derive (formally) the optimal nonlinear shrinkage function \eqref{eq:RIE_optimal} for the bulk eigenvalues of the measurement model \eqref{eq:model_freemult} which generalizes the case of  
sample covariance matrices (see Section \ref{sec:MP}). 

To that end, let us define $\M \deq \C^{1/2} \b \Omega \B \b \Omega^* \C^{1/2}$ where $\B$ is a $N \times N$ symmetric rotational invariant noise term and $\b \Omega$ is a $N \times N$ random rotation matrix that is distributed according to the Haar measure. One can easily check from Eq.\ \eqref{eq:global_law_freemult} that 
\begin{equation}
%\label{oracle_MP_tmp}
\Tr \left[{\G}_{\M}(z) \C\right] = N (z\stj_{\M}(z) - 1) \str_{\B}(z \stj_{\M}(z) - 1)\,.
\end{equation}
Using the analyticity of the $\str$-transform, we define the function $\gamma_{\B}$ and $\omega_{\B}$ such that:
\begin{equation}
\label{eq:s_transform_decomposition}
\underset{z \rightarrow \lambda - \ii0^{+}}{\lim}  \str_{\B}(z\stj_{\M}(z) - 1) := \gamma_{\B}(\lambda) + \ii\pi\rho_{\M}(\lambda) \omega_{\B}(\lambda)\,,
\end{equation}
and as a result, the optimal RIE for bulk eigenvalues of the free multiplicative noise model \eqref{eq:model_freemult} may be inferred from \eqref{eq:RIE_resolvent_relation}:
\begin{equation}
\label{eq:RIE_freemult}
{\xi}^{\text{ora.}}_i \sim F_2(\lambda_i); \qquad F_2(\lambda)= \lambda \gamma_{\B}(\lambda) + (\lambda \hil_{\M}(\lambda) - 1)\omega_{\B}(\lambda)\,.
\end{equation}
Note that one retrieves the estimator \eqref{eq:RIE_optimal} by plugging Eqs.\ \eqref{eq:S_transform_MP} and \eqref{eq:s_transform_decomposition} into Eq.\ \eqref{eq:RIE_freemult}. We omit details, which can be found in \cite{bun2015rotational}, and we conclude  that the formula \eqref{eq:RIE_freemult} indeed generalizes Eq.\ \eqref{eq:RIE_optimal}. Again, we see that the final solution does not depend explicitly on $\C$ but somehow requires a prior on the spectral distribution of the matrix $\B$. It would be quite satisfying to find models in which we may obtain an explicit formula for Eq.\ \eqref{eq:RIE_freemult} (see Chapter \ref{chap:conclusion} for some relevant applications of this model).

We emphasize in passing that we may also derive the mean squared overlap \eqref{eq:overlap} in the bulk of the distribution using Eq.\ \eqref{eq:global_law_freemult}. To that end, we invoke the relation \eqref{eq:resolvent_overlap} and Eq.\ \eqref{eq:global_law_freemult} to obtain \cite{bun2015rotational}:
\begin{equation}
\label{eq:overlap_freemult}
\mso(\lambda, \mu) = \frac{\mu \beta_m(\lambda)}{(\lambda - \mu \alpha_m(\lambda))^2 + \pi^2 \mu^2 \beta_m(\lambda)^2 \rho_{\M}(\lambda)^2}\,,
\end{equation}
where we defined the functions $\alpha_m$ and $\beta_m$ as
\begin{equation}
\label{decomposition_S}
\begin{dcases}
\alpha_{m}(\lambda) :=  \underset{z \rightarrow \lambda - i0^{+}}{\lim} \re\left[ \frac{1}{\str_{\B}(z \stj_{\M}(z) - 1)} \right]  \\
\beta_{m}(\lambda) :=  \underset{z \rightarrow \lambda - i0^{+}}{\lim} \im\left[ \frac{1}{\str_{\B}(z \stj_{\M}(z) - 1)} \right] \frac{1}{\pi \rho_{\M}(\lambda)}\,,
\end{dcases}
\end{equation}
and the subscript $m$ stands for ``multiplication''.

We conclude this technical section by mentioning one open problem which is the extension of these results in the presence of outliers. Indeed, it would be interesting to see whether the optimal RIE formula \eqref{eq:RIE_freemult} remains \emph{universal} (as we believe it is) in the sense that the cleaning formula for bulk eigenvalues and outliers is identical. The block matrix representation \eqref{eq:SCM_resolvent_block} might be useful in that respect. 

\end{changemargin} 
\normalsize

%---------  Applications part ---------------- %

\clearpage%!TEX root = RMT_Covariance_Review.tex
\section{Application: Markowitz portfolio theory and previous ``cleaning'' schemes}
\label{chap:application}

\subsection{Markowitz optimal portfolio theory}
\label{sec:markowitz}

%\subsubsection{Mean-Variance optimization and optimal portfolio} 

For the reader not familiar with Markowitz's optimal portfolio theory \cite{markowitz1952portfolio}, we recall in this section some of the most important results. 
Suppose that an investor wants to invest in a portfolio containing $N$ different assets, with optimal ``weights'' to be determined. An intuitive strategy is the so-called mean-variance optimization: the investor seeks an allocation such that the overall quadratic risk of the portfolio is minimized given an expected return target. It is not hard to see that this mean-variance optimization can be translated into a simple quadratic optimization program with a linear constraint. Before going into more mathematical details, let us introduce some notations that will be used in the following. We suppose that we observe the return time series of $N$ different stocks. For each stock, we observe a time series of size $T$, where $T$ is often larger than $N$ in practice. This yields the (normalized) $N \times T$ return matrix $\b Y = (Y_{it}) \in \R^{N\times T}$ whose true correlation matrix is defined by
\begin{equation}
\langle Y_{it} Y_{jt'} \rangle = \C_{ij} \delta_{tt'},
\end{equation}
where the absence of correlations in the time direction is a only a first approximation since weak, but persistent linear correlations are known to exist in stock markets.

As natural in the present ``Big Data'' era, we place ourselves in the high-dimensional regime $N, T \rightarrow \infty$ with a finite ratio $q = N/T$. Markowitz's optimal portfolio amounts to solving the following quadratic optimization problem  
\begin{equation}
\label{quad_prog}
\left\{
  \begin{array}{lr}
    \min_{\b w \in \mathbb{R}^{N}} \frac12 \b w^{*} \C \b w \\
    \text{s.t.} \; \b w^* \b g \ge {\cal G}
  \end{array}
\right.
\end{equation}
where $\b g$ is a $N$-dimensional vector of predictors (assumed to be deterministic and given by, e.g. in depth analysis of economic data) and $\cal G$ is the expected gain. This mathematical problem can be easily solved by introducing a Lagrangian multiplier $\gamma$ to rewrite this constrained optimization problem 
as a unconstrained one\footnote{One can check that the so-called Karush-Kuhn-Tucker conditions are satisfied.}:
\begin{equation}
\min_{\b w \in \mathbb{R}^{N}} \frac12 \b w^{*} \C \b w - \gamma \b w^{*} \b g.
\end{equation}
Assuming that $\C$ is invertible, it is not hard to find the optimal solution and the value of $\gamma$ such that overall expected return is exactly 
$\cal G$. It is given by
\begin{equation}
\label{eq:markowitz_weight}
\b w_{\C} = {\cal G} \frac{\C^{-1} \b g}{\b g^{*} \C^{-1} \b g},
\end{equation}
that requires the knowledge of both $\C$ and $\b g$, which are \textit{a priori} unknown. As mentioned above, forming expectations of future returns is the job of the investor or of the financial analyst, based on his/her information and anticipations, so we assume that $\b g$ is given. Even if these predictions were completely wrong, it would still make sense to look for the minimum risk portfolio consistent with these expectations. We are still left with the problem of estimating $\b C$, or maybe $\b C^{-1}$ before applying Markowitz's formula, Eq. \eqref{eq:markowitz_weight}. We will see below why one should actually find the best estimator of $\b C$ itself before inverting it and determining the weights.

What is the \emph{minimum} risk associated to this allocation strategy, measured as the variance of the returns of the portfolio?\footnote{An equivalent risk measure is the volatility which is simply the square root of the variance of the portfolio strategy.} If one knew the population correlation matrix, $\C$, the \emph{true} optimal risk associated $\b w_\C$ would be given by  
\begin{equation}
\label{eq:true_risk}
{\cal R}_{\text{true}}^{2} \deq \scalar{\b w_{\C}}{\C \b w_{\C}} = \frac{{\cal G}^2}{\b g^{*} \C^{-1} \b g}.
\end{equation}
However, the optimal strategy \eqref{eq:markowitz_weight} is not attainable in practice as the matrix $\C$ is unknown. What can one do then, and how badly is
the realized risk of the portfolio estimated?
% and its risk is measured by the total variance, that is to say
% \begin{equation}
	% \label{eq:markowitz_risk}
	% \cal R^2_{\text{true}} \deq \b w_\C^{*} \C \b w_\C.
% \end{equation}

\subsubsection{Predicted and realized risk}

\begin{figure}[h]
  \begin{center}
   \includegraphics[scale = 0.5]{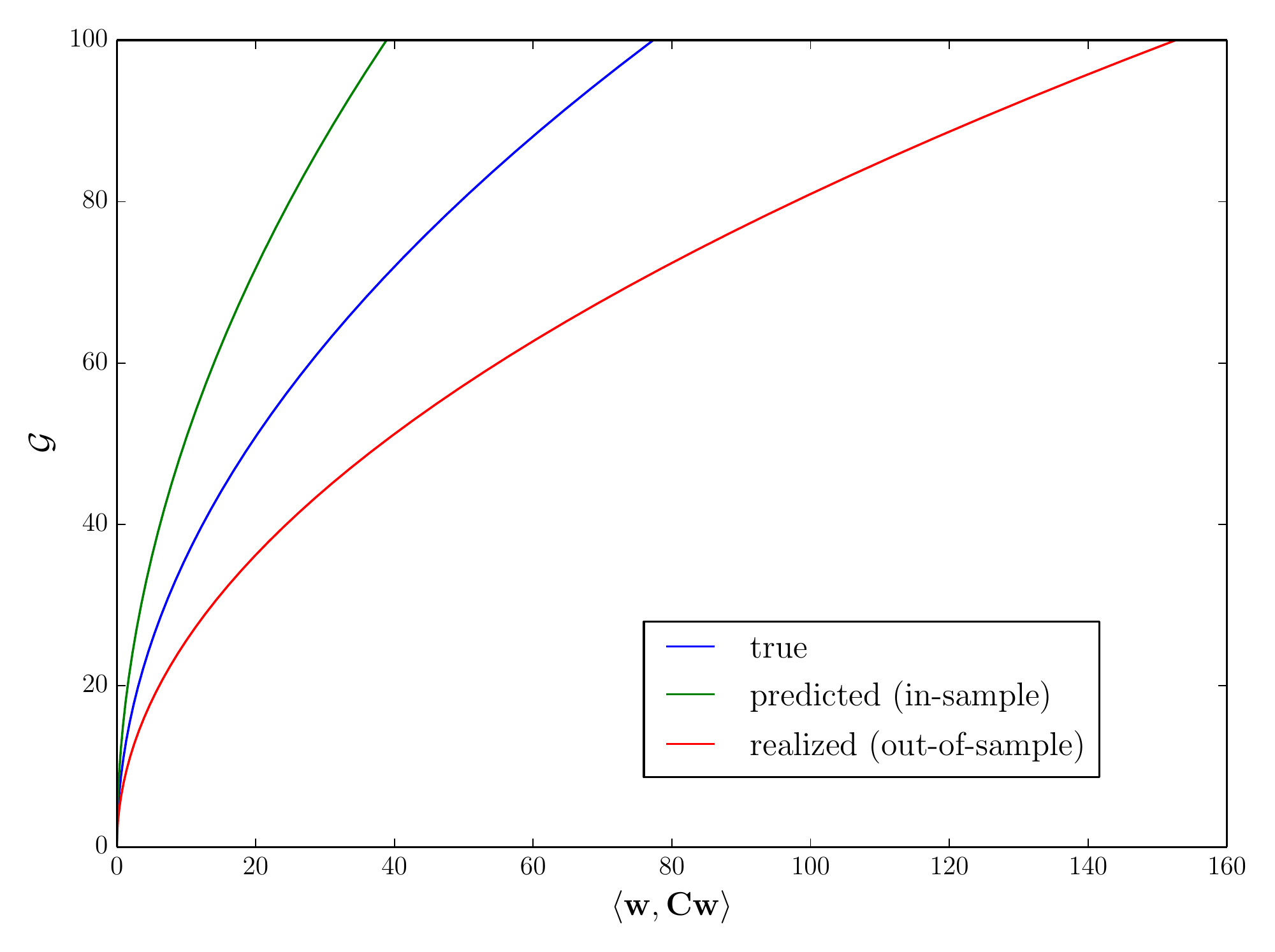} 
   \end{center}
   \caption{Efficient frontier associated to the mean-variance optimal portfolio \eqref{eq:markowitz_weight} for $\b g = (1,\dots,1)^*$ and $\C$ a shifted GOE
    around the identity matrix, with $\sigma=0.2$ and for $q=0.5$. The blue line depicts the expected gain as a function of the \emph{true} optimal risk \eqref{eq:true_risk} in percentage. 
    The green line the predicted (in-sample) risk while the red line gives the realized (out-of-sample) risk, which is well above the true risk.}
   \label{fig:efficient_curve}
\end{figure}

 One very naive way to use the Markowitz optimal portfolio is to apply \eqref{eq:markowitz_weight} using the empirical matrix $\E$ instead of $\C$. Recalling the results of Chapter \ref{chap:spectrum} and \ref{chap:eigenvectors}, it is not hard to see that this strategy should suffer from strong biases whenever $T$ is not sufficiently large compared to $N$, which is precisely the case we consider here. Notwithstanding, the optimal investment weights using the empirical matrix $\E$ read: 
\begin{equation}
\b w_{\E} = {\cal G} \frac{\E^{-1} \b g}{\b g^{*} \E^{-1} \b g},
\end{equation}
and the minimum risk associated to this portfolio is thus given by
\begin{equation}
\label{eq:in_sample_risk}
{\cal R}_{\text{in}}^{2} = \scalar{\b w_{\E}}{\E \,\b w_{\E}} = \frac{{\cal G}^2}{\b g^{*} \E^{-1} \b g},
\end{equation}
which is known as the ``in-sample'' risk, or the \emph{predicted} risk. Let us assume for a moment that $\b g$ is independent from $\C$ (and hence, from $\E$). Then, using the convexity with respect to $\E$ of $\b g^{*} \E^{-1} \b g$ we find from Jensen inequality that 
\begin{equation}
	\label{eq:in_true_jensen}
	\mathbb{E}[{\b g^{*} \E^{-1} \b g}] \geq \b g^{*} \mathbb{E}\qb{{\E}}^{-1} \b g = \b g^{*} {\C^{-1}} \b g
\end{equation} 
because $\E$ is an unbiased estimator of $\C$. Hence, we conclude that the in-sample risk is lower than the `true' risk and therefore, our optimal portfolio 
suffers from an in-sample bias: its predicted risk underestimates the true optimal risk, and even more so the future \textit{out-of-sample} or \emph{realized} risk, that is the risk realized in the period subsequent to the estimation period. Let us denote by $\E'$ the empirical matrix of this out-of-sample period; the \emph{out-of-sample} risk is then naturally defined by:
\begin{equation}
\label{eq:out_sample_risk}
{\cal R}_{\text{out}}^{2} = \scalar{\b w_{\E}}{\E' \,\b w_{\E}} = \frac{{\cal G}^2 \b g^{\dag} \E^{-1} \E' \E^{-1} \b g}{(\b g^{\dag} \E^{-1} \b g)^2}.
\end{equation}
For large matrices, we expect the result to be self-averaging and given by its expectation. Since the noise in $\b w_\E$ can be assumed to be independent from that in $\E'$, we get for large $N$ \cite{pafka2003noisy}:
\begin{equation}
\label{out_sample_risk_ave}
w_{\E}^{*} \E' w_{\E} \approx w_{\E}^{*} \C w_{\E}
\end{equation}
and one readily obtains, from the fact that Eq.\ \eqref{eq:true_risk} is the minimum possible risk, the following inequality: ${\cal R}_{\text{true}}^{2} \leq {\cal R}_{\text{out}}^{2}$. We plot in Figure \ref{fig:efficient_curve} an illustration of these inequalities using the so-called efficient frontier where we assumed that $\b g = (1,\dots,1)^*$. For a given $\C$ (here a shifted GOE around the identity matrix, with $\sigma=0.2$), we build $\b w_\C$ and $\b w_\E$ and compare Eqs.\ \eqref{eq:true_risk}, \eqref{eq:in_sample_risk} and \eqref{eq:out_sample_risk} for $q=0.5$. We see that using $\b w_\E$ is clearly overoptimistic and can potentially lead to disastrous results in practice. We emphasize that this conclusion holds for different risk measures as well\cite{caccioli2015portfolio,ciliberti2007feasibility}.

\subsubsection{The case of high-dimensional random predictors}

In the limit of large matrices and with some assumptions on the structure $\b g$, we can make these inequalities more precise using tools from RMT. In particular, we will show that we can link the true and the realized risk using the Mar{\v c}enko-Pastur equation and free probability theory. Let us suppose for simplicity that
\begin{equation}
	\label{eq:mkw_gaussian_predictor}
	\b g \sim \cal N_N(0, \In),
\end{equation}
but the result holds for any vector $\b g$ whose direction is independent of $\C$ or $\E$, such that $\b g$ is normalized as $\b g^* \b g = N$, i.e. each component of $\b g$ is of order unity. We emphasize that these assumptions are not necessarily realistic (predictors can be biased along the principal components of $\C$) but allow us to quantify more precisely the relation between the in/true/out of sample risk. The suboptimal returns that follow the use ``bad'' predictors $\b g$ is outside of the scope of this review. Let $\M$ be a positive definite matrix which is independent from the vector $\b g$, then we have in the large $N$ limit,
\begin{equation}
	 \frac{\b g^* \M \b g}{N} \;=\; \frac1N \tr[\b g\b g^* \M]\; \underset{\text{freeness}}{=} \; \frac{\b g^* \b g}{N} \varphi(\M)\,
\end{equation}
where we recall that $\varphi$ is the normalized trace operator. Thus, from our assumption \eqref{eq:mkw_gaussian_predictor} we easily deduce,
% Then, we have \cite[Lemma 3.1]{silverstein1995empirical}
% **this formula looks wrong** \cob [If I understand correctly the result of Bai and Silverstein, the following formula is correct and one has to rescaled by $N^{-1}$ to have an almost sure convergence, see \eqref{eq:quad_form_trace} below.] \nc
% \begin{equation}
% \mathbb{E} \lvert \b g^{*} \M \b g - \Tr \M \lvert^{6} \le K \| \M \| N^{3}
% \end{equation}
% for $\M$ an arbitrary $N \times N$ matrix and $K$ a constant that is independent of $N$, $\M$ and $g$. Therefore, we deduce that
% It is easy to get from the Central Limit Theorem that for any $\b g$ independent from $\M$:**
% So why do we need the above ?** It could also be mentionned that $\b g^{\dag} \M \b g = \Tr \b g  \b g^{\dag} \M$ and by the freeness argument this is $\Tr \M$**
\begin{equation}	
	\label{eq:quad_form_trace}
\frac{\b g^{*} \M \b g}{N} - \varphi(\M) \underset{N \rightarrow \infty}{\rightarrow} 0.
\end{equation}
Now setting $\M = \{ \E^{-1}, \, \C^{-1} \}$, we apply Eq.\ \eqref{eq:quad_form_trace} to Eqs.\ \eqref{eq:in_sample_risk}, \eqref{eq:true_risk} and \eqref{eq:out_sample_risk} respectively, to find
\begin{eqnarray}
\label{eq:markowitz_risk_RMT_tmp}
{\cal R}_{\text{in}}^{2}   & \rightarrow & \frac{{\cal G}^{2}}{N \varphi(\E^{-1})}, \nonumber \\
{\cal R}_{\text{true}}^{2} & \rightarrow & \frac{{\cal G}^{2}}{N \varphi(\C^{-1})}, \nonumber \\
{\cal R}_{\text{out}}^{2}  & \rightarrow & \frac{{\cal G}^{2} \varphi(\E^{-1} \C \E^{-1})}{N \varphi^2(\E^{-1})},
\end{eqnarray}
where we recall that $\varphi$ is the normalized trace operator defined in Eq.\ \eqref{eq:trace_matrix}. Let us focus on the first two terms above. For $q < 1$, we have shown above that in the high-dimensional regime one has $\varphi(\C^{-1}) = (1-q) \varphi( \E^{-1})$ -- see Eq.\ \eqref{eq:SCM_inverse_moment_1}. As a result, we have, for $N \to \infty$
\begin{equation}
	\label{eq:in_true_risk}
	{\cal R}_{\text{in}}^{2}  = (1-q) {\cal R}_{\text{true}}^{2}. 
\end{equation}
Hence, for any $q \in (0,1)$, we see that the in-sample risk associated to $\b w_\E$ always provides an over-optimistic estimator. Even better, we are able to quantify exactly the risk underestimation thanks to \eqref{eq:in_true_risk}. 

Next we would like to find the same type of relation for the ``out-of-sample'' risk. We recall that under the framework of Chapter \ref{chap:spectrum}, we may always rewrite $\E = \C^{1/2} \Wishart \C^{1/2}$ where $\Wishart$ is a white Wishart matrix of parameter $q$ independent from $\C$. Hence, we have for the out-of-sample risk 
\begin{equation*}
	{\cal R}_{\text{out}}^{2} = \frac{{\cal G}^{2} \varphi(\C^{-1} \Wishart^{-2})}{N \varphi^2(\E^{-1})}
\end{equation*}
when $N \to \infty$. Then, the trick is to notice that in the limit of large matrices, $\Wishart$ and $\C$ are \emph{asymptotically free}. This allows us to conclude from the freeness relation \eqref{eq:mixed_moment1} that
\begin{equation}
	\label{eq:trWishart_minus2}
	\varphi(\C^{-1} \Wishart^{-2}) = \varphi (\C^{-1}) \, \varphi (\Wishart^{-2}),
\end{equation}
% $N^{-1} \Tr(AB) = N^{-1} \Tr A \Tr B$ if $A$ and $B$ are free. Hence, in our case, we have:
% \begin{equation}
% \frac1N \Tr \C^{1/2} \X \X^{\dag} \C^{1/2} = \frac1N \Tr \C \X \X^{\dag} = \frac1N \Tr \C \Tr W
% \end{equation}
% with $W = \X \X^{\dag}$ and can be thought as a 'white' Wishart matrix. Putting this result in the out-of-sample risk formula leads to:
% \begin{equation}
% {\cal R}_{\text{out}}^{2} = \frac{{\cal G}^{2} \Tr(\C^{-1} W^2)}{(\Tr \E^{-1})^2} = \frac{{\cal G}^{2} \Tr\C^{-1} \Tr W^{2}}{(\Tr \E^{-1})^2},
% \end{equation}
Hence, using the asymptotic relation \eqref{eq:SCM_inverse_moment_1}, we find:
\begin{equation}
{\cal R}_{\text{out}}^{2} = {\cal G}^2 (1-q)^2 \frac{ \varphi(\Wishart^{-2})}{N \varphi(\C^{-1})} ,
\end{equation}
Finally, one can readily compute $\varphi(\Wishart^{-2})$ by performing the large $z \to 0$ expansion of the Stieltjes transform of the Mar\v{c}enko-Pastur density given Eq.\ in \eqref{eq:SCM_inverse_moment_1} by replacing $\C$ with $\In$, that is to say $\varphi( \Wishart^{-2} )= (1-q)^{-3}$ for $q < 1$. We finally get:
\begin{equation}
{\cal R}_{\text{out}}^{2} = \frac{{\cal R}_{\text{true}}^{2}}{1-q}.
\end{equation}
All in all, we obtained the following asymptotic relations: 
\begin{equation}
	\label{eq:markowitz_risk_RMT}
	\frac{{\cal R}_{\text{in}}^{2}}{1-q} = {\cal R}_{\text{true}}^{2} = (1-q) {\cal R}_{\text{out}}^{2},
\end{equation}
which holds for a completely general $\C$. Note that similar results have been obtained in a slightly different context in \cite{pafka2003noisy} for $\C = I_N$ and later in \cite{collins2013compound}. Hence, if one invests with the ``naive'' weights $\b w_\E$, it turns out that the predicted risk underestimate the realized risk by a factor $(1-q)^2$ and in the extreme case $N = T$ or $q = 1$, the in-sample risk is equal to zero while the out-of-sample risk diverges. We thus conclude that, as announced, the use of the sample covariance matrix $\E$ for the Markowitz optimization problem can lead to disastrous results. This suggests that we should have a more reliable estimator of $\C$ in order to control the out-of-sample risk. 

\subsubsection{Out-of-sample risk minimization}
\label{sec:markowitz_opt_risk}

We insisted throughout the last section that the relevant quantity to control in portfolio management is the realized, out-of-sample risk. It is also clear from Eq.\ \eqref{eq:markowitz_risk_RMT} that using the sample estimate $\E$ is a very bad idea and hence, it is natural to ask: which estimator of $\C$ should one use to minimize the out-of-sample risk? The Markowitz formula \eqref{eq:markowitz_weight} naively suggests that one should look for a faithful estimator of the so-called precision matrix $\C^{-1}$. But in fact, since the expected out-of-sample risk involves the matrix $\C$ linearly, it is that matrix that should be estimated.
There are two different approaches to argue that the oracle estimator indeed yields the optimal out-of-sample risk.

The first approach consists in rephrasing the Markowitz problem in terms of conditional expectation. Indeed, the Markowitz problem can be thought as the minimization of the expected future risk given the observations available at the investment date. More formally, it can be written as\footnote{Recall that we neglect the 
expected return $\b g$ in the calculation of the variance, since the latter is usually small compared to the volatility.}
\begin{equation}
	\label{eq:mkw_bayes}
	\left\{
  \begin{array}{lr}
    \min_{\b w} \mathbb{E}\qBB{ \frac{1}{T_{\text{out}}} \pB{\sum_{t'=t+1}^{t+T_{\text{out}}} \scalar{\b w}{\b r_{t'}}}^2 \Bigg\lvert \cal F(t) }\,, \\
    \text{s.t.} \; \b w^* \b g \ge {\cal G}\,,
  \end{array}
\right.
\end{equation}
where $\cal F(t)$ is all the information available at time $t$ (the investment data), $T_{\text{out}}$ is the out-of-sample period, and $\b r$ is the vector of returns of the $N$ stocks in our portfolio. Assuming iid returns means that the optimal weights are independent from the future realizations of $\b r$. Moreover, we assume that $\cal P(\b r_{t'}) \propto \cal P(\b r_{t'} | \C) \cal P_0(\C)$ for $t' > t$, where $\cal P_0(\C)$ is an (arbitrary) prior distribution on the population covariance matrix $\C$.  One then has:
\begin{eqnarray}
	\mathbb{E}\qBB{ \frac{1}{T_{\text{out}}} \pB{\sum_{t'=t+1}^{t+T_{\text{out}}} \scalar{\b w}{\b r_{t'}}}^2 \Bigg\lvert \cal F(t) }\,, & = & \scalarBB{\b w}{\frac{1}{T_{\text{out}}} \sum_{t'} \mathbb{E}\qB{ \b r_t \b r_t^* \Big\lvert \cal F(t) } \; \b w  }\,, \nonumber \\
	& = & \scalarBB{\b w}{ \mathbb{E}\qB{ \C \Big\lvert \cal F(t)} \; \b w  }\,.
\end{eqnarray}
Recalling the results from Chapter \ref{chap:bayes}, we see that $\mathbb{E}[\C \lvert \cal F(t)] = \avg{\C}_{\cal P(\C |\E)}$ under a  multivariate Gaussian assumption on the returns\footnote{We expect this result to hold also for the multivariate Student, see Section \ref{sec:SCM_entries}.} (see Eq.\ \eqref{eq:gaussian_likelihood}). Therefore, using the result Eq.\ \eqref{eq:bayes_MMSE_SCM}, we can conclude that the oracle estimator is the one that minimizes the out-of-sample risk in that specific framework. 

There exists another, perhaps more direct derivation of the same result that we shall now present. It is based on the relation \eqref{eq:out_sample_risk}. Let us show this explicitly in the context of rotationally invariant estimators, that we considered in Chapter \ref{chap:bayes} and \ref{chap:RIE}. Let us define our RIE as 
\begin{equation*}
	\Xi = \sum_{i=1}^{N} \xi(\lambda_i) \b u_i \b u_i^*,
\end{equation*}
where we recall that $[\b u_i]_{i \in \qq{1,N}}$ are the sample eigenvectors and $\xi(\cdot)$ is a function that has to be determined. Suppose that we construct our portfolio $\b w_{\Xi}$ using this RIE, that we assume to be independent of the prediction vector $\b g$. Again, we assume for simplicity that $\b g$ is a Gaussian vector with zero mean and unit variance. Consequently, the estimate \eqref{eq:quad_form_trace} is still valid, such that the realized risk associated to the portfolio $\b w_{\Xi}$ reads for $N \to \infty$:
\begin{equation}
	\label{eq:out_sample_risk_rie}
	\cal R_{\text{out}}^2(\Xi) = \cal G^2 \frac{\tr \pB{\Xi^{-1} \C \Xi^{-1}}}{\pB{\tr \Xi^{-1}}^2}.
\end{equation}
using the spectral decomposition of $\Xi$, we can rewrite the numerator as 
\begin{equation}
	\label{eq:trace_C_rie_inv}
	\tr \pB{\Xi^{-1} \C \Xi^{-1}} = \sum_{i=1}^{N} \frac{\scalar{\b u_i}{\b C \b u_i}}{\xi^2(\lambda_i)}.
\end{equation}
On the other hand, one can rewrite the denominator of Eq.\ \eqref{eq:out_sample_risk_rie} as
\begin{equation}
	\label{eq:trace_rie_inv_squared}
	\pb{\tr \Xi^{-1}}^2 = \pBB{\sum_{i=1}^{N} \frac{1}{\xi(\lambda_i)}}^2.
\end{equation}
Regrouping these last two equations allows us to rewrite Eq.\ \eqref{eq:out_sample_risk_rie} as
\begin{equation}
	\label{eq:markowitz_out_risk_rie}
	\cal R_{\text{out}}^2(\Xi) = \cal G^2  \sum_{i=1}^{N} \frac{\scalar{\b u_i}{\b C \b u_i}}{\xi^2(\lambda_i)} \pBB{\sum_{i=1}^{N} \frac{1}{\xi(\lambda_i)}}^{-2}.
\end{equation}
Our aim is to find the optimal shrinkage function $\xi(\lambda_j)$ associated to the sample eigenvalues $[\lambda_j]_{j\in \qq{1,N}}$, such that the out-of-sample risk is minimized. This can be done by solving, for a given $j$, the following first order condition:
\begin{equation}
	\frac{\partial \cal R_{\text{out}}^2(\Xi)}{\partial \xi(\lambda_j)} = 0.
\end{equation}
By performing the derivative with respect to $\xi(\lambda_j)$ in \eqref{eq:markowitz_out_risk_rie}, one obtains 
\begin{equation}
	- 2 \frac{\scalar{\b u_j}{\C \b u_j} \xi'(\lambda_j)}{\xi^3(\lambda_j)} \pBB{\sum_{i=1}^{N} \frac{1}{\xi(\lambda_i)}}^{-2} + 2\frac{\xi'(\lambda_j)}{\xi^2(\lambda_j)} \pBB{\sum_{i=1}^{N} \frac{\scalar{\b u_i}{\C \b u_i}}{\xi^2(\lambda_i)}} \pBB{\sum_{i=1}^{N} \frac{1}{\xi(\lambda_i)}}^{-3} = 0,
\end{equation}
and one can check that the solution is precisely given by 
\begin{equation}
	\xi(\lambda_j) = \scalar{\b u_j}{\C \b u_j}  := \xi^{\text{ora.}}_j,
\end{equation}
which is the oracle estimator that we have studied in  chapters \ref{chap:bayes} and \ref{chap:RIE}. Note that this result has been obtained in \cite{ledoit2014nonlinear} where the authors also showed that this estimator maximizes the Sharpe ratio, i.e., the expected return of the strategy divided by its volatility. 

As a conclusion, the optimal RIE \eqref{eq:RIE_optimal} actually minimizes the out-of-sample risk under the class of rotationally invariant estimators under some distribution assumptions. Moreover, the corresponding ``optimal'' realized risk is given by
\begin{equation}
	\label{eq:out_sample_risk_rie_opt}	
	\cal R_{\text{out}}^2(\Xi^{\text{ora.}})  = \frac{\cal G^2}{\tr \qb{(\Xi^{\text{ora.}})^{-1}}},
\end{equation}
where we used the notable property that for any $n \in \mathbb{Z}$:
\begin{equation}
	\tr[ (\Xi^{\text{ora.}})^{n} \C] = \tr[ (\Xi^{\text{ora.}})^{n+1}],
\end{equation}
which directly follows from the general formula \eqref{eq:oracle}. 

\subsubsection{Optimal in and out-of-sample risk for an Inverse Wishart prior}

In this section, we specialize the result \eqref{eq:out_sample_risk_rie_opt} to the case when $\C$ is an Inverse-Wishart matrix with parameter $\kappa > 0$, corresponding to the simple linear shrinkage optimal estimator. Notice that we shall assume throughout this section that there are no outliers ($r=0$). Firstly, we infer from Eq.\ \eqref{eq:stieltjes_invWishart} by $z \to 0$ that
\begin{equation}
	\label{eq:normTrace_trueInv_iw}
	\varphi(\C^{-1}) = - \stj_\C(0) = 1 + \frac{1}{2\kappa}\,,
\end{equation}
so that we get from Eq.\ \eqref{eq:markowitz_risk_RMT_tmp} that in the large $N$ limit:
\begin{equation}
	\label{eq:true_risk_invWishart}
	\cal R^2_{\text{true}} = \frac{\cal G^2}{N} \frac{2 \kappa}{1+2\kappa}\,.
\end{equation}

Next, we see from Eq.\ \eqref{eq:out_sample_risk_rie_opt} that the optimal out-of-sample risk requires the computation of $\varphi((\Xi^{\text{ora.}})^{-1})$. In general, the computation of this normalized is highly non-trivial but we shall show that some genuine simplifications appear when $\C$ is an inverse Wishart. In the LDL, the final result, whose derivation is postponed at the end of this section, reads:
\begin{equation}
	\label{eq:normTrace_oracleInv_iw}
	\varphi((\Xi^{\text{ora.}})^{-1}) = - (1+2q\kappa) \stj_\E(-2q\kappa) = 1 + \frac{1}{2\kappa(1+q(1+2\kappa))}\,,
\end{equation}
and therefore we have from Eq.\ \eqref{eq:out_sample_risk_rie_opt} 
\begin{equation}
	\label{eq:oracle_realized_risk_invWishart}
	\cal R_{\text{out}}^2(\Xi^{\text{ora.}}) = \frac{\cal G^2}{N}\frac{2\kappa(1+q(1+2\kappa))}{1+2\kappa(1+q(1+2\kappa))}\,,
\end{equation}
from which it is clear from Eqs.\ \eqref{eq:oracle_realized_risk_invWishart} and \eqref{eq:true_risk_invWishart} that for any $\kappa > 0$:
\begin{equation}
	\label{eq:oracle_ratio_out_true_risk}
	\frac{\cal R_{\text{out}}^2(\Xi^{\text{ora.}})}{\cal R^2_{\text{true}}} = 1 + q \frac{2\kappa}{1+2\kappa(1+q(1+2\kappa))} \geq 1 \,,
\end{equation}
where the last inequality becomes an equality only when $q = 0$, as it should.

It is also interesting to evaluate the in-sample risk associated to the oracle estimator. It is defined by
\begin{equation}
	\label{eq:oracle_in_risk}
	\cal R^2_{\text{in}}(\Xi^{\text{ora.}}) = \cal G^2 \frac{\tr\qb{(\Xi^{\text{ora.}})^{-1}\E (\Xi^{\text{ora.}})^{-1}}}{N\varphi^2((\Xi^{\text{ora.}})^{-1})}\,,
\end{equation}
where the most challenging term is the numerator. As above, the computation of this term is, to our knowledge, not trivial in the general case but using the fact that the eigenvalues of $\Xi^{\text{ora.}}$ are given by \eqref{eq:RIE_lin}, we can once again find a closed formula. 
As above, we relegate the derivation at the end of this section and the result reads:
\begin{equation}
	\label{eq:normTrace_inrisk_num_iw}
	\varphi\pb{(\Xi^{\text{ora.}})^{-1}\E (\Xi^{\text{ora.}})^{-1}} = -(1-z)^2 \qb{\stj_\E(z) + z\stj_\E'(z)}\bigg\lvert_{z=-2q\kappa} = \frac{(1+2\kappa)(1+2q\kappa)^3}{2\kappa (1+q(1+2\kappa))^3} \,.
\end{equation}
Hence by plugging Eqs.\ \eqref{eq:normTrace_inrisk_num_iw} and \eqref{eq:normTrace_oracleInv_iw} into Eq.\ \eqref{eq:oracle_in_risk}, we obtain
\begin{equation}
	\label{eq:oracle_in_risk_invWishart}
	\cal R^2_{\text{in}}(\Xi^{\text{ora.}}) = \frac{\cal G^2}{N} \frac{2\kappa(1+2q\kappa)}{(1+2\kappa)(1+q(1+2\kappa))}\,, 
\end{equation}
and we therefore deduce with Eq.\ \eqref{eq:true_risk_invWishart} that for any $\kappa > 0$:
\begin{equation}
	\label{eq:oracle_ratio_in_true_risk}
	\frac{\cal R^2_{\text{in}}(\Xi^{\text{ora.}})}{\cal R^2_{\text{true}}} = 1 - \frac{q}{1+q(1+2\kappa)} \leq 1\,,
\end{equation}
where the inequality becomes an equality for $q = 0$ as above. 

Finally, one may easily check from Eqs.\ \eqref{eq:markowitz_risk_RMT}, \eqref{eq:oracle_ratio_out_true_risk} and \eqref{eq:oracle_ratio_in_true_risk},  that
\begin{equation}
	\cal R^2_{\text{in}}(\Xi^{\text{ora.}}) - \cal R_{\text{in}}^2(\E) \geq 0, \qquad \cal R^2_{\text{out}}(\Xi^{\text{ora.}}) - \cal R_{\text{out}}^2(\E) \leq 0\,,
\end{equation}
showing explicitly that we indeed reduce the over-fitting by using the oracle estimator instead of the sample covariance matrix in the high dimensional framework. 

\begin{changemargin}{1.0cm}{1.0cm} 
\footnotesize

The aim of this technical section is to derive the results \eqref{eq:normTrace_oracleInv_iw} and \eqref{eq:normTrace_inrisk_num_iw}. We begin with Eq.\ \eqref{eq:normTrace_oracleInv_iw} and we use that the eigenvalues of the oracle estimator converge to Eq.\ \eqref{eq:RIE_lin} when $N \to \infty$. $\C$ is assumed to be an inverse Wishart of parameter $\kappa > 0$. Hence, one has 
\begin{equation}
	\varphi((\Xi^{\text{ora.}})^{-1}) = \frac1N \sum_{i=1}^{N} \frac{1}{1+\alpha_s(\lambda_i - 1)} = \frac{1}{\alpha_s} \frac{1}{N} \sum_{i=1}^{N} \frac{1}{\frac{1-\alpha_s}{\alpha_s} +\lambda_i}\,,
\end{equation}
and using Eq.\ \eqref{eq:linear_shrinkage}, we also have
\begin{equation*}
	\frac{1}{\alpha_s} = 1+2q\kappa, \quad\text{and}\quad \frac{1-\alpha_s}{\alpha_s} = 2q\kappa\,.
\end{equation*}
We may conclude that
\begin{equation}
	\label{eq:stj_normTrace_oracleInv_iw}
	\varphi((\Xi^{\text{ora.}})^{-1}) \sim (1+ 2q\kappa) \stj_{\E}(-2q\kappa)\,,
\end{equation}
where we emphasize that the Stieltjes transform is analytic since its argument is non-positive for any $\kappa > 0$. This is the first equality of Eq.\ \eqref{eq:normTrace_oracleInv_iw} that relates the computation of the normalized trace with the Stieltjes transform of $\E$. When $\C$ is an Inverse Wishart, we know that $\stj_\E$ is explicit and given by \eqref{eq:stieltjes_IW_MP}. Nonetheless, it seems that Eq.\ \eqref{eq:stieltjes_IW_MP} is diverging for $z = -2q\kappa$ so that one has to be careful in the evaluation of $\stj_{\E}(-2q\kappa)$. To that end, we fix $z = -2q\kappa + \e$ with $\e > 0$ and expand the numerator of Eq.\ \eqref{eq:stieltjes_IW_MP} as a power of $\e$ to find:
\begin{equation*}
	\stj_\E(z) = \frac{q-z}{z(1+q-z)} + \cal O(\e),
\end{equation*}
meaning that for $\e = 0$, we obtain
\begin{equation}
	\stj_\E(-2q\kappa) = - \frac{1+2\kappa}{2\kappa(1+q(1+2\kappa))}.
\end{equation}
It is then easy to deduce Eq.\ \eqref{eq:normTrace_oracleInv_iw} from this last equation and Eq.\ \eqref{eq:stj_normTrace_oracleInv_iw}.

The computation of Eq.\ \eqref{eq:normTrace_inrisk_num_iw} is a bit more tedious but very similar to the derivation of the previous paragraph. Indeed, using that $(\Xi^{\text{ora.}})^{-1} \E (\Xi^{\text{ora.}})^{-1}$ share the same eigenbasis, we have thanks to Eq.\ \eqref{eq:RIE_lin}:
\begin{equation}
	\varphi((\Xi^{\text{ora.}})^{-1} \E (\Xi^{\text{ora.}})^{-1}) = \frac{1}{N} \sum_{i=1}^{N} \frac{\lambda_i}{(1+\alpha_s(\lambda_i - 1))^2}\,,
\end{equation}
which gives after some simple manipulations:
\begin{equation}
	\varphi((\Xi^{\text{ora.}})^{-1} \E (\Xi^{\text{ora.}})^{-1}) = \frac{1}{\alpha_s} \frac{1}{N} \sum_{i=1}^{N} \qBB{ \frac{1}{1+\alpha_s(\lambda_i - 1)} -  \frac{1-\alpha_s}{(1+\alpha_s(\lambda_i - 1))^2}}  \,.
\end{equation}
Defining $z = - 2q\kappa < 0$, one can deduce the first equality of Eq.\ \eqref{eq:normTrace_inrisk_num_iw} using the same identification with the Stieltjes transform (and its derivative with respect to $z$) as above. The derivative of Eq.\ \eqref{eq:stieltjes_IW_MP} reads:
\begin{equation}
	\stj_\E'(z) = \frac{1}{z^2(z+2q\kappa)^2}\qBB{ z (2 \kappa q+z) \left(1+\kappa -\frac{\kappa (\kappa (q-z+1)+1)}{\sqrt{\kappa^2(z+q-1)^2 - 2\kappa z(1+2\kappa)}}\right)- 2(q\kappa+z) \beta(z)},
\end{equation}
where $\beta(z)$ is defined by
\begin{equation}
	\beta(z) \;\deq\; z(1+\kappa) - \kappa(1-q) + \sqrt{\kappa^2(z+q-1)^2 - 2\kappa z(1+2\kappa)}\,,
\end{equation}
which is the denominator of Eq.\ \eqref{eq:stieltjes_IW_MP}. We omit further details as the proof of the second equality of Eq.\ \eqref{eq:normTrace_inrisk_num_iw} relies on a Taylor expansion around $-2q\kappa$ in the same spirit than in the previous paragraph. This regularizes the Stieltjes transform and its derivative and one eventually obtains:
\begin{equation}
	-2q\kappa\stj_\E'(-2q\kappa) = \frac{q(1+2\kappa)\qb{q + 2(1+\kappa+ 2q\kappa(1+\kappa))}}{2\kappa(1+q(1+2\kappa))^3}\,
\end{equation}
and we find the desired result by plugging this last equation into Eq.\ \eqref{eq:normTrace_inrisk_num_iw}. 

\end{changemargin} 
\normalsize

\subsection{A short review on previous cleaning schemes} 
\label{sec:past_cleaning}

In this section, we give a short survey of the many attempts in the literature to circumvent the above ``in-sample'' curse by \textit{cleaning} the covariance matrix before using it for e.g. portfolio construction. 
Even if most of the recipes considered below are not optimal (in a statistical sense), a lot of interesting ideas have been proposed to infer the statistical properties of the unknown population matrix. 
As we shall see, most of the methods appeared after the seminal work of Mar{\v c}enko \& Pastur \cite{marchenko1967distribution}. We nonetheless stress that the literature on estimating large covariance matrices 
is so large that it is impossible to make justice to all the available results here. We will only consider methods for which RMT results offer interesting insights and refer to, e.g. \cite{bouchaud2009financial,bartz2015advances,paul2014random} for complementary sources of information. 

We shall present four different classes of estimators. The first one is the \textbf{linear shrinkage} method. This estimator has been studied in details in Chapters \ref{chap:bayes} and \ref{chap:RIE} but here, we focus on the estimation of the shrinkage intensity. As we will see, RMT will provide very simple methods to estimate parameters from the data. 

Then we will present the \textbf{eigenvalues clipping} method of \cite{laloux2000random, plerou2002random} where the aim is to separate ``trustworthy'' eigenvalues from ``noisy'' ones. The basic idea of this method is the spiked covariance matrix model that we presented in Section \ref{chap:spectrum} where the true eigenvalues consist in a finite number $r$ of spikes and one degenerate eigenvalue $\approx 1 - \cal O(r/N)$, with multiplicity $N-r$. 

The third method, that we name \textbf{eigenvalues substitution}, consists in solving the inverse Mar{\v c}enko-Pastur problem (see Section \ref{chap:spectrum}). Roughly speaking, in the presence of a very large number of eigenvectors, one can discretize the Mar{\v c}enko-Pastur equation and solve the inverse problem using either a parametric \cite{bouchaud2009financial} or non-parametric approach \cite{el2008spectrum}. 

The last method concerns \textbf{factors models}, or structured covariance estimators, where one tries to explain the correlation matrix through a simplified model of the underlying structure of the data. This is a very popular approach in finance and economics, and we will see how RMT has allowed some recent progress.  

All these methods will be tested using real financial data in the next chapter. 

\subsubsection{Linear Shrinkage}
\label{sec:linear_shrinkage}
We recall that the linear shrinkage is given by
\begin{equation}
	\Xi^{\text{lin}} = \alpha_s \E + (1-\alpha_s) \In, \qquad \alpha \in [0,1].
\end{equation}
As discussed in Chapter \ref{chap:bayes}, this estimator has a long history in high-dimensional statistics \cite{haff1980empirical,ledoit2004well} as it provides a simple proof that the sample estimator $\E$ is 
inconsistent whenever $N$ and $T$ are both large. A very exhaustive presentation of the properties of this estimator in the high-dimensional regime can be found in \cite{ledoit2004well} or in \cite{karoui2011geometric} in a more RMT oriented standpoint. It is easy to see that $\Xi^{\text{lin}}$ shares the same eigenbasis as the sample estimator $\E$, and is thus a rotationally invariant estimator with  
\begin{equation}
	\label{eq:linear_shrinkage_eigen}
	\Xi^{\text{lin}} = \sum_{i=1}^{N} \xi^{\text{lin}} \b u_i \b u_i^*, \qquad  \xi^{\text{lin}} = 1 + \alpha_s(\lambda_i - 1) 
\end{equation}
We already emphasized that this estimator exhibits all the expected features: the small eigenvalues are shifted upwards (compared to the sample eigenvalues) while the top eigenvalues are pulled downwards (see Figure \ref{fig:cleaning_old}). As alluded to above, this estimator has been fully investigated in \cite{ledoit2004well}. Most notably, the authors were able to determine an asymptotic optimal formula to estimate $\alpha_s$ directly from the data. Keeping the notations of Section \ref{chap:spectrum}, our data set is $\Y = (\b y_1, \dots, \b y_T) \in \R^{N\times T}$ and we assume that $\mathbb{E} [Y_{it}] = 0$ and $\mathbb{E} [Y_{it}^2] = T^{-1}$ for all $i \in \qq{1,N}$. 
Defining:
\begin{eqnarray}
	\label{eq:linear_shrinkage_auxvar}
	\beta & \deq & \frac1N \tr\left[ (\E - \In)(\E - \In)^* \right] \nonumber \\
	\gamma & \deq & \max\left( \beta, \frac{1}{T^2} \sum_{k=1}^{T} \frac1N \tr\left[ (\b y_k \b y_k^* - \E)(\b y_k \b y_k^* - \E)^*\right]  \right),
\end{eqnarray}
then 
\begin{equation}
	\label{eq:shrinkage_intensity_LW}
	\wh \alpha_s = 1 - \frac{\beta}{\gamma},
\end{equation}
is a consistent estimator of $\alpha_s$ in the high-dimensional regime \cite{ledoit2004well}. 

Using tools from RMT, and more precisely the result of Sections \ref{chap:spectrum} and \ref{chap:eigenvectors}, we can find another consistent estimators of $\alpha_s$ which uses the fact that linear shrinkage implicitly 
assumes the underlying correlation matrix to be an Inverse-Wishart matrix with parameter $\kappa$, from which $\alpha_s$ is deduced as $\alpha_s = (1 + 2q\kappa)^{-1}$. The value of $\kappa$ can be extracted from the 
data using the relation (valid for $q < 1$): 
\begin{equation}
	\stj_\C(0) = (1-q) \stj_\E(0) = 1 + 2 \kappa.
\end{equation}
where the last equality can be deduced from \eqref{eq:stieltjes_invWishart} and \eqref{eq:SCM_inverse_moment_1}. Therefore, we obtain a simple estimate for $\kappa$ from the trace of $\E^{-1}$ as:
\begin{equation}
	\label{eq:shrinkage_intensity_trace}
		\kappa = \frac12\pBB{(1-q)\frac{\tr\E^{-1}}{N} - 1 }.
\end{equation}
However, this estimate is only reliable when $\kappa$ is not too large, i.e. when $\C$ is significantly different from the identity matrix (in the opposite case, $(1-q)\tr\E^{-1} \approx N$ so that one can obtain negative 
values for $\kappa$). A more robust alternative to estimate $\kappa$ is the ``two-sample'' test introduced in Chapter \ref{sec:mso_two_sample}, see Eqs \eqref{eq:mso_SCM_bulk} and \cite{bun2016overlaps}.

\subsubsection{Eigenvalues clipping}

This method is perhaps the first RMT-based estimator for large covariance matrices. It has been investigated in several papers \cite{laloux1999noise, laloux2000random,plerou2002random} where the Mar{\v c}enko-Pastur distribution is used in a very intuitive way to correct the sample eigenvalues. The idea of the method is as follows: all the eigenvalues that are beyond the largest expected eigenvalue of the empirical matrix $\lambda_{+} = (1+\sqrt{q})^2$ (within a null hypothesis) are interpreted as signal while the others are pure noise (see Figure \ref{fig:eigen_justMP}). An alternative interpretation would be that outliers are true factors while the others are meaningless. 
% We plot in Fig. \ref{eigen_justMP} an illustration of this method where all outliers contains true signal. 

% \begin{figure}
% 	\begin{center}
%    \includegraphics[scale = 0.4]{Figures/oldCleaning/eigen_justMP} 
%    \end{center}
%    \caption{Test of the null hypothesis on the empirical correlation matrix $\E$ using US stocks' data.}
%    \label{eigen_justMP}
% \end{figure}

In a recent paper \cite{bloemendal2014principal}, this idea has been made rigorous in the sense that if we suppose that $\C$ is a finite rank perturbation of $\In$ as defined in \eqref{eq:spikeless_population_cov}, then the reference matrix of the bulk eigenvalues of $\E$ simply corresponds to the (isotropic) Wishart matrix $\Wishart$. Differently said, for this specific model, these bulk eigenvalues should be seen as pure noise, and the right edge $(1+\sqrt{q})^2$ can be interpreted as the threshold between noise and signal.

\begin{figure}[!ht]
	 \begin{center}
	\includegraphics[scale = 0.5]{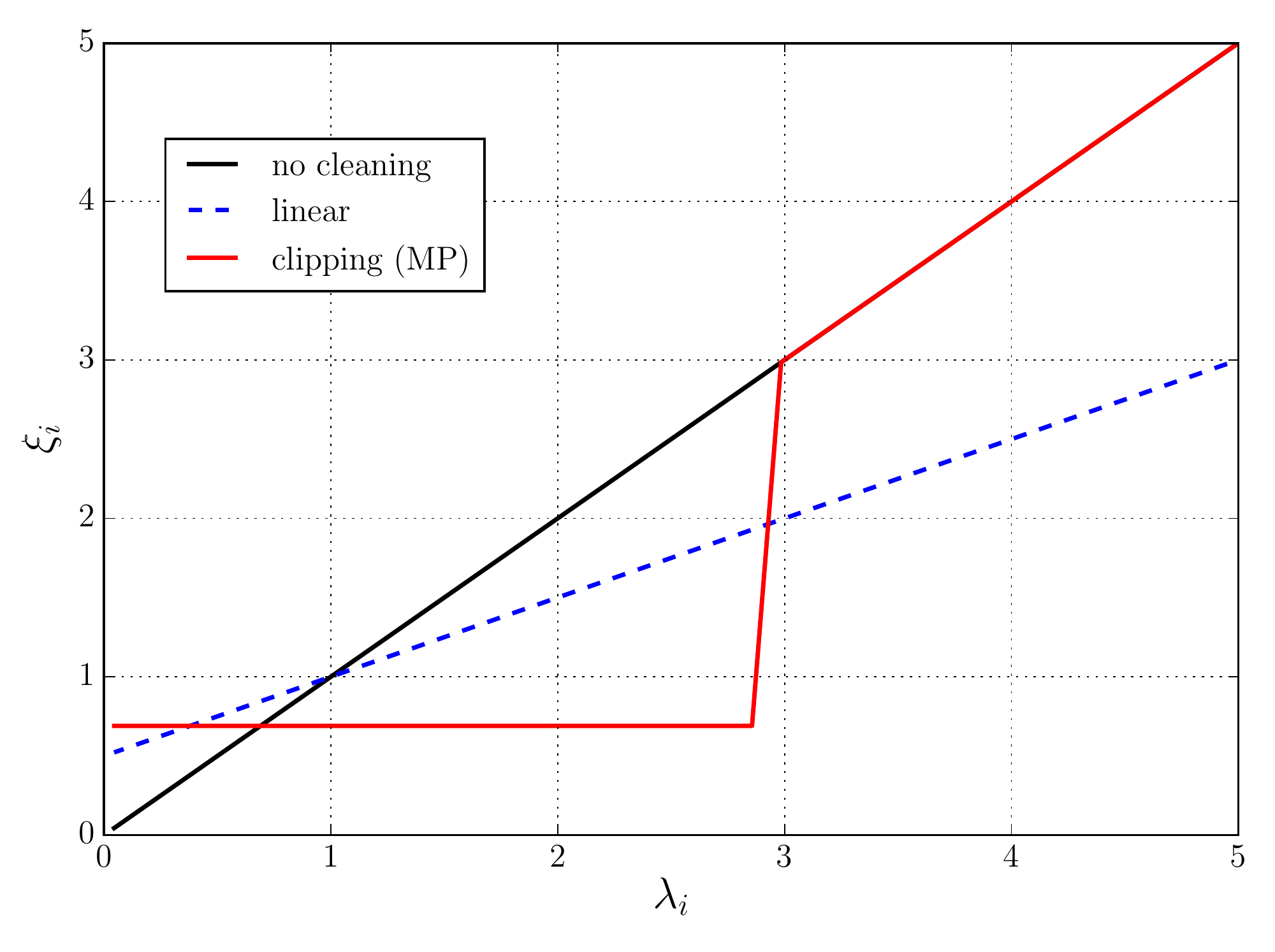} 
	\end{center}
	\caption{Impact on sample eigenvalues of the eigenvalues clipping \eqref{clipping_scheme} (red plain line) with a threshold given by $(1+\sqrt{q})^2$ with $q = 0.5$ and the linear shrinkage \eqref{eq:linear_shrinkage_eigen} (blue dashed line) with intensity $\alpha_s = 0.5$. We see that the lowest eigenvalues are shifted upward. }
	\label{fig:cleaning_old}
\end{figure}

Endowed with a simple rule to isolate the signal eigenvalues, how should one clean the noisy ones? Laloux et al. \cite{laloux2000random} proposed the following rule: first diagonalize the matrix $\E$ and keep the eigenvectors unchanged. Then apply the following scheme in order to denoise the sample eigenvalues:
\begin{equation}
\label{clipping_scheme}
\Xi^{\text{clip.}} \;\deq\; \sum_{i=1}^{N} \xi_i^{\text{c}} \b  u_i \b u_i^*, \qquad
\xi_i^{\text{clip.}} \;=\; \begin{dcases}
\lambda_{i} & \text{if } \lambda_{i} \geq (1+\sqrt{q})^2 \\
\bar{\lambda} & \text{otherwise},
\end{dcases}
\end{equation}
where $\bar{\lambda}$ is chosen such that $\Tr\Xi^{\text{clip.}} = \Tr\E$. Roughly speaking, this method simply states that the noisy eigenvalues are shrunk toward a (single) constant such that the trace is preserved. 
This procedure is known as \textit{clipping} and Figure \ref{fig:cleaning_old} shows how it shifts upwards the lowest eigenvalues in order to avoid {\it a priori} abnormal low variance modes.

Nonetheless, the method suffers from several separate problems. First, one often observes empirically, especially with financial data, that the value of $q=N/T$ that is fixed by the dimensionality of the matrix and the length of the
time series is significantly different from the ``effective'' value $q_{\text{eff}}$ that allows one to fit best the empirical spectral density \cite{laloux2000random}. This effect can be induced either by small temporal 
autocorrelation in the time series \cite{burda2004spectral,burda2004free,bartz2014covariance} and/or by the inadequacy of the null hypothesis $\C = \In$ for the bulk of the distribution. In any case, a simple recipe would be to use a corrected upper edge $\lambda_+ =
(1+\sqrt{q_{\text{eff}}})^2$ for the threshold separating wheat from chaff. Another possibility, proposed in \cite{bouchaud2009financial}, is to introduce a fine-tuning parameter $\alpha_{c} \in [0,1]$ such that the $\lceil N \alpha_{c} \rceil$ largest eigenvalues are kept unaltered while the others are still replaced by a common $\bar{\lambda}$. It is easy to see that for $\alpha_c = 1$, we get the empirical covariance matrix while for $\alpha_c = 0$, we get the identity matrix. So $\alpha_c$ plays the role of the upper bound $\lambda_+$ of the Mar{\v c}enko-Pastur density, and allows one to interpolate between $\E$ and the null hypothesis $\In$, much like linear shrinkage. Nevertheless, the calibration of the parameter $\alpha_c$ is not based on any theoretical rule. 

Another concern about this method is that we know from section \ref{sec:RIE_prop} that the optimal estimator of the large outliers is {\it not} their bare empirical value $\lambda_i$. Rather, one should shift them downwards even when far from the bulk, by a quantity 
equal to $-2q$ (in the limit $\lambda_i \gg 1$). Hence, at the very least, such a shift should be included in the eigenvalue clipping scheme from Eq.\ \eqref{clipping_scheme} (see \cite{bartz2013directional} for a related discussion). 

\subsubsection{Eigenvalue substitution}

The main idea behind the eigenvalue substitution method is also quite intuitive and amounts to replacing the sample eigenvalues by their corresponding ``true'' values obtained by inverting the Mar{\v c}enko-Pastur equation \eqref{eq:MP_equation_stieltjes}. More formally, we seek the set of true eigenvalues $[\mu_j]_{j \in \qq{1,N}}$ that solve Eq.\ \eqref{eq:MP_equation_stieltjes} for a \emph{given} set of sample eigenvalues $[\lambda_j]_{j \in \qq{1,N}}$. As for the eigenvalues clipping procedure, this technique can be seen a nonlinear shrinkage function and has the advantage to lean upon a more robust theoretical framework than the clipping ``recipe''. However, as we emphasized in Section \ref{sec:MP}, inverting the Mar{\v c}enko-Pastur equation is quite challenging in practice. In this section, we present several possibilities to achieve this goal in the limit of large dimensions.

\paragraph{Parametrization of Mar{\v c}enko-Pastur equation}

One way to think about the inverse Mar{\v c}enko-Pastur problem is to adopt a Bayesian viewpoint (like in Chapter \ref{chap:bayes}). More specifically, we assume that $\C$ belongs to a rotationally invariant ensemble -- so that there is no a priori knowledge about the eigenvectors -- and assume a certain structure on the LSD $\rho_\C(\mu)$, parameterized by one or several numbers. The optimal values of these parameters (and the corresponding optimal $\widehat \rho_\C$) are then fixed by e.g. a maximum likelihood procedure on the associated $\rho_\E$, obtained from the direct Mar\v{c}enko-Pastur equation. Once the fit is done, the \emph{substitution} cleaning scheme reads
\begin{equation}
\label{eq:eigenvalues_sub_param}
\lambda_i \rightarrow \widehat\mu_i \quad\text{such that}\quad \frac{i}{N} = \int_{\widehat\mu_i}^{\infty} \widehat\rho_{\C}(x)\dd x.
\end{equation}
Note that under the transformation \eqref{eq:eigenvalues_sub_param}, we assume that the eigenvalues of $\C$ are allocated smoothly according to the quantile of the limiting density $\widehat\rho_{\C}$. 
%Note that there is no guarantee that the trace is perfectly preserved by this operation, so some final rescaling might be necessary.

\begin{figure}
	 \begin{center}
	\includegraphics[scale = 0.4]{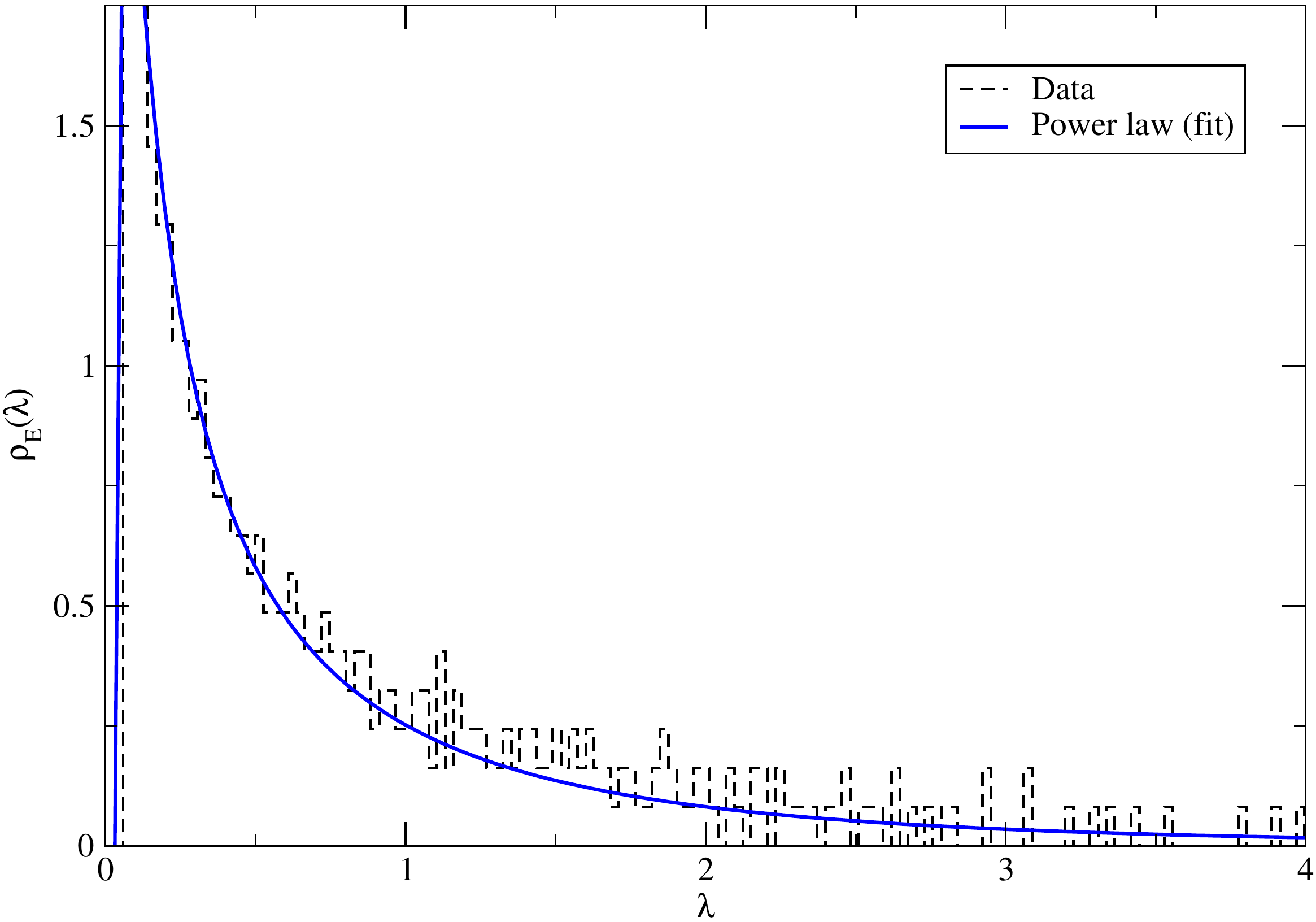} 
	\end{center}
	\caption{Fit of the power law distribution \eqref{eq:density_powerlaw} on the sample eigenvalues of the 450 most liquid assets of the S\&P index from 2006 to 2010 using the Mar{\v c}enko-Pastur equation \eqref{eq:MP_equation_stieltjes}. The fit has been performed using a maximum likelihood procedure and yields $\alpha \approx 0.3$. The black dashed histogram represents the empirical spectral density. 
	}
	\label{Estimation_PL_2006_bulk}
\end{figure}

As an illustration of this parametric substitution method, let us consider a power law density \eqref{eq:density_powerlaw} as the prior for $\rho_\C(\mu)$. Such a probabilistic model for the population eigenvalues density is thought to be plausible for financial markets, and reflect the power-law distribution of sector sizes in the economy \cite{bouchaud2009financial,marsili2002dissecting}. In that case, the parametric substitution turns out to be explicit in the limit of large dimension. Moreover, the estimation of the unique parameter $\lambda_0$ in this model can be done using e.g.\ maximum likelihood, as we can compute exactly $\rho_\E$ on $\mathbb{R}^+$ using \eqref{eq:stieltjes_powerlaw} and \eqref{eq:MP_equation_dual_inverse}. This then yields a parameter $\widehat\lambda_0$ and hence $\widehat\rho_\C$ as well. As a result, the substitution procedure \eqref{eq:eigenvalues_sub_param} becomes for $N \to \infty$ \cite{bouchaud2009financial}:
\begin{equation}
\label{eigenvalues_sub_powerlaw}
\mu_i = -\wh\lambda_0 + \frac{(1 + \wh\lambda_{0})}{2} \sqrt{\frac{N}{i}} \qquad i \in \qq{1,N}\,.
\end{equation}
We present such a procedure in Fig. \ref{Estimation_PL_2006_bulk} using US stocks data. We conclude from this figure that the fit is indeed fairly convincing, i.e. that a power-law density for the eigenvalues 
of $\C$ is a reasonable assumption. 

\paragraph{Discretization of Mar{\v c}enko-Pastur equation}
\label{sec:NEK}

Interestingly, a ``quasi'' non-parametric procedure is possible under some smoothness assumption on the density $\rho_{\C}$. This algorithm is due to N. El Karoui \cite{el2008spectrum} who proposed to solve an approximate form of the  Mar{\v c}enko-Pastur inverse problem. The starting point is to notice that each eigenvalue of $\E$ satisfies:
\begin{equation*}
\left\{ z_j = \frac{1}{\stj_{\S}(z_j)} \left[ 1-q +q \int \frac{\rho_{\C}(\mu)d\mu}{1-\mu\, \stj_{\S}(z_j)} \right], \quad\text{with}\quad z_j = \lambda_j - i\eta \right\}_{j=1}^{N}
\end{equation*}
that follows from Eq.\ \eqref{eq:MP_equation_dual_inverse} and where we recall that $\S$ is the $T\times T$ dual matrix of $\E$ defined in \eqref{eq:SCM_dual}. The main assumption of this method is to decompose the density of states $\rho_{\C}$ as a weighted sum of Dirac masses:
\begin{equation}
\label{NEK_discret_density}
\rho_{\C}(\mu) = \sum_{k=1}^{N} \widehat{w}_k \delta(\mu - \mu_k), \quad\text{such that }  \sum_{k=1}^{N} \widehat{w}_k = 1 \quad\text{and } \widehat{w}_k \ge 0, \, \forall \, k \in [\![1,N]\!].
\end{equation}
Note that this decomposition simply use the discreteness of the eigenvalues that follows from the very definition of an ESD where each eigenvalues are associated with a weight equals to $N^{-1}$. One notices that there are two different sources of uncertainty: the ``true'' eigenvalues $\mu_j$ and their corresponding weights $\wh w_j$ so that the parametrization looks inextricably complex. In \cite{el2008spectrum}, the author suggested to fix the 
positions $[\mu_j]_{j \in \qq{1,N}}$ {\it a priori} such that we are left with the weights $[\widehat{w}_j]_{j \in \qq{1,N}}$ as the only unknown variables in the problem. Within this framework, the author then proposed to obtain the optimal weights through the following optimization program:
\begin{equation}
\label{NEK_optim_pbm}
[\widehat{w}_j]_{j \in \qq{1,N}} =
\begin{dcases} 
		\argmin_{\{ w_i\}_{i=1}^{N}} {\cal L}\left( \left\{ \frac{1}{\stj_{\S}(z_j)} \left[ 1-q +q \sum_{k=1}^{N} \frac{w_k}{1-\mu_k \, \stj_{\S}(z_j)} \right] - z_j \right\}_{j = 1}^{N} \right) \\
		\text{subject to } \sum_{k=1}^{N} w_k = 1, \quad\text{and }  w_k \ge 0 \quad \forall \, k \in [\![1,N]\!],
\end{dcases}
\end{equation}
where ${\cal L}$ is a certain loss function and $z_j = \lambda_j - i\eta$. In addition to the error we make by approximating the true density by a sum of weighted Dirac masses, there are at least two others sources of errors: 
\begin{itemize}
  \item[1.] The approximation $\stj_{\E}(z_j) \approx N^{-1} \Tr(z_j \In - \E)^{-1}$;
  \item[2.] The position of the eigenvalues $[\mu_j]_{j \in \qq{1,N}}$ that have to be chosen.
\end{itemize}
In the large $N$ limit, the first approximation is fairly accurate (see Section \ref{chap:application}). However, the second is much more difficult to handle especially in the case of a very diluted spectrum. Note that if we define $e_j$ as the error we make term in \eqref{NEK_optim_pbm} for each $\lambda_j$, then the consistency of the algorithm has been showed in \cite{el2008spectrum} under the norm $L_{\infty} = \max_{j=1,\dots,N} \max(  | \re(e_j) |, | \im(e_j)|)$. Once we get the optimal weight $[\widehat w_j]_{j \in \qq{1,N}}$, the cleaning procedure is immediate
\begin{equation}
\label{eq:substitution_estimator}
\lambda_{i} \rightarrow \widehat{\mu}_i \quad\text{where } \quad \widehat{\mu}_i = \min\left\{ x \in \mathbb{R}^{+} :  \sum_{k=1}^{N} \widehat{w}_{k} \Theta(\mu_k -  x)  \ge \frac{i}{N} \right\}
\end{equation}
where we have used the approximation
\begin{equation*}
\int_{x}^{\infty} \rho_{\C}(u)du \approx \sum_{k=1}^{N} \widehat{w}_{k} \Theta(\mu_k -  x) ,
\end{equation*}
with $\Theta(x)$ that denotes the Heaviside step function. 

While the method is backed by a theoretical framework, it turns out that the error source \# 2. above is a strong limitation in practice. A recent proposal to invert the Mar{\v c}enko-Pastur equation by optimizing directly the eigenvalues $[\mu_j]_{j \in \qq{1,N}}$ has therefore been proposed in \cite{ledoit2013spectrum}. This alternative 
method, called QuEST, turns out to be much more robust numerically (see \cite{ledoit2016numerical} and Chapter \ref{chap:numerical} for an extended discussion and some applications). 

As a conclusion, we see that it is possible to solve (approximately) the inverse Mar{\v c}enko-Pastur equation in a quite general fashion, meaning that we might 
indeed be able to find an estimator of the true eigenvalues $\widehat \mu_i$ for all $i =1,\dots,N$. As a result, the eigenvalue substitution estimator is then obtained as
\begin{equation}
	\label{eq:sub_estimator}
\Xi^{\text{sub}} = \sum_{k=1}^{N} \widehat{\mu}_k \b u_k \b u_k^*.
\end{equation}
However, even when a perfect estimation of the true density $\rho_\C$ is feasible, we see that this estimator does not take into account the fact that the sample eigenvectors are not consistent estimators of the true ones, as shown in Chapter \ref{chap:eigenvectors}. Therefore, for covariance matrices estimation, it is not advised to use the substitution \eqref{eq:sub_estimator} since this is not the optimal solution. However, it can be used to compute the optimal RIE \eqref{eq:RIE_optimal} and we refer to Section \ref{sec:QuEST} for more details.

\subsection{Factor models}

The main idea behind linear factor models is quite simple: the (normalized) data $Y_{it}$ is represented as a linear combination of $M$ common factors $F$
\begin{equation}
\label{eq:factor_model}
Y_{it} = \sum_{k=1}^{M} \beta_{ik} f_{kt} + \e_{it} 
%\equiv \b \beta F + \b \e
\end{equation}
where the $\beta_{ik}$ are the linear exposures of the variable $i$ to the factors $k=1,\dots, M$ at time $t$ and the $N \times T$ matrix $\e_{it}$ is the idiosyncratic part of $Y_{it}$ (or the residual in Statistics), assumed to be of zero mean. The model \eqref{eq:factor_model} in matrix form reads
\begin{equation}
\label{eq:factor_model_matrix}
\b Y = \b \beta \b F + \b{\cal E},
%\equiv \b \beta F + \b \e
\end{equation}
which is known as \emph{Generalized Linear Model} \cite{mccullagh1989generalized}. It is often assumed that the residuals are i.i.d. across $i$ with $t$ fixed (see e.g. \cite{chamberlain1982arbitrage} for an application in Finance). It is not hard to see that the covariance matrix under the model \eqref{eq:factor_model}, is given by
\begin{equation}
\label{linear_model_true_covar}
\C = \beta \Sigma_{F} \beta^* + \Sigma_{\e}
\end{equation}
where $\Sigma_{F}$ is the covariance matrix of size $M \times M$ of the factors $F$ -- which can be chosen, without loss of generality, to be proportional to the identity matrix -- and $\Sigma_{\e}$ is the $N \times N$ covariance matrix of the residuals $\e$, which is simply the identity in the simplest framework. Within the linear decomposition \eqref{eq:factor_model}, we see that we have generically a number of parameters to estimate of order $\cal O(NM)$ out of datasets of size $\cal O(NT)$. Hence, we see that the curse of dimensionality disappears as soon as $M \ll N,T$ which implies that the empirical estimate
\begin{equation}
\label{linear_model_empirical_covar}
\E = \frac1T (\b \beta \b F + \b{\cal E})(\b \beta \b F + \b{\cal E})^*,
\end{equation}
becomes more accurate. This is a simple way of cleaning high-dimensional covariance matrices within factor models. 

However, this cleaning scheme leaves open at least one question of practical use. How should the number of factor $M$ be chosen? In the case where one has \textit{ a priori} information on the factors $F$, we are just left with the estimation of $\b \beta$ and $\b{\cal E}$. But in the general case, this question is still an open problem. Let us treat the general case, in which several authors considered tools from RMT to choose the number of factor $M$. 

In \cite{kapetanios2004new}, the author assumes that the empirical estimator of $\Sigma_{\e}$ is given by an isotropic Wishart matrix for which the upper bounds of the spectrum is exactly known. Hence, if there were no tangible factor in the data, one should observe that largest eigenvalues of the matrix $\E$ defined in \eqref{linear_model_empirical_covar} cannot exceed 
\begin{equation}
\lambda_{+}^{\text{eff}}(q) := (1+\sqrt{q})^2 + \delta(q,N)
\end{equation}
where the last term $\delta$ is a suitably defined constant as to reflect the width of the Tracy-Widom tail, i.e. $\delta(q,N) \sim N^{-2/3}$ \cite{kapetanios2004new}. If however one observes that the largest sample eigenvalue $\lambda_1$ exceeds $\lambda_{+}^{\text{eff}}$, then a true factor probably exists. In that case, the procedure suggested  in \cite{kapetanios2004new} is to extract the corresponding largest component from the data:
\begin{equation*}
Y_{it}^{(1)} = Y_{it} - \beta_{1t} f_{1t},
\end{equation*}
which is the residual from a regression of the data on the first principal component. Next, we compare the largest eigenvalue of $\Y^{(1)} \Y^{(1) \, *}/T$ against the new threshold $\lambda_{+}^{\text{eff}}(q'=q-1/T)$ and iterate the procedure until $\Y^{(M)}\Y^{(M)\,*}/T$ has all its eigenvalues within the Mar{\v c}enko-Pastur  sea. This approach has been generalized in \cite{onatski2010determining} to the case where the empirical estimator of the $\Sigma_{\e}$ is an \textit{anisotropic} Wishart matrix for which one has several results concerning the spectrum (see Chapter \ref{chap:spectrum}). The procedure is similar to the one above: the author proposed an algorithm to detect outliers for this anisotropic Wishart matrix using the results of Ref. \cite{paul2009no}. We refer to \cite{onatski2010determining} for more details. We can therefore see that RMT allows one to derive some rigorously based heuristics to determine the number of true factors $M$, which are quite
similar in spirit to the eigenvalue clipping method described above. 

It is also possible that one has some \textit{a priori} insight on the structure of the relevant factors. This for instance is a standard state of affairs in theoretical finance, where the so-called Capital Asset Pricing Model (CAPM) \cite{merton1973intertemporal} assumes a unique factor 
corresponds to the market portfolio, or its extension to three factors model by Fama-French \cite{fama1993common} (see \cite{tanskanen2016random} for further more recent extensions). In that case, one can simplify the problem to the estimation of the $\beta$ by assuming that the factors $f_k$ and the residuals $\e_i$ are linearly uncorrelated:
\begin{equation}
\langle f_k f_l \rangle = \delta_{kl} \quad,\quad \langle \e_i \e_j \rangle = \delta_{ij} \left( 1 - \sum_{l} \beta^2_{li} \right) \quad\text{and}\quad \langle f_k \e_l \rangle = 0,
\end{equation}
such that the true correlation becomes: 
\begin{equation*}
C_{ij} = \sum_{k=1}^{M} \beta_{ki} \beta_{kj} + \delta_{ij} \left(1 - \sum_{l=1}^{M} \beta^2_{li} \right)
\end{equation*}
that is to say
\begin{equation}
\label{eq:correl_factor_models}
C_{ij} = \begin{dcases}
1 \quad \text{if} \quad i = j \\
\left( \b\beta \b\beta^{*} \right)_{ij} \quad \text{otherwise}.
\end{dcases}
\end{equation}
Again, we emphasize that we are reduced to the estimation of only $N \times M$ parameters out of $N \times T$ points. We now give an insight on how one can estimate the coefficients of $\beta$ using the sample data, which is due to the recent paper \cite{chicheportiche2015nested}. Note that the eigenvalue clipping  \eqref{eq:correl_factor_models} can recovered by setting $\beta \equiv \beta_{PCA}$ where
\begin{equation}
\label{eq:beta_clipping}
\beta_{PCA} \;\deq\; \b U_{\lvert M} \b \Lambda^{1/2}_{\lvert M}\,,
\end{equation}
with $\b U$ the sample eigenvectors, $\Lambda$ the $N\times N$ diagonal matrix with the sample eigenvalues and the subscript $\mbox{}_{\lvert M}$ denotes that only the $M$ largest components are kept, where $M$ is such that $\lambda_i > (1+\sqrt{q})^2$ for any $i \le M$. The method of \cite{chicheportiche2015nested} suggests finding the $\b\beta$s such that:
\begin{equation}
\label{beta_chicheportiche}
\widehat{\b\beta} \;\deq\; \underset{\beta}{\argmin} \, {\cal L}\left( \bigg\| \frac1T \Y \Y^{*} - \b{\beta} \b{\beta}^* \bigg\|_{\text{off-diag}}\right),
\end{equation}
with ${\cal L}$ a given loss function and ``off-diag'' to denote the off-diagonal elements. (The diagonal elements are all equal to unity by construction). Numerically, the authors solve the latter equation in the vicinity of the PCA beta's \eqref{eq:beta_clipping} and with a quadratic norm ${\cal L}$. We refer the reader to \cite{chicheportiche2015nested} for more details on the procedure and its implementation, as well as an extension of the model to non-linear (volatility) dependencies.

\clearpage%!TEX root = RMT_Covariance_Review.tex 
\section{Numerical Implementation and Empirical results}
\label{chap:numerical}

This chapter aims at putting all the above ideas into practice in a financial context, the final goal being to achieve minimum out-of-sample, or forward looking risk. As we have seen above, the Rotationally Invariant Estimator
framework is promising in that respect. Still, as one tries to implement this method numerically, some problems arise. For example, we saw in Section \ref{sec:RIE_simulations} that the discrete version \eqref{eq:RIE_optimal_obs} of the optimal RIE \eqref{eq:RIE_optimal} deviates systematically from its limiting value for small eigenvalues. But as we discussed in Section \ref{chap:application}, the estimation of these small eigenvalues is particularly 
important since Markowitz optimal portfolios tend to overweight them and hence, inadequate estimators of these small eigenvalues may lead to disastrous results. We will therefore first discuss two different 
regularization schemes that appeared in the recent literature (see \cite{ledoit2016numerical} and \cite{bun2016cleaning}) that attempt to correct this systematic underestimation of the small eigenvalues. We will then turn to
numerical experiments on synthetic and real financial data and test the quality of the regularized RIE for real world applications.  

\subsection{Finite $N$ regularization of the optimal RIE \eqref{eq:RIE_optimal_obs}}
\label{sec:estimation}

\subsubsection{Why is there a problem for small-eigenvalues?}
\label{sec:rie_denoise}

\begin{figure}[h]
  \begin{center}
   \includegraphics[scale = 0.5]{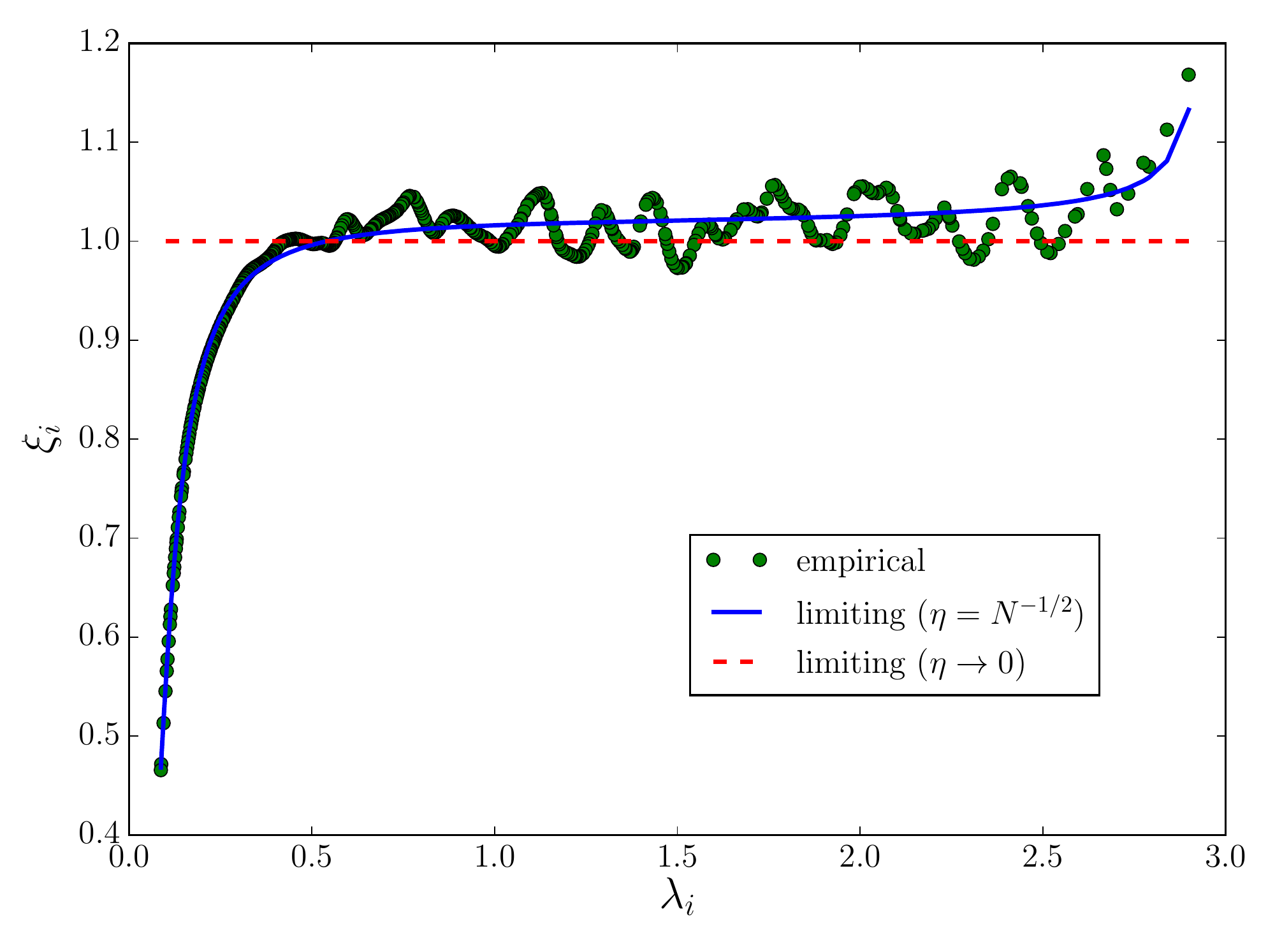} 
   \end{center}
   \caption{Evaluation of the empirical RIE \eqref{eq:RIE_optimal_obs} (green points) for $\C = \In$ with $N=500$. The matrix $\E$ is generated using Wishart matrix with parameter $q = 0.5$. We compare the result with its limiting value for $\eta = N^{-1/2}$ (blue line) and $\eta \to 0^{+}$ (red dotted line). }
   \label{fig:rie_emp_null}
\end{figure}

The small eigenvalue bias can be best illustrated using the null hypothesis on the sample covariance matrix. Indeed, we know that for $\C = \b I_N$, the optimal RIE \eqref{eq:RIE_optimal} should yield $\widehat\xi(\lambda_i) = 1$ exactly  as $N \to \infty$ (see Eq.\ \eqref{eq:RIE_nullH}). We therefore compare the observable shrinkage function $\widehat\xi^N$ \eqref{eq:RIE_optimal_obs} for finite $N$ with its limiting value $\widehat\xi=1$. The results are reported in Figure \ref{fig:rie_emp_null} where the observable estimator Eq.\ \eqref{eq:RIE_optimal_obs} appears as green points while the limiting value is given by the red dotted line. We see that the bulk and the right edge are relatively well estimated, but this is clearly not the case for the left edge, below which the estimated
eigenvalues dive towards zero instead of remaining close to unity. This highlights, as stated in \cite{bun2016optimal} or \cite{knowles2014anisotropic}, that the behavior for small eigenvalues is more difficult to handle compared to the rest of the spectrum. 

This underestimation can be investigated analytically. With $z = \lambda - \ii \eta$, we actually see from the Figure \ref{fig:rie_emp_null} that the discrete RIE $\widehat\xi^N$ is a very good approximation of the limiting quantity $\widehat\xi(z)$, i.e., with $\eta = N^{-1/2}$ (blue plain line). Hence, the deviation at the left edge is \emph{systematic} for any finite $N$ and only disappears as $N \to \infty$ ($\eta \to 0^+$). This finite size effect is due to the \emph{hard} left edge as eigenvalues are confined to stay on $\mathbb{R}^+$. Let us illustrate this: under the one-cut assumption, we can always decompose the Stieltjes transform as (see Eq.\ \eqref{eq:potential_stieltjes})
\begin{equation}
	\stj_\E(z) = \hil(z) + Q(z) \sqrt{d_+(z)}\sqrt{d_-(z)}, \qquad d_\pm(z) \;\deq\; z - \lambda_\pm
\end{equation}
where $\hil(z)$ is the Hilbert transform of $\rho_\E$ and $Q(z)$ is a given function that we assumed be smoothed on $\mathbb{C}^+$. We place ourselves in the situation where  $d_-(\lambda) = \e \ll \eta$, i.e. the eigenvalue $\lambda$ is very close to zero. Then, we have
\begin{eqnarray}
	\label{eq:Stieltjes_finite_N_left_edge}
	\stj_\E(z) & = &  \hil(z) + Q(z) \sqrt{- \ii\eta} \sqrt{d_+(\lambda) - \ii\eta} + \cal O(\e) \nonumber \\
	& = & \hil(z) - (1+\ii) Q(z)\sqrt{\frac{\eta \abs{d_+(\lambda)}}{2}} + \cal O(\e)\,.
\end{eqnarray}
Specializing this last equation to the null hypothesis $\C = \In$, one infers from Eq.\ \eqref{eq:stieltjes_isotropic_wishart} that $1/Q(z) = 2qz$ and $\hil(z) = Q(z)(z+q-1)$. Then plugging \eqref{eq:Stieltjes_finite_N_left_edge} into \eqref{eq:RIE_optimal} yields, at the left edge:
\begin{equation}
	\label{eq:RIE_finite_N}
	\widehat\xi(\lambda_- - \ii\eta) = 1 - \sqrt{\frac{2\eta\sqrt{q}}{(1-\sqrt{q})^2}} + \cal O(\eta),
\end{equation}
that is to say, there is a finite size ``correction'' to the asymptotic result $\widehat\xi(z)=1$ of order $N^{-1/4}$ when $\eta = N^{-1/2}$.  This correction is therefore quite significant if $N$ is not large enough. One tempting solution would be to decrease the value of $\eta$ to be arbitrarily small. However, we know that the empirical Stieltjes transform is only a good approximation of the limiting value up to an error of order $(T\eta)^{-1}$, so that $\eta$ cannot be too small either \cite{knowles2014anisotropic}. We conclude that the underestimation effect we observe in Figures \ref{fig:rie_emp_null} and \ref{fig:rie_examples} is purely due to a finite size effect and would furthermore occur for any model of $\rho_\C$ (see Fig.\ \ref{fig:rie_examples}). We emphasize that this effect is different from the phase transition affecting left outliers, as displayed in Fig.\ \ref{fig:rie_left_edge_pt}.

\subsubsection{Regularizing the empirical RIE \eqref{eq:RIE_optimal_obs}}
\label{sec:rie_denoise}

There are two ways to address this problem. The first one is to use a simple ad-hoc de-noising procedure that we shall now explain; the second is a more sophisticated scheme recently proposed by Ledoit and Wolf (see below). 

Firstly, using the fact that the finite size corrections are rather harmless for large eigenvalues (see Figure \ref{fig:rie_emp_null}), we can focus on small sample eigenvalues only. The idea is to use a regularization that would be exact if the true correlation matrix was of the Inverse-Wishart type, with $\rho_\C$ to be given by Eq. \eqref{eq:IMP_density}, for which we know that the associated optimal RIE is the linear shrinkage \eqref{eq:RIE_lin}.\footnote{A yet simpler solution, proposed in \cite{bun2016cleaning} is to consider a rescaled Mar{\v c}enko-Pastur's spectrum 
in such a way to fit the smallest eigenvalue $\lambda_N$. This is indistinguishable from the IW procedure when $\kappa$ is large enough, and provides very accurate predictions for US stocks return \cite{bun2016cleaning}. Nevertheless, in the presence of very small ``true'' eigenvalues, corresponding to of e.g.\ very strongly pairs of correlated financial contracts, this simple recipe fails.} Within this specification, the parameter $\kappa$ allows one to interpolate $\rho_\C$ between the infinitely wide measure on $\mathbb{R}^{+}$ ($\kappa \to 0^+$) and the null hypothesis ($\kappa \to \infty$). 

\begin{figure}[h]
  \begin{center}
   \includegraphics[scale = 0.4]{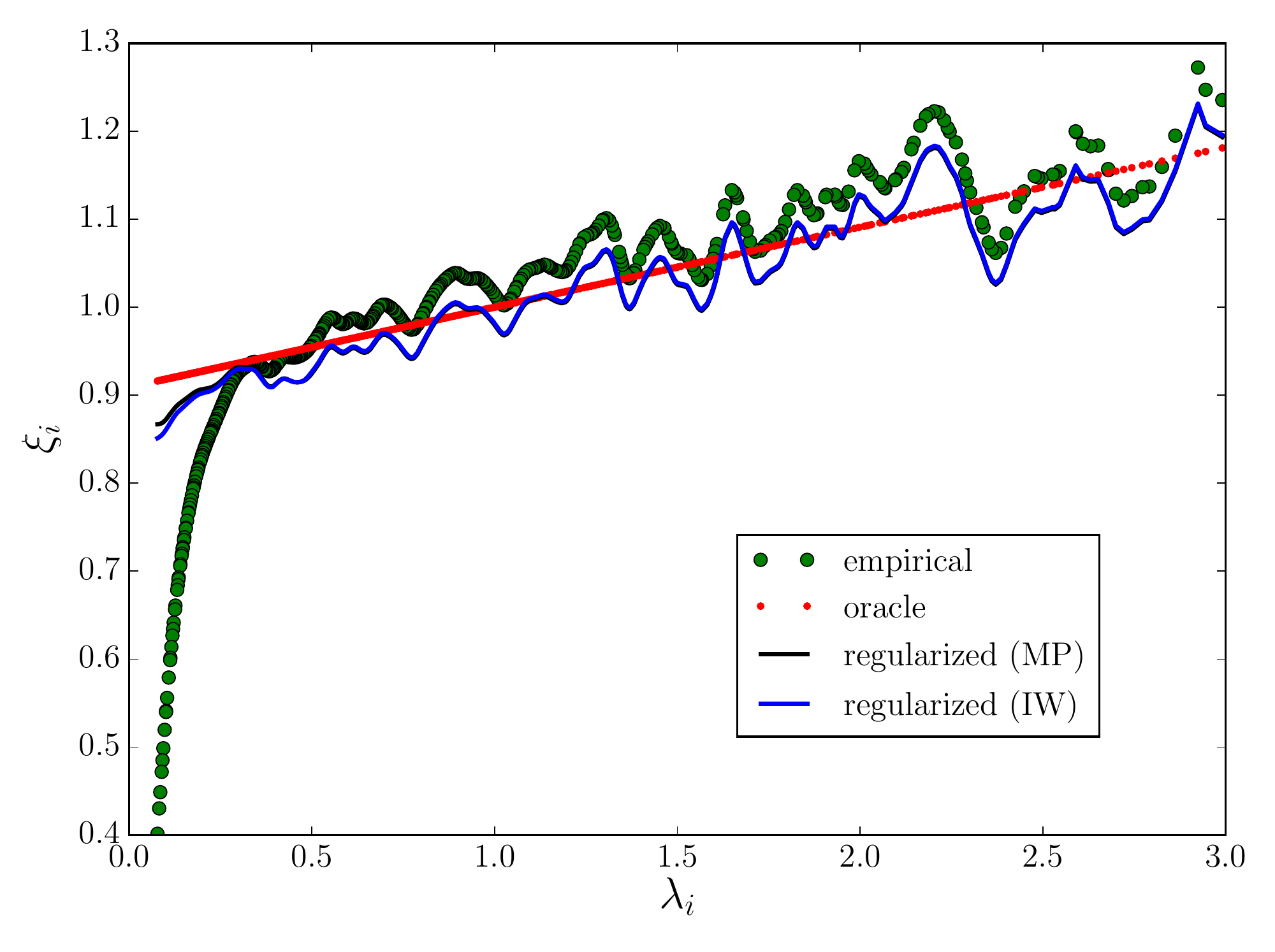} 
   \end{center}
   \caption{We apply the IW-regularization $\widehat\xi_i^{\text{reg}}$ with $z = \lambda - \ii N^{-1/2}$ in the case where $\C$ is an Inverse-Wishart matrix with $\kappa=10$ and $q=0.5$. The finite size effect of the empirical RIE \eqref{eq:RIE_optimal_obs} (green points) is efficiently corrected. The red points correspond to the Oracle estimator which is, in this case, the linear shrinkage procedure. We also compare the result of a ``rescaled''  Mar{\v c}enko-Pastur
   spectrum, as proposed in \cite{bun2016cleaning}.}
   \label{fig:rie_reg_iw}
\end{figure}

Our procedure, {\it for the only purpose of regularization}, is to calibrate $\kappa$ such that the lower edge $\lambda_{-}^{\text{iw}}$ of the corresponding empirical spectrum (and given in Eq.\ \eqref{eq:stieltjes_IW_MP}), coincides with the observed smallest eigenvalue $\lambda_N$. We then rescale the 
smallest eigenvalue using the exact factor that would be needed if $\C$ was indeed an Inverse-Wishart matrix, i.e.:
\begin{equation}
	\label{eq:rie_reg_invW}
	\widehat{\xi}_i^{\,\text{reg}} = \widehat\xi_i^N \times \max(1,\Gamma^{\text{iw}}_i), \qquad \Gamma^{\text{iw}}_i = \frac{\abs{1-q+qz_i \stj_{\E}^{\text{iw}}(z_i)}^2}{\lambda_i/(1 + \alpha_s(\lambda_i - 1))}, 
	\quad z_i = \lambda_i - \ii N^{-1/2},
\end{equation}
where $\alpha_s = 1/(1+2q\kappa)$ and $\stj_\E^{\text{iw}}$ is given in Eq.\ \eqref{eq:stieltjes_IW_MP}. We give a more precise implementation of this ``IW-regularization'' in the Algorithm \ref{algo:rie_iw}, and a numerical illustration for an Inverse Wishart matrix \eqref{eq:inverse_wishart_distribution} with parameter $\kappa = 10$ and $q=0.5$, for which $\alpha_s \approx 0.09$. The results are plotted in Figure \ref{fig:rie_reg_iw} where the empirical points come from a single simulation with $N=500$.

\begin{algorithm}
\caption{IW-regularization of the empirical RIE \eqref{eq:RIE_optimal_obs}}
\label{algo:rie_iw}
\begin{algorithmic}
\Function{g\_iw}{$z, q, \kappa$}:
	\State $\lambda_{\pm} \gets \qb{ (1+q) \kappa + 1 \pm \sqrt{(2\kappa +1)(2q\kappa+1)}\;} / \kappa$;
	\State \Return $\qb{ z(1+\kappa) - \kappa(1-q) - \sqrt{z-\lambda_+}\sqrt{z-\lambda_-}}/(z(z+2q\kappa))$; 
\EndFunction
\State
\Function{rie}{$z, q, g$}:
	\State \Return $\re[z]/\abs{1-q+qzg}^2$; 
\EndFunction
\State
\Function{denoising\_rie}{$N, q, \{\lambda_i\}_{i=1}^{N}$}: \qquad //$\lambda_1 \geq \lambda_2\geq\dots\geq \lambda_N$
	\State $\kappa \gets 2\lambda_N/\pb{ (1-q-\lambda_N)^2 - 4q\lambda_N }$;
	\State $\alpha \gets 1/(1+2q\kappa)$;
	\For {$i = 1 \text{ to } N$}
	\State $z \gets \lambda_i - \ii N^{-1/2}$;
	\State $g \gets \pb{\sum_{j\neq i}^{N} 1/(z - \lambda_j)}/(N-1)$;
	\State $\hat\xi_i \gets \Call{rie}{z,q,g}$;
	\State $g \gets \Call{g\_iw}{z,q,\kappa}$
	\State $\Gamma_i \gets (1+\alpha(\lambda_i - 1))/\Call{rie}{z,q,g}$;
	\If {$\Gamma_i > 1$ and $\lambda_i < 1$}
    	\State $\hat\xi_i \gets \Gamma_i \hat\xi_i$;
	\EndIf
	\EndFor
	\State $s \gets \sum_i \lambda_i /\sum_i \hat\xi_i; \qquad\qquad \text{//preserving the trace}$
	\State \Return $\{s\times\hat\xi_i\}_{i=1}^{N}$
\EndFunction
\end{algorithmic}
\end{algorithm}

\begin{figure}[!ht]
\begin{subfigure}{.5\textwidth}
  \centering
  \includegraphics[width=.9\linewidth]{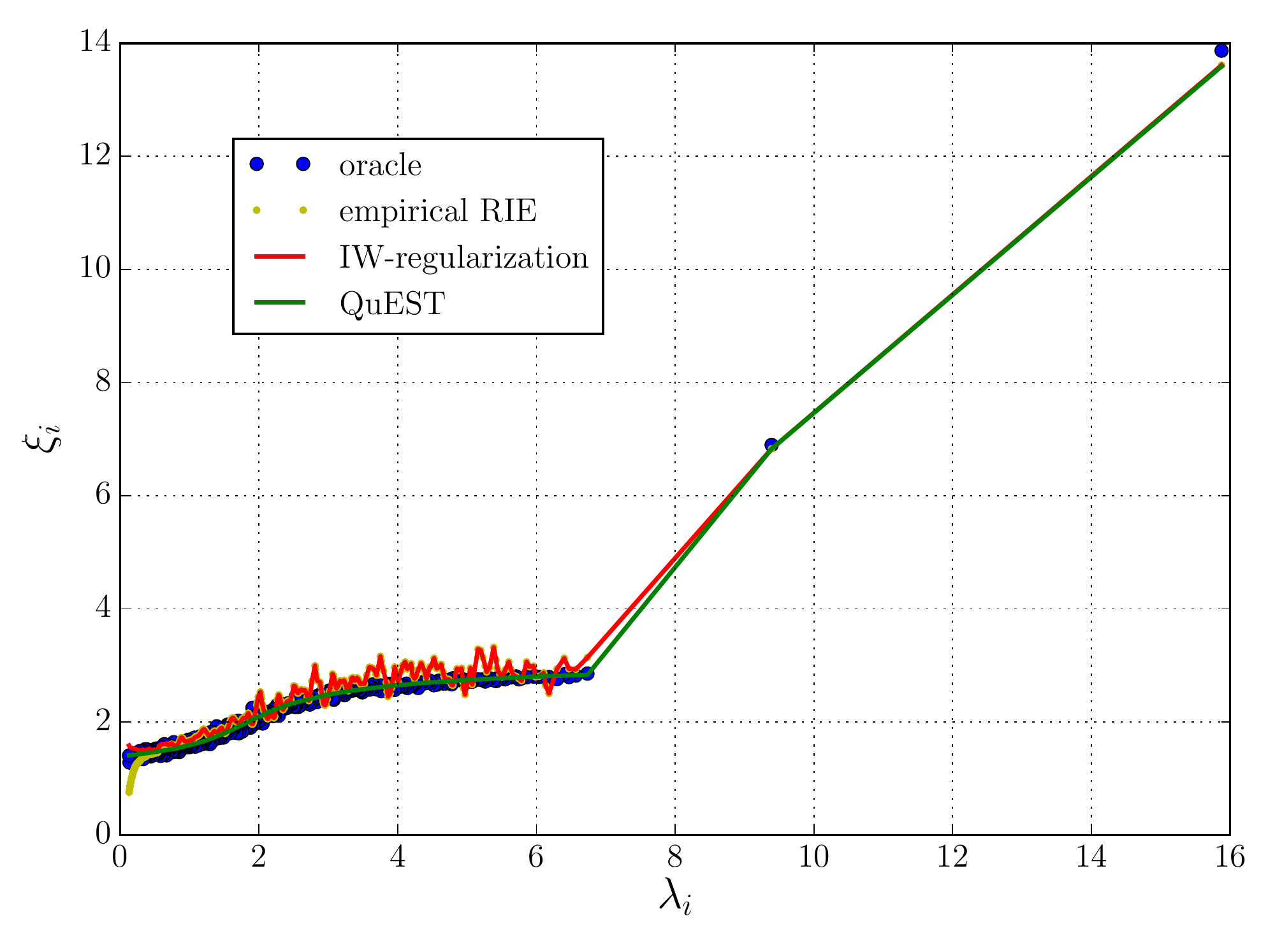}
  \caption{Multiple sources (case (i)).}
  \label{fig:multiple_debias}
\end{subfigure}%
\begin{subfigure}{.5\textwidth}
  \centering
  \includegraphics[width=.9\linewidth]{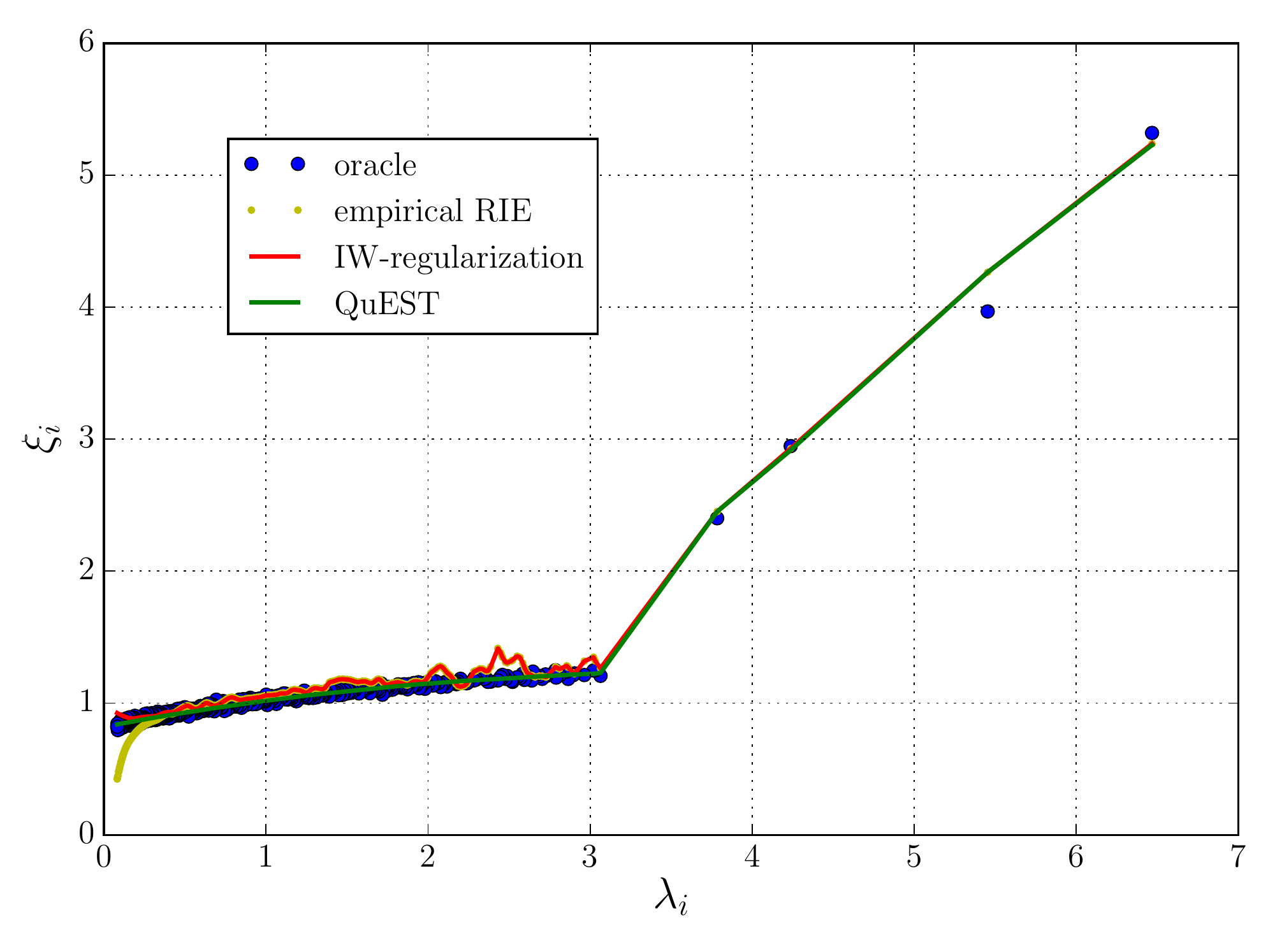}
  \caption{deformed GOE (case (ii)).}
  \label{fig:dGOE_debias}
\end{subfigure}\\
\begin{subfigure}{.5\textwidth}
  \centering
  \includegraphics[width=.9\linewidth]{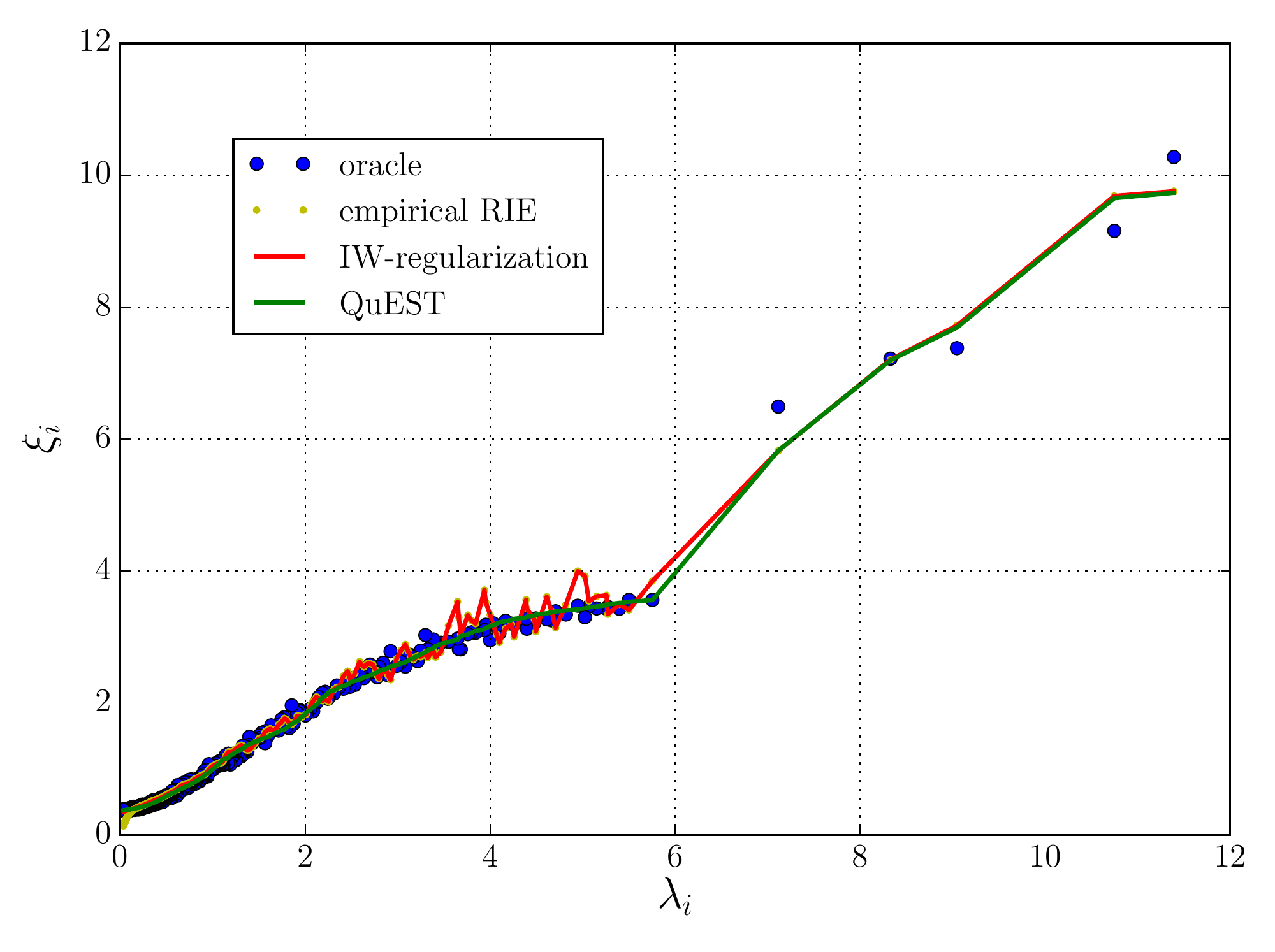}
  \caption{Toeplitz (case (iii))}
  \label{fig:toeplitz_debias}
\end{subfigure}%
\begin{subfigure}{.5\textwidth}
  \centering
  \includegraphics[width=.9\linewidth]{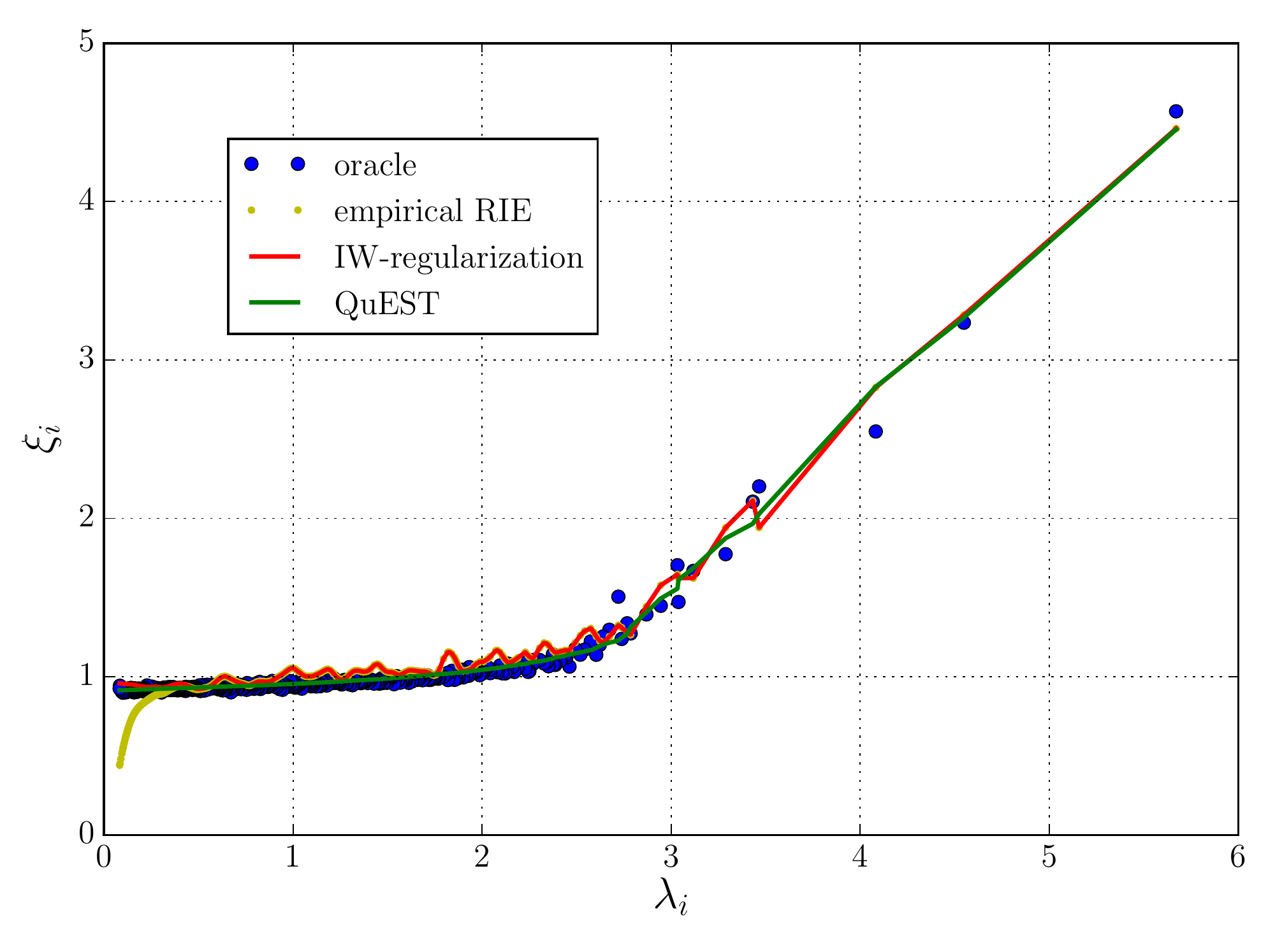}
  \caption{Power law (case (iv))}
  \label{fig:powerlaw_debias}
\end{subfigure}
\caption{Comparison of the IW-regularization \eqref{eq:RIE_optimal_obs} (red line) with the empirical RIE \eqref{eq:RIE_optimal_obs} (yellow dots) and the Oracle estimator \eqref{eq:oracle} (blue points) for the four cases presented at the beginning of Section \ref{sec:RIE_simulations} with $N = 500$ and $T = 1000$. We also plot the estimation we get using QuEST estimator \eqref{eq:QuEST_rie} (green line). The results a generated with a single realization of $\E$ using a multivariate Gaussian measurement process, and the four specifications of Section \ref{sec:RIE_simulations}.
%v The X-axis denotes the sample eigenvalues and the Y-axis the \emph{cleaned} eigenvalues.
}
\label{fig:rie_examples_debias}
\end{figure}

We now reconsider the numerical examples given in Section \ref{sec:RIE_simulations}, for which we apply the IW-regularization algorithm \eqref{algo:rie_iw}. The results are plotted in Figure \ref{fig:rie_examples_debias} and we observe that this IW-regularization works perfectly for all four population eigenvalues we consider in our simulations. Indeed, if we look at the left edge region, the regularized eigenvalues have been shifted upwards to coincide with the Oracle estimator (blue points) while one observes a significant discrepancy for the empirical, bare estimator (green dots). Hence, the IW-regularization (Algorithm \ref{algo:rie_iw}) provides a very simple way to correct this systematic downside bias which is of crucial importance whenever we need to invert the covariance matrix. Note that we can further improve the result by sorting the regularized eigenvalues. This is justified by the fact that we expect the RIE to be monotone with respect to the sample eigenvalues in the limit $N \to \infty$. We will investigate this point numerically in the next section (see Table \ref{table:regularization_accuracy}). 

%\begin{document}

\subsubsection{Quantized Eigenvalues Sampling Transform (QuEST)}
\label{sec:QuEST}

An alternative method, recently proposed by Ledoit and Wolf \cite{ledoit2016numerical} to approximate numerically the optimal RIE \eqref{eq:RIE_optimal}, is to work with the Mar\v{c}enko-Pastur equation \eqref{eq:MP_equation_stieltjes}. It is somewhat similar to the numerical scheme proposed by N. El Karoui (see Section \ref{sec:NEK}) to solve the indirect problem of the Mar\v{c}enko-Pastur equation. 

\begin{changemargin}{1.0cm}{1.0cm}
\footnotesize
The method, named as QuEST (Quantized Eigenvalues Sampling Transform), is based on a quantile representation of the eigenvalues. More formally, the key assumption is that the empirical eigenvalues are allocated smoothly according to the quantile of the spectral distribution, i.e.
\begin{equation}
	\label{eq:QuEST}
	\frac{i}{N} = \int_{-\infty}^{\lambda_i} \rho_\E(x)\dd x,
\end{equation}
and the aim is to find the quantile, as a function of the population eigenvalues $[\mu_i]_{i \in \qq{1,N}}$, such that \eqref{eq:QuEST} holds. Note that the representation \eqref{eq:QuEST} is the definition of the classical location of the \emph{bulk} eigenvalues, encountered in Eq.\ \eqref{eq:classical_location_bulk}. Hence, for $N \to \infty$, this method does not seem to be appropriate for outliers as we know that the spectral density $\rho_\E$ puts no weights on these outliers. Nevertheless, for constructing RIEs, this might not be that important since, roughly speaking, all we need to know is the Stieltjes transform of the spikeless covariance matrix $\ul\E$ (see Section \ref{sec:rie_outlier}). That being said, the ``quantized'' eigenvalues, expected to
be close to the empirical eigenvalues, are defined as
\begin{equation}
	\label{eq:QuEST_classical_location}
	\tilde\gamma_{i}(\b \mu) \;\deq\; N \int_{(i-1)/N}^{i/N} F_\E^{-1}(p) \dd p, \qquad i \in \qq{1,N}, \; p \in [0,1 ],
\end{equation}
where $\b\mu = (\mu_1,\dots, \mu_N)$, and
\begin{eqnarray}
	\label{eq:QuEST_quantile}
	F_\E^{-1}(p) & \deq & \sup \hB{x \in \mathbb{R}\,:\, F_\E(x) \leq p}, \nonumber\\
	F_\E(x) & \deq & \begin{cases}
		\max\pB{1-1/q, N^{-1} \sum_{i=1}^{N} \delta_0(\mu_i)} & \text{if } x = 0,\\
		\int_0^{x} \rho_\E(u) \dd u, & \text{otherwise},
	\end{cases}
\end{eqnarray}
with $\rho_\E(u) = \lim_{\eta\downarrow 0}\im \stj_\E^N(u - \ii\eta)$ and $\stj_\E^N$ is the unique solution in $\mathbb{C}^+$ of the discretized Mar{\v c}enko-Pastur equation \eqref{eq:MP_equation_int}
\begin{equation}
	\label{eq:QuEST_MP}
	\stj_\E^N(z) = \frac1N \sum_{i=1}^{N} \frac{1}{z - \mu_i(1-q+q z \stj_\E^N(z))}.
\end{equation}
Even if the numerical scheme seems quite intricate, all these quantities are simply a discretized version of the Mar{\v c}enko-Pastur equation. Indeed, Eq.\ \eqref{eq:QuEST} is equivalent to Eq.\ \eqref{eq:MP_equation_int} for large $N$ and \eqref{eq:QuEST_classical_location} is nothing but a discrete estimator of Eq.\ \eqref{eq:classical_location_bulk}. 

Finally, the optimization program reads
\begin{equation}
	\label{eq:QuEST_optim}
	\b{\tilde\mu} \;\deq\; \begin{cases}
	\argmin_{\b \mu \in \mathbb{R}_{+}^{N}} \sum_{i=1}^{N} \qB{ \tilde\gamma_i(\b\mu) - \lambda_i}^2, \\
	\text{s.t. } \; \tilde\gamma_i(\b\mu) \; \text{ satisfies Eqs. \eqref{eq:QuEST_classical_location}, \eqref{eq:QuEST_quantile} and \eqref{eq:QuEST_MP}}.
	\end{cases}
\end{equation}
From there, the regularization scheme of the empirical RIE \eqref{eq:RIE_optimal_obs} reads
\begin{equation}
	\label{eq:QuEST_rie}
	\xi_i^{\,\text{QuEST}} = \frac{\lambda_i}{\abs{1-q+q\lambda_i \lim_{\eta\downarrow 0} \tilde\stj_\E^N(\lambda_i - \ii\eta)}^2},
\end{equation}
where $\tilde\stj_\E^N(z) \in \mathbb{C}^+$ is the unique solution of 
\begin{equation}
	 \tilde\stj_\E^N(z) = \frac1N \sum_{i=1}^{N} \frac{1}{z - \tilde\mu_i(1-q+q z \tilde \stj_\E^N(z))}.
\end{equation}

\end{changemargin}
\normalsize

We see that the above regularization scheme allows one  to estimate -- in principle -- the limiting RIE \eqref{eq:RIE_optimal} since we can now set $\eta$ to be arbitrarily small. This means that, contrary to the empirical estimate \eqref{eq:RIE_optimal_obs}, the QuEST procedure should not suffer from a systematic underestimation at the left edge. The main advantage of this method is that it also allows us to estimate the population eigenvalues, which can be useful in some particular cases. However, from a numerical standpoint, this algorithm is far more complicated to implement than the above IW-regularization (Algorithm \eqref{algo:rie_iw}). Indeed, we see that the starting point of the optimization \eqref{eq:QuEST_optim} is the vector of population eigenvalues, which can be problematic for very ``diluted'' spectrum. Moreover, the algorithm might suffer from instabilities in the presence of very large and isolated eigenvalues. Note that a detailed presentation of the implementation of QuEST is given in \cite{ledoit2016numerical}, where the authors advise to sort the cleaned eigenvalues $[\xi_i^{\text{QuEST}}]_{i \in \qq{1,N}}$ since, as said above, we expect the optimal cleaned eigenvalues to be monotonic with respect to the sample eigenvalues. 

\subsubsection{Empirical studies}

We compare in Figure \ref{fig:rie_examples_debias} the above QuEST numerical scheme with the simple IW-regularization of Section \ref{sec:rie_denoise}. The eigenvalues coming from the QuEST regularization are shown as green lines and we see that the results are very satisfactory. In particular, it indeed does not suffer from the systematic bias in the left edge and seems to handle efficiently outliers even if the formula \eqref{eq:QuEST} is a priori not valid for isolated eigenvalues in the large $N$ limit. We nonetheless notice that the algorithm suffers sometimes from instabilities in the presence of ``clustered'' outliers as in the power law example (see Figure \ref{fig:powerlaw_debias}). On the other hand, and perhaps surprisingly, the much simpler, somewhat ad-hoc IW-regularization given in Algorithm \ref{algo:rie_iw} provides very similar results. However, the QuEST method requires solving a nonlinear and non-convex optimization problem (see Eq.\ \eqref{eq:QuEST_optim}) which implies heavy numerical computations that may not even converge to the global minimum (when it exists). 

% \begin{figure}[h]
% \begin{subfigure}{.5\textwidth}
%   \centering
%   \includegraphics[width=.9\linewidth]{Figures/numerical/rie_final_multiple}
%   \caption{Multiple sources (case (i)).}
%   \label{fig:multiple_final}
% \end{subfigure}%
% \begin{subfigure}{.5\textwidth}
%   \centering
%   \includegraphics[width=.9\linewidth]{Figures/numerical/rie_final_sw}
%   \caption{deformed GOE (case (ii)).}
%   \label{fig:dGOE_debias}
% \end{subfigure}\\
% \begin{subfigure}{.5\textwidth}
%   \centering
%   \includegraphics[width=.9\linewidth]{Figures/numerical/rie_final_toeplitz}
%   \caption{Toeplitz (case (iii))}
%   \label{fig:toeplitz_final}
% \end{subfigure}%
% \begin{subfigure}{.5\textwidth}
%   \centering
%   \includegraphics[width=.9\linewidth]{Figures/numerical/rie_final_power}
%   \caption{Power law (case (iv))}
%   \label{fig:powerlaw_final}
% \end{subfigure}
% \caption{Comparison of the IW-regulation \eqref{eq:RIE_optimal_obs} (red line) with the Oracle estimator \eqref{eq:Oracle} (blue points) for the four cases presented at the beginning of Section \ref{sec:RIE_simulations} with $N = 500$ and $T = 1000$. The results come from a single realization of $\E$ using a multivariate Gaussian measurement process.\cob **To be merged with the previous graph and suppressed the legend is wrong anyway**\nc
% %v The X-axis denotes the sample eigenvalues and the Y-axis the \emph{cleaned} eigenvalues.
% }
% \label{fig:rie_examples_final}
% \end{figure}

We want to further investigate the efficiency of these two regularizations. One direction is to change the number of variables $N$ with $q=0.5$ fixed. This allows us to assess the finite size performance of the two algorithms. The second direction is to fix $N = 500$ and vary the observation ratio $q$.  We shall consider three different regularizations in the following: (i) IW-regularization (Algorithm \ref{algo:rie_iw}), (ii) IW-regularization + sorting (name ``IWs regularization'' in the following) and (iii) QuEST procedure. Note that we will focus our study on the power law example of Figure \ref{fig:powerlaw_debias} since this simple prior allows use to generate very complex spectrum with possibly ``clustered'' outliers, similar to financial data. We emphasize again the regularization scheme (ii) is justified by the fact that we expect the estimator to preserve the monotonicity of the sample eigenvalues.

To measure the accuracy and the stability of each algorithm, we characterize the deviation between a given estimator and the Oracle \eqref{eq:oracle}. Using the mean squared error (MSE), we may also analyze the relative performance (RP) in percentage compared to the sample covariance. This is given by 
\begin{equation}
	\label{eq:rpi}
	\text{RP}(\Xi) \;\deq\; 100\times\pBB{1- \frac{\mathbb{E}\norm{ \Xi -\Xi^{\text{ora.}}}_2}{\mathbb{E}\norm{ \E - \Xi^{\text{ora.}}}_2}},
\end{equation}
where $\Xi \equiv \Xi(\E)$ is a RIE of $\C$ and $\Xi^{\text{ora.}}$ is the Oracle estimator. We also report in each case the average computational time needed to perform the estimation\footnote{Simulations were implemented in Python and based on an $\text{Intel}^{\textregistered}$ Core\texttrademark$\,$ i7-4700HQ and CPU of 8 $\times$ 2.40 GHz processor.}. 

First, let us assess the usefulness of sorting the cleaned eigenvalues. We report in Table \ref{table:regularization_accuracy} the performance we obtained for $N=500$ and $q=0.5$ fixed over 100 realizations of $\E$ (which is a Wishart matrix with population covariance matrix $\C$). We conclude from Table \ref{table:regularization_accuracy} that it is indeed better to sort the eigenvalues when using the IW-regularization \eqref{eq:rie_reg_invW} as the difference is statistically significant, while being nearly equally efficient in terms of computational time. For large $N$, the QuEST procedure yields the best accuracy score but the difference with the IWs eigenvalues is not statistically significant and the QuEST requires much more numerical operations than the ad-hoc IWs algorithm. Note that the performance improvement over to the sample covariance matrix is very substantial. 

\begin{table}[h]
\caption{We reconsider the setting of Figure \ref{fig:powerlaw_debias} and check the consistency over $100$ samples. The population density $\rho_\C$ is drawn from \eqref{eq:power law proxy} with $\lambda_0 = -0.6$ and $N = 500$ and the sample covariance matrix is obtained from the Wishart distribution. MSE stands for the mean squared error with respect to the Oracle estimator \eqref{eq:oracle}, stdev stands for the standard deviation of the squared error and the RP defined in Eq.\ \eqref{eq:rpi}. Running time shows the average time elapsed for the cleaning of one sample set of eigenvalues of size $N$. }
\label{table:regularization_accuracy}
%{\bf should we use $\sqrt{R^2}$, units would be easier}
\centering

\begin{tabular}{|*{5}{c|}}
    \hline
    Method  & MSE & stdev & RP  & Running time (sec)  \\
    \hline
    IW-regularization  & 0.64 & 0.13  & 99.69  & 0.02   \\
    \hline
    IWs-regularization  & 0.45 & 0.12  & 99.78  & 0.03  \\
    \hline
    QuEST  & 0.44 & 0.15  & 99.79  & 33.5 \\
    \hline
\end{tabular}
%\bigskip
%\subcaption*{}

\end{table}

We now investigate how these conclusions change when $N$ varies with $q = 0.5$ fixed. The results are given in Table \ref{table:regularization_N}. First, we stress that the RP with respect to the sample covariance matrix is already greater than $98\%$ for $N=100$ which is why we did not report these values in the table. As above, for $N \geq 100$, sorting the eigenvalues improves significantly the mean squared error with respect to the Oracle estimator. We also emphasize that for $N = 1000$, it takes $0.06$ seconds to get the regularized RIE while the QuEST algorithm requires $80$ seconds on average. We see that as the size $N$ grows to infinity, the high degree of complexity needed to solve the nonlinear and non-convex optimization \eqref{eq:QuEST_optim} becomes very restrictive, while improvement over the simple IWs method is no longer significant.

\begin{table}[!ht]
\caption{Check of the consistency of the three regularizations with respect to the dimension $N$. The population density $\rho_\C$ is drawn from \eqref{eq:power law proxy} with $\lambda_0 = -0.6$ and the sample covariance matrix is obtained from the Wishart distribution with $T=2N$. We report in the table the mean squared error with respect to the Oracle estimator \eqref{eq:oracle} and the standard deviation in parenthesis as a function of $N$. }
\label{table:regularization_N}
%{\bf should we use $\sqrt{R^2}$, units would be easier}
\centering

\begin{tabular}{|*{7}{c|}}
    \hline
    Method  & $N=100$ & $N=200$ & $N=300$  & $N=400$ & $N=500$  & $N=1000$ \\
    \hline
    IW-regularization  & 0.53 (0.17) & 0.56 (0.15)  & 0.64 (0.16)  & 0.65 (0.14) & 0.64 (0.14) & 0.74 (0.14)   \\
    \hline
    IWs-regularization  & 0.35 (0.14) & 0.39 (0.14)  & 0.45 (0.14)  & 0.45 (0.13) & 0.46 (0.12) & 0.53 (0.12) \\
    \hline
    QuEST  & 0.26 (0.16) & 0.33 (0.15)  & 0.39 (0.15)  & 0.4 (0.15) & 0.44 (0.15) & 0.5 (0.13) \\
    \hline
\end{tabular}
\end{table}

We now look at the second test in which $N = 500$ is fixed and we vary $q = 0.25, 0.5, 0.75, 0.95$. For each $q$, we perform the same procedure as in Table \ref{table:regularization_N} and the results are reported in Table \ref{table:regularization_q}. It is easy to see that the conclusions of the first consistency test are still valid for the three regularization schemes as a function of $q$ with $N= 500$. Note that we do not consider here the case $q \geq 1$ which is less immediate since $\E$ generically possess $(N-T)$ zero eigenvalues. Both regularization schemes, IWs-regularization and QuEST algorithm, fail to handle this case and we shall come back to this problem in Chapter \ref{chap:conclusion}. 

\begin{table}[!ht]
\caption{Check of the consistency of the three regularizations with respect to the dimension ratio $q$. The population density $\rho_\C$ is drawn from \eqref{eq:power law proxy} with $\lambda_0 = -0.6$ and $N = 500$ and the sample covariance matrix is obtained from the Wishart distribution with parameter $T=N/q$. We report in the table the mean squared error with respect to the Oracle estimator \eqref{eq:oracle} and the standard deviation in parenthesis as a function of $q$. }
\label{table:regularization_q}
%{\bf should we use $\sqrt{R^2}$, units would be easier}
\centering

\begin{tabular}{|*{5}{c|}}
    \hline
    Method  & $q=0.25$ & $q=0.5$ & $q=0.75$  & $q = 0.95$ \\
    \hline
    IW-regularization & 0.31 (0.06) & 0.65 (0.14) &   1.2 (0.18) & 1.78 (0.44) \\
    \hline
    IWs-regularization  & 0.28 (0.05) & 0.46 (0.12) & 0.71 (0.17) & 0.94 (0.39) \\
    \hline
    QuEST  & 0.25 (0.05) & 0.45 (0.15)  & 0.72 (0.17) & 0.98 (0.35) \\
    \hline
\end{tabular}
\bigskip
%\subcaption*{}

\end{table}

To conclude, we observed using synthetic data that we are able to estimate accurately the Oracle estimator for finite $N$ both for small eigenvalues and outliers. The QuEST procedure is found to behave efficiently for any $N$ and any $q < 1$, and allows one to estimate both the population eigenvalues and the limiting Stieltjes transform with high precision. However, as far as the estimation of large sample covariance matrices is concerned, the improvement obtained by solving the nonlinear and non-convex optimization problem \eqref{eq:QuEST_optim} becomes insignificant as $N$ increases (see Tables \ref{table:regularization_N} and \ref{table:regularization_q}). Furthermore, the computational time of the QuEST algorithm increases considerably as $N$ grows. We shall henceforth use the IWs RIE as our estimator of $\C$ for the applications below. Nonetheless, whenever $N$ is not very large, the QuEST procedure is clearly advised as it yields a significant improvement with an acceptable computational time.

\subsection{Optimal RIE and out-of-sample risk for optimized portfolios}

% \subsubsection{Setting the stage} 
% \label{sec:setting}

As alluded above (see Section \ref{sec:markowitz}), the concept of correlations between different assets is the cornerstone of Markowitz' optimal portfolio theory, and more generally for risk management purposes \cite{markowitz1968portfolio}. It is therefore of crucial importance to use a correlation matrix that faithfully represents {\it future} risks, and not past risks -- otherwise the over-allocation on spurious low risk
combination of assets might prove disastrous. In that respect, we saw in Section \ref{sec:markowitz_opt_risk} that the best estimator inside the space of estimators restricted to possess the sample eigenvectors is precisely the Oracle estimator \eqref{eq:oracle} which is not observable a priori. However, if the number of variables is sufficiently large, we know -- thanks to the numerical study of the previous section -- that it is possible to estimate very accurately the Oracle estimator using only observable variables. The main objective in the present section is to investigate the IWs RIE procedure for financial stock market data. 

Let us now explain the construction of our test. We consider a universe made of $N$ different financial assets -- say stocks -- that we observe at -- say -- the daily frequency, defining a vector of returns $\b r_t = (r_{1t}, r_{2t}, \dots, r_{Nt})$ for each day $t=1,\dots,T$. 
It is well known that volatilities of financial assets are heteroskedastic \cite{bouchaud2003theory} and we therefore focus specifically on {\it correlations} and not on volatilities in order to study the systemic risk. To that end, we standardize these returns as follows: (i) we remove the sample mean of each asset; (ii) we normalize each return by an estimate $\widehat \sigma_{it}$ of its daily volatility: $\widetilde r_{it} = r_{it}/\widehat \sigma_{it}$. There are many possible choices for $\widehat \sigma_{it}$, based e.g. on GARCH or FIGARCH models of historical returns, or simply implied volatilities from option markets, and the reader can choose his/her favorite estimator which can easily be combined with the correlation matrix cleaning schemes discussed below. For simplicity, we have chosen here the 
cross-sectional daily volatility, that is 
\begin{equation}
	\label{eq:cross_sec_vol}
	\widehat \sigma_{it}\;\deq\;\sqrt{\sum_{j} r^2_{jt}}\,,
\end{equation}
to remove a substantial amount of non-stationarity in the volatilities. The final standardized return matrix $\b Y = (Y_{it}) \in \R^{N\times T}$ is then given by $Y_{it} \deq \widetilde r_{it}/\sigma_i$ where $\sigma_i$ is the sample estimator of the $\widetilde r_i$ which is now, to a first approximation, stationary.

We may now compute the sample covariance matrix $\E$ as in Eq.\ \eqref{eq:SCM}. We stress that the Mar{\v c}enko and Pastur result does not require multivariate normality of the returns, which can have fat-tailed distributions. In fact, the above normalization by the cross-sectional volatility can be seen as a proxy for a robust estimator of the covariance matrix \eqref{eq:Maronna_SCM} with $U(x) = x^{-1}$ which can be studied using the tools of Chapters \ref{chap:spectrum} and \ref{chap:eigenvectors} (see Section \ref{sec:SCM_entries} for a discussion on this point). All in all, we are able to construct the optimal RIE either using IWs-regularization (Algorithm \ref{algo:rie_iw} + sorting) or the QuEST regularization, the latter allowing us to estimate the population eigenvalue spectrum as well.

\begin{figure}[!ht]
	\begin{center}
   \includegraphics[scale = 0.45]{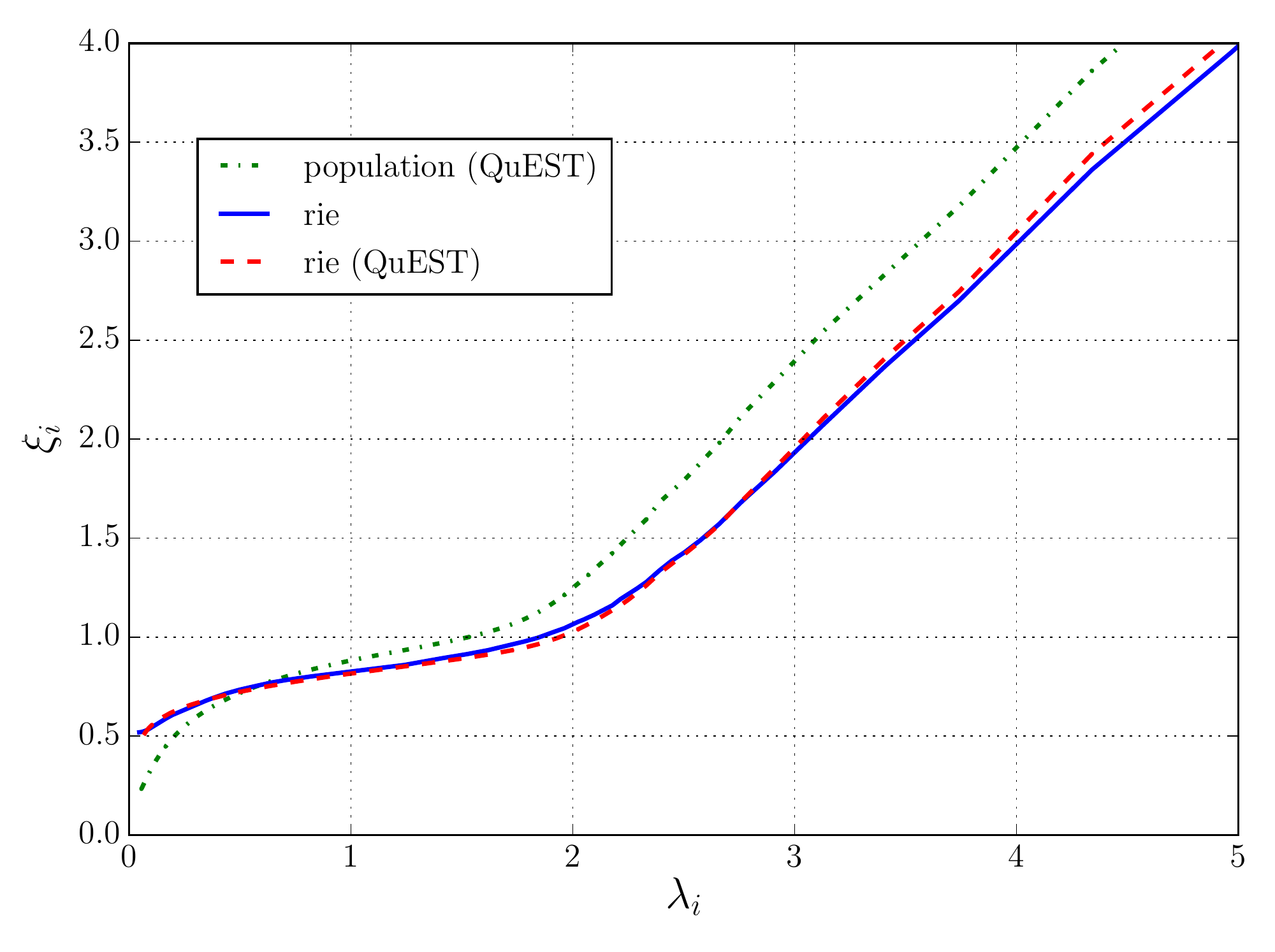} 
   \end{center}
   \caption{Comparison of the IWs-regularization \eqref{eq:rie_reg_invW} (blue) with the QuEST procedure \eqref{eq:QuEST_rie} (red dashed line) using 500 US stocks from 1970 to 2012. The agreement between those two regularizations is quite remarkable. We also provide the estimation of the population eigenvalues obtained from \eqref{eq:QuEST_optim} (green dashed-dotted line). }
   \label{fig:mkw_rie_SPX}
\end{figure}

For our simulations, we consider an international pools of stocks with daily data:
\begin{enumerate}
%{\bf check numbers: 500 or 450 out of 500?}
	\item US: $500$ most liquid stocks during the training period of the S\&P 500 from 1966 until 2012;
	\item Japan: $500$ most liquid stocks during the training period of the all-shares TOPIX index from 1993 until 2016;
	\item Europe: $500$ most liquid stocks during the training period of the Bloomberg European 500 index from 1996 until 2016.
\end{enumerate}
We chose $T = 1000$ (4 years) for the training period, i.e.\ $q = 0.5$, and $T_{\text{out}} = 60$ (three months) for the out-of-sample test period. Let us first analyze the optimal RIE for US stocks. We plot in Figure \ref{fig:mkw_rie_SPX} the average nonlinear shrinkage curve for the IWs-regularization (blue line) and for the QuEST regularization (red dashed line) -- where we sorted the eigenvalues in both cases -- and compare it with the estimated population eigenvalues obtained from \eqref{eq:QuEST_optim}. We see that IWs-regularization and QuEST still yield very similar results. Furthermore, we notice that the spectrum of the cleaned eigenvalues is, as expected, narrower than the spectrum of the (estimated) population matrix.

\begin{figure}[!ht]
	\begin{center}
   \includegraphics[scale = 0.45]{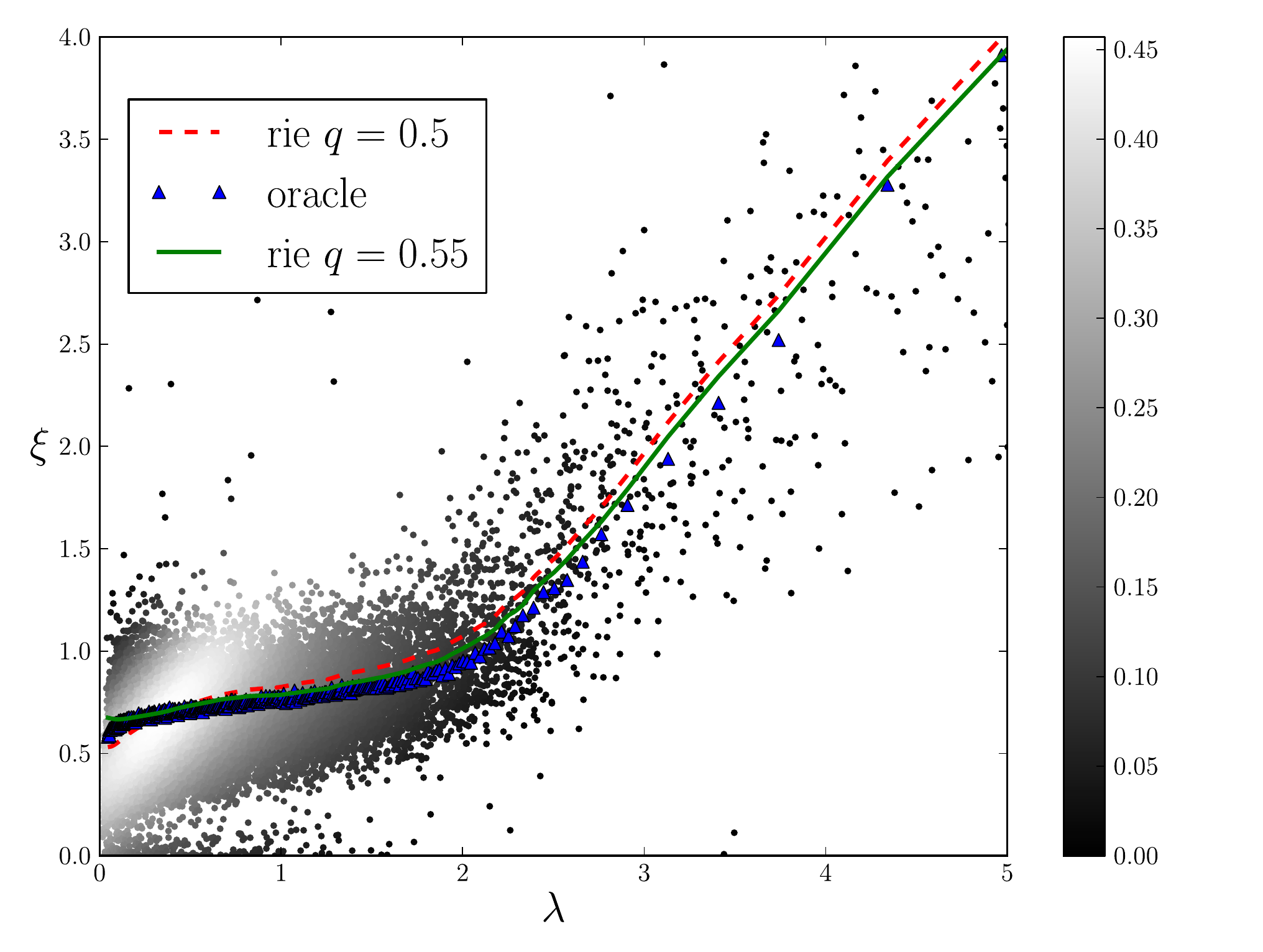} 
   \end{center}
   \caption{Comparison of the IWs-regularized RIE \eqref{eq:rie_reg_invW} with the proxy \eqref{eq:approxoracle} using 500 US stocks from 1970 to 2012. The points represent the density map of each realization of \eqref{eq:approxoracle} and the color code indicates the density of data points. The average IWs-regularized RIE is plotted with the red dashed line and the average realized risk in blue. We also provide the prediction of the IWs-regularized RIE with an effective observation ratio $q_{\text{eff}}$ which is slightly bigger than $q$ (green plain line). The agreement between the green line and the average Oracle estimator (blue triangle) is quite remarkable.}
   \label{fig:scatter_RIE_SPX_IDX}
\end{figure}

Interestingly, the Oracle estimator \eqref{eq:oracle} can be estimated empirically and used to directly test the accuracy of the IWs-regularized RIE \eqref{eq:rie_reg_invW}. The trick is to remark that the Oracle eigenvalues \eqref{eq:oracle} can be interpreted as the ``true'' (out-of-sample) risk associated to a portfolio whose weights are given by the $i$-th eigenvector. 
Hence, assuming that the data generating process is stationary, we estimate the Oracle estimator through the realized risk associated to such eigen-portfolios \cite{pafka2003noisy}. More
precisely, we split the total length of our time series $T_{\text{tot}}$ into $n$ consecutive, non-overlapping samples of length $T_{\text{out}}$. The ``training'' period has length $T$, so
$n$ is given by:
\begin{equation}
	n \deq \lfloor \frac{T_{\text{tot}} - T - 1}{T_{\text{out}}} \rfloor.
\end{equation}
The Oracle estimator \eqref{eq:oracle} is then computed as:
\begin{equation}
	\label{eq:approxoracle}
	\hat\xi_i^{\text{ora.}} \approx \frac{1}{n} \sum_{j=0}^{n-1} {\cal R_{\text{out}}^2}(t_j,\b u_i) \quad i=1,\dots,N,
\end{equation}
for $t_j = T + j \times T_{\text{out}} + 1$ and ${\cal R}(t,\b w)$ denotes the out-of-sample variance of the returns of portfolio $\b w$ built at time $t$, that is to say 
\begin{equation}
	\label{eq:risk_os}
	\cal R_{\text{out}}^2(t,\b w) \deq \frac{1}{T_\text{out}} \sum_{\tau = t + 1}^{t + T_{\text{out}}} \left( \sum_{i=1}^{N} \b w_i Y_{i\tau} \right)^2,
\end{equation}
where $Y_{i\tau}$ denotes the rescaled realized returns. Again, as we are primarily interested in estimating correlations and 
not volatilities, both our in-sample and out-of-sample returns are made approximately stationary and normalized. This implies that $\sum_{i=1}^{N} \cal R_{\text{out}}^2(t, \b u_i) = N$ for any time $t$. We plot our results for the estimated Oracle estimator \eqref{eq:approxoracle} using US data in Fig \ref{fig:scatter_RIE_SPX_IDX}, which we compare with the IWs-regularized RIE. The results are, we believe, quite remarkable:  the RIE formula \eqref{eq:rie_reg_invW} (red dashed line) tracks very closely the average realized risk (blue triangles), especially in the region where there is a lot of eigenvalues.

\begin{figure}[!ht]
\begin{subfigure}{.5\textwidth}
  \centering
  \includegraphics[width=.9\linewidth]{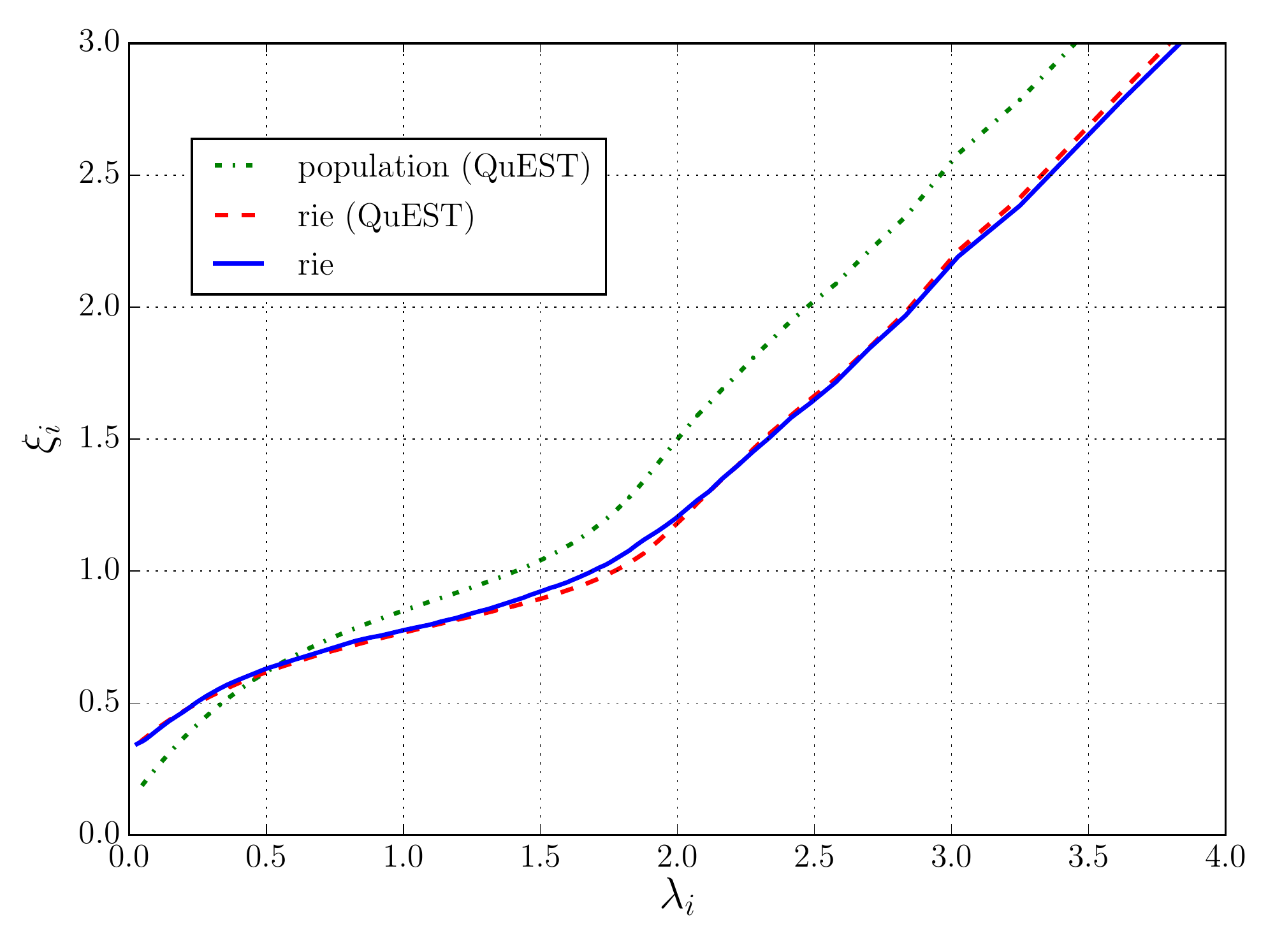}
  \caption{Population and optimal RIE bulk eigenvalues.}
  \label{fig:mkw_rie_TPX}
\end{subfigure}%
\begin{subfigure}{.5\textwidth}
  \centering
  \includegraphics[width=.9\linewidth]{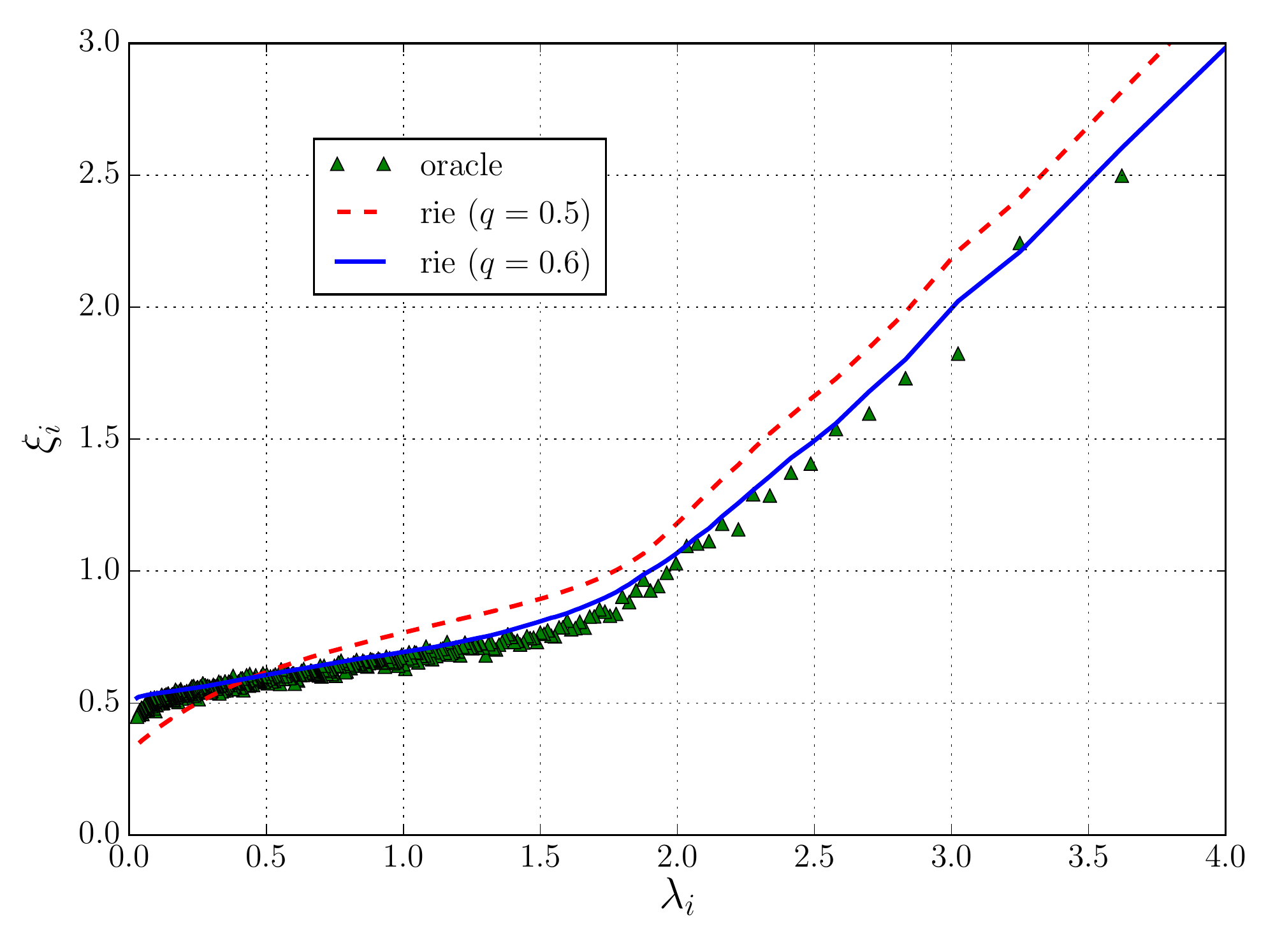}
  \caption{Comparison with Oracle estimator \eqref{eq:approxoracle}.}
  \label{fig:mkw_Oracle_TPX}
\end{subfigure}\\
\caption{Left figure: analysis of the population (green dashed line) and optimal RIE bulk eigenvalues (red dashed line for Eq.\ \eqref{eq:QuEST_rie} and blue plain line for the IWs-regularization) using the $500$ most liquid stocks during the training period of the all-shares TOPIX index from 1993 until 2016. Right figure: Comparison between the IWs-regularized RIE (red dashed line) with the Oracle estimator \eqref{eq:approxoracle} (green triangle). We also provide the plot of the IWs-regularized RIE with an effective observation ratio (blue line).
%v The X-axis denotes the sample eigenvalues and the Y-axis the \emph{cleaned} eigenvalues.
}
\label{fig:mkw_TPX}
\end{figure}

We may now repeat the analysis for the other pools of stocks as well. We begin with the TOPIX where we plot in Figure \ref{fig:mkw_rie_TPX} the estimation of the population eigenvalues (using Eq.\ \eqref{eq:QuEST_optim}) and the regularized RIE (using Algorithm \ref{algo:rie_iw} or Eq.\ \eqref{eq:QuEST_rie}). Again, the results we get from the simple IWs-regularization and QuEST procedure are nearly indistinguishable.  This is another manifestation of the robustness of both algorithms at a finite $N$. We then plot in Figure \ref{fig:mkw_Oracle_TPX} the comparison between the IWs-regularized RIE (red dashed line) and the Oracle estimator, approximated by \eqref{eq:approxoracle} (green triangles). We observe that the overall estimation is not as convincing as for US stocks (Figure \ref{fig:scatter_RIE_SPX_IDX}) but as above, the deviation can be explained by the presence of weak autocorrelations in the return time series (more on this below). Indeed, there  exists an effective ratio $q_{\text{eff}} = 1.2 q$ such that the estimation is extremely good (see blue line in Figure \ref{fig:mkw_Oracle_TPX}). 

\begin{figure}[!ht]
\begin{subfigure}{.5\textwidth}
  \centering
  \includegraphics[width=.9\linewidth]{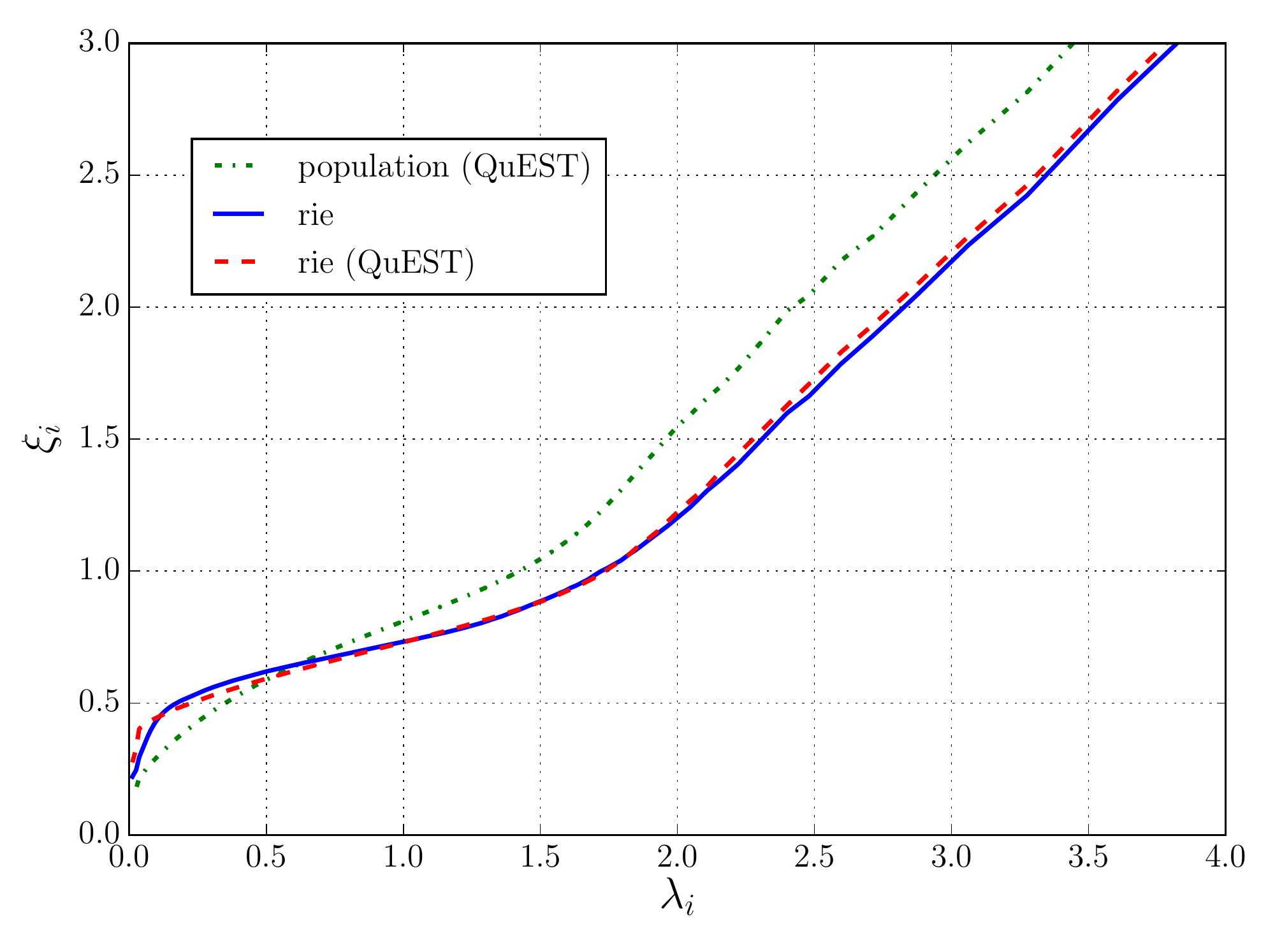}
  \caption{Population and optimal RIE bulk eigenvalues.}
  \label{fig:mkw_rie_EUR}
\end{subfigure}%
\begin{subfigure}{.5\textwidth}
  \centering
  \includegraphics[width=.9\linewidth]{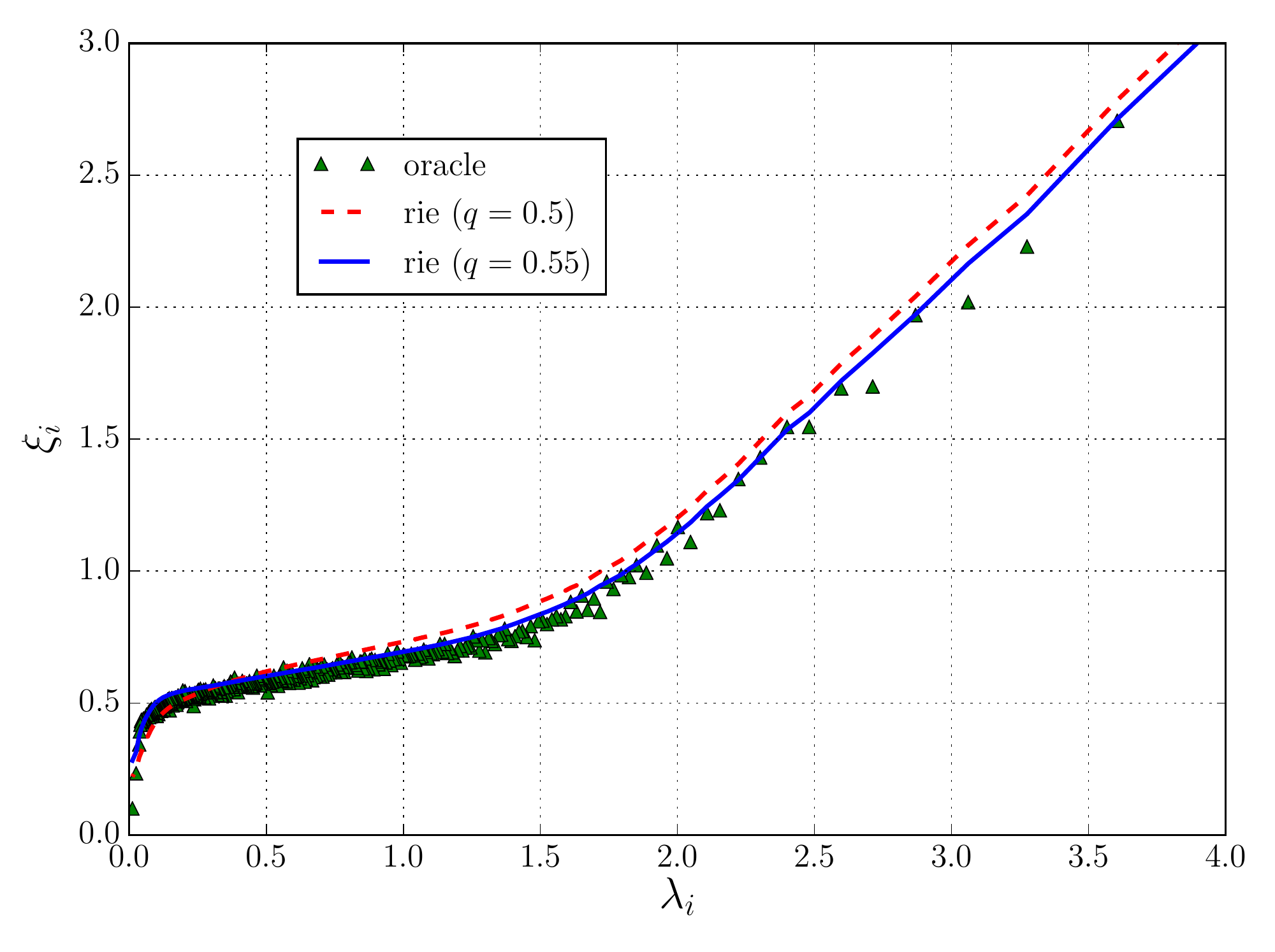}
  \caption{Comparison with Oracle estimator \eqref{eq:approxoracle}.}
  \label{fig:mkw_Oracle_EUR}
\end{subfigure}\\
\caption{Left figure: analysis of the population (green dashed line) and optimal RIE bulk eigenvalues (red dashed line for Eq.\ \eqref{eq:QuEST_rie} and blue plain line for the IWs-regularization) using the $500$ most liquid stocks during the training period of the Bloomberg European 500 index from 1996 until 2016. Right figure: Comparison between the IWs-regularized RIE (red dashed line) with the Oracle estimator \eqref{eq:approxoracle} (green triangle). We also provide the plot of the IWs-regularized RIE with an effective observation ratio (blue line)
%v The X-axis denotes the sample eigenvalues and the Y-axis the \emph{cleaned} eigenvalues.
}
\label{fig:mkw_TPX}
\end{figure}

Finally we look at European stocks where the conclusion are similar than for the US stocks. In particular, we notice in Figure \ref{fig:mkw_Oracle_EUR} that the estimation we obtained for the IWs-regularized RIE with the observed $q=0.5$ (red dashed line) yields a very good approximation of the Oracle estimator (green triangle). We can nonetheless improve the estimation with an effective ratio $q_{\text{eff}} = 1.1 q$ (blue plain line). 

All in all, we see that both the simple IWs-regularization and the QuEST regularization allow one to estimate accurately the (approximated) Oracle estimator using only observables quantities. This study highlights that the optimal RIE is robust with respect to the data generating process, as financial stock markets are certainly not Gaussian. The cross sectional volatility estimator \eqref{eq:cross_sec_vol} does not remove entirely heteroskedastic effects, nor the temporal dependence of the variables since it appears that one can choose an \emph{effective} observation ratio $q_{\text{eff}} > q$ for which the IWs-regularized RIE and the Oracle estimate nearly coincide. This effect may be understood by the presence of autocorrelations in the
stock returns that are not taken into account in the model of $\b E$. The presence of autocorrelations has been shown to widen the spectrum of the sample matrix $\b E$ \cite{burda2004spectral}. We shall come back to the open problem of calibrating $q_{\text{eff}}$ on empirical data in the Chapter \ref{chap:conclusion}.  It would be interesting to quantify the information kept by the optimal RIE compared to other estimators using e.g.\ the Kullback-Leibler distance as in \cite{tumminello2007kullback,biroli2007student}.

\subsection{Out-of-sample risk minimization}

\begin{figure}[!ht]
	%\begin{center}
	\centering
   \includegraphics[scale = 0.43]{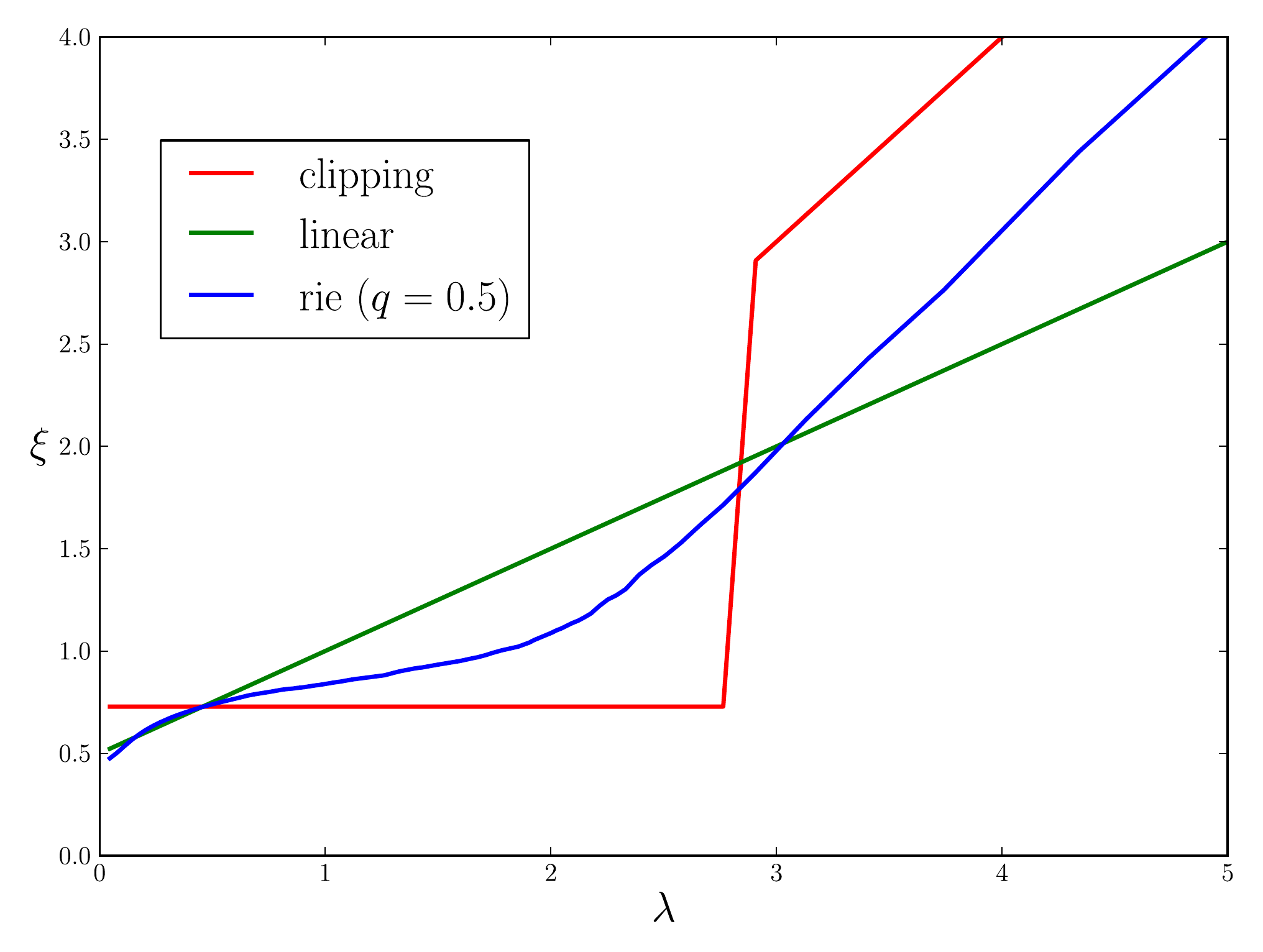} 
   %\end{center}
   \caption{Comparison of the de-biased RIE \eqref{eq:rie_reg_invW} (blue line) with clipping at the edge of the Mar{\v c}enko-Pastur (red dashed line) 
   and the linear shrinkage with $\alpha = 0.5$ (green dotted line). We use here the same data set as in Figure \ref{fig:scatter_RIE_SPX_IDX}. }
   \label{fig:RIE_SPX}
\end{figure}

It is interesting to compare the different shrinkage functions that map the empirical eigenvalues $\lambda_i$ onto their ``cleaned'' counterparts $\hat\xi_i$. We show these functions in Figure \ref{fig:RIE_SPX} for the three schemes we retained here, i.e.\ linear shrinkage, clipping and RIE, using the same data set as in Figure \ref{fig:scatter_RIE_SPX_IDX}. This figure clearly reveals the difference
between the three schemes. For clipping (red dashed line), the intermediate eigenvalues are quite well estimated but the convex shape of the optimal shrinkage function for larger $\lambda_i$'s is not captured. Furthermore, the larger eigenvalues are systematically overestimated. For the linear shrinkage (green dotted line), it is immediate from Figure \ref{fig:RIE_SPX} why this method is not optimal for any shrinkage parameters $\alpha_s \in [0,1]$ (that fixes the slope of the line).

We now turn to optimal portfolio construction using the above three cleaning schemes, with the aim of comparing the (average) realized risk of optimal Markowitz portfolios constructed as:
\begin{equation}
	\label{eq:markowitz var}
	 \b w \deq \frac{\widehat{\b \Sigma}^{-1} \b g}{\b g^* \widehat{\b \Sigma}^{-1} \b g},
\end{equation}
where $\b g$ is a vector of \emph{predictions} and $\widehat{\b \Sigma}$ is the cleaned covariance matrix $\widehat \Sigma_{ij} \deq \sigma_i \sigma_j \widehat \Xi_{ij}$ for $i,j \in \qq{1,N}$. Note again that we consider
here returns normalized by an estimator of their volatility: $\widetilde r_{it} = r_{it}/\widehat \sigma_{it}$. This means that our tests are immune against an overall increase or decrease of the volatility
in the out-of-sample period, and are only sensitive to the quality of the estimator of the correlation matrix itself.  

In order to ascertain the robustness of our results in different market situations, we consider the following four families of predictors $\b g$:
\begin{enumerate}
\item The minimum variance portfolio, corresponding to $g_i = 1, \, \forall i \in \qq{1,N}$
\item The omniscient case, i.e. when we know exactly the realized returns on the next out-of-sample period for each stock. This is given by $g_i = {\cal N} \, \tilde r_{i,t}(T_{\text{out}})$ 
where $r_{i,t}(\tau) = (P_{i,t+\tau} - P_{i,t})/P_{i,t}$ with $P_{i,t}$ the price of the $i$th asset at time $t$ and $\wt r_{it} = r_{it}/\wh \sigma_{it}$.
\item Mean-reversion on the return of the last day: $g_i = - {\cal N} \, \widetilde r_{it} \; \forall i \in \qq{1,N}$. 
\item Random long-short predictors where $\b g = {\cal N} \, \b v$ where $\b v$ is a random vector uniformly distributed on the unit sphere.
\end{enumerate} 
The normalisation factor ${\cal N} := \sqrt{N}$ is chosen to ensure $\b w_i \sim \cal O(N^{-1})$ for all $i$. The out-of-sample risk $\cal R^2$ is obtained from Eq.\ \eqref{eq:risk_os} by replacing the matrix $\b X$ by the normalized return matrix $\wt{\b R}$ defined by $\wt{\b R} \deq (\wt r_{it}) \in \mathbb{R}^{N \times T}$. We report the average out-of-sample risk for these various portfolios in Table \ref{table:results}, for the three above cleaning schemes and the three geographical zones, keeping the same value of $T$ (the learning period) and $T_{\text{out}}$ (the out-of-sample period) as above. The linear shrinkage estimator uses a shrinkage intensity $\alpha$ estimated from the data following \cite{ledoit2003improved} (LW). The eigenvalues clipping procedure uses the position of the Mar{\v c}enko-Pastur edge, $(1+\sqrt{q})^2$, to discriminate between meaningful and noisy eigenvalues.  The second to last line gives the result obtained by taking the identity matrix (\emph{total} shrinkage, $\alpha_s=0$) and the last one is obtained by taking the uncleaned, in-sample correlation matrix ($\alpha_s=1$).

\begin{table}[!ht]
\caption{Annualized average volatility (in \%) of the different strategies. Standard deviations are given in bracket.}
\label{table:results}
%{\bf should we use $\sqrt{R^2}$, units would be easier}
\small
\centering
\subcaption*{Minimum variance portfolio}
\begin{tabular}{|*{4}{c|}}
    \hline
    $\langle {\cal R} \rangle_e$  & US & Japan  & Europe  \\
    \hline
     RIE (IWs)  & \b{10.4} (0.12)  & 30.0 (2.9)  & \b{13.2} (0.12)  \\
    \hline
    Clipping MP  & 10.6 (0.12)  & 30.4 (2.9)  & 13.6 (0.12)  \\
    \hline
     Linear LW  & 10.5 (0.12) & \b{29.5} (2.9) \nc & 13.2 (0.13)  \\
	\hline
    Identity $\alpha_s=0$ & 15.0 (0.25) & 31.6 (2.92) & 20.1 (0.25)  \\
	\hline
    In sample $\alpha_s=1$ & 11.6 (0.13) & 32.3 (2.95) & 14.6 (0.2)  \\
     \hline
\end{tabular}
\bigskip
\subcaption*{Omniscient predictor}
\begin{tabular}{|*{4}{c|}}
    \hline
    $\langle {\cal R} \rangle_e$  & US & Japan  & Europe  \\
    \hline
     RIE (IWs) &  \b{10.9} (0.15)   & \b{12.1} (0.18)  & \b{9.38} (0.18)	 \\
    \hline
     Clipping MP  & 11.1 (0.15)  & 12.5 (0.2)  & 11.1 (0.21)  \\
    \hline
     Linear LW  & 11.1 (0.16) & 12.2 (0.18) & 11.1 (0.22) \\
	\hline
    Identity $\alpha_s=0$ & 17.3 (0.24) & 19.4 (0.31) & 17.7 (0.34) \\
    \hline
	In sample $\alpha_s=1$ & 13.4 (0.25) & 14.9 (0.28) & 12.1 (0.28)  \\
     \hline
\end{tabular}

\bigskip
\subcaption*{Mean reversion predictor}
\begin{tabular}{|*{4}{c|}}
    \hline
    $\langle {\cal R} \rangle_e$  & US & Japan  & Europe  \\
    \hline
     RIE (IWs) & \b{7.97} (0.14)  & \b{11.2} (0.20)  & \b{7.85} (0.06) \\
    \hline
     Clipping MP  & 8.11 (0.14)  & 11.3 (0.21)  & 9.35 (0.09) \\
    \hline
     Linear LW  & 8.13 (0.14) & 11.3 (0.20) & 9.26 (0.09) \\
	\hline
    Identity $\alpha_s=0$ & 17.7 (0.23) & 24.0 (0.4) & 23.5 (0.2) \\
    \hline
	In sample $\alpha_s=1$ & 9.75 (0.28) & 15.4 (0.3) & 9.65 (0.11) \\
    \hline
\end{tabular}

\bigskip
\subcaption*{Uniform predictor}
\begin{tabular}{|*{4}{c|}}
    \hline
    $\langle {\cal R} \rangle_e$  & US & Japan  & Europe  \\
    \hline
     RIE  & \b{1.30} (8e-4)  & \b{1.50} (1e-3)  & \b{1.23} (1e-3) \\
    \hline
     Clipping MP  & 1.31 (8e-4)  & 1.55  (1e-3) & 1.32 (1e-3)  \\
    \hline
     Linear LW  & 1.32 (8e-4) & 1.61 (1e-3) & 1.27 (1e-3) \\
	\hline
    Identity $\alpha_s=0$ & 1.56 (2e-3) & 1.86 (2e-3) & 1.69 (2e-3)  \\
    \hline
	In sample $\alpha_s=1$ & 1.69 (1e-3) & 2.00 (2e-3) & 2.7 (0.01)  \\
     \hline
\end{tabular}
\normalsize
\end{table}

These tables reveal that: (i) it is always better to use a cleaned correlation matrix: the out-of-sample risk without cleaning is, as expected, always higher than with any of the cleaning schemes, even 
with four years of data. This is in agreement with previous work of Pantaleo et al.\ \cite{pantaleo2011improved}; (ii) in all cases but one (Minimum risk portfolio in Japan, where the LW linear shrinkage outperforms), the regularized RIE is providing the lowest out-of-sample
risk, independently of the type of predictor used. Note that these results are statistically significant everywhere, except perhaps for the minimum variance strategy with Japanese stocks: see the standard errors that are given between parenthesis in Table \ref{table:results}. Finally, we test the robustness in the dimension $N$ by repeating the same test for $N=\{100,200,300\}$. We focus on relatively small values of $N$ as the conclusions are 
valid in all cases as soon as $N \geq 300$. We see that apart from some fluctuations for $N=100$, the result for out-of-sample test with the RIE is robust to the dimension $N$ as indicated in the Table \ref{table:r_out_dimension}.

\begin{table}[!ht]
\caption{Annualized average volatility (in \%) of the different strategies as a function of $N$ with $q = 0.5$. We report the standard deviation in parenthesis. We highlight the smallest annualized average volatility amongst all estimators in bold. }
\label{table:r_out_dimension}
\small
%{\bf should we use $\sqrt{R^2}$, units would be easier}
\subcaption*{Minimum variance portfolio}
\begin{tabular}{*{10}{c|}}
    %\multirow{2}{*}{} &
    \cline{2-10} 
    & \multicolumn{3}{c|}{US} & 
    \multicolumn{3}{c|}{Japan} & 
    \multicolumn{2}{c}{Europe} &  \\
    %\hline
    \thickline
 	\multicolumn{1}{|c|}{$N$} & 100 & 200  & 300 &  100 & 200  & 300 &  100 & 200  & 300 \\
    \thickline
    \multicolumn{1}{|c|}{RIE (IWs)} & \bf 12.1 (0.1) & \bf  11.0 (0.2)  & \bf  10.4 (0.1) & 28.7 (2.7) & 28.2 (2.7)  & 27.8 (2.7) & 15.3 (0.2) & \bf  13.5 (0.1)  & \bf  13.4 (0.1) \\
    \hline
    \multicolumn{1}{|c|}{Clipping} & 12.2 (0.2) & 11.0 (0.2)  & 10.5 (0.1) & 28.7 (2.7) & 28.5 (2.7)  & 28.1 (2.8) & \bf 15.0 (0.2) & 13.7 (0.1)  & 13.8 (0.1)  \\
    \hline
    \multicolumn{1}{|c|}{Linear} & 12.3 (0.2) & 11.3 (0.2)  & 10.6 (0.1) & \bf 28.6 (2.7) & \bf 28.0 (2.7) & \bf 27.7 (2.8) & 15.4 (0.2) & 13.7 (0.1)  & 13.5 (0.2) \\
    \hline
    \multicolumn{1}{|c|}{Identity} & 16.4 (0.3) & 15.7 (0.3)  & 15.3 (0.3) & 31.3 (2.7) & 31.0 (2.7)  & 31.0 (2.8)  & 20.4 (0.3) & 20.1 (0.4)  & 20.2 (0.4) \\
    \hline
    \multicolumn{1}{|c|}{In sample} & 14.6 (0.2) & 13.1 (0.2)  & 12.3 (0.2) & 32.0 (2.8) & 31.3 (2.8)  & 31.0 (2.8) & 18.2 (0.2) & 16.6 (0.2)  & 18.2 (0.4) \\
    \hline
 %     RIE  & \bf 21.9 \nc  & \bf 11.7 \nc& \bf 10.0 \nc & \bf 8.51 \nc \\
 %    \hline
 %    Clipping MP  & 22.0  & 11.9  & 10.1 & 8.62	\\
 %    \hline
 %     Linear LW  & 22.6 &  12.1  & 10.3 & 8.74 \\
	% \hline
 %    Identity $\alpha_s=0$  & 43.2 & 27.3 & 21.1 & 19.3 \\
	% \hline
 %    In sample $\alpha_s=1$  & 30.0 & 15.7 & 13.5 & 11.4 \\
 %     \hline
\end{tabular}

\bigskip
\subcaption*{Mean reversion predictor}
\begin{tabular}{*{10}{c|}}
    %\multirow{2}{*}{} &
    \cline{2-10} 
    & \multicolumn{3}{c|}{US} & 
    \multicolumn{3}{c|}{Japan} & 
    \multicolumn{2}{c}{Europe} &  \\
    %\hline
    \thickline
 	\multicolumn{1}{|c|}{$N$} & 100 & 200  & 300  & 100 & 200  & 300  & 100 & 200  & 300  \\
    \thickline
    \multicolumn{1}{|c|}{RIE (IWs)} &  \bf  21.9 (0.3) & \bf  11.8 (0.07)  & \bf  10.0 (0.1) & \bf{24.5} (0.4) & \bf  13.8 (0.1)  & \bf  12.5 (0.2)  & \bf  26.4 (0.8) & \bf  15.4 (0.3)  & \bf  10.0 (0.1)  \\
    \hline
    \multicolumn{1}{|c|}{Clipping} & 22.1 (0.3) & 11.9 (0.08)  & 10.2 (0.1)  & 25.2 (0.4) & 14.3 (0.1)  & 13.2 (0.4)  & 27.3 (0.9) & 15.9 (0.2)  & 10.1 (0.1) \\
    \hline
    \multicolumn{1}{|c|}{Linear} & 22.6 (0.4) & 12.1 (0.08)  & 10.3 (0.1)  & 25.5 (0.5) & 14.2 (0.1)  & 12.8 (0.3)  & 27.3 (0.9) & 16.1 (0.3)  & 10.3 (0.2)  \\
    \hline
    \multicolumn{1}{|c|}{Identity} & 43.2 (2.5) & 27.3 (0.6)  & 21.1 (0.3)  & 64.0 (4.6) & 43.9 (3.9)  & 41.3 (5.2)  & 66.2 (2.5) & 42.2(1.7)  & 31.2 (0.7) \\
    \hline
    \multicolumn{1}{|c|}{In sample} & 30.0 (0.6) & 15.7 (0.2)  & 13.5 (0.2) & 31.7 (0.4) & 18.5 (0.3)  & 15.8 (0.5)  & 34.5 (1.2) & 20.0 (0.4)  & 11.4 (0.1)   \\
    \hline
 
\end{tabular}

\bigskip
\subcaption*{Omniscient predictor}
\begin{tabular}{*{10}{c|}}
    %\multirow{2}{*}{} &
    \cline{2-10} 
    & \multicolumn{3}{c|}{US} & 
    \multicolumn{3}{c|}{Japan} & 
    \multicolumn{2}{c}{Europe} &  \\
    %\hline
    \thickline
 	\multicolumn{1}{|c|}{$N$} & 100 & 200  & 300  & 100 & 200  & 300 & 100 & 200  & 300 \\
    \thickline
    \multicolumn{1}{|c|}{RIE (IWs)} & \bf 13.6 (0.2) & \bf 11.1 (0.2)  & \bf 11.7 (0.2)  & \bf 12.1 (0.2) & \bf 11.2 (0.1)  & \bf 12.2 (0.2) & \bf 10.2 (0.1) & \bf 9.9 (0.2)  & \bf 9.82 (0.2) \\
    \hline
    \multicolumn{1}{|c|}{Clipping} & 13.8 (0.2) & 11.2 (0.2)  & 11.9 (0.2) & 12.3 (0.2) & 11.4 (0.1)  & 12.7 (0.2) & 10.4 (0.1) & 11.3 (0.2) & 9.91 (0.2)  \\
    \hline
    \multicolumn{1}{|c|}{Linear} & 13.9 (0.2) & 11.5 (0.2)  & 12.0 (0.2) & 12.3 (0.2) & 11.4 (0.1)  & 12.5 (0.2) & 10.6 (0.1) & 11.3 (0.2)  & 9.87 (0.2) \\
    \hline
    \multicolumn{1}{|c|}{Identity} & 19.4 (0.5) & 16.4 (0.4)  &  16.3 (0.3)  & 20.7 (0.5) & 19.1 (0.3)  & 22.6 (0.9)  & 18.5 (0.3) & 18.4 (0.4)  & 18.3 (0.5) \\
    \hline
    \multicolumn{1}{|c|}{In sample} & 16.7 (0.4) &  13.7 (0.3) & 14.6 (0.3) & 14.0 (0.3) & 14.7 (0.3)  & 15.0 (0.3) & 11.0 (0.1) & 10.5 (0.2)  & 11.4 (0.2)  \\
    \hline

\end{tabular}

\bigskip
\footnotesize
\subcaption*{Uniform predictor}
\begin{tabular}{*{10}{c|}}
    %\multirow{2}{*}{} &
    \cline{2-10} 
    & \multicolumn{3}{c|}{US} & 
    \multicolumn{3}{c|}{Japan} & 
    \multicolumn{2}{c}{Europe} &  \\
    %\hline
    \thickline
 	\multicolumn{1}{|c|}{$N$} & 100 & 200  & 300 & 100 & 200  & 300 & 100 & 200  & 300 \\
    \thickline
    \multicolumn{1}{|c|}{RIE (IWs)} & \bf 2.72 (3e-3) & \bf 1.91 (2e-3)  & \bf 1.57 (1e-3) & \bf 3.06 (4e-3) & \bf 2.16 (2e-3)  & \bf 1.73 (1e-3)  & \bf 2.85 (5e-3) & \bf 2.01 (4e-3)  & \bf 1.58 (1e-3) \\
    \hline
    \multicolumn{1}{|c|}{Clipping} & 2.77 (3e-3) & 1.94 (2e-3)  & 1.59 (1e-3) & 3.19 (5e-3) & 2.2 (2e-3)  & 1.80 (1e-3)  & 2.96 (6e-3) & 2.16 (4e-3)  & 1.63 (1e-3)  \\
    \hline
    \multicolumn{1}{|c|}{Linear} & 2.74 (3e-3) & 1.93 (2e-3)  & 1.61 (1e-3)  & 3.07 (4e-3) & 2.18 (2e-3)  & 1.75 (1e-3) & 2.90 (5e-3) & 2.03 (3e-3)  & 1.6 (1e-3)  \\
    \hline
    \multicolumn{1}{|c|}{Identity} & 3.25 (6e-3) & 2.36 (3e-3)  &  1.85 (2e-3) & 4.82 (3e-2) & 3.23 (1e-2)  & 3.13 (2e-2) & 3.71 (7e-3) & 3.01 (8e-3)  & 2.3 (5e-3) \\
    \hline
    \multicolumn{1}{|c|}{In sample} & 3.71 (7e-3) & 2.56 (3e-3)  & 2.12 (2e-3) & 4.11 (8e-3) & 3.0 (4e-3)  & 2.38 (3e-2) & 3.69 (9e-3) & 3.13 (2e-2)  & 2.33 (9e-3) \\
    \hline
 
\end{tabular}
\normalsize

\end{table}

\subsection{Testing for stationarity assumption}
\label{sec:MP_2sample_test}

In this section, we investigate in more details the stationarity assumption underlying the Mar{\v c}enko-Pastur framework, i.e. that the future (out-of-sample) is statistically identical to the past (in-sample), in the 
sense that the empirical correlation matrices $\E_{\text{in}}$ and $\E_{\text{out}}$ are generated by the same underlying statistical process characterized by a unique correlation matrix $\C$. We will use the two-sample 
eigenvector test introduced in Section \ref{sec:mso_two_sample}. 

Let us reconsider the two-sample self-overlap formula \eqref{eq:mso_SCM_bulk_same_q_eig} for which the key object is the \emph{limiting} Stieltjes transform \eqref{eq:m0}. As we saw in Section \ref{sec:rie_denoise}, using the ``raw'' empirical Stieltjes transform yields a systematic bias for small eigenvalues which can be problematic when applying Eq.\ \eqref{eq:mso_SCM_bulk_same_q_eig}. Hence, we shall split the numerical computation of the overlap formula \eqref{eq:mso_SCM_bulk} or \eqref{eq:mso_SCM_bulk_same_q_eig} into two steps. The first step is to estimate the population eigenvalues using the QuEST method of Ledoit and Wolf (see Section \ref{sec:QuEST}). Since these eigenvalues are designed to solve the Mar{\v c}enko-Pastur equation, the second step consists in extracting from Eq.\ \eqref{eq:QuEST_MP} an estimation of the Stieltjes transform of $\E$ for an arbitrarily small imaginary part $\eta$, that we denote by $\widehat \stj_\E(z)$ for any $z \in \mathbb{C}_-$. Using  $\widehat \stj_\E(z)$ in Eq.\ \eqref{eq:m0} allows us to obtain the overlaps. 

\subsubsection{Synthetic data}

We test this procedure on synthetic data first. Our numerical procedure is as follows. As in Section \ref{sec:mso_two_sample}, we consider $100$ independent realizations of the Wishart noise ${\Wishart}$ with parameter $T$ and covariance $\C$. Then, for each pair of samples, we compute the smoothed overlaps as:
\begin{equation}
	\label{eq:mso_empirical_proc}
	\scalar{\b u_i}{ \tilde{\b u}_i}^2 = \frac{1}{Z_i} \sum_{j = 1}^{N} \frac{\scalar{\b u_i}{\tilde{\b u}_j}^2}{(\lambda_i - \tilde{\lambda}_j)^2 + \eta^2},
\end{equation}
with $Z_i = \sum_{k=1}^{N} ( (\lambda_i - \tilde{\lambda}_k)^2 + \eta^2)^{-1}$ the normalization constant and $\eta$ the width of the Cauchy kernel, that we 
choose to be $N^{-1/2}$ in such a way that $N^{-1} \ll \eta \ll 1$. We then average this quantity over all sample pairs for a given label $i$ to obtain 
$[ \scalar{\b u_i}{ \tilde{\b u}_i}^2 ]_e$, which should be a good approximation of Eq.\ \eqref{eq:mso} provided that we have enough data. 

\begin{figure}[!ht]
\begin{subfigure}{.5\textwidth}
  \centering
  \includegraphics[width=.9\linewidth]{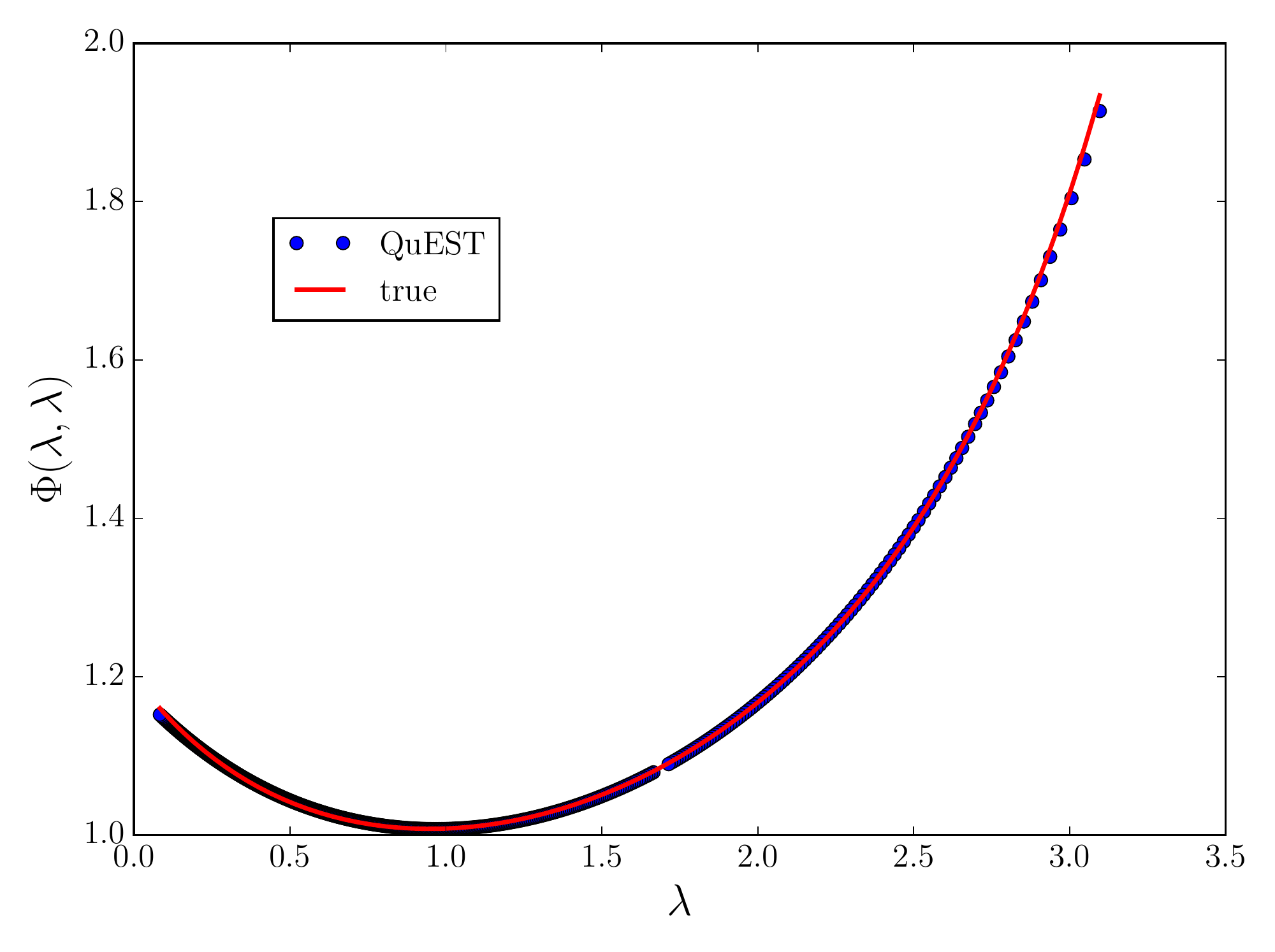}
  \caption{Inverse Wishart ($\kappa = 10$).}
  \label{fig:multiple_final}
\end{subfigure}%
\begin{subfigure}{.5\textwidth}
  \centering
  \includegraphics[width=.9\linewidth]{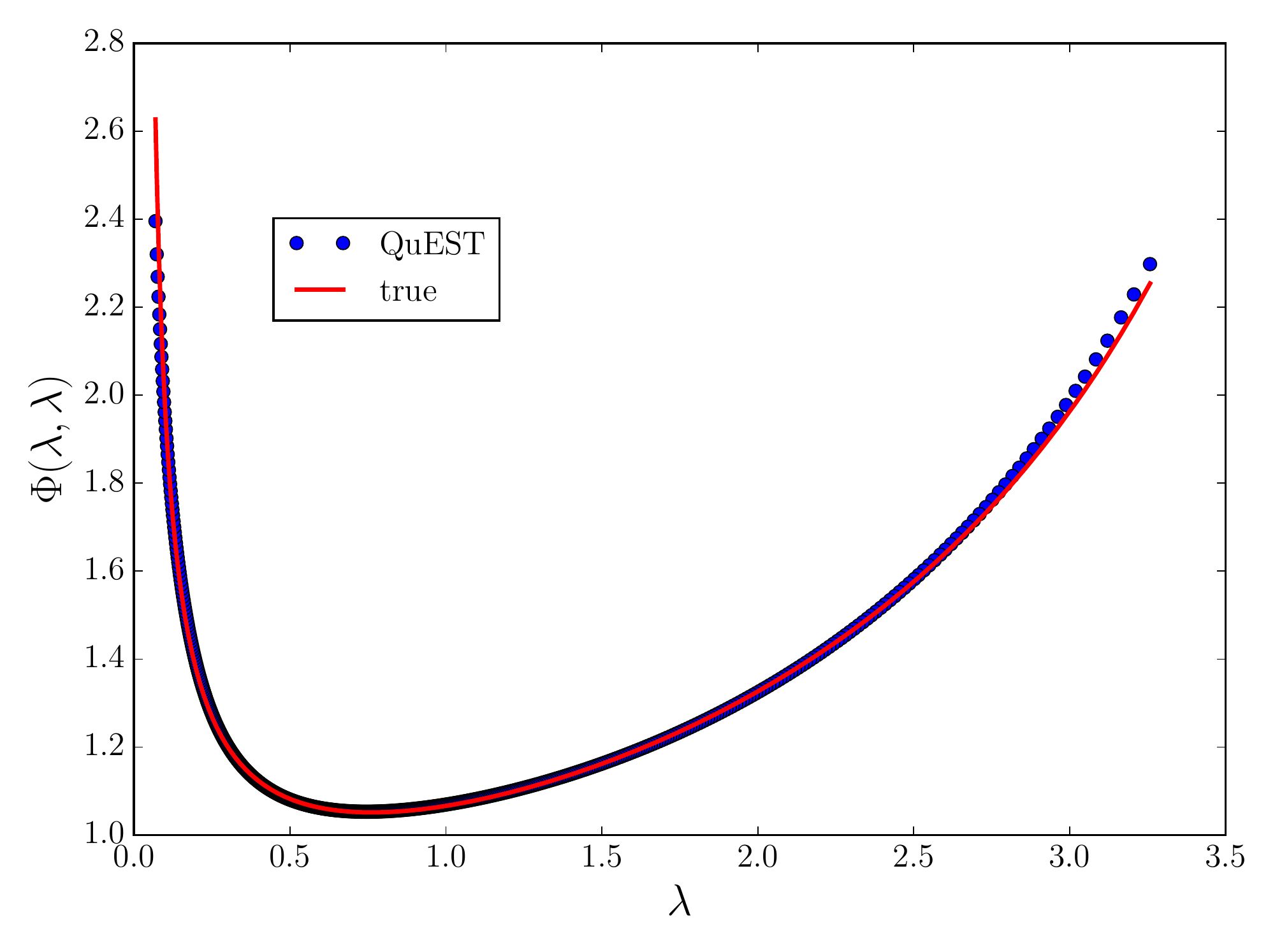}
  \caption{$\In +$ GOE ($\sigma = 0.35$).}
  \label{fig:dGOE_debias}
\end{subfigure}\\
   \caption{Evaluation of the self-overlap $\Phi(\lambda,\lambda)$ as a function of the sample eigenvalues $\lambda$ when $\C$ is an inverse Wishart of parameter $\kappa = 10$ (left) and $\C$ is a GOE centered around the identity with $\sigma = 0.35$ (right). In both cases, we compute the self-overlap \eqref{eq:mso_SCM_bulk_same_q_eig} using analytical solution (red line) and the estimated from the sample eigenvalues using QuEST algorithm (blue points).}
   %as a function the typical $[\lambda_i]_e$
   \label{fig:mso_iw_goe}
\end{figure}

We consider two simple synthetic cases. Let us assume that $\C$ is an inverse Wishart with parameter $\kappa = 10$. We generate one sample of $\E \sim \text{Wishart}(N,T, C^{-1}/T)$ with $N = 500$, $T=2N$ and we can compute the self-overlap \eqref{eq:mso_SCM_bulk_same_q_eig} using the sample eigenvalues. We compare in Figure \ref{fig:mso_iw_goe} the estimation that we get using QuEST algorithm (blue points) with the limiting ``true'' analytical solution \eqref{eq:mso_same_q_eig_invW} (red line) and we see that the fit is indeed excellent. The same conclusion is reached when $\C$ is a GOE centered around the identity matrix.

\begin{figure}[!ht]
  \begin{center}
   \includegraphics[scale = 0.45]{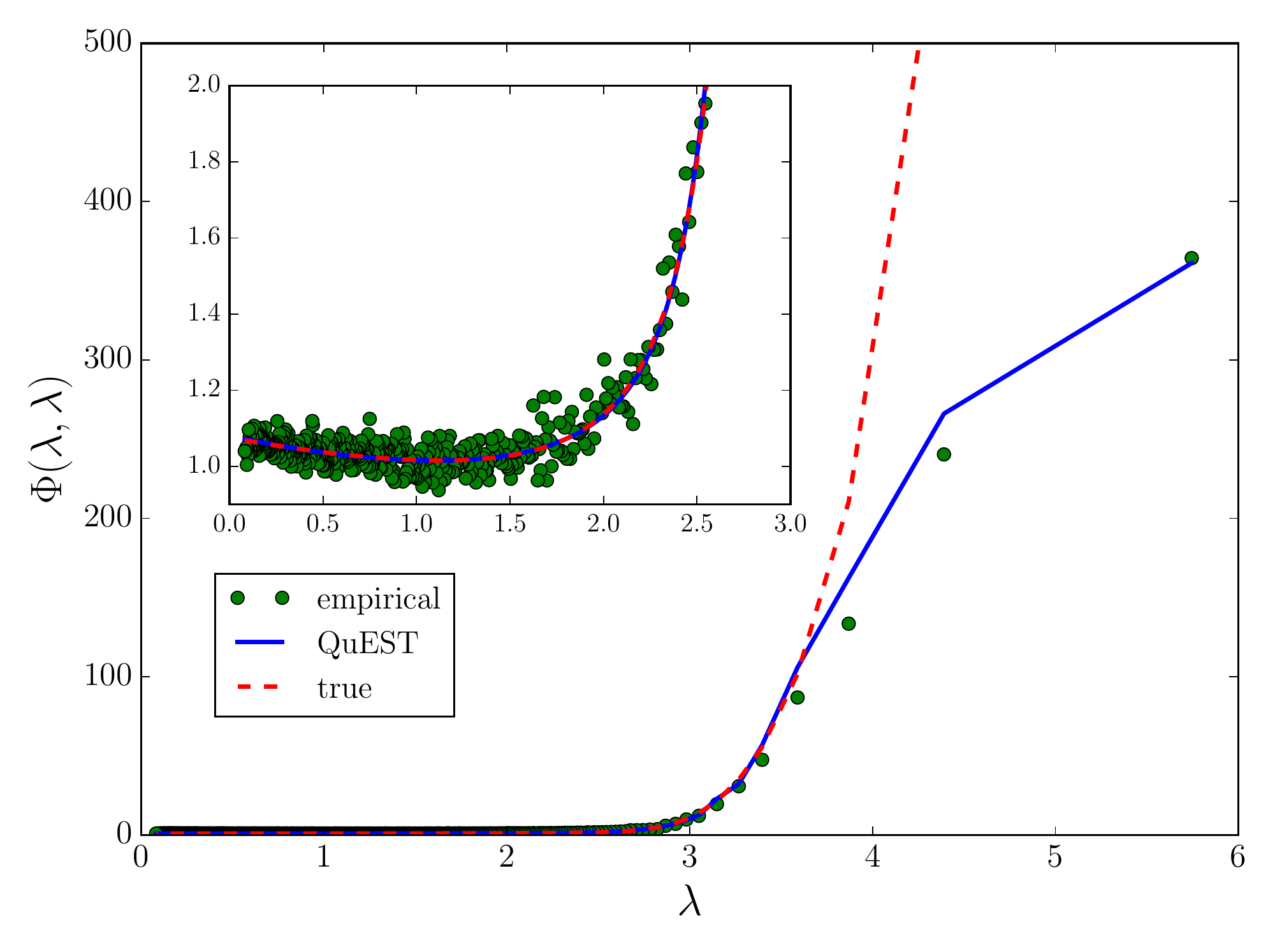} 
   \end{center}
   \caption{Main figure: Evaluation of the self-overlap $\Phi(\lambda,\lambda)$ as a function of the sample eigenvalues $\lambda$ when $\rho_\C$ is obtained from the power law proxy \eqref{eq:power law proxy} with $\lambda_0 = 0.8$. We compare the analytical true solution using Eq.\ \eqref{eq:stieltjes_powerlaw} (red dashed line) with the QuEST estimation (blue plain line) and also an empirical estimate over $100$ realizations of $\E$ using Eq.\ \eqref{eq:wishart_distribution} (green points). Inset: zoom in the bulk region of the main figure. }
   %as a function the typical $[\lambda_i]_e$
   \label{fig:mso_powerlaw}
\end{figure}

Next, we proceed to the same test using the power law distribution proxy \eqref{eq:power law proxy} for $\rho_\C$ with $\lambda_0 = -0.6$ (see Eq.\ \eqref{eq:density_powerlaw} for the precise definition of $\lambda_0$). We emphasize again that this model is quite complex since it naturally generates a finite number of outliers. The result is reported in Figure \ref{fig:mso_powerlaw} where we plotted the self-overlap obtained by the limiting exact spectral density using Eq.\ \eqref{eq:stieltjes_powerlaw} (red dashed line), the QuEST algorithm (blue plain line) and the empirical estimate \eqref{eq:mso_empirical_proc} over $100$ realizations of $\E$ (green points). Quite surprisingly, we see that the estimation obtained from the QuEST algorithm remains accurate for the outliers while the analytical solution becomes inaccurate for $\lambda \gtrsim 3.5$. This can be understood by the fact that the discrete approximation of the density \eqref{eq:QuEST} in QuEST yields a Dirac mass of weight of order $\cal O(N^{-1})$ (with $N$ finite numerically) while the limiting continuous density $\rho_\E(\lambda)$ becomes arbitrarily small for  large eigenvalues. 

\subsubsection{Financial data}

We now investigate an application to real data, in the case of stock markets and using a bootstrap technique to generate different samples. Indeed, the difficulty here is to measure the empirical mean squared overlaps between the two sample correlation matrices $\E$ and $\E'$, as in Eq. \eqref{eq:mso_empirical_proc}, because we do not have enough data points to evaluate accurately an average over the noise as required in Eq.\ \eqref{eq:mso}. To bypass this problem, we use a Bootstrap procedure to increase the size of the data.\footnote{This technique is especially useful in machine learning and we refer the reader to e.g. \cite[Section 7.11]{friedman2001elements} for a more detailed explanation.}  Specifically, we take a total period of 2400 business days from 2004 to 2013 for the same three pools of assets that we split into two non-overlapping subsets of same size of 1200 days, corresponding to 2004 to 2008 and 2008 to 2013. Then, for each subset and of each 
Bootstrap sample $b \in \{1,\dots, B\}$, we select randomly $T=600$ distinct days for $N = 300$ stocks returns such that we construct two independent sample correlation matrices 
$\E_b$ and $\E^{\prime}_{b}$, with $q=N/T=0.5$. Note that we restrict to $N = 300$ stocks such that all of them are present throughout the whole period from 2004 to 2013. We then compute the empirical mean squared overlap \eqref{eq:mso} and also the theoretical limit \eqref{eq:mso_SCM_bulk} -- using QuEST algorithm -- from these $B$ bootstrap data-sets. 

\begin{figure}[!ht]
% \begin{subfigure}{.5\textwidth}
%   \centering
%   \includegraphics[width=1\linewidth]{Figures/numerical/mso_SPX_1999}
%   \caption{1999-2005}
%   \label{fig:mso_SPX_1998}
% \end{subfigure}%
% \begin{subfigure}{.5\textwidth}
%   \centering
%   \includegraphics[width=1\linewidth]{Figures/numerical/mso_SPX}
%   \caption{2006-2012}
%   \label{fig:mso_SPX_2006}
% \end{subfigure}\\
	\begin{center}
   \includegraphics[scale = 0.45]{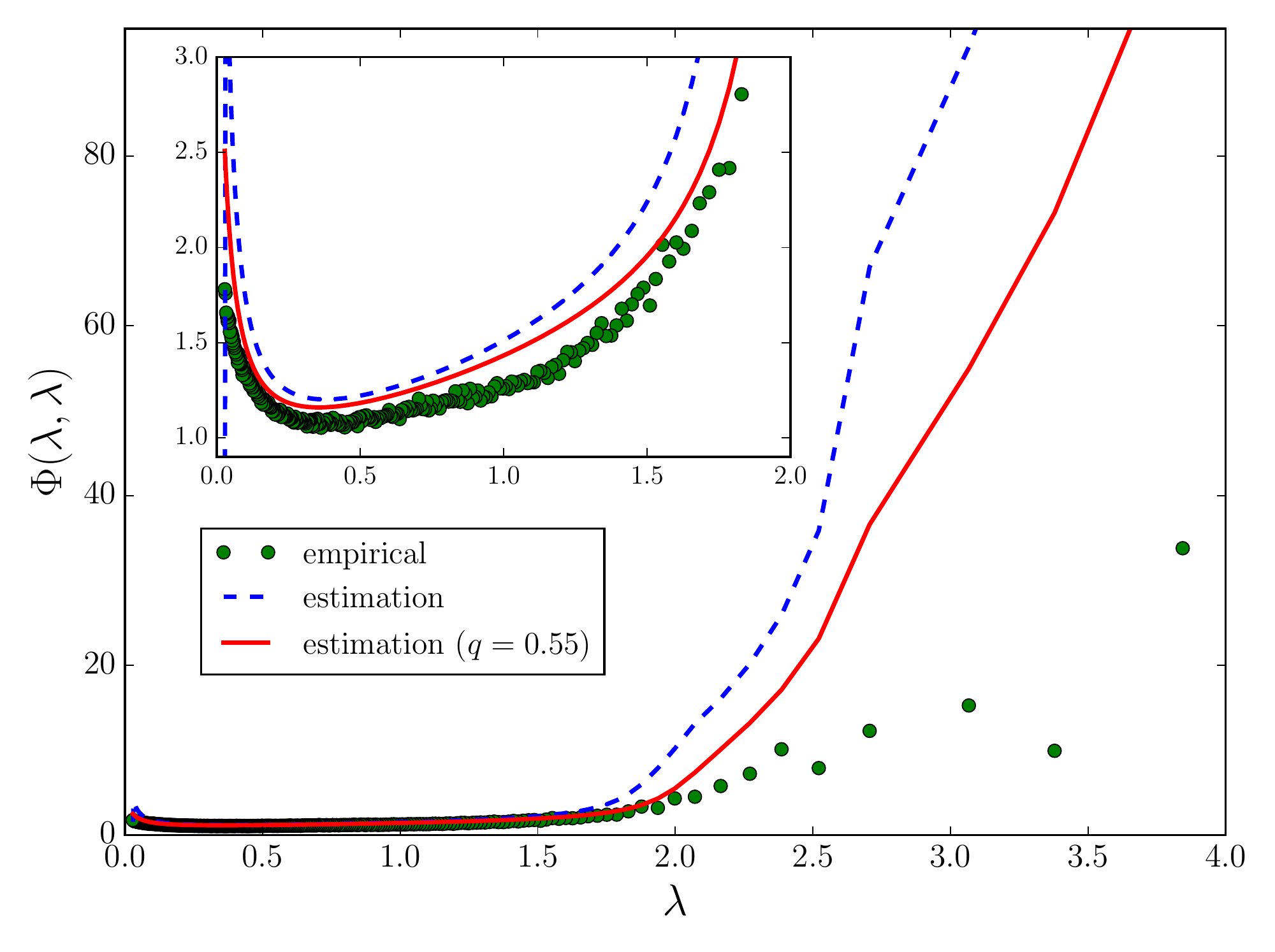} 
   \end{center}
   \caption{Evaluation of the self-overlap $\Phi(\lambda,\lambda)$ as a function of the sample eigenvalues $\lambda$ using the $N=300$ most liquid US equities from 2004 to 2013. 
   We split the data into two non-overlapping period with same sample size 1200 business days. For each period, we randomly select $T=600$ days and we repeat $B=100$ bootstraps of the original data. The empirical self-overlap is computed using Eq.\ \eqref{eq:mso_empirical_proc} over these 100 bootstraps (green points) and the limiting formula \eqref{eq:mso_SCM_bulk_same_q_eig} is estimated using QuEST algorithm with $q=0.5$ (blue dashed line). We also provide the estimation we get using the same effective observation ratio $q_{\text{eff}}$ than in Figure \ref{fig:scatter_RIE_SPX_IDX}. Inset: focus in the bulk of eigenvalues.}
   %as a function the typical $[\lambda_i]_e$
   \label{fig:mso_SPX}
\end{figure}

For our simulations, we set $B = 100$ and plot in Figure \ref{fig:mso_SPX} the resulting estimation of Eq.\ \eqref{eq:mso} we get from the QuEST algorithm (blue dashed line) and the empirical 
bootstrap estimate \eqref{eq:mso_empirical_proc} (green points) using US stocks. We also perform the estimation with an effective observation ratio $q_{\text{eff}}$ (red plain line) where we use for each market the values of $q_{\text{eff}}$ obtained above (see Figures \ref{fig:scatter_RIE_SPX_IDX}-\ref{fig:mkw_Oracle_TPX}-\ref{fig:mkw_Oracle_EUR}). Note that the behaviour in bulk is quite well estimated by the asymptotic prediction Eq.\ \eqref{eq:mso_SCM_bulk_same_q_eig} for both periods. This is consistent with the conclusion of Figure \ref{fig:scatter_RIE_SPX_IDX}.

 %These observations are robust in time since the results we obtained over 1999 to 2005 and 2006 to 2012 are quite similar even though the bulk region is slightly more peaked in the second period. 

It is however clear from Figure \ref{fig:mso_SPX} that the eigenvectors associated to large eigenvalues are not well described by the theory: we notice a discrepancy between the (estimated) theoretical curve and the empirical data even with an effective ratio $q_{\text{eff}}$. The difference is even worse for the market mode (data not shown). This is presumably related to the fact that the largest eigenvectors are expected to genuinely evolve with time, as
argued in  \cite{allez2012eigenvector}. Note also the strong at the left edge between the theoretical and empirical data in Figure \ref{fig:mso_SPX}, which can be partly corrected using the effective ratio $q_{\text{eff}}$. This suggests that one can still improve the Mar{\v c}enko-Pastur framework by adding e.g. autocorrelation or heavy tailed entries which allows one to widen the LSD of $\E$ (see e.g.\ \cite{burda2004spectral,bartz2014covariance} for autocorrelation and \cite{biroli2007student,burda2004free,el2009concentration} for heavy tailed entries).

\begin{figure}[!ht]
\begin{subfigure}{.5\textwidth}
  \centering
  \includegraphics[width=1\linewidth]{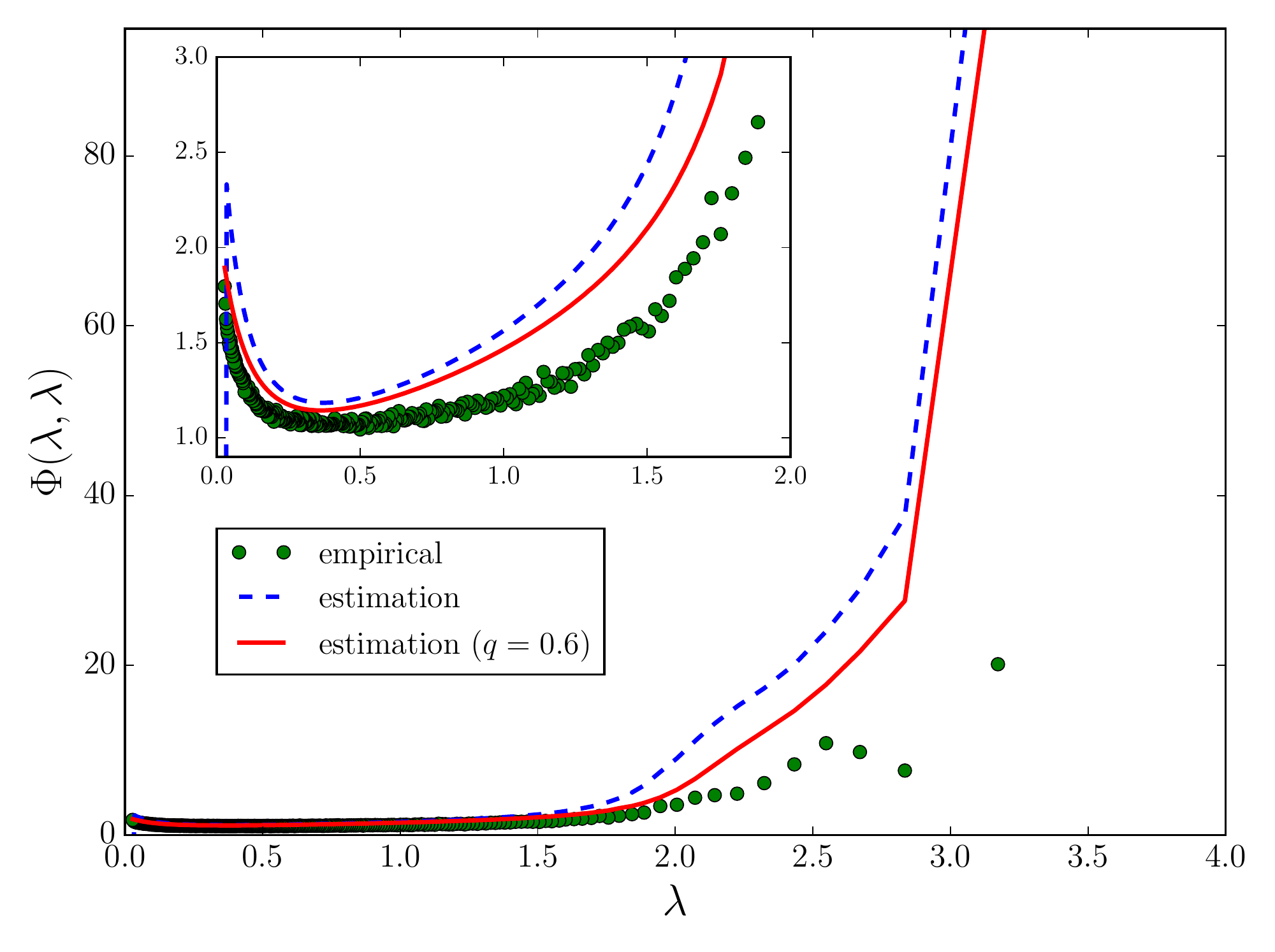}
  \caption{TOPIX (Japan)}
  \label{fig:mso_TPX}
\end{subfigure}%
\begin{subfigure}{.5\textwidth}
  \centering
  \includegraphics[width=1\linewidth]{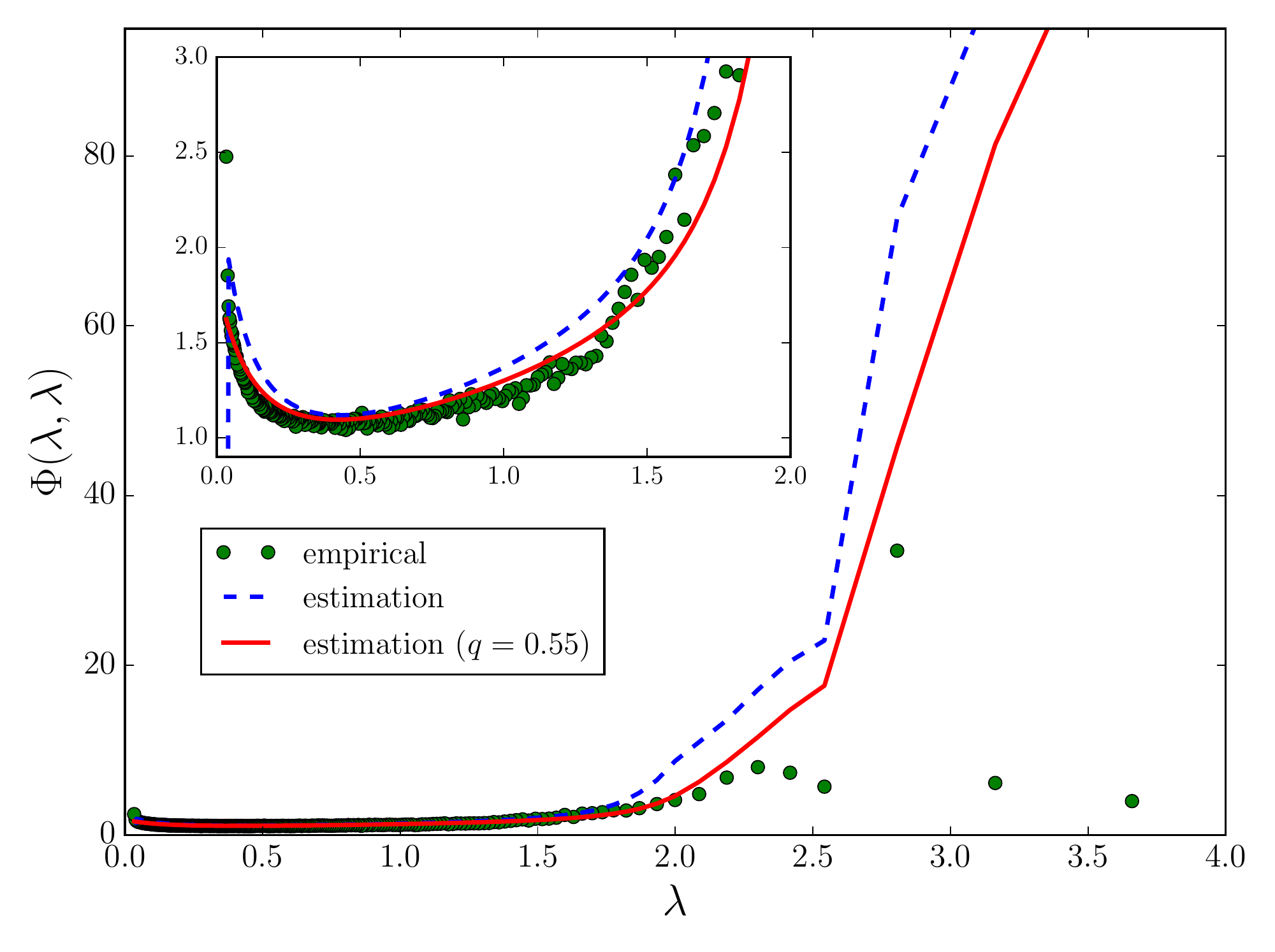}
  \caption{Bloomberg 500 (Europe)}
  \label{fig:mso_EUR}
\end{subfigure}\\
   \caption{Evaluation of the self-overlap $\Phi(\lambda,\lambda)$ as a function of the sample eigenvalues $\lambda$ using the $N=300$ most liquid equities from the Japanese TOPIX (left) and the European Bloomberg 500 index (right) from 2004 to 2013. For each case, we split the data into two non-overlapping period with same sample size $T=1200$ business days. For each period, we randomly select $600$ realizations of the returns and we repeat $B=100$ bootstraps of the original data. The empirical self-overlap is computed using Eq.\ \eqref{eq:mso_empirical_proc} over these 100 bootstraps (green points) and the limiting formula \eqref{eq:mso_SCM_bulk_same_q_eig} is estimated using QuEST algorithm with $q=0.5$ (blue dashed line). We also provide the estimation we get using the same effective observation ratio $q$ than in Figure \ref{fig:scatter_RIE_SPX_IDX}. Inset: focus in the bulk of eigenvalues.}
   %as a function the typical $[\lambda_i]_e$
   %\label{fig:mso_TPX}
\end{figure}

All the above results can be extended and confirmed in the case of Japanese and European stocks, for which the results are plotted respectively in Figures \ref{fig:mso_TPX} and \ref{fig:mso_EUR}.

To conclude, these observations suggest further improvements upon the time independent framework of Mar{\v c}enko and Pastur, that would allow one to account for some ``true'' dynamics of the underlying correlation matrix. That such dynamics exist for eigenvectors corresponding to the largest eigenvalues is intuitively reasonable, and empirically confirmed by the analysis of Ref. \cite{allez2012eigenvector}. The full correlation matrix might in fact evolve and jump between different ``market states'', as suggested in various recent papers of the Guhr group (see e.g.\ \cite{schmitt2013non,wang2016average} and references therein). Extending the present framework to these cases is quite interesting and would shed light on the optimal value of the observation ratio $q_{\text{eff}}$ which was systematically found to be larger than $q=N/T$. This could be an indication of non-stationarity effects. This is particularly apparent for the Japanese stocks (see e.g.\ Fig.\ \ref{fig:mso_TPX}) where the theoretical prediction deviates significantly from the empirical one even if we calibrate the effective quality ratio $q_{\text{eff}}$. The case of eigenvectors associated to the small eigenvalues is particularly striking and probably need further scrutiny, in particular in the case of futures markets where the presence of very strongly correlated contracts (i.e. two different maturities for the same underlying) leads to very small true eigenvalues of the correlation matrix, for which the above IW-regularizing scheme is probably inadequate. We leave these issues, as well as several others alluded to in the following concluding chapter, for further investigations.

% The results are 
% reported in Figure \ref{fig:mso_spx_inset} where we only focus on ``bulk'' eigenvalues (the top eigenvectors are governed by non trivial dynamics, see e.g. \cite{allez2012eigenvector}). We conclude from Figure \ref{fig:mso_spx_inset} that bulk eigenvalues have a non trivial structure since 
% the mean squared overlaps strongly depart from the null hypothesis and in fact cannot be estimated using the universal formula Eq.\ \eqref{eq:mso_same_q_eig_invW_exp} as a parabolic fit is clearly unwarranted. A fit with the
% full inverse Wishart formula \eqref{eq:mso_same_q_eig_invW} leads to $\kappa \approx 0.7$. \footnote{The calibration of $\kappa$ is performed by least squares using Eq.\ \eqref{eq:mso_same_q_eig_invW} and where we removed the 10 largest eigenvalues.}. A much better fit could be achieved by choosing another prior for the 
% spectrum of $\b C$, for example a power-law as in \cite{bouchaud2009financial}; we however leave this issue for later investigations. The conclusion of this 
% study is that the cross-sample eigenvector overlaps provide precious insights about the structure of the true eigenvalue spectrum, much beyond the
% information contained in the empirical spectrum itself.

%---------  Conclusion ---------------- %

\clearpage%!TEX root = RMT_Covariance_Review.tex
\section{Conclusion and perspectives}
\label{chap:conclusion}

In this review, we have discussed some of the most advanced techniques in RMT and their usefulness for estimating large correlation matrices, in particular within a rotational invariant framework. Moreover, we showed through an extended empirical analysis that these estimators can be of great interest in real world situations. Instead of repeating the main messages emphasized in the previous sections, we want to end this review with an (incomplete) list of potentially interesting open problems that represent natural extensions of the results obtained above.

\subsection{Extension to more general models of covariance matrices}

One important assumption of the sample covariance matrix model \eqref{eq:SCM} is the absence of temporal correlations and/or temporal structure in the data. However, this assumption does not hold in most real life applications (see e.g.\ Section \ref{sec:MP_2sample_test}). It is thus natural to extend the present work to estimators that account for some temporal dependence. The simplest case is 
when some \emph{autocorrelations} are present. A standard assumption is that of an exponential autocorrelation of the form \cite{burda2004spectral,burda2004free,bartz2014covariance}:
\begin{equation}
	\label{eq:expo_autocorrel}
	\mathbb{E}[ Y_{it} Y_{jt'}] = C_{ij} \exp\qb{- \abs{t-t'}/\tau},
\end{equation}
where $\tau$ controls the range of the time correlations. 

Another frequent situation is when covariances are measured through an \emph{Exponential Weighted Moving Average} (EWMA)\cite{burda2004free,pafka2004exponential}:\footnote{We denote in the following the different estimators of $\C$ by $\M$ to avoid confusion with Pearson's sample estimator $\E = \X \X^*/T$.}
\begin{equation}
	\label{eq:EWMA}
	M_{ij}(\tau,T) = (1-\alpha)  \sum_{t=0}^{T} \alpha^{t} Y_{i,\tau-t} Y_{j,\tau-t},
\end{equation}
where $\tau$ is the last estimation date available, $\alpha \in (0,1)$ is a constant and $T$ is the total size of the time series. Roughly, the idea of this estimator is that old data become gradually obsolete so that they should contribute less than more recent information. We see that the estimator \eqref{eq:EWMA} can be rewritten as
\begin{equation}
	\label{eq:SCM_EWMA}
	M_{ij}(\tau) \;=\; \sum_{t=0}^{T} H_{it} H_{jt}, \quad\text{with}\quad \mathbb{E}\qb{H_{it} H_{it'}} = \delta_{tt'} (1-\alpha) \alpha^t,
\end{equation}
i.e. the variance of the random variables have an explicit time dependence.

Another interesting way to generalize the Mar{\v c}enko-Pastur framework concerns the distribution of the entries. An important assumption for the Mar{\v c}enko-Pastur equation to be valid is that each entry $Y_{it}$ possesses a finite fourth moment. Again, this assumption may not be satisfied in real dataset, especially in finance \cite{chicheportiche2013non}. As alluded to in Section \ref{sec:SCM_entries}, a more robust estimate of the covariance matrix in then needed \cite{maronna2006robust}. Let us assume that we can rewrite the observations as $Y_{it} = \sigma_{t} \C^{1/2} X_{it}$ for any $i \in \qq{1,N}$ and $t \in \qq{1,T}$, where $\sigma_t$ is a fluctuating global volatility that sets the overall scale of the returns, and $\X$ are IID Gaussian variables. 
In that particular context, 
the sample covariance matrix is obtained as the solution of the fixed-point equation \cite{maronna2006robust}:
\begin{equation*}
		\M \deq \frac{1}{T} \sum_{t=1}^{T} U\pB{\frac1N \b y_t^* \M^{-1} \b y_t} \b y_t \b y_t^*,
\end{equation*}
where $U$ is a non-increasing function. As mentioned in Section \ref{sec:SCM_entries}, it is possible to show that for the $U(x) = x^{-1}$, one has $\M \to \E$ in the large $N$ limit \cite{biroli2007student,el2009concentration,zhang2014marchenko,couillet2016second}, where $\E =  \C^{1/2} {\b{\cal W}} \C^{1/2}$ and ${\b{\cal W}}$ is a Wishart matrix. 
However, the asymptotic limit is more complex for general $U$'s and reads:
\begin{equation}
	\label{eq:Maronna_asymp_limit}
	\M \to \C^{1/2} \X \B \X^* \C^{1/2}\,,
\end{equation}
where $\B$ is a \emph{deterministic} diagonal $T\times T$ matrix where each entry is a functional of the $\{\sigma_{t}\}_t$ and the function $U$ (see e.g.\ \cite{couillet2016second} for the exact expression of the matrix $\B$).

Interestingly, all the above models, \eqref{eq:expo_autocorrel}, \eqref{eq:SCM_EWMA} and \eqref{eq:Maronna_asymp_limit}, can be wrapped into a general multiplicative framework that reads:
\begin{equation}
	\label{eq:multiplicative_model}
	\M \;\deq\; \C^{1/2} \X \B \X^* \C^{1/2},
\end{equation}
where $\X \deq (X_{it}) \in \mathbb{R}^{N\times T}$ is a random matrix with zero mean and variance $T^{-1}$ IID entries and $\B = (B_{tt'}) \in \mathbb{R}^{T\times T}$ is fixed matrix, independent from $\C$. Indeed, for \eqref{eq:expo_autocorrel}, we have $B_{tt'} = \exp[-\abs{t-t'}/\tau]$ while we set $B_{tt'} = \delta_{tt'} (1-\alpha) \alpha^t$ for \eqref{eq:SCM_EWMA}. 

The optimal RIE for this model has been briefly mentioned in Section \ref{sec:RIE_freemult} and can be found in exquisite details in \cite{bun2015rotational}. We saw that the oracle estimator associated to the model \eqref{eq:multiplicative_model} converges -- at least for bulk eigenvalues --  to a limiting function that does not depend explicitly on the spectral density of $\C$ (see Eq.\ \eqref{eq:RIE_freemult}). It is thus interesting to see whether one of the aforementioned models can be solved in full generality using e.g.\ the results of \cite{burda2004spectral} for the model \eqref{eq:expo_autocorrel} and whether one can explain the appearance of an effective ratio $q_{\text{eff}} > q$, as encountered in Chapter \ref{chap:numerical}. Furthermore, another important result would be to see whether the estimator \eqref{eq:RIE_freemult} is also valid for outliers, as is the case for the time-independent sample covariance matrices. 

\subsection{Singular Value Decomposition}

A natural extension of the work presented in this review is to consider rectangular correlation matrices. This is particularly useful when one wishes to  measure the correlation between $N$ \emph{inputs} variables $\b x \deq (x_1, \dots, x_N)$ and $M$ \emph{outputs} variables $\b y \deq (y_1, \dots, y_M)$. The vector $\b x$ and the $\b y$ may be completely different
from one another (for example, $\b x$ could be production indicators and $\b y$ inflation indexes) or it also could be
the same set of observables but observed at different times (\emph{lagged} correlation matrix \cite{bouchaud2009financial}). The cross-correlations is thus characterized by a rectangular $N \times M $ matrix $\b{\cal{C}}$ defined as:
\begin{equation}
	\label{eq:population_correl_XY_SVD}
		\cal C_{ia}\;\deq\; \mathbb{E}[x_i y_a],
\end{equation}
where we assumed that both quantities have zero mean and unit variance. 

What can be said about the structure of this rectangular and non symmetric correlation matrix \eqref{eq:population_correl_XY_SVD}? 
The answer is obtained from the singular value decomposition (SVD) in the following sense: what is the (normalized) linear combination of $\b x$'s on the one
hand, and of $\b y$'s on the other hand, that have the strongest mutual correlation? In other words, what is the best
pair of predictor and predicted variables, given the data? The largest singular value -- say $c_{1} \in (0,1)$ and its corresponding left
and right eigenvectors answer precisely this question: the eigenvectors tell us how to construct these optimal linear
combinations, and the associated singular value gives us the strength of the cross-correlation. We may then repeat this operation on the $N-1$ and $M-1$ dimensional sub-spaces orthogonal to the two eigenvectors for both input and output variables. This yields a list of singular values $\{c_{i}\}_i$ that represent the prediction power of the corresponding linear combinations (in decreasing order). This is called \emph{Canonical Correlation Analysis} (CCA) in the literature and has (see \cite{hotelling1936relations} or \cite{johnstone2008multivariate,yang2015independence} for more recent works). 

In order to study the singular values and the associated left and right eigenvectors, we consider the $N \times N$
matrix $\b{\cal C}\b{\cal C}^*$, which is now symmetric and has $N$ non negative eigenvalues. Indeed, the trick behind this change of variable is that the eigenvalues of $\b{\cal C}\b{\cal C}^*$ are equal to the square of a singular value of $\b{\cal C}$ itself. Then, the eigenvectors give us the weights of the linear combination of the $\b x$'s that construct the \emph{best} predictors in the above sense. In order to obtain the right eigenvectors of $\b{\cal C}$, one forms the $M\times M$ matrix $\b{\cal C}^* \b{\cal C}$ that has exactly the same non zero eigenvalues as $\b{\cal C}\b{\cal C}^*$; the corresponding eigenvectors now give us the weights of the linear combination of the $\b y$'s that construct the \emph{best} predictees. If $M > N$, the matrix $\b{\cal C}^* \b{\cal C}$ has $M-N$ additional zero eigenvalues; whereas in the other case, it is $\b{\cal C} \b{\cal C}^*$
that has an excess of $N-M$ zero eigenvalues. 

However, as for standard correlation matrices, the knowledge of the true population matrix Eq.\ \eqref{eq:population_correl_XY_SVD} is unavailable. Hence, one resorts to an empirical determination of $\b{\cal C}$ that is strewn with measurement noise, as above. We expect to be able to use tools from RMT to understand the how the true singular values are dressed by the measurement noise. To that end, suppose that we have a total of $T$ observations of both quantities that we denote by $[X_{it}]_{t}$ and $[Y_{at}]_t$. Then, the empirical estimate of $\b{\cal C}$ is given by
\begin{equation}
	\label{eq:sample_correl_XY_SVD}
		\cal E_{ia}\;\deq\; \frac1T \sum_{t=1}^{T} X_{it} Y_{at}\,,
\end{equation}
and the aim is to study the singular values of this matrix. Indeed, as in Chapter \ref{chap:spectrum}, we expect the measurement noise to affect the accuracy of the estimation in the limit $N,M,T \to \infty$ with $n=N/T$ and $m=M/T$ finite, which we will assume to be both smaller than unity in the following. As explained in the previous paragraph, a convenient way to perform this analysis is to consider the eigenvalues of $\b{\cal E} \b{\cal E}^*$ (or $\b{\cal E}^* \b{\cal E}$). Using tools from Appendix \ref{app:linear_algebra}, especially Eq.\ \eqref{eq:sylvester}, we see that 
\begin{equation*}
	\det(\b{\cal E} \b{\cal E}^* - z \b I_N) = \det\pBB{\b S_\X \b S_\Y - z \b I_T}\,, \qquad \b S_\X \;\deq\; \frac{\X^* \X}{T}, \quad \b S_\Y \;\deq\; \frac{\Y^* \Y}{T}
\end{equation*}
so that $\b{\cal E} \b{\cal E}^*$ shares the same non-zero eigenvalues than the product of the dual $T\times T$ samples covariance matrix $\b S_\X$ and $\b S_\Y$. 

It is easy to see that when $\X$ and $\Y$ are uncorrelated, i.e. $\b{\cal C} = \b 0$, one can compute the spectral density of $\b S_\X \b S_\Y$ using the free multiplication formula \eqref{eq:free_mult}. 
However, the result depends in general on the correlation structure of the input variables, $\C_X$, and of the output variables $\C_Y$. A way to obtain a universal result is to consider the 
exact normalized PCA's of the $\X$ and of the $\Y$, that we call $\hat \X$ and $\hat \Y$, such that $\b S_{\hat \X}$ has $N$ eigenvalues equal to $1$ and $T - N$ eigenvalues equal to zero, while 
$\b S_{\hat \Y}$ has $M$ eigenvalues equal to $1$ and $T - M$ eigenvalues equal to zero. In this case, the limiting spectrum of singular values can be found explicitly (see 
\cite{bouchaud2007large} and \cite{wachter1980limiting} for an early derivation without using free probability methods), and is given by:
\begin{equation}\label{resultSVD}
\rho(c) = \max(m+n-1,0) \delta(c-1) + \Re \frac{\sqrt{(c^2-\gamma_-)(\gamma_+-c^2)}}{\pi c(1 - c^2)},
\end{equation}
where $\gamma_\pm$ are given by: 
\begin{equation}
\gamma_{\pm} = n + m - 2mn \pm 2\sqrt{mn(1 - n)(1 - m)}, \quad 0 \leq \gamma_{\pm} \leq 1
\end{equation}
The allowed $c$'s are all between $0$ and $1$, as they
should since these singular values can be interpreted as correlation coefficients. In the limit $T \to \infty$ at fixed $N$, $M$, all singular values collapse to zero, as they should
since there is no true correlations between $X$ and $Y$. The allowed band in the limit $n,m \to 0$ becomes:
\begin{equation*}
c \in \left[|\sqrt{m}-\sqrt{n}|,{\sqrt{m}+\sqrt{n}}\right],
\end{equation*}
showing that for fixed $N,M$, the order of magnitude of allowed singular values decays as $T^{-1/2}$. The above result allows one devise precise statistical tests, see \cite{johnstone2008multivariate,bouchaud2007large,yang2015independence}.

The general case where when $\X$ and $\Y$ are correlated, i.e. $\b{\cal C} \neq \b 0$, is, to our knowledge, unknown. This is particularly relevant for practical cases since one might expect some true correlations between the input and output variables. It would be interesting to characterize how the noise distorts the ``true'' cross-correlations between $\X$ and $\Y$, as the analogue of the Mar{\v c}enko-Pastur equation \eqref{eq:MP_equation_stieltjes}. Moreover, an analysis of the left and right eigenvectors like in Chapter \ref{chap:eigenvectors} would certainly be of interest in many real life problems (see e.g.\ \cite{friedman2001elements,klema1980singular,furnas1988information,alter2000singular} for standard applications). Note that the case of outlier singular values and vectors of rectangular random matrices subject to a low rank perturbation has been considered \cite{benaych2012singular}.

\subsection{Estimating the eigenvectors}

As indicated by its name, the optimal RIE is optimal under the assumption that we have no prior insights on the true components, i.e.\ the eigenvectors of the population covariance matrix $\C$. However, in some problems we expect these eigenvectors to have some specific, non isotropic structure. One possible solution to this problem is to formulate prior structures for these eigenvectors through factor models \cite{fama1993common,chicheportiche2015nested}, ultra-metric tree models (\emph{eigenvector clustering}) \cite{tumminello2010correlation,dimov2012hidden}, or constraints on the participation ratios 
\cite{monasson2015estimating}.

Very recently, an attempt to ``clean'' empirical outlier eigenvectors was formulated in \cite{monasson2015estimating}. Let us focus for example on the top eigenvector; the prior is then defined as a weighted sum of the sample eigenvectors:
\begin{equation}
	\label{eq:top_eigenvector_estimator}
	\wh{\b v}_1 = \sqrt{\mso(\mu_1, \lambda_1)} \, \b u_1 + \sum_{j=2}^{N} \e_{j} \sqrt{\mso(\mu_1, \lambda_j)} \b u_j\,,
\end{equation}
where the bivariate mean squared overlap $\mso$ is defined in Eq.\ \eqref{eq:overlap} and the $\{\e_j\}_{j\geq 2}$ is a set of i.i.d.\ Gaussian random variables with zero mean and unit variance, that must 
be determined in such a way that $\wh{\b v}_1$ is, for example, as ``localized'' as possible. One notices that the first term in the RHS of Eq.\ \eqref{eq:top_eigenvector_estimator} can be computed using Eq.\ \eqref{eq:overlap_outlier_outlier} and the second one can be inferred from Eq.\ \eqref{eq:overlap_outlier_bulk}. On average, we see that $\langle \wh{\b v}_1 \rangle_\e \cdot \b u_1 = \sqrt{\mso(\mu_1, \lambda_1)}$, as it should. While this prior requires some knowledge about the number of outliers -- which is still an open question -- it is shown in \cite{monasson2015estimating} that this method improves the accuracy of the estimation on synthetic data. It would be interesting to make use of some of these ideas in the context financial data. 

\subsection{Cleaning recipe for $q > 1$}

\begin{figure}[h]
  \begin{center}
   \includegraphics[scale = 0.5]{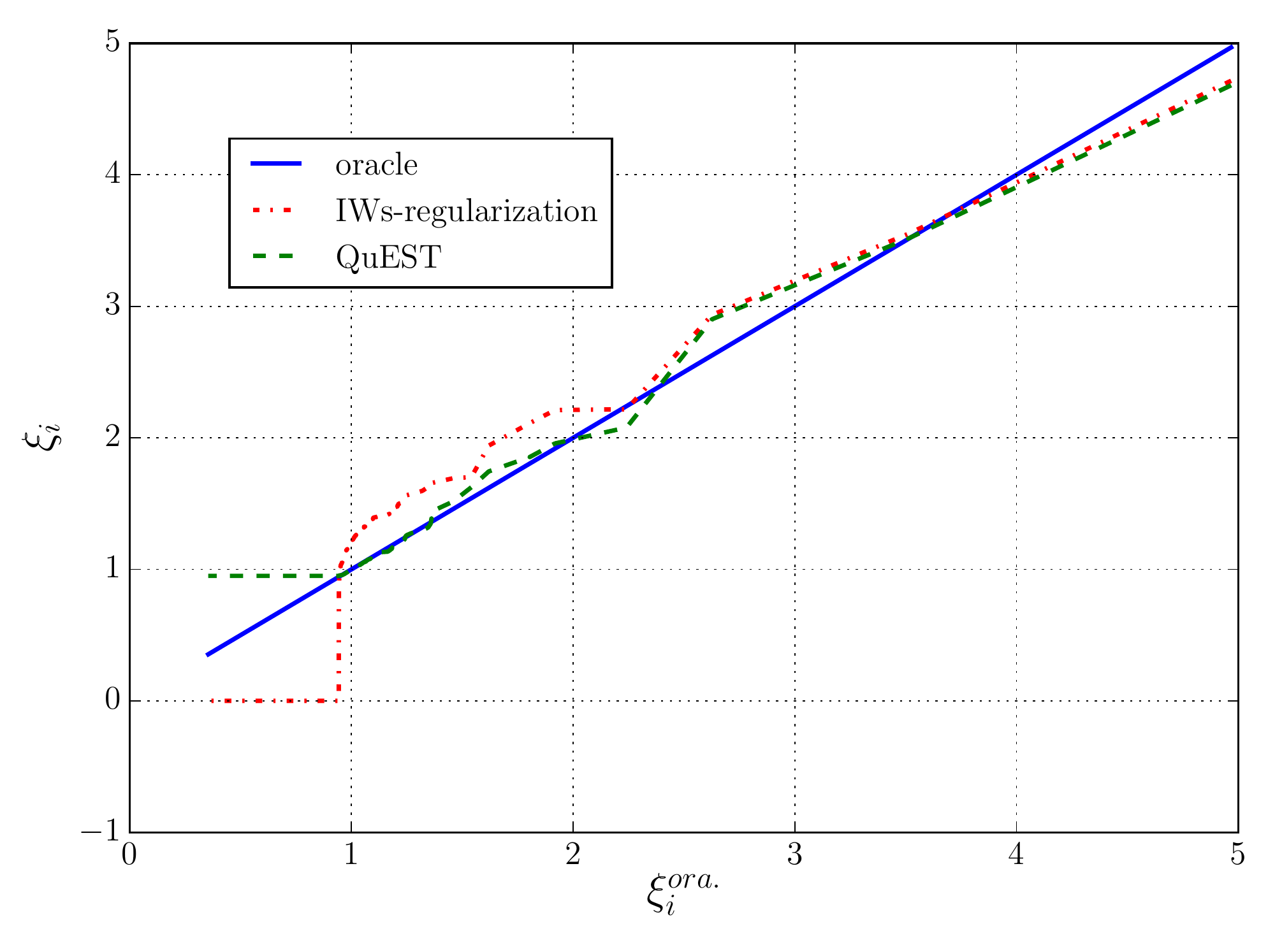} 
   \end{center}
   \caption{We apply the IWs (red dash-dotted line) and QuEST (green dashed line) regularization of Chapter \ref{chap:numerical} as a function of the oracle estimator \eqref{eq:oracle} with $\rho_\C$ given by Eq.\ \eqref{eq:power law proxy} with $\lambda_0 = 0.8$ and $N = 1000$. The sample covariance matrix $\E$ is a Wishart matrix with $q=2$. We see that both regularizations provide results that are far from the optimal solution (blue plain line). }
   \label{fig:rie_q_2}
\end{figure}

As observed in Chapter \ref{chap:numerical}, the optimal RIE \eqref{eq:rie_reg_invW} returns very satisfactory results in terms of estimating the oracle estimator either with synthetic or real data when the sample size is greater than the number of variables. However, it may happen in practice that one is confronted to the case where $N > T$ in which the sample covariance matrix $\E$ has generically $N-T$ zero eigenvalues. The main difficulty is to interpret these null eigenvalues since they could either be due to the fact we do not have enough data points, or else that $\C$ has some exact zero modes. It is therefore not surprising that both regularizations schemes of Chapter \ref{chap:numerical} fail to estimate correctly the small eigenvalues in this case (see Figure \ref{fig:rie_q_2}). However, they fail in different ways: the IWs-regularization leaves zero eigenvalues unaltered while the QuEST algorithm shrinks the small eigenvalues upwards too much. 

A naive and ad-hoc approach to this problem when $\C$ has \emph{no} zero mode is to rescale the $N-T$ zero eigenvalues of the IWs-regularization by a constant so that the trace of the estimator is equal to $N$, as it should be. This is similar to the clipping procedure of Section \ref{sec:past_cleaning}. We see that the main problem with this simple recipe is that when $\C$ has some exact zero modes, then we will always overestimate the volatility of these zero risk modes. Hence, at this stage, it seems that there are no satisfactory systematic cleaning recipe when $q > 1$, in the absence of 
some information about the possibility of true zero modes. 

\subsection{A Brownian Motion model for correlated Wishart matrices}

We present in Appendix \ref{app:addition} that Dyson's Brownian Motion that offers a nice physical interpretation of dynamics of the sample eigenvalues and eigenvectors in the case of 
an {\it additive} noise. It also provides a straightforward tool to compute the dynamics of the resolvent of the sample matrix; Eq. \eqref{eq:dGOE_resolvent_DBM} is quite remarkable in that eigenvectors' overlaps may be easily inferred. 

We are not aware of a similar result in the multiplicative case, with sample covariance matrices in mind, although Eq.\ \eqref{eq:MP_equation_diff} suggest that such a process should
exist. In the case where $\C=\In$, Bru's Wishart process \cite{bru1991wishart} allows one to obtain many interesting properties about both the eigenvalues and eigenvectors -- see \cite{allez2012invariant,bourgade2013eigenvector}, but time in this case is not related to the quality parameter $q$, as one would like it to be. This question is quite fundamental and also has practical applications, as it would for example allow to understand the overlap of the eigenvectors of $\E$ at different ``times'' (see e.g.\ \cite{allez2013eigenvectors,allez2012eigenvector} for a related question in the additive model). As this review was being completed, we managed to characterize this process, and the reader is referred to \cite{bun2016dyson} for details.

\section*{Acknowledgment}

We want to warmly thank all our collaborators on the different topics considered in this review, in particular Romain Allez, Antti Knowles and Satya N. Majumdar. We also acknowledge insightful discussions with Marc Abeille, Jean-Yves Audibert, Yanis Bahroun, Florent Benaych-Georges, Rapha{\"e}l Benichou, Alexios Beveratos, Giulio Biroli, R{\'e}my Chicheportiche, Benoit Collins, Romain Couillet, Noureddine El Karoui, Sandrine P{\'e}ch{\'e}, Adam Rej, Emmanuel S{\'e}ri{\'e}, Guillaume Simon and Denis Ullmo.

\clearpage\section{References}
\bibliography{RMT_Covariance_Review}

\begin{thebibliography}{100}
\expandafter\ifx\csname url\endcsname\relax
  \def\url#1{\texttt{#1}}\fi
\expandafter\ifx\csname urlprefix\endcsname\relax\def\urlprefix{URL }\fi
\expandafter\ifx\csname href\endcsname\relax
  \def\href#1#2{#2} \def\path#1{#1}\fi

\bibitem{van2000asymptotic}
A.~W. Van~der Vaart, Asymptotic statistics, Vol.~3, Cambridge university press,
  2000.

\bibitem{amemiya1985advanced}
T.~Amemiya, Advanced econometrics, Harvard University Press, 1985.

\bibitem{hansen1982large}
L.~P. Hansen, Large sample properties of generalized method of moments
  estimators, Econometrica: Journal of the Econometric Society (1982)
  1029--1054.

\bibitem{friedman2001elements}
J.~Friedman, T.~Hastie, R.~Tibshirani, The elements of statistical learning,
  Vol.~1, Springer series in statistics Springer, Berlin, 2001.

\bibitem{markowitz1952portfolio}
H.~Markowitz, Portfolio selection, The journal of finance 7~(1) (1952) 77--91.

\bibitem{caccioli2015portfolio}
F.~Caccioli, I.~Kondor, G.~Papp, Portfolio optimization under expected
  shortfall: contour maps of estimation error, arXiv preprint arXiv:1510.04943.

\bibitem{ciliberti2007feasibility}
S.~Ciliberti, I.~Kondor, M.~M{\'e}zard, On the feasibility of portfolio
  optimization under expected shortfall, Quantitative Finance 7~(4) (2007)
  389--396.

\bibitem{wishart1928generalised}
J.~{W}ishart, The generalised product moment distribution in samples from a
  normal multivariate population, Biometrika (1928) 32--52.

\bibitem{anderson1984introduction}
T.~W. Anderson, An introduction to multivariate statistics, Wiley, New York,
  1984.

\bibitem{stein1956inadmissibility}
C.~Stein, Inadmissibility of the usual estimator for the mean of a multivariate
  normal distribution, in: Proceedings of the Third Berkeley symposium on
  mathematical statistics and probability, Vol.~1, 1956, pp. 197--206.

\bibitem{efron1977stein}
B.~Efron, C.~N. Morris, Stein's paradox in statistics, WH Freeman, 1977.

\bibitem{james1961estimation}
W.~James, C.~Stein, Estimation with quadratic loss, in: Proceedings of the
  fourth Berkeley symposium on mathematical statistics and probability, Vol.~1,
  1961, pp. 361--379.

\bibitem{efron1976multivariate}
B.~Efron, C.~Morris, Multivariate empirical {B}ayes and estimation of
  covariance matrices, The Annals of Statistics (1976) 22--32.

\bibitem{haff1977minimax}
L.~Haff, Minimax estimators for a multinormal precision matrix, Journal of
  Multivariate Analysis 7~(3) (1977) 374--385.

\bibitem{haff1980empirical}
L.~Haff, Empirical {B}ayes estimation of the multivariate normal covariance
  matrix, The Annals of Statistics (1980) 586--597.

\bibitem{ledoit2004well}
O.~Ledoit, M.~Wolf, A well-conditioned estimator for large-dimensional
  covariance matrices, Journal of multivariate analysis 88~(2) (2004) 365--411.

\bibitem{marchenko1967distribution}
V.~A. Marchenko, L.~A. Pastur, Distribution of eigenvalues for some sets of
  random matrices, Matematicheskii Sbornik 114~(4) (1967) 507--536.

\bibitem{anderson1963asymptotic}
T.~W. Anderson, Asymptotic theory for principal component analysis, Annals of
  Mathematical Statistics (1963) 122--148.

\bibitem{yin1986limiting}
Y.~Q. Yin, Limiting spectral distribution for a class of random matrices,
  Journal of multivariate analysis 20~(1) (1986) 50--68.

\bibitem{silverstein1995strong}
J.~W. Silverstein, Strong convergence of the empirical distribution of
  eigenvalues of large dimensional random matrices, Journal of Multivariate
  Analysis 55~(2) (1995) 331--339.

\bibitem{sengupta1999distributions}
A.~Sengupta, P.~P. Mitra, Distributions of singular values for some random
  matrices, Physical Review E 60~(3) (1999) 3389.

\bibitem{laloux1999noise}
L.~Laloux, P.~Cizeau, J.-P. Bouchaud, M.~Potters, Noise dressing of financial
  correlation matrices, Physical review letters 83~(7) (1999) 1467.

\bibitem{plerou2002random}
V.~Plerou, P.~Gopikrishnan, B.~Rosenow, L.~A.~N. Amaral, T.~Guhr, H.~E.
  Stanley, Random matrix approach to cross correlations in financial data,
  Physical Review E 65~(6) (2002) 066126.

\bibitem{bouchaud2003theory}
J.-P. Bouchaud, M.~Potters, Theory of financial risk and derivative pricing:
  from statistical physics to risk management, Cambridge university press,
  2003.

\bibitem{johnstone2001distribution}
I.~M. Johnstone, On the distribution of the largest eigenvalue in principal
  components analysis, Annals of statistics (2001) 295--327.

\bibitem{tracy1994level}
C.~A. Tracy, H.~Widom, Level-spacing distributions and the {A}iry kernel,
  Communications in Mathematical Physics 159~(1) (1994) 151--174.

\bibitem{laloux2000random}
L.~Laloux, P.~Cizeau, M.~Potters, J.-P. Bouchaud, Random matrix theory and
  financial correlations, International Journal of Theoretical and Applied
  Finance 3~(03) (2000) 391--397.

\bibitem{bouchaud2009financial}
J.-P. Bouchaud, M.~Potters, Financial applications of random matrix theory: a
  short review, in: The Oxford handbook of Random Matrix Theory, Oxford
  University Press, 2011, pp. 824--850.

\bibitem{silverstein1995analysis}
J.~W. Silverstein, S.-I. Choi, Analysis of the limiting spectral distribution
  of large dimensional random matrices, Journal of Multivariate Analysis 54~(2)
  (1995) 295--309.

\bibitem{mestre2008improved}
X.~Mestre, Improved estimation of eigenvalues and eigenvectors of covariance
  matrices using their sample estimates, Information Theory, IEEE Transactions
  on 54~(11) (2008) 5113--5129.

\bibitem{yao2012eigenvalue}
J.~Yao, A.~Kammoun, J.~Najim, Eigenvalue estimation of parameterized covariance
  matrices of large dimensional data, Signal Processing, IEEE Transactions on
  60~(11) (2012) 5893--5905.

\bibitem{el2008spectrum}
N.~El~Karoui, Spectrum estimation for large dimensional covariance matrices
  using random matrix theory, The Annals of Statistics (2008) 2757--2790.

\bibitem{silverstein1986eigenvalues}
J.~W. Silverstein, Eigenvalues and eigenvectors of large dimensional sample
  covariance matrices, Contemporary Mathematics 50 (1986) 153--159.

\bibitem{silverstein1989eigenvectors}
J.~W. Silverstein, On the eigenvectors of large dimensional sample covariance
  matrices, Journal of multivariate analysis 30~(1) (1989) 1--16.

\bibitem{paul2007asymptotics}
D.~Paul, Asymptotics of sample eigenstructure for a large dimensional spiked
  covariance model, Statistica Sinica (2007) 1617--1642.

\bibitem{ledoit2011eigenvectors}
O.~Ledoit, S.~P{\'e}ch{\'e}, Eigenvectors of some large sample covariance
  matrix ensembles, Probability Theory and Related Fields 151~(1-2) (2011)
  233--264.

\bibitem{bun2015rotational}
J.~Bun, R.~Allez, J.-P. Bouchaud, M.~Potters, Rotational invariant estimator
  for general noisy matrices, arXiv preprint arXiv:1502.06736.

\bibitem{bun2016optimal}
J.~Bun, A.~Knowles, An optimal rotational invariant estimator for general
  covariance matrices, in preparation.

\bibitem{benaych2011eigenvalues}
F.~Benaych-Georges, R.~R. Nadakuditi, The eigenvalues and eigenvectors of
  finite, low rank perturbations of large random matrices, Advances in
  Mathematics 227~(1) (2011) 494--521.

\bibitem{monasson2015estimating}
R.~Monasson, D.~Villamaina, Estimating the principal components of correlation
  matrices from all their empirical eigenvectors, EPL (Europhysics Letters)
  112~(5) (2015) 50001.

\bibitem{brezin1978planar}
E.~Br{\'e}zin, C.~Itzykson, G.~Parisi, J.-B. Zuber, Planar diagrams,
  Communications in Mathematical Physics 59~(1) (1978) 35--51.

\bibitem{wigner1951statistical}
E.~P. {W}igner, On the statistical distribution of the widths and spacings of
  nuclear resonance levels, in: Mathematical Proceedings of the Cambridge
  Philosophical Society, Vol.~47, Cambridge Univ Press, 1951, pp. 790--798.

\bibitem{voiculescu1985symmetries}
D.~Voiculescu, Symmetries of some reduced free product {C}*-algebras, Springer,
  1985.

\bibitem{voiculescu1991limit}
D.~Voiculescu, Limit laws for random matrices and free products, Inventiones
  mathematicae 104~(1) (1991) 201--220.

\bibitem{edwards1976eigenvalue}
S.~Edwards, R.~C. Jones, The eigenvalue spectrum of a large symmetric random
  matrix, Journal of Physics A: Mathematical and General 9~(10) (1976) 1595.

\bibitem{mezard1987spin}
M.~M{\'e}zard, M.~A. Virasoro, G.~Parisi, Spin glass theory and beyond, World
  Scientific, 1987.

\bibitem{morone2014replica}
F.~Morone, F.~Caltagirone, E.~Harrison, G.~Parisi, Replica theory and spin
  glasses, arXiv preprint arXiv:1409.2722.

\bibitem{weidenmuller2009random}
H.~Weidenm{\"u}ller, G.~Mitchell, Random matrices and chaos in nuclear physics:
  Nuclear structure, Reviews of Modern Physics 81~(2) (2009) 539.

\bibitem{akemann2011oxford}
G.~Akemann, J.~Baik, P.~Di~Francesco, The Oxford handbook of random matrix
  theory, Oxford University Press, 2011.

\bibitem{dean2006large}
D.~S. Dean, S.~N. Majumdar, Large deviations of extreme eigenvalues of random
  matrices, Physical review letters 97~(16) (2006) 160201.

\bibitem{majumdar2006random}
S.~Majumdar, Random matrices, the {U}lam problem, directed polymers and growth
  models, and sequence matching, chapter 4, Les Houches-Session LXXXV. Elsevier
  (2006) 179--216.

\bibitem{anderson2010introduction}
G.~W. Anderson, A.~Guionnet, O.~Zeitouni, An introduction to random matrices,
  Cambridge University Press, 2010.

\bibitem{terence2012topics}
T.~Tao, Topics in random matrix theory, Vol. 132, American Mathematical Soc.,
  2012.

\bibitem{tulino2004random}
A.~M. Tulino, S.~Verd{\'u}, Random matrix theory and wireless communications,
  Communications and Information theory 1~(1) (2004) 1--182.

\bibitem{bai2009spectral}
Z.~Bai, J.~W. Silverstein, Spectral analysis of large dimensional random
  matrices, Springer, 2009.

\bibitem{couillet2011random}
R.~Couillet, M.~Debbah, et~al., Random matrix methods for wireless
  communications, Cambridge University Press Cambridge, MA, 2011.

\bibitem{hotelling1936relations}
H.~Hotelling, Relations between two sets of variates, Biometrika 28~(3/4)
  (1936) 321--377.

\bibitem{wachter1980limiting}
K.~W. Wachter, The limiting empirical measure of multiple discriminant ratios,
  The Annals of Statistics (1980) 937--957.

\bibitem{bouchaud2007large}
J.-P. Bouchaud, L.~Laloux, M.~A. Miceli, M.~Potters, Large dimension
  forecasting models and random singular value spectra, The European Physical
  Journal B 55~(2) (2007) 201--207.

\bibitem{tao2011random}
T.~Tao, V.~Vu, Random matrices: universality of local eigenvalue statistics,
  Acta mathematica 206~(1) (2011) 127--204.

\bibitem{bordenave2012}
B.~Charles, D.~Chafai, Around the circular law, Probability Surveys 9 (2012)
  1--89.

\bibitem{voiculescu1992free}
D.~Voiculescu, K.~Dykema, A.~Nica, Free random variables, American Mathematical
  Soc., 1992.

\bibitem{speicher2009free}
R.~Speicher, Free probability theory, in: The Oxford handbook of Random Matrix
  Theory, Oxford University Press, 2011, pp. 452--470.

\bibitem{zee1996law}
A.~Zee, Law of addition in random matrix theory, Nuclear Physics B 474~(3)
  (1996) 726--744.

\bibitem{burda2013free}
Z.~Burda, Free products of large random matrices--a short review of recent
  developments, in: Journal of Physics: Conference Series, Vol. 473, IOP
  Publishing, 2013, p. 012002.

\bibitem{burda2004signal}
Z.~Burda, A.~G{\"o}rlich, A.~Jarosz, J.~Jurkiewicz, Signal and noise in
  correlation matrix, Physica A: Statistical Mechanics and its Applications 343
  (2004) 295--310.

\bibitem{guhr2010supersymmetry}
T.~Guhr, Supersymmetry, in: The Oxford handbook of Random Matrix Theory, Oxford
  University Press, 2011, pp. 135--154.

\bibitem{mehta2004random}
M.~L. Mehta, Random matrices, Vol. 142, Academic press, 2004.

\bibitem{dean2008extreme}
D.~S. Dean, S.~N. Majumdar, Extreme value statistics of eigenvalues of
  {G}aussian random matrices, Physical Review E 77~(4) (2008) 041108.

\bibitem{zuber2012introduction}
J.-B. Zuber, Introduction to random matrices (2012).

\bibitem{allez2012invariant}
R.~Allez, J.-P. Bouchaud, S.~N. Majumdar, P.~Vivo, Invariant $\beta$-{W}ishart
  ensembles, crossover densities and asymptotic corrections to the
  {M}ar{\v{c}}enko--{P}astur law, Journal of Physics A: Mathematical and
  Theoretical 46~(1) (2012) 015001.

\bibitem{cizeau1994theory}
P.~Cizeau, J.-P. Bouchaud, Theory of {L}{\'e}vy matrices, Physical Review E
  50~(3) (1994) 1810.

\bibitem{arous2008spectrum}
G.~Ben~Arous, A.~Guionnet, The spectrum of heavy tailed random matrices,
  Communications in Mathematical Physics 278~(3) (2008) 715--751.

\bibitem{tarquini2016level}
E.~Tarquini, G.~Biroli, M.~Tarzia, Level statistics and localization
  transitions of {L}{\'e}vy matrices, Physical Review Letters 116~(1) (2016)
  010601.

\bibitem{bai1993convergence}
Z.~D. Bai, Convergence rate of expected spectral distributions of large random
  matrices. part i. {W}igner matrices, The Annals of Probability (1993)
  625--648.

\bibitem{bai2003convergence}
Z.~Bai, B.~Miao, J.-F. Yao, Convergence rates of spectral distributions of
  large sample covariance matrices, SIAM journal on matrix analysis and
  applications 25~(1) (2003) 105--127.

\bibitem{dwyer1967some}
P.~S. Dwyer, Some applications of matrix derivatives in multivariate analysis,
  Journal of the American Statistical Association 62~(318) (1967) 607--625.

\bibitem{haff1979identity}
L.~Haff, An identity for the {W}ishart distribution with applications, Journal
  of Multivariate Analysis 9~(4) (1979) 531--544.

\bibitem{speicher1994multiplicative}
R.~Speicher, Multiplicative functions on the lattice of non-crossing partitions
  and free convolution, Mathematische Annalen 298~(1) (1994) 611--628.

\bibitem{hooft1974planar}
G.~Hooft, A planar diagram theory for strong interactions, Nuclear Physics B
  72~(3) (1974) 461--473.

\bibitem{khorunzhy1994eigenvalue}
A.~M. Khorunzhy, L.~Pastur, On the eigenvalue distribution of the deformed
  {W}igner ensemble of random matrices, Advances in Soviet Mathematics 19
  (1994) 97--127.

\bibitem{brezin1995universalb}
E.~Br{\'e}zin, S.~Hikami, A.~Zee, Universal correlations for deterministic plus
  random {H}amiltonians, Physical Review E 51~(6) (1995) 5442.

\bibitem{zinn1999adding}
P.~Zinn-Justin, Adding and multiplying random matrices: a generalization of
  voiculescu’s formulas, Physical Review E 59~(5) (1999) 4884.

\bibitem{burda2011multiplication}
Z.~Burda, R.~Janik, M.~Nowak, Multiplication law and {S} transform for
  non-hermitian random matrices, Physical Review E 84~(6) (2011) 061125.

\bibitem{burda2004spectral}
Z.~Burda, J.~Jurkiewicz, B.~Wac{\l}aw, Spectral moments of correlated {W}ishart
  matrices, Physical Review E 71~(2) (2005) 026111.

\bibitem{parisi1980sequence}
G.~Parisi, A sequence of approximated solutions to the sk model for spin
  glasses, Journal of Physics A: Mathematical and General 13~(4) (1980) L115.

\bibitem{harish1957differential}
Harish-Chandra, Differential operators on a semisimple lie algebra, American
  Journal of Mathematics (1957) 87--120.

\bibitem{itzykson1980planar}
C.~Itzykson, J.-B. Zuber, The planar approximation. ii, Journal of Mathematical
  Physics 21 (1980) 411--421.

\bibitem{talagrand2006parisi}
M.~Talagrand, The parisi formula, Annals of Mathematics 163 (2006) 221--263.

\bibitem{benaych2016lectures}
F.~Benaych-Georges, A.~Knowles, Lectures on the local semicircle law for
  {W}igner matrices, arXiv preprint arXiv:1601.04055.

\bibitem{kargin2015subordination}
V.~Kargin, Subordination for the sum of two random matrices, The Annals of
  Probability 43~(4) (2015) 2119--2150.

\bibitem{paul2014random}
D.~Paul, A.~Aue, Random matrix theory in statistics: a review, Journal of
  Statistical Planning and Inference 150 (2014) 1--29.

\bibitem{vivo2007large}
P.~Vivo, S.~N. Majumdar, O.~Bohigas, Large deviations of the maximum eigenvalue
  in {W}ishart random matrices, Journal of Physics A: Mathematical and
  Theoretical 40~(16) (2007) 4317.

\bibitem{majumdar2012number}
S.~N. Majumdar, P.~Vivo, Number of relevant directions in principal component
  analysis and {W}ishart random matrices, Physical review letters 108~(20)
  (2012) 200601.

\bibitem{perret2015finite}
A.~Perret, G.~Schehr, Finite n corrections to the limiting distribution of the
  smallest eigenvalue of {W}ishart complex matrices, Random Matrices: Theory
  and Applications (2015) 1650001.

\bibitem{wirtz2013distribution}
T.~Wirtz, T.~Guhr, Distribution of the smallest eigenvalue in the correlated
  {W}ishart model, Physical review letters 111~(9) (2013) 094101.

\bibitem{bloemendal2014principal}
A.~Bloemendal, A.~Knowles, H.-T. Yau, J.~Yin, On the principal components of
  sample covariance matrices, arXiv preprint arXiv:1404.0788.

\bibitem{huber2011robust}
P.~J. Huber, Robust statistics, Springer, 2011.

\bibitem{maronna2006robust}
R.~A. Maronna, D.~R. Martin, V.~J. Yohai, Robust Statistics: Theory and
  Methods, John Wiley and Sons, 2006.

\bibitem{biroli2007student}
G.~Biroli, J.-P. Bouchaud, M.~Potters, The {S}tudent ensemble of correlation
  matrices: eigenvalue spectrum and kullback-leibler entropy, arXiv preprint
  arXiv:0710.0802.

\bibitem{el2009concentration}
N.~El~Karoui, Concentration of measure and spectra of random matrices:
  applications to correlation matrices, elliptical distributions and beyond,
  The Annals of Applied Probability 19~(6) (2009) 2362--2405.

\bibitem{chicheportiche2013non}
R.~Chicheportiche, Non-linear dependences in finance, arXiv preprint
  arXiv:1309.5073.

\bibitem{couillet2015random}
R.~Couillet, F.~Pascal, J.~W. Silverstein, The random matrix regime of
  {M}aronna’s {M}-estimator with elliptically distributed samples, Journal of
  Multivariate Analysis 139 (2015) 56--78.

\bibitem{tyler1987distribution}
D.~E. Tyler, A distribution-free $ m $-estimator of multivariate scatter, The
  Annals of Statistics 15~(1) (1987) 234--251.

\bibitem{zhang2014marchenko}
T.~Zhang, X.~Cheng, A.~Singer, Marchenko-{P}astur law for {T}yler's and
  {M}aronna's {M}-estimators, arXiv preprint arXiv:1401.3424.

\bibitem{couillet2016second}
R.~Couillet, A.~Kammoun, F.~Pascal, Second order statistics of robust
  estimators of scatter. application to glrt detection for elliptical signals,
  Journal of Multivariate Analysis 143 (2016) 249--274.

\bibitem{silverstein1995empirical}
J.~W. Silverstein, Z.~Bai, On the empirical distribution of eigenvalues of a
  class of large dimensional random matrices, Journal of Multivariate analysis
  54~(2) (1995) 175--192.

\bibitem{mingo2004annular}
J.~A. Mingo, A.~Nica, Annular noncrossing permutations and partitions, and
  second-order asymptotics for random matrices, International Mathematics
  Research Notices 2004~(28) (2004) 1413--1460.

\bibitem{ledoit2013spectrum}
O.~Ledoit, M.~Wolf, Spectrum estimation: A unified framework for covariance
  matrix estimation and pca in large dimensions, Tech. rep., Working Paper
  Series, Department of Economics, University of Zurich (2013).

\bibitem{bun2016dyson}
J.~Bun, J.-P. Bouchaud, M.~Potters, A {D}yson {B}rownian motion for the
  resolvant of {W}igner and {W}ishart random matrices, in preparation.

\bibitem{knowles2014anisotropic}
A.~Knowles, J.~Yin, Anisotropic local laws for random matrices, arXiv preprint
  arXiv:1410.3516.

\bibitem{biroli2007top}
G.~Biroli, J.-P. Bouchaud, M.~Potters, On the top eigenvalue of heavy-tailed
  random matrices, EPL (Europhysics Letters) 78~(1) (2007) 10001.

\bibitem{bowick1991universal}
M.~J. Bowick, {\'E}.~Br{\'e}zin, Universal scaling of the tail of the density
  of eigenvalues in random matrix models, Physics Letters B 268~(1) (1991)
  21--28.

\bibitem{majumdar2014top}
S.~N. Majumdar, G.~Schehr, Top eigenvalue of a random matrix: large deviations
  and third order phase transition, Journal of Statistical Mechanics: Theory
  and Experiment 2014~(1) (2014) P01012.

\bibitem{nadal2011simple}
C.~Nadal, S.~N. Majumdar, A simple derivation of the tracy--widom distribution
  of the maximal eigenvalue of a {G}aussian unitary random matrix, Journal of
  Statistical Mechanics: Theory and Experiment 2011~(04) (2011) P04001.

\bibitem{ramirez2011beta}
J.~Ramirez, B.~Rider, B.~Vir{\'a}g, Beta ensembles, stochastic airy spectrum,
  and a diffusion, Journal of the American Mathematical Society 24~(4) (2011)
  919--944.

\bibitem{johansson2000shape}
K.~Johansson, Shape fluctuations and random matrices, Communications in
  mathematical physics 209~(2) (2000) 437--476.

\bibitem{baik2005phase}
J.~Baik, G.~Ben~Arous, S.~P{\'e}ch{\'e}, Phase transition of the largest
  eigenvalue for nonnull complex sample covariance matrices, Annals of
  Probability (2005) 1643--1697.

\bibitem{peche2009universality}
S.~P{\'e}ch{\'e}, Universality results for the largest eigenvalues of some
  sample covariance matrix ensembles, Probability Theory and Related Fields
  143~(3-4) (2009) 481--516.

\bibitem{peche2003universality}
S.~P{\'e}ch{\'e}, Universality of local eigenvalue statistics for random sample
  covariance matrices, Ph.D. thesis, EPFL (2003).

\bibitem{hachem2015survey}
W.~Hachem, A.~Hardy, J.~Najim, A survey on the eigenvalues local behavior of
  large complex correlated {W}ishart matrices, ESAIM: Proceedings and Surveys
  51 (2015) 150--174.

\bibitem{allez2014eigenvectors}
R.~Allez, J.~Bun, J.-P. Bouchaud, The eigenvectors of {G}aussian matrices with
  an external source, arXiv preprint arXiv:1412.7108.

\bibitem{allez2013eigenvectors}
R.~Allez, J.-P. Bouchaud, Eigenvector dynamics under free addition, Random
  Matrices: Theory and Applications 03 (2014) 1450010.

\bibitem{bun2016overlaps}
J.~Bun, J.-P. Bouchaud, M.~Potters, On the overlaps between eigenvectors of
  correlated random matrices, arXiv preprint arXiv:1603.04364.

\bibitem{hoyle2003limiting}
D.~Hoyle, M.~Rattray, Limiting form of the sample covariance eigenspectrum in
  pca and kernel pca, in: Advances in Neural Information Processing Systems,
  2003, p. None.

\bibitem{weyl1949inequalities}
H.~Weyl, Inequalities between the two kinds of eigenvalues of a linear
  transformation, Proceedings of the national academy of sciences 35~(7) (1949)
  408--411.

\bibitem{deutsch1991quantum}
J.~M. Deutsch, Quantum statistical mechanics in a closed system, Physical
  Review A 43~(4) (1991) 2046.

\bibitem{gelman2014bayesian}
A.~Gelman, J.~B. Carlin, H.~S. Stern, D.~B. Rubin, {B}ayesian data analysis,
  Vol.~2, Taylor \& Francis, 2014.

\bibitem{takemura1983orthogonally}
A.~Takemura, An orthogonally invariant minimax estimator of the covariance
  matrix of a multivariate normal population, Tech. rep., DTIC Document (1983).

\bibitem{karoui2011geometric}
N.~El~Karoui, H.~K{\"o}sters, Geometric sensitivity of random matrix results:
  consequences for shrinkage estimators of covariance and related statistical
  methods, arXiv preprint arXiv:1105.1404.

\bibitem{dicker2016ridge}
L.~H. Dicker, et~al., Ridge regression and asymptotic minimax estimation over
  spheres of growing dimension, Bernoulli 22~(1) (2016) 1--37.

\bibitem{wiener1949extrapolation}
N.~Wiener, Extrapolation, interpolation, and smoothing of stationary time
  series, Vol.~2, MIT press Cambridge, MA, 1949.

\bibitem{pafka2003noisy}
S.~Pafka, I.~Kondor, Noisy covariance matrices and portfolio optimization ii,
  Physica A: Statistical Mechanics and its Applications 319 (2003) 487--494.

\bibitem{collins2013compound}
B.~Collins, D.~McDonald, N.~Saad, Compound {W}ishart matrices and noisy
  covariance matrices: Risk underestimation, arXiv preprint arXiv:1306.5510.

\bibitem{ledoit2014nonlinear}
O.~Ledoit, M.~Wolf, Nonlinear shrinkage of the covariance matrix for portfolio
  selection: {M}arkowitz meets {G}oldilocks, Available at SSRN 2383361.

\bibitem{bartz2015advances}
D.~Bartz, Advances in high-dimensional covariance matrix estimation, Technische
  Universit{\"a}t Berlin, Doctoral Thesis.

\bibitem{burda2004free}
Z.~Burda, J.~Jurkiewicz, M.~A. Nowak, G.~Papp, I.~Zahed, Free {L}{\'e}vy
  matrices and financial correlations, Physica A: Statistical Mechanics and its
  Applications 343 (2004) 694--700.

\bibitem{bartz2014covariance}
D.~Bartz, K.-R. M{\"u}ller, Covariance shrinkage for autocorrelated data, in:
  Advances in Neural Information Processing Systems, 2014, pp. 1592--1600.

\bibitem{bartz2013directional}
D.~Bartz, K.~Hatrick, C.~W. Hesse, K.-R. M{\"u}ller, S.~Lemm, Directional
  variance adjustment: Bias reduction in covariance matrices based on factor
  analysis with an application to portfolio optimization, PloS one 8~(7) (2013)
  e67503.

\bibitem{marsili2002dissecting}
M.~Marsili, Dissecting financial markets: sectors and states, Quantitative
  Finance 2~(4) (2002) 297--302.

\bibitem{ledoit2016numerical}
O.~Ledoit, M.~Wolf, Numerical implementation of the quest function, Tech. rep.,
  Department of Economics-University of Zurich (2016).

\bibitem{mccullagh1989generalized}
P.~McCullagh, J.~A. Nelder, Generalized linear models, Vol.~37, CRC press,
  1989.

\bibitem{chamberlain1982arbitrage}
G.~Chamberlain, M.~Rothschild, Arbitrage, factor structure, and mean-variance
  analysis on large asset markets (1982).

\bibitem{kapetanios2004new}
G.~Kapetanios, A new method for determining the number of factors in factor
  models with large datasets, Tech. rep., Working Paper, Department of
  Economics, Queen Mary, University of London (2004).

\bibitem{onatski2010determining}
A.~Onatski, Determining the number of factors from empirical distribution of
  eigenvalues, The Review of Economics and Statistics 92~(4) (2010) 1004--1016.

\bibitem{paul2009no}
D.~Paul, J.~W. Silverstein, No eigenvalues outside the support of the limiting
  empirical spectral distribution of a separable covariance matrix, Journal of
  Multivariate Analysis 100~(1) (2009) 37--57.

\bibitem{merton1973intertemporal}
R.~C. Merton, An intertemporal capital asset pricing model, Econometrica:
  Journal of the Econometric Society (1973) 867--887.

\bibitem{fama1993common}
E.~F. Fama, K.~R. French, Common risk factors in the returns on stocks and
  bonds, Journal of financial economics 33~(1) (1993) 3--56.

\bibitem{tanskanen2016random}
A.~Tanskanen, J.~Lukkarinen, K.~Vatanen, Random factor approach for large sets
  of equity time-series, arXiv preprint arXiv:1604.05896.

\bibitem{chicheportiche2015nested}
R.~Chicheportiche, J.-P. Bouchaud, A nested factor model for non-linear
  dependencies in stock returns, Quantitative Finance 15~(11) (2015) 1--16.

\bibitem{bun2016cleaning}
J.~Bun, J.-P. Bouchaud, M.~Potters, Cleaning correlation matrices, Risk
  magazine.

\bibitem{markowitz1968portfolio}
H.~M. Markowitz, Portfolio selection: efficient diversification of investments,
  Vol.~16, Yale University Press, 1968.

\bibitem{tumminello2007kullback}
M.~Tumminello, F.~Lillo, R.~N. Mantegna, Kullback-leibler distance as a measure
  of the information filtered from multivariate data, Physical Review E 76~(3)
  (2007) 031123.

\bibitem{ledoit2003improved}
O.~Ledoit, M.~Wolf, Improved estimation of the covariance matrix of stock
  returns with an application to portfolio selection, Journal of Empirical
  Finance 10~(5) (2003) 603--621.

\bibitem{pantaleo2011improved}
E.~Pantaleo, M.~Tumminello, F.~Lillo, R.~N. Mantegna, When do improved
  covariance matrix estimators enhance portfolio optimization? an empirical
  comparative study of nine estimators, Quantitative Finance 11~(7) (2011)
  1067--1080.

\bibitem{allez2012eigenvector}
R.~Allez, J.-P. Bouchaud, Eigenvector dynamics: general theory and some
  applications, Physical Review E 86~(4) (2012) 046202.

\bibitem{schmitt2013non}
T.~A. Schmitt, D.~Chetalova, R.~Sch{\"a}fer, T.~Guhr, Non-stationarity in
  financial time series: Generic features and tail behavior, EPL (Europhysics
  Letters) 103~(5) (2013) 58003.

\bibitem{wang2016average}
S.~Wang, R.~Sch{\"a}fer, T.~Guhr, Average cross-responses in correlated
  financial market, arXiv preprint arXiv:1603.01586.

\bibitem{pafka2004exponential}
S.~Pafka, M.~Potters, I.~Kondor, Exponential weighting and
  random-matrix-theory-based filtering of financial covariance matrices for
  portfolio optimization, arXiv preprint cond-mat/0402573.

\bibitem{johnstone2008multivariate}
I.~M. Johnstone, Multivariate analysis and jacobi ensembles: Largest
  eigenvalue, tracy--widom limits and rates of convergence, Annals of
  statistics 36~(6) (2008) 2638.

\bibitem{yang2015independence}
Y.~Yang, G.~Pan, et~al., Independence test for high dimensional data based on
  regularized canonical correlation coefficients, The Annals of Statistics
  43~(2) (2015) 467--500.

\bibitem{klema1980singular}
V.~C. Klema, A.~J. Laub, The singular value decomposition: Its computation and
  some applications, Automatic Control, IEEE Transactions on 25~(2) (1980)
  164--176.

\bibitem{furnas1988information}
G.~W. Furnas, S.~Deerwester, S.~T. Dumais, T.~K. Landauer, R.~A. Harshman,
  L.~A. Streeter, K.~E. Lochbaum, Information retrieval using a singular value
  decomposition model of latent semantic structure, in: Proceedings of the 11th
  annual international ACM SIGIR conference on Research and development in
  information retrieval, ACM, 1988, pp. 465--480.

\bibitem{alter2000singular}
O.~Alter, P.~O. Brown, D.~Botstein, Singular value decomposition for
  genome-wide expression data processing and modeling, Proceedings of the
  National Academy of Sciences 97~(18) (2000) 10101--10106.

\bibitem{benaych2012singular}
F.~Benaych-Georges, R.~R. Nadakuditi, The singular values and vectors of low
  rank perturbations of large rectangular random matrices, Journal of
  Multivariate Analysis 111 (2012) 120--135.

\bibitem{tumminello2010correlation}
M.~Tumminello, F.~Lillo, R.~N. Mantegna, Correlation, hierarchies, and networks
  in financial markets, Journal of Economic Behavior \& Organization 75~(1)
  (2010) 40--58.

\bibitem{dimov2012hidden}
I.~I. Dimov, P.~N. Kolm, L.~Maclin, D.~Y. Shiber, Hidden noise structure and
  random matrix models of stock correlations, Quantitative Finance 12~(4)
  (2012) 567--572.

\bibitem{bru1991wishart}
M.-F. Bru, {W}ishart processes, Journal of Theoretical Probability 4~(4) (1991)
  725--751.

\bibitem{bourgade2013eigenvector}
P.~Bourgade, H.-T. Yau, The eigenvector moment flow and local quantum unique
  ergodicty, arXiv preprint arXiv:1312.1301.

\bibitem{tao-blog}
T.~Tao,
  http://terrytao.wordpress.com/2013/02/08/the-harish-chandra-itzykson-zuber-integral-formula/
  (2013).

\bibitem{matytsin1994large}
A.~Matytsin, On the large-n limit of the {I}tzykson-{Z}uber integral, Nuclear
  Physics B 411 (1994) 805--820.

\bibitem{bun2014instanton}
J.~Bun, J.~P. Bouchaud, S.~N. Majumdar, M.~Potters, Instanton approach to large
  $n$ harish-chandra-itzykson-zuber integrals, Phys. Rev. Lett. 113 (2014)
  070201.

\bibitem{guionnet2004asymptotic}
A.~Guionnet, M.~Ma{\"i}da, A {F}ourier view on the {R}-transform and related
  asymptotics of spherical integrals, Journal of Functional Analysis 222~(2)
  (2005) 435 -- 490.

\bibitem{zuber2008large}
J.-B. Zuber, The large-n limit of matrix integrals over the orthogonal group,
  Journal of Physics A: Mathematical and Theoretical 41~(38) (2008) 382001.

\bibitem{guionnet2002large}
A.~Guionnet, O.~Zeitouni, Large deviations asymptotics for spherical integrals,
  Journal of functional analysis 188~(2) (2002) 461--515.

\bibitem{collins2009asymptotics}
B.~Collins, A.~Guionnet, E.~Maurel-Segala, Asymptotics of unitary and
  orthogonal matrix integrals, Advances in Mathematics 222~(1) (2009) 172--215.

\bibitem{tanaka2008asymptotics}
T.~Tanaka, Asymptotics of {H}arish-{C}handra-{I}tzykson-{Z}uber integrals and
  free probability theory, in: Journal of Physics: Conference Series, Vol.~95,
  IOP Publishing, 2008, p. 012002.

\bibitem{marinari1994replica}
E.~Marinari, G.~Parisi, F.~Ritort, Replica field theory for deterministic
  models. ii. a non-random spin glass with glassy behaviour, Journal of Physics
  A: Mathematical and General 27~(23) (1994) 7647.

\bibitem{erdHos2011universality}
L.~Erd{\H{o}}s, Universality of {W}igner random matrices: a survey of recent
  results, Russian Mathematical Surveys 66~(3) (2011) 507.

\bibitem{beenakker1997random}
C.~W. Beenakker, Random-matrix theory of quantum transport, Reviews of modern
  physics 69~(3) (1997) 731.

\bibitem{ithier2015thermalisation}
G.~Ithier, F.~Benaych-Georges, Thermalisation of a quantum system from first
  principles, arXiv preprint arXiv:1510.04352.

\bibitem{nandkishore2015many}
R.~Nandkishore, D.~A. Huse, Many-{Body} {Localization} and {Thermalization} in
  {Quantum} {Statistical} {Mechanics}, Annual Review of Condensed Matter
  Physics 6~(1) (2015) 15--38.

\bibitem{eisert2015quantum}
J.~Eisert, M.~Friesdorf, C.~Gogolin, Quantum many-body systems out of
  equilibrium, Nature Physics 11~(2) (2015) 124--130.

\bibitem{dyson1962brownian}
F.~J. Dyson, A brownian-motion model for the eigenvalues of a random matrix,
  Journal of Mathematical Physics 3 (1962) 1191--1198.

\bibitem{shlyakhtenko1996random}
D.~Shlyakhtenko, Random {G}aussian band matrices and freeness with
  amalgamation, International Mathematics Research Notices 1996~(20) (1996)
  1013--1025.

\end{thebibliography}

%\bibliographystyle{amsplain}

%\bibliography{mybibfile}

\appendix

\clearpage%!TEX root = RMT_Covariance_Review.tex
\section{Harish-Chandra--Itzykson-Zuber integrals}
\label{app:HCIZ}

\subsection{Definitions and results}

The (generalized) Harish-Chandra-Itzykson-Zuber (HCIZ)
integral \cite{harish1957differential,itzykson1980planar} ${\cal I}_\beta(\AAA,\BBB)$ is defined as:
\begin{equation}\label{eq:HCIZ-def}
{\cal I}_\beta(\AAA,\BBB) = \int_{G(N)} {\cal D} \b\Omega \,\, e^{\frac{\beta N}{2} \Tr  \AAA \b\Omega \BBB \b\Omega^{\dag}},
\end{equation}
where the integral is over the (flat) Haar measure of the compact group $\b\Omega \in \b G(N)=\b O(N), \b U(N)$ or $Sp(N)$ in $N$ dimensions and $\AAA,\BBB$ are arbitrary 
$N \times N$ symmetric (hermitian or symplectic) matrices. The parameter $\beta$ is the usual Dyson ``inverse temperature'', with $\beta=1,2,$ or $4$, respectively for the three groups. This integral has found several applications in many different fields, including Random Matrix Theory, disordered systems or quantum gravity (for a particularly insightful introduction, see \cite{tao-blog}). In RMT, this integral naturally appears in many problems, e.g. the derivation of the free addition and multiplication or the evaluation of eigenvalues density of states of a partition function whose potential is subject to a multiplicative external field.

In the unitary case 
$\b G(N)=\b U(N)$ and $\beta=2$, it turns out that the HCIZ integral can be expressed exactly, for all $N$, as the ratio of determinants that depend on $\AAA,\BBB$, and additional $N$-dependent prefactors:
\begin{equation}
{\cal I}_{\beta=2}(\AAA,\BBB)=\frac{c_{N}}{N^{(N^2 - N)/2}} \frac{\det\left( (e^{N a_{i} b_{j}})_{1 \le i,j \le N}\right)}{\Delta(\AAA) \Delta(\BBB)}
\end{equation}
with $\{a_i\}$, $\{b_i\}$ the eigenvalues of $\AAA$ and $\BBB$, $\Delta(\AAA) = \prod_{i<j} |a_{i} - a_{j}|$ the Vandermonde determinant of $\AAA$ 
[and, similarly, for $\Delta(\BBB)$], and $c_{N} = \prod_{i}^{N} i!$. Finding the expression of $\beta = 1$ or $\beta = 4$ is still an open problem.

Also, as is well known, determinants contain $N!$ terms of alternating signs, which makes their order of magnitude 
very hard to estimate {\it a priori}. This difficulty appears clearly when one is interested in the large $N$ asymptotic of HCIZ integrals, for which one 
would naively expect to have a simplified, explicit expression as a functional $F_2(\rho_\AAA,\rho_\BBB) = \lim_{N \to \infty} N^{-2} \ln {\cal I}_{\beta=2}(\AAA,\BBB)$ 
of the eigenvalue densities $\rho_{\AAA,\BBB}$ of $\AAA,\BBB$ \cite{matytsin1994large}. Using Dyson's Brownian motion, one can finds \cite{bun2014instanton,guionnet2004asymptotic}:
$F_{\beta=2}(\AAA,\BBB) = \lim_{N \to \infty} N^{-2} \ln \, {\cal I}_2(\AAA,\BBB)$:
\begin{equation}
\label{HCIZ_largeN} 
\nonumber F_{2}(\AAA,\BBB) = -\frac34-S_2(\AAA,\BBB) + \frac{1}{2} \int {\rm d}x\, x^2 (\rho_\AAA(x)+\rho_\BBB(x)) - \frac{1}{2} \int {\rm d}x {\rm d}y\, [\rho_\AAA(x) \rho_\AAA(y)+\rho_\BBB(x) \rho_\BBB(y)] \ln |x - y|, 
\end{equation}
where 
\begin{equation}
S_{2}(\AAA,\BBB) = \frac12 \int \dd t \int d\lambda \, \rho(\lambda,t) \left\{ v^2(\lambda,t) + \frac{\pi^2}{3} \rho^2(\lambda, t) \right\}
\end{equation}
with $\rho(\lambda,t)$ and $v(\lambda,t)$ solution of the following Euler equation
\begin{equation}
\begin{dcases}
\partial_t \rho(\lambda,t) + \partial_{\lambda} [\rho(\lambda,t)v(\lambda,t)] = 0, \\
\partial_t v(\lambda,t) + v(\lambda,t) \partial_{\lambda} v(\lambda,t) = \frac{\pi^2}{2} \partial_{\lambda} \rho^2(\lambda, t), \\
\text{with } \rho(\lambda, 0) = \rho_{\AAA}(\lambda) , \, \text{and } \, \rho(\lambda,1) = \rho_{\B}(\lambda).
\end{dcases}
\end{equation}
In fact, this result can be extended to arbitrary value of $\beta$ with the final (simple) result $F_{\beta}(\AAA,\BBB)=\beta F_{2}(\AAA,\BBB)/2$. This coincides with the result obtained by Zuber in the orthogonal case $\beta=1$ \cite{zuber2008large} 
(see also \cite{guionnet2002large,collins2009asymptotics,tanaka2008asymptotics} for arbitrary $\beta$).

Nonetheless, explicit results concerning the asymptotic of this integral are scarce. When $\AAA$ and $\BBB$ are both Wigner matrices, the Euler-Matytsin system of equation can be solved explicitly \cite{bun2014instanton}. Another soluble case is when one of the two matrix has a Flat distribution \cite{guionnet2002large}. Last but not least, a beautiful explicit result is available when one of the matrices has lower rank $n \ll N$. Precisely, let us assume that $\AAA$ has $n$ eigenvalues $a_1, a_2,\dots,a_n$ and $N-n$ zero eigenvalues. Then we have \cite{marinari1994replica,guionnet2004asymptotic,tanaka2008asymptotics}:
\begin{equation}
	\label{eq:HCIZ_rankn}
	{\cal I}_{\beta}(\AAA,\BBB) = \exp\left[ \frac{N\beta}{2} \sum_{i=1}^{n} \wtr_{\BBB}(a_i) \right],
\end{equation}
where $\wtr_{\BBB}$ is the primitive of the $\rtr$-transform of $\BBB$. This result is of particular importance when we do Replica analysis since we introduce a finite number $n$ of ``replicas'' (see Section \ref{sec:replica}). We provide hereafter a complete derivation with elementary calculus in the rank-one case in the following section and explain how to generalize it to the rank-$n$ case. 

% one can notice that Eq. (\ref{eq:replica_free_add}) is the Orthogonal low-rank version of the Harish-Chandra-Itzykson-Zuber integrals (\cite{harish1957differential}, \cite{itzykson1980planar}). The result is known for all symmetry groups (\cite{marinari1994replica, tanaka2008asymptotics} or \cite{guionnet2004asymptotic} 

\subsection{Derivation of \eqref{eq:HCIZ_rankn} in the Rank-1 case}
\label{appsec:HCIZ_rank1}

This section is devoted to the derivation of the result \eqref{eq:HCIZ_rankn} in the sample case where $\AAA = \diag(a_1, 0, \dots, 0)$ and $\BBB = \diag(b_1, \dots, b_N)$. Firstly, we rewrite \eqref{eq:HCIZ-def} (we set $\beta = 1$ for simplicity):
\begin{equation}
	\label{eq:HCIZ_rank1_integral}
	{\cal I}_{1}(\AAA,\BBB) = \frac{1}{\cal Z} \int \pBB{\prod_{k=1}^{N} \dd \Omega_{1k}} \exp\qBB{\frac{N}{2} a_1 \sum_{k=1}^{N} \Omega_{1k}^2 b_k } \delta\pBB{\sum_{k=1}^{N} \Omega_{1k}^2 - 1},
\end{equation}
where the Dirac delta function enforces the orthogonality and $\cal Z$ is normalization constant defined as:
\begin{equation}
	\label{eq:Haar_partition_fct}
	\cal Z \;\deq\; \int \pBB{\prod_{k=1}^{N} \dd \Omega_{1k}} \delta\pBB{\sum_{k=1}^{N} \Omega_{1k}^2 - 1}\;,
\end{equation}
which allows us to omit constant variables in the following. We then use the following integral representation of the delta function:
\begin{equation}
	\delta\pBB{\sum_{k=1}^{N} \Omega_{1k}^2 - 1} =  \frac{1}{2\pi} \int \exp\qBB{\ii\zeta\pB{\sum_{k=1}^{N} \Omega_{1k}^2 - 1}} \dd\zeta,
\end{equation}
so that we have (after renaming $\zeta \to -2\ii\zeta/N$)
\begin{eqnarray}
	\label{eq:HCIZ_rank1_free_energy}
	{\cal I}_{1}(\AAA,\BBB) & \propto & \frac{N}{4\pi} \int_{-\ii\infty}^{\ii\infty} \dd\zeta \int \pBB{\prod_{k=1}^{N} \dd \Omega_{1k}} \exp\qBB{\frac{N}{2} \pBB{ a_1 \sum_{k=1}^{N} \Omega_{1k}^2 b_k  + \zeta\pB{\sum_{k=1}^{N} \Omega_{1k}^2 - 1}} } \nonumber \\
	& = & \frac{N}{4\pi} \int_{-\ii\infty}^{\ii\infty} \dd\zeta \exp\qBB{\frac{N\zeta}{2}} \int \pBB{\prod_{k=1}^{N} \dd \Omega_{1k}} \exp\qBB{- \frac{N}{2} \sum_{k=1}^{N} \Omega_{1k}^2  \pB{\zeta - a_1 b_k} } \nonumber \\
	& = & \frac{N}{4\pi} \int_{-\ii\infty}^{\ii\infty} \exp\qBB{- \frac{N}{2} \pBB{ \frac{1}{N} \sum_{k=1}^{N} \log(\zeta - a_1 b_k)  - \zeta} }  \dd\zeta.
\end{eqnarray}
Since we consider $N \to \infty$, the integral over $\zeta$ is performed by a saddle-point method, leading to the following equation:
\begin{equation}
	\frac1N \sum_{k=1}^{N} \frac{1}{\zeta - a_1 b_k} = 1,
\end{equation}
which is equivalent to 
\begin{equation}
	\stj_{\BBB}(\zeta/a_1) = a_1.
\end{equation}
We therefore find that 
\begin{equation}
	\zeta = a_1 \btr_{\BBB}(a_1) = a_1 \rtr_\BBB(a_1) + 1.
\end{equation}
By plugging this solution into \eqref{eq:HCIZ_rank1_free_energy}, we obtain 
\begin{equation}
	\frac{2}{N} \log {\cal I}_{1}(\AAA,\BBB) \sim  a_1 \rtr_{\BBB}(a_1) - \frac{1}{N} \sum_{k=1}^{N} \log\pB{1 + a_1 (\rtr_\BBB(a_1 - b_k))}.
\end{equation}
One can then check, by taking the derivative of both sides, that 
\begin{equation}
	a_1 \rtr_{\BBB}(a_1) - \frac{1}{N} \sum_{k=1}^{N} \log\pB{1 + a_1 (\rtr_\BBB(a_1 - b_k))} = \wtr_{\BBB}(a_1),
\end{equation}
where $\wtr_\BBB$ is the primitive integral of the $\rtr$-transform of $\BBB$ satisfying $\wtr_\BBB^{\,\prime}(\omega) = \rtr_\BBB(\omega)$. We therefore conclude that
\begin{equation}
	\frac{2}{N} \log {\cal I}_{1}(\AAA,\BBB) \sim \wtr_\BBB(a_1),
\end{equation}
which is the claim.

Let us now explain briefly how to extend this derivation to the rank-$n$ case. Formally, the integral reads
\begin{equation}
	\label{eq:HCIZ_rankn_integral}
	{\cal I}_{1}(\AAA,\BBB) = \frac{1}{\cal Z} \int \pBB{\prod_{i=1}^{n} \prod_{k=1}^{N} \dd \Omega_{ik}} \exp\qBB{\frac{N}{2} \sum_{i=1}^{n} a_i  \sum_{k=1}^{N} \Omega_{ik}^2 b_k } \prod_{i,j=1}^{n} \delta\pB{ \sum_{k=1}^N \Omega_{ik} \Omega_{jk} - \delta_{ij} },
\end{equation}
where the normalization $\cal Z$ is easily deduced from \eqref{eq:Haar_partition_fct}, and $\AAA = \diag(a_1, a_2, \dots, a_n, \dots, 0)$. When $n = \cal O(N)$, i.e. when $\AAA$ has close to full rank, the orthogonality 
constraint $\sum_{k=1}^N \Omega_{ik} \Omega_{jk} = 0$ for $i \neq j$ becomes dominant and makes the calculation difficult. However, when $n \ll N$, this constraint is nearly automatically satisfied since two random unit vectors in $N$
dimensions have naturally a scalar product of order $1/\sqrt{N}$. In this limit, only the normalization constraint is operative, i.e.  $\sum_{k=1}^N \Omega_{ik}^2 = 1$, $\forall i = 1,\dots, n$. But one then easily sees that the
above integral factorizes into $n$ independent integrals of the type we considered above, hence leading to result \eqref{eq:HCIZ_rankn} above. For a more rigorous proof that this result holds as long as $n \ll \sqrt{N}$, see \cite{guionnet2004asymptotic}.

\clearpage%!TEX root = RMT_Covariance_Review.tex
\section{Reminders on linear algebra}
\label{app:linear_algebra}

\subsection{Schur complement}

The derivation of recursion relation mostly relies on linear algebra. More specifically, let us define the $(N+M)\times(N+M)$ matrix $\b M$ by
\begin{equation}
\b M  \;\deq\;  
\begin{pmatrix}
\b A & \b B
\\
\b C & \b D
\end{pmatrix}\,,
\end{equation}
where the matrices $\b A, \b B, \b C$ and $\b D$ are respectively of dimension $N\times N, N\times M, M \times N$ and $M\times M$. Suppose that $\b D$ is invertible, then the \emph{Schur complement} of the block $\b D$ of the matrix $\M$ is given by the $N \times N$ matrix
\begin{equation}
	\label{eq:Schur_D}
	\M / \b D = \b A - \b B \b D^{-1} \b C.
\end{equation}
Using it, one obtains after using block Gaussian elimination (or LU decomposition) that the determinant of $\M$ can be expressed as
\begin{equation}
	\label{eq:det_schur_blockD}
	\det(\M) = \det(\b D) \det(\M / \b D).
\end{equation}
Moreover, one can write the inverse matrix $\M^{-1}$ in terms of $\b D^{-1}$ and the inverse of the Schur complement \eqref{eq:Schur_D}
\begin{equation}
	\label{eq:inv_schur_blockD}
	\b M^{-1} = 
	\begin{pmatrix}
	(\M / \b D)^{-1}  & - (\M / \b D)^{-1} \b B \b D^{-1}
\\
- \b D^{-1} \b C (\M / \b D)^{-1} & \b D^{-1} + \b D^{-1} \b C (\M / \b D)^{-1} \b B \b D^{-1}
\end{pmatrix}\,.
	% \begin{pmatrix}
	% (\b A - \b B \b D^{-1} \b C)^{-1} & - (\b A - \b B \b D^{-1} \b C)^{-1} \b B \b D^{-1}
	% \\
 %    - \b D^{-1} \b C (\b A - \b B \b D^{-1} \b C)^{-1} & \b D^{-1} + \b D^{-1} \b C (\b A - \b B \b D^{-1} \b C)^{-1} \b B \b D^{-1}
	% \end{pmatrix}\,
\end{equation}

Similarly, if $\b A$ is invertible, the Schur complement of the block $\b A$ of the matrix $\M$ is given by the $M \times M$ matrix
\begin{equation}
	\label{eq:Schur_A}
	\M / \b A = \b D - \b C \b A^{-1} \b B.
\end{equation}
One easily obtains $\det(\M)$ in terms of $\b A$ and $\M / \b A$ from \eqref{eq:det_schur_blockD} by replacing $\b D$ by $\b A$
\begin{equation}
	\label{eq:det_schur_blockA}
	\det(\M) = \det(\b A) \det(\M / \b A).
\end{equation}
The inverse matrix $\M^{-1}$ can also be written in terms of $\b A^{-1}$ and the inverse of the Schur complement \eqref{eq:Schur_A}
\begin{equation}
	\label{eq:inv_schur_blockA}
	\b M^{-1} = 
	\begin{pmatrix}
	\b A^{-1} + \b A^{-1} \b B (\M / \b A)^{-1} \b C \b A^{-1}  & - \b A^{-1} \b B (\M / \b A)^{-1} 
	\\
    - (\M / \b A)^{-1} \b C \b A^{-1}  & (\M / \b A)^{-1}
\end{pmatrix}\,.
\end{equation}

\subsection{Matrix identities}

There are several useful identities that can be inferred from Schur complement formula. Firstly, using \eqref{eq:inv_schur_blockD} and \eqref{eq:inv_schur_blockA}, we may immediately deduce the so-called \emph{Woodbury} matrix identity 
\begin{equation}
	\label{eq:woodbury}
	(\b A + \b B \b D^{-1} \b C)^{-1} = \b A^{-1} - \b A^{-1} \b B (\b D + \b C \b A^{-1} \b B)^{-1} \b C \b A^{-1}.
\end{equation}
Moreover, if  $\b D = I_M$, we get the \emph{matrix determinant lemma} from \eqref{eq:det_schur_blockD}  and \eqref{eq:det_schur_blockA} 
\begin{equation}
	\label{eq:mat_det_lemma}
	\det(\b A - \b B \b C) = \det(\b A) \det(\b I_M - \b C \b A^{-1} \b B),
\end{equation}
and if $\b A = \b I_N$ in addition, one gets \emph{Sylvester's determinant identity} 
\begin{equation}
	\label{eq:sylvester}
	\det(\b I_N - \b B \b C) = \det(\b I_M - \b C \b B).
\end{equation}
Now, assuming that both $\b B$ and $\b C$ are column vectors, one readily find from \eqref{eq:woodbury} the \emph{Sherman-Morrison} formula. 

\subsection{Resolvent identities}
\label{app:resolvent_identities}

Another useful application of Schur complement formula concerns the resolvent. We keep the notations of Section \ref{section:RMT_transforms} and thus
\begin{equation}
	\label{eq:resolvent_identities_model}
	\G(z) = \b H^{-1}(z), \qquad \b H(z) \;\deq\; z\In - \M\,,
\end{equation}
with $\G$ a $N \times N$ symmetric matrix. We now rewrite $\b H(z)$ as a block matrix:
\begin{equation}
	\label{eq:resolvent_block_matrix}
	\b H(z) \;=\;
\begin{pmatrix}
	\b A  & \b B 
	\\
    \b B^* & \b C 
\end{pmatrix}\,,
\end{equation}
where the matrices $\b A, \b B$ and $\b C$ are respectively of dimension $K\times K, \; K\times M$ and $M\times M$ with $N=K+M$. Next, we define from \eqref{eq:Schur_D} the Schur complement $\b D \deq \b A - \b B \C^{-1} \b B^*$. In the following, we consider $K = 2$ for simplicity. We have for any $i,j \in \{1,2\}$, we have from \eqref{eq:inv_schur_blockD}:
\begin{equation}
	\label{eq:schur_resolvent_ij}
	G_{ij} = (\b D^{-1})_{ij}.
\end{equation}
As a warm-up exercise, let us first consider the simplest case $i=j$ ($K=1$) and we set without loss of generality that $i = 1$. Then $\AAA$ becomes a scalar and so is $\b D$. Using Eq.\ \eqref{eq:resolvent_identities_model}, one obtains $\AAA = z - M_{11}$, $\b B = [ M_{12}, \dots, M_{1N}]$ and $\C = \b H^{(1)}(z)$ where $\b H^{(i)}$ denotes the ``minor'' of $\b H$, i.e. $\b H^{(i)} \deq \pb{H_{st} : s,t \in \qq{1,N} \backslash\{i\}}$. Hence, it is easy to see from the very definition of $\b D$ that
\begin{equation}
	\label{eq:schur_resolvent_ii_complement}
	\b D \equiv D_{11} = z - M_{11} - \sum_{\alpha,\beta}^{(1)} M_{1\alpha} G^{(1)}_{\alpha,\beta} M_{\beta 1}, 
\end{equation}
where and we used the abbreviation
\begin{equation}
	\label{eq:partial_sum}
	\sum_{\alpha,\beta}^{(i)} \;\equiv\; \sum_{\alpha,\beta \in \qq{1,N} \backslash\{i\}}.
\end{equation} 
Therefore, we deduce from \eqref{eq:schur_resolvent_ij} that
\begin{equation}
	\label{eq:resolvent_schur_diag}
	G_{11}(z) = \frac{1}{z - M_{11} - \sum_{\alpha\beta}^{(1)} M_{1\alpha} G^{(1)}_{\alpha,\beta} M_{\beta 1}}.
\end{equation}
This last result holds for any other diagonal term of the resolvent $\b G$. 

Next, we consider the general case $K=2$ so that $\b D$ is a $2 \times 2$ matrix. Again, using the block representation \eqref{eq:resolvent_block_matrix} and Eq.\ \eqref{eq:resolvent_identities_model}, one deduces that:
\begin{equation}
	\label{eq:schur_resolvent_ij_complement}
	D_{kl} = z\delta_{kl} - M_{kl} - \sum_{\alpha,\beta}^{(kl)} M_{k\alpha} G^{(kl)}_{\alpha,\beta} M_{\beta l}, \quad k,l \in \qq{i,j}.
\end{equation}
It is not hard to see that $D_{kk}$ yields Eq.\ \eqref{eq:schur_resolvent_ii_complement} as it should. Using that \eqref{eq:schur_resolvent_ij_complement} is a $2 \times 2$ matrix, one can readily invert the matrix $\b D$ to obtain the relation
\begin{equation}
	\label{eq:schur_Gij_minor}
	G_{ij} - G_{ij}^{(m)} = \frac{G_{im} G_{mj}}{G_{mm}},
\end{equation}
for any $i,j \in \qq{1,K}$ and $m \in \qq{1,N}$ with $i, j \neq m$. This last equation allows one to write a recursion relation on the entries of the resolvent (see the following appendix).

\clearpage%!TEX root = RMT_Covariance_Review.tex
\section{Self-consistent relation for Green's function and Central Limit Theorem}
\label{app:recursion}

We focus in this section on another frequently used analytical tool in RMT based on recursion relation for the resolvent of a given matrix $\M$. This technique has many advantages compared to the method compared to the Replica analysis: (i) the entries of the matrix need not to be identically distributed, (ii) no ansatz is required to perform the calculations. In the limit of $N \to \infty$, an interesting application of the Central Limit Theorem (CLT) concerns the spectral properties of random matrices. Precisely, we shall see that relations like that of Eq.\ \eqref{eq:global_law_SCM} are actually a consequence of the CLT. 

\subsection{Wigner matrices}
\label{sec:schur_wigner}

As a warm-up exercise, we consider the simplest ensemble of random matrices where all elements of the matrix $\M$ are iid random variables, with the only constraint that the matrix be symmetrical. This is the well-known Wigner ensemble where we assume that 
\begin{equation}
	\label{eq:moments_Wigner}
	\mathbb{E}[ M_{ij} ]	= 0 , \quad \mathbb{E} [ M_{ij}^{2}] = \frac{\sigma^2}{N},
\end{equation}
for any $i,j \in \qq{1,N}$. Note that the scaling with $N^{-1}$ for the variance comes from the fact that we want the eigenvalues of $\M$ to stay bounded when $N \to \infty$. This allows to conclude that $M_{ij} \sim 1/\sqrt{N}$ for any $i,j \in \qq{N}$. 

In order to derive a self-consistent equation for the resolvent of $\M$, we use \eqref{eq:moments_Wigner} and Wick's theorem into \eqref{eq:schur_resolvent_ij_complement} and one can check that
\begin{eqnarray}
	\mathbb{E} \qBB{\;\sum_{\alpha,\beta}^{(kl)} M_{k\alpha} G^{(kl)}_{\alpha\beta} M_{\beta l}} & = & \delta_{kl}  \frac{\sigma^2}{N} \sum_{\alpha}^{(k)} G^{(k)}_{\alpha\alpha} \nonumber \\
	\mathbb{V} \qBB{\; \sum_{\alpha,\beta}^{(kl)} M_{k\alpha} G^{(kl)}_{\alpha\beta} M_{\beta l}} & \sim & \frac{\sigma^4}{N}. \nonumber \\
\end{eqnarray}
Consequently, using the Central Limit Theorem, we conclude that for Wigner matrices, \eqref{eq:schur_resolvent_ij_complement} converges for large $N$ towards
\begin{equation}
	\label{eq:matrix_D_wigner}
	D_{kl} = \delta_{kl} \pBB{z - \frac{\sigma^2}{N} \sum_{\alpha}^{(k)} G^{(k)}_{\alpha\alpha}} + O(N^{-1/2}) \quad k,l \in \{i,j\},
\end{equation}
from which one deduces that $G_{ij} \sim N^{-1/2}$ using \eqref{eq:schur_resolvent_ij}. Moreover, we may consistently check  that $G_{\ell\ell}^{(k)} \sim G_{\ell\ell} + O(N^{-1})$ for any $\ell \in \qq{1,N}$ thanks to \eqref{eq:schur_Gij_minor} and  we therefore obtain for any $i \in \qq{1,N}$:
\begin{equation}
	G_{ii} \sim \frac1{z - \sigma^2 \stj(z)} + O(N^{-1/2}).
\end{equation}
By taking the normalized trace in this last equation, we obtain at leading order the equation of the semi-circle law's Stieltjes transform 
\begin{equation}
	\stj(z) = \frac{1}{z - \sigma^2\stj(z) } ,
\end{equation}
so that we conclude 
\begin{equation}
	\label{eq:wigner_green_fct}
	G_{ij}(z) \sim \delta_{ij} \stj(z) + O(N^{-1/2}).
\end{equation}
This result has been extended in a much more general framework -- see e.g. the recent reviews \cite{benaych2016lectures,erdHos2011universality}. In particular, it is possible to show that the error term we obtain in Eq.\ \eqref{eq:wigner_green_fct} is quite similar to \eqref{eq:global_law_SCM_error_term} and reads for $\eta = \wh \eta N$ with $\wh\eta \gg 1$:
\begin{equation}
  \label{eq:GOE_error_term}
  \Psi_{\text{GOE}}(z) \;\deq\; \sqrt{\frac{\im \stj_\S(z)}{\widehat \eta}} + \frac{1}{\widehat \eta}\,,
\end{equation}
provided that $N$ is large enough. We illustrate this ergodic behavior for the GOE in Figure \ref{fig:resolvent_entries}, and we see the agreement is excellent and each diagonal entry indeed converges to the semicircle law. 

\begin{figure}[!]
\begin{subfigure}{.5\textwidth}
  \centering
  \includegraphics[width= 1\linewidth]{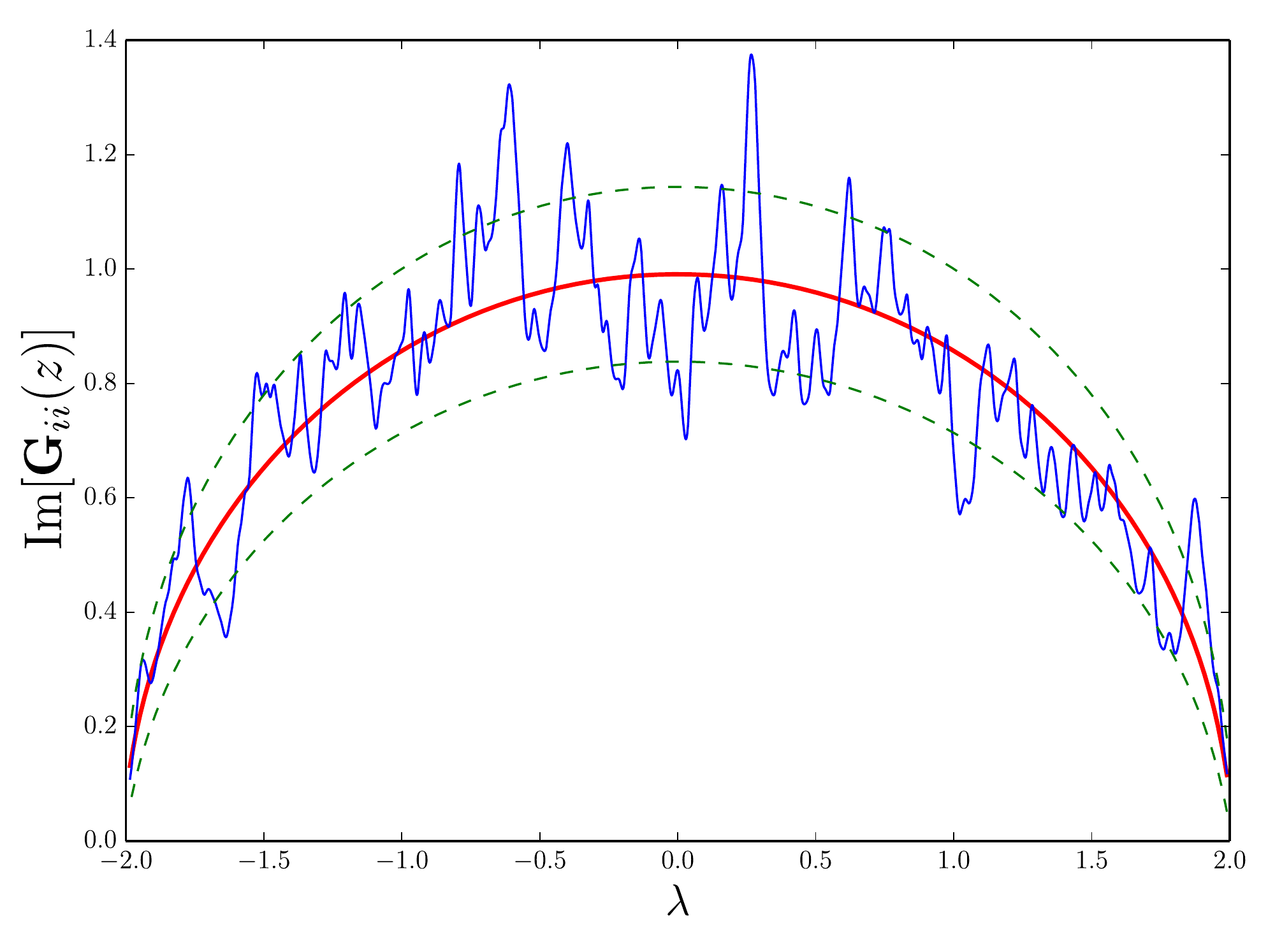}
  \caption{Diagonal entry of $\im[\G_\E(z)]$ with $i = 1$. }
  \label{fig:multiple}
\end{subfigure}%
\begin{subfigure}{.5\textwidth}
  \centering
  \includegraphics[width= 1.0\linewidth]{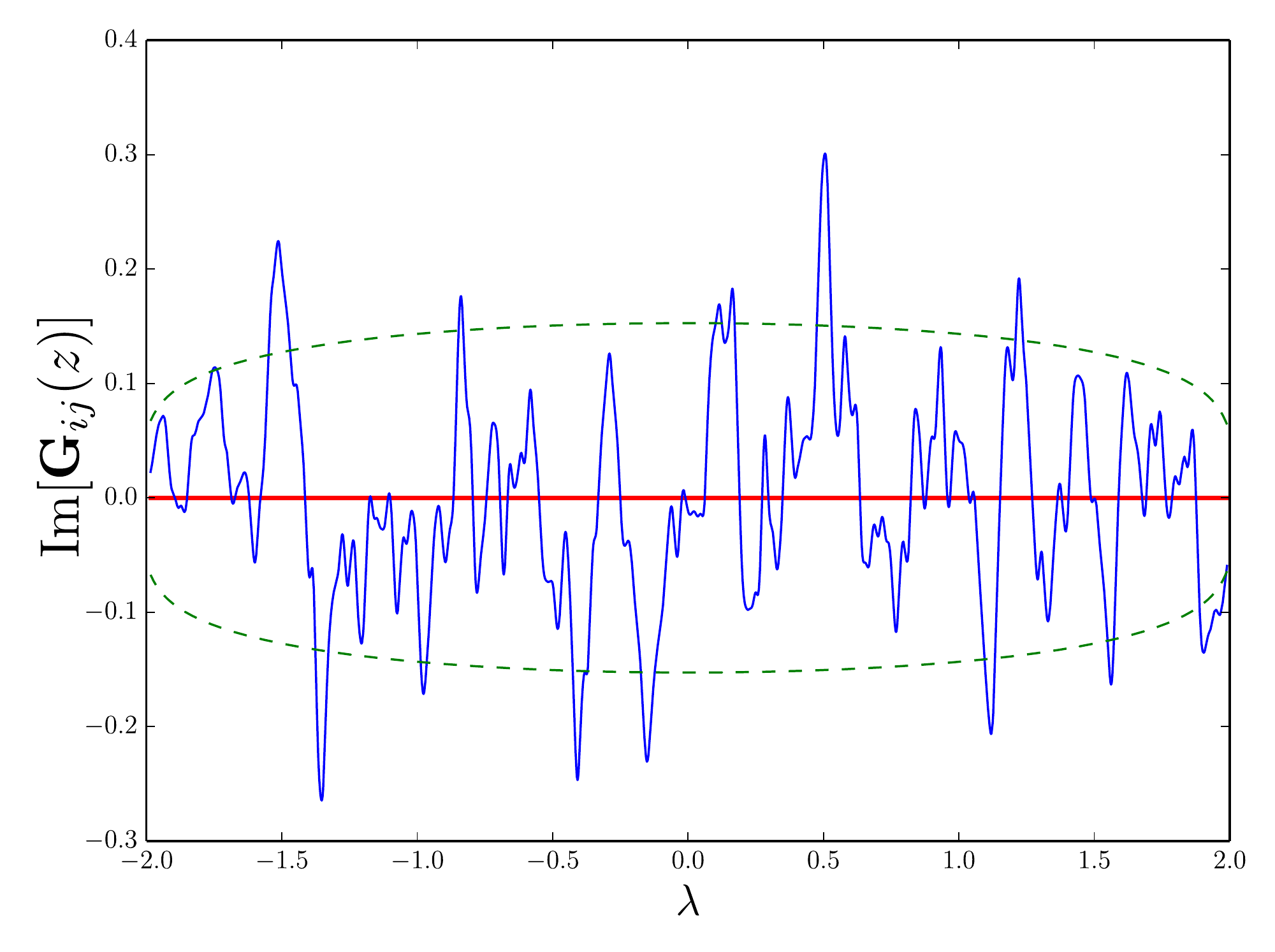}
  \caption{Off diagonal entry of $\im[\G_\E(z)]$ with $i = 1$ and $j = 2$. }
  \label{fig:dGOE}
\end{subfigure}\\
\caption{Illustration of the imaginary part of Eq.\ \eqref{eq:wigner_green_fct} with $N = 1000$. The empirical estimate of $\G_\E(z)$ (blue line) is computed for any $z = \lambda_i - \ii N^{-1/2}$ with $i \in \qq{1,N}$ and comes from one sample. The theoretical one (red line) is given by the RHS of Eq.\ \eqref{eq:wigner_green_fct}. The green dotted corresponds to the confidence interval whose formula is given by Eq.\ \eqref{eq:GOE_error_term}.
}
\label{fig:resolvent_entries_GOE}
\end{figure}

\subsection{Sample covariance matrices}
\label{sec:schur_SCM}

We now want to derive \eqref{eq:global_law_SCM} using the same type of arguments than in the previous section. Suppose that $\E$ is defined as in \eqref{eq:SCM} and we denote by $\b G(z)$ its resolvent. Let us assume for simplicity that $\C = \diag(\mu_1, \mu_2, \dots, \mu_N)$. Since $\E$ is a product of two rectangular matrices, it is convenient to introduce the $(N+T)\times(N+T)$ block matrix $\b R \;\deq\; (R_{ij}) \in \mathbb{R}^{(N+T)\times(N+T)}$ defined as:
\begin{equation}
	\label{eq:SCM_resolvent_block}
	\b R(z) \deq \b H^{-1}(z), \quad \b H(z) \deq
	\begin{pmatrix}
	\b \C^{-1}  & \b \X
	\\
    \b \X^* & z\,\b I_T
\end{pmatrix}\,.
\end{equation}
To simplify the notations, we introduce the set of indexes $\cal I_N \deq \qq{1,N}$ and $\cal I_T \deq \qq{1,T}$. Then using \eqref{eq:inv_schur_blockD} and \eqref{eq:inv_schur_blockA}, we see that 
\begin{equation}
	\label{eq:R_ij}
	R_{ij}(z) = z (\C^{1/2} \G_\E(z) \C^{1/2})_{ij}, \quad i,j \in \cal I_N,
\end{equation}
where $\E$ is the sample covariance matrix defined in Eqs.\ \eqref{eq:SCM} and \eqref{eq:SCM_entries}, but also
\begin{equation}
	\label{eq:R_munu}
	R_{\alpha\beta}(z) = (\b G_\S(z))_{\alpha\beta}, \qquad \alpha,\beta \in \cal I_T,
\end{equation}
where the $T \times T$ matrix $\S$ is defined in Eq.\ \eqref{eq:SCM_dual}.

We are interested in the computations of $R_{ij}$ for $i,j \in \cal I_N$ and this can be done using \eqref{eq:schur_resolvent_ij} and \eqref{eq:schur_resolvent_ij_complement}. Note that one can finds $R_{\alpha\beta}$ by proceeding in the same way. We obtain from \eqref{eq:schur_resolvent_ij} and \eqref{eq:schur_resolvent_ij_complement} that
\begin{equation}
	\label{eq:R_ij_schur}
	R_{ij}(z) = (\b D^{-1})_{ij}, \qquad D_{kl} \;\deq\; \frac{\delta_{kl}}{\mu_k} - \sum_{\alpha,\beta \in \cal I_T} X_{k\alpha} R^{(kl)}_{\alpha\beta} X_{l\alpha}.
\end{equation}
for any $k,l \in \{i,j\}$. Using that $\mathbb{E}[X_{it}] = 0$ and $\mathbb{E} [X_{it}^2] = T^{-1}$ from \eqref{eq:SCM_entries_X}, we remark thanks to Wick's theorem that the sum in the term $D_{kl}$ obeys
\begin{eqnarray}
	\mathbb{E} \qBB{\;\sum_{\alpha,\beta \in \cal I_T} X_{k\alpha} R^{(kl)}_{\alpha\beta} X_{l\alpha} } & = & \frac{\delta_{kl}}{T} \sum_{\alpha}^{(k)} R^{(k)}_{\alpha\alpha} \nonumber \\
	\mathbb{V} \qBB{\; \sum_{\alpha,\beta \in \cal I_T} X_{k\alpha} R^{(kl)}_{\alpha\beta} X_{l\alpha}} & \sim & \frac{1}{T}, \nonumber \\
\end{eqnarray}
where we used the notation \eqref{eq:partial_sum} for the sum. Invoking once again the CLT, we find that the entry $D_{kl}$ converges for large $N$ towards
\begin{equation}
	\label{eq:Schur_complement_SCM}
	D_{kl} \sim \delta_{kl}\pBB{\frac{1}{\mu_k} - \frac{1}{T} \sum_{\alpha \in \cal I_T} R^{(k)}_{\alpha\alpha}} +  O(T^{-1/2}),
\end{equation}
so that we may conclude from \eqref{eq:R_ij_schur} that $R_{ij} \sim O(T^{-1/2})$ for $i\neq j$. Note that one may repeat the same arguments for $R_{\alpha\beta}$ with $\alpha,\beta \in \cal I_T$ to obtain
\begin{equation}
	\label{eq:Schur_complement_SCM_dual}
	D_{\alpha\beta} \sim \delta_{\alpha\beta}\pBB{z - \frac{1}{T} \sum_{k \in \cal I_N} R^{(\alpha)}_{kk}} + \cal O(T^{-1/2}),.
\end{equation}
Let us now investigate $R^{(k)}_{\alpha\alpha}$ which can be rewritten thanks to \eqref{eq:schur_Gij_minor} as:
\begin{equation}
	\label{eq:schur_Rij_minor}
	R^{(k)}_{\alpha\alpha} = R_{\alpha\alpha} - \frac{R_{k\alpha}R_{\alpha k}}{R_{kk}}\,.
\end{equation}
We deduce from \eqref{eq:Schur_complement_SCM} that $R_{kk} \sim  O(1)$. We will now show that $R_{k\alpha}$ (and $R_{\alpha k}$) are vanishing as $T^{-1/2}$. To that end, we apply \eqref{eq:inv_schur_blockA} to \eqref{eq:SCM_resolvent_block} to find
\begin{equation}
	R_{k\alpha} \;=\;  - \pb{\C \X \G_\S}_{k\alpha} = - \mu_k \sum_{\beta \in \cal I_T} X_{k\beta} (\G_\S)_{\beta\alpha}. 
\end{equation}
Using Eqs.\ \eqref{eq:R_munu}, \eqref{eq:Schur_complement_SCM_dual} and that $X_{k\beta} \sim T^{-1/2}$, one can self-consistently check that $R_{k\alpha} \sim T^{-1/2}$. This is also true for $R_{\alpha k}$. Hence, if we plug this into Eq.\ \eqref{eq:schur_Rij_minor}, we see that for $N\to\infty$:
\begin{equation}
	\frac{1}{T} \sum_{\alpha}^{(k)} R^{(k)}_{\alpha\alpha} = \frac{1}{T} \sum_{\alpha}^{(k)} R_{\alpha\alpha} +  O(T^{-1}) = \stj_\S(z) + O(T^{-1})\,,
\end{equation}
and we therefore have from Eqs.\ \eqref{eq:Schur_complement_SCM} and \eqref{eq:R_ij_schur}: 
\begin{equation}
	\label{eq:SCM_green_fct_tmp}
	R_{ij}(z) = \delta_{ij}\pBB{\frac{\mu_k}{1- \mu_k \stj_S(z)}} +  O(T^{-1/2}).
\end{equation}
Finally, recalling that $\stj_\S(z) = q \stj_\E(z) + (1-q)/z$ from Eq.\ \eqref{eq:stieltjes_E_dual} and $R_{ii} = z \mu_i G_{ii}$ from Eq.\ \eqref{eq:R_ij}, we conclude that 
\begin{equation}
	\label{eq:SCM_green_fct}
	(\b G_\E(z))_{ij} = \delta_{ij}\pBB{\frac{1}{z - \mu_k (1-q+qz \stj_\E(z))}} +  O(T^{-1/2}), \qquad i,j \in \qq{1,N},
\end{equation}
which is the prediction obtained in \eqref{eq:global_law_SCM_diag} with the Replica method. Similarly, we obtain for the $T\times T$ block that: 
\begin{equation}
	%\label{eq:SCM_dual_green_fct}
	(\b G_\S(z))_{\alpha\beta} = \frac{\delta_{\alpha\beta}}{z - \frac{1}{T} \sum_{k \in \cal I_N} (\b G_\E(z))_{ij}}\, + O(T^{-1/2})\,.
\end{equation}
Moreover, by using \eqref{eq:SCM_green_fct_tmp} and \eqref{eq:MP_equation_BaiSilverstein}, we see that for $N\to\infty$
\begin{equation}
	z - \frac{1}{T} \sum_{k \in \cal I_N} (\b G_\E(z))_{kk} = \frac{1}{\stj_\S(z)},
\end{equation}
so that we may conclude
\begin{equation}
	\label{eq:SCM_dual_green_fct}
	(\b G_\S(z))_{\alpha\beta} = \delta_{\alpha\beta} \stj_\S(z) + O(T^{-1/2})\,.
\end{equation}
This last result highlights that it is often easier to work with the $T \times T$ sample covariance matrix $\S$ rather than with the $N \times N$ matrix $\E$ since the resolvent can be approximated simply by its normalized trace. All these results can be found in a much more general and rigorous context in \cite{knowles2014anisotropic}.

\clearpage%!TEX root = RMT_Covariance_Review.tex
\section{Additive noise model}
\label{app:addition}

In this review, we mainly focus on sample covariance matrices which is a particular case of models of random matrices with multiplicative noise. In this appendix, we consider the case of an additive external noise which 
can also be important in many situations, in particular in quantum chaos and quantum transport \cite{beenakker1997random}, with renewed interest coming from problems of quantum ergodicity 
(``eigenstate thermalisation'') \cite{deutsch1991quantum,ithier2015thermalisation}, entanglement and dissipation (for recent reviews see \cite{nandkishore2015many,eisert2015quantum}). We will show briefly here 
how we can extend the results of Chapter \ref{chap:eigenvectors}  to this specific case with a special focus to the overlaps \eqref{eq:overlap}.

As above, we shall denote the $N \times N$ real symmetric population matrix, i.e. the one we wish to infer, by $\C$ and to avoid confusion, we denote  by $\M$ the sample matrix that is the matrix we measure with the data. Throughout this section, we deal with models of the form
\begin{equation}
	\label{eq:additive_model_app}
	\M = \C + \b\Omega {\bf B} \b\Omega^{*},
\end{equation}
where ${\bf B}$ is a fixed matrix with eigenvalues $b_1 > b_2 > \dots > b_N$, spectral $\rho_{\BBB}$, and $\b\Omega$ is a random matrix chosen in the Orthogonal group $\b O(N)$ according to the Haar measure. Clearly, the noise term is invariant under rotation so that we expect the resolvent of $\M$ to be (for large $N$) in the same basis 	as $\C$. We therefore posit without loss of generality that $\C$ is diagonal. 
The most common example of such models in the literature \cite{brezin1995universalb} is the case where $\BBB$ belongs to the GOE but for now, we do not specify any distribution or structure assumption on the fixed matrix $\BBB$. We first present this simple model and then show that we can generalize it to the general case \eqref{eq:additive_model_app}. We shall also provide an elementary derivation of the free addition in the limit $N \to \infty$. 

\subsection{Gaussian external noise}

In order to give some insights about the general model \eqref{eq:additive_model_app}, we focus first on the case where the external noise $\BBB$ belongs to the GOE with a variance of $\sigma^2$. More formally, we consider $\BBB$ to be a $N\times N$ real symmetric matrix with Gaussian entries that satisfies
\begin{equation}
	\label{eq:B_is_GOE}
	\mathbb{E} [B_{ij}] = 0\, \qquad \mathbb{E} [B_{ij}^2] \;=\; 
	\begin{cases}
		2\sigma^2/N & \text{if } i = j\,, \\
		\sigma^2/N & \text{otherwise} .
	\end{cases}
\end{equation}
In the case where $\BBB$ satisfies \eqref{eq:B_is_GOE}, we say that $\M$ defined as \eqref{eq:additive_model_app} is a \emph{deformed GOE} matrix. As usual, all the information about the eigenvalues and eigenvectors of $\M$ can be analyzed through the resolvent. In fact, as for sample covariance matrices, it is possible to show that each entry of the resolvent $\G_\M$ converges to a deterministic limit for $N \to \infty$. There are a lot of different mathematical methods to prove this last assertion and we shall cover only two of them: the first method is to use a straightforward generalization of the arguments of Section \ref{sec:schur_wigner} above. The second method is based on the representation of a GOE matrix as a (dynamical) stochastic process, known as \emph{Dyson's Brownian motion}. As we shall see below, this second approach provides insightful physical interpretation about the behavior of $\M$.

\subsubsection{Schur complement arguments}

Let us start with the first method. We expect the resolvent of $\M$ to be in the same basis than $\C$, at least in the limit $N \to \infty$, meaning that we can work in the basis where $\C$ is diagonal. Moreover, since matrix $\C$ is deterministic, one may easily repeat the arguments of Section \ref{sec:schur_wigner} to generalize Eq.\ \eqref{eq:matrix_D_wigner} to:
\begin{equation}
	\label{eq:matrix_D_dWigner}
	D_{kl} = \delta_{kl} \pBB{z - \mu_k - \frac{\sigma^2}{N} \sum_{\alpha}^{(k)} G^{(k)}_{\alpha\alpha}} + \cal O(N^{-1/2})\,, \qquad k,l \in \{i,j\}\,.
\end{equation}
As above, we can consistently check that $G_{ij} \sim N^{-1/2}$ using \eqref{eq:schur_resolvent_ij}. Moreover, we also obtain that $G_{\ell\ell}^{(k)} \sim G_{\ell\ell} + \cal O(N^{-1})$ for any $\ell \in \qq{1,N}$ thanks to \eqref{eq:schur_Gij_minor}. Therefore, we obtain for any $i \in \qq{1,N}$:
\begin{equation}
	\label{eq:dGOE_resolvent_schur}
	G_{ii} \sim \frac1{z - \sigma^2 \stj(z) - \mu_i} + \cal O(N^{-1/2})\,,
\end{equation}
which the result obtained in e.g. \cite{allez2014eigenvectors,knowles2014anisotropic} using more rigorous arguments.

\subsubsection{Dyson Brownian Motion}
\label{app:DBM}

Since the seminal paper of Dyson in 1962 \cite{dyson1962brownian}, it is well known that the spectrum induced by the addition of free random matrices in the Gaussian orthogonal ensemble\footnote{All these results may be easily extended to the Hermitian case.} can be investigated through the evolution of a time-dependent real symmetric $N \times N$ Brownian motion. More precisely, let us introduce a fictitious time $t$ and rewrite the model \eqref{eq:additive_model_app} as :
\begin{equation}
	\label{eq:dGOE_BM}
	\M(t) = \C + \BBB(t)
\end{equation}
with
\begin{equation}
	\label{eq:GOE_entries}
	B_{ii}(t) \;=\; \sqrt{\frac{2 \sigma^2}{N}} W_{ii}(t), \qquad B_{ij}(t) \;=\; \sqrt{\frac{\sigma^2}{N}} W_{ij}(t) \quad (i\neq j),
\end{equation}
where the $W_{ij}(t)$, $i \leq j$ are independent and identically distributed real Brownian motions. We see that $\BBB(t)$ is an external noise whose variance increases as the time $t$ grows. We suppose that the eigenvalues of $\C$ are all distinct and satisfy $\mu_1 \geq \mu_2 \geq \dots \mu_N$. Then, the dynamics of the eigenvalues of $\M(t)$ may also be characterized by a stochastic differential equation (SDE), known as \emph{Dyson's Brownian motion}:
\begin{eqnarray}
	\label{eq:DBM_eigenvalues}
	\dd\lambda_i(t) & = & \sqrt{\frac{2 \sigma^2}{N}} \dd b_{i}(t) + \frac1N \sum_{j\neq i}^{N} \frac{\dd t}{\lambda_i(t) - \lambda_j(t)}, \nonumber \\
	\lambda_i(0) & = & \mu_i, 
\end{eqnarray}
for any $i = 1,\dots,N$, and where the $b_i(t)$ are independent real Brownian motions. We observe that the eigenvalues of $\M(t)$ defines Dyson's Coulomb gas model that describes positively charged particles on a line interacting via a logarithmic potential and subject to a thermal noise $\dd b_i(t)$. 

Conditionally to the eigenvalues paths, the trajectories of the associated eigenvectors $\b u_i(t)$ can also be characterized by a SDE:
\begin{eqnarray}
	\label{eq:DBM_eigenvectors}
	\dd\b u_i(t) & = & \frac{1}{\sqrt{N}} \sum_{k\neq i} \frac{\dd w_{ik}(t)}{\lambda_i(t) - \lambda_k(t)} \b u_k(t) - \frac{1}{2N} \sum_{k\neq i}  \frac{\dd t}{(\lambda_i(t) - \lambda_k(t))^2} \b u_i(t) , \nonumber \\
	\b u_i(0) & = & \b v_i,
\end{eqnarray}
where the family of independent (up to symmetry) of Brownian motions $\{w_{ij}\}$ is independent from the Brownian motions $\{b_i\}$ that drive the eigenvalues trajectories. As a result, in order to study the dynamics of the eigenvectors, we may always freeze the eigenvalues paths and work conditionally to the realized trajectories. This is the approach used in \cite{allez2014eigenvectors,allez2013eigenvectors,bourgade2013eigenvector} in order to study the mean squared overlap \eqref{eq:overlap} in this additive model. 

In this appendix, we present an alternative approach that considers directly the time evolution of the full resolvent, which we have not seen in the literature before. To that end, we define
\begin{equation}
	\label{eq:resolvent_BM}
	\G(z,t) \;\deq\; \b H^{-1}(z,t), \qquad \b H(z,t) \;\deq\; z \b I_N - \M(t).
\end{equation}
Using It{\^o} formula and the fact that $\dd M_{kl} = \dd B_{kl}$, one has
\begin{eqnarray}
	\label{eq:G_DBM_step1}
	\dd G_{ij}(z,t)  & = &  \sum_{k,l=1}^{N} \frac{\partial G_{ij}}{\partial M_{kl}} \dd B_{kl} + \frac12 \sum_{k, l,m,n=1}^{N} \sum_{m, n=1}^{N} \frac{\partial^2 G_{ij}}{\partial M_{kl} \partial M_{mn}} \dd \qb{B_{kl} B_{mn}},
\end{eqnarray}
Next, we compute the derivatives:
\begin{equation}
	\label{eq:resolvent_derivatives_1}
	\frac{\partial G_{ij}}{\partial M_{kl}} = \frac12 \left[ G_{ik} G_{jl}+G_{jk} G_{il} \right],  
\end{equation}
from which we deduce the second derivatives
\begin{equation}
	\label{eq:resolvent_derivatives_2}
	\frac{\partial^2 G_{ij}}{\partial M_{kl} \partial M_{mn}} = \frac14 \left[ \left(G_{im} G_{kn} + G_{im} G_{kn} \right) G_{jl} + ... \right],
\end{equation}
where we have not written the other 6 $GGG$ products. Now, using \eqref{eq:GOE_entries}, the quadratic co-variation reads
\begin{equation}
	\label{eq:quadratic_covar}
	\dd\qb{B_{kl} B_{mn}} =  \frac{\sigma^2\dd t}{N} \pBB{ 2\delta_{k=l=m=n} + \delta_{k=m}\delta_{l=n}+\delta_{k=n}\delta_{l=m} }
	%\dd \qb{W_{kk}^2} = \frac{2\sigma^2}{N} \dd t, \qquad \dd \qb{W_{kl}^2} = \dd \qb{W_{kl}, W_{lk}} \dd t = \frac{\sigma^2}{N} \quad (k\neq l)\,,
\end{equation}
so that we get from \eqref{eq:G_DBM_step1} and taking into account symmetries:
\begin{equation}
	\label{eq:G_DBM_step3}
	\dd G_{ij}(z,t)  =  \sum_{k,l=1}^{N} G_{ik} G_{jl} \dd B_{kl} + \frac{\sigma^2}{N} \sum_{k,l=1}^{N} \pB{G_{ik} G_{lk} G_{lj} + G_{ik} G_{kj} G_{ll}} \dd t\,.
\end{equation}
As above, we expect the entries of $\G$ to be self-averaging. Hence, we consider the average with respect to the Brownian motion $W_{kl}$ defined in Eq.\ \eqref{eq:GOE_entries}, we find the following evolution for the average resolvent:
\begin{equation}
	\label{eq:G_DBM_avg}
	\partial_t \mathbb{E} [\b G(z,t)] \;=\; \sigma^2 \stj(z,t) \, \mathbb{E} [\G^2(z,t)] +  \frac{1}{N} \mathbb{E} [\b G^{3}(z,t)].
\end{equation}
Now, one can notice that:
\begin{equation}
\G^2(z,t) = - \partial_z \G(z,t); \qquad \G^3(z,t) =  \partial^2_{zz} \G(z,t),
\end{equation}
which hold even before averaging. By sending $N\to\infty$, we obtain the following matrix PDE for the resolvent:
\begin{equation}
	\label{eq:G_DBM_avg_asymp}
	\partial_t \mathbb{E} [\b G(z,t)] \;=\; - \sigma^2 \stj(z,t) \, \partial_z \mathbb{E} [\b G(z,t)] \,, \quad\text{with}\quad 
	\mathbb{E} [\G(z,0)] \; = \; \G_\C(z)\,.
\end{equation}
Taking the trace of this equation immediately leads to a Burgers equation for the Stieltjes transform \cite{allez2014eigenvectors,allez2013eigenvectors}:
\begin{equation}
	\label{eq:St_DBM_avg_asymp}
	\partial_t \stj(z,t) \;=\; - \sigma^2 \stj(z,t) \, \partial_z \stj(z,t) \,, \quad\text{with}\quad 
	\stj(z,0) \; = \; \stj_\C(z)\,.
\end{equation}
Its solution can be found using the method of characteristics and reads:
\begin{equation}
	\label{eq:dGOE_resolvent_DBM}
	\stj(z,t) = \stj_\C(Z(z,t)), \qquad Z(z,t)\;\deq\; z - \sigma^2 t \stj(z,t).
\end{equation}
The solution of Eq.\ \eqref{eq:G_DBM_avg_asymp} then reads \cite{allez2014eigenvectors,shlyakhtenko1996random}:
\begin{equation}
	\label{eq:dGOE_resolvent_DBM}
	\mathbb{E} [\G(z,t)] = \G_\C(Z(z,t)),
\end{equation}
and is exactly equivalent to \eqref{eq:dGOE_resolvent_schur} except that the variance here is given by $\sigma^2 t$. 
 
Note that we can then easily study from \eqref{eq:dGOE_resolvent_DBM} the mean squared overlap between the \emph{perturbed} eigenvectors $\b u_{i}(t)$ and the \emph{pure} ones $\b u_{j}(0) = \b v_j$ for any $i,j \in \qq{1,N}$. 
Indeed, it suffices to consider in the basis where $\C$ is diagonal the following projection $\scalar{\b v_j}{G_{ii}(z,t) \b v_j}$ with $z = \lambda_i -\text{i} \eta$ as in Chapter \ref{chap:eigenvectors} and we finally obtain
\begin{equation}
	\label{eq:overlap_DBM}
	N \mathbb{E} \qb{\scalar{\b u_i(t)}{\b v_j}^2} = \frac{\sigma^2 t}{\absb{\lambda_i(t) - \sigma^2 t \stj_\M(z,t) - \mu_j }^2}.
\end{equation}

\subsection{Extension to an arbitrary rotational invariant noise}

\subsubsection{An elementary derivation of the free addition formula}

We now turn on the general case where the noise term $\B$ is a (asymptotically) rotational invariant random matrix. We saw in Section \ref{sec:free_probability} that the limiting spectrum of such models can be investigated using the free probability formalism. The first part of this section is dedicated to a formal but elementary derivation of Voiculescu's free addition \eqref{eq:free_add_formula} \cite{voiculescu1991limit} by following the arguments of \cite{bun2015rotational}. From this result, we will be able to derive the asymptotic behavior of the resolvent of the model \eqref{eq:additive_model_app} using the Replica formalism of Section \ref{sec:replica}.

As in Section \ref{subsubsec:freemult}, the starting point is to notice that since the noise is rotationally invariant, we can always work in the basis where the matrix $\C$ is diagonal. Thus, we may specialize the Replica formalism \eqref{eq:resolvent_replica} for the resolvent of \eqref{eq:additive_model_app} which yields\footnote{One may also use the Replica formalism for the Stieltjes transform as well.} 
\begin{equation}
\label{eq:replica_free_add}
{\b G}_{\M}(z)_{i,j} = \int \left(\prod_{\alpha = 1}^{n} \prod_{k=1}^{N} d\eta_k^{\alpha}\right) \eta_i^1 \eta_j^1 \prod_{\alpha = 1}^{n} e^{-\frac{1}{2} \sum_{k=1}^{N} (\eta_k^{\alpha})^2 (z - c_k) } \left\langle e^{-\frac{1}{2} \sum_{k,l=1}^{N} \eta_k^{\alpha} (\b\Omega\BBB \b\Omega^{*})_{k,l} \eta_l^{\alpha}} \right\rangle_{\b\Omega}.
\end{equation}
One recognizes that the average value in the RHS of the latter equation is again the finite rank version of HCIZ integrals studied in details in Section \ref{appsec:HCIZ_rank1}. Hence, one deduces from \eqref{eq:HCIZ_rankn} that 
\begin{equation}
\label{free_convolution_HCIZ_rank_one}
\cal I_1\pBB{\sum_{\alpha=1}^{n} \eta^{\alpha}\pb{\eta^{\alpha}}^*, \BBB} = \exp\left[ {\frac{N}{2} \sum_{\alpha = 1}^{n} \wtr_{\BBB}\left(\frac1N(\eta^{\alpha})^{\dag}\eta^{\alpha}\right)} \right], 
\end{equation}
with $\wtr_{\BBB}^{\,\prime}(.) = \cal R_{\BBB}(.)$ the primitive of the $\rtr$-transform of $\BBB$. As a result, the computation of the resolvent (\ref{eq:replica_free_add}) becomes
\begin{equation}
\G_{\M}(z)_{i,j} = \int \left(\prod_{k=1}^{N} d\eta_k\right) \eta^{1}_i \eta^{1}_j  \exp\left\{ \frac{N}{2}\sum_{\alpha=1}^{n} \left[ \wtr_{\BBB}\left(\frac1N(\eta^{\alpha})^{\dag}\eta^{\alpha} \right) -\frac{1}{2} \sum_{k=1}^{N} (\eta_k^{\alpha})^2 (z - \mu_k) \right] \right\},
\end{equation}
and by introducing a Lagrange multiplier $p^{\alpha} \deq \frac1N (\eta^{\alpha})^{\dag}\eta^{\alpha}$, we obtain using Fourier transform (and renaming $\zeta^{\alpha} \to -2i\zeta^{\alpha}/N$)
\begin{eqnarray}
\G_{\M}(z)_{i,j} & \propto & \int \int \left( \prod_{\alpha = 1}^{n} dp^{\alpha} d\zeta^{\alpha} \right) \exp\left\{\frac{N}{2} \sum_{\alpha = 1}^{n} \left[ \wtr_{\BBB}(p^{\alpha}) - p^{\alpha}\zeta^{\alpha} \right] \right\} \nonumber \\
& & \times \int \left(\prod_{\alpha = 1}^{n}\prod_{k=1}^{N} d\eta_k^{\alpha} \right) \eta^{1}_i \eta^{1}_j  \exp\left\{ -\frac{1}{2} \sum_{\alpha = 1}^{n} \sum_{k=1}^{N} (\eta_k^{\alpha})^2 (z - \zeta^{\alpha} - \mu_k) \right\} \nonumber.
\end{eqnarray}
One can readily find 
% This additional constraint allows one to retrieve a Gaussian integral over the $\{\eta_j\}$ which can be computed exactly. Ignoring normalization terms, we obtain
\begin{equation}
\label{eq:resolvent_free_add_tmp}
\G_{\M}(z)_{i,j} \propto \int \int \left( \prod_{\alpha=1}^{n}  dp^{\alpha} d\zeta^{\alpha}\right) \frac{\delta_{ij}}{z + \zeta^{1} - \mu_i} \exp\left\{ -\frac{Nn}{2} F_0(p^{\alpha}, \zeta^{\alpha}) \right\},
\end{equation}
where the `free energy' $F_0$ is given by
\begin{equation}
\label{eq:free_energy_addition}
F_0(p^\alpha, \zeta^\alpha) = \frac{1}{Nn} \sum_{\alpha = 1}^{n} \left[ \sum_{k=1}^{N} \log(z - \zeta^{\alpha} - \mu_k) - \wtr_{\BBB}(p^{\alpha}) + p^{\alpha}\zeta^{\alpha} \right].
\end{equation}
As in Section \ref{subsubsec:freemult}, the integral \eqref{eq:resolvent_free_add_tmp} can be evaluated by considering the saddle-point of the free energy $F_0$ as the other term is obviously sub-leading. Moreover, we use the \textit{replica symmetric} ansatz that tells us if the free energy is invariant under the action of the symmetry group $\b O(N)$, then we expect a saddle-point which is also invariant. This implies that we have at the saddle-point
\begin{equation}
p^{\alpha} = p \quad \text{ and } \quad \zeta^{\alpha} = \zeta, \qquad \forall\, \alpha \in \{1, \dots, n\},
\end{equation}
from which, we obtain the following set of equations:
\begin{equation}
	\label{eq:free_add_saddle_pts}
	\zeta^* = \rtr_{\BBB}(p^*) \qquad\text{and}\qquad p^* = \stj_{\C}(z - \zeta^*).
\end{equation}
If we apply the Blue transform of $\C$ on the second equation of \eqref{eq:free_add_saddle_pts}, we obtain
\begin{equation}
	\label{eq:free_add_saddle_pts_2}
	z = \btr_\C(p^*)  + \rtr_\BBB(p^*) \equiv \rtr_{\C}(p^*) + \rtr_{\BBB}(p^*) - \frac{1}{p^*}.
\end{equation}
On the other hand, we see that the resolvent \eqref{eq:resolvent_free_add_tmp} is given in the large $N$ limit and the limit $n \rightarrow 0$ by 
\begin{equation}
	\label{eq:resolvent_free_add_tmp_2}
	\G_{ij}(z) \sim \frac{\delta_{ij}}{z - \rtr_\BBB(p^*) - \mu_i}.
\end{equation}
The trick is to see that we can get rid of one variable by taking the normalized trace in this later equation as it yields
\begin{equation}
	\label{eq:p_star_add}
	\stj_\M(z) = \stj_\C(z - \rtr_{\BBB}(p^*)) = p^*
\end{equation}
where the last equation follows from \eqref{eq:free_add_saddle_pts}. Therefore, we conclude by plugging this last equation into \eqref{eq:free_add_saddle_pts_2} that
\begin{equation*}
	z - \frac{1}{\stj_\M(z)} =  \rtr_{\C}(\stj_\M(z)) + \rtr_{\BBB}(\stj_\M(z)),
\end{equation*}
from which one can check by renaming $z = \btr_\M(\omega)$ that
\begin{equation}
	\rtr_\M(\omega) = \rtr_{\C}(\omega) + \rtr_{\BBB}(\omega),
\end{equation}
which is exactly the free addition formula \eqref{eq:free_add_formula}. 

\subsubsection{Asymptotic resolvent of \eqref{eq:additive_model_app}}

A trivial application of the result above is the evaluation of the resolvent entry-wise for the general model \eqref{eq:additive_model_app}. Indeed, we see by plugging Eq.\ \eqref{eq:p_star_add} into Eq.\ \eqref{eq:resolvent_free_add_tmp_2} that
\begin{equation}
	\G_\M(z)_{ij} \sim \frac{\delta_{ij}}{z - \rtr_\BBB(\stj_\M(z)) - \mu_i},
\end{equation}
which is equivalent to
\begin{equation}
	\label{eq:resolvent_free_add}
	\G_\M(z)_{ij} = \G_\C(Z(z))_{ij}, \qquad Z(z) \;\deq\; z - \rtr_\BBB(\stj_\M(z)).
\end{equation}
One notices that this formula is indeed the generalization of the formula \eqref{eq:stieltjes_free_addition} as a matrix. Moreover, we see that in the large $N$ limit, each entry of the random resolvent of $\M$ converges to a deterministic quantity that lies in the basis of $\C$. We moreover see that the additive case is even simpler than the multiplicative one as expected. It also means that all the computations we considered in Section \ref{chap:eigenvectors} can be performed nearly verbatim for the additive model \eqref{eq:additive_model_app} and the exact results can be found in \cite{bun2015rotational}.

\subsection{Overlap and Optimal RIE formulas in the additive case}

\subsubsection{Mean squared overlaps}

We were able to show that each entries of the resolvent of $\M$ in the general additive model \eqref{eq:additive_model_app} converges to a deterministic limit that is given in Eq.\ \eqref{eq:resolvent_free_add}. We see that this matrix relation can be simplified when written in the basis where $\C$ is diagonal, since
in this case ${\G}_{\C}(Z)$ is also diagonal. Therefore, the evaluation of the mean squared overlap between a given sample and true eigenvectors, denoted as $\mso(\lambda, \mu)$, is straightforward using the same techniques as in Section \ref{sec:eigenvectors_sample_true_bulk}. We omit details that may be found in \cite{bun2015rotational} and one finds that the overlap for the free additive noise is given by:
\begin{equation}
\label{overlap_addition}
\mso(\lambda, \mu) = \frac{\beta_1(\lambda)}{(\lambda - c - \alpha_{a}(\lambda))^2 + \pi^2 \beta_a(\lambda)^2 \rho_{\M}(\lambda)^2},
\end{equation}
where $\mu$ is the corresponding eigenvalue of the true matrix $\C$, and where we defined: 
\begin{equation}
\label{decomposition_R}
\begin{dcases}
\alpha_a(\lambda) \;\deq\;  \re[\rtr_{\B} \left( \hil_{\M}(\lambda) + \ii \pi \rho_{\M}(\lambda) \right)],  \\
\beta_a(\lambda) \;\deq\; \frac{\im[\rtr_{\B} \left( \hil_{\M}(\lambda) + \ii \pi \rho_{\M}(\lambda) \right)]}{\pi \rho_{\M}(\lambda)}\,.
\end{dcases}
\end{equation}

\begin{figure}[!ht]
	\begin{center}
   \includegraphics[scale = 0.4]{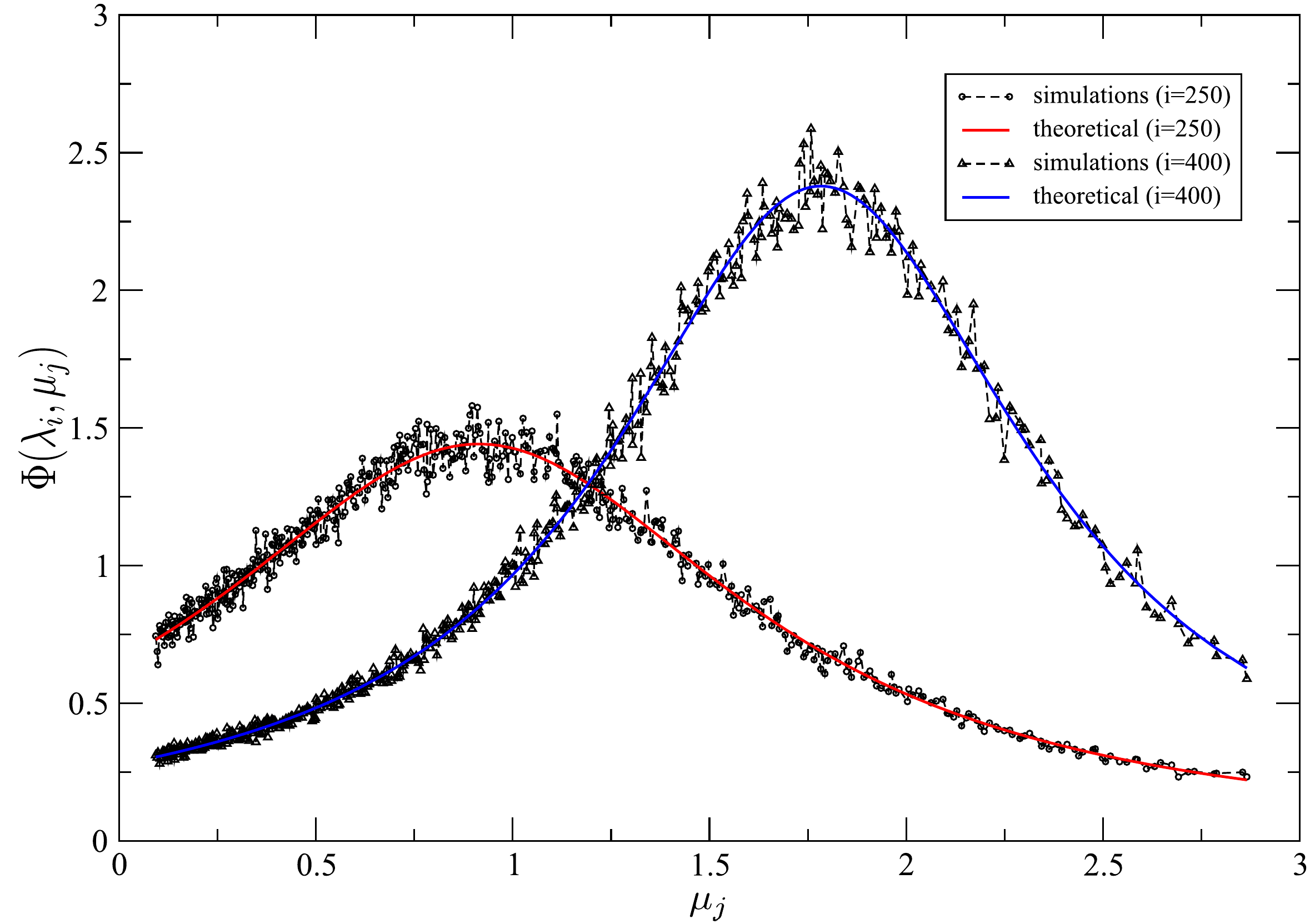} 
   \end{center}
   \caption{Computations of the rescaled overlap $\mso(\lambda, \mu)$ as a function of $\mu$ in the free addition perturbation. 
   We chose $i = 250$, $\C$ a Wishart matrix with parameter $q = 0.5$ and $\B$ a Wigner matrix with $\sigma^2 = 1$. The black dotted points are computed using numerical simulations and the plain red curve is the theoretical predictions Eq. \eqref{overlap_addition}. The agreement is excellent. For $i = 250$, we have $\mu_i \approx 0.83$ and we see that the peak of the curve is in that region. The same observation holds for $i = 400$ where $\mu_i \approx 1.66$. {
   The numerical curves display the empirical mean values of the overlaps 
   over 1000 samples of $\M$ given by Eq. \eqref{eq:additive_model_app} with $\C$ fixed}.}
   \label{Overlaps_add}
\end{figure}

As a simple consistency check, we specialize our result to the case where $\b\Omega\BBB \b\Omega^{*}$ is a GOE matrix such that the entries have a variance equal to $\sigma^2/N$. Then, one has $\rtr_{\B}(z) = \sigma^2 z$ meaning that $Z(z)$ of Eq. \eqref{eq:resolvent_free_add} simply becomes $Z(z) = z - \sigma^2 \stj_{\M}(z)$. This allows us to get a simpler expression for the overlap:
\begin{equation}
\label{eq:overlap_addition_gaussian}
\mso(\lambda, \mu)= \frac{\sigma^2}{(c -\lambda + \sigma^2 \hil_{\M}(\lambda))^2 + \sigma^4 \pi^2 \rho_{\M}(\lambda)^2},
\end{equation}
which is exactly the result obtained in Eq.\ \eqref{eq:overlap_DBM}. In Fig.\ \ref{Overlaps_add}, we illustrate this formula in the case where $\C = \Wishart$ with parameter $q$. We set $N = 500$, $T = 1000$, and take $\b\Omega{\bf B}\b\Omega^{*}$ as a GOE matrix with variance $1/N$. For a fixed $\C$, we generate 200 samples of $\M$ given by Eq. (\ref{eq:additive_model_app}) for which we can measure numerically the overlap \eqref{eq:overlap}. We see that the theoretical prediction (\ref{eq:overlap_addition_gaussian}) agrees remarkably with the numerical simulations.

\subsubsection{Optimal RIE}

Since the overlaps are explicit in this general model, it is easy to compute the asymptotic limit of the oracle estimator \eqref{eq:oracle} for the bulk eigenvalues in the model \eqref{eq:additive_model_app}. Indeed, it is easy to see from Eqs.\ \eqref{eq:resolvent_decomposition} and (\ref{eq:oracle}) that:
\begin{equation}
\xi^{\text{ora.}}_i \sim \frac{1}{\pi \rho_{\M}(\lambda_i)} 
\lim_{z \rightarrow \lambda_i - \ii 0^{+}} \im\qBB{\int \frac{\mu\, \rho_{\C}(\mu)}{Z(z) - \mu} \, \dd \mu}
= \frac{1}{N\pi \rho_{\M}(\lambda_i)} \underset{z \rightarrow \lambda_i - \ii 0^{+}}{\lim} \, \im \Tr \left[ {\G}_{\M} (z) \C \right],
\end{equation}
where $Z(z)$ is given by Eq. \eqref{eq:resolvent_free_add}. From Eq. \eqref{eq:resolvent_free_add} one also has 
$\Tr [{\G}_{\M}(z) \C] = N (Z(z) \stj_{\M}(z) - 1)$, and using Eqs. \eqref{eq:resolvent_free_add} and \eqref{decomposition_R}, 
we end up with:
\begin{equation*}
\underset{z \rightarrow \lambda - \ii 0^{+}}{\lim} \, \im \Tr \left[ {\G}_{\M}(z) \C \right] = N \pi \rho_{M}(\lambda) \left[ \lambda - \alpha(\lambda) - \beta(\lambda) \hil_{\M}(\lambda) \right].
\end{equation*}
We therefore find the following optimal RIE nonlinear ``shrinkage'' function $F_a$:
\begin{equation}
\label{eq:oracle_free_add}
\xi^{\text{ora.}}_i \sim F_a(\lambda_i); \qquad F_a(\lambda)= \lambda - \alpha_a(\lambda) - \beta_a(\lambda) \hil_{\M}(\lambda),
\end{equation}
where $\alpha_a, \beta_a$ are defined in Eq.\ (\ref{decomposition_R}). 
This result states that if we consider a model where the signal $\C$ is perturbed by an additive noise 
(that is free with respect to $\C$), 
the optimal way to 'clean' the eigenvalues of $\M$ in order to get $\widehat{\Xi}(\M)$ is to keep the eigenvectors of $\M$ and apply the nonlinear shrinkage formula \eqref{eq:oracle_free_add}. We see that the non-observable oracle estimator converges in the limit $N \to \infty$ towards a deterministic function of the observable eigenvalues.

As usual, let us consider the case where $\B$ is a GOE  matrix in order to give more intuitions about \eqref{eq:oracle_free_add}. Using the definition of $\alpha_a$ and $\beta_a$ given in Eq. (\ref{decomposition_R}), the nonlinear shrinkage function is given by
\begin{equation}
\label{oracle_gaussian}
F_a({\lambda}) = \lambda - 2 \sigma^2 \hil_{\M}(\lambda).
\end{equation}
Moreover, suppose that $\C$ is also a GOE matrix so that $\M$ is a also a GOE matrix with variance 
$\sigma^2_{\M} = \sigma^2_{\C} + \sigma^2$. As a consequence, the Hilbert transform of $\M$ can be computed straightforwardly from the Wigner semicircle law and we find
\begin{equation*}
\hil_{\M}(\lambda) = \frac{\lambda}{2\sigma_{\M}^2}\,.
\end{equation*}
The optimal cleaning scheme to apply in this case is then given by: 
\begin{equation}
\label{oracle_gaussian_Wigner}
F_a({\lambda})= \lambda \left( \frac{\sigma_{\C}^2}{\sigma^2_{\C} + \sigma^2} \right)\,,
\end{equation}
where one can see that the optimal cleaning is given by rescaling the empirical eigenvalues by the signal-to-noise ratio. This result is expected in the sense that we perturb a Gaussian signal by adding a Gaussian noise. We know in this case that the optimal estimator of the signal is given, element by element, by the Wiener filter \cite{wiener1949extrapolation}, and this is exactly the result that we have obtained with \eqref{oracle_gaussian_Wigner}. We can also notice that the ESD of the cleaned matrix is narrower than the true one. Indeed, let us define the signal-to-noise ratio $\text{SNR} = \sigma_{\C}^{2}/\sigma_{\M}^{2} \in [0,1]$, and it is obvious from \eqref{oracle_gaussian_Wigner} that $\widehat{\Xi}(\M)$ is a Wigner matrix with variance $\sigma_{\Xi}^{2} \times \text{SNR}$ which leads to 
\begin{equation}
\sigma_{\M}^{2} \ge \sigma_{\C}^{2} \ge \sigma_{\C}^{2} \times \text{SNR}\,,
\end{equation}
as it should be.

\begin{figure}[ht]
	\begin{center}
   \includegraphics[scale = 0.4]{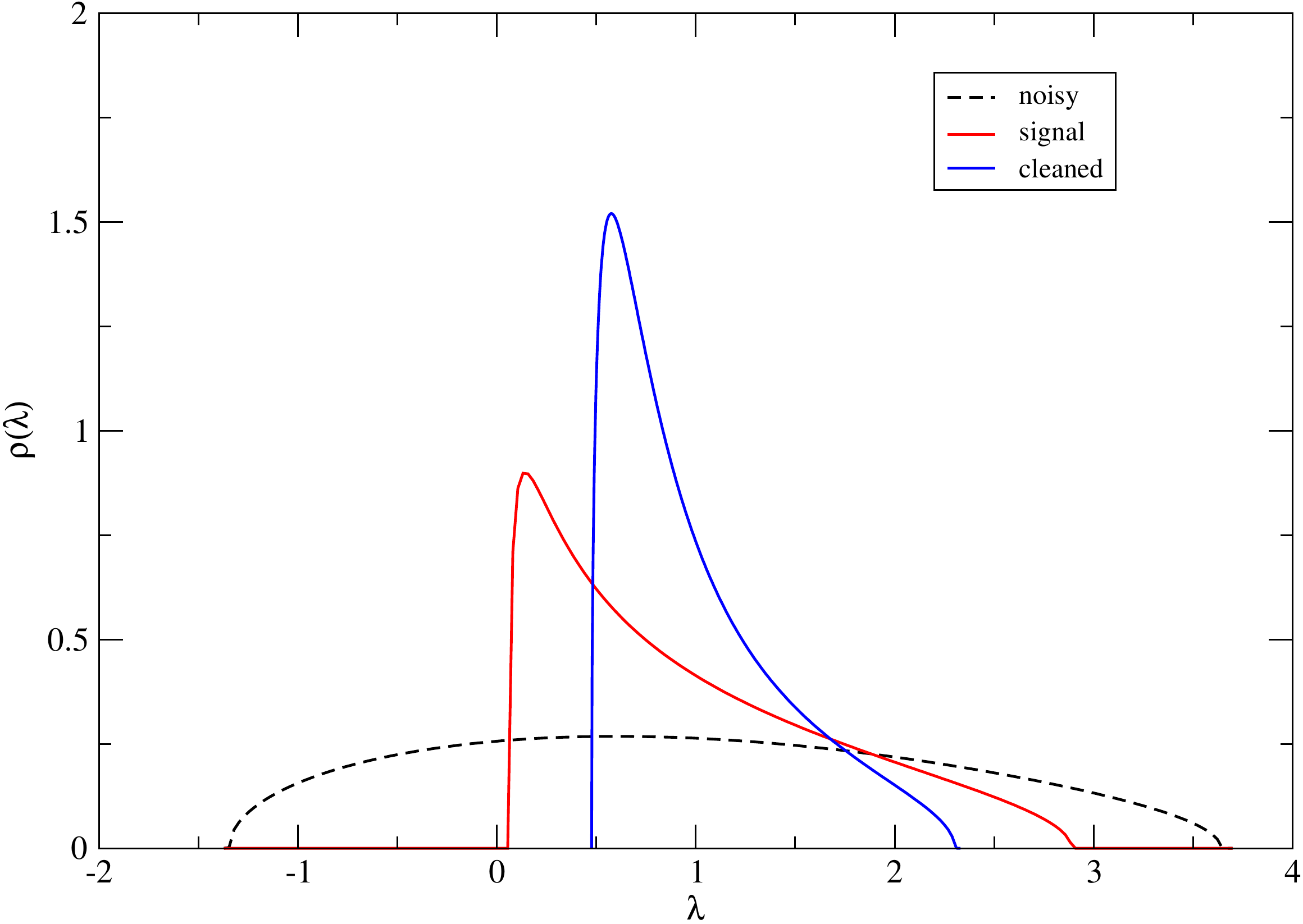} 
   \end{center}
   \caption{Eigenvalues of the noisy measurement $\M$ (black dotted line) compared to the true signal $\C$ drawn from a $500 \times 500$ Wishart matrix of parameter $q_0 = 0.5$ (red line). We have corrupted the signal by adding a GOE matrix with radius 1. The eigenvalues density of $\M$ allows negative values while the true one has only positive values. The blue line is the LSD of the optimally cleaned matrix. We clearly notice that the cleaned eigenvalues are all positive and its spectrum is narrower than the true one, while preserving the trace.}
   \label{fig:oracle_addition_density}
\end{figure}

\begin{figure}[ht]
	\begin{center}
   \includegraphics[scale = 0.4]{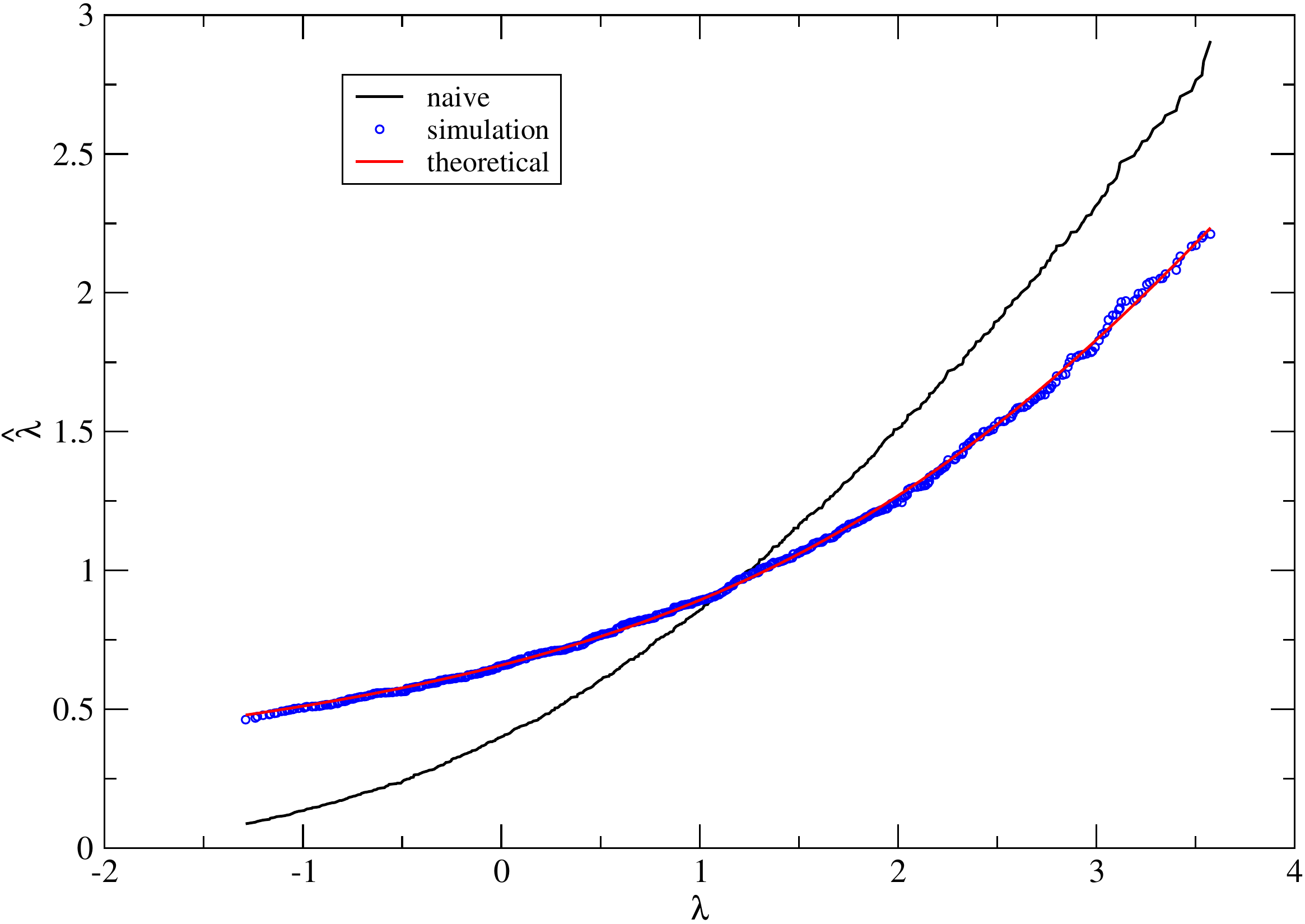} 
   \end{center}
   \caption{Eigenvalues according to the optimal cleaning formula (\ref{oracle_gaussian_Wigner}) (red line) as a function of the observed noisy eigenvalues $\lambda$. The parameter are the same as in Fig. \ref{fig:oracle_addition_density}. We also provide a comparison against the naive eigenvalues substitution method (black line) and we see that the optimal cleaning scheme indeed narrows the spacing between eigenvalues. }
   \label{fig:oracle_addition}
\end{figure}

As a second example, we now consider a less trivial case and suppose that $\C$ is a 
white 
Wishart matrix with parameter $q_0$. For any $q_0 > 0$, it is well known that the Wishart matrix has non-negative eigenvalues. However, we expect that the noisy effect coming from the GOE matrix pushes some true eigenvalues towards the negative side of the real axis. In Fig.\ \ref{fig:oracle_addition_density}, we clearly observe this effect and a good cleaning scheme should bring these negative eigenvalues back to positive values.
In order to use Eq. (\ref{oracle_gaussian}), we invoke once again the free addition formula to find the following equation for the Stieltjes transform of $\M$:
\begin{equation*}
- q_0\sigma^2\stj_{\M}(z)^3 + (\sigma^2 +q_0\,z)\stj_{\M}(z)^2 + (1-q_0-z) \stj_{\M}(z) + 1 = 0,
\end{equation*}
for any $z = \lambda - \ii\eta$ with $\eta \rightarrow 0$. It then suffices to take the real part of the Stieltjes transform $\stj_{\M}(z)$ that solves this equation\footnote{We take the solution which has a strictly non-negative imaginary part} to get the Hilbert transform. {In order to check formula Eq.\ \eqref{eq:oracle_free_add} using numerical simulations, 
we have generated a matrix of $\M$ given by Eq.\ \eqref{eq:additive_model_app} with $\C$ a fixed white Wishart matrix with parameter $q_0$ and $\b \Omega\B \b \Omega^{*}$  a GOE matrix with radius 1. As we know exactly $\C$, we can compute numerically
the oracle estimator as given in \eqref{eq:oracle} for each sample. In Fig.\ \ref{fig:oracle_addition}, we see that our theoretical prediction in the large $N$ limit 
compares very nicely with the mean values of the empirical oracle estimator computed from the sample}. 
We can also notice in Fig. \ref{fig:oracle_addition_density} that the spectrum of the cleaned matrix 
(represented by the ESD in green) is narrower 
than the standard Mar{\v c}enko-Pastur density. This confirms the observation made in Chapter \ref{chap:RIE}.

\clearpage%!TEX root = RMT_Covariance_Review.tex
\section{Conventions, notations and abbreviations}

\begin{center}
\textbf{Conventions}
\end{center}
We use bold capital letters for matrices and bold lowercase letters for vectors, which we regard as $N \times 1$ matrices. The superscript $*$ denotes the transpose operator. We use the abbreviations $\qq{a,b} \deq [a,b] \cap \N$ and $\qq{a} \equiv \qq{1,a}$ for $a,b \in \N$. 

\begin{center}
\textbf{Mathematical symbols}
\end{center}
We list here some of the most important notations of the review. 
\begin{longtable}[l]{p{50pt} p{200pt}} 
\textbf{Symbol}	& \textbf{Description} \\ 
$\btr_{\M}$ & Blue transform of $\M$ \eqref{eq:blue} \\
$\C$ & Population/True covariance matrix \eqref{eq:population_covmat} \\
$\ul{\C}$ & Spikeless version of $\C$ \eqref{eq:spikeless_population_cov} \\
$\mathbb{C}_{\pm}$ & Complex upper/lower half plane \\
$\E$ & Sample/Empirical covariance matrix \eqref{eq:SCM} \\
$\mathbb{E}$ & Expectation value over the noise \\
$\b G_{\M}$ & Resolvent of $\M$, \eqref{resolvent} \\
$\stj^N_{\M}$ & Empirical Stieltjes transform of $\rho_\M$ \eqref{eq:stieltjes_emp} \\
$\stj_{\M}$ & Stieltjes transform of $\rho_\M$ \eqref{eq:stieltjes} \\
$\ii$ & $\sqrt{-1}$ \\
$i$ & integer index \\
$N$	& Number of variables \\ 
$\b O(N)$ & Orthogonal group on $\mathbb{R}^{N\times N}$ \\
$\cal O$	& Big $O$ notation \\
${\cal P}(\cdot)$ & Probability density function \\
${\cal P}(\cdot | \cdot)$ & Conditional probability measure \\
$q$	& Observation ratio ($N/T$) \\
$r$ & Number of outliers \\
$\rtr_{\M}$ & R-transform of $\M$ \eqref{eq:r_transform} \\
$\cal R^2_{\text{in}}$ & In-sample/predicted  risk \eqref{eq:in_sample_risk} \\
$\cal R^2_{\text{out}}$ & Out-of-sample/realized risk \eqref{eq:out_sample_risk} \\
$\cal R^2_{\text{true}}$ & True risk \eqref{eq:true_risk} \\
$\S$ & ``Dual'' sample covariance matrix \eqref{eq:SCM_dual} \\
$\str_{\M}$ & S-transform of $\M$ \eqref{eq:s_transform} \\
$T$	& Sample size \\ 
$\ttr_{\M}$ & T-transform of $\M$ \eqref{eq:T_transform} \\
$\b u_i$ & Sample eigenvector associated to $\lambda_i$ \\
$\b v_i$ & Population eigenvector associated to $\mu_i$ \\
$\mathbb{V}$ & Variance of a random variable \\
$\wtr_{\M}$ & Primitive of the $\rtr$-Transform of $\M$ \eqref{eq:W_transform} \\
$\Y$ & $N \times T$ normalized data matrix \\
$\alpha_s$ & Linear shrinkage intensity \eqref{eq:linear_shrinkage}\\
$\lambda_i$ & $i$th sample eigenvalue \\
$\mu_i$ & $i$th population (true) eigenvalue \\
$\Xi^{\text{lin.}}$ & Linear Shrinkage estimator \eqref{eq:linear_shrinkage} \\
$\hat\Xi(\E)$ & Optimal RIE of $\C$ depending on $\E$ \\
$\Xi^{\text{ora.}}$ & Oracle estimator \eqref{eq:oracle} \\
$\Xi(\E)$ & RIE of $\C$ depending on $\E$ \\
$\rho_{\M}^{N}$ & Empirical spectral density of $\M$ \eqref{ESD} \\
$\rho_{\M}$ & Limiting spectral density of $\M$ \eqref{average_ESD} \\
$\Phi$ & Rescaled mean squared overlap \eqref{eq:overlap} and \eqref{eq:mso} \\
$\varphi(\M)$ & Normalized trace of $\M$ \eqref{eq:trace_matrix}\\
$\b\Omega$ & Rotation matrix \\
$\langle \cdot \rangle_{\M}$ & Expectation value with respect to ${\cal P}(\M)$ \\
$\langle \, , \, \rangle$ & inner product \\

\end{longtable} 

\begin{center}
\textbf{Abbreviations}
\end{center}
\begin{longtable}[l]{p{50pt} p{200pt}} 
\textbf{Symbol}	& \textbf{Description} \\
CCA & Canonical Correlation Analysis \\
ESD & Empirical Spectral Density \\
GOE & Gaussian Orthogonal Ensemble \\
HCIZ & Harish-Chandra--Itzykson-Zuber \\
IW & Inverse Wishart \\
IWs & Inverse Wishart + sorting \\
LDL & Large dimension limit \\
LHS & Left Hand Side \\
LSD & Limiting Spectral Density \\
MMSE & Minimum Mean Squared Error  \\
MSE & Mean Squared Error  \\
MP & Mar{\v c}enko-Pastur  \\
PCA & Principal Component Analysis \\
PDE & Partial Differential Equation \\
PDF & Probability Density Function \\
RHS & Right Hand Side \\
QuEST & Quantized Eigenvalues Sampling Transform \\
RI & Rotational Invariance \\
RIE & Rotational Invariant Estimator \\
RP & Relative Performance \\
RMT & Random Matrix Theory \\
%SCM & Sample Covariance Matrix \\
SVD & Singular Value Decomposition \\
\end{longtable} 

%ora. & Oracle estimator \eqref{eq:oracle}\\

\end{document}